\documentclass[rmp,aps,amsfonts,amssymb,nofootinbib,floats,twocolumn,eqsecnum,longbibliography]{revtex4-2}
\pdfoutput=1 
\usepackage{graphics}
\usepackage{epsfig}
\usepackage{xcolor}
\usepackage{mathtools}
\usepackage[colorlinks, linkcolor=blue]{hyperref}
\usepackage{mathrsfs,bm}
\usepackage{changes}

\definechangesauthor[color=magenta]{OP}

\usepackage{braket}
\usepackage{amsmath}

\def \q{{\vec q}}

\renewcommand{\vec}[1]{\boldsymbol{#1}}
\def \beq{\begin{eqnarray}}
\def \eeq{\end{eqnarray}}
\def \bit{\begin{itemize}}
\def \eit{\end{itemize}}
\def\ds{\displaystyle}
\def \r {{\vec r}}
\def \k {{\vec k}}
\def \P {{\mathcal{P}}}
\newcommand{\beqn}{\begin{equation}}
\newcommand{\eeqn}{\end{equation}}
\newcommand{\nn}{\nonumber \\}
\def\bea{\begin{eqnarray}}
\def\eea{\end{eqnarray}}
\def \tn{\textnormal}

\newcommand{\dos}{{\cal N}} % Many-Body density of states

\begin{document}
~\vskip 2in
\begin{flushright}
\includegraphics[width=8cm]{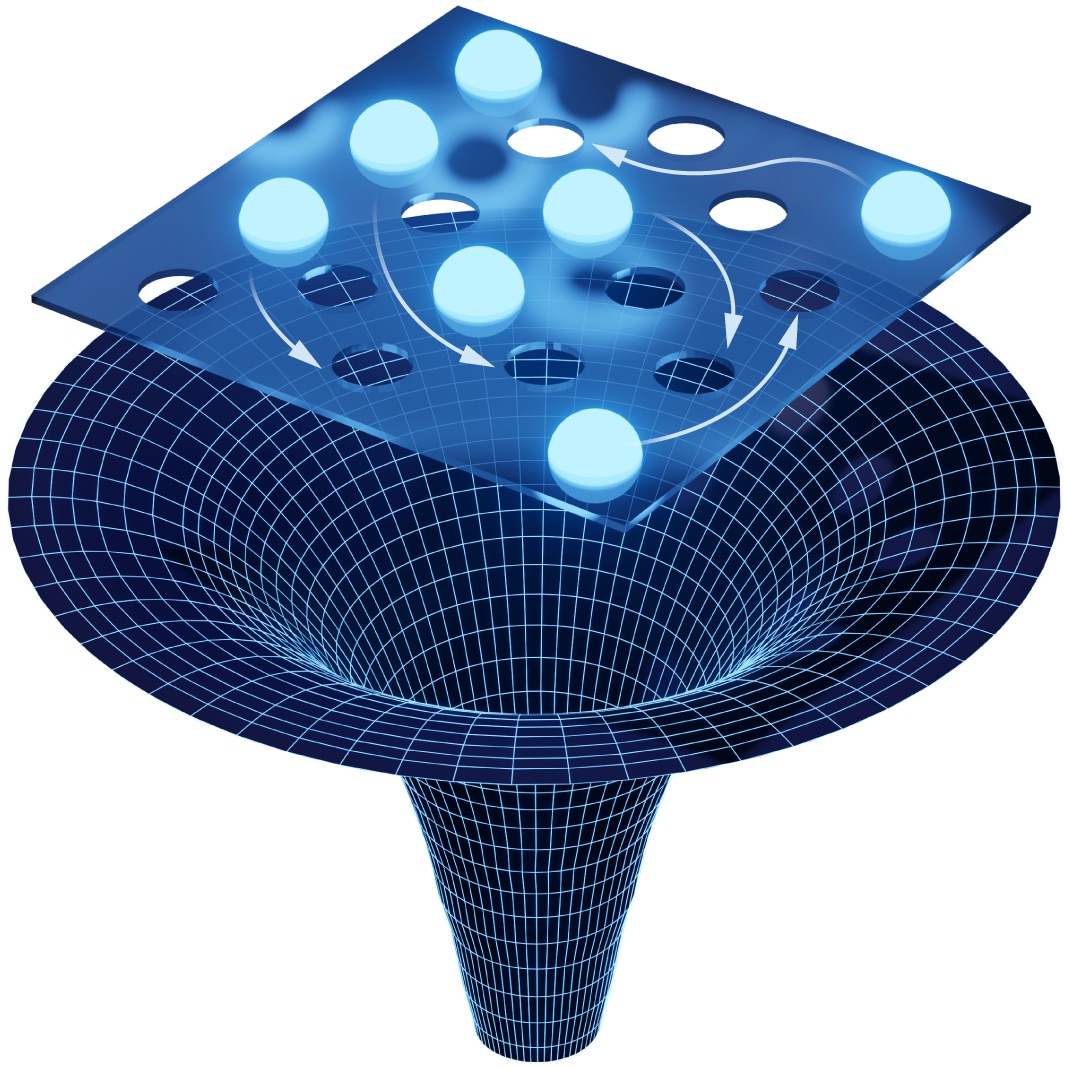}\\
{\small Figure by Lucy Reading-Ikkanda, Simons Foundation}
\end{flushright}
\vskip -6.5in
\title{Sachdev-Ye-Kitaev Models and Beyond: A Window into Non-Fermi Liquids}
\author{Debanjan Chowdhury}
\email{debanjanchowdhury@cornell.edu}
\affiliation{Department of Physics, Cornell University, Ithaca NY-14853, USA}

\author{Antoine Georges}
\email{ageorges@flatironinstitute.org}
\affiliation{Coll{\`e}ge de France, 11 place Marcelin Berthelot, 75005 Paris, France}
\affiliation{Center for Computational Quantum Physics, Flatiron Institute, New York, New York, 10010, USA}
\affiliation{CPHT, CNRS, Ecole Polytechnique, IP Paris, F-91128 Palaiseau, France}
\affiliation{DQMP, Universit{\'e} de Gen{\`e}ve, 24 quai Ernest Ansermet, CH-1211 Gen{\`e}ve, Suisse}

\author{Olivier Parcollet}
\email{oparcollet@flatironinstitute.org}
\affiliation{Center for Computational Quantum Physics, Flatiron Institute, New York, New York, 10010, USA}
\affiliation{Universit{\'e} Paris-Saclay, CNRS, CEA, Institut de physique th\'eorique, 91191, Gif-sur-Yvette, France}
\author{Subir Sachdev}
\email{sachdev@g.harvard.edu}
\affiliation{Department of Physics, Harvard University, Cambridge MA-02138, USA}
\affiliation{School of Natural Sciences, Institute for Advanced Study, Princeton, NJ-08540, USA}
\date{\today} 
\begin{abstract}
We present a review of the Sachdev-Ye-Kitaev (SYK) model of compressible quantum many-body systems without quasiparticle excitations, and its connections to various theoretical studies of non-Fermi liquids in condensed matter physics. The review is placed in the context of numerous experimental observations on correlated electron materials.
Strong correlations in  metals are often associated with their proximity to a Mott transition to an insulator created by the local Coulomb repulsion between the electrons. We explore the phase diagrams of a number of models of such local electronic correlation, employing a dynamical mean field theory in the presence of random spin exchange interactions. Numerical analyses and analytical solutions, using renormalization group methods and expansions in large spin degeneracy, lead to critical regions which display SYK physics. The models studied include the single-band Hubbard model, the $t$-$J$ model 
and the two-band Kondo-Heisenberg model in the presence of random spin exchange interactions.
We also examine non-Fermi liquids obtained by considering each SYK model with random four-fermion interactions to be a multi-orbital atom, with the SYK-atoms arranged in an infinite lattice. We connect to theories of sharp Fermi surfaces without any low-energy quasiparticles in the absence of spatial disorder, obtained by coupling a Fermi liquid to a gapless boson; a systematic large $N$ theory of such a critical Fermi surface, with SYK characteristics, is obtained by averaging over an ensemble of theories with random boson-fermion couplings. Finally, we present an overview of the links between the SYK model and quantum gravity and end with an outlook on open questions. 
\end{abstract}
\maketitle

\tableofcontents

\bigskip

\section{Introduction}
\label{sec:intro}

The discovery of high temperature superconductivity in the cuprate compounds in 1986 posed numerous challenges to quantum theories of electronic matter.
The biggest mystery, as became evident early on, was the unusual metallic state of these materials, above the superconducting critical temperature. This `strange metal' as it has since come to be called, displayed unusual temperature and frequency dependencies in its properties, which indicated that the strange metal was an entangled many-body quantum state without `quasiparticles'. 
Almost all of quantum condensed matter physics is built on the idea of quasiparticles: it allows us to account for the Coulomb interactions between electrons, by assuming their main effect is to renormalize each electron with a cloud of electron-hole pairs, after which we can treat each electron as a nearly independent quasiparticle. This decomposition of the excitations of a many-body system into a composite of simple quasiparticle excitations is an assumption so deeply engrained in the theoretical framework that it is usually left unstated. 

The aim of this review is to present some recent advances in describing quantum phases of matter that do not host {\it any} quasiparticle excitations. Much has been understood theoretically in recent years about the properties of a solvable model of a many-body quantum system without quasiparticle excitations in the regime of strong interactions: the Sachdev-Ye-Kitaev (SYK) model. We will review some of these advances in this article, along with a discussion of the application of these advances to more realistic models of quantum matter without quasiparticles.

The idea of employing a quasiparticle description of a macroscopic many-particle system can be traced back to Boltzmann \cite{boltzmann}. Boltzmann was thinking of a {\it dilute classical\/} gas of molecules, as found in the atmosphere. In 1872, he introduced an equation which described the time evolution of the observable properties of a dilute gas in response to external forces. He applied Newton's laws of motion to individual molecules, and obtained an equation for $f_{\bm p}$, the density of particles with momentum ${\bm p}$. In a spatially uniform situation, Boltzmann's equation takes the form
\begin{equation}
\frac{\partial f_{\bm p}}{\partial t} + {\bm F} \cdot \nabla_{\bm p} f_{\bm p} = \mathcal{C}[f]\,, \label{be}
\end{equation}
where $t$ is time, and ${\bm F}$ is the external force. The left-hand-side of Eq.~(\ref{be}) is just a restatement of Newton's laws for individual molecules. Boltzmann's innovation was the right-hand-side, which describes collisions between the molecules. Boltzmann introduced the concept of `molecular chaos', which asserted that in a sufficiently dilute gas successive collisions were statistically independent. With this assumption, Boltzmann showed that
\begin{equation}
\mathcal{C}[f] \propto -\int_{{\bm p}_{1,2,3}} \cdots \left [f_{\bm p} f_{{\bm p}_1} - f_{{\bm p}_2} f_{{\bm p}_3} \right] \label{coll1}
\end{equation}
for molecules with momenta ${\bm p},~ {\bm p}_1$ colliding to momenta ${\bm p}_{2},~{\bm p}_{3}$. The statistical independence of collisions is reflected in the products of the densities in Eq.~(\ref{coll1}), and the second term represents the time-reversed collision.

The remarkable fact is that Boltzmann's equation also applies, with relatively minor modifications, to the {\it dense quantum\/} gas of electrons found in ordinary metals, as was argued in Landau's Fermi liquid theory \cite{landau}. Individual electrons move in Bloch waves \cite{Bloch29} characterized by a crystal momentum ${\bm p}$.
Now collisions become rare because of Pauli's exclusion principle, and the statistical independence of collisions is assumed to continue to apply. The main modification is that the collision term in Eq.~(\ref{coll1}) is replaced by
\begin{align}
\mathcal{C}[f] \propto - \int_{{\bm p}_{1,2,3}} \cdots &\left [f_{\bm p} f_{{\bm p}_1}(1-f_{{\bm p}_2})(1-f_{{\bm p}_3}) \right. \nonumber \\
& ~~~\left. - f_{{\bm p}_2} f_{{\bm p}_3} (1-f_{{\bm p}})(1-f_{{\bm p}_1})\right] \,,\label{coll2}
\end{align}
where the additional $(1-f)$ factors ensure that the final states of collisions are not occupied. Now the $f_{\bm p}$ measure the distribution of electronic quasiparticles, and a cloud of particle-hole pairs around each electron only renormalizes the microscopic scattering cross-section.
Such a quantum Boltzmann equation is the foundation of the quasiparticle theory of the electron gas in metals, superconductors, semiconductors, and insulators, and indeed almost all of condensed matter physics before the 1980's. 

Our interest here is in quantum materials in which the description in terms of a quasiparticle distribution function $f_{\bm p}$ obeying a quantum Boltzmann equation breaks down. The time between collisions becomes so short that the quantum interference between successive collisions cannot be ignored, and the collisions cannot be treated as statistically independent. Landau's Fermi liquid theory has the feature that the quasiparticles are essentially dressed electrons, but there are situations in which the quasiparticles are emergent excitations of the many-body system with no simple relation to the bare electrons; such systems can be treated by extensions of Landau's approach, and these will also not be of interest to us.

Given a quantum many-body system, how do we ascertain the absence of low-energy quasiparticles in {\it any\/} basis and the associated universal diagnostics, if any? 
The simplest diagnostic we might consider for detecting the presence of electronic quasiparticles is via poles 
in the single-particle Green's function (sharp peaks in the spectral function).
%in a sharp spectral function. 
However, the existence of a broad electron spectral function is, by itself, not sufficient to conclude that there are no quasiparticle excitations. 
After all, interacting electrons in one dimension 
have broad electron spectral functions~\cite{giamarchi}. This is understood in Luttinger liquid theory, using a description in terms of a different set of quasiparticles: linearly dispersing 
bosons associated with collective excitations. 
The electron operator is an exponential of the boson operator, and this leads to the broad spectral functions. The bosonic quasiparticles describe {\it all\/} the many-body eigenstates, but the electron operator has a quite complicated form in this representation. Similarly, while the electron spectral function in certain fractional quantum Hall phases and paramagnetic Mott insulators \cite{QSL} can be complicated, at low-energies they might host {\it emergent} quasiparticle excitations that are well defined but impossible to diagnose using a two-point spectral function, as the latter quantity is not even a gauge-invariant observable. These examples illustrate that the electron spectral function is not a universal diagnostic for detecting quasiparticles; it is useful when the overlap between the wavefunction of the low-energy quasiparticle and the physical electron is non-zero ({\it e.g.\/}, as in a Landau Fermi liquid \cite{AGD}). On the other hand, when the two are orthogonal, as in the examples highlighted above, the diagnostic fails and the spectral function is ill equipped to analyze the fate of quasiparticles. 
A further weakness in the spectral function diagnostic is apparent when we consider disordered systems (e.g., even a disordered Fermi liquid). Electronic quasiparticles are well defined in such systems \cite{Abrahams81}, but they are not apparent in electronic spectral functions unless the spatial form of the quasiparticle wavefunction is already known: they are not plane waves, as in Fermi liquids in clean crystals. 

These considerations make it clear that a system with quasiparticle excitations 
is best characterized by an extension of the original Landau perspective \cite{landau}: 
the low energy states of a many body system can be decomposed into composites of single quasiparticle states, 
and the energies of these states are functionals of the densities of individual quasiparticle states. 
In other words, quasiparticles are {\it additive} excitations of a many-body system. 
Analyzing the spectrum of low-lying eigenstates of a many-body quantum systems for a large but finite volume 
therefore provides a useful diagnostic of the validity of a quasiparticle description or of its 
failure. We will use this `spectral fingerprint' in several places in this review, see {\it e.g.\/} Sec.~\ref{sec:matrix2}. 

With this perspective, in a many-body quantum system without quasiparticle excitations, it is not possible to decompose 
the low-lying states into any basis of quasiparticle excitations. 
This is however a practical definition only when the full low-lying spectrum is available. Furthermore, 
it may be possible to exclude a candidate quasiparticle basis but it is often difficult to exclude them all.
For a more positive and practical definition, we  consider the approach of a quantum many-body system to local thermal equilibrium at a temperature $T$ after the action of a local perturbation. In a system with quasiparticle excitations, such as a Fermi liquid, solution of the quantum Boltzmann equation shows that this will happen in a time that is at least as long as $ \sim 1/ T^2$ as $T \rightarrow 0$. This long time is required for  individual quasiparticles to collide with each other. In a system without quasiparticles, we expect the time to be much shorter. But how short can the local equilibration time get as $T \rightarrow 0$ ? Studies of numerous model systems without quasiparticle excitations, some of which are described in the present review, show that the time is never shorter than a time of order the `Planckian time', $\hbar /(k_B T)$,  
{\it i.e.\/} the minimum time associated with an energy of order $k_B T$ according to the Heisenberg uncertainty principle. 
On the other hand, it is clear from a study of systems with quasiparticles, that such systems can never equilibrate as quickly as the Planckian time, as long as quasiparticles are well-defined. So we reach the proposal that many-body quantum systems without quasiparticles are those that locally equilibrate in a time of order $\hbar/ (k_B T)$, and no system can equilibrate any faster \cite{QPT,Hartnoll:2016apf}. 

Our focus in this article will primarily be on metallic quantum many-body systems without quasiparticle excitations {\it i.e.\/} non-Fermi liquids. Section~\ref{sec:typology} presents a general  perspective on non-Fermi liquids, with a summary of some of their experimental signatures and an overview of some theoretical ideas and their relationship to the SYK models presented in this review. 
A detailed outline of the perspective of this paper appears in Section~\ref{sec:outline}. Readers wishing to focus on the SYK viewpoint can skip ahead directly to Section~\ref{sec:outline} and then to Section~\ref{sec:matrix}.
In Sec.~\ref{sec:bad_planck}, we discuss qualitatively the properties of `bad metals' and `Planckian metals', 
two forms of unconventional transport often encountered in non-Fermi liquids. In Section~\ref{sec:matrix}, as a warmup, we first review the random matrix model for non-interacting fermions that realizes a Fermi liquid with quasiparticles.
The SYK model system is introduced and reviewed in Section~\ref{sec:SYK}. The insights gained from this study are then applied to several extensions thereof in Sections~\ref{sec:rqm}, \ref{sec:tJU}, \ref{sec:KH},  \ref{sec:lattice}, and \ref{sec:cfs}, with an eye towards capturing certain universal phenomenological aspects of quantum materials with strong electronic correlations. There are also remarkable connections between the SYK model and quantum theories of Einstein gravity in black holes, and these will be reviewed in Section~\ref{sec:qg}. In recent years, precise diagnostics of a class of non-quasiparticle systems have appeared by introducing ideas from quantum chaos and quantum gravity which will be discussed briefly in Section~\ref{sec:otoc}.

\section{Typology of non-Fermi liquids }
\label{sec:typology}

Numerous strongly correlated systems, {\it e.g.\/} materials with partially filled $d$- or $f$-shell orbitals and more recently in moir\'e systems, 
display a phenomenology which, while metallic, can drastically deviate from the predictions of the standard Fermi liquid (FL) theory of metals.
These {\it Non Fermi liquids} (nFL) raise a series of central challenges in condensed matter physics, both experimentally and theoretically.
As they are defined by what they are not, they
constitute a rich and very diverse family of systems.
Conceptually, they are not characterized by a few universal experimental traits, unlike Fermi liquids. 
In practice, they can not always be clearly identified using simple response functions,
unlike other familiar phases of quantum matter with or without spontaneously broken symmetries (e.g. superconductor, antiferromagnet, quantum Hall insulator). 

The family of SYK models discussed in this review  constitute a solvable theoretical route to study a class of nFL behaviour, 
as they have some of the major characteristics of nFL metals. 
In particular, we will discuss their relation with {\it Planckian metals}, characterized by a linear dependence of resistivity with temperature and a characteristic scattering rate $\sim k_BT/\hbar$.
In order to set the stage for this review, we therefore start in this section 
by discussing a selection of the most important nFL behavior encountered experimentally (Section \ref{sec:typology}A).
We then introduce the main theoretical routes which have been proposed to characterise and explain them (Section \ref{sec:typology}B), 
along with their connections to the aspects of SYK physics discussed in later sections.
Finally, in Section~\ref{sec:outline}, we present the general perspective of this review article and provide a detailed outline.
Readers wishing to go directly to the theoretical models of this paper can skip ahead to Section~\ref{sec:outline}.

\subsection{Experimental signatures of non-Fermi liquids}
\label{subsec:defn}

We start by discussing a few experimental signatures of nFL, based on a variety of spectroscopic and transport measurements.
Since {\it d.c.} transport can be difficult to interpret, it is important not to rely only on it exclusively to characterize nFL behaviour.
The various signatures include:

\begin{itemize}
\item ``Short'' single particle lifetimes for excitations near the Fermi surface, as deduced e.g. from spectroscopic measurements such as angle resolved photoemission spectroscopy (ARPES) \cite{zxs}. In FL metals,  the inverse quasiparticle lifetime (i.e. the scattering rate) scales as, $\Gamma_{\tn{sp}}\equiv g^2 W \left(k_B T/E_F^*\right)^2$, where $g$ is a dimensionless electron-electron interaction strength, 
$W$ is a bare electronic energy scale (bandwidth or hopping) 
and $E_F^*$ is a characteristic energy scale below which coherent 
long-lived quasiparticle excitations emerge. 
$E_F^*$ can be viewed as a degeneracy scale for the Fermi gas of quasiparticles. 
In contrast, a strong departure from the above form that persists over a large range of energy scales is an indication of breakdown of FL behavior. In a number of experimental systems that display nFL behavior, $\Gamma_{\tn{sp}}(\omega,T)\sim \tn{max}(\omega,k_BT/\hbar)$ \cite{Valla99,johnson}.

\item A power-law temperature dependence of the dc resistivity deviating from the expected FL form $\sim T^2$ (due to umklapp scattering) over a broad range of temperatures, without any signs of crossovers or saturation. One of the most commonly reported behaviors is $\rho = \rho_0 + AT$, over an extended range $T_{\tn{coh}} < T < T_{\tn{uv}}$ \cite{Mackenzie_rev}; see Sec.~\ref{subsec:planckian}. 
However, other power-laws $\Delta\rho~(\equiv\rho - \rho_0) \sim T^\alpha$, have also been observed \cite{allen,Eom02}. Identifying a material as a nFL on the basis of observation of $T-$linear resistivity above $T_{\tn{coh}}$ requires special care, since electron-phonon scattering in conventional metals leads to a trivial example of the same \cite{ziman}. However, $T-$linear resistivity presents a clear indication of behavior at odds with Boltzmann theory of FL transport in examples where $T_{\tn{coh}}$ is significantly low compared to the Debye (or Bloch-Gr\"uneissen) scale, the linearity persists without any crossovers across multiple phonon energy scales, and there are no obvious collective-modes to which a similar phonon-type argument can be applied directly. We return to a discussion of the physical significance of $T_{\tn{coh}}$, in subsequent sections. It is also worth noting that in some materials, such as optimally doped cuprates \cite{boebinger}, certain heavy-fermion materials \cite{Stewart} and twisted bilayer graphene \cite{efetov21}, this behavior persists down to a low $T_{\tn{coh}}\rightarrow0$.

\item Bad metallic behavior~\cite{emery_kivelson_prl_1995,GunnarssonRMP,Hussey04}
with a resistivity that is an increasing function of temperature with 
$\rho\gtrsim \rho_Q$ ($\rho_Q=h/e^2[a]^{d-2}$, where $a$ is a microscopic length scale 
and $h/e^2\simeq 25.8\mathrm{k}\Omega$ the quantum of resistance) is also indicative of nFL behavior. A majority of the systems of interest to us are quasi two-dimensional (with appreciable transport anisotropy in the $ab-$plane vs. along the $c-$axis) and it is thus useful to quote the results for the sheet-resistivities in units of $h/e^2$. While bad-metals can arise at very high temperatures for rather simple reasons, the key puzzle is often related to their smooth evolution into a low-temperature regime without any characteristic crossovers that defies Fermi liquid behavior.  
In the literature, the expressions {\it bad}, or, {\it strange} are often used to refer to certain nFL metals. 
In this review, we reserve the term {\it bad metals} to designate systems in which the resistivity is larger than the Mott-Ioffe-Regel value and {\it strange metals} to materials with a resistivity smaller than this value but displaying a set of behavior incompatible with the quasiparticle-based framework of Fermi liquid theory.
 We discuss bad metallic transport in the high-temperature regime in more detail in Sec.~\ref{sec:badmetal} below.

\item An anomalous power-law dependence of the optical conductivity, $\sigma(\omega)\sim 1/\omega^\gamma$, 
over an extended range of frequencies, 
%and without a clear 
differing from conventional Drude behavior. 
This is observed in cuprates ~\cite{Baraduc_1996,ElAzrak_1994,Hwang_2007,Marel2003,Schlesinger_1990} and has also been 
reported in other materials~\cite{Kostic_1998,Limelette_2013,Phanindra_2018,Schwartz_1998,Mena_2003,Dodge_2000}. 
This is also often accompanied by $(\omega/T)-$scaling as a function of temperature, 
i.e. $\sigma(\omega,T)\sim1/\omega^\gamma F(\omega/T)$~\cite{Marel2003,Eom02,Limelette_2013,vdM_2006,Michon2022,Heumen22}. 
At higher energy or temperature, a transfer of spectral weight over energy scales larger (sometimes much larger) than $k_BT$ are also typically observed as temperature is varied~\cite{Rozenberg_1996,DMFT,Basov_RMP}. 
A simultaneous analysis of both dc transport and optical conductivity (or other frequency-dependent response functions) is often crucial 
in reaching an understanding of the nFL phenomenology in a specific material.

\item An unconventional charge-density response, exemplified by a featureless continuum extending over a broad range of energy scales, as measured in Raman scattering experiments \cite{Bozovic87,Ginsberg91}. Recent measurements using momentum-resolved electron energy loss spectroscopy have further revealed a featureless two-particle continuum and an overdamped plasmon excitation \cite{Abbamonte1,Abbamonte2,Abbamonte3}, that is strikingly at odds with the expectations in a Fermi liquid metal.  

\end{itemize}

\subsection{Theoretical models of non-Fermi liquids}
\label{subsec:class}

Classifying insulating {\it gapped} phases of matter in terms of their symmetry and topological properties, using the lens of many-body entanglement, has been a remarkably successful venture \cite{XGW17}. On the other hand, classifying {\it gapless} phases of matter, and non-Fermi liquids in particular, remains an outstanding challenge. We shall not attempt to embark on such an endeavor here. This review will focus on a few distinct classes of nFL without quasiparticles, that can be described using various generalizations of the solvable SYK model. We find it useful nevertheless to first provide a broader overview of some of the theoretical frameworks and routes that lead to examples of non-Fermi liquids in clean crystalline systems without disorder.

\begin{figure}
\begin{center}
\includegraphics[scale=0.32]{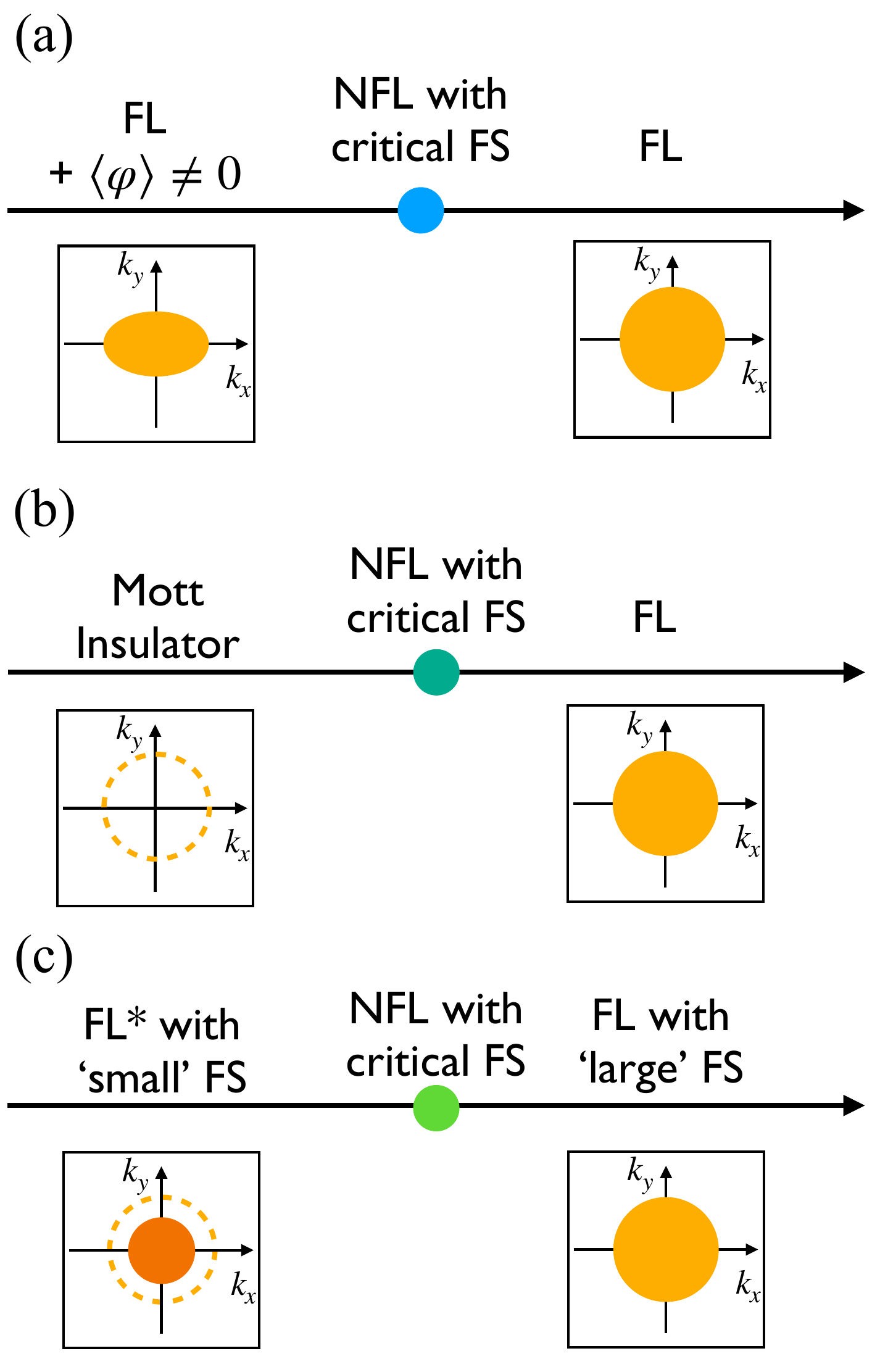}
\end{center}
\caption{(a) A nFL obtained by coupling a critical boson (e.g. nematic order with $\vec{Q}=0$) to an electronic Fermi surface. 
(b) A bandwidth-tuned metal to paramagnetic Mott insulator transition.   
The Mott insulator hosts a neutral Fermi surface (dashed circle) of fractionalized degrees of freedom coupled to an emergent gauge field. 
(c) A Fermi volume changing transition between two distinct metals across a `Kondo-breakdown' 
quantum critical point.  
The quantum critical point hosts a critical Fermi surface of {\it electrons} in all the examples. 
The Mott insulator and the FL* phases host a critical Fermi surface of `spinons' in (b) and (c), respectively. }
\label{QCNFL}
\end{figure}

\begin{itemize}
\item 
A class of models involves the quantum-critical fluctuations of a bosonic degree of freedom 
coupled to an electronic Fermi surface \cite{rmpqcp}. These fluctuations are associated with 
the order-parameter corresponding to the spontaneous breaking of a point-group (`nematic’), 
translational (spin/charge density-wave) or spin-rotation (ferromagnetism) symmetry. 
In the absence of any other instability, e.g. to pairing, the resulting ground state is a nFL that controls the properties of the system in a range of temperatures above the critical point. The nature of the low-energy excitations near the Fermi surface are clearly different depending on whether the order parameter carries zero, or a finite center-of-mass momentum, $\vec{Q}$. This framework of an electronic Fermi surface coupled to the low-energy fluctuations of a Landau order-parameter often goes under the name of Hertz-Millis-Moriya 
criticality~\cite{Moriya1985,Millis1993,QPT}. A critical boson with $\vec{Q}=0$ (e.g. nematic order) can destroy electronic quasiparticles around the entire Fermi surface (Fig.~\ref{QCNFL}a). At the critical point, the resulting state realizes a classic example of a critical Fermi surface \cite{metlitski1,mross} and provides an ideal setting for studying the interplay of nFL physics and superconductivity \cite{MM15,Chubukov16,BergAR19}. The low-energy field theory for such metallic criticality in $(2+1)-$dimensions presents a significant theoretical challenge \cite{sungsik1}. The insights provided by the solvable SYK model into such systems are reviewed in Section~\ref{sec:cfs}. 
A critical boson with $\vec{Q}\neq0$ (e.g. density-wave order) destroys electronic quasiparticles near only certain special points on the Fermi surface (`hot-spots') as it gets reconstructed into pockets, while much of the Fermi surface continues to host long-lived quasiparticles. See \cite{lee_review,BergAR19} for some recent complementary theoretical progress into both classes of such order-parameter based metallic criticality.

\item 
A different form of quantum criticality leading to nFL behavior is associated with the disappearance of entire electronic Fermi surfaces \cite{Colemanetal}. Prominent examples of such criticality include continuous metal-insulator transitions between a FL metal and a paramagnetic Mott insulator at fixed density (Fig. \ref{QCNFL}b) \cite{Florens2004,TS08,TSmott}; see also 
\cite{kotliar_largeN}, and a Kondo breakdown transition in a heavy Fermi liquid to a fractionalized FL (Fig. \ref{QCNFL}c) \cite{Coleman2000,Si1,Si2,FLSPRL,TSFL04,Norman07,Norman08,Norman13,Burdin2002,Colemanetal}. The critical point across both of these transitions also hosts an electronic critical Fermi surface without low-energy Landau quasiparticles \cite{TSFL04,TS08}. All currently known low-energy theories for describing such continuous transitions involve {\it fractionalized} degrees of freedom coupled to emergent dynamical gauge fields. Most theoretical descriptions of these continuous transitions have a remnant Fermi surface of the fractionalized degrees of freedom (and {\it not} of electrons) coupled to dynamical gauge fields on one side of the critical point; we will continue to refer to these as critical Fermi surfaces in this article. Continuous metal-insulator transitions without {\it any} remnant Fermi surface of even fractionalized degrees of freedom provide examples of a new form of `deconfined' metallic quantum criticality; see \cite{LZDC20,YZSS20} for some recent progress in describing such transitions. In particular, all of these transitions fall beyond the order-parameter based Hertz-Millis-Moriya framework described above. The insights of SY model with random exchange interactions in the presence of a uniform Kondo exchange for two-orbital models will be applied to study a special case of such abrupt Fermi-volume changing transitions in Section~\ref{sec:KH}.

\item 
In contrast to the examples above that arise at certain $T=0$ quantum critical points, a nFL can arise as a stable phase at zero temperature. One of the most well-known examples of such nFL behavior is found in a two-dimensional electron gas at high magnetic fields at a filling factor,  $\nu=1/2$. The metallic nFL state is compressible and otherwise known as the composite Fermi liquid (CFL); it hosts a sharp Fermi surface but the low-energy excitations are not electrons, but composite fermions (CF) \cite{JJCF}. The low-energy theory for the CFL is described in terms of a CF Fermi-sea coupled to a dynamical gauge-field \cite{HLR,Son15}. Other examples of nFL phases at $T=0$ have also been observed in numerical studies of lattice models \cite{mishmash}. 

We note that there are {\it insulating} (and {\it incompressible}) phases of matter that are expected to arise in a class of paramagnetic Mott insulators, where fractionalized degrees of freedom (e.g. spinons) form a Fermi surface and are coupled minimally to an emergent gauge-field \cite{PALee89,AIM}. The low-energy field theory for such phases shares similarities with the theory for the CFL, but there are important conceptual differences. A theoretical description of the low-energy field theory for the Fermi surface of spinons coupled to a dynamical gauge field suffers from the same problem that was noted earlier \cite{sungsik1}; the solvable SYK model of Section~\ref{sec:cfs} offers a controlled complementary understanding of this problem.
\item 
For sufficiently strong interactions and over a range of intermediate temperatures, it is possible that nFL behavior emerges generically and is not controlled by the proximity to a quantum critical point (or phase). Moreover, the nFL regime appears only as a crossover regime at intermediate temperatures while the ground state is a conventional phase (such as a FL, superconductor etc.). These nFL regimes can be described as ``infra-red (IR) incomplete", unlike the examples described earlier which are, in principle, controlled by $T=0$ fixed-points. Some prominent and well understood examples of such IR incomplete behavior include the classic electron-phonon system above the Debye temperature \cite{ziman}, spin-incoherent Luttinger liquids \cite{sill}, generic lattice models with a finite bandwidth at high temperatures \cite{Oganesyan} (see also \cite{lindner}), and certain holographic non-Fermi liquids \cite{Liu1,Liu2}.
Interestingly, a number of theoretical examples of such IR-incomplete behavior are accompanied by an extensive residual entropy, obtained from an extrapolation to the limit of $T\rightarrow 0$; the excess entropy is then relieved below the crossover to the conventional phase. Our treatment of such systems will appear in the discussion on lattice models of one and two-band models of SYK atoms in Section~\ref{sec:lattice}.
\end{itemize}

\subsection{Perspective of this review}
\label{sec:outline}

An important idea in  our approach is that it is possible to make progress on many intractable problems in the theory of non-Fermi liquids by considering models with random interactions. At first sight, this appears counter-intuitive, because spatial randomness introduces new phenomena associated with localization which are not of interest to us here. 
However, most of the models considered below live on fully connected lattices on which disorder-induced localization cannot take place. 
Indeed, the local electronic properties are strongly self-averaging, and the observable properties of a single sample with disorder are 
indistinguishable from the average of an ensemble of samples in the infinite-volume limit. 
Furthermore, one can argue that the strong incoherence associated with the absence of quasiparticles also removes localization effects 
which require quantum coherence and interference processes~\cite{LeeTVRRMP}. 
A non-Fermi liquid system without disorder thermalizes in the shortest possible time, and this implies chaotic behavior in which the memory of the initial conditions is rapidly lost. Consequently, it is possible to view averaging over disorder 
as a technical tool which allows access to the collective properties of a system with strong many-body quantum chaos. 

We can also restrict the disorder exclusively to a flavor space, and so study non-Fermi liquids with full translational symmetry, as we shall do in Sections~\ref{sec:lattice} and \ref{sec:cfs}. Here, the idea is that, after some renormalization group flow, a large set of theories flow to the same universal low energy behavior. And we find that it is easier to access the universal theory by averaging over a suitable set of microscopic couplings.

Indeed, the idea of using an average over random systems to understand quantum chaos has long been present in the theory of single-particle quantum chaos. We will discuss this in Section~\ref{sec:matrix}, where we will review the random matrix theory of non-interacting fermions: this has been a successful model of the quantum theory of particles whose classical dynamics is chaotic.

Section~\ref{sec:SYK} introduces the SYK model of fermions with random two-body interactions with $N$ single particle states. 
We will present the exact solution of the many-body system without quasiparticle excitations obtained in the $N \rightarrow \infty$ limit. Much is also understood about the finite $N$ fluctuations, including some results with a remarkable accuracy of $\exp(-N)$. This fluctuation theory relies on a mapping to a low energy effective theory of time reparameterization fluctuations (which is also the theory of a `boundary graviton' in the quantum theory of certain black holes of Einstein-Maxwell theory of gravity and electromagnetism, as will be discussed in Section~\ref{sec:qg}).

Section~\ref{sec:rqm} turns to a quantum generalization of the thoroughly studied Sherrington-Kirkpatrick model of a classical spin glass with Ising spins $\sigma_i = \pm 1$ ($i = 1 \ldots N \rightarrow \infty$) with random and all-to-all interactions $J_{ij}$ with zero mean. The quantum model replaces $\sigma_i$ with quantum $S=1/2$ SU(2) spins $\vec{S}_i$, which have random Heisenberg interactions $J_{ij}$. We will review a variety of studies of this model, involving numerical exact diagonalization, renormalization group, and large $M$ expansions of models with SU($M$) spin symmetry. These results show that the $S=1/2$ SU(2) model has spin glass order similar to that of the classical Sherrington-Kirkpatrick model. However, the spin glass order parameter is quite small, and for a wide range of intermediate frequencies, the dynamical spectrum of the SU(2) model matches that of the SYK model (obtained here in the large $M$ limit).

Sections~\ref{sec:tJU} and \ref{sec:KH} discuss the familiar and intensively studied single-band Hubbard and two-band Kondo-Heisenberg 
models, respectively, of strong electronic correlations. We will consider models with an {\it additional} random exchange interaction $J_{ij}$, 
which can be used to justify an extended dynamic mean field theory with self-consistency conditions on both the single electron and spin correlators. 
Such theories apply also to models with non-random single-particle dispersion, but it is useful to focus on a simplified limit with random and all-to-all single electron hopping $t_{ij}$. We will use methods similar to those in Section~\ref{sec:rqm} to show that these models exhibit quantum phase transitions between two metals: a metallic spin glass and a Fermi liquid. In the quantum critical region, we find a non-Fermi liquid with SYK-like correlations. Section~\ref{sec:numerical_methods} will present an overview of recent advances in the numerical methods employed for the analyses in Sections~\ref{sec:tJU} and \ref{sec:KH}.

Section~\ref{sec:lattice} presents a different approach towards generalizing SYK models to lattice systems. We consider a lattice of `SYK-atoms', where each lattice site has $N$ orbitals, and the intra-atomic electronic interactions are assumed to have the random SYK form. We will consider the case where all SYK atoms are identical (so that there is lattice translational symmetry) vs. the case where the interactions are different random instances on each site, and comment on their similarities and differences. These models can be used to realize non-Fermi liquids with a SYK character and no singular spatial correlations, but with a bad metallic resistivity. Generalizations of these models to include additional orbitals, in the spirit of two-band models of heavy-fermion materials, lead to strange metals with $T-$linear resistivity, critical Fermi surfaces and a marginal Fermi liquid behavior \cite{Varma89}.

Section~\ref{sec:cfs} returns to models of Fermi surfaces coupled to critical bosons, which we introduced earlier in Section~\ref{subsec:class}.
We describe how a systematic large $N$ theory of a class of non-Fermi liquids can be obtained by applying SYK-like approaches to these well-studied models.
We generalize the models to $\sim N$ flavors of fermions and bosons, with a random Yukawa coupling between the fermions and bosons.
The randomness can be independent of space, so that the models have translational symmetry. 

Section~\ref{sec:qg} explores the remarkable connections between the SYK model and the quantum theory of black holes. We will highlight some 
recent developments, particularly those we think are of interest to condensed matter physicists.

We end with a brief outlook on open questions in Section~\ref{sec:outlook}.

\section{Bad Metals and Planckian Metals}
\label{sec:bad_planck}

As emphasized above, a prime signature of nFL behaviour is unconventional transport. In this section, we provide a 
qualitative discussion contrasting high-temperature `bad metallic' 
behaviour to `Planckian transport' persisting 
down to low-$T$. This review focuses mostly on solvable models aiming at providing insight into the latter. 

\subsection{Bad Metals: Mott-Ioffe-Regel criterion and a high-temperature perspective}
\label{sec:badmetal}

In considering transport in semiconductors, Ioffe and Regel \cite{IR} and Mott \cite{mott} 
argued that metallic transport in the conventional sense requires that the mean-free path, $\ell$, 
of quasiparticles should be longer than the typical lattice spacing, $a$. 
For a quasi two-dimensional conductor with a single parabolic band and a simple cylindrical Fermi surface of 
radius $k_F$, the Drude expression for conductivity $\sigma=ne^2\tau/m$ can be rewritten as: 
\begin{equation}
    \sigma\,=\,\frac{e^2}{h} \frac{1}{c}\, k_F\ell,
    \label{eq:conductivity2D}
\end{equation}
in which $c$ is the interlayer distance. 
Hence, when the sheet conductance becomes smaller than the 
conductance quantum $e^2/h$, the Mott-Ioffe-Regel (MIR) criterion is violated and this 
suggests that a Drude-Boltzmann description of transport is no longer valid. 
The criterion itself is not a quantitatively precise one, depending 
on whether $\ell$ is compared to $a$, or to the Fermi wavelength $\lambda_F=2\pi/k_F$. 

`Good' metals typically have resistivities that are much smaller than $\rho_Q$ and correspondingly $\ell\gg a$. In the context of unconventional metallic transport, the physical significance of the MIR criterion 
has been a confusing issue for quite a while, as reviewed e.g. in \cite{Hussey04,GunnarssonRMP}. Some materials, such as the A15 compounds~\cite{Fisk76}, display a resistivity saturation as the 
MIR value is approached, leading to the speculation that resistivity saturation should perhaps be a general fact. It is worth noting that there is no fundamental theoretical understanding for resistivity saturation in metals.~\footnote{Recent work has analyzed resistivity saturation, and lack thereof, in solvable models of electrons coupled to a large number of phonon modes \cite{Werman,Werman2}.}
Moreover, a wealth of experimental data collected on materials with strong electronic correlations, 
most notably transition-metal oxides, came in to contradict the very notion of resistivity saturation. Indeed, resistivity in many such 
materials can increase significantly above the MIR value without any trend towards saturation or even any 
characteristic feature signalling this crossover in the temperature dependence of $\rho$. 
The term `bad' metal was coined to highlight this behavior \cite{emery_kivelson_prl_1995}. 
A material displaying bad metallic behavior at a high temperature can become a good Fermi liquid at a low temperature with long-lived coherent quasiparticles, an outstanding example being 
Sr$_2$RuO$_4$~\cite{Tyler1998}. Low-carrier density materials such as doped 
SrTiO$_3$ also have bad metallic behavior at high-$T$ \cite{BehniaPRX} while displaying quantum oscillations 
and coherent transport at low-$T$ \cite{behnia_review}.

Recent studies \cite{Deng2013,Deng2014} have considerably clarified the physical significance 
of the MIR criterion. It is now understood that the temperature $T_{\mathrm{MIR}}$ at which the 
resistivity becomes of the order of the MIR value corresponds to the complete disappearance of quasiparticles. Typically, in systems which become FL at low-$T$, 
the scale $T_F^*$ below which long-lived coherent (Landau) quasiparticles with  $\Gamma_{\tn{sp}}\sim T^2$ are observed is significantly smaller than  $T_{\mathrm{MIR}}$. 
In Sr$_2$RuO$_4$ for example, $T_F^*\simeq 30$~K, while $T_{\mathrm{MIR}}$ is several hundreds 
degrees Kelvin. Studies of the doped Hubbard model in the dynamical mean-field theory (DMFT) framework 
have documented this interpretation in a precise manner. There, $T_{\mathrm{MIR}}$ was found to 
be of the order of the Brinkman-Rice scale $\sim p.t$ (with $p$ the doping level and $t$ the 
typical hopping or bare Fermi energy), while a much lower scale is associated with $T_F^*$ 
--- for a renormalization group interpretation of that scale, see~\cite{Held2013}. 
It was shown that `resilient quasiparticles' exist in the intermediate 
regime $T_F^*<T<T_{\mathrm{MIR}}$: 
the spectral function displays a broadened but well-defined peak 
and transport can still be described in terms of these 
excitations, reminiscent of the notions introduced~\cite{Prange}
for electron-phonon scattering. It was also shown~\cite{Deng2014} that the quasiparticle lifetime 
follows a $1/T^2$ law up to a higher temperature than the transport lifetime itself, and hence than the 
temperature at which the resistivity deviates from $T^2$.

Considerable insight in interpreting transport results can be gained by simultaneously considering spectroscopy experiments, most notably optical conductivity, and the corresponding transfers of 
spectral weight upon changing temperature. 
In studies of the doped Hubbard model~\cite{Deng2013}, it 
was shown that these transfers are limited to the low-energy region between the Drude peak and 
the mid-infrared range for $T_F^*<T<T_{\mathrm{MIR}}$, while the MIR crossover is signalled by 
spectral weight transfers over a much larger energy range, leading to a broad featureless 
optical conductivity for $T > T_{\mathrm{MIR}}$. 

At temperatures exceeding the (finite) bandwidth for lattice fermions, it is natural to 
find bad metallic transport with a resistivity scaling linearly with temperature. We briefly 
review here the physical nature of this high-$T$ 
regime which is by now well understood. One approach to this regime is to start from the Kubo formula 
for the optical conductivity:
\beq
\sigma(\omega,T) = \pi \frac{1-e^{-\beta \hbar\omega}}{\omega Z}\sum_{n,m} e^{-\beta E_n} |J_{nm}|^2~\delta(E_n-E_m-\hbar\omega) \nonumber\\
\label{eq:Kubo}
\eeq
where $n,m$ label the eigenstates of the generic many-body Hamiltonian with energies $E_n,~E_m$, respectively. The matrix elements of the total current operator between the two states are denoted $J_{nm}$ and $Z = \sum_n e^{-\beta E_n}$ is the partition function. 
When $T$ is the largest energy scale in the problem, this expression reduces to
\beq
\sigma(\omega,T) = \frac{\pi}{k_B T} \frac{1}{Z}
\sum_{n,m} |J_{nm}|^2~\delta(E_n-E_m-\hbar\omega).
\label{eq:HighT}
\eeq
For generic lattice models, and more generally any system for which the sum is finite in the thermodynamic limit,
Eq.~(\ref{eq:HighT}) implies that $T$-linear 
resistivity is expected in the high-$T$ regime; importantly for generic non-integrable models the matrix-elements $J_{nm}$ are expected to have a `random-matrix' form even in the absence of any randomness \cite{Oganesyan}. 
This analysis has recently been extended to study several interacting models over a wider range of temperatures~\cite{Changlani}. 
The expression for the conductivity in Eq.~(\ref{eq:Kubo}) looks deceptively simple but usually presents a significant computational challenge when evaluated for the entire many-body spectrum. 

The origin of 
$T$-linear resistivity (and deviations thereof at lower $T$) can also be approached from a systematic high-$T$ expansion of the optical conductivity \cite{Perepelitsky2016,lindner}. 
Computational investigations of transport in two-dimensional Hubbard models 
in the high-$T$ regime have appeared recently, using 
quantum Monte Carlo~\cite{Huang19} and the finite-temperature Lanczos 
method~\cite{Vucicevic2019,Vranic2020}.

Complementary and model-independent insights into this high-$T$ regime can be obtained by considering the Einstein-Sutherland relation relating the dc conductivity $\sigma_{dc}$, charge diffusion 
coefficient $D_c$, and charge compressibility $\chi_c$~\cite{GunnarssonRMP}; see also~\cite{Hartnoll2014,Perepelitsky2016}. 
 
When thermoelectric effects can be neglected, this relation reads: 
\beq
\sigma_{dc} = \chi_c\,D_c\,\,\,,\,\,\,
\chi_c = \frac{\partial n}{\partial\mu},
\label{eq:Einstein}
\eeq
with $n$ the average density and $\mu$ the chemical potential. In the high-temperature limit, where the gas of Fermi particles is non-degenerate, the origin of $\sigma_c\sim 1/T$ is tied simply to the 
thermodynamic property $\chi_c\sim 1/T$ rather than to the $T-$dependence of $D_c$ 
(or equivalently, of the scattering rate). 
Hence, in that regime, bad metallic transport does correspond to a saturation phenomenon, although not of the resistivity itself but rather of the 
diffusion constant or scattering rate. Indeed, in a lattice model, it is natural that 
the minimum possible value of the diffusion constant 
should be of order $D_c \sim a^2/\tau_0$ with $a$ the lattice spacing and 
the microscopic time-scale $\tau_0 \sim \hbar/t$ with $t$ the bare hopping. 

\begin{figure*}[!htbp]
\begin{center}
\includegraphics[scale=0.32]{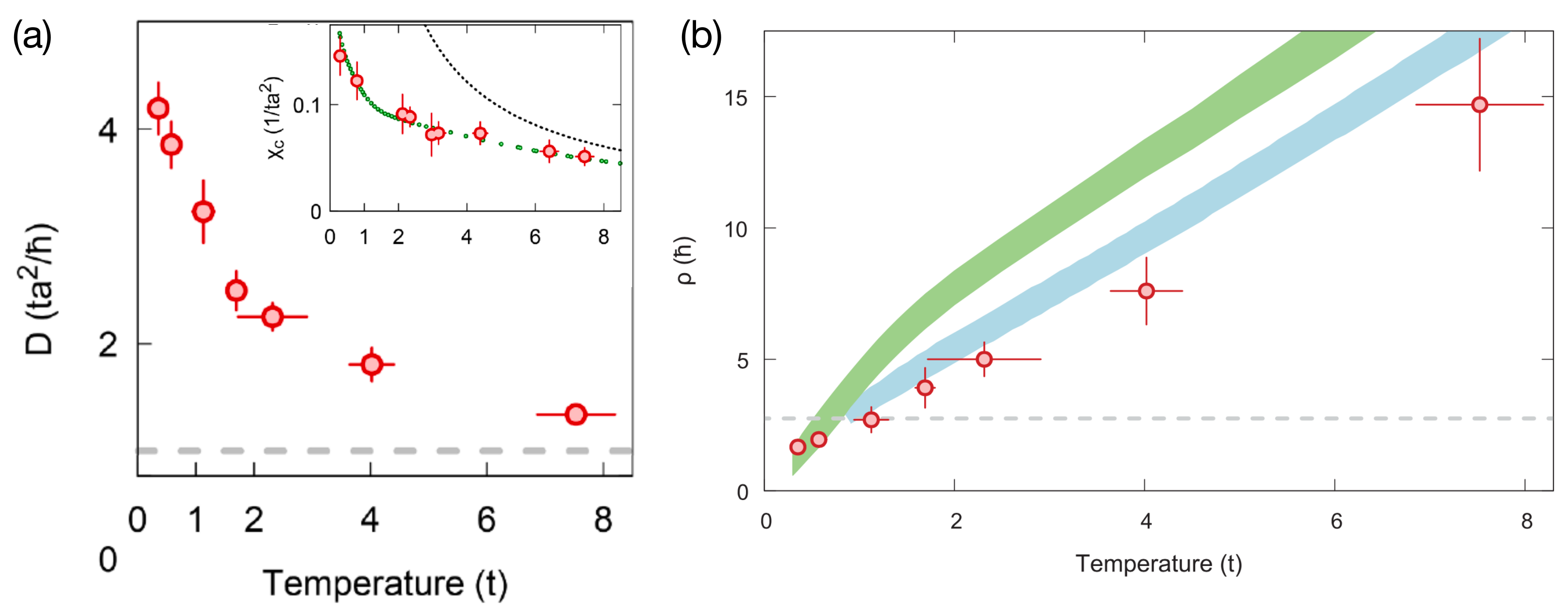}
\end{center}
\caption{Measurement of the diffusion constant (a) and compressibility ((a)-inset) 
for a gas of ultra-cold $^6$Li atoms in an optical lattice, realizing 
a two-dimensional Fermi-Hubbard model with $U/t\simeq 7.5$ at a density $n\simeq 0.825$. 
(b) Reconstructed `resistivity' using Einstein-Sutherland relation. Grey horizontal dashed line represents the estimated MIR value. Theoretical calculations using DMFT (in green) and the finite-$T$ Lanczos method (in blue) are shown; the band representation indicates estimated error bars. Adapted from \cite{Bakr}. 
\label{coldatoms}
}
\end{figure*}

In the solid-state context, probing experimentally the regime where $T$ is comparable to the hopping amplitude is challenging, except in flat-band materials, but is usually complicated by the intervening role of phonons and other remote dispersive bands. From that perspective, cold atomic gases in optical lattices offer an 
ideal platform for studying transport in `hot' or intermediate temperature regimes, as 
documented by recent experimental investigations~\cite{Bakr,Marco,Anderson2019}. 
Fig.~\ref{coldatoms}(a)-(b) displays the measured diffusion constant and compressibility, 
and the `resistivity' calculated using the Einstein-Sutherland relation for two-component fermions in an optical lattice realizing a two-dimensional Hubbard model,  
measured as a function of temperature in the range $T/t = 0.3 - 8$~\cite{Bakr}. 
It is seen that the regime dominated by thermodynamics $\chi_c\sim 1/T, ~D_c\sim\mathrm{const.}$ is indeed observed at 
the highest temperatures, crossing over into a regime at lower $T$ in which both 
the diffusion constant and compressibility exhibit $T$-dependent crossovers. 
Correspondingly, the resistivity as given by Eq.~(\ref{eq:Einstein}) becomes smaller 
than the MIR value at the lowest temperature while exhibiting a $T-$linear behavior
without any noticeable feature or change of slope across the crossover. 

The high-$T$ mechanism for $T$-linear bad-metallic transport should be contrasted with 
the `Planckian regime'~\cite{Zaanen04}, discussed in more details below, in which   
the diffusion constant (or, scattering time) is temperature dependent, 
$D_c\sim a^2\, \hbar/k_B T$, 
while the compressibility is temperature independent~\cite{Hartnoll2014}. 
In most of the low-temperature nFL exhibiting $T-$linear resistivity, it is widely believed that it is the scattering rate that is temperature dependent and not the compressibility. However, establishing this is, in general, difficult in the solid-state setting. Recent experimental progress has allowed for direct measurements of the electronic compressibility in two-dimensional gate-tunable materials~\cite{Zondiner2020}, indeed demonstrating that the Planckian regime of low-temperature transport in magic-angle twisted bilayer graphene~\cite{Cao20,Young19,efetov21} corresponds to $D_c\sim 1/T$~\cite{ParkDiffusivity2020}. 
It should be noted that Planckian behavior and bad metallic behavior are not mutually exclusive: indeed we shall discuss in Sections~\ref{sec:tJPG} and \ref{sec:lattice} models in which $D_c\sim 1/T$ while the resistivity is larger 
than the MIR value.

In the remainder of this review, we continue to refer to `bad' metals as systems 
with a resistivity larger than the MIR value. We reserve the term `strange' metal to 
systems or regimes with a resistivity smaller than the MIR value 
but having an unconventional power-law behavior at odds with expectations in a Fermi liquid.   
This article devotes special interest to the latter, 
only occasionally discussing bad metals when relevant. 

\subsection{Planckian relaxation: unity in diversity?}
\label{subsec:planckian}

\begin{figure*}[!ht]
\begin{center}
\includegraphics[scale=0.3]{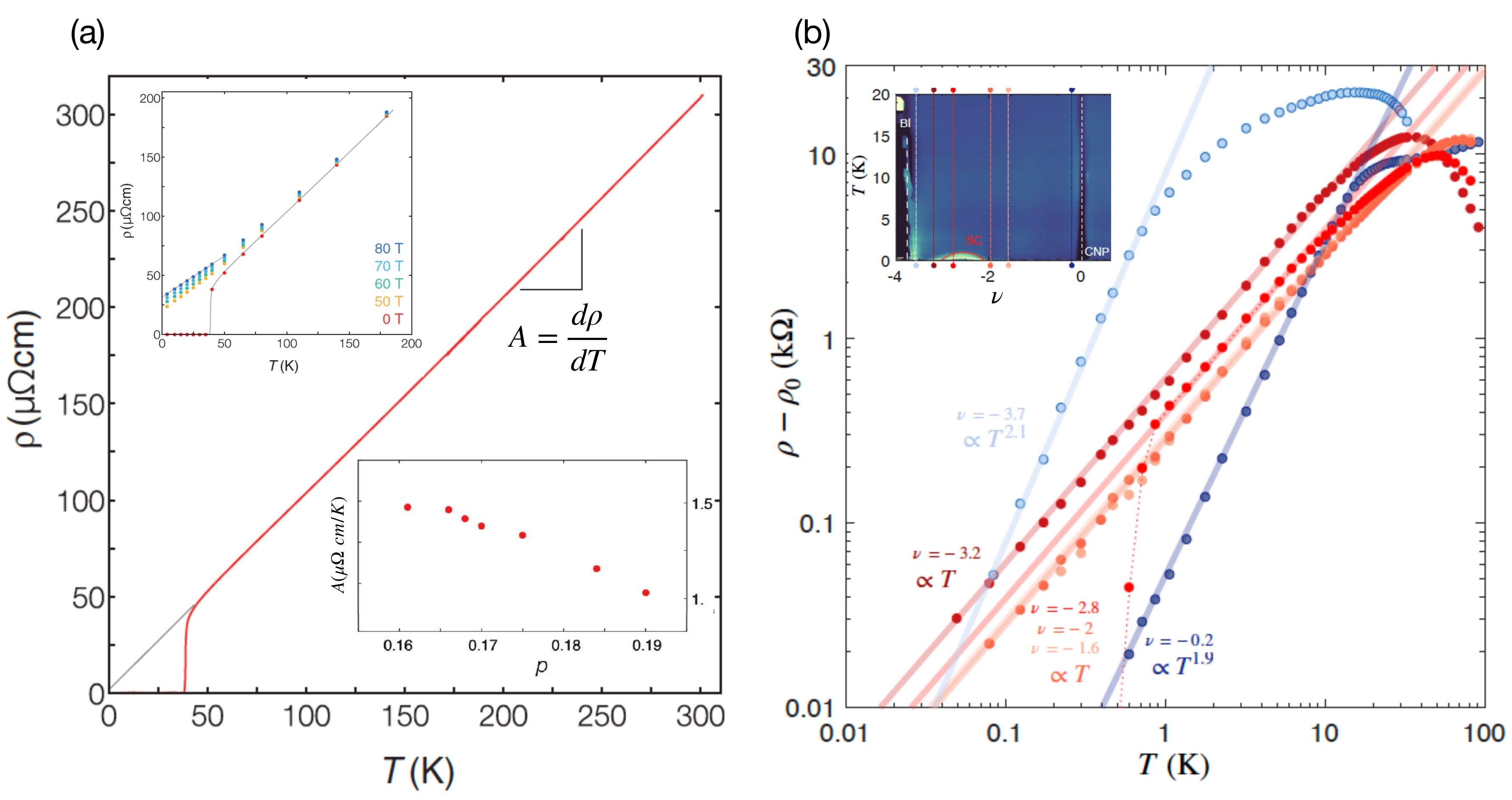}
\end{center}
\caption{Examples of $T-$linear resistivity extending over a wide range of temperature scales in (a) hole-doped La$_{2-x}$Sr$_x$CuO$_4$ (LSCO) near optimal doping (adapted from \cite{boebinger}), and (b) magic-angle twisted bilayer graphene (MATBG) near $\nu\approx -2$, relative to charge neutrality, $\nu=0$ (adapted from \cite{efetov21}). In LSCO, $T_{\tn{coh}}$ can be inferred to be much lower than any characteristic energy scales by turning on a magnetic field and accounting for the finite magnetoresistance ((a)-top inset); the variation of the slope ($A$) on hole-doping is shown in (a)-bottom inset. In MATBG, the linearity for a range of dopings near $\nu\approx-2$ ((b)-inset) persists down to $\sim 40$ mK. Both family of materials also display a Planckian form of $\Gamma_\tn{dc}$ (Eq.~\ref{eq:Planckian}).} 
\label{fig:Tlinear}
\end{figure*}

Carrier numbers and effective masses may be very different from one material to another, and thus it is often not a meaningful exercise to compare the actual values of the resistivity across different materials. 
Instead, comparing the relaxation timescales associated with transport can shed interesting light on the universal mechanisms that govern nFL properties. Unfortunately, obtaining a transport lifetime from measurements of a dc resistivity is not a straightforward exercise.

We focus here on instances in which a resistivity depending linearly on temperature 
$\rho=\rho_0+A\,T$ is observed --- see Fig.~\ref{fig:Tlinear} for some examples and 
\cite{Hussey2008,Proust2019,Varma2020,Mackenzie_rev} for reviews. 
A particular procedure that has been adopted to extract a temperature 
dependent transport scattering rate, $\Gamma_{\tn{dc}}$ in such materials~\cite{Bruin13} 
relies on a ``Drude'' fit,{\footnote{The dc transport need not have a Drude-like form in generic nFL metals.}} where one expresses $\rho = m^*\Gamma_{\tn{dc}}/n_c e^2$. Assuming that the effective mass, $m^*$, and carrier concentration, $n_c$, are temperature independent, one writes 
\beq
\Gamma_\tn{dc}\equiv \alpha \frac{k_BT}{\hbar},~~\alpha = \frac{\hbar}{k_B}\frac{e^2n_c}{m^*} A.
\label{eq:Planckian}
\eeq
In the experimental analysis, $m^*$ and $n_c$ are typically extracted from low-temperature measurements (i.e. $n_c \equiv n_c(T\rightarrow0),~m^* \equiv m^*(T\rightarrow0)$), which 
does not always coincide with the regime in which the clearest signature of an extended $T-$linear resistivity is observed. The above analysis becomes especially difficult in multi-orbital systems and the effective masses are often extracted from quantum oscillations, or, specific heat; it is far from being clear why this is a relevant quantity that should determine the momentum relaxation rate even within Drude theory.

Nevertheless, it is quite remarkable that for a number of metals exhibiting a broad regime of $T-$linear resistivity including the cuprates, pnictides, ruthenates, organics and rare-earth element materials, the above ``operational" definition of a scattering rate leads to $\alpha\approx1$ \cite{Bruin13}. 
A similar analysis in magic-angle twisted bilayer graphene near half-filling of the electron and hole-like flat-bands \cite{Cao20,efetov21}, in twisted transition metal dichalcogenides \cite{AP21}, 
several cuprates over an extended range of doping levels~\cite{Legros19} 
and a non-superconducting iron-pnictide \cite{Paglione} have also found indication of a Planckian scattering rate with $\alpha\approx 1$. Recent measurements of angle-dependent magnetoresistance (ADMR) near the pseudogap critical point in Nd-LSCO also reveal a Fermi surface with an isotropic Planckian scattering set by $\alpha\approx 1$ \cite{Grissonnanche2020}. 
Note that this conclusion holds in the latter case 
provided a $T$-independent effective mass associated 
with intermediate energy scales (and consistent with ARPES and ADMR) is used, rather than the 
thermodynamic effective mass associated with specific heat which displays a logarithmic $T$-dependence.

It is important to note that a $T-$linear resistivity with a Planckian scattering rate (Eqn.~\ref{eq:Planckian}) is observed in conventional metals like copper, gold etc. 
This is not a surprise and as noted earlier, the behavior is associated with electron-phonon scattering where the phonons are in a classical equipartition regime. 
There have been discussions \cite{Sadovskii2020,Sadovskii2021} of a possible rationale 
for $\alpha\simeq 1$ in regimes where electron-phonon and electron-electron interactions 
contribute to $T$-linearity on a similar footing.
However, Planckian scattering that persists down to extremely low temperatures \cite{boebinger,Cao20,efetov21} in nFL that are not low-density materials, and where the behavior persists across multiple phonon frequencies without any crossovers presents a challenge to theory. A more in-depth discussion of Planckian timescales across solid-state materials has appeared in a recent review \cite{Mackenzie_rev}.

We end this section by noting that there does {\it not} exist a universal definition 
of a ``transport scattering rate'', making it difficult to formulate 
a precise theoretical Planckian bound. Even experimentally, as seen above, the procedure used most often to extract a scattering rate relies on a number of approximations. 
In that sense, the use of Einstein-Sutherland relation to extract a diffusion constant, 
combined with the recent progress in measuring electronic compressibility discussed above, may be a safer route to follow whenever possible.

Optical spectroscopy measurements of the complex conductivity are often parametrized in terms of a frequency and temperature-dependent optical time scale and effective mass enhancements 
as~\cite{Basov_RMP}: 
$4\pi\sigma(\omega)/\omega_p^2=\left[1/\tau_{\mathrm{opt}}(\omega)-i\omega m^*_{\mathrm{opt}}(\omega)/m)\right]^{-1}$, which can be directly determined from experimental data 
as $1/\tau_{\mathrm{opt}}=\omega_p^2/4\pi\,\mathrm{Re}[1/\sigma]$, 
$m^*_{\mathrm{opt}}/m=\omega_p^2/4\pi\, \mathrm{Im}[1/\sigma]$ once a normalisation of the spectral weight $\omega^2_p/4\pi$ has been chosen. 

In a subset of the nFL metals highlighted above, including optimally doped cuprates~\cite{Marel2003}, 
the low-frequency limit of 
$1/\tau_{\mathrm{opt}}$ was also shown to have a Planckian form and 
$\omega/T$ scaling was observed. 

In later sections of this review, we will discuss a number of recent studies that have demonstrated the existence of a Planckian timescale for transport in solvable models of correlated electrons.

\section{Random matrix model: free fermions} 
\label{sec:matrix}

In the study of charge transport in mesoscopic structures, much experimental effort has focused on electrons moving through `quantum dots' \cite{alhassid}. We can idealize a quantum dot as a `billiard', a cavity with irregular walls. The electrons scatter off the walls, before eventually escaping through the leads. If we treat the electron motion classically, we can follow a chaotic trajectory of particles bouncing off the walls of the billiard. Much mathematical effort has been devoted to the semi-classical quantization of such non-interacting particles: the `quantum billiard' problem. The Bohigas-Giannoni-Schmidt conjecture~\cite{BGS} states that many statistical properties of this quantum billiard can be described by a model in which the electrons 
hop on a random matrix; there has been recent progress \cite{Altland09,anantharaman2011} towards establishing this conjecture. It is this random matrix problem that we will describe in this section. 

Many properties of the random-matrix model are similar to a model of a disordered metal in which the electrons occupy plane wave eigenstates which scatter off randomly placed impurities with a short-range potential. However, unlike the random impurity case, there is no regime in which the eigenstates of a random matrix can be localized. As every site is coupled to every other site, there is no sense of space or distance along which the eigenstate can decay exponentially. 
The absence of localization also extends to non-fully 
connected lattices with infinite 
connectivity, such as a regular hybercubic lattice in $d$-dimensions in the $d\rightarrow\infty$ limit. 
Indeed, it can be shown that in this limit the local density of states 
self-averages (see below), 
which implies the absence of Anderson localisation~\cite{Dobro1997}.

\subsection{Green's function}

We consider electrons $c_i$ (assumed spinless, for simplicity) hopping between sites labeled $i=1, \ldots, N$, with a hopping matrix element $t_{ij}/\sqrt{N}$:
\begin{subequations}
\begin{align}
&~~~H_2 = \frac{1}{(N)^{1/2}} \sum_{i,j=1}^N  t_{ij} c_{i}^\dagger  c_{j} - \mu \sum_i c_i^\dagger c_i \label{rmt1}\\
& c_{i} c_{j} + c_{j } c_{i} = 0 \quad, \quad c_{i }^{\vphantom \dagger} c_{j }^\dagger + c_{j }^\dagger 
c_{i }^{\vphantom \dagger} = \delta_{ij} \\
&~~~~~~~~~~~~~~~\frac{1}{N} \langle\sum_i c_{i }^\dagger c_{i } \rangle = \mathcal{Q}.
\end{align}
\end{subequations}

The  $t_{ij}$ are chosen to be {\it independent\/} random complex numbers with $t_{ij} = t_{ji}^\ast$, $\overline{t_{ij}} = 0$ and $\overline{|t_{ij}|^2} = t^2$.
The $1/\sqrt{N}$ scaling of the hopping has been chosen so that the bandwidth of the single electron eigenstates will be of order unity in the $N \rightarrow \infty$ limit, and therefore (as there are $N$ eigenstates) the spacing between the successive eigenvalues will be of order $1/N$.
We have also included a chemical potential so that the average density of electrons on each site is $\mathcal{Q}$. The subscript (`$2$') in the Hamiltonian, $H_2$, denotes that it only includes two electron operators. 

For a given set of $t_{ij}$, one can numerically diagonalize
the $N \times N$ matrix $t_{ij}$ to solve this problem.
We denote 
by $\{\ket{\lambda}$, $\epsilon_\lambda\}$ the spectrum of eigenstates of the matrix $t_{ij}$ 
for a given realisation. 
%However, in the limit of large $N$, it turns out that certain quantities self-average. 
%In other words, certain observables take the same value on every site, and that value is realized 
%for a given sample with probability $1$ in the $N \rightarrow \infty$ limit. 
%We will only be interested in such observables here. 

However, in the limit of large $N$, it turns out that certain quantities are self-averaging. 
This means that, for a given sample $t_{ij}$, their value converges with probability one 
in the $N \rightarrow \infty$ limit 
to their averaged value over all samples. 
We will only be interested in such observables here.

We define as usual the single-particle Green's function as: 
\beqn
G_{ij} (\tau) =  - \left\langle T_\tau c_i (\tau) c_j^\dagger (0) \right\rangle,
\eeqn
with $\tau$ the imaginary time and $G_{ij}(\tau+\beta)=-G_{ij}(\tau)$.
For a given sample, we can expand this function in terms of the one-particle eigenstates as: 
\beqn
G_{ij}(z) = \frac{1}{N}\sum_\lambda \braket{i|\lambda} \frac{1}{z+\mu-\epsilon_\lambda} 
\braket{\lambda|j},
\eeqn
where $z$ denotes a complex frequency, for example the Matsubara frequencies 
$\omega_n=(2n+1)\pi/\beta$. 
%In the limit of large $N$, and the sites $i,j$ being fixed, the Green's function self-averages:
%\beqn
%G_{ij} (\tau) \quad \rightarrow \quad G(\tau) \delta_{ij}\,.
%\eeqn

In the limit of large $N$, for a given site $i$, the local Green's function self-averages:
\beqn
G_{ii} (\tau) \quad \rightarrow \quad G(\tau)\,
\eeqn
with $G=1/N\sum_i G_{ii}$, also identical to the average over all samples $\overline{G_{ii}}$. 
In contrast, $G_{i\neq j}$ is of order $1/\sqrt{N}$ for a given pair of sites $i,j$ and depends on the 
specific sample. 

\begin{figure}
\begin{center}
\includegraphics[scale=0.15]{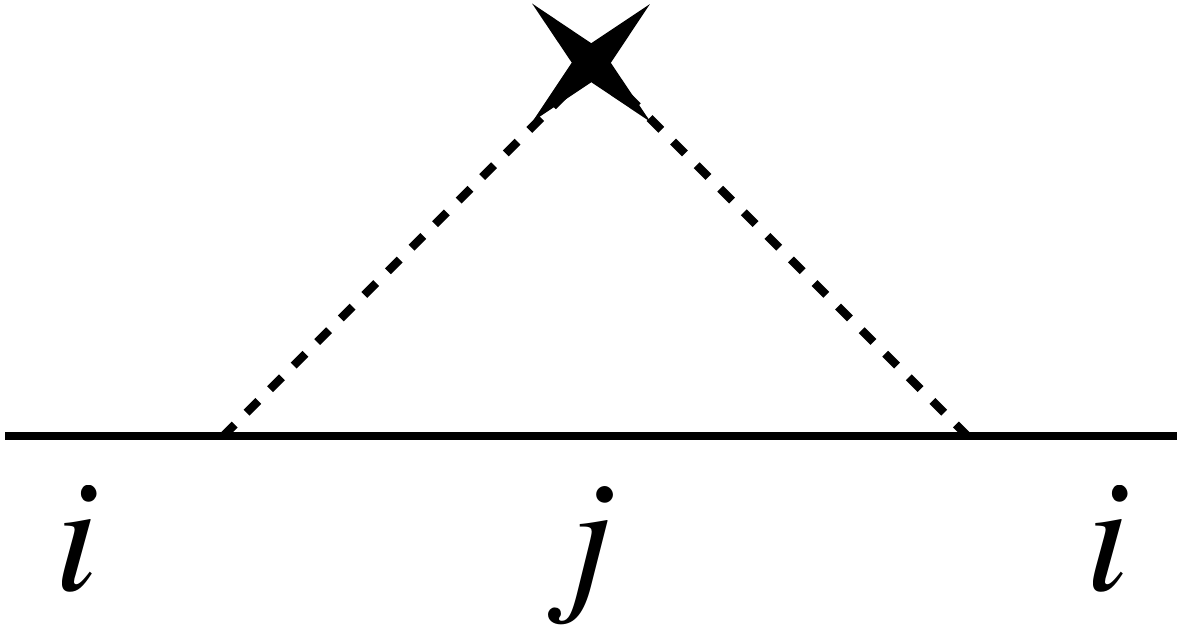}
\end{center}
\caption{The graph for the electron self-energy, $\Delta(\tau)$, in Eq. (\ref{rm2}). Solid lines denote fully dressed electron Green's functions. The dashed line represents the disorder averaging associated with $\overline{|t_{ij}|^2}$. }
\label{rmtse}
\end{figure}

The simplest way to establish this result consists in evaluating averages of $G_{ij}$ order-by-order in a perturbation theory in $t_{ij}$.
At zeroth-order, the Green's function is simply
\beqn
G^0_{ij} (i \omega_n) = \frac{\delta_{ij}}{i \omega_n + \mu}.
\eeqn
The Feynman graph expansion consists of a single particle line, with an
infinite set of possible products of $G_{ij}^0$ and $t_{ij}$. We then average
each graph over the distribution of $t_{ij}$. In the $N \rightarrow
\infty$ limit, only a simple set of graphs survive (Fig.~\ref{rmtse}) and the average
Green's function is a solution of the following set of equations
\begin{subequations}
\begin{align}
G(i\omega_n) &= \frac{1}{i \omega_n + \mu - \Delta (i\omega_n)} \\
\Delta (\tau) &=  t^2 G (\tau)  \label{rm2} \\
G(\tau = 0^-) &= \mathcal{Q}. \label{rm1}
\end{align}
\end{subequations}

The solution of Eq.~(\ref{rm2}) reduces to solving a quadratic equation for $G(z)$, 
and so we obtain for a complex frequency $z$
\beqn
G(z) = \frac{1}{2t^2} \left[ z + \mu \pm \sqrt{(z+\mu)^2 - 4t^2} \right]. 
\label{GFsoln}
\eeqn
The sign in front of the square root ($=\mathrm{sign}[\mathrm{Im}(z+\mu)]$) 
is to be chosen such that $G(z)$ has the correct analytic properties:
\begin{itemize}
\item $G(|z| \rightarrow \infty) = 1/z$,
\item $\mbox{Im} \, G(\omega+i0^+) < 0$ for real $\omega$,
\item $\mbox{Im} \, G(\omega+i0^-) > 0$ for real $\omega$.
\end{itemize}
All of these constraints can be obtained from the spectral representation of the Green's function. 
We can also define the density of single-particle states as
\beqn
\rho (\omega) = - \frac{1}{\pi} \mbox{Im} \, G(\omega-\mu+i0^+) = \frac{1}{2 \pi t^2} \sqrt{4 t^2 - \omega^2} \label{rm3a}
\eeqn
for $\omega \in [-2t, 2t ]$, and $\rho(\omega) = 0$ otherwise. This is the famous Wigner 
semi-circle density of states for the random matrix \cite{Mehta}.

The chemical potential is fixed by requiring that Eq. (\ref{rm1}) is satisfied, which can be written as
\beqn
\int_{-2t}^{2t} d \omega \, \rho (\omega) f(\omega-\mu) = \mathcal{Q}\,, \label{rm3}
\eeqn
where $f(\varepsilon) = 1/(e^{\varepsilon/T} + 1)$ is the Fermi function. Performing a Sommerfeld expansion of the left-hand side
for $T \ll t$, we obtain
\beqn
\int_{-2t}^{\mu} d \omega \,  \rho(\omega)  + \frac{\pi^2 T^2}{6} \rho' (\mu) = \mathcal{Q}\,.
\eeqn
where $\rho' (\omega) = d \rho/d \omega$. 
In order to satisfy this equation for all $T$ in the low-$T$ regime, 
%This equation must be satisfied at all $T$, and depending upon the particular ensemble, 
%it requires variation of $\mu$ or $\mathcal{Q}$ with $T$. 
$\mu$ or alternatively $\mathcal{Q}$ must depend on $T$ (depending upon the particular ensemble). 
In particular, if we keep $\mathcal{Q}$ fixed and vary $T$, then
\beqn
\mu(T) = \mu_0 - \frac{\rho' (\mu_0)}{\rho(\mu_0)} \frac{\pi^2 T^2}{6} 
\eeqn 
where $\mu_0 = \mu(T=0)$. 

An alternative way to prove the self-averaging properties is to use the `cavity' construction, 
which is also a useful method to establish the local effective action 
associated with interacting models considered later in this article.  
In a nutshell (see e.g. \cite{DMFT} for details), this consists of integrating over all sites $i=2,\cdots,N$ except site 
$i=1$, and noting that the term 
$\sum_{i>1} c^\dagger_i (t_{i1} c_1)$ can be viewed as a 
source term coupling to $c^\dagger_i$. Performing the integration over sites is a Gaussian problem in this non-interacting case and leads to the following effective action for site $1$: 
\begin{equation}
    \int d\tau \int d\tau' c^\dagger_1(\tau) \left[\delta(\tau-\tau')(\partial_\tau-\mu) + 
    \Delta_1(\tau-\tau')\right] c_1(\tau'),
\end{equation}
with 
\begin{equation}
    \Delta_1(z) = \frac{1}{N}\sum_{i\neq 1} t_{1i}^2 G_{ii}^{[1]}(z)+
    \frac{1}{N}\sum_{i\neq j, i,j >1} t_{1i} t_{1j} G_{ij}^{[1]}(z).
\end{equation}
In the above expression, $G_{ij}^{[1]}(z)$ denotes the Green's function of the lattice with one less site  
(site $1$ removed, $N-1$ sites), also removing all connections to that site. 
We see that the sum over $i$ in the first term amounts to a statistical 
average as $N\rightarrow\infty$ and we note, importantly, that $G_{ii}^{[1]}(z)$ does not depend on $t_{1i}$. 
Hence the two terms under the sum can be averaged independently, yielding $t^2 G$. 
A similar reasoning shows that the second term vanishes since the average of the $t_{ij}$'s are zero. 
This proves the self-averaging of the local Green's function, $G_{11}$. Inverting the quadratic kernel leads to Eq. (\ref{rm2}), $G^{-1}(z) = z+\mu-t^2 G(z)$. 
This also proves that the local one-particle density of states for a given sample 
$\frac{1}{N}\sum_\lambda |\braket{i|\lambda}|^2 \delta(\epsilon-\epsilon_\lambda)$ converges with probability 
1 to the Wigner semi-circular law in the thermodynamic limit $N\rightarrow\infty$. For a given single-particle energy $\epsilon$ within this distribution, one can consider the energy-resolved Green's function: 
\begin{equation}
\label{eq:G0_energyresolved}
    G(\epsilon,i\omega_n) = \frac{1}{i\omega_n+\mu-\epsilon}
\end{equation}
which allows to locate the position $\epsilon=\mu$ of the Fermi energy of this random but self-averaging model, 
and corresponding energy distribution of particles $\theta(\mu-\epsilon)$ at $T=0$. 

\subsection{Many-body density of states}
\label{sec:matrix2}

A quantity that will play an important role in our subsequent discussion of the SYK model is the {\it many-body} density of states, $\dos(E)$. 
Unlike the single particle density of states, $\rho(\omega)$, this is not an intensive quantity, but is typically exponentially large in $N$, because there is an exponentially large number of ways of making states within a small window of an energy $E \sim N$. 
In the grand canonical ensemble, we can relate the grand potential $\Omega (T)$ to $\dos(E)$ via an expression for the grand partition function
\beqn
Z = \exp \left(-\frac{\Omega(T)}{T} \right) = \int_{-\infty}^{\infty} dE ~\dos(E) e^{-E/T}. \label{rm4}
\eeqn
Note that we have absorbed a contribution $-\mu N \mathcal{Q}$ into the definition of the (grand) energy $E$, as is frequently done in Fermi liquid theory. So we can obtain $\dos(E)$ by an inverse Laplace transform of $\Omega(T)$. 

\begin{figure}[t]
\begin{center}
\includegraphics[width=3.3in]{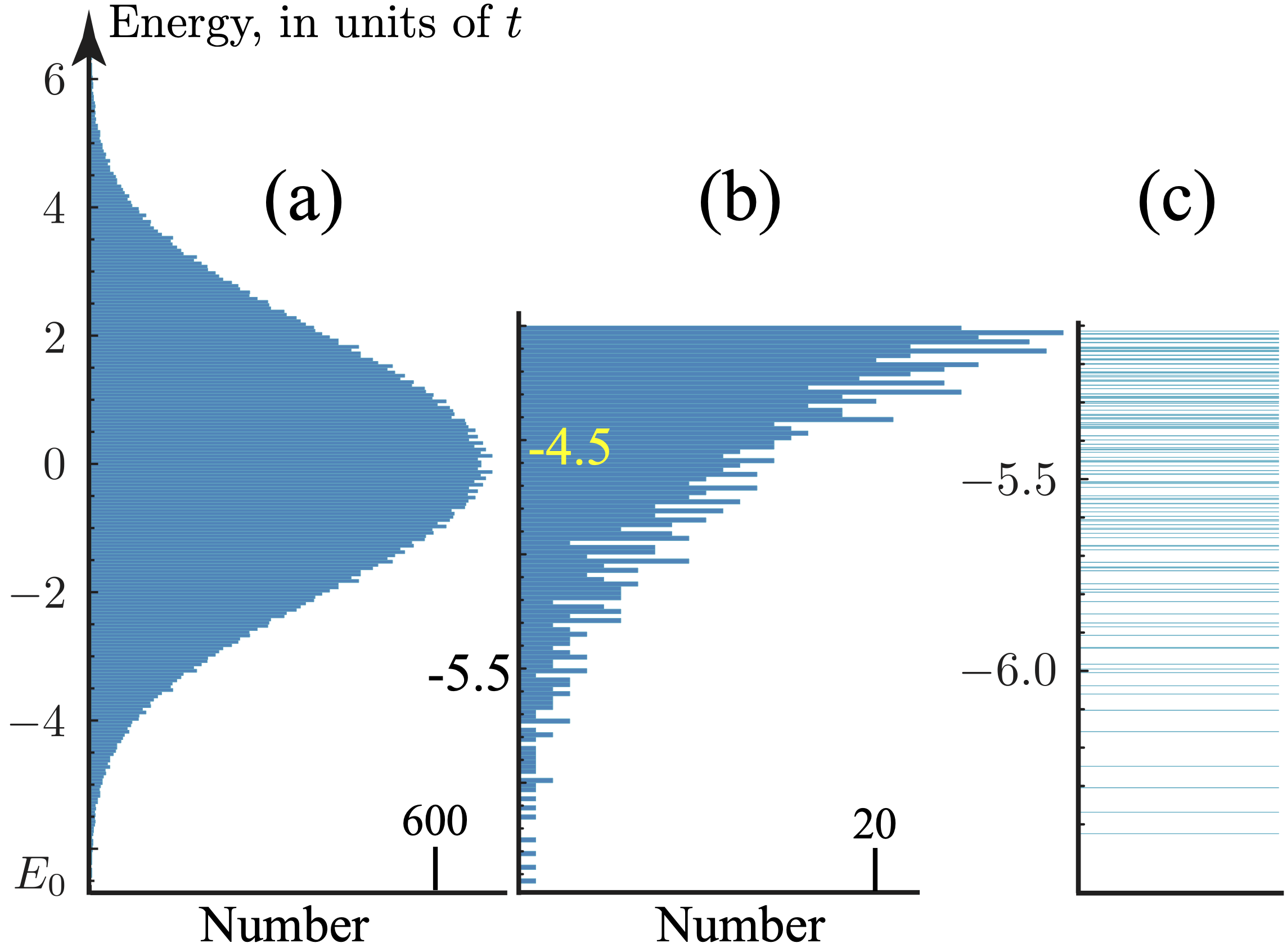}
\end{center}
\caption{65536 many-body eigenvalues of a $N=32$ Majorana matrix model with random $q=2$ fermion terms. $\mathcal{N}(E)$ is plotted in (a) and (b) in 200 and 100 bins, (b) and (c) zoom into the bottom of the band. Individual energy levels are shown in (c), and these are expected to have spacing $1/(N \rho(\mu))$ at the bottom of the band as $N \rightarrow \infty$. }
\label{fig:specq2}
\begin{center}
\includegraphics[width=3.3in]{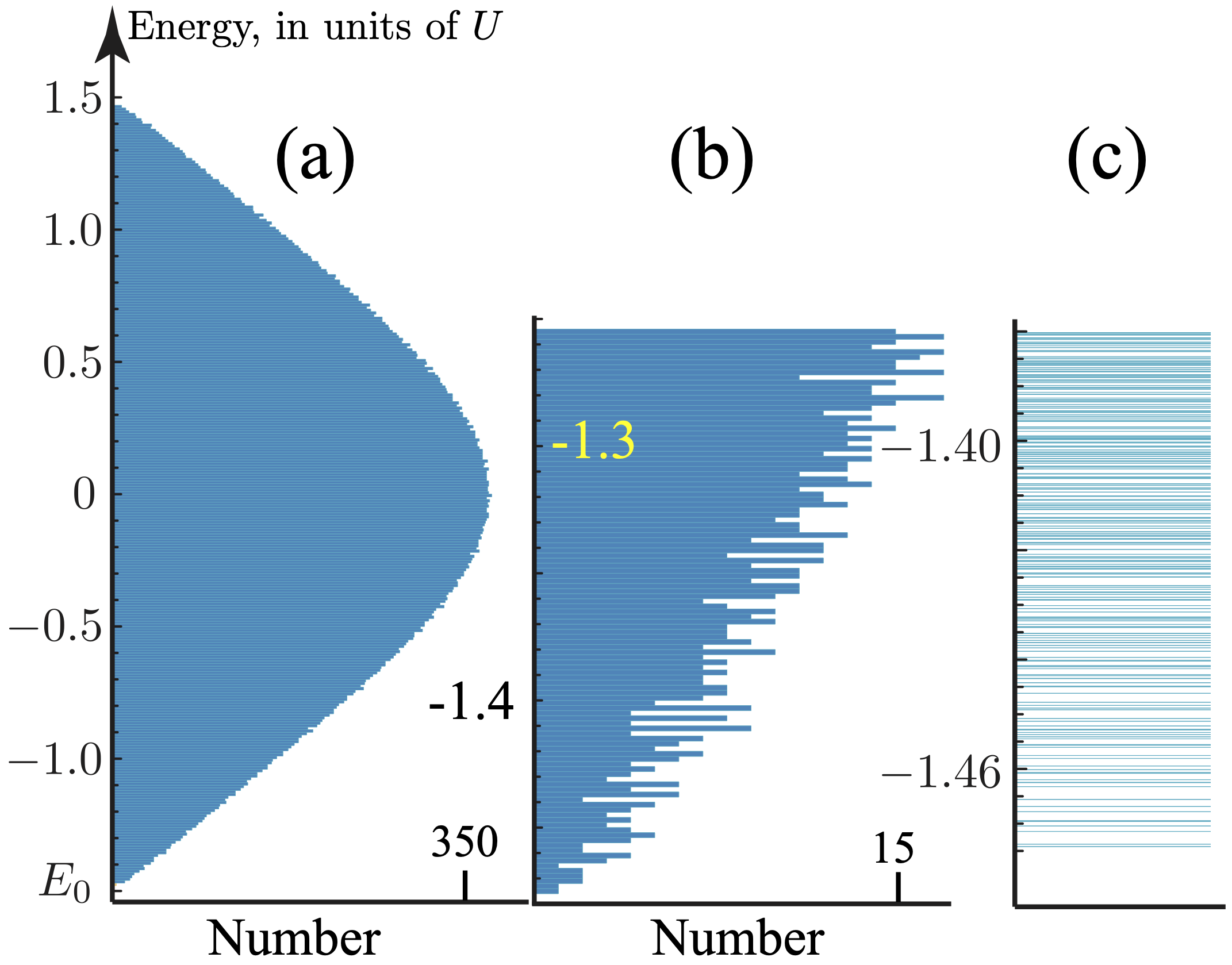}
\end{center}
\caption{65536 many-body eigenvalues of a $N=32$ Majorana SYK Hamiltonian with random $q=4$ fermion terms. $\mathcal{N}(E)$ is plotted in (a) and (b) in 200 and 100 bins, (b) and (c) zoom into the bottom of the band. Individual energy levels are shown in (c), and these are expected to have spacing $e^{- N \mathcal{S}}$ at the bottom of the band as $N \rightarrow \infty$. Compare to Fig.~\ref{fig:specq2} for the random matrix model, which has a much sparser spacing $\sim 1/N$ at the bottom of the band.}
\label{fig:specq4}
\end{figure}
First, let us evaluate $\Omega(T)$. By the standard Sommerfeld expansion for free fermions, we have
\bea
\Omega (T) &=& - N T \int_{-2t}^{2t} d \omega \rho(\omega)  \ln \left( 1 + e^{-(\omega-\mu)/T} \right) \nonumber \\
&=& N \int_{-2t}^{\mu} d \omega (\omega-\mu) \rho(\omega) - \frac{N\pi^2 T^2}{6} \rho (\mu) \nonumber \\ 
&\equiv& E_0 - \frac{N \pi^2 T^2}{6} \rho (\mu).
\label{rm5}
\eea
We now have to insert Eq. (\ref{rm5}) into Eq. (\ref{rm4}) and determine $\dos(E)$. Rather than perform the inverse Laplace transform, we make a guess of the form of $\dos(E)$. First, it is not difficult to see that $\dos(E<E_0) = 0$. Next, we expect $\dos(E)$ to be exponentially large in $N$ when $E-E_0 \sim N$. So we make a guess
\beqn
\dos(E)  \sim \exp \left( a N [(E-E_0)/N]^b \right) \quad, \quad E> E_0 \label{rm6}
\eeqn
for some constants $a$ and $b$. Then we insert Eq. (\ref{rm6}) into Eq. (\ref{rm4}), and perform the integral over $E$ by steepest descent method in the large $N$ limit. Matching the result to the left hand side of Eq. (\ref{rm4}), we obtain the main result of this section
\bea
\dos(E) &\sim& \exp\left(S(E)  \right) \label{rm7} \\
S(E) &=& \left\{ \begin{array}{ccc}
\pi \sqrt{2 N \rho (\mu) (E - E_0)/3}   &,&  E>E_0  \\
0 &,& E < E_0 \end{array} \right. , \nonumber
\eea
where $S(E)$ is the entropy as a function of the energy.
Consideration of the derivation shows that this result is valid for
\beqn
1 \ll \rho (\mu) (E-E_0) \ll N\,, \label{rm8}
\eeqn
in the limit of large $N$. Note that the entropy vanishes as $E \searrow E_0$ in Eq. (\ref{rm7}).
We show numerical results for $\dos(E)$ for a closely related random Majorana fermion model
in Fig.~\ref{fig:specq2}. When $E-E_0 \sim N$, the entropy $S(E)$ is extensive, the energy level spacing is exponentially small $\sim e^{-a N}$ with $a>0$, and $\dos(E) \sim e^{a N}$ is exponentially large. However, when $E - E_0 \sim 1/N$, we expect the energy levels to be few particle excitations with energies $\sim 1/(N \rho (\mu))$, and so $\dos (E) \sim N$. This rapid decrease of $\dos (E)$ near the bottom of the band is clearly evident in Fig.~\ref{fig:specq2}a from the `tails' in the density of states. 
A more complete analysis of the finite $N$ corrections is needed to understand the behavior of the $\dos (E)$ at low energy, along the lines of recent analyses \cite{Liao:2020lac,Liao:2021ofk}.

We also show in Fig.~\ref{fig:specq4} the corresponding results for the Majorana SYK model. These results will be discussed further in Section~\ref{sec:schwarzian}, but for now the reader should note
the striking absence of the tails in $\dos (E)$ in Fig.~\ref{fig:specq4}a in comparison to Fig.~\ref{fig:specq2}a.

There is an interesting interpretation of Eq. (\ref{rm7}) which gives us some insight into the structure of the random matrix eigenenergies, and also highlights a key characteristic of many body systems {\it with} quasiparticle excitations. It is known that the eigenvalues of a random matrix undergo level repulsion and their spacings obey Wigner-Dyson statistics \cite{Mehta}. For a zeroth order picture, let us assume that the random matrix eigenvalues are rigidly equally spaced, with energy level spacing (near the chemical potential) of $1/(N \rho(\mu))$. Now we ask for the number of ways to create a many body excitation with energy $E-E_0$. 
With the simplifying assumption that we made on the one-particle spectrum, each many-body eigenstate can be described 
by a unique set of particle-hole excitations, each one of them having an excitation energy which is an 
integer $n_i$ times the level spacing $1/(N \rho(\mu))$. 
This mapping is the essence of bosonization in one dimension, see e.g. \cite{giamarchi,QPT}. 
Hence the excitation energy reads:
%This many body excitation energy is the sum of particle-hole excitations, 
%each of which has an energy equal to an integer times the level spacing $1/(N \rho(\mu))$: 
\beqn
N \rho(\mu)(E- E_0) = n_1 + n_2 + n_3 + n_4 + \ldots
\eeqn
where the $n_i$ are the excitation numbers of the particle-hole excitations. %(this mapping is the essence of bosonization in one dimension).
So we estimate that the number of such excitations is equal to the number of partitions of the integer $N \rho(\mu) (E-E_0)$. Now we use the Hardy-Ramanujan result that the number of partitions of an integer $n$ is $p(n) \sim \exp(\pi \sqrt{2n/3})$ at large $n$. This immediately yields Eq. (\ref{rm7}). Note that the special case with exactly equally spaced quasiparticle levels (which is the case for the linearly-dispersing free Fermi gas in one dimensions) has many body levels with a spacing $\sim 1/N$ but an exponentially large degeneracy; in contrast, the generic random matrix case has no degeneracy but an exponentially small many-body level spacing.

This argument highlights a key feature of the many-body spectrum: it is just the sum of single particle excitation energies. 
We expect that if we add four fermion interactions to the random matrix model, we will obtain quasiparticle 
excitations in a Fermi liquid state whose energies add to give many-particle excitations. 
This can be checked for weak interactions by a perturbative calculation, in SYK models with random hopping \cite{Parcollet1,Balents}, and also holds non-perturbatively as 
shown by dynamical mean-field theory~\cite{DMFT}, which is exact for the random matrix Hubbard model with a local interaction.
Therefore, we expect the general form of Eq. (\ref{rm7}) to continue to hold even with interactions. 
However, we will see at the end of Section~\ref{sec:schwarzian} that such a decomposition into quasiparticle excitations does not hold for the SYK model. 

We can also estimate the lifetime of the quasiparticles at weak coupling by a perturbative computation based on Fermi's Golden Rule: 
we obtain $1/\tau \sim U^2 T^2/t^3$ at low $T$ with $U$ the strength of the local interaction. 
As this is parametrically smaller than a thermal excitation energy $\sim T$, quasiparticles remain well-defined excitations. 
The existence of such quasiparticles can be diagnosed from the poles of the energy-resolved Green's function to be presented in Eq.~(\ref{eq:G_energyresolved}), 
supplemented by the self-energy as defined in Sec.~\ref{sec:EDMFT} to account for interactions, while 
the energy integrated (local) Green's function Eq.~(\ref{GFsoln}) yields the disorder-averaged total density of states.

\section{The SYK model}
\label{sec:SYK}

As in the random matrix model, we consider electrons (assumed spinless for simplicity) which occupy sites labeled $i=1,2 \ldots N$. However, instead of a random one-particle hopping $t_{ij}$, we now have only a random two-particle interaction $U_{ij;k\ell}$:
\begin{subequations}
\begin{align}
&H_4 = \frac{1}{(2 N)^{3/2}} \sum_{ijk\ell=1}^N U_{ij;k\ell} \, c_i^\dagger c_j^\dagger c_k^{\vphantom \dagger} c_\ell^{\vphantom \dagger} 
-\mu \sum_{i} c_i^\dagger c_i^{\vphantom \dagger}  \label{syk1} \\
& ~~~~~~c_i c_j + c_j c_i = 0 \quad, \quad c_i^{\vphantom \dagger} c_j^\dagger + c_j^\dagger c_i^{\vphantom \dagger} = \delta_{ij}  \\
&~~~~~~~~~~~~~~~~~~~~\mathcal{Q} = \frac{1}{N} \sum_i \langle c_i^\dagger c_i^{\vphantom \dagger}\rangle. 
\end{align}
\end{subequations}
We choose the couplings $U_{ij;k\ell}$ to be {\it independent} random variables with zero mean $\overline{U_{ij;k\ell}} = 0$, while satisfying $U_{ij;k\ell} = -U_{ji;k\ell} = - U_{ij;\ell k} = U_{k\ell; ij}^\ast$. All the random variables have the same variance $\overline{|U_{ij;k\ell}|^2} = U^2$.

A model similar to $H_4$ appeared in nuclear physics,
where it was called the two-body random ensemble \cite{Bohigas71,French81}, and studied numerically. The existence and structure of the large $N$ limit was understood \cite{SY,Parcollet1,GPS1,GPS2} in the context of a closely related model that we will examine in Section~\ref{sec:rqm}. More recently, a Majorana version was introduced \cite{kitaev_talk}, and the large $N$ limit of $H_4$ was obtained \cite{SS15}.

\begin{figure}[h]
\begin{center}
\includegraphics[scale=0.2]{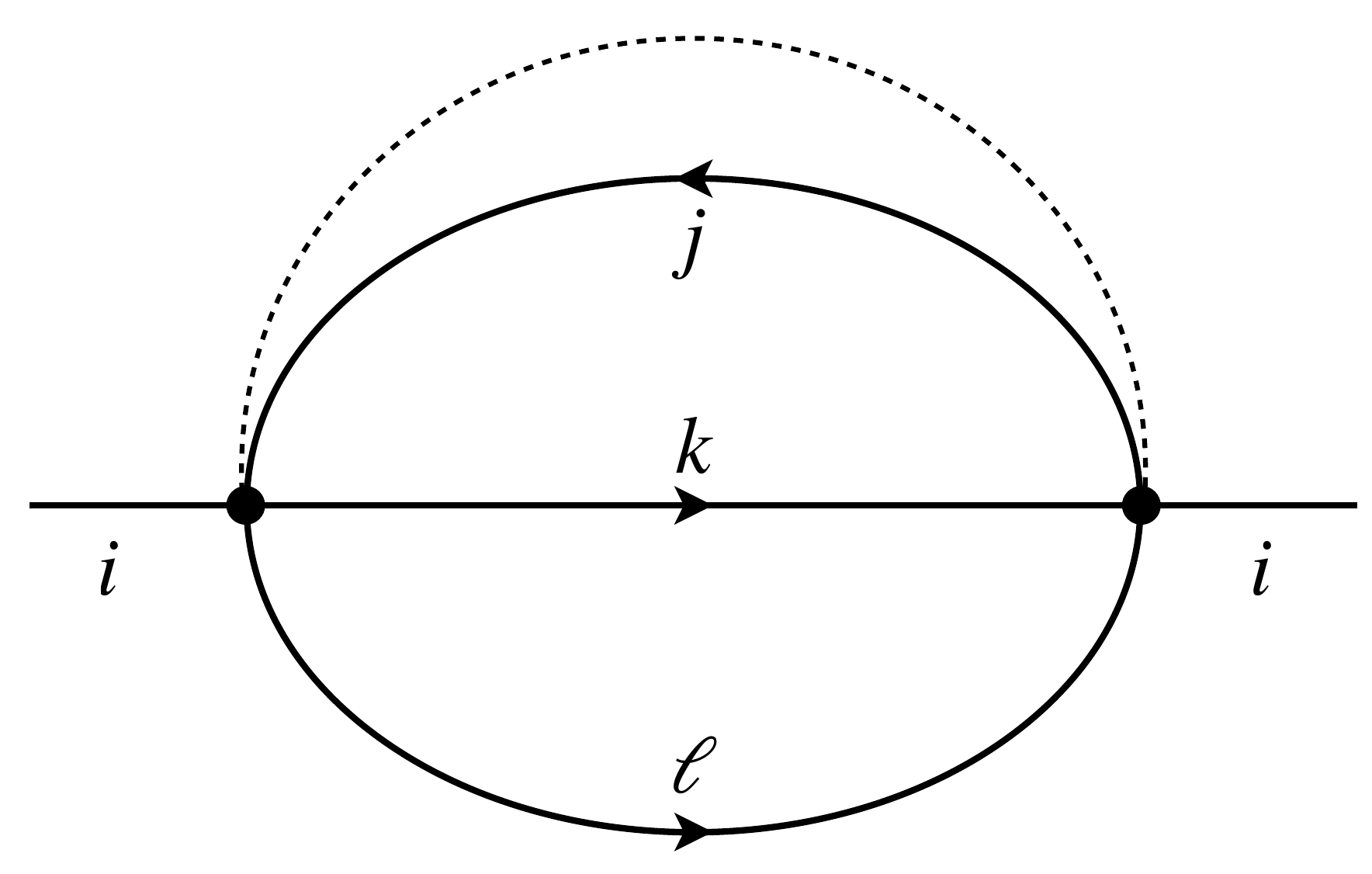}
\end{center}
\caption{The `melon graph' for the electron self-energy, $\Sigma (\tau)$, in Eq. (\ref{syk2b}). Solid lines denote fully dressed electron Green's functions. The dashed line represents the disorder averaging associated with the interaction vertices (denoted as solid circles), $\overline{|U_{ij;k\ell}|^2}$. }
\label{melon}
\end{figure}

The useful self-averaging properties of the random matrix model as $N \rightarrow \infty$ also apply to the SYK model Eq. (\ref{syk1}). Indeed, the self-averaging properties are {\it much\/} stronger, as the average takes place over the many-body Hilbert space of size $e^{\alpha N}$, rather than the single-particle Hilbert space of size $N$. Proceeding just as in the random matrix model, we perform a 
Feynman graph expansion in $U_{ij;k\ell}$, and then average graph-by-graph. In the large $N$ limit, only the so-called `melon graphs' survive (Fig. \ref{melon}), and the determination of the on-site Green's function reduces to the solution of the following equations
\begin{subequations}\label{eq:SYK}
\begin{align}
G(i\omega_n) & = \frac{1}{i \omega_n + \mu - \Sigma (i\omega_n)} \label{syk2a} \\ 
\Sigma (\tau) &= -  U^2 G^2 (\tau) G(-\tau) \label{syk2b} \\
G(\tau = 0^-) & = \mathcal{Q}. \label{syk2c}
\end{align}
\end{subequations}
Unlike the random matrix equations, these equations cannot be solved analytically as a result of their non-linearity, and a full solution can only be obtained numerically. However, it is possible to make significant analytic progress at frequencies and temperatures much smaller than $U$, as we shall describe in the following subsections.

Before embarking on a general low energy solution of Eq. (\ref{syk2a}-\ref{syk2c}), let us note a remarkable feature that can be deduced on general grounds \cite{SY}:
any non-trivial solution ({\it i.e.\/} with $\mathcal{Q} \neq 0,1$) must be gapless. Let us suppose otherwise, and assume there is a gapped solution with $\mbox{Im}\, G(\omega) = 0$ for $|\omega| < E_G$. Then, by an examination of the spectral decomposition of the equation for the self energy in Eq. (\ref{syk2b}), we can establish that $\mbox{Im}\, \Sigma(\omega) = 0$ for $|\omega| < 3E_G$. Inserting this back into Dyson's equation Eq. (\ref{syk2a}), we obtain the contradictory result that $\mbox{Im}\, G(\omega) = 0$ for $|\omega| < 3E_G$. So the only possible value is $E_G = 0$. 

\subsection{Low energy solution at $T=0$}
\label{sec:sykT0}

Knowing that the solution must be gapless, let us assume that we have a power-law singularity at zero frequency. So we assume \cite{SY}
\beqn
G(z) = C \frac{e^{- i (\pi \Delta +  \theta)}}{z^{1-2 \Delta}} \quad , \quad \mbox{Im} (z) > 0,~ |z| \ll U\,. \label{syk3}
\eeqn
We have a prefactor $C>0$, a power-law singularity determined by the exponent $\Delta > 0$, and a spectral asymmetry angle $\theta$ which yields distinct density of states for particle and hole excitations. We now have to insert the ansatz Eq. (\ref{syk3}) into Eqs.~(\ref{syk2a},\ref{syk2b}) and find the values of $C$, $\Delta$ and $\theta$ for which there 
is a self-consistent solution. Of course, the solution also has to satisfy the constraint arising from the spectral representation, $\mbox{Im} \, G(\omega +i 0^+) < 0$; for Eq.~(\ref{syk3}) this translates to 
\beqn
-\pi \Delta < \theta < \pi \Delta\,. \label{syk4}
\eeqn

We now wish to obtain the Green's function as a function of imaginary time $\tau$. For this purpose, we write the spectral representation using the density of states $\rho (\Omega)  = - (1/\pi) \mbox{Im} \, G(\omega +i 0^+) > 0$, so that
\beqn
G(z) = \int_{-\infty}^{\infty} d \Omega \frac{\rho (\Omega)}{z - \Omega}. \label{syk5}
\eeqn
We can take a Fourier transform and obtain
\bea 
G(\tau) = \left\{
\begin{array}{ccc}
\displaystyle - \int_0^\infty d \Omega ~\rho (\Omega) e^{- \Omega \tau} &,& \mbox{for $\tau > 0$ } \\[1em] 
\displaystyle\int_0^\infty d \Omega ~\rho (-\Omega) e^{ \Omega \tau} &,& \mbox{for $\tau < 0$ }
\end{array} \right. . \label{syk6}
\eea
Using Eq.~(\ref{syk6}) we obtain in $\tau$ space
\bea 
G (\tau) = \left\{
\begin{array}{ccc}
\displaystyle - \frac{C \Gamma (2 \Delta) \sin (\pi \Delta + \theta)}{\pi |\tau|^{2 \Delta}} &,& \mbox{for $\tau \gg 1/U$ } \\[1em] 
\displaystyle \frac{C \Gamma (2 \Delta) \sin (\pi \Delta - \theta)}{\pi |\tau|^{2 \Delta}} &,& \mbox{for $\tau \ll -1/U$ }
\end{array} \right. . \label{syk7}
\eea
corresponding to the low-frequency behaviour of the spectral function: 
\bea 
\rho(\Omega) = \left\{
\begin{array}{ccc}
\displaystyle \sin(\pi \Delta + \theta)\frac{C}{\pi|\Omega|^{1-2\Delta}} 
&,& \mbox{for $0<\Omega\ll U$ } \\[1em] 
\displaystyle \sin(\pi \Delta - \theta)\frac{C}{\pi|\Omega|^{1-2\Delta}} 
&,& \mbox{for $-U\ll\Omega<0$ }
\end{array} \right. . \label{syk7spectral}
\eea
This expression makes it clear that $\theta$ determines the particle-hole asymmetry, associated with the fermion propagation 
forward and backward in time (positive/negative frequencies). 
For our later purpose, it is also useful to parametrize the asymmetry in terms 
of a real number $- \infty < \mathcal{E} < \infty$ so that
\bea
G (\tau) \sim \left\{
\begin{array}{ccc}
\displaystyle - \frac{e^{\pi \mathcal{E}}}{|\tau|^{2 \Delta}} &,& \mbox{for $\tau \gg 1/U $ } \\[1em] 
\displaystyle \frac{e^{-\pi \mathcal{E}}}{|\tau|^{2 \Delta}} &,& \mbox{for $\tau \ll - 1/U$ }
\end{array} \right. , \label{syk7a}
\eea
and then we have
\beqn
e^{2 \pi \mathcal{E}} = \frac{ \sin (\pi \Delta + \theta)}{\sin (\pi \Delta - \theta)}\,, \label{syk7b}
\eeqn
and $\mathcal{E} = \theta = 0$ is the particle-hole symmetric case. 
This spectral asymmetry plays a key role in the physics of the complex SYK model, 
as well as in the large-$M$ solution of multichannel Kondo models \cite{ParcolletKondo}, where the notation $\alpha=2\pi\mathcal{E}$ was used. 

We also use the spectral representation for the self energy 
\beqn
\Sigma (z) = \int_{-\infty}^{\infty} d \Omega \frac{\sigma (\Omega)}{z - \Omega}. \label{syk8}
\eeqn
Using Eqs.~(\ref{syk2b}) and (\ref{syk7}) to obtain $\Sigma (\tau)$, and performing the inverse Laplace transform as for $G(\tau)$, we obtain 
\bea 
\sigma (\Omega) = \left\{
\begin{array}{c}
\displaystyle \Upsilon(\Delta)
\left[\sin(\pi \Delta + \theta) \right]^{2} \left[\sin(\pi \Delta - \theta) \right]
|\Omega|^{6 \Delta - 1}  \\
~~~~~~~~~~~~~~ \mbox{for $\Omega > 0$} \\[1em] 
\displaystyle \Upsilon(\Delta)
\left[\sin(\pi \Delta + \theta) \right] \left[\sin(\pi \Delta - \theta) \right]^{2}
|\Omega|^{6 \Delta - 1} \\
~~~~~~~~~~~~~~ \mbox{for $\Omega < 0$}
\end{array} \right.  \label{syk9}
\eea
where $\Upsilon(\Delta)=[{U^2}/{\Gamma (6 \Delta)}] [{C \Gamma (2 \Delta)}/{\pi}]^{3} $.
Finally we have to insert the $\Sigma (i \omega_n)$ obtained from Eqs.~(\ref{syk8}) and (\ref{syk9}) back into Eq.~(\ref{syk2a}). 
To understand the structure of the solution, let us first assume that $0 < 6 \Delta -1 < 1$; we will find soon that this is indeed the case, and no other solution is possible.  Then as $|\omega_n| \rightarrow 0$, the frequency dependence in $\Sigma(i\omega_n)$ is much larger than that from the $i \omega_n$ term in Eq. (\ref{syk2a}). Also, we have $1-2 \Delta > 0$, and so $G(z)$ in Eq. (\ref{syk3}) diverges as $|z| \rightarrow 0$. So we find that a solution of Eq. (\ref{syk2a}) is only possible under two conditions:
\bea
\mu - \Sigma (0) &=& 0 \, , ~~\mbox{and} \nonumber \\
1 - 2 \Delta &=& 6 \Delta - 1 \quad \Rightarrow \quad \Delta = \frac{1}{4}\,. \label{syk10}
\eea
Matching the divergence in the coefficient of $G(z)$ as $z \rightarrow 0$, we also obtain the value of $C$:
\beqn
C = \left(\frac{\pi}{U^2 \cos(2 \theta)} \right)^{1/4}\,. \label{syk11}
\eeqn

The value of asymmetry angle, $\theta$, remains undetermined by the solution Eqs. (\ref{syk2a}) and (\ref{syk2b}). As we will see in Section~\ref{sec:syklutt}, the value of $\theta$ is fixed by a generalized Luttinger's theorem, which relates it to the value of the fermion density $\mathcal{Q}$~\cite{GPS2}. 
But without further computation we can conclude that at the particle-hole symmetric point with $\mathcal{Q} = 1/2$, we have $\mathcal{E} = \theta = 0$. 

The main result of this section is therefore summarized in Eq. (\ref{syk7a}). The fermion has `dimension' $\Delta =1/4$ and its two-point correlator decays as $1/\sqrt{\tau}$; there is 
a particle-hole asymmetry determined by $\mathcal{E}$ 
(which is unknown at this stage, but determined in the next section). 
This should be contrasted with the corresponding features of the random matrix model with a Fermi liquid ground state: the two-point fermion correlator decays as $1/\tau$, and the leading decay is particle-hole symmetric.

\subsection{Luttinger's theorem}
\label{sec:syklutt}

In Fermi liquid theory, Luttinger's theorem relates an equal time property --- the total electron density--- to a low energy property, the Fermi wavevector which is the location of zero energy excitations. There turns out to be a similar low-to-high energy mapping that can be made in a `generalized' Luttinger theorem for the SYK model, relating the angle $\theta$ characterizing the particle-hole asymmetry at long times in Eq. (\ref{syk3}), to the fermion density $\mathcal{Q}$ \cite{GPS2}. As in the conventional Luttinger analysis, we start by manipulating the expression for $\mathcal{Q}$ into 2 terms
\bea
\mathcal{Q}-1 &=& \int_{-\infty}^{\infty} \frac{d \omega}{2 \pi} G(i \omega) e^{-i \omega 0^+} = I_1 + I_2, \nn
I_1 &=& i \int_{-\infty}^{\infty} \frac{d \omega}{2 \pi} \frac{d}{d \omega} \ln \left[ G(i \omega) \right] e^{-i \omega 0^+} \nn
I_2 &=& -i \int_{-\infty}^{\infty} \frac{d \omega}{2 \pi} G(i \omega) \frac{d}{d \omega} \Sigma(i \omega)  e^{-i \omega 0^+}. \label{sykl1}
\eea
In Fermi liquid theory, $I_2$ vanishes because of the existence of the Luttinger-Ward functional~\cite{Luttinger_Ward_1960,AGD}, 
while $I_1$ is easily evaluated because it is a total derivative, and this yields the Luttinger theorem. 
The situation is more complicated for the SYK model because of the singular nature of $G(\omega)$ as $|\omega| \rightarrow 0$. Indeed, both $I_1$ and $I_2$ are logarithmically divergent at small $|\omega|$, although, naturally, their sum is well defined. Nevertheless the separation of $\mathcal{Q}$ into $I_1$ and $I_2$ is useful, because it allows us to use the special properties of the Luttinger-Ward functional to account for the unknown high frequency behavior of the Green's function. We shall define $I_{1,2}$ by a regularization procedure, and it is then important that the same regularization be used for both $I_1$ and $I_2$. We employ the symmetric principal value, with 
\beq
 \int_{-\infty}^{\infty} d \omega \Rightarrow
\lim_{\eta \rightarrow 0} \left[ \int_{-\infty}^{-\eta} d \omega + \int_{\eta}^{\infty} d \omega \right]. \label{sykl2}
\eeq

Now we evaluate $I_1$ using the usual procedure: we distort the contour of integration to the real frequency axis 
and have to evaluate:
\bea
%I_1 &=& 
&i& \lim_{\eta \rightarrow 0} \int_{0}^{\infty} \frac{d \omega}{2 \pi} \frac{d}{d \omega} \ln \left[\frac{G(\omega + i \eta)}{G (\omega - i \eta)} \right] \nn
&=&- \frac{1}{\pi} \lim_{\eta \rightarrow 0} \left[ 
\mbox{arg} \, G(\infty + i \eta) -  \mbox{arg} \, G(i \eta) \right]\,. \label{sykl3}
\eea
In a Fermi liquid, this is the only contribution to $I_1$, which
evaluates to unity outside the Fermi surface, and vanishes inside the Fermi surface. 
In the present case however, the imaginary frequency integral (\ref{sykl2}) differs from the real frequency integral (\ref{sykl3}) because of the singularity at $\omega=0$, for 
which a small contour encircling the origin must be introduced, finally leading via
%In the present case, 
Eq. (\ref{syk3}) to:
\beq
I_1 = - \frac{1}{2} - \frac{\theta}{\pi}. \label{sykl4}
\eeq
Note that this yields a contribution $1/2-\theta/\pi$ to $\mathcal{Q}$ which does obey 
$\mathcal{Q}\rightarrow 1-\mathcal{Q}$ under $\theta\rightarrow -\theta$ as dictated by particle-hole symmetry, but does 
not have the expected limits $\mathcal{Q}\rightarrow 0,1$ as $\theta\rightarrow\pm\pi/4$. This is already a hint that 
$I_2$ must yield a non-zero contribution.

In the evaluation of $I_2$ we must substitute the expression Eq. (\ref{syk2b}) for $\Sigma$ into $I_2$, because then we ensure cancellations at high frequencies arising from the existence of the Luttinger-Ward functional:
\beq
\Phi_{LW}[G] = - \frac{U^2}{4} \int d\tau \, G^2 (\tau) G^2(-\tau)\,. \label{sykl5}
\eeq
Using $\Sigma = \delta \Phi_{LW}/\delta G$, and ignoring the singularity at $\omega=0$, we obtain, as in Fermi liquid theory, $I_2 = -i \int_{-\infty}^{\infty} d \omega (d/d \omega) \Phi_{LW} = 0$. So the entire contribution to $I_2$ arises from the regularization of singularity near $\omega =0$. We can therefore evaluate $I_2$ by using Eq. (\ref{syk2b}) for $\Sigma$, the regularization in Eq. (\ref{sykl2}), and the low frequency spectral density in Eq. (\ref{syk9}), and ignore the high frequency contribution to $I_2$. 
The explicit evaluation of the integral is somewhat involved~\cite{GPS2,GKST}. The 
result can be guessed however from a heuristic argument~\cite{GPS2}, which can also be 
generalized to the SYK model with $q$-fermion interactions~\cite{SS17}. 
The low-energy contribution to $I_2$ involves a product of $G$ and $\Sigma$ and must be a homogeneous polynomial of degree $4$ in the two coefficients that 
enter the low-energy behaviour of $G$, Eq.~(\ref{syk7spectral}). Using particle-hole 
symmetry, and imposing the absence of singularity as $\theta\rightarrow\pm\pi/4$, it is seen that only the combination 
$C^4\left[\sin^3(\frac{\pi}{4}+\theta)\sin(\frac{\pi}{4}-\theta)
-\sin^3(\frac{\pi}{4}-\theta) \sin(\frac{\pi}{4}+\theta)\right]
\propto \sin 2\theta$ is allowed. The proportionality coefficient is fixed by imposing that 
$\mathcal{Q}=+1$ for $\theta=-\pi/4$ in Eq.~(\ref{sykl1}). Indeed, the explicit evaluation yields:
\beq
I_2 = - \frac{\sin (2 \theta)}{4}\,. \label{sykl6}
\eeq
Combining Eqs. (\ref{sykl1},\ref{sykl4},\ref{sykl6}), we obtain our generalized 
Luttinger theorem \cite{GPS2,SS17,GKST},
\beqn
\mathcal{Q} = \frac{1}{2} - \frac{\theta}{\pi} -  \frac{\sin(2 \theta)}{4} \,. \label{sykl7}
\eeqn
This expression evaluates to the limiting values $\mathcal{Q}=1,0$ for the limiting values of $\theta = -\pi/4,\pi/4$ in Eq. (\ref{syk4}), and decreases monotonically in between; $\mathcal{Q}$ is also a monotonically decreasing function between these limits of $-\infty<\mathcal{E}<\infty$, via Eq. (\ref{syk7b}). 

All our results have so far been obtained by an analytic analysis of the low energy behavior. A numerical analysis is needed to ensure that such low energy solutions have high energy continuations which also obey Eqs.~(\ref{syk2a},\ref{syk2b}). Such analyses show that complete solutions only exist for a range of values around $\mathcal{Q}=1/2$ \cite{Azeyanagi:2017drg}; for values of $\mathcal{Q}$ close to 0,1 there is phase separation into the trivial $\mathcal{Q}=0,1$ state, and densities closer to half-filling. However, this conclusion is only for the specific microscopic Hamiltonian in Eq. (\ref{syk1}): other Hamiltonians, with additional $q$-fermion terms (see Appendix \ref{app:SYKq}), with $q>4$, could have solutions with the same low energy behavior described so far for a wider range of $\mathcal{Q}$, because these higher $q$ terms are irrelevant at low energy.

\subsection{Non-zero temperatures}
\label{sec:SYKT}

It turns out to be possible to extend the solutions for $T=0$ Green's functions obtained so far to non-zero $T \ll U$ by employing a subtle argument involving conformal invariance. 
However, let us first take a simple minded approach to look for a solution directly from 
Eq. (\ref{syk2a}) and Eq. (\ref{syk2b}), and show that we can {\it guess\/} a solution.

We initially limit consideration to the particle-hole symmetric case with $\mathcal{Q} = 1/2$ and $\theta=0$.
We use the similarity to multichannel Kondo problems \cite{ParcolletKondo}, 
to generalize the $\tau$ dependence of the Green's function in Eq. (\ref{syk7}) to \cite{Parcollet1} 
\beqn
G(\tau) = B\,  \mbox{sgn}(\tau) \left| \frac{\pi T}{\sin (\pi T \tau)} \right|^{1/2}\,, \quad T, |\tau|^{-1} \ll U\,, \label{syk12}
\eeqn
where $B$ is some $T$-independent constant. 
Making contact with the notations of Sec.~\ref{sec:sykT0}, 
we have $-B=C\Gamma(1/2)\,\sin (\pi/4)/\pi=C/\sqrt{2\pi}$, with 
$C^4=\pi/U^2$ for this case with $\theta=0$.
Note that Eq. (\ref{syk12}) reduces to Eq. (\ref{syk7}) for $1/U \ll |\tau| \ll 1/T$.
Then the self-energy is
\begin{displaymath}
\Sigma (\tau) = U^2 B^3 \mbox{sgn} (\tau) \left| \frac{\pi T}{\sin (\pi T \tau)} \right|^{3/2}\,, \quad T, |\tau|^{-1} \ll U\,.
\end{displaymath}
Taking Fourier transforms, we have as a function of the Matsubara frequency $\omega_n$
\begin{subequations}
\begin{eqnarray}
G (i\omega_n) &=& \left[ i  B  \right]  \frac{ T^{-1/2} \,
 \Gamma \left( \displaystyle  \frac{1}{4}  + \frac{\omega_n}{2 \pi T} \right)}{
 \Gamma \left( \displaystyle \frac{3}{4} + \frac{\omega_n}{2 \pi T} \right)  }, \label{syk12pa}
\\
\Sigma_{\rm sing} (i\omega_n ) &=& \left[ i  4 \pi U^2 B^3\right]  \frac{  T^{1/2} \,
 \Gamma \left( \displaystyle  \frac{3}{4}  + \frac{\omega_n}{2 \pi T} \right)}{
 \Gamma \left( \displaystyle \frac{1}{4} + \frac{\omega_n}{2 \pi T} \right)  } \,,~~~~~~~~~~\label{syk12pb}
\end{eqnarray}
\end{subequations}
where we have subtracted $\Sigma (\omega=0,T=0)$ in $\Sigma_{\rm sing} (i\omega_n)$.
Now the singular part of Dyson's equation is 
\beqn
G(i\omega_n) \Sigma_{\rm sing} (i\omega_n) = -1. \label{syk13}
\eeqn
Remarkably, the $\Gamma$ functions in Eqs. (\ref{syk12pa}) and (\ref{syk12pb}) appear with just the right arguments, 
so that they can obey Eq. (\ref{syk13}) for all $\omega_n$, 
and the amplitude indeed obeys $4\pi U^2B^4=1$.

\begin{figure}
\begin{center}
\includegraphics[height=6cm]{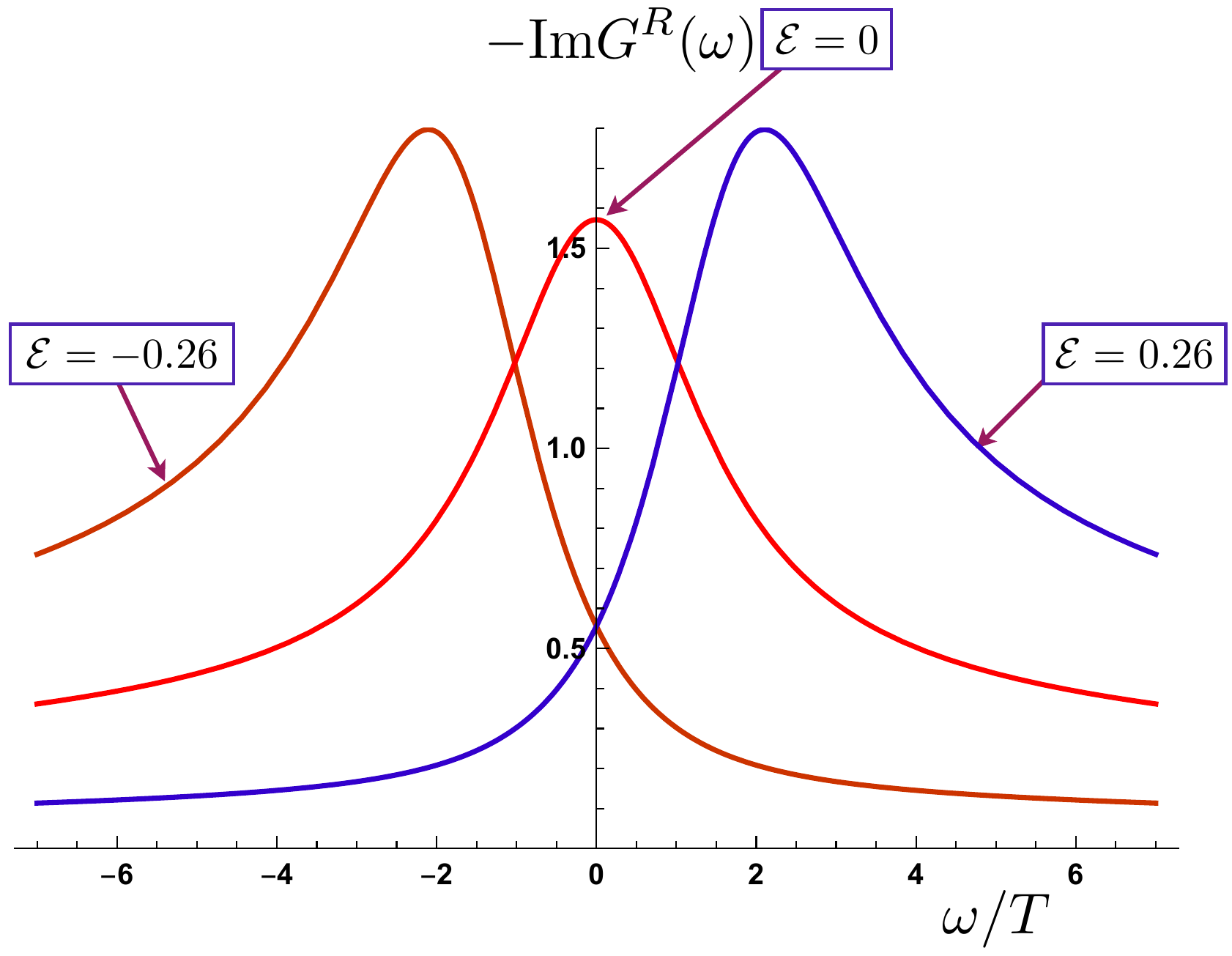}
\end{center}
\caption{Plot of the electron spectral density in the SYK model, obtained from imaginary part of Eq. (\ref{syk18}). }
\label{fig:specsyk}
\end{figure}

A deeper understanding of the origin of Eq. (\ref{syk12}), and its generalization to the particle-hole asymmetric case, can be obtained by analyzing the low energy limit of the original saddle point equations Eqs.~(\ref{syk2a}) and (\ref{syk2b}). These equations are characterized by a remarkably large set of emergent symmetries, which we describe in Appendix \ref{sec:timerepar}. The final result for the Green's function in imaginary time away from the particle-hole symmetric point is
\beqn
G(\tau) =  - C \frac{e^{-2 \pi \mathcal{E} T \tau}}{\sqrt{1 + e^{-4 \pi \mathcal{E}}}} \left( \frac{T}{\sin (\pi T \tau)} \right)^{1/2} \,.
 \label{syk17}
\eeqn
for $0< \tau < \frac{1}{T}$.
This can be extended to all real $\tau$ using the KMS condition Eq.~(\ref{KMS}). Performing a Fourier transform, and analytically continuing to real frequencies leads to the Green's function \cite{Parcollet1,SS15}
\beqn
G (\omega+i0^+) = \frac{-i C e^{-i \theta}}{(2 \pi T)^{1/2}}
\frac{\Gamma \left( \displaystyle \frac{1}{4} + i \mathcal{E} - \frac{i \omega }{2 \pi  T}  \right)}
{\Gamma \left(  \displaystyle \frac{3}{4} + i \mathcal{E} - \frac{i  \omega }{2 \pi  T} \right)}. \label{syk18}
\eeqn
We show a plot of the imaginary part of the Green's function in Fig.~\ref{fig:specsyk}.

For later comparison with other models, let us note that these results imply that the singular part of the electron self energy in Eq. (\ref{syk12pb}) obeys the scaling form
\beq
\Sigma (\omega, T) = U^{1/2} T^{1/2} \Phi \left( \frac{\hbar\omega}{k_B T} \right)
\label{syk18a}
\eeq
where $\Phi$ is a universal scaling function with a known dependence on the particle-hole asymmetry parameter $\mathcal{E}$. 

The universal dependence of the self energy on the Planckian ratio, $\hbar\omega/(k_B T)$, implies the absence of electronic quasiparticles \cite{QPT}: the characteristic lifetime of the excitations $\sim \hbar/(k_B T)$ is of the same order as their energy $\sim \hbar \omega$, and so quasiparticles are not well defined. This behaviour is very different from the random matrix model 
studied in Section~\ref{sec:matrix2} where the self-energy was negligible at low $T$.

A Planckian lifetime has also been obtained by non-equilibrium studies of SYK models, which display a recovery of thermal Green's functions in a time that is independent of $U$, and proportional to the inverse final temperature \cite{Eberlein:2017wah,Jatkar,Almheiri:2019jqq,Zhang:2019fcy,Dhar:2018pii,Kourkoulou:2017zaj,Lensky:2020fqf,Rossini:2019nfu,Samui:2020jli}; for closely related and complementary insights, see also \cite{Sonner2017,Cheipesh20,Banerjee20,Knap20,Haque19,Schiro21,Bandyopadhyay:2021wpy}. 

\subsection{Computation of the $T \rightarrow 0$ entropy}
\label{sec:SYKentropy}

We have now presented detailed information on the nature of the Green's function of the SYK model at low $T$. We will proceed next to use this information to compute some key features of the low $T$ thermodynamics. 

First, we establish some properties of the behavior of the chemical potential, $\mu$, as $T \rightarrow 0$ at fixed $\mathcal{Q}$. 
Recall that for the random matrix model, and more generally for any Fermi liquid, there was a $\sim T^2$ correction to the chemical potential, which depended upon the derivative of the density of single particle states.
For the SYK model, the leading correction is much stronger; the correction $\sim T$, which is universally related to parameters in the Green's function \cite{GPS2}.

A simple way to determine the linear $T$ dependence of $\mu$ is to examine the particle-hole asymmetry of the Green's function at $T>0$. From Eqs.~(\ref{syk7a}), (\ref{syk17})
this is given by the ratio
\beq
\lim_{T \rightarrow 0} \frac{G ( \tau )}{G (1/T - \tau)} =  e^{2 \pi \mathcal{E}}\,, \label{syk21a}
\eeq
where the limit is taken at a fixed $\tau \gg 1/U$.
We now use a crude picture of the low energy physics, and imagine that all the low-energy degrees of freedom are essentially at zero energy, compared to $U$. So we compare Eq.~(\ref{syk21a}) with the corresponding ratio for a zero energy fermion whose chemical potential has been shifted by $\delta \mu$
\beq
\frac{G_0 ( \tau )}{G_0 (1/T - \tau)} =  e^{-(\delta\mu)(1/T-2\tau)}\,.
\eeq
From this comparison, we conclude that there is a linear-in-$T$ dependence of the chemical potential that keeps the particle-hole asymmetry fixed as $T \rightarrow 0$:
\bea
\mu - \mu_0 &=& \delta \mu =  - 2 \pi \mathcal{E} T \nonumber \\
&+& \mbox{terms vanishing as $T^p$ with $p>1$}\,, \label{syk22}
\eea
with $\mu_0$ a non-universal constant. Note that the density of the zero energy fermion $= 1/(e^{-\delta\mu/T} + 1)$ remains fixed
as $T \rightarrow 0$, and so Eq.~(\ref{syk22}) applies at fixed $\mathcal{Q}$.

A more formal analysis \cite{ParcolletKondo,GPS2,SS15}, leading to the same result for the $T$ dependence of $\mu$,
relates the long-time conformal Green's function (valid for $\tau \gg 1/U$) to its short-time behavior. In particular at $|\omega_n| \gg U$ we have
\beqn
G(i \omega_n) = \frac{1}{i \omega_n} - \frac{\mu}{(i \omega_n)^2} + \ldots
\eeqn
which implies for the spectral density of the Green's function, $\rho (\Omega)$
\beqn
\mu = -\int_{-\infty}^{\infty} d \Omega \, \Omega \rho (\Omega),
\eeqn
which makes it evident that $\mu$ depends only upon the particle-hole asymmetric part of the spectral density.
Next, using the spectral relations we can relate the $\Omega$ integrals to the derivative of the imaginary time correlator
\beqn
\mu = - \partial_\tau G (\tau = 0^+) - \partial_\tau G(\tau = (1/T)^-).
\eeqn
We pull out an explicitly particle-hole asymmetric part of $G(\tau)$ by defining
\beqn
G(\tau) \equiv e^{-2 \pi \mathcal{E} T \tau} G_c(\tau) \quad,\quad 0 < \tau < \frac{1}{T}.
\eeqn
where $G_c$ will be given by a particle-hole symmetric conformal form at low $T$ and low $\omega$. Then we obtain
\begin{eqnarray}
\mu &=& 2 \pi \mathcal{E} T \left[ G (\tau = 0^+) + G(\tau = (1/T)^-) \right]  \nonumber \\
&~&~~~~~~~~~~+ \mbox{terms dependent on $G_c$} \nonumber\\
&=& - 2 \pi \mathcal{E} T + \mbox{terms dependent on $G_c$} \nonumber
\end{eqnarray}
It can be shown that all the terms dependent upon $G_c$ have a $T$ dependence that is weaker than linear in $T$ provided $\mathcal{Q}$ is held fixed. Hence, we obtain Eq.~(\ref{syk22}). 

\begin{figure}[t]
\centering
%\begin{center}
\includegraphics[width=0.9\columnwidth]{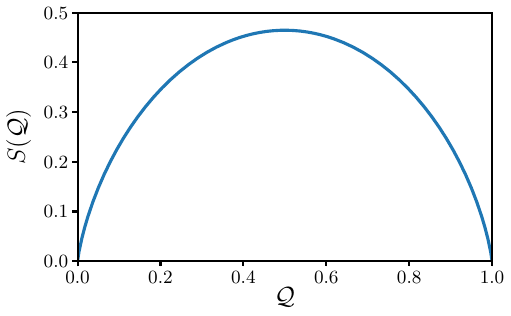}
%\end{center}
\caption{$T=0$ entropy density $\mathcal{S}$ vs. $\mathcal{Q}$ \cite{GPS2}. }
\label{fig:entropy}
\end{figure}

Now we can deduce the $T$ dependence of the entropy by the Maxwell relation
\beqn
\left( \frac{\partial \mu}{\partial T} \right)_{\mathcal{Q}} = - \frac{1}{N}\left( \frac{\partial S}{\partial \mathcal{Q}} \right)_{T}, \, \label{syk22a}
\eeqn
where the $1/N$ is needed because we define $S$ to be the total extensive entropy, and so we must use the total number $N \mathcal{Q}$ in the Maxwell relation.
Applying this to Eq. (\ref{syk22}) we obtain
\beqn
\frac{1}{N}\left( \frac{\partial S}{\partial \mathcal{Q}} \right)_{T} = 2 \pi \mathcal{E} \neq 0 ~~\mbox{as $T \rightarrow 0$}. \label{syk23}
\eeqn
In Section~\ref{sec:syklutt}, we obtained an `extended' Luttinger relationship between the density $\mathcal{Q}$ and the particle-hole asymmetry parameter $\mathcal{E}$. Assuming that $S=0$ at $\mathcal{Q}=0$, we can now integrate Eq. (\ref{syk23}) to obtain
for the entropy $S$ \cite{GPS2}
\beqn
S(T \rightarrow 0) = N \mathcal{S}  \quad,  \quad \mathcal{S} = 2 \pi \int_0^{\mathcal{Q}} d \tilde{\mathcal{Q}} \mathcal{E} ( \tilde{\mathcal{Q}} ). \label{syk24}
\eeqn
which can be rewritten, using Eqs.~(\ref{syk7b}) and (\ref{sykl7}) in the following parametric form:
\begin{eqnarray}
\mathcal{S}(\Theta)&=&\int_{-\pi/4}^\Theta d\theta\, \ln\frac{\sin (\pi/4+\theta)}{\sin(\pi/4-\theta)}\,\frac{\partial\mathcal{Q}}{\partial\theta} \\ 
\mathcal{Q}(\Theta)&=&\frac{1}{2}-\frac{\Theta}{\pi}-\frac{\sin 2\Theta}{4}
\end{eqnarray}
Fig.~\ref{fig:entropy} displays the entropy density vs. $\mathcal{Q}$ obtained from this expression.

The remarkable feature of this result is that the entropy $S$ is extensive, {\it i.e.} proportional to $N$, as $T \rightarrow 0$. Specifically, we have
\beqn
\lim_{T \rightarrow 0} \lim_{N \rightarrow \infty} \frac{S}{N} \neq 0\,.
\eeqn
The order of limits is crucial here; the above order of limits defines the zero temperature entropy density, in which the thermodynamic limit is taken before the zero temperature limit. If we had taken the other order of limits, we would obtain the ground state entropy density, which does indeed vanish.

\subsection{Corrections to scaling}
\label{sec:corrscaling}

All of our low energy results for the SYK model have so far been obtained in a scaling limit in which the $i \omega_n$ term in the Green's function in Eq.~(\ref{syk2a}) was neglected, as discussed above Eq.~(\ref{syk10}). This subsection will consider the structure of the corrections that arise when this $i \omega_n$ term is included. We emphasize that all of the computations here are in the $N = \infty$ limit, and we are computing corrections to the low energy approximation to the saddle point equations. A significant result of our computations will be $T$-dependent corrections to the entropy in Eq.~(\ref{syk24}); these will continue to be proportional to $N$. We will consider finite $N$ corrections to such saddle point results in Section~\ref{sec:fluctuations}.

To understand the structure of the possible corrections, we postulate that the low energy corrections can be computed from an effective action of the following form:
\beq
I = I_\ast + \sum_h g_h  \int_0^\beta d \tau\, O_h (\tau) \label{syks1}
\eeq
where $O_h$ are a set of scaling operators with scaling dimension $h$. One of our tasks for the subsection is to determine the possible values of $h$, and we will accomplish this shortly. The term $I_\ast$ is the leading critical theory which leads to the results described so far; in particular to the Green's function in 
Eqs.~(\ref{syk3}) and (\ref{syk18}), and the entropy in Eq.~(\ref{syk24}). We normalize the perturbing operators by the two-point correlator
\beq
\left \langle O_h (\tau) O_h (0) \right\rangle = \frac{1}{|\tau|^{2h}},\label{syks2}
\eeq
then the co-efficient $g_h$ is fully specified. In general, the $g_h$ are a set of non-universal numbers of order $U^{1-h}$,  whose precise values depend upon the details of the underlying theory {\it e.g.\/} on possible higher-order fermion interaction terms we can add to the SYK Hamiltonian.

Given Eq.~(\ref{syks1}), we can easily estimate the form of the corrections to the grand potential $\Omega (T)$. We expect that 
\beq
\left \langle O_h \right\rangle_{T \ast} = \Omega_h T^h , \label{syks4}
\eeq
where the expectation value is evaluated at a temperature $T$ in $I_\ast$, and the $T$-dependence follows from the scaling dimension of $O_h$. Taking the expectation value of the action, we obtain
\beq
\Omega(T) = E_0 - N \mathcal{S} T + \sum_h g_h \Omega_h T^{h} \label{syks3}
\eeq
where $E_0$ is the ground state energy, $\mathcal{S}$ is the entropy in Eq.~(\ref{syk24}), and the set of co-efficients $\Omega_h$ were specified in Eq.~(\ref{syks4}). Similarly, we can write the corrections to the Green's function in Eq.~(\ref{syk7})
from the $O_h$ perturbations:
\beq
G(\tau) = G_\ast (\tau) \left( 1 + \sum_h \frac{g_h \alpha_h}{|\tau|^{h-1}} \right)\,,  \label{syks5}
\eeq
where we now use $G_\ast$ to denote the leading order result in Eq.~(\ref{syk12}),
and we have used $\mbox{dim}[g_h] = 1-h$ from Eq.~(\ref{syks1}).
Here, and below, we will limit ourselves to the particle-hole symmetric case with $\theta=0$, $\mu=0$, $\mathcal{E}=0$, and refer to Ref.~\cite{Tikhanovskaya:2020elb} for the general case. The co-efficients $\Omega_h$ and $\alpha_h$ are universal dimensionless numbers.

\begin{figure}
\includegraphics[width=3in]{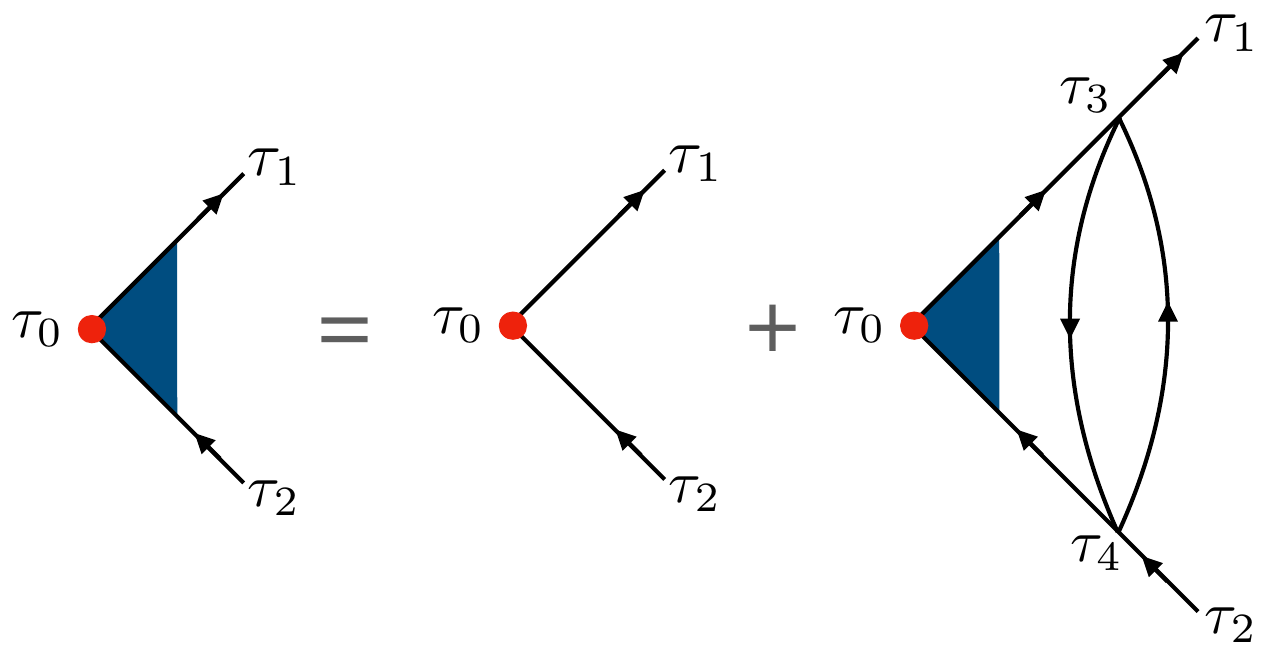}
\caption{Large $N$ equation satisfied by the three point correlator in Eq.~(\ref{threepoint}). The red circle represents the operator $O_h$.}
\label{fig:Ohfig}
\end{figure}

Our remaining task here is to determine the allowed values of $h$. We only consider \cite{Gross:2016kjj,Klebanov:2016xxf,Klebanov:2018fzb} the `antisymmetric' operators $O_h$ which are represented at short times by $O_{h_{n}}=c^{\dag}_{i}\partial_{\tau}^{2n+1}c_{i}$ with $n=0,1,2,\dots$.  The needed information is contained in the three point functions
  \begin{equation}
   v_{h}(\tau_{1},\tau_{2},\tau_{0})=
   \langle c(\tau_1)c^{\dagger}(\tau_2)O_{h}(\tau_0)\rangle \,. \label{threepoint}
  \end{equation}
In the large $N$ limit, this three point function obeys the integral equation shown in Fig.~\ref{fig:Ohfig}. 
In the long time scaling limit, we can drop the bare first time on the right hand side, and then Fig.~\ref{fig:Ohfig} reduces to the eigenvalue equation
\cite{Gross:2016kjj}
 \begin{equation}
 \label{eq:DS3ptfa}
  k(h) v(\tau_{1},\tau_{2},\tau_{0})=   \int d\tau_{3}d\tau_{4} K(\tau_{1},\tau_{2};\tau_{3},\tau_{4})v_{h}(\tau_{3},\tau_{4},\tau_{0})\,,
  \end{equation}
 where the kernel $K$ is
  \begin{equation}
 K (\tau_{1},\tau_{2};\tau_{3},\tau_{4}) =  -3U^{2}G_\ast (\tau_{13})G_\ast (\tau_{24})G_\ast (\tau_{34})^{2}\,, \label{syks10}
 \end{equation}
 with $\tau_{ij} \equiv \tau_i - \tau_j$, and we have introduced an eigenvalue $k(h)$ by hand which must obey
 \beq
  k(h) = 1\,. \label{syks12}
 \eeq
For our purposes, it is sufficient to solve Eq.~(\ref{eq:DS3ptfa}) in the limit $\tau_0 \rightarrow \infty$. Then, we can use the operator product expansion to write
\begin{equation}
c(\tau_1) c^\dagger (\tau_2) \sim \mbox{sgn} (\tau_{12})  \sum_h \frac{c_h}{|\tau_{12}|^{1/2 - h}} O_h (\tau_1) + \ldots
\end{equation}
for some constants $c_h$, where the sum over $h$ now includes the identity operator with $h=0$.
Inserting this into Eq.~(\ref{threepoint}), we conclude that $v \sim \textrm{sgn}(\tau_{12})/|\tau_{12}|^{1/2-h}$ as $\tau_0 \rightarrow \infty$. Then Eq.~(\ref{eq:DS3ptfa}) yields the eigenvalue
%
% The solution of Eq.~(\ref{eq:DS3ptfa}) is aided by an assumption of conformal symmetry, aspects of which will be discussed in Appendix~\ref{sec:timerepar}. With these assumptions, the three point functions are expected to obey the functional form \cite{Gross:2016kjj}
%\begin{equation}
%     v(\tau_{1},\tau_{2},\tau_{0})  = \frac{c_{h} B\,\textrm{sgn}(\tau_{12})}{ |\tau_{12}|^{1/2-h}|\tau_{10}|^{h}|\tau_{20}|^{h}}\,, \label{syks11}
% \end{equation}
%where we have introduced a set of dimensionless `structure constants' $c_h$.   Inserting Eqs.~(\ref{syks10}) and (\ref{syks11}) into Eq.~(\ref{eq:DS3ptfa}), and evaluating the integrals over $\tau_3$ and $\tau_4$ in the limit $\tau_0 \rightarrow \infty$, we find that Eq.~(\ref{eq:DS3ptfa}) yields the eigenvalue 
\beq
k(h) = - \frac{3 \tan(\pi h/2 - \pi/4)}{2h-1}\,. \label{syks13}
\eeq
The solution of Eqs.~(\ref{syks12}) and (\ref{syks13}) finally yields the needed values of $h$. There are an infinite number of solutions, and the lowest values are 
\beq
h=2,~ 3.77354\ldots, ~ 5.567946 \ldots, ~ 7.63197\ldots.~~ 
\eeq
Only the lowest value $h=2$ is an integer, and all higher values are irrational numbers. 

We will have a particular interest here in the $h=2$ operator. This plays a special role in the finite $N$ fluctuations, and leads eventually to a violation of scaling, as will be discussed in Section~\ref{sec:fluctuations}. At $N=\infty$, it is also the lowest dimension operator, and so yields the most important corrections to Eqs.~(\ref{syks3}) and (\ref{syks5}). For the entropy at fixed $\mathcal{Q}$, we can take a derivative of Eq.~(\ref{syks3}), and write the correction to Eq.~(\ref{syk24}) as \cite{kitaevsuh,Maldacena_syk,GKST}
\beqn
S (T \rightarrow 0, \mathcal{Q}) = N\left[ \mathcal{S} + \gamma T \right]\,, \label{syk36c}
\eeqn
where $\gamma \sim 1/U$ is the non-universal co-efficient of the linear-in-$T$ specific heat at fixed $\mathcal{Q}$, a quantity familiar from Fermi liquid theory. The SYK non-Fermi liquid has a similar specific heat; but note the presence of the residual entropy $\mathcal{S}$ which vanishes in a Fermi liquid. We will see in Section~\ref{sec:fluctuations} that $\gamma$ also appears as the co-efficient of the Schwarzian effective action for finite $N$ fluctuations.

\subsection{Finite $N$ Fluctuations}
\label{sec:fluctuations}

This section will turn to a theory of the fluctuations about the large $N$ saddle point examined so far. We will focus on the corrections to the result for the entropy in Eqs.~(\ref{syk24},\ref{syk36c}). The dominant finite $N$ corrections arise from a universal, exactly soluble theory for the low energy fluctuations about the large $N$ saddle point. Along the way, we will also obtain an example of the corrections discussed in Section~\ref{sec:corrscaling} associated with irrelevant operators in the $N=\infty$ saddle point theory. This will lead to the $T$-dependent correction in Eq.~(\ref{syk36c}), and allow us to identify $\gamma$ with a coupling in the effective action.

We begin with a path integral representation of the underlying SYK Hamiltonian Eq.~(\ref{syk1}). To treat the random couplings, we need to perform a quenched average using the replica method. However, the strongly self-averaging properties we shall compute below do not depend upon the replica structure, and so we will simply ignore these technicalities, and work directly with the averaged theory.
So after averaging over the $U_{ijk\ell}$, the path integral becomes
\begin{align}
\overline{\mathcal{Z}} &= \int \mathcal{D} c_{i}(\tau) \exp \left[ - \sum_{i} \int_0^\beta d\tau \, c_{i}^\dagger
\left( \frac{\partial}{\partial \tau} - \mu \right) c_{i} \right. \nonumber \\
& ~~~~~~~~~\left. + \frac{U^2}{4 N^3} \int_0^\beta 
d\tau d \tau' \left| \sum_i c_{i }^\dagger (\tau) c_{i } (\tau') \right|^4 \right]\,, \label{syk30}
\end{align}
where $\beta=1/T$.
We now introduce the following `trivial' identity in the path integral, 
\begin{align}
1 &= \int \mathcal{D} G(\tau_1, \tau_2) \mathcal{D} \Sigma(\tau_1, \tau_2) \nonumber \\
&~~~~~~~\times \exp \Biggl[ - N \int_0^\beta d\tau_1 d\tau_2 \Sigma(\tau_1, \tau_2) \Biggl( G(\tau_2, \tau_1) \nonumber \\
&~~~~~~~~~~~~~~+ \frac{1}{N} \sum_i c_{i} (\tau_2) c_i^\dagger (\tau_1) \Biggr) \Biggr] \,. \label{syk31}
\end{align}
and interchange orders of integration.
Then the partition function can be written as a `$G-\Sigma$' theory, 
a path integral with an action $I[G, \Sigma]$ for the Green's function and the self energy analogous to a Luttinger-Ward functional \cite{GPS2,kitaevsuh,Maldacena_syk}. 
\begin{align}
& \overline{\mathcal{Z}} = \int \mathcal{D} G(\tau_1, \tau_2) \mathcal{D} \Sigma (\tau_1, \tau_2) \exp (-N I[G, \Sigma]) \nonumber \\
& I[G, \Sigma] = -\ln \det \left[ (\partial_{\tau_1} - \mu)\delta(\tau_1 - \tau_2)  + \Sigma (\tau_1, \tau_2) \right] \nonumber \\
&~~~~~~~~~~~~ - \mbox{Tr} \left( \Sigma \cdot G \right) - \frac{U^2}{4} 
\mbox{Tr} \left( G^2 \cdot G^2 \right)\,.  \label{syk31a} 
\end{align}
We have integrated over the fermions to obtain the $\ln \det$ term. This is an exact representation of the averaged partition function. Notice that it involves $G$ and $\Sigma$ as bilocal fields that depend upon two times, and we have introduced a compact notation for such fields:
\beqn
\mbox{Tr} \left( f \cdot g \right) \equiv \int d \tau_1 d \tau_2 \,  f(\tau_2, \tau_1) g(\tau_1, \tau_2)\,. \label{syk31b}
\eeqn
See Appendix ~\ref{sec:timerepar} for a discussion of the symmetries of the bilocal fields, where we also show that after ignoring the explicit time derivative in Eq. (\ref{syk31a}), the action is invariant under time reparametrization and gauge symmetries  (Eq. (\ref{syk15})).

The path integral in Eq. (\ref{syk31a}) is complicated to be evaluated, in general. We now make a low energy approximation by integrating only along directions in the vast $(G,~\Sigma)$ space where the variation $S[G, \Sigma]$ is small at low energies \cite{kitaevsuh,Maldacena_syk}. Given the unimportance of the $\partial_\tau$ in Eq. (\ref{syk31a}), and the resulting symmetries of the action, a powerful conclusion is that we need only perform the path integral along trajectories where the Green's function obeys Eq. (\ref{syk19}) (and similarly for the self energy). In this manner, we formally convert the $G$-$\Sigma$ path integral into a path integral over the time reparameterization $f(\tau)$ and the gauge transformation $\phi (\tau)$ \cite{kitaevsuh,Maldacena_syk,SS17,GKST}.
\beqn
\overline{\mathcal{Z}} \approx e^{-E_0/T + N \mathcal{S}} \int \mathcal{D}f(\tau) \mathcal{D} \phi (\tau) \exp\left( - I_{\rm eff} [f, \phi] \right) \,, \label{syk32}
\eeqn
where $E_0 \propto N$ is the ground state energy (including the $-\mu \mathcal{Q} N$ contribution). 
We will shortly deduce the form of $I_{\rm eff} [f, \phi]$ from symmetry arguments. But before we turn to that, let us note that the 
combination of Eqs.~(\ref{syk19}) and (\ref{syk32}) also yield the most important contributions to the fluctuation corrections to the Green's function
\bea
&& G(\tau_1 -  \tau_2) = \frac{e^{-E_0/T + N \mathcal{S}}}{\overline{\mathcal{Z}}}  \int \mathcal{D}f(\tau) \mathcal{D} \phi (\tau) \nonumber \\
&&~~~~~~~\times \exp\left( - I_{\rm eff} [f, \phi] \right) [f'(\tau_1) f'(\tau_2)]^{1/4} \nonumber \\
&&~~~~~~~\times G_c (f(\tau_1) - f(\tau_2)) e^{i \phi (\tau_1) - i \phi (\tau_2)}\,, \label{syk33}
\eea
where the conformal saddle-point Green's function $G_c (\tau)$  is given by Eq.~(\ref{syk17}). 

Now our task is to determine the action $I_{\rm eff} [f, \phi]$, and then evaluate the path integrals in Eqs.~(\ref{syk32}) and (\ref{syk33}). It turns out the partition function for the free energy in Eq.~(\ref{syk32}) can be evaluated exactly. The consequences of the path integral in Eq.~(\ref{syk33}) for the long-time behavior of $G(\tau)$ have also been investigated \cite{kitaev_talk,Maldacena_syk,kitaevsuh,Bagrets:2016cdf,Bagrets:2017pwq,Altland:2019czw,Kruchkov:2019idx,Kobrin:2020xms}: they lead to a violation of scaling at times of order $N/U$, but we will not describe this further here.

The form of $I_{\rm eff} [f, \phi]$ is strongly constrained by the requirement that $I$ vanish for the case where $f(\tau)$ and $\phi(\tau)$ are given by Eq.~(\ref{syk20}). This follows immediately from the fact that Eq.~(\ref{syk20}) leads to no changes in the from of the saddle point Green's function when inserted into Eq.~(\ref{syk19}).  As the action was originally a functional of the Green's function, it can also not change. The action 
$I_{\rm eff} [f, \phi]$ with the smallest number of derivatives that satisfies this requirement is \cite{kitaevsuh,Maldacena_syk,SS17,GKST}
\bea
I_{\rm eff} [f, \phi] &=& \frac{NK}{2} \int_0^{\beta} d \tau \left(\frac{\partial \phi}{\partial \tau} + i (2 \pi \mathcal{E} T) \frac{\partial f}{\partial \tau}\right)^2 \nonumber \\
&~&-  \frac{N\gamma}{4 \pi^2} \int_0^{\beta} d \tau \, \{ \tan (\pi T f(\tau)), \tau\}\,. \label{syk34}
\eea
The curly brackets in Eq.~(\ref{syk34}) represent a Schwarzian derivative:
\beqn
\{g, \tau\} \equiv \frac{g'''}{g'} - \frac{3}{2} \left( \frac{g''}{g'} \right)^2 .
\eeqn
This has the defining property that
\beqn
\{ \frac{a \tau +b}{c \tau + d} , \tau \} = 0\,,
\eeqn
which ensures that $I_{\rm eff} [f, \phi]$ vanishes for Eq.~(\ref{syk20}). (We note that there are coupled SYK models which are not described by a Schwarzian effective action \cite{JMDS16b,Milekhin:2021cou}.) 

For the origin of $f(\tau)$ and $\phi(\tau)$ as time reparameterization and gauge transformations of the Green's function, we must also place some constraints on the nature of the path integral over them. The function $f(\tau)$ must be monotonic, and obey
\beqn
f(\tau + \beta) = f(\tau) + \beta\,. \label{syk37}
\eeqn
Moreover, we should sum over all possible phase windings with
\beqn
\phi(\tau + \beta) = \phi (\tau) + 2 \pi n \label{syk38}
\eeqn
where $n$ is an integer. 

The action in Eq.~(\ref{syk34}) has 2 dimensionful coupling constants, $K$ and $\gamma$. By dimensional analysis, we can see that $K \sim \gamma \sim 1/U$, the only energy scale at $T=0$. Determining the precise values of $K$ and $\gamma$ is not simple, and requires a numerical study of the higher energy properties of the SYK model. We will now relate the values of $K$ and $\gamma$ to thermodynamic observables of the $N=\infty$ theory, and numerical computation of these observables is usually the simplest way to determine $K$ and $\gamma$. 

At $T=0$, the action for $\phi$ represents the path integral of a particle of mass $NK$ moving on a ring of circumference $2 \pi$. So the energy levels are $\ell^2/(2 N K)$, where the integer $\ell$ measures the total fermion number. With a chemical potential shift $\delta\mu$, the energy levels will shift as $\ell^2/(2NK) - \delta\mu \ell$. From the optimum value of this function for different $\ell$, we conclude that $K$ is the just the compressibility
\beqn
K = \frac{d \mathcal{Q}}{d \mu} \quad, \quad T=0\,. \label{syk35a}
\eeqn

Turning to the value of $\gamma$, note that the action $I_{\rm eff} [f, \phi]$ does not vanish at the $N=\infty$ saddle point $f (\tau) = \tau$. 
Evaluating Eq.~(\ref{syk34}) for this value of $f(\tau)$, and setting $\phi=0$, we obtain the grand potential at $N=\infty$ for small $T>0$:
\beqn
\Omega (T) = E_0 - N \mathcal{S} T - \frac{1}{2} N (\gamma + 4 \pi^2 \mathcal{E}^2 K) T^2. \label{syk35}
\eeqn
Taking the $T$ derivative, we obtain the leading low temperature correction to the entropy in Eq.~(\ref{syk24})
\beqn
S (T \rightarrow 0, \mu) = N\left[ \mathcal{S} + \left(\gamma + 4 \pi^2 \mathcal{E}^2 K \right) T \right]\,. \label{syk36}
\eeqn
As denoted above, this the entropy at a fixed chemical potential. We can use standard thermodynamic relations to compute the entropy at fixed $\mathcal{Q}$, using the thermodynamic relations Eqs.~(\ref{syk23}) and (\ref{syk35a}), and obtain Eq.~(\ref{syk36c}); indeed the co-efficient of the Schwarzian was chosen so that the entropy would obey the form in Eq.~(\ref{syk36c}). 
The $T$-dependent corrections in Eq.~(\ref{syk36}) and Eq.~(\ref{syk36c}) are proportional to $N$, and so constitute corrections from irrelevant operators which were studied in Section~\ref{sec:corrscaling}, and identify the Schwarzian as representing the corrections arising from the $h=2$ operator. 

In the remainder of our discussion of the SYK model, we will evaluate the path integral in Eq.~(\ref{syk32}), and so obtain the finite $N$ corrections to the free energy and entropy in Eqs.~(\ref{syk35}) and (\ref{syk36}). These results will also allow us to compute the many-particle density of states $D(E)$. 

A key observation in the evaluation of Eq.~(\ref{syk32}) is that the path integrals factorize.  To establish this, we use the boundary conditions in Eqs.~(\ref{syk37}) and (\ref{syk38}) to parameterize
\begin{eqnarray}
    f(\tau) &=& \tau + \epsilon (\tau) \nonumber \\
    \phi (\tau) &=& 2 \pi n T \tau + \bar{\phi} (\tau)\,, \label{syk36a}
\end{eqnarray}
where the `winding number' $n$ is an integer, and $\epsilon$ and $\bar{\phi}$ are then periodic functions of $\tau$ with period $\beta$. In the first term in the action Eq.~(\ref{syk34}), we can absorb $\epsilon$ by a shift in $\bar{\phi}$; then the remaining dependence on $\epsilon$ is only in the Schwarzian. In this manner, we can write Eq.~(\ref{syk32}) as \cite{GKST}
\beqn
\overline{\mathcal{Z}} = e^{-E_0/T} \, \mathcal{Z}_\mathcal{Q}  \, \mathcal{Z}_{\rm Sch} \label{syk36b}
\eeqn
and we will evaluate $\mathcal{Z}_{\mathcal{Q}}$ and $\mathcal{Z}_{\rm Sch}$ in the following subsections.

\subsubsection{Rotor path integral}
\label{sec:sykrotor}

The partition function $\mathcal{Z}_{\mathcal{Q}}$ represents fluctuations in the total charge on the SYK dot. It is expressed as a path integral over the co-ordinates of a particle moving on a unit circle (a `rotor')
\bea
  \mathcal{Z}_{\mathcal{Q}} &=&  \left( \, \sum_{n=-\infty}^{\infty} \exp \left[ - 2 \pi^2 N K T (n + i \mathcal{E})^2 \right]  \right) \nonumber \\
  &~&\times \int \frac{\mathcal{D} \bar{\phi}}{U(1)} \exp \left[- \frac{NK}{2} \int_0^{\beta} d \tau \left(\bar{\phi}'(\tau)  \right)^2 \right] \,. \label{syk40}
\eea
The second term is just the imaginary time amplitude for a `free particle' of mass $NK$ to return to its starting point in a time $\beta$, divided by the volume ($=2\pi$) of the U(1) group (because a $\tau$ independent $\phi$ does not make any changes to the Green's function in Eq.~(\ref{syk19})).
So we obtain an expression for $\mathcal{Z}_{\mathcal{Q}}$ which is useful at temperatures $T \gg 1/(NK)$. 
\beqn
\mathcal{Z}_{\mathcal{Q}} =  \left( \, \sum_{n=-\infty}^{\infty} \exp \left[ - 2 \pi^2 N K T (n + i \mathcal{E})^2 \right]  \right) \sqrt{ \frac{N K T}{2 \pi}}\,. \label{syk41}
\eeqn
For lower temperatures, $T \ll  1/(NK)$, we can apply the Poisson summation formula to Eq.~(\ref{syk41}) and obtain
\beqn
\mathcal{Z}_{\mathcal{Q}} = \frac{1}{2 \pi} \sum_{p=-\infty}^{\infty}  \exp \left[ - \frac{p^2}{2 N K T} - 2 \pi \mathcal{E} p \right]. \label{syk42}
\eeqn
We note, however, that both expressions Eqs.~(\ref{syk41}) and (\ref{syk42}) are convergent and exact at all $T$ \cite{GKST}.

The physical interpretation of Eq.~(\ref{syk42}) is especially transparent. It describes a `quantum dot' with equilibrium charge $N \mathcal{Q}$, which has fluctuations to states with charge $N \mathcal{Q} + p$. The energy of a charge $p$ fluctuation is determined by a `capacitance' $NK$, and a temperature-dependent chemical potential $-2 \pi \mathcal{E} T$. Note that the chemical potential shift is exactly that appearing in Eq.~(\ref{syk22}), and indeed the present analysis can be viewed as another derivation of Eq.~(\ref{syk22}). Recall that the key relation for the entropy in Eq.~(\ref{syk33}) followed after applying a Maxwell thermodynamic relation to Eq.~(\ref{syk22}). 

The above physical interpretation also indicates that in a fixed $\mathcal{Q}$ ensemble, we should take $\mathcal{Z}_{\mathcal{Q}} = 1$.
That turns out to be not quite correct, and a more careful analysis of finite $N$ corrections shows that $\mathcal{Z}_{\mathcal{Q}} \sim 1/N^2$.

\subsubsection{Schwarzian path integral}
\label{sec:schwarzian}

The other component of Eq.~(\ref{syk36b}) is the Schwarzian path integral
\beqn
\mathcal{Z}_{\rm Sch} = e^{N \mathcal{S}} \int \frac{\mathcal{D} f(\tau)}{{\rm SL(2,R)}} \exp \left[ \frac{N\gamma}{4 \pi^2} \int_0^{\beta} d \tau \, \{ \tan (\pi T f(\tau)), \tau\} \right] \label{syk43}
\eeqn
We have normalized the path integral by the (infinite) volume of the non-compact group SL(2,R) because, as we argued earlier, the action must vanish under SL(2,R) transformations. This quotient will be crucial in obtaining a well-defined answer.

It was shown \cite{Stanford:2017thb} that the path integral in Eq.~(\ref{syk43}) can be evaluated exactly. The key to their result is the remarkable fact that a Gaussian approximation to the path integral is in fact exact. We will exploit this by just evaluating Eq.~(\ref{syk43}) in the Gaussian approximation.

To this end, we expand the Schwarzian action in dimensional Fourier coefficients of $\epsilon (\tau)$ in Eq.~(\ref{syk36a})
\beqn
\epsilon(\tau) = \frac{1}{T} \sum_{n=-\infty}^{\infty} \epsilon_n e^{-2 \pi i n T \tau} \,, \label{syk44b}
\eeqn
and obtain  
\beqn
I_{\rm eff} [ \epsilon ] =  - \frac{N\gamma T}{2} + 2 \pi^2 N \gamma T \sum_{n} n^2 (n^2 - 1)|\epsilon_n|^2  \,. \label{syk44}
\eeqn

Now notice that $I_{\rm eff} [ \epsilon ]$ vanishes for the smallest 3 Matsubara frequencies $\omega_n = 0, \pm 2 \pi T$. 
Indeed, the action was designed to vanish for any time reparameterization which belongs to SL(2,R), a three-dimensional non-compact space. And here we have discovered three Fourier components which cause no variation in the action to second order: clearly, we can identify the frequency components at $n = 0, \pm 1$ as the infinitesimal limit of the  SL(2,R) transformations. At Gaussian order, the path integral over these action-free normal modes therefore cancels against the volume of SL(2,R) in Eq.~(\ref{syk43}). Actually, this cancellation happens also for large SL(2,R) transformations, but that we do not prove here.

Performing the Gaussian integral over the remaining modes, we obtain for the logarithm of the partition function
\beqn
\ln \mathcal{Z}_{\rm Sch} = N \mathcal{S} + \frac{N\gamma T}{2} - \frac{1}{2} \sum_{n\neq 0, \pm 1} \ln \left[ 2 \pi^2 N \gamma T n^2 (n^2 - 1) \right]. \label{syk44a}
\eeqn
The sum over the Matsubara frequency $\omega_n$ is clearly divergent, and should be cutoff at a frequency $|\omega_n| \sim U$, above which our low energy Schwarzian theory does not apply. We describe the regulation of the divergence in Appendix~\ref{app:schwarzian}: there is a contribution $\sim U/T$, but this can be absorbed into a redefinition of $E_0$ in Eq.~(\ref{syk36b}). The needed subleading term is $\sim \ln (U/T)$, and an important result is that the co-efficient of the $\ln (T)$ term is universal; we find for $T \ll U$ \cite{Maldacena_syk,kitaevsuh,Stanford:2017thb}
\beqn
\ln \mathcal{Z}_{\rm Sch}  = N \mathcal{S} + \frac{N \gamma T}{2} - \frac{3}{2} \ln \left( \frac{U}{T} \right) \,. \label{syk45}
\eeqn
Apart from the finite $N$ corrections in the rotor components (which had a simple physical interpretation), we have now obtained our first non-trivial finite $N$ correction to the SYK model: the $-(3/2) \ln (1/T)$ correction to the logarithm of the partition function. Note that the logarithm in Eq.~(\ref{syk45}) becomes as large as the leading term only at an exponentially low $T \sim U e^{-N}$,  below which the large $N$ theory does not apply. 

It is also useful to compare Eq.~(\ref{syk45}) to our earlier large $N$ result for $-T \ln Z$ in the random matrix model in Eq.~(\ref{rm5}). That had a leading $N \gamma T/2$ term, but there was no $T$-independent term proportional to $N$, as the random matrix model does not have an extensive entropy in the zero temperature limit.

The $-(3/2) \ln (1/T)$ correction to Eq.~(\ref{syk45}) has important consequences for the many-body density of states,  $\dos_{\rm Sch} (E)$. We define this by
\beqn
\mathcal{Z}_{\rm Sch} (T) = \int_0^{\infty} dE \, \dos_{\rm Sch} (E) e^{-E/T}\,. \label{syk46}
\eeqn
As we have absorbed the $\sim 1/T$ term in Eq.~(\ref{syk45}) into a redefinition of $E_0$ in Eq.~(\ref{syk36b}), we can assume in Eq.~(\ref{syk46}) that 
$\dos_{\rm Sch} (E)$ vanishes for $E < 0$.
It turns out to be possible to determine $\dos_{\rm Sch} (E)$ by performing the inverse Laplace transform exactly using the value in Eq.~(\ref{syk45}). 
This yields \cite{Bagrets:2017pwq,kitaevsuh,Stanford:2017thb,Garcia-Garcia:2017pzl,Cotler:2016fpe}
\beqn
\dos_{\rm Sch} (E) \propto e^{N \mathcal{S}} \sinh \left( \sqrt{2 N \gamma E } \right) \,. \label{syk47}
\eeqn
It is easier to insert the result Eq.~(\ref{syk47}) into Eq.~(\ref{syk46}), perform the $E$ integral, and verify that we obtain Eq.~(\ref{syk45}).

The result Eq.~(\ref{syk47}) is accurate for $E \ll NU$, and even down to $E \sim U/N$. Near the lower bound it predicts a many-body density of states $\sim e^{ N\mathcal{S}}$, in sharp contrast to the random matrix model of Section~\ref{sec:matrix} which did not have an exponentially large density of states at such low energies.
We showed numerical plots of the many-body density of states \cite{FuSS,Cotler:2016fpe,ShenkerRMT} for a closely related Majorana fermion model in Fig.~\ref{fig:specq4}. Notice the much larger density of states, and much smaller level spacing near the bottom of the band, in comparison to the free fermion random matrix model in Fig.~\ref{fig:specq2} of the same size. This is also evident from a comparison of the Schwarzian result in Eq.~(\ref{syk47}), with the free fermion result in Eq.~(\ref{rm7}): the most important difference is the presence of the prefactor of $e^{N \mathcal{S}}$ in Eq.~(\ref{syk47}). 

We now recall our discussion at the end of Section~\ref{sec:matrix2} where we argued that the low-lying many-body eigenstates at excitation energies of order $1/N$ could be interpreted as the sums of quasiparticle energies. In the SYK model we have order $\sim e^{N \mathcal{S}}$ energy levels even within energy $\sim 1/N$ above the many-body ground states. It is impossible to construct these many-body eigenstates from order $\sim N$ quasiparticle states. This is therefore strong evidence that there is no quasiparticle decomposition of the many-body eigenstates of the SYK model. Note that the presence of an extensive entropy as $T \rightarrow 0$ (the non-zero value of $\mathcal{S}$) is a {\it sufficient\/}, but not a {\it necessary\/}, condition for the absence of quasiparticles: the models we shall study in Section~\ref{sec:cfs} do not have quasiparticles, but do not have an extensive entropy as $T \rightarrow 0$ as described in more detail in Section~\ref{sec:cfsthermo}.

Finally, we combine the results for the rotor and Schwarzian partition functions, and obtain corresponding results for the SYK model \cite{GKST}.
Using the $n=0$ term in Eqs.~(\ref{syk41}) and (\ref{syk45}) in Eq.~(\ref{syk36b}), we obtain for $U/N \ll T \ll U$
\begin{equation}
    \Omega = E_0 - N \mathcal{S} T - \frac{N(\gamma + 4 \pi^2 \mathcal{E}^2 K) T^2}{2} 
    + 2 T \ln \left( \frac{U}{T} \right) + \ldots\,. \label{syk48}
\end{equation}
This contains the $1/N$ correction to the result Eq.~(\ref{syk35}) for the grand partition function: the $2 T \ln (1/T)$ term.
As for the random matrix model, we can invert Eq.~(\ref{syk48}) to obtain the many body density of states in the grand canonical ensemble for grand energies $U/N \ll E \ll NU$
\bea
\dos(E) &\sim& \exp\left(S(E)  \right) \nonumber \\
S(E) &=& 
N \mathcal{S} + \sqrt{2 N  (\gamma + 4 \pi^2 \mathcal{E}^2 K) (E - E_0)}  \label{syk49}
\eea
for $E>E_0$, and $S(E)=0$ for $E<E_0$.
Comparing this to the random matrix model, we find that $S(E)$ has a similar functional form of $E$, but without the leading $N \mathcal{S}$ term.

%%%%%%%%%%%%%%%%%%%%
%%%%%%% Quantum MAGNET
%%%%%%%%%%%%%%%%%%%%%%%%%%

\section{Random exchange quantum magnets}
\label{sec:rqm}

The SYK model discussed so far provides valuable insight into quantum systems
without quasiparticle excitations. However, the microscopic Hamiltonian in
Eq.~(\ref{syk1}) has a short-coming, namely that strong local ({\it i.e.\/} on-site)
interactions are absent. As a result, there are no Mott insulating phases at
any commensurate density in the large $N$ limit. Such local correlations are
clearly important for understanding the interplay of electron itinerancy and the tendency for interaction-induced localization in numerous correlated electron materials.

We now turn to a number of random and fully connected models which restore ``Mottness''. We refer to Mottness here as a generic term to indicate qualitatively the tendency of electrons to localize due to strong repulsive interactions in the vicinity of a Mott transition.
In the present section we discuss the original SY model \cite{SY}, a pure spin model in which explicit on-site charge fluctuations are absent.
In Section~\ref{sec:tJU}, we will introduce charge fluctuations and consider itinerant electron models, with a strong, on-site, repulsive interaction $U$.
Notably, we will find substantial evidence that near critical points and/or over significant intermediate energy scales, 
these correlated models exhibit singular behavior which connects to the critical properties of the SYK model. Section~\ref{sec:KH} will extend the present random quantum magnet in a different manner, by adding a second band of free electrons (similar to that in Section~\ref{sec:matrix}), and so describe a random exchange Kondo-Heisenberg model.

Unlike the SYK model, the models mentioned above are not solvable analytically in the limit of a large number of sites $N$.
We will follow two routes towards understanding their phase diagram.
First, analytical results can be obtained by extending the spin symmetry from SU(2) to SU($M$) and taking the large-$M$ limit or, 
in the SU(2) case by using renormalization group methods in the vicinity of specific fixed points.
Secondly, modern computational algorithms now provide a controlled numerical solution of such models in the SU(2) case directly, even close to quantum critical points.
Some algorithms are briefly reviewed in Sec.~\ref{sec:numerical_methods}.

This section applies the above approaches to random exchange quantum magnets. We consider insulating quantum magnets with a Hamiltonian of the form
\beqn
H_J = \frac{1}{\sqrt{N}}
\sum_{1\leq i < j \leq N} %\sum_{i<j=1}^{N} 
J_{ij} \vec{S}_i \cdot \vec{S}_j, \label{rqm1}
\eeqn
where $\vec{S}_i$ are quantum spin operators on site $i$, and $J_{ij}$ are independent random variables with vanishing mean and variance $J$. In the most important case, the spins belong to the SU(2) algebra, and we have $S=1/2$ states on each site. As noted above, we will also consider generalizations to SU($M$) spins. 

Models like Eq.~(\ref{rqm1}) with {\it classical} spins have served as the foundations of spin glass theory, and more generally of optimization problems and also of neural networks \cite{mezard1987spin}. Here, we will see that such models are also a valuable starting point for understanding correlated electron systems without quasiparticle excitations.

\subsection{SU($M$) symmetry with $M$ large}
\label{sec:SY}

As stated above, Eq.~(\ref{rqm1}) is not analytically solvable for the SU(2) case, even in the limit of $N \rightarrow \infty$. We will return to the SU(2) case in Section~\ref{sec:SYRG}, but here we consider the extension to SU($M$) spin symmetry, with $M$ large, that was originally examined by Sachdev and Ye \cite{SY}. We will see that the limit $N \rightarrow \infty$ followed by the limit $M \rightarrow \infty$ leads to the same saddle point equations and $G$-$\Sigma$ action as the SYK model of Section~\ref{sec:SYK}.

For the SU($M$) case, we employ the representation of spin using fermionic spinons $f_{i,\alpha}$, $\alpha = 1 \ldots M$. These fermions obey the constraint
\beqn
\sum_{\alpha=1}^M f_{i \alpha}^\dagger f_{i \alpha} = \kappa M
\label{rqm2}
\eeqn
on each site $i$, where $0 < \kappa < 1$. The SU(2) case corresponds to $M=2$ and $\kappa = 1/2$. Then, we can write the spin operators as $S_{i, \alpha\beta} = f_{i \alpha}^\dagger f_{i\beta}$, and generalize Eq.~(\ref{rqm1}) to 
\beqn
H_{J} = \frac{1}{\sqrt{NM}} \sum_{\alpha,\beta=1}^M 
\sum_{1\leq i < j \leq N}%\sum_{i<j=1}^{N} 
J_{ij} f_{i \alpha}^\dagger f_{i \beta} f_{j \beta}^\dagger f_{j \alpha}. 
\label{rqm3}
\eeqn
This fermionic spinon representation has {\it fractionalized\/} the spin operator, where the U(1) gauge transformation $f_{i\alpha} \rightarrow e^{i \phi_i} f_{i\alpha}$ leaves the spin operator invariant. Below we will see that, in the large $M$ limit, the $f_\alpha$ form a SYK state: in the context of the random quantum magnet, this state is a critical, gapless, {\it spin liquid\/}. In the present large $N,M$ expansion, the Lagrange multiplier $\lambda_i$, introduced below, plays the role of an emergent gauge field in this spin liquid.

We proceed \cite{SY} with an analysis of Eq.~(\ref{rqm3}) similar to that presented for Eq.~(\ref{syk1}). We average over $J_{ij}$, and obtain the averaged partition function analogous to Eq.~(\ref{syk30}):
\begin{align}
& \overline{\mathcal{Z}} = \int \mathcal{D} f_{i\alpha}(\tau) \mathcal{D} \lambda_i (\tau)
e^{- \mathcal{S}_B - \mathcal{S}_J} \label{rqm4} \\
&\mathcal{S}_B =  \sum_{i} \int_0^\beta d\tau \left\{ f_{i\alpha}^\dagger
\left( \frac{\partial}{\partial \tau} + i \lambda \right) f_{i\alpha} - i \lambda \kappa M \right\} \nonumber \\
& \mathcal{S}_J = - \frac{J^2}{4 N M} \int_0^\beta 
d\tau d \tau' \left| \sum_i f_{i \alpha }^\dagger (\tau) f_{i \beta } (\tau) f_{i \gamma}^\dagger (\tau') f_{i \delta} (\tau') \right|^2 \,. \nonumber
\end{align}
In the large $N$ limit, we assume self-averaging among the sites, and in the large $M$ limit, we can replace the quartic operator of fermions by the product of Green's functions of the $f$ fermions:
\beqn
f_{\alpha }^\dagger (\tau) f_{\beta } (\tau) f_{\gamma}^\dagger (\tau') f_{ \delta} (\tau') = \delta_{\alpha\delta} \delta_{\beta\gamma} G(\tau,\tau') G(\tau', \tau)\,. \label{rqm5}
\eeqn
Then, the analysis proceeds just as for the SYK model, and we obtain an expression for the $G$-$\Sigma$ action nearly identical to that in Eq.~(\ref{syk31a}) but with a prefactor of $N$ replaced by $NM$:
\begin{align}
& I[G, \Sigma,\lambda] = -\ln \det \left[ (\partial_{\tau_1} + i \lambda (\tau_1)) \delta(\tau_1 - \tau_2)  + \Sigma (\tau_1, \tau_2) \right] \nonumber \\
&- \mbox{Tr} \left( \Sigma \cdot G \right) - \frac{J^2}{4} 
\mbox{Tr} \left( G^2 \cdot G^2 \right) - i \kappa \int_0^\beta \!\! d \tau\, \lambda (\tau)  \,. \label{rqm6}
\end{align}
Consequently the subsequent results for the fermion Green's function and the large $NM$ thermodynamics are identical to those in Section~\ref{sec:SYK} after the replacement $U \rightarrow J$ and $\mathcal{Q} \rightarrow \kappa$.

The local spin-spin correlation can also be obtained as
\bea
Q(\tau) &=& \frac{1}{M^2} \left\langle f_{\alpha}^\dagger (\tau) f_\beta(\tau) f_{\beta}^\dagger (\tau') f_\alpha (\tau') \right\rangle \nonumber \\
&=& \frac{C^2 e^{- 2 \pi \mathcal{E}}}{(1 + e^{- 4 \pi \mathcal{E}})} \frac{T}{ \sin (\pi T \tau)}\,, \quad 0 < \tau < \frac{1}{T}, \label{rqm7}
\eea
which has been obtained from Eq.~(\ref{syk17}).
We can obtain the spin spectral density, $\rho_Q$, by a Fourier transform, which yields \cite{Parcollet1}
\beqn
\rho_Q (\omega) \sim \tanh \left( \frac{\omega}{2T} \right)\,. \label{rqm7a}
\eeqn
At $T=0$, this corresponds to a spin density of states $\sim \mbox{sgn} (\omega)$, which is a starting assumption in the original theory of the marginal Fermi liquid \cite{Varma89}.

Recent work \cite{Tikhanovskaya:2020elb} has obtained corrections to the correlators of the quantum magnet $H_J$ from perturbations of the critical theory by leading irrelevant operators described in Section~\ref{sec:corrscaling}.
The most important corrections arise from the operator with scaling dimension $h=2$, and similar to Eq.~(\ref{syks3}) and Eq.~(\ref{syks5}), we obtain
\beqn
\rho_Q (\omega) \sim \tanh \left( \frac{\omega}{2T} \right) \left[
1 - \mathcal{C} \gamma \, \omega \tanh \left( \frac{\omega}{2T} \right) \right]\,, \label{rqm10}
\eeqn
where $\gamma \sim 1/J$ is the co-efficient of the Schwarzian in Eq.~(\ref{syk34}), and also the linear-in-$T$ coefficient of the specific heat in Eq.~(\ref{syk36c}). The dimensionless number $\mathcal{C}$ is universal,
\beq
\mathcal{C} = \frac{24}{\pi \left[ 2 \cos (2 \theta) + 3 \pi \cos^2 (2 \theta) \right]}\,.
\eeq
Here $\theta$ is the spectral asymmetry angle which appeared in Eq.~(\ref{syk3}), and which is related by the Luttinger theorem in Eq.~(\ref{sykl7}) to $\kappa$ in Eq.~(\ref{rqm2}). 
We will compare Eq.~(\ref{rqm10}) with numerical studies of the SU(2) magnet in Section~\ref{sec:SYSU2}. 
%%%%%%%%%%%%%%%%%%%%%%

\subsection{SU(2) model}
\label{sec:SYSU2}

We now return to the original model in Eq.~(\ref{rqm1}), and examine it for the physically important case with SU(2) symmetry. We proceed as in the analyses of classical spin glass problems by introducing replicas and then averaging  over the replicated partition function. This yields a self-consistent problem of a single quantum spin with replica indices~\cite{BrayMoore}. The replica structure is important for the spin-glass phase~\cite{GPS1,GPS2, Biroli_OP2002} 
but in this article we will mostly focus on the disordered paramagnetic phase 
above the spin-glass ordering temperature or on quantum critical points corresponding to the destruction 
of spin-glass order at $T=0$ (Secs.~\ref{sec:ChaHubbard} and \ref{sec:tJnumerics})
In these cases, it is permissible at large $N$ to ignore the replica indices and consider the following path integral for a single quantum spin $S=1/2$
\begin{eqnarray}\label{eq:BrayMoore}
\mathcal{Z}_J &=& \int \mathcal{D} \vec{S} (\tau) \delta( \vec{S}^2 - 1) e^{-\mathcal{S}_B - \mathcal{S}_J} \label{ZJ} \\
\mathcal{S}_B &=& \frac{i}{2} \int_0^{1} du \int d \tau ~\vec{S} \cdot \left(\frac{\partial \vec{S}}{\partial \tau} \times \frac{\partial \vec{S}}{\partial u} \right) \nonumber \\
\mathcal{S}_J &=&   - \frac{J^2}{2} \int d\tau d \tau' Q (\tau - \tau') \vec{S} (\tau) \cdot \vec{S} (\tau')  \nonumber \,. \nonumber
\end{eqnarray} 
This is a coherent state path integral, and $\mathcal{S}_B$ is the geometric Berry phase, closely connected to the spin commutation relations. The spin has a temporal self-interaction with itself, represented by the function $Q(\tau)$. The value of $Q(\tau)$ is to be determined self-consistently by computing the correlator,
\begin{eqnarray}
\overline{Q} (\tau - \tau') &\equiv&  \frac{1}{3} \left\langle \vec{S} (\tau) \cdot \vec{S} (\tau') \right\rangle_{\mathcal{Z}_J}, \label{defQb}
\end{eqnarray}
and then imposing the self-consistency condition, 
\beqn
Q(\tau) = \overline{Q} (\tau). \label{defQ}
\eeqn

A major difference with the $SU(M)$ model in the fermionic large-$M$ limit is that the $SU(2)$ model 
has a spin-glass phase at low temperature. A semi-classical picture of this phase is that of local moments 
pointing randomly in all directions so that the global magnetisation vanishes but the variance of the 
distribution of local magnetisations $\frac{1}{N}\sum_i m_i^2 = q_{EA}$ is non-zero. The latter is the 
Edwards-Anderson order parameter of the spin-glass phase~\cite{mezard1987spin}. 
A hallmark of the spin-glass phase is also 
that local quantities (starting with the local magnetisation itself) are no-longer self-averaging. 

The existence of a spin-glass phase in the $SU(2)$ case can be 
%Contrary to the large-$M$ limit, the SU(2) random Heisenberg model orders in a spin-glass at low temperatures.
%This 
can be established in two ways.
First, the replica diagonal effective action Eq.~(\ref{eq:BrayMoore}) for the disordered averaged 
Green functions can be solved numerically exactly using Quantum Monte Carlo methods in the paramagnetic phase \cite{GrempelRozenberg98}.
At low temperature, the spin-glass susceptiblity diverges at $T= T_{SG}\approx 0.14J$ at the boundary of the spin glass phase.

\begin{figure}
\begin{center}
\includegraphics[width=3.5in]{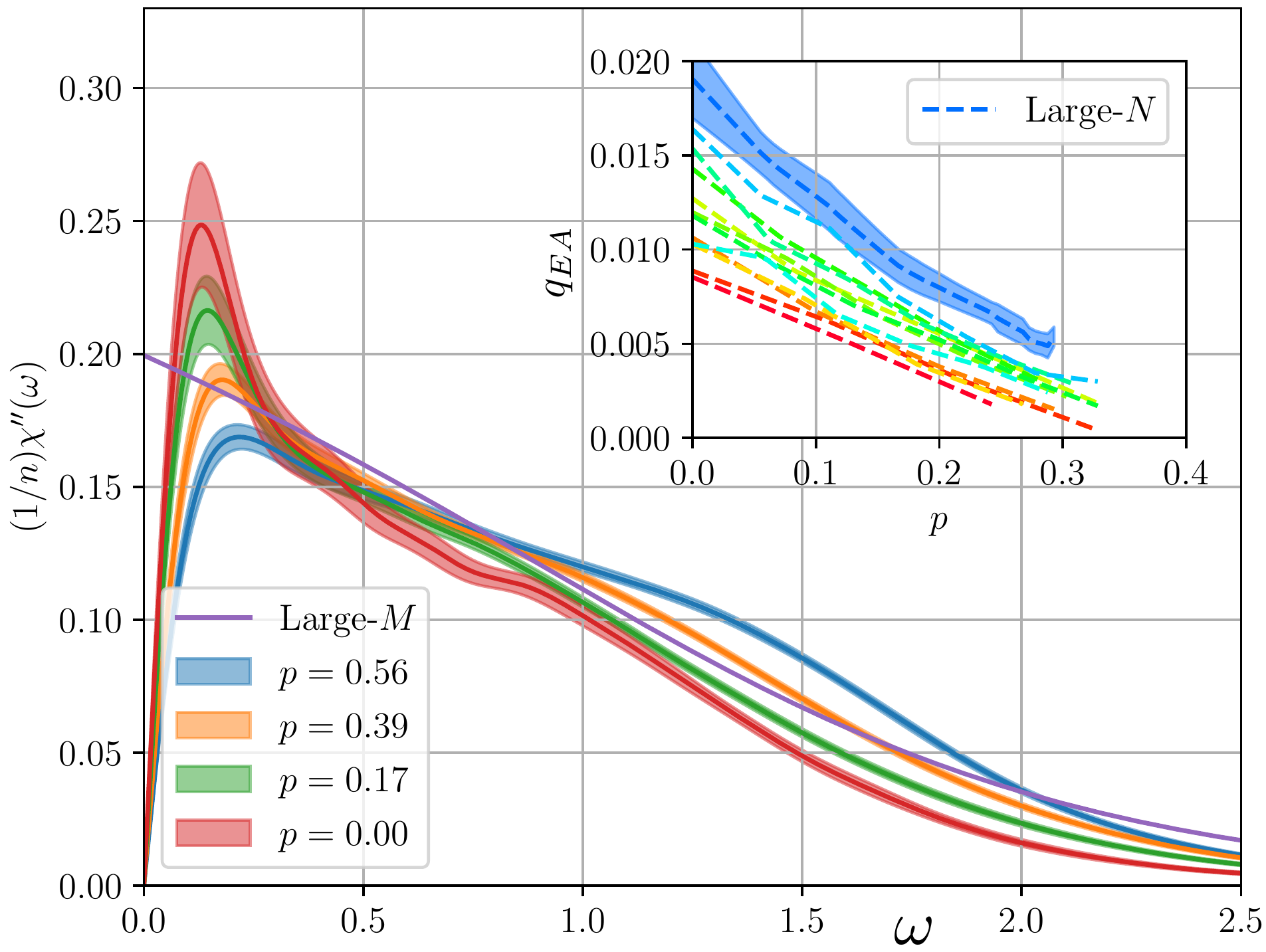}
\end{center}
\caption{\cite{Shackleton2021} Local spin response function for the spin-$1/2$ doped random exchange t-J model, as obtained by exact diagonalization of 
an $N=18$ site cluster, averaged over $100$ disorder realizations, for $t=J=1$. $n= 1-p$ is the particle density.
\label{fig:tJ_ED_chi} 
}
\end{figure}

Second, exact diagonalization of finite size systems have been performed directly in the spin glass phase for many realizations ($10^3$ to $10^5$) of the quenched disorder
\cite{ArracheaRozenbergSG2002,Shackleton2021}.
The local dynamical spin susceptibility $\chi''_\text{loc}(\omega)$ was computed from both full diagonalization of small systems at finite $T$,
and Lanczos method at $T=0$.
From a finite size scaling analysis, the $T=0$ disordered averaged susceptibility in the thermodynamic limit 
is of the form
\begin{equation}\label{eq:ChiQEADelta}
\chi''_\text{loc}(\omega) = q_{EA} \pi \beta \omega \delta(\omega) + \chi''_\text{reg}(\omega),
\end{equation}
where $q_{EA} \approx 0.02$ is the Edwards-Anderson parameter \cite{Shackleton2021} and $\chi''_\text{reg}$ is the regular part.
Fig.~\ref{fig:tJ_ED_chi}, presents numerical results for $\chi''_\text{loc}(\omega)$ for the $t$-$J$ model 
for various dopings $p$; the present discussion is for $p=0$, and the doped cases will be discussed in Section~\ref{sec:SYSU2}.
Apart from the delta function spin glass contribution at very low frequencies, the structure of $\chi''_\text{reg}(\omega)$ is notable. Specifically, the theory of the gapless spin fluid phase studied in Section~\ref{sec:SY} predicts quite generally that
\beq
\chi''_\text{reg}(\omega) = C_1 \mbox{sgn} (\omega) \left[ 1 - C_2 |\omega| + \ldots \right] ,~ T=0.~~ \label{chireg}
\eeq
Formally, this result follows from taking the $T \rightarrow 0$ limit of the large $M$ result in Eq.~(\ref{rqm10}). However, the structure of Eq.~(\ref{chireg}) is much more general: the $\mbox{sgn}(\omega)$ is linked to the exact SU(2) exponent we will obtain below in Eq.~(\ref{SStau}). And the $|\omega|$ correction term is similarly robust, and is related to the Schwarzian operator with $h=2$, as in 
Eq.~(\ref{syks5}) (in Section~\ref{sec:bh_fluc}, we will relate this $h=2$ mode to the boundary graviton in the holographic dual.)
As shown in Fig.~\ref{fig:tJ_ED_chi}, the form in Eq.~(\ref{chireg}) provides a good fit to the numerical susceptibility of the $p=0$ SU(2) model, apart from the low frequency peak associated with spin glass order. We can therefore conclude that the spin glass order $q \approx 0.02$ is weak, and there is clear evidence of the SY spin liquid behavior at intermediate energy scales in the SU(2) random exchange quantum magnet.

A theory for the quantum spin glass state can be obtained using bosonic spinons, and the spin glass order appears when the bosonic spinons condense \cite{GPS1,GPS2}. Such a theory is applicable when $q_{EA}$ is large, and yields $\chi''_\text{reg}(\omega) \sim \omega$ in Eq.~(\ref{eq:ChiQEADelta}) at small $|\omega|$ after an assumption of marginal stability in the replica symmetry breaking structure. More recently, the onset of spin glass order has been studied \cite{Christos:2021wno} using the fermionic spinon large $M$ theory of Section~\ref{sec:SY}. Such a theory yields an estimate of the critical temperature to spin glass order
\beq
T_{\rm sg} \sim J \exp \left(- \sqrt{M\pi}\right)\,, \label{Tsg}
\eeq
and also has $\chi''_\text{reg}(\omega) \sim \omega$ for $|\omega| < \omega_{\ast}$. The fermionic spinon theory describes the crossover above the frequency $\omega_\ast = J q_{EA}$ to the spin liquid spectrum in Eqs.~(\ref{chireg}) or (\ref{rqm10}). The exponential factor in Eq.~(\ref{Tsg}) is small even for $M=2$, $e^{- \sqrt{2 \pi}} = 0.0815\ldots$, and this could be the justification for the applicability of the large $M$ theory to the SU(2) case.

%%%%%%%%%%%%%%%%%%%%%%

\subsection{RG analysis of the SU(2) model}
\label{sec:SYRG}

We now turn to an analytic study of the SU(2) model, as this will help us understand the structure of non-zero frequency spin susceptibility observed in the numerics, as described by Eq.~(\ref{chireg}).

We present here a systematic RG procedure to analyze the problem defined by Eqs.~(\ref{ZJ}) - (\ref{defQ}). We begin by assuming that there is a critical solution in which $Q(\tau)$ has a power-law decay in time. Notice that this is similar to the assumption made for the SYK model in Eq.~(\ref{syk3}): in that case we were able to solve the self-consistency problem exactly at low energies. That will not be possible here, and we will have to introduce an $\epsilon$-expansion defined below.
We assume the power-law decay 
\begin{equation}
Q(\tau) \sim \frac{\gamma^2}{|\tau|^\alpha}\,, \label{Qgamma}
\end{equation}
and postpone consideration of the self-consistency condition. 
Then, we have to solve the well-defined problem of computing $\overline{Q}(\tau)$ from Eq.~(\ref{defQb}), given the $Q(\tau)$ in Eq.~(\ref{Qgamma}). 

This problem can be reduced to the solution of a quantum impurity problem, sometimes called the Bose-Kondo problem \cite{Sengupta2000,Beccaria:2022bcr,Cuomo:2022xgw,Nahum:2022fqw,Weber:2022ada}. We begin by decoupling the $\vec{S}(\tau) \cdot \vec{S}(0)$ interaction in Eq.~(\ref{ZJ}) with a 
bosonic field $\phi_a$, $a = 1 \ldots 3$. We assume that there is a bosonic `bath' field that lives in $d$ spatial dimensions, $\phi_a(x, \tau)$, and the decoupling field is $\phi_a (x=0, \tau)$. Then the path integral for $\mathcal{Z}_J$ in Eq.~(\ref{ZJ}) reduces to the solution of the following Bose-Kondo Hamiltonian of a $S=1/2$ spin $S_a$ coupled to a bosonic scalar field $\phi_a (x, \tau)$:
\begin{eqnarray}
H_{\rm imp} & =&  \gamma S_a \, \phi_a (0)  + \frac{1}{2} \int d^d x \left[ \pi_a^2 + (\partial_x \phi_a)^2 \right]\,. \label{SY30}
\end{eqnarray}
Here $\pi_a$ is canonically conjugate to the field $\phi_a$, and
$\phi_a (0) \equiv \phi_a (x=0)$. We identify $Q (\tau)$ with temporal correlator of $\phi_a (0)$, and then from Eq.~(\ref{Qgamma}) we conclude that we need $\alpha = d-1$.

We now wish to determine the properties of the theory $H_{\rm imp}$ in a renormalized perturbation expansion in the coupling $\gamma$. A simple determination of scaling dimensions at tree level shows that $\gamma$ has scaling dimension $(3-d)/2$, and so an expansion in powers of $\gamma$ is equivalent to an RG expansion in
\beqn
\epsilon = 3-d=2-\alpha\,. \label{epsalpha}
\eeqn
Such a computation can be performed \cite{SBV1999,Sengupta2000,Vojta2000,Sachdev2001,Si0,Beccaria:2022bcr,Cuomo:2022xgw,Nahum:2022fqw,Weber:2022ada} while imposing the fermion constraint in Eq.~(\ref{rqm2}) for SU(2) {\em exactly}, and yields the two-loop $\beta$ function,
\begin{align}
\beta (\gamma) &= - \frac{\epsilon}{2} \gamma + \gamma^3 - \gamma^5 + \ldots \label{betagamma}
\end{align}
This has a stable fixed point at $\gamma^{\ast 2} = \epsilon/2 + \epsilon^2/4  + \ldots$ which provides the needed critical theory of $\mathcal{Z}_J$ with the interaction in Eq.~(\ref{Qgamma}). 

To solve the self-consistent theory, we need to compute $\overline{Q}(\tau)$ in Eq.~(\ref{defQb}) at this fixed point. The scaling dimension of the spin operator $\mbox{dim}[\vec{S}]$ can be computed by standard RG methods order-by-order in $\epsilon$, but we encounter an unexpected simplification. Because of the quantized Berry phase (Wess-Zumino-Witten) term, the renormalization of the coupling $\gamma$ is given only by the wavefunction renormalization, and this fixes the scaling dimension of the spin-operator at the non-trivial fixed point of the $\beta$ function: we find \cite{Vojta2000,Sachdev2001}
\beqn
\mbox{dim}[\vec{S}] =\epsilon/2, \label{dimS}
\eeqn
exact to {\em all\/} orders in $\epsilon$. 
This implies the correlator
\begin{align}
\overline{Q} (\tau) = \frac{1}{3} \left\langle \vec{S} (\tau) \cdot \vec{S}(0) \right\rangle &\sim \frac{1}{|\tau|^{2-\alpha}} \,. \label{Qexact}
\end{align}

Finally, we impose the self-consistency condition in Eq.~(\ref{defQ})
at least at the level of the exponent. 
Comparing Eqs.~(\ref{Qgamma}) and (\ref{Qexact}), we conclude that the self-consistent value is $\alpha=1$. Note that this value is well outside the domain of applicability of the $\epsilon$ expansion, given Eq.~(\ref{epsalpha}). Nevertheless, given that Eq.~(\ref{dimS}) has been obtained to all orders in $\epsilon$, the only requirement for the validity of Eq.~(\ref{Qexact}) is the continued existence of the non-trivial fixed point of the $\beta$ function at $\epsilon$ of order unity.
The self-consistent spin correlator is therefore
\begin{align}
\left\langle \vec{S} (\tau) \cdot \vec{S}(0) \right\rangle &\sim \frac{1}{|\tau|} \,. \label{SStau}
\end{align}
Comparing with the large $M$ result in Eq.~(\ref{rqm7}), we find perfect agreement between the large $M$ and RG exponents.

As discussed in Section~\ref{sec:SYSU2}, the ground state of Eq.~(\ref{rqm1}) is actually a spin glass for SU(2) spins. The
above analysis obtaining the result in Eq.~(\ref{Qexact}) is certainly correct
for SU(2), and applies exactly to the Bose-Kondo impurity model defined by Eq.~(\ref{SY30}) for small $\epsilon$.
Recent studies have shown \cite{Beccaria:2022bcr,Cuomo:2022xgw,Nahum:2022fqw,Weber:2022ada} that the fixed point is not present at large $\epsilon$, and this is consistent with appearance of spin glass order.

Despite the direct inapplicability of the RG to the SU(2) model in Eq.~(\ref{rqm1}), the analysis presented here turns out to be very useful. A closely related RG applies to the SU($M$) generalization considered in Section~\ref{sec:SY} \cite{Joshi:2019csz}, and from this we can conclude that there are no corrections to the exponent in Eq.~(\ref{dimS}) (which is related to the exponent in Eq.~(\ref{rqm7})) at all orders in $1/M$.
Moreover extensions of the RG of the Bose-Kondo model obtained here will apply to the correlated electrons models to be considered in the following sections: to the superspin Bose-Fermi-Kondo model in Section~\ref{sec:tJRG}, and the Bose-Fermi-Kondo model in Section~\ref{sec:KHRG}.

%%%%%%%%%%%%%%%%%%%%%%%%%%%%%%%%%%%%%%%%%%%%%
%               RANDOM TJU
%%%%%%%%%%%%%%%%%%%%%%%%%%%%%%%%%%%%%%%%%%%%%

\section{Random exchange $t$-$U$-$J$ Hubbard models}
\label{sec:tJU}

In the following, we will consider models of itinerant electrons on a fully connected lattice with a strong local interaction 
and random exchange constants. One such example is the `$t$-$U$-$J$' model in which random $J_{ij}$'s are 
added to the Hubbard model with random hoppings: 
\begin{align}
   \label{eq:tUJmodel}
& H_{tUJ} = - \frac{1}{\sqrt{N}} \sum_{i,j=1}^N \sum_{\alpha=1}^{M} 
t_{ij} \, c_{i \alpha}^\dagger c_{j \alpha}  - \mu \sum_{i\alpha} c_{i \alpha}^\dagger c_{i \alpha} \\
& + \frac{U}{2} \sum_i \left(\sum_\alpha c_{i \alpha}^\dagger c_{i \alpha} - M/2  \right)^2 
+ \frac{1}{\sqrt{N}} \sum_{1\leq i < j \leq N} J_{ij} \, \vec{S}_i \cdot \vec{S}_j \,. \nonumber
\label{eq:tUJ}
\end{align}
In this expression, we have introduced $M$ `colors' of fermions, so that the model has $U(M)=U(1)\times SU(M)$ symmetry, 
corresponding to an extension of the spin symmetry to SU($M$). The usual SU(2), $S=1/2$ Hubbard model corresponds to 
$M=2$ ($\alpha=\uparrow,\downarrow$). The electron spin operators, $\vec{S}_i=\sum_{\alpha\beta} c^\dagger_\alpha (\vec{\sigma}_{\alpha\beta}/2) c_\beta$, with $\vec{\sigma}/2$ the $M^2-1$ generators of SU($M$) ($\vec{\sigma}$ are the Pauli matrices 
for $M=2$). 
As before, the $t_{ij}$'s and $J_{ij}$'s are drawn from distributions with zero mean and 
variances $\overline{t_{ij}^2}=t^2$, $\overline{J_{ij}^2}=J^2$; however, as we will note below, closely related results also apply to the case where the $t_{ij}$ are non-random, and determine an electronic dispersion $\epsilon_{\bf k}$.
Note a change in notation from the SYK model above: $U$ 
designates here the on-site repulsion while the variance $J$ of the random bonds is more directly analogous to the 
variance of the random SYK interactions.
Also note that the chemical potential $\mu$ is defined with reference to the half-filled case ($M/2$ electrons per site). 

We can also consider the $t$-$J$ limit of this model~\cite{Parcollet1}, which reads: 
\begin{eqnarray}
&& H_{tJ} = - \frac{1}{\sqrt{N}} \sum_{i,j=1}^N \sum_\alpha t_{ij} \, 
\P c_{i \alpha}^\dagger c_{j \alpha} \P  - \mu \sum_{i\alpha} c_{i \alpha}^\dagger c_{i \alpha} \nonumber \\
&&~~~~~~~~+ \frac{1}{\sqrt{N}} \sum_{1\leq i < j \leq N}  J_{ij} \, \vec{S}_i \cdot \vec{S}_j \, 
\label{tJ1}
\end{eqnarray}
in which the operator $\P$ enforces a Gutzwiller-type projection such that the total number of fermions on each site is at most $M/2$: 
\begin{equation}
  \sum_\alpha c_{i \alpha}^\dagger c_{i \alpha} \leq \frac{M}{2}\,\,\,,\,\,\,\forall ~i.
\end{equation}
At half-filling ($\mu=0$) this reduces to the random-bond Heisenberg (SY) model of the previous section. 

\subsection{Effective local action}
\label{sec:EDMFT}

In the thermodynamic limit $N\rightarrow\infty$, the calculation of the single-particle 
Green's function and self-energy of this model, as well as that of the local spin-spin correlator, 
reduces to a local effective action subject to a 
self-consistency condition. This corresponds to the (extended) dynamical mean-field theory construction 
(EDMFT)~\cite{SG95,Si_1996,smith_si_prb_2000,Chitra2000,DMFT}, 
which is exact for these random fully-connected models. 
The term `extended' is commonly used to indicate that the mapping 
involves a self-consistency over both single-particle and two-particle correlation functions. 
When considering the system outside the spin-glass phase, all local correlators are self-averaging and 
this mapping is most easily derived following the cavity construction, similar to Sec.~\ref{sec:matrix}. 
We skip the details here, since the reasoning is completely analogous to the one in that section.
One obtains the single-site effective action: 
\begin{eqnarray}
\mathcal{S}_{\mathrm{tUJ}} &=& 
\int d \tau \sum_\alpha c_\alpha^\dagger (\tau) \left( \frac{\partial}{\partial \tau} 
-\mu \right) c_\alpha (\tau)  \nonumber \\
&+&\frac{U}{2}\int d\tau \Bigl(\sum_\alpha c_{\alpha}^\dagger c_{\alpha} - M/2 \Bigr)^2  \nonumber \\
   &+& \int d\tau d \tau' \Delta (\tau - \tau') \sum_\alpha c_\alpha^\dagger (\tau) c_\alpha (\tau') \nonumber \\ 
 &-& \frac{1}{2}\int d\tau d \tau' {\cal J} (\tau - \tau') \vec{S} (\tau) \cdot \vec{S} (\tau').
\label{eq:Seff_tUJ}
\end{eqnarray} 
From this action we have to determine the Green's function and spin correlator:
\begin{eqnarray}
G(\tau - \tau') &\equiv& - \frac{1}{M}\sum_\alpha \left\langle c_{\alpha}(\tau) c^{\dagger}_\alpha (\tau') 
\right\rangle_{\mathcal{S}_{tUJ}} \nonumber \\
\chi (\tau - \tau') &\equiv&  \frac{1}{M^2-1} 
\left\langle \vec{S} (\tau) \cdot \vec{S} (\tau') \right\rangle_{\mathcal{S}_{tUJ}}\,, 
\label{eq:def_local_correlators}
\end{eqnarray}
and impose the self-consistency condition that results from the cavity construction: 
\begin{equation}
    \Delta(\tau)=t^2\,G(\tau)\,\,\,,\,\,\,
    {\cal J}(\tau) = J^2\,\chi(\tau).
    \label{eq:EDMFT}
\end{equation}
The electronic self-energy can be defined by reference to the non-interacting system $U=J=0$ (the random 
matrix model of Sec.~\ref{sec:matrix}) as $G^{-1}_{ij}(i\omega_n)=i\omega_n+\mu-t_{ij}-\Sigma_{ij}$, for a 
given sample $\{t_{ij}\}$. 
In the infinite-volume limit $N\rightarrow\infty$, the self-energy becomes local $\Sigma_{ij}=\Sigma_{ii}\delta_{ij}$ 
and self-averaging, when not in the spin-glass phase. The local Green's function $G_{ii}$ is also self-averaging and is 
related to $\Sigma$ by: 
\begin{eqnarray}
    G_{ii}(i\omega_n)&=&\sum_\lambda |\langle i|\lambda\rangle|^2 G(i\omega_n,\varepsilon_\lambda) \nonumber \\
    &\rightarrow& \int \rho_0(\varepsilon) G(i\omega_n,\varepsilon) = G(i\omega_n),
\end{eqnarray}
with $\rho_0$ the semi-circular density of states defined in Sec.~\ref{sec:matrix} and 
\begin{equation}
G(i\omega_n,\varepsilon)\,=\,\frac{1}{i\omega_n+\mu-\varepsilon-\Sigma(i\omega_n)}
\label{eq:G_energyresolved}
\end{equation}
is the Green's function in the basis of the single-particle states of the free system at an energy $\varepsilon$. 
The self-energy $\Sigma$ coincides with that of the effective action Eq.~(\ref{eq:Seff_tUJ}) and hence reads:
\begin{equation}
    \Sigma(i\omega_n)=i\omega_n+\mu-t^2\Delta(i\omega_n)-G^{-1}(i\omega_n).
\end{equation}
Substituting this expression into Eq.~(\ref{eq:G_energyresolved}) and performing the Hilbert transform of $\rho_0$, one recovers  
the self-consistency condition $\Delta=t^2 G$ \cite{DMFT}. 

When a spin-glass phase exists, self-averaging of the local observables does not hold inside the ordered phase. 
A mapping onto a local effective action still applies however after introducing $n$ replicas and performing the 
average of $(Z^n-1)/n$ over the $t_{ij}$ and $J_{ij}$ random variables, after which the $n\rightarrow 0$ limit must be taken also  
allowing for the possibility of replica symmetry breaking. We do not write these equations in detail here, and 
refer the reader to Refs.~\cite{GPS1,GPS2}. 

In order to make contact with the (E)DMFT literature, we have used in this section 
notations that are rather standard in this field. In particular $\Delta(\tau)$ is the dynamical mean-field 
(quantum generalization of the Weiss field), describing the hybridisation between a local site and 
its self-consistent bath. In the following, we will often use a somewhat different notations which are 
more commonly used in the SY/SYK literature, such as $\Delta\rightarrow t^2\overline{R}$, 
$G\rightarrow \overline{R}$, ${\cal J}\rightarrow J^2Q$ and $\chi\rightarrow \overline{Q}$. 

We also note that the single-site effective action permits a spin glass phase after we include replica off-diagonal components of the correlators \cite{GPS1,GPS2}. In the replica diagonal components ${\cal J} (\tau \rightarrow \infty) \neq 0$ at zero temperature. Naively, such a non-zero limit signals a problem in the replica diagonal action in  Eq.~(\ref{eq:Seff_tUJ}), as the expectation value of the last term in the action diverges as $\sim \beta^2$ as $\beta \rightarrow \infty$, implying a divergent ground state energy. However, this problem is cured upon including the replica off-diagonal components and taking the replica $n \rightarrow 0$ limit \cite{RSY95}. This issue highlights the difficulty in interpreting the EDMFT framework in the magnetically ordered phase for non-random systems \cite{Si1,Si2,SiRMP,Pankov2002}.
%%%%%%%%%%%%%%%%%%%%%%%%%%%%%%%%%%

\subsection{The SU(2) Hubbard model at half-filling}
\label{sec:ChaHubbard}

\begin{figure}
\begin{center}
\includegraphics[width=3.5in]{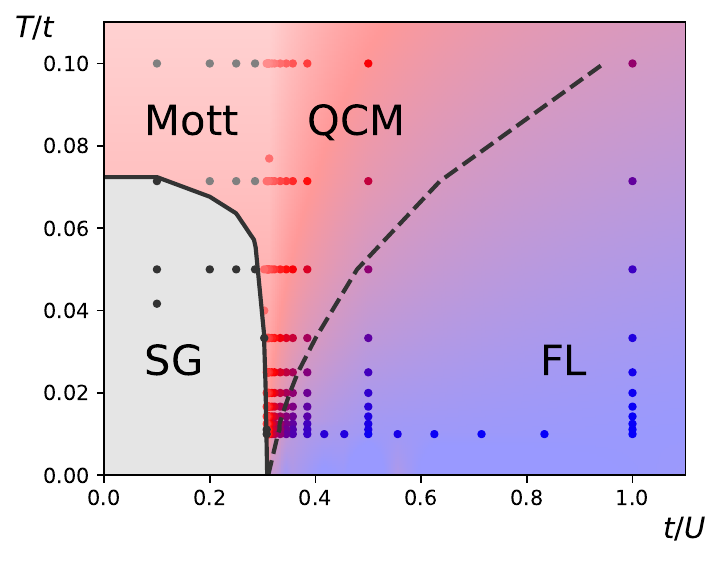}
\end{center}
\caption{Phase diagram of the spin-$1/2$ half-filled random exchange $t$-$J$-$U$ model. 
At low temperature, a quantum critical point separates the spin glass phase (SG) from a Fermi liquid phase (FL).
The background color corresponds to the fitted power-law exponent of the local spin correlation function $\chi(\tau)\sim 1/\tau^{2\Delta}$, 
with $2\Delta\simeq 1$ in the quantum critical metal (QCM) (red), and  $\Delta= 1$ in the Fermi liquid (blue). At high temperature and $U$, one obtains a Mott insulator.
Reproduced from \cite{Cha_2020}.
\label{fig:tJU_phasediagram_vsU}
}
\end{figure} 

The SU(2) $t$-$U$-$J$ model in Eq.~(\ref{eq:tUJmodel}) was studied at half-filling in the EDMFT framework \cite{Cha_2020} 
--- for a previous study in the large-$M$ limit, see \cite{Florens2013}.
The phase diagram is reproduced in Fig. \ref{fig:tJU_phasediagram_vsU} as a function of $t/U$ and temperature, as obtained by a quantum Monte Carlo solution of the EDMFT equations (cf Section \ref{sec:numerical_methods}).
A quantum critical point (QCP) at $U=U_c$ separates a Fermi liquid phase at small $U$ from an insulator at large $U$, 
which orders into a spin glass phase at low temperature.
At the quantum critical point, the spin correlations decays as $\chi(\tau)\sim 1/\tau$, as in the large-$M$ limit of the SY model, 
while it is  the expected $\chi(\tau)\sim 1/\tau^2$ in the Fermi liquid phase.

The electronic self-energy $\Sigma$ is strongly affected by the QCP. While it takes its regular form in the Fermi liquid,  the coherence temperature vanishes at the QCP,
where a linear temperature behaviour $\mbox{Im}\, \Sigma(\omega =0, T) \propto T$ is found numerically.
As detailed below in Sec.~\ref{sec:transport}, this behaviour leads at the QCP 
to a $T$-linear dependence of the resistivity, which is smaller than the MIR value.
Furthermore, in the accessible range of temperatures, the frequency dependence of the self-energy is compatible with a Marginal Fermi liquid form.
Finally, we note a theoretical study \cite{Tarnopolsky:2020spd} analyzing the metal-insulator transition at half-filling, related to the finite doping theoretical models that are described in Section~\ref{sec:dopedtJ}.

%%%%%%%%%%%%%%%%%%%%%%%%%%%%%%5

\subsection{The SU(2) Hubbard model away from half-filling}
\label{sec:tJnumerics}

Section~\ref{sec:ChaHubbard} has shown that the Hubbard model exhibits a novel phase transition at half-filling: between a Fermi liquid at small $U/t$, and a metallic spin glass at large $U/t$. Next we turn to the case with hole-doping $p$ away from half-filling. Here, we assume throughout that $U/t$ is large, so that at $p=0$ we obtain the insulating spin glass phase which was described in Section~\ref{sec:SYSU2}. We review numerical studies \cite{Otsuki2013,Shackleton2021,Dumitrescu2021} showing that the spin glass order survives in a metallic state up to a critical doping $p=p_c$, and that there is a Fermi liquid for $p > p_c$. (We note an exactly diagonalization study \cite{Kumar21} which presents evidence for the spin glass transition from quasiparticle spectra.) The critical point at $p=p_c$ displays a SYK-like criticality, with some similarities to  the $U=U_c$ critical point at $p=0$ described in Section~\ref{sec:ChaHubbard}. Analytic analyses of the $p>0$ Hubbard model appear next in Section~\ref{sec:dopedtJ}.

A recent study \cite{Shackleton2021}  approached the large $U$ and $p \geq 0$ Hubbard model in the $t$-$J$ model framework by performing exact diagonalizations of fully connected 
clusters of $N$ sites, up to $N=18$, for a fixed sample of random hopping amplitudes and exchange constants, then taking averages or 
histograms over samples. This study confirms the existence of a spin-glass phase at low doping, which survives up to 
$p_c\simeq 0.3$ (in agreement with earlier analytic arguments \cite{Joshi:2019csz} to be presented in Sections~\ref{sec:JoshiM} and \ref{sec:tJRG}). Their result for the local spin response function 
$\chi{''}(\omega)$ was displayed in Fig.~\ref{fig:tJ_ED_chi}. 
The spin-glass phase is signalled by a sharp low frequency peak in $\chi{''}(\omega)$, which is absent for $p>p_c$, and the 
spin-fluctuation spectrum close to the critical point is seen to be well approximated by the large-$M$ SYK theory of Section~\ref{sec:SY}. 
These authors also computed thermodynamic properties (entropy, specific heat and entanglement entropy) as a function of temperature 
and found that the specific heat coefficient $\gamma=C/T$ displays a maximum as a function of doping for $p\simeq p_c$.

\begin{figure}[t]
\begin{center}
\includegraphics[width=3.5in]{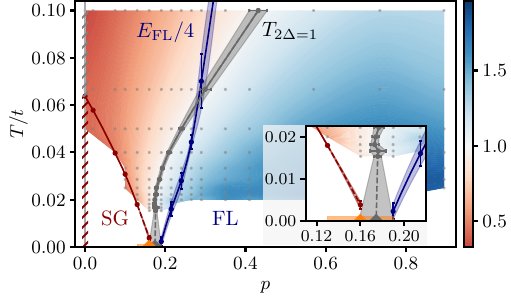}
\end{center}
\caption{Phase diagram \cite{Dumitrescu2021} of the spin-$1/2$ doped random exchange $t$-$U$-$J$ model, as obtained by a quantum Monte Carlo solution of the EDMFT equations. 
FL indicates Fermi liquid, while SG is a metallic spin glass for $p \neq 0$.
The background color corresponds to the fitted power-law exponent 
of the local spin correlation function $\chi(\tau)\sim 1/\tau^{2\Delta}$ (color scale on the right).
Along the dashed grey line, SYK behaviour $2\Delta\simeq 1$ is found. 
A linear-in-$T$ resistivity is obtained in the quantum-critical region with a resistivity which becomes lower than the MIR resistivity.
(Inset) Zoom close to the quantum critical point.
\label{fig:tJU_phasediagram}
}
\end{figure}

\begin{figure}[t]
\begin{center}
\includegraphics[width=3.5in]{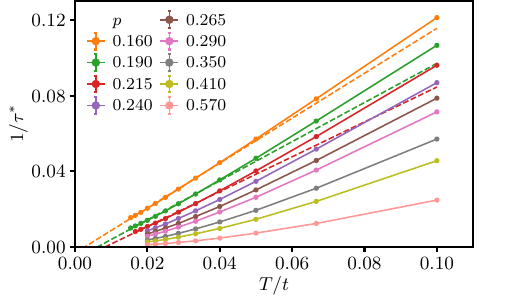}
\end{center}
\caption{Inverse single electron excitations lifetime $1/\tau^*$ as a function of temperature $T$
  in the spin-$1/2$ doped random exchange $t$-$U$-$J$ model, for different doping $p$. A Planckian behaviour 
  (\ref{tau_star_Planckian}) is observed close to the quantum critical point.
\label{fig:tJU_tau_star}
}
\end{figure}

A different and complementary approach \cite{Otsuki2013} was used recently \cite{Dumitrescu2021}. The 
EDMFT equations of Section~\ref{sec:dopedtJ} were solved using the Quantum Monte Carlo algorithms reviewed 
in Section~\ref{sec:numerical_methods}, corresponding to a direct solution in the thermodynamic limit 
$N=\infty$ for disorder-averaged observables. The model considered \cite{Dumitrescu2021} is actually a finite-$U$ random 
exchange model, with $U/t$ large enough so that the physics of a doped Mott insulating spin-glass is being captured. 
The phase diagram obtained in this study is displayed on Fig.~\ref{fig:tJU_phasediagram}. 
The spin-glass phase itself (requiring replica off-diagonal terms) was not studied in this work, but the location of 
the critical boundary in the $T$-$U$ plane was identified from the criterion $J\chi_{\mathrm{loc}}=1$. The $T=0$ critical doping was 
found to be at $p_c\simeq 0.17$ for the finite value of $U/t$ studied, in contrast to the higher value $p_c\simeq 0.3$ for the $U=\infty$ model. 
Consistently with the exact diagonalization study \cite{Shackleton2021}, the local spin dynamics at the critical point is of SYK type with $\chi(\tau)\propto 1/\tau$. 
The self-energy obeys interesting scaling properties near the critical point: the imaginary-time data for different temperatures can be collapsed onto: 
\begin{equation}
  \frac{\Sigma(\tau)}{\Sigma(\beta/2)}=\frac{e^{2\pi\mathcal{E}(\tau/\beta-1/2)}}{(\sin \pi\tau/\beta)^\nu}
\end{equation}
corresponding to the conformally invariant scaling form for the real-frequency scattering rate:
\begin{eqnarray} \label{eq:skewed_sigma}
    &-\frac{1}{\pi}\mathrm{Im}\Sigma(\omega+i0^+)=\lambda T^\nu \Phi_{\nu,\mathcal{E}}\left(\frac{\omega}{T}\right),\\
    &\Phi_{\nu,\mathcal{E}}\left(x\right)=\cosh\frac{x}{2}\,\Bigl |\Gamma\left[\frac{1+\nu}{2}+ i\frac{x}{2\pi}+i\mathcal{E}\right] \Bigr|^2. \nonumber
\end{eqnarray}
The exponent $\nu$ at criticality was estimated to be in the range $\nu\simeq 0.6 - 0.8$. Note that a value of $\nu$ smaller than unity implies 
that the lifetime of single-electron excitations (inverse width of the spectral function) satisfies Planckian $T$-linear behaviour:
\begin{equation}\label{tau_star_Planckian}
    \frac{1}{\tau^*} \equiv -Z\, \mathrm{Im}\Sigma(i0^+) = c \frac{\hbar}{k_B T}
\end{equation}
since, as detailed below in Sec.~\ref{sec:transport}, $Z=\left(1-\partial\mathrm{Re}\Sigma(\omega)/\partial\omega)\big|_{\omega=0}\right)^{-1}$ vanishes as $Z\propto \lambda T^{1-\nu}$ at low-$T$. 
The overall coupling constant $\lambda$ cancels in the expression of $\tau^*$ to dominant order, hence the prefactor $c$ is generically of order unity. 
This quantity is displayed on Fig. \ref{fig:tJU_tau_star}.
The spectral asymmetry $\mathcal{E}$ was found to be non-zero but temperature-dependent over some extended range of $T$.  
Whether there is an intrinsic particle-hole asymmetry of the scaling function at criticality down to $T=0$ is an open question. 

The metallic state is a Fermi liquid for $p>p_c$, satisfying Luttinger theorem with a large Fermi energy associated with a fermion density of  $1-p$, see Section~\ref{sec:Kondolut} for a discussion of the Luttinger theorem in disordered systems; for the present system, it is expressed by the relation $\mu-\mathrm{Re}\Sigma(0)=\varepsilon_F$ at $T=0$, with $\varepsilon_F$ the Fermi energy of 
the non-interacting system (random matrix model) for a density $n=1-p$. 
When solving the EDMFT equations without allowing for spin-glass ordering, a sudden breakdown 
of this relation is found for $p<p_c$~\cite{Otsuki2013,Dumitrescu2021}, signalling a breakdown of the Luttinger theorem. 
These solutions correspond to a metastable state with unquenched local magnetic moments. 
These local moments order into a spin-glass which is the actual stable phase. The finite size exact diagonalisation 
results~\cite{Shackleton2021} suggest that the Fermi energy may collapse to a small one of volume $p$ in this 
metallic spin-glass phase. This fascinating possibility awaits confirmation from an infinite-volume solution of the EDMFT 
equations inside the 
spin-glass phase. 

%%%%%%%%%%%%%%%%%%%%%%%%%%%%%%%%%%

\subsection{Doped $t$-$J$ model: analytical insights}
\label{sec:dopedtJ}

We now extend the analytic considerations of Sections~\ref{sec:SY} and \ref{sec:SYRG} from the undoped quantum magnet at $p=0$ to the non-zero doping $t$-$J$ model with $p \neq 0$. This will provide insight to the numerical results presented in Section~\ref{sec:tJnumerics} for the doped Hubbard model. This analysis will be carried out in the $U \rightarrow \infty$ limit, employing the  $t$-$J$ model in Eq.~(\ref{tJ1}).

In the SU(2) ($M=2$) case, the Hilbert space of the $tJ$ model on each site consists of 3 states
\beq
\left| 0 \right\rangle, \quad c_\uparrow^\dagger\left| 0 \right\rangle, \quad c_\downarrow^\dagger \left| 0 \right\rangle\,. \label{tJ2}
\eeq
We will treat these 3 states in close analogy to the 2 spin states of the random magnet in Eq.~(\ref{rqm1}) \cite{VojtaFritz2004,FritzVojta2004}. Apart from the increase in the number of states, a crucial difference is the Fermi statistics of the electron operator, which requires that the 3 states are components of a `{\em superspin}'. However, there remains a choice on whether the spinful or spinless component of the superspin is fermionic. In an exact treatment of the problem, either choice is permitted and should lead to equivalent results; but in approximate treatments, one or the other choice may be superior, and it is often useful to exploit this freedom. For now, we will present our discussion by representing the superspin by a spinless boson $b$ (the holon) and a spinful fermion $f_\alpha$ (the spinon):
\beqn
\left| 0 \right\rangle \Rightarrow b^\dagger \left|v \right \rangle \quad , \quad c_\alpha^\dagger \left| 0 \right\rangle \Rightarrow f^\dagger_\alpha \left|v \right \rangle\,.
\label{tJ3}
\eeqn
The physical states are obtained when the constraint 
\beqn
f_\alpha^\dagger f_\alpha + b^\dagger b = 1 \label{tJ4}
\eeqn
is obeyed (we implicitly sum over SU(2) indices in this discussion for $M=2$). Hence, the physical states are invariant under the U(1) gauge transformation which generalizes that in Section~\ref{sec:SY} $f_\alpha \rightarrow f_\alpha e^{i \phi}$, $b \rightarrow be^{i \phi}$, while individual spinon and holon excitations carry U(1) gauge charges. At the moment, the fractionalized representation, and associated emergent gauge symmetry, is just a convenient exact description of the Hilbert space. But we will see later in Sections~\ref{sec:JoshiM} and \ref{sec:tJRG} that the fractionalized operators yield a simple way to understand the exponents at a non-Fermi liquid critical point as a realization of a critical doped spin liquid.

The physical electron ($c_\alpha$) and spin ($\vec{S}$) operators can be viewed as rotation operators of the superspin:
\beq
c_\alpha = b^\dagger f_\alpha^{}, \qquad \vec{S} = \frac{1}{2} f_\alpha^\dagger \vec{\sigma}_{\alpha\beta} f_\beta^{} \,. 
\label{defcS1}
\eeq
If we combine these operators with an operator $V$ which measures the electron density
\bea
V &=& b^\dagger b^{} + \frac{1}{2}f^\dagger_\alpha f_\alpha^{} \nonumber \\
&=& 1 - \frac{1}{2} c_\alpha^\dagger c_\alpha \,, \label{tJ5}
\eea
we obtain all the generators of the supergroup SU($1|2$). The notation indicates that this group acts on a superspin with 1 bosonic component, $b^\dagger \left|v \right \rangle$, and 2 fermionic components, $f^\dagger_\alpha \left|v \right \rangle$. These generators realize
the superalgebra SU($1|2$), which is

\begin{subequations}
\begin{align}
 \label{super} 
[S^a, S^b] &= i \epsilon_{abc} S^c  \\ 
\{c_\alpha , c_\beta \} &= 0 \\ 
   \{c_\alpha , c_\beta^\dagger\} &= \delta_{\alpha\beta} V +  \sigma^a_{\alpha \beta} S^a  \\ 
   [ S^a , c_\alpha ] &= - \frac{1}{2} \sigma^a_{\alpha\beta} c_\beta  \\ %[ S^a , c_\alpha^\dagger ] = \frac{1}{2} \sigma^a_{\beta\alpha} c_\beta^\dagger  \nonumber \\
  [S^a , V] &= 0  \\ 
  [V, c_\alpha ] &= \frac{1}{2} c_\alpha %  \qquad  [V, c_\alpha^\dagger ] = -\frac{1}{2} c_\alpha^\dagger\,. \nonumber \\ 
\end{align}
\end{subequations}
If we had made the opposite choice of using spinful bosonic spinons and spinless fermionic holons, we would have obtained the superalgebra SU($2|1$), which is isomorphic to SU($1|2$).

The effective local action associated with this model along the lines of Sec.~\ref{sec:EDMFT} can be viewed as 
that of a single SU($1|2$) superspin, in complete analogy with 
Eqs.~(\ref{ZJ})-(\ref{defQ}) for the self-consistent dynamics of a single SU(2) spin.
The local effective action can be written in terms of the spinon and holon fields as:
\begin{eqnarray}
&& \mathcal{Z}_{tJ} = \int \mathcal{D} f_\alpha (\tau) \mathcal{D} b (\tau) \mathcal{D} \lambda (\tau) e^{-\mathcal{S}_B -\mathcal{S}_{tJ}} \label{ZtJ}  \\
&& \mathcal{S}_B = \int d \tau \left[ f_\alpha^\dagger (\tau) \left( \frac{\partial}{\partial \tau} + i \lambda \right) f_\alpha (\tau) \right. 
\nonumber \\
&~&~~~~~~~~~~~~~~~\left. + b^\dagger (\tau) \left( \frac{\partial}{\partial \tau} + i \lambda \right) b (\tau) - i \lambda \right] \nonumber \\
&& \mathcal{S}_{tJ} = s_0\int d\tau f_\alpha^\dagger f_\alpha  %\nonumber\\ 
%&~& 
- \frac{J^2}{2} \int d\tau d \tau' Q (\tau - \tau') \vec{S} (\tau) \cdot \vec{S} (\tau') \nonumber \\
&&- t^2 \int d\tau d \tau' R (\tau - \tau') f_\alpha^\dagger (\tau) b(\tau) b^\dagger(\tau') f_\alpha (\tau') + \mbox{H.c.} \nonumber
%&~&   - t^2 \int d\tau d \tau' R (\tau - \tau') c_\alpha^\dagger (\tau) c_\alpha (\tau') + \mbox{H.c.} \,. \nonumber
\end{eqnarray} 
The action $\mathcal{S}_B$ is the Berry phase of a SU($1|2$) superspin, which we have expressed as the path integral over canonical bosonic and fermionic fields while imposing the constraint Eq.~(\ref{tJ4}) with the field $\lambda (\tau)$. The chemical potential $\mu$ of the $t$-$J$ Hamiltonian is now represented by the coupling $s_0$.
From this action we have to determine the correlators
\begin{eqnarray}
\overline{R}(\tau - \tau') &=& - \frac{1}{2} \left\langle c^{}_{\alpha} (\tau) c^{\dagger}_\alpha (\tau') \right\rangle_{\mathcal{Z}_{tJ}} \nonumber \\
\overline{Q} (\tau - \tau') &=&  \frac{1}{3} \left\langle \vec{S} (\tau) \cdot \vec{S} (\tau') \right\rangle_{\mathcal{Z}_{tJ}}\,, \label{defRQb}
\end{eqnarray}
analogous to Eq.~(\ref{defQb}). 
And then we impose the self-consistency conditions in Eqs.~(\ref{eq:def_local_correlators}) and (\ref{eq:EDMFT}), which take the form
\beqn
R(\tau) = \overline{R}(\tau), \qquad Q(\tau) = \overline{Q} (\tau)\,, \label{defRQ}
\eeqn
analogous to Eq.~(\ref{defQ}).

It is not possible to solve the self-consistent single-site quantum problem defined by Eqs.~(\ref{ZtJ})-(\ref{defRQ}) exactly. The following subsections will describe various theoretical expansions and numerical results, analogous to those discussed in Section~\ref{sec:rqm} for the random quantum magnet.

\subsubsection{SU($M$) symmetry: Fermi liquid large-$M$ limit}
\label{sec:tJPG}

A first approach \cite{Parcollet1} is to extend the SU($M$) large $M$ model of Section~\ref{sec:SY} by using fermionic spinons $f_\alpha$ with an index $\alpha = 1 \ldots M$, while the bosonic holons $b$ have no index. In this case, the constraints Eqs.~(\ref{rqm2}) and (\ref{tJ4}) become
\beqn
\sum_{\alpha=1}^M f_{i \alpha}^\dagger f_{i \alpha} + b_i^\dagger b_i = \frac{M}{2}\,
\label{tJ20}
\eeqn
on each site $i$; we are restricting to the case with self-conjugate representations of SU($M$) at half-filling, with $\kappa=1/2$. We also fix the doping density $p$ by
\beq
\frac{1}{N} \sum_i b_i^\dagger b_i = \frac{M p}{2}. \label{tJ21}
\eeq

This particular large $M$ limit is similar to that employed for non-random $t$-$J$ models \cite{LeeRMP,kotliar_largeN}, and has the crucial feature that the bosonic holons are strongly condensed at $T=0$.
Indeed, in the large $M$ limit, we may replace the boson by a number $b_i = \sqrt{Mp}$ obtained from the constraint in Eq.~(\ref{tJ21}). Then the fermions $f_\alpha$ have the same quantum numbers as an electron, with spin $S=1/2$ and charge $-1$. The effective theory of these electrons is a sum of the random matrix Hamiltonian $H_2$ in Eq.~(\ref{rmt1}), and the SYK Hamiltonian $H_4$ in Eq.~(\ref{syk1}). We will discuss very similar Hamiltonians in a different context in Section~\ref{sec:lattice} and defer a complete discussion until then. 

\begin{figure}
\begin{center}
\includegraphics[width=3.25in]{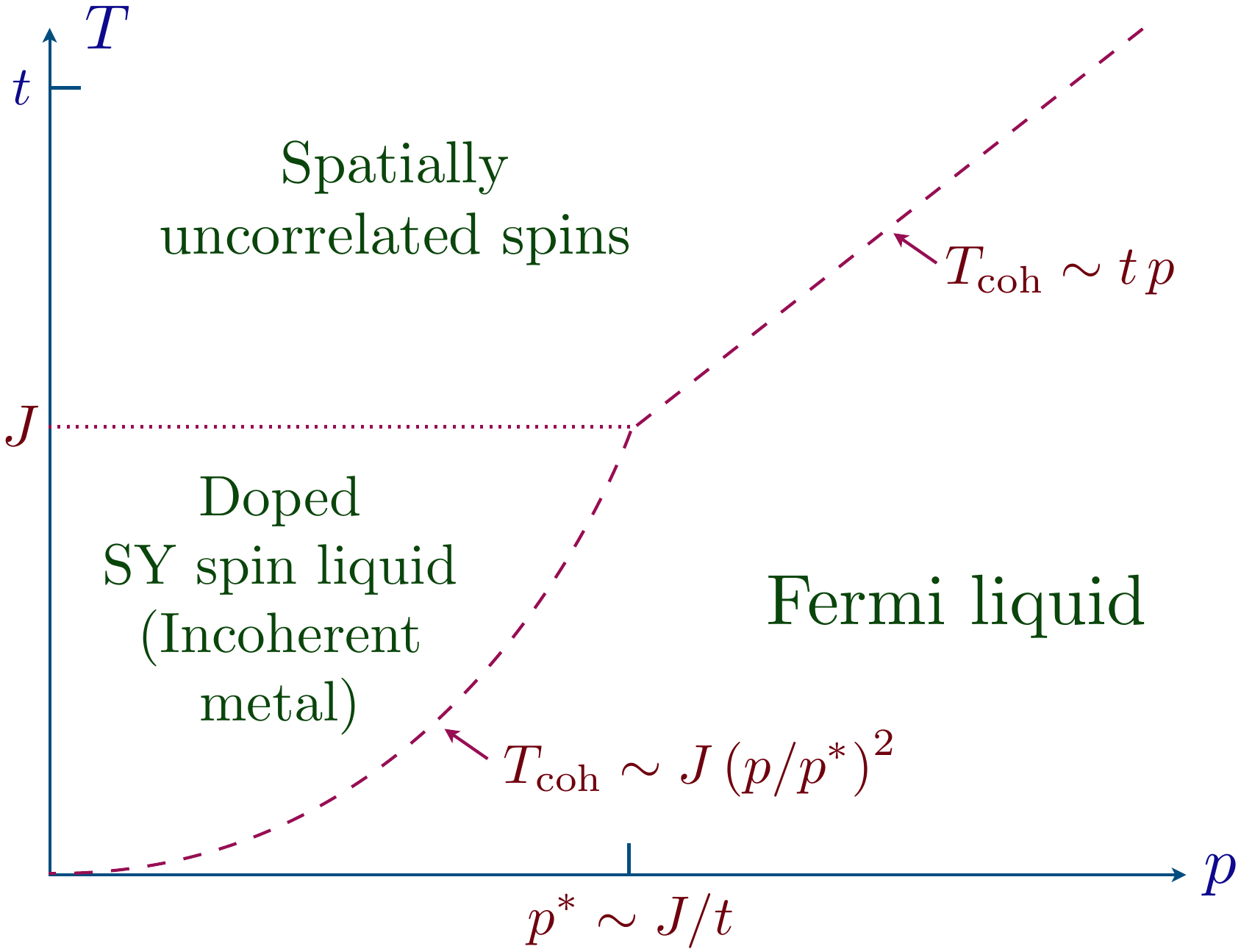}
\end{center}
\caption{Phase diagram \cite{Parcollet1} of the doped $t$-$J$ model in the large-$M$ limit with a condensed bosonic holon $b$. The SY spin liquid (incoherent metal) displays linear-in-$T$ resistivity with a large `bad metal' resistivity.  
\label{fig:tJlargeM}
}
\end{figure}

For now, we note a few important features of this large $M$ limit. 
The phase diagram \cite{Parcollet1} is displayed on Fig.~\ref{fig:tJlargeM}. 
At $p=0$, we have the SYK spin liquid state described in Section~\ref{sec:rqm}. At any non-zero $p$, because of the condensation of the holons $b$, 
we obtain a disordered Fermi liquid ground state, with quasiparticles moving with an effective hopping $tp$. These quasiparticles are present at a {\it large\/} Fermi energy below which there are states of $(1-p)/2$ electrons per spin. 
There is a characteristic doping $p^*\sim J/t$ which separates two different regimes with a distinct 
doping-dependence of the effective mass enhancement and spectral weight $Z$ of these quasiparticles. 
For $p>p^*$, the usual Brinkman-Rice \cite{BrinkmanRice} behaviour $m^*/m=1/Z\propto 1/p$ is recovered, as in the absence of random exchange couplings. 
In contrast for $p<p^*$, much heavier quasiparticles are found with $m^*/m=1/Z\propto (p^*/p)^2$. Correspondingly, the 
Fermi liquid coherence scale is $T_{\mathrm{coh}}\sim (pt)^2/J$ in this regime. 
Hence, the random exchanges strongly modify the usual Brinkman-Rice behaviour of the 
doped Mott insulator at low doping. 
For $p<p^*$, there is an interesting crossover at $T\gtrsim T_{\mathrm{coh}}$, above which non-Fermi liquid behaviour with spin-liquid 
local correlations of the SYK type are recovered (Fig.~\ref{fig:tJlargeM}). 
This regime corresponds to a `bad metal' with a resistivity larger than the MIR limit and, 
interestingly, depending linearly on temperature; see Section \ref{sec:transport} for a discussion of related models for which we define a proper notion of transport. 
The mechanism for this $T$-linear dependence is unusual. Indeed, in this regime the single particle scattering rate has the $\mathrm{Im}\Sigma\propto \sqrt{\omega}, \sqrt{T}$ 
dependence of the spinon self-energy characteristic of the SYK regime. Despite this, the resistivity is found to be linear in $T$, because the dispersion of the 
quasiparticles is negligible as compared to this large scattering rate, so that the conductivity as obtained from the Kubo formula is 
proportional to $1/(\mathrm{Im}\Sigma)^2\propto 1/T$. 

These conclusions can be drawn by examining the large-$M$ equation for the spinon Green's function $G_f$, which read~\cite{Parcollet1}: 
\begin{equation}
    G_f^{-1}=i\omega_n+\mu-\overline{\lambda}-(pt)^2 G_f - \Sigma_f(i\omega_n)
    \label{eq:G_tJlargeM}
\end{equation}
where $i \lambda = \overline{\lambda}$ at the saddle point, and  $\Sigma_f(\tau)=-J^2G_f^2(\tau)G_f(-\tau)$ as in the large-$M$ SY model. It is easy to see that the doping-induced term $(pt)^2 G_f$ 
is a singular perturbation that cuts-off the SYK behaviour. Indeed, substituting $\Sigma_f\propto\sqrt{J\omega}$ in the equation above, 
corresponding to $G_f\propto 1/\sqrt{J\omega}$, we see that a stable solution of this type can only exist for 
$(pt)^2/\sqrt{J\omega} \lesssim \sqrt{J\omega}$ which yields $\omega\gtrsim (pt)^2/J \sim T_{\mathrm{coh}}$, corresponding to the crossover regime 
described above. For $T,\omega\lesssim T_{\mathrm{coh}}$, the consistent solution of (\ref{eq:G_tJlargeM}) is a Fermi liquid.

\subsubsection{SU($M$) symmetry: non-Fermi liquid large-$M$ limit}
\label{sec:JoshiM}

We know from the numerical studies of the random quantum magnet discussed in Section~\ref{sec:SYSU2} that the actual ground state of the undoped model, $p=0$, is a spin glass, in contrast to the spin liquid appearing in Section~\ref{sec:tJPG}. It is reasonable to expect that this spin glass state survives for a range of non-zero $p$, and this has been confirmed by numerical studies discussed in Section~\ref{sec:tJnumerics}.
In the large $M$ method of Section~\ref{sec:tJPG} the boson $b$ condenses at any non-zero doping, and so the correlated spin liquid (or its associated spin glass state) is absent at $T=0$ away from the insulator. In this section, we will discuss an alternative large $M$ approach in which the boson need not condense at non-zero doping, and can instead form a SYK-like critical state.

We consider a large $M$ theory of a SU($M'|M$) superspin, in which large $M$ and $M'$ limit is taken with $k = M'/M$ fixed \cite{Joshi:2019csz,Tikhanovskaya:2020zcw}. This requires a theory of fermionic spinons $f_\alpha$, $\alpha = 1 \ldots M$, just as in Section~\ref{sec:tJPG}. However, the bosonic holons $b_\ell$ now have an additional `orbital' index $\ell = 1 \ldots M'$.
The electrons $c_{\ell \alpha}$ also have an additional orbital index $\ell$, and are related to the spinons $f_\alpha$ and holons $b_\ell$ by
\begin{eqnarray}
&&~~~c_{\ell\alpha} = f_\alpha b_\ell^\dagger \nonumber \\ 
&& \sum_{\alpha=1}^M f_\alpha^\dagger f_\alpha + \sum_{\ell = 1}^{M'} b_\ell^\dagger b_\ell = \frac{M}{2}\,. \label{tJ22}
\end{eqnarray}
The doping density $p$ is given by
\beqn
\frac{1}{N} \sum_{i \ell} b_{i \ell}^{\dagger} b_{i \ell} = M' p \label{tJ23}
\eeqn
The physical case corresponds to $M=2$, $M'=1$, and $k=1/2$.

We can now take the large $M$ limit in a manner which closely parallels Section~\ref{sec:rqm}. Then we obtain SYK-like equations for the boson and fermion Green's functions, now describing a critical doped spin liquid:
\begin{eqnarray}
G_b ( i \omega_n) &=& \frac{1}{i \omega_n + \mu_b - \Sigma_b (i \omega_n) } \nonumber\\
 \Sigma_b (\tau) &=& - t^2 G_f(\tau) G_f(-\tau) G_b (\tau) \nonumber \\
G_f (i \omega_n) &=& \frac{1}{i \omega_n + \mu_f - \Sigma_f (i \omega_n)} \label{tJ24} \\
 \Sigma_f (\tau) &=& - J^2 G_f^2 (\tau) G_f(-\tau)
+ k \, t^2 G_f (\tau) G_b (\tau) G_b (-\tau). \nonumber
\end{eqnarray}
These equations share some similarities with the ones introduced in a study \cite{Haule2002} of the 
non-random $t-J$ model using the non-crossing approximation in the EDMFT framework.
They can be obtained from a $G$-$\Sigma$ action which generalizes those in Eqs.~(\ref{syk31a}), (\ref{rqm6}), and (\ref{kh23})
\begin{align}
 I[G, \Sigma] =& -\ln \det \left[ (\partial_{\tau} -\mu_f )\delta(\tau_1 - \tau_2)  + \Sigma_f (\tau_1, \tau_2) \right] \nonumber \\
 & + k \ln \det \left[ (\partial_{\tau} -\mu_b) \delta(\tau_1 - \tau_2)  + \Sigma_b (\tau_1, \tau_2) \right] \nonumber \\
 & + k \mbox{Tr} \left( \Sigma_b \cdot G_b \right) + \frac{k t^2}{2} 
\mbox{Tr} \left( [G_f G_b] \cdot [G_f G_b] \right) \nonumber \\
& - \mbox{Tr} \left( \Sigma_f \cdot G_f \right) - \frac{J^2}{4} 
\mbox{Tr} \left( G_f^2 \cdot G_f^2 \right) \,. \label{tj10}
\end{align}
Here $\mu_f$ and $\mu_b$ are chemical potentials chosen to satisfy
\beqn
\left\langle f^\dagger f \right\rangle = \frac{1}{2}-k p \quad, \quad  \left\langle b^\dagger b \right\rangle = p \,. 
\eeqn
As for the SYK model, we search for solutions of Eq. (\ref{tJ24}) with the following low energy critical behavior
\begin{eqnarray}
&& G_f (z) = C_f \frac{e^{-i (\pi \Delta_f + \theta_f)}}{z^{1-2 \Delta_f}}, \quad
\mbox{Im}(z) >0 \nonumber \\ 
&& G_b (z) = C_b \frac{e^{-i (\pi \Delta_b + \theta_b)}}{z^{1-2 \Delta_b}} , \quad
\mbox{Im}(z) >0 \label{tJ25} \\
&&~~~~~~~~~~~~ \frac{\theta_f}{\pi} + \left( \frac{1}{2} -\Delta_f \right) \frac{\sin (2 \theta_f)}{\sin (2\pi \Delta_f)} = k p \nonumber \\
&&~~~~~~~~~~~~ \frac{\theta_b}{\pi} + \left( \frac{1}{2} -\Delta_b \right) \frac{\sin (2 \theta_b)}{\sin (2\pi \Delta_b)} = \frac{1}{2} + p \,. \nonumber
\end{eqnarray}
The last two equations follow from Luttinger theorems, similar to those discussed in Section~\ref{sec:syklutt} \cite{GPS2,GKST}.
Inserting this ansatz into Eq.~(\ref{tJ24}), we find that self consistency of the terms involving the hopping $t$ leads to the following constraint on the scaling dimensions of the fermion ($\Delta_f$) and boson ($\Delta_b$)
\beqn
\Delta_f + \Delta_b = \frac{1}{2} \,. \label{tJ25a}
\eeqn

Inserting the ansatzes for $G_b$ and $G_f$ into the correlation functions for the electron and spin operators (as in Eq.~(\ref{rqm7})), we obtain for the gauge-invariant observables
\begin{align}
\left\langle c_\alpha (\tau) c_\alpha^\dagger (0) \right\rangle & \sim 
\left\{
\begin{array}{ccc} \displaystyle \frac{A_{+}}{|\tau|} &,& \tau > 0 \\
~&~&~\\
\displaystyle -\frac{A_{-}}{|\tau|} &,& \tau < 0
\end{array}
\right. \nonumber \\
  \left\langle \vec{S} (\tau) \cdot \vec{S} (0) \right\rangle & \sim \frac{1}{|\tau|^{4 \Delta_f}} \,. \label{tJ26}
\end{align}
The electron Green's function is similar to that of a Fermi liquid, with the difference that the present large $M$ limit allows solutions with a particle-hole asymmetry with $A_+ \neq A_-$, whereas a Fermi liquid always has $A_+=A_-$. 
We note that this is a rather unusual situation in which the $T=0$ spectral function is discontinuous at $\omega=0$; the electron Green's function obtained from the RG analysis below in Section~\ref{sec:tJRG} does not share this feature.
A Fermi liquid would also have a spin correlation function with a $1/\tau^2$ decay, which is potentially different from the $1/|\tau|^{4 \Delta_f}$ decay above.

Our discussion so far has been rather general, but the nature of the state obtained depends crucially on the values of the exponents $\Delta_f$ and $\Delta_b$. Determining their values requires further analysis of Eq.~(\ref{tJ24}), and we now describe the 3 distinct possibilities.

\paragraph{$\Delta_b = \Delta_f =1/4$: doped SY spin liquid.}
In such a solution, the $J$ terms in Eq.~(\ref{tJ24}) also contribute to determining the parameters in the scaling ansatz in Eq.~(\ref{tJ25}). 
The scaling dimension of the spinons and the spin operator are the same as those in the insulating SY spin liquid described in Section~\ref{sec:SY}.
Numerical analyses of Eq.~(\ref{tJ24}) at all energies \cite{Tikhanovskaya:2020zcw} show that such solutions do indeed exist, but only at very small values of the doping $p$.

\paragraph{$\Delta_b =0$, $\Delta_f = 1/2$: disordered Fermi liquid.}
This state is the same as that obtained in the large $M$ limit of Section~\ref{sec:tJPG}, but it turns out not to be a valid solution of the saddle point equations in Eq.~(\ref{tJ24}) of the present large $M$ limit \cite{Christos:toappear}.  If $\Delta_b =0 $, we have $b$ condensate with $\langle b (\tau \rightarrow \infty) b^\dagger (0) \rangle \neq 0$ at $T=0$. Inserting this condensate in the equation for $\Sigma_b$ in Eq.~(\ref{tJ24}), we find a contribution $\Sigma_b (\omega) \sim |\omega|$ from the fermion polarizability, which leads to a $\ln (1/\tau)$ contribution to $G_b (\tau)$, inconsistent with presence of a $b$ condensate.

\begin{figure}
\begin{center}
\includegraphics[width=3.25in]{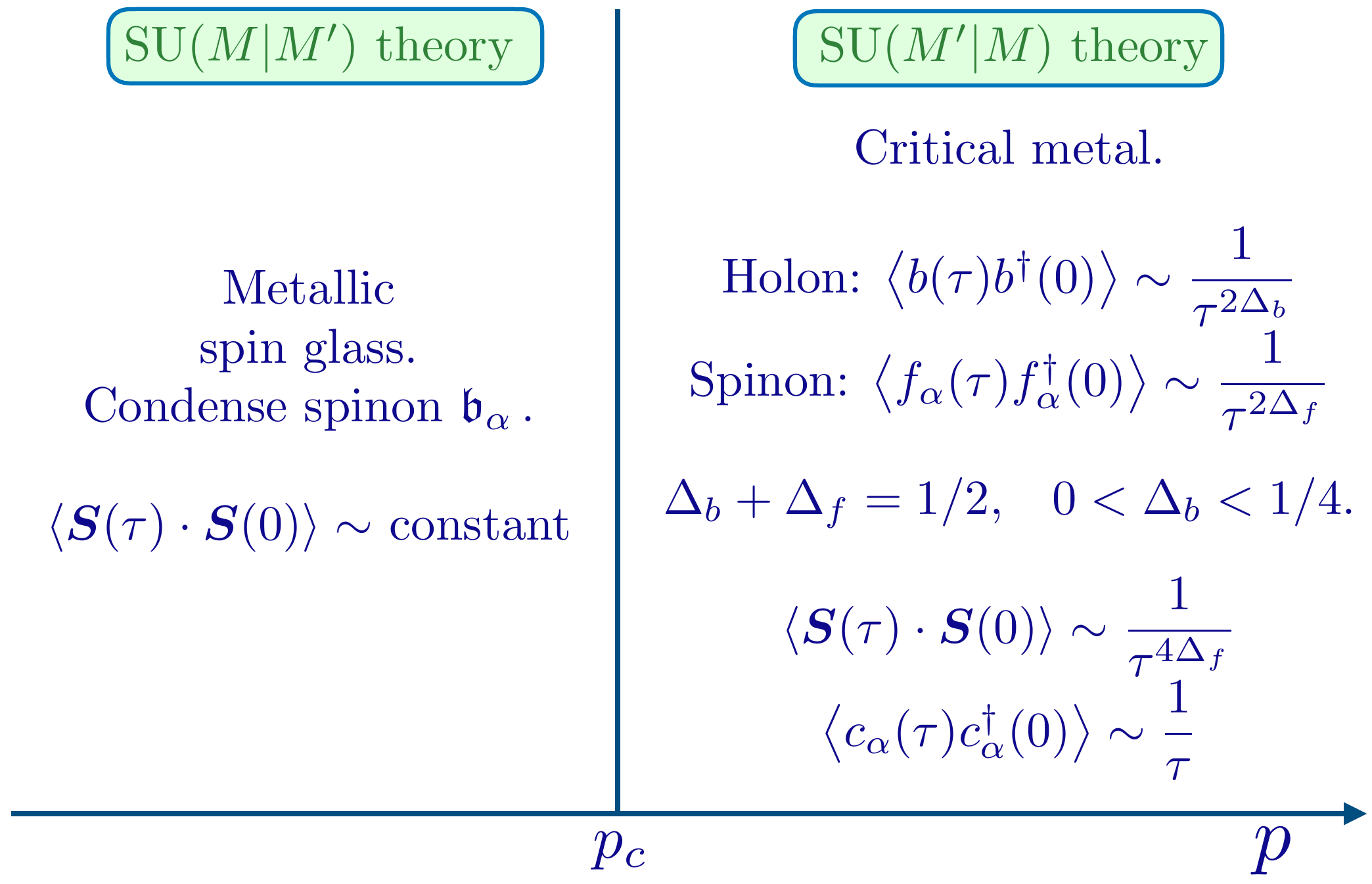}
\end{center}
\caption{Schematic phase diagram of the $t$-$J$ model in the non-Fermi liquid $M$ limit of Section~\ref{sec:JoshiM} \cite{Christos:toappear}. In the critical metal phase, the exponents obey $0 < \Delta_b < 1/4$, $\Delta_f = 1/2 - \Delta_b$, and $\Delta_b$ decreases monotonically towards 0 (the Fermi liquid value) with increasing $p$.
\label{fig:phasediag_nfl}
}
\end{figure}
\paragraph{$0 < \Delta_b < 1/4$, $\Delta_f = 1/2 - \Delta_b$: critical metal.}
Numerical analyses \cite{Christos:toappear} of Eq.~(\ref{tJ24}) shows that this is indeed a valid solution for a wide range of doping $p$. 
The $J$ terms in Eq.~(\ref{tJ24}) are subdominant to the critical ansatz at low energies, but they do contribute at higher energies.
The exponents in this critical metal vary continuously as a function of the doping and $J/t$, and can be determined by demanding numerically that Eq.~(\ref{tJ24}) apply at all energies. For finite $M$, the critical metal can be {\it stable \/} to spin glass order at $T=0$ for $\Delta_f < 1/4$, unlike the finite $T$ instability in (\ref{Tsg}) for the SY spin liquid. There can be an instability to a metallic spin glass below a critical doping $p_c$ \cite{Christos:toappear}, and that is indicated in the schematic phase diagram in Fig.~\ref{fig:phasediag_nfl}. This spin glass phase can be described \cite{Christos:toappear} using a theory of bosonic spinons $\mathfrak{b}_\alpha$ similar to that used for the insulating spin glass \cite{GPS1,GPS2} in a related large $M$ limit of a SU($M|M'$) superspin \cite{Tikhanovskaya:2020zcw}, as 
indicated in Fig.~\ref{fig:phasediag_nfl}.

\subsubsection{RG analysis for SU(2) symmetry}
\label{sec:tJRG}

This section returns to the original $t$-$J$ model with SU(2) spin symmetry, 
as defined by Eqs.~(\ref{ZtJ})-(\ref{defRQ}). We will describe a RG treatment similar to that for the insulating quantum magnet presented in Section~\ref{sec:SYRG}. The RG finds a critical point with one relevant direction, which is naturally identified with the deviation of the doping density $p$ from the critical density $p_c$. Moreover, the theory of this critical point turns out to be very similar to the large $M$ theory just described in Section~\ref{sec:JoshiM}. 

We proceed \cite{Joshi:2019csz} in a manner that parallels Section~\ref{sec:SYRG}.
First, we assume power-law decays for the correlators in the action in Eq.~(\ref{ZtJ}),
\begin{equation}
Q(\tau) \sim \frac{1}{|\tau|^{d-1}} \quad, \quad R(\tau) \sim \frac{\mbox{sgn}(\tau)}{|\tau|^{r+1}}\,, \label{tJ27}
\end{equation}
and ignore the self-consistency condition Eq.~(\ref{defRQ}) to begin with. We decouple the $J^2$ and $t^2$ terms in the action  by introducing 
bosonic ($\phi_a$, $a = 1 \ldots 3$) and fermionic ($\psi_\alpha$) baths. Then the problem reduces to solving the impurity Hamiltonian
\begin{eqnarray}
H_{\rm imp} && = (s_0 + \lambda) f_\alpha^\dagger f_\alpha + \lambda \, b^\dagger b + g_0 \left( f^\dagger_\alpha b \, \psi_\alpha (0) + \mbox{H.c.} \right) \nonumber \\
&&~~~~~~~~~~+ \gamma_0 f_\alpha^\dagger \frac{\sigma^a_{\alpha\beta}}{2} f_\beta \, \phi_a (0) \label{tJ28} \\
&&~~+ \int |k|^r dk \, k \, \psi_{k\alpha}^\dagger \psi_{k \alpha} + \frac{1}{2} \int d^d x \left[ \pi_a^2 + (\partial_x \phi_a)^2 \right] \nonumber
\end{eqnarray}
where the constraint in Eq.~(\ref{tJ4}) is imposed exactly by taking $\lambda \rightarrow \infty$ \cite{FritzVojta2004},  $a = (x,y,z)$, $\sigma^a$ are Pauli matrices, $\pi_a$ is canonically conjugate to the field $\phi_a$, and
$\phi_a (0) \equiv \phi_a (x=0)$, $\psi_\alpha (0)  \equiv \int |k|^r dk \, \psi_{k \alpha}$.
We identify $Q (\tau)$ with temporal correlator of $\phi_a (0)$, and $R(\tau)$ with the temporal correlator of $\psi_\alpha (0)$, and it can be verified that these correlators decay as in Eq.~(\ref{tJ27}).

So we have reduced the problem to an impurity Hamiltonian of a SU($1|2$) superspin interacting with separate bosonic and fermionic baths. 
By analogy with the Bose-Kondo model in Eq.~(\ref{SY30}), we can identify it as a superspin Bose-Fermi-Kondo model, where both the fermionic and bosonic baths have to be determined self-consistently.  
Such a model can be analyzed by a RG computation which performs the exact path integral over the superspin space {\it i.e.} imposes the constraint in Eq.~(\ref{tJ4}) exactly. The methods are similar to those used for the insulating spin problem that were used to obtain Eq.~(\ref{betagamma}), which ultimately only depended upon the spin commutation relations. In a similar manner, the RG results follow in a similar manner from the SU($1|2$) commutation relations in Eq.~(\ref{super}). We also note that the same RG equations would have been obtained from the commutation relations of the isomorphic SU($2|1$) algebra {\it i.e.\/} we get the same results from the formulation in terms of either the bosonic spinons or the fermionic spinons.

\begin{figure}
\begin{center}
\includegraphics[width=3.5in]{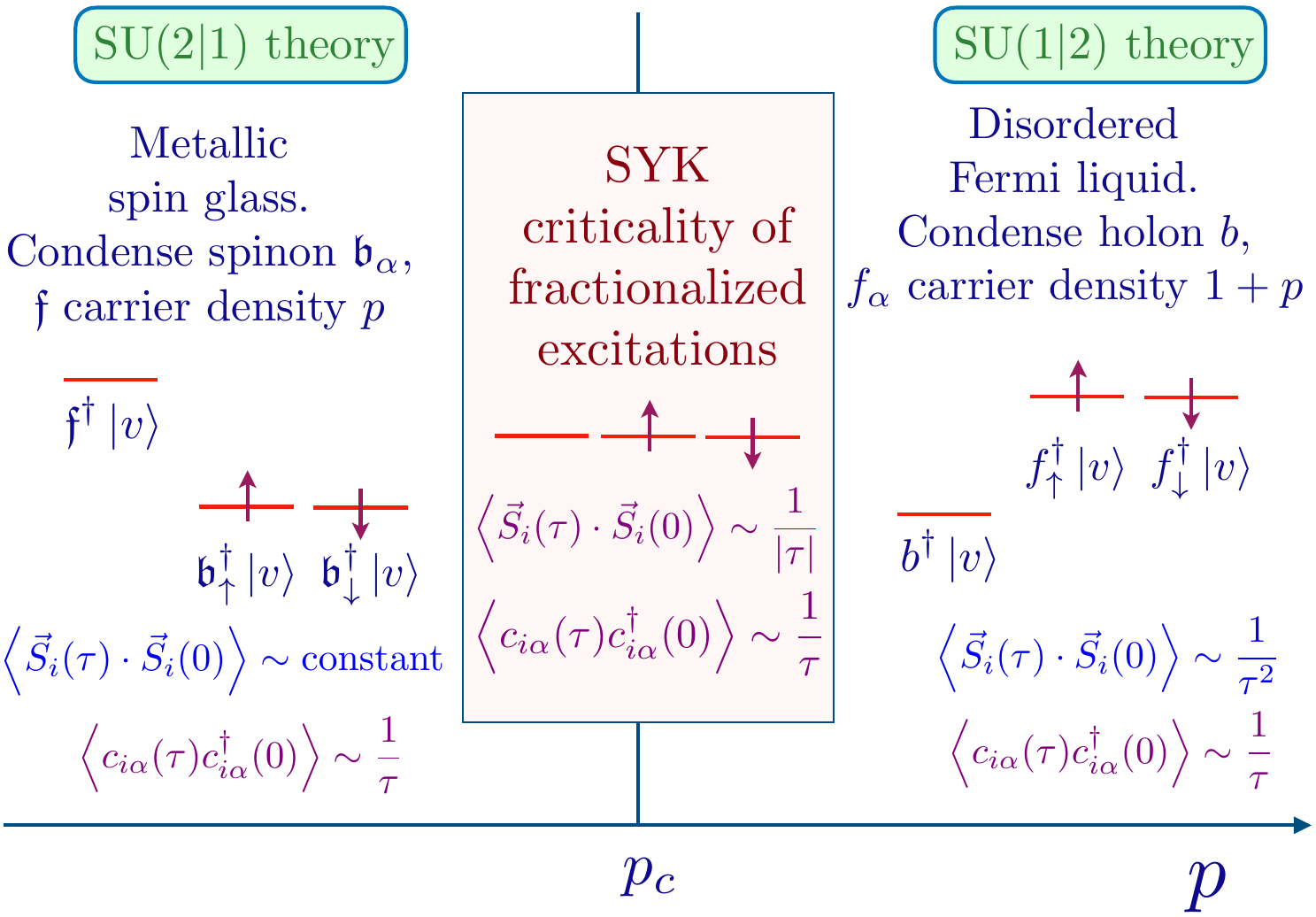}
\end{center}
\caption{Schematic phase diagram of the RG analysis of the random, fully-connected $t$-$J$ model \cite{Joshi:2019csz}. The spinon and holon states are nearly degenerate in the critical spin liquid theory, while the holon (spinon) states have lower energy for $p>p_c$ ($p< p_c$).}
\label{fig:Joshi}
\end{figure} 
The impurity has 3 coupling constants, and we represent their renormalized values by $\gamma$, $g$, and $s$. The coupling $\gamma$ measures the coupling to the bosonic bath, just as in Eq.~(\ref{SY30}). Similarly, $g$ is the coupling to the fermionic bath. We will see shortly that $g$ and $\gamma$ can be chosen to be nearly marginal, by appropriate choices of the exponents in Eq.~(\ref{tJ27}). 
The coupling $s$ tunes the relative energies of the spin and holon states, as is clear from Eq.~(\ref{tJ28}). This is the relevant perturbation mentioned at the start of this subsection, and its flow leads to the phase diagram in Fig.~\ref{fig:Joshi}. For $s \rightarrow + \infty$, the energy of the holon is much lower, and we expect the holon $b$ to condense, leading to a disordered Fermi liquid. Conversely for $s \rightarrow -\infty$, the spinons will condense (in the SU($2|1$) formulation, as in Fig.~\ref{fig:Joshi}), leading to a spin glass. And in between, at some $s=s_c$ we will have the fixed point which describes the critical theory we are interested in. To zeroth order in $g, \gamma$ the critical point is at $s_c = 0$: this corresponds a 3-fold degeneracy in the 3 states of the superspin (see Fig~\ref{fig:Joshi}), and a doping density $p_c = 1/3$. So we have the interesting prediction that the critical doping density of the fully-connected random $t$-$J$ model is close to $p=1/3$, a result that is indeed supported by the numerical results~\cite{Shackleton2021} reviewed in Section~\ref{sec:tJnumerics}.

\begin{figure}
\centering
\includegraphics[width=3.2in]{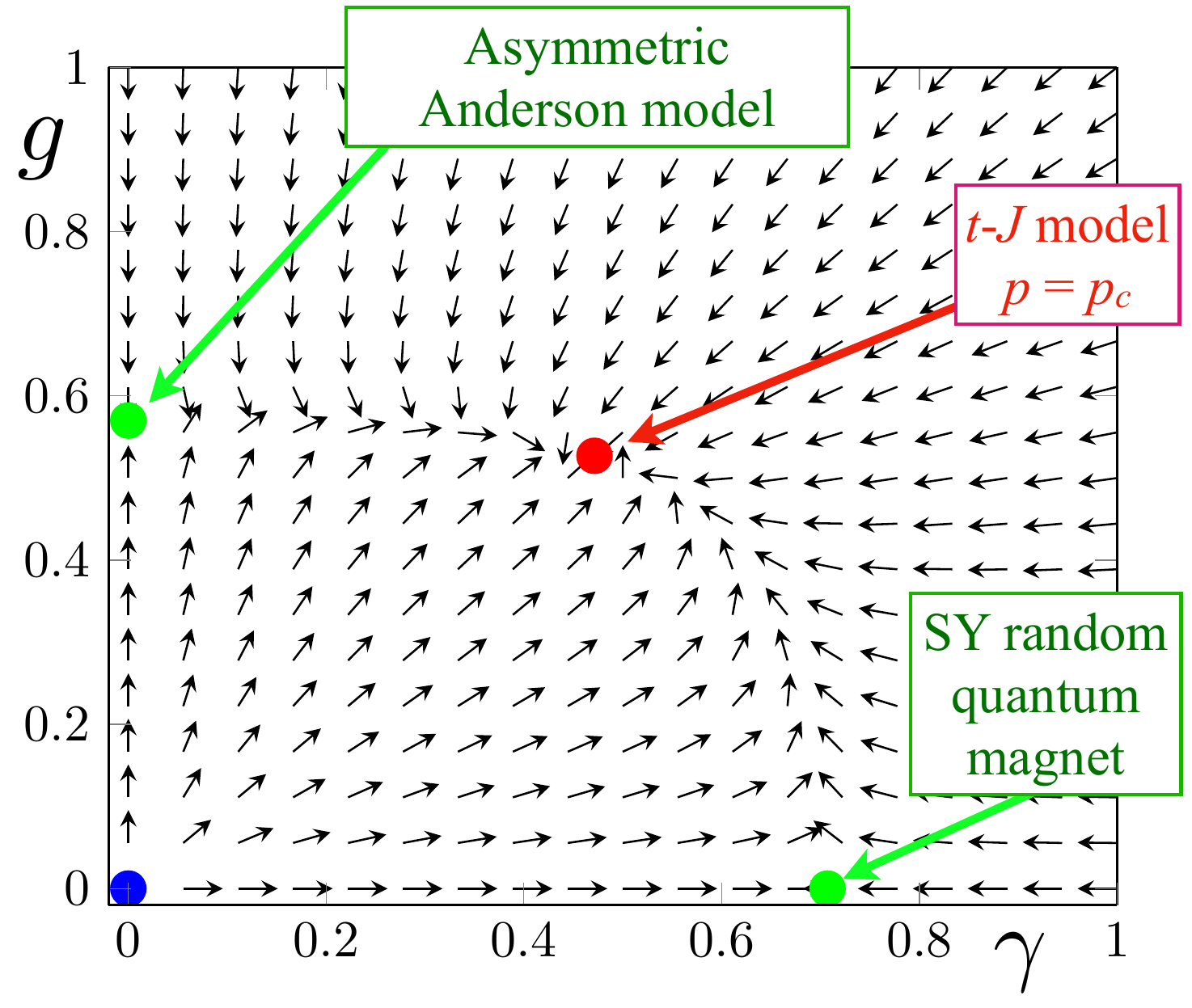}
\caption{RG flow \cite{Joshi:2019csz} of Eq.~(\ref{tJbeta}) in the $\gamma$-$g$ plane plotted for $\epsilon = 1$ and $\bar{r}=0.5$. The red point is the stable fixed point in this plane, which is unstable only to flows predominantly in the $s$ direction out of the plane; this fixed point describes the $p=p_c$ critical state in Fig.~\ref{fig:Joshi}, and $p-p_c$ tunes the co-efficient of the relevant perturbation (not shown), which presumably drives the system into the $p>p_c$ and $p<p_c$ phases shown in Fig.~\ref{fig:Joshi}.
}
\label{fig:tJrg}
\end{figure}

The one loop RG equations are \cite{Joshi:2019csz}
\begin{align}
\beta (g)  &= - \bar{r} g + \frac{3}{2} g^{3} + \frac{3}{8} g \gamma^{2} \nonumber \\
\beta (\gamma) &= - \frac{\epsilon}{2} \gamma + \gamma^{3} + g^{2} \gamma \nonumber \\
\beta (s) &= -s + 3 g^{2} s - g^{2} + \frac{3}{4} \gamma^{2}  \,. \label{tJbeta}
\end{align}
We have introduced the variables
\beqn
\epsilon = 3-d \quad , \quad \bar{r}=(1-r)/2 \label{tJ29}
\eeqn
and it is clear from Eq.~(\ref{tJbeta}) that the fixed points at small $\epsilon$ and $\bar{r}$ are under perturbative control in powers of $\epsilon$ and $\bar{r}$. The RG flows in the $g,\gamma$ plane are shown in Fig.~\ref{fig:tJrg}: there is an attractive fixed point in this plane at $g^{\ast 2}$, $\gamma^{\ast 2}$ of order $\epsilon, \bar{r}$. The relevant perturbation $s$ induces flows away from this fixed point in a direction which is predominantly transverse to the $g, \gamma$ plane. 
There are also fixed points in Fig.~\ref{fig:tJrg} along the $g=0$ line, corresponding to the fixed point of the insulating magnet in Eq.~(\ref{betagamma}); and along the $\gamma=0$ line, corresponding to the fixed point of the asymmetric pseudogap Anderson impurity \cite{VojtaFritz2004,FritzVojta2004}, and has properties similar to the large $M$ critical metal solution of Section~\ref{sec:JoshiM} in Fig.~\ref{fig:phasediag_nfl}.

Finally, we can compute the scaling dimensions of the electron and spin operators at the red fixed point of Fig.~\ref{fig:tJrg}. As in Section~\ref{sec:SYRG}, these scaling dimensions are protected by the Berry phase term in Eq.~(\ref{ZtJ}) that imposes the SU($1|2$) commutation relations at any fixed at non-zero $g^\ast$ and $\gamma^\ast$. So we are able to compute the exponents in Eq.~(\ref{defRQb}) to all loop order; we find 
\begin{align}
\left\langle c_\alpha (\tau) c_\alpha^\dagger (0) \right\rangle & \sim \frac{\mbox{sgn}(\tau)}{|\tau|^{1-r}}  \nn 
\left\langle \vec{S} (\tau) \cdot \vec{S} (0) \right\rangle & \sim \frac{1}{|\tau|^{3-d}} \,. \label{tJ30}
\end{align}
We now restore the self-consistency condition in Eq.~(\ref{defRQ}), and find the self-consistent values $r=0$ and $d=2$. These self-consistent exponents are the same as those obtained for the doped SY spin liquid case in the large $M$ computation of Eq.~(\ref{tJ26}).
There is however an interesting difference that will benefit from further study: the electron correlator in Eq.~(\ref{tJ26}) is allowed to have particle-hole asymmetry with $A_+ \neq A_-$, but that is not the case for the present RG analysis.

\subsection{Transport in random exchange $t$-$U$-$J$ models}
\label{sec:transport}

Discussing conductivity requires a slightly different setup than a fully-connected lattice in order to properly define transport and the current operator. One possibility is to consider the model on the Bethe lattice 
with non-random hopping amplitudes $t_{ij}=t/\sqrt{z}$, with $z$ the connectivity of 
the lattice. In the limit $z\rightarrow\infty$, the self-energy and local 
Green's function obey the same equations as the model with random $t_{ij}$ \cite{DMFT}. 
Another possibility is to consider a translationally invariant lattice of 
fully connected dots, as in Section~\ref{sec:lattice}.  

The conductivity is given by the Kubo formula: 
\begin{equation}
\label{eq:kubo_bubble}
\sigma_{\rm dc} =
\frac{2\pi e^2}{\hbar}\int d\omega \frac{\beta}{4\cosh^2(\beta\omega/2)}\,
\int d\epsilon\, \phi(\epsilon) A(\epsilon,\omega)^2.
\end{equation}
In this expression, $\epsilon$ is the energy of a bare single-particle state within the band  
and $A(\epsilon,\omega)=-(1/\pi)\mbox{Im}\, G^R(\epsilon, \omega)$ 
is the energy (momentum-) resolved spectral function. 
The transport function $\phi(\epsilon)$ is defined on a Bravais lattice by: 
\beq
\phi(\epsilon) = \int \frac{d^dk}{(2\pi)^d}\, v_{kx}^2\delta(\epsilon-\epsilon_{\mathbf{k}})\,,
\eeq
in which $v_{kx}=(\nabla_{\mathbf{k}}\epsilon_{\mathbf{k}})_x/\hbar$ is the velocity in the considered direction. 
On the infinite-connectivity Bethe lattice $\phi(\epsilon)=\phi(0)[1-\left(\epsilon/2t\right)^2]^{3/2}$ \cite{DMFT}. 
Here, we have assumed that the self-energy as well as the 2-particle vertex function only depends on frequency. 
As a result, because the current vertex is odd in momentum, vertex corrections to the conductivity vanish and 
the full Kubo formula reduces to the fermionic bubble in Eq.~(\ref{eq:kubo_bubble})~\cite{Khurana1990}. 
Note that this is not the case for other correlation functions which are even-parity 
(such as charge or spin)~\cite{DMFT}. 

Let us discuss the behaviour of the resistivity following from Eq.~(\ref{eq:kubo_bubble}) in two different situations. 
First, we consider a case in which $\mathrm{Im}\Sigma$ is much larger than the 
dispersion of the band itself (i.e the range over which $\epsilon$ varies in the integral). Then the 
dispersion can be entirely neglected and we obtain: 
$\sigma_{\rm dc}  \propto \int d\omega  \left[ {\beta}/({4\cosh^2(\beta\omega/2)}) \right]\, 
\left(\mathrm{Im} 1/\Sigma\right)^2$. This applies for example to the large-$M$ limit of the 
random exchange $t$-$J$ model discussed in Sec.\ref{sec:tJPG} in the SYK regime where $T>T_{\rm coh}$. 
In that case, $\mathrm{Im}\Sigma \propto \sqrt{JT} f(\omega/T)$, where $f(...)$ is a scaling function. Inserting this into the expression above 
leads to $\rho(T)/\rho_Q \propto T/T_{\rm coh}$, 
i.e. a resistivity which is $T$-linear but larger than the MIR value (introduced in Section \ref{sec:bad_planck}). 
This bad metallic behaviour does correspond however to a Planckian regime with 
a diffusion constant $\propto 1/T$ since the compressibility is temperature independent. 
Interestingly, the conductivity is proportional to the 
{\it square} of the transport scattering rate in this regime; the latter is $T$-linear while 
the single-particle scattering rate is $\propto \sqrt{JT}$. This mechanism for 
a Planckian bad metal with $T$-linear resistivity was first discussed in \cite{Parcollet1}. 

In the second case $\mathrm{Im}\Sigma$ is, in contrast, smaller than the band dispersion. 
This applies, for example, in the low-$T$ limit of most of the models discussed in this review.
The integral in Eq.~(\ref{eq:kubo_bubble}) can then be approximated as:
$\int d\epsilon\, 
\phi(\epsilon) A(\epsilon,\omega)^2\,\sim\,\phi[\omega+\mu-\mathrm{Re}\Sigma(\omega)]
/(2\pi|\mathrm{Im}\Sigma(\omega)|)$. 
Due to the derivative of the Fermi function, one can set $\omega=0$ in the numerator. 
Defining the renormalized Fermi level as $\epsilon_F=\mu-\mathrm{Re}\Sigma(0,0)$ (which 
coincides with the bare Fermi energy when Luttinger's theorem is satisfied), one obtains: 
\begin{equation}
\sigma_{\rm dc}\,\simeq\,\frac{e^2\phi(\epsilon_F)}{\hbar}\int d\omega \frac{\beta}{4\cosh^2(\beta\omega/2)}\frac{1}{|\mathrm{Im}\Sigma(\omega,T)|}.
\label{eq:quasiBoltzmann}
\end{equation}
This expression is similar to Drude-Boltzmann theory, but we emphasize that it is valid even when 
the scattering rate has a non-Fermi liquid form. 
For example when $\mathrm{Im}\Sigma = T^\nu f(\omega/T)$, we obtain $\rho \propto T^\nu$; 
$\nu=1$ corresponds to a Planckian metal. Evidence for such nFL behavior of the 
scattering rate was discussed above in the quantum critical regime of the random bond $t$-$J$ 
and Hubbard models~\cite{Cha_2020,Dumitrescu2021}.  
We also emphasize, as is well known from transport theory, that the wave function normalisation 
$Z(T)\propto (1-\left.\partial\mathrm{Re}\Sigma(\omega, T)/\partial\omega\right|_{\omega=0})^{-1}$ does not enter 
the expression of the conductivity, in contrast to the width of the one-electron spectral function 
which is $\propto Z|\mathrm{Im}\Sigma|$ (and can be interpreted as the inverse of the 
quasiparticle lifetime in a Fermi liquid). 
It is interesting to note that, for a nFL with $\mathrm{Im}\Sigma \propto T^\nu$ and $\nu<1$, the 
latter always displays Planckian behavior $\propto T$, independently 
of the value of the exponent $\nu$ since $Z(T,\omega=0)$ vanishes as $Z(T)\propto T^{1-\nu}$. 
Indeed, the real part of the self-energy is related to the imaginary part by 
$\mathrm{Re}\Sigma(\omega) = - \int d\omega^\prime \frac{\mathrm{Im}\Sigma(\omega')/\pi}{\omega-\omega'}$, from which it follows 
for $\nu<1$ that $\mathrm{Re}\Sigma(\omega)=T^\nu \tilde{f}(\omega/T)$ 
and hence $1/Z=1-\partial_\omega\mathrm{Re}\Sigma=1-T^{\nu-1}\tilde{f}'(\omega/T)$; 
see \cite{Georges2021skewed} for details. 

We note that Eq.~(\ref{eq:kubo_bubble}) also applies to non-random models in the (DMFT) limit of infinite connectivity. 
An interesting connection was recently noted~\cite{Cha_Patel_2020} between the $T$-linear behavior of the resistivity 
in such models in the high-$T$ bad metal regime~\cite{Palsson_1998,Perepelitsky2016} discussed in Sec.~\ref{sec:badmetal} 
and the SYK equations for the self-energy. 
Whether such a connection also exists in the lower temperature regime is an interesting open question - for a recent 
study of $T$-linear resistivity in the non-random Hubbard model using cluster extensions of DMFT, 
see \cite{Tremblay21}. 
Possible connections between the SYK model and nFL regimes of non-random multi-orbital models have also been 
pointed out~\cite{Werner_2018,Tsuji_2019}. Relevance of SYK criticality to possible instabilities of `Luttinger surfaces' has 
also been discussed~\cite{Setty20,Setty21}.

Thermoelectric transport has also been analyzed in random-exchange and SYK models. 
It was pointed out~\cite{SS17,Kruchkov:2019idx} that the thermopower of a lattice of SYK islands is directly related to the 
spectral asymmetry parameter ${\cal E}$ introduced in Eq.~(\ref{syk17}), and hence offers a possible 
probe of the residual $T=0$ entropy. That relation may be more involved in general however~\cite{Kruchkov:2019idx,Kiselev21}. 
Recently, \cite{Georges2021skewed} emphasized that the intrinsic particle-hole asymmetry of the 
$\omega/T$ scaling function in Eq.~(\ref{eq:skewed_sigma}), characteristic of `skewed' Planckian (or sub-Planckian) metals, 
has remarkable consequences for the sign and $T$-dependence of the thermopower down to low-$T$, 
even in the presence of additional elastic scattering. Possible relevance to Seebeck measurements on 
cuprate superconductors was explored~\cite{Gourgout2021}. 

\subsection{General mechanism for $T$-linear resistivity as $T \rightarrow 0$ from time reparameterization}
\label{sec:linearT}

The quantum-critical $T$-linear resistivity computed numerically in Section~\ref{sec:tJnumerics}
(and also in Section~\ref{sec:ChaHubbard}) is somewhat mysterious when compared with the analytical results. 
Recall that we found a leading Fermi liquid-like behavior in the electron Green's function at the quantum critical point in the non-Fermi liquid large $M$ limit in Eq.~(\ref{tJ26}), and also in the RG analysis for $M=2$ in Eq.~(\ref{tJ30}). The RG analysis also makes clear that this Fermi liquid exponent for the electron operator is likely exact to all orders in $1/M$. Inserting such an electron spectral density in Eq.~(\ref{eq:kubo_bubble}), we obtain temperature {\it independent\/} residual resistivity as $T \rightarrow 0$, $\rho (0)$. We note that this large residual resistivity, present even for a large dimension lattice without hopping disorder, appears to be an artifact of the non-Fermi liquid large $M$ limit of Section~\ref{sec:JoshiM} \cite{Guo:2020aog}. The Fermi liquid large $M$ limit of Section~\ref{sec:tJPG} has vanishing residual resistivity \cite{Parcollet1}, and this also appears to be the case in the numerical study in the SU(2) limit \cite{Dumitrescu2021}. It is possible that the non-Fermi liquid large $M$ limit of Section~\ref{sec:JoshiM} has a crossover in the residual resistivity at a frequency which vanishes as $M$ becomes large.

We obtain a $T$-dependence to the resistivity as $T \rightarrow 0$ by considering corrections to scaling for the electron operator in the $N=\infty$ theory. The structure of these corrections can be easily deduced from the theory described in Section~\ref{sec:corrscaling}, which generalizes directly to the $t$-$J$ model \cite{Tikhanovskaya:2020zcw}. As for the entropy in Eq.~(\ref{syk36c}), and the spin spectral density in Eq.~(\ref{rqm10}), we consider the corrections due to $h=2$ operator. The scaling dimension of this operator is also `protected' at $h=2$, given its connection to the Schwarzian theory in Section~\ref{sec:fluctuations} {\it i.e.\/} it is the `time reparameterization' operator, and the `boundary graviton' in the holographic theory to be discussed in Section~\ref{sec:bh_fluc}. 
Therefore, we expect that the $h=2$ scaling dimensions does not acquire any $1/M$ corrections. By the same arguments leading to Eq.~(\ref{rqm10}) for the spin spectral density, we now obtain for the temperature dependence for the 
resistivity \cite{Guo:2020aog}
\beq
\rho (T) = \rho (0) \left[ 1 + \mathcal{C}_\rho \gamma\, T + \ldots \right]\,.
\eeq
The linear $T$ dependence is the power $T^{h-1}$, which is related to that in Eq.~(\ref{syks5}), for the time reparameterization mode with $h=2$. The parameter $\gamma$ is the same as that in the entropy in Eq.~(\ref{syk36c}), and $\mathcal{C}_\rho$ is a dimensionless universal number similar to $\mathcal{C}$ in Eq.~(\ref{rqm10}). The value of $\mathcal{C}_\rho$ can be computed in the large $M$ limit of the $t$-$J$ model \cite{Guo:2020aog}. 
While the co-efficient of linear $T$ resistivity is controlled by the residual resistivity in this large $M$ computation, that is not the case for the numerical SU(2) computation in Fig.~\ref{fig:tJU_tau_star}, with the corresponding phase diagram in Fig.~\ref{fig:tJU_phasediagram} \cite{Dumitrescu2021}. We also note that the large $M$ theory of the doped $t$-$J$ model has operators with $h<2$; but the scaling dimension of these operators is not protected, and their contribution to the resistivity is numerically small in the large $M$ theory \cite{Tikhanovskaya:2020zcw}.

%%%%%%%%%%%%% EXPERIMENTAL %%%%%%%%%%%%%%%%%%%%%

\subsection{Experimental relevance}
\label{sec:exprel}

The models described in this section are, of course, not meant to be microscopically realistic models of 
materials displaying nFL behaviour, such as the cuprate strange metal. 
Nonetheless, as we now discuss, the physics of the doped Hubbard and $t$-$J$ models with random exchange couplings exposed above 
present rather striking similarities with some of the salient phenomenology of the cuprates and can serve as a building block for capturing certain universal aspects of nFL behavior in general. 
We recall two of the most fundamentally intriguing phenomena observed in these materials:

\begin{itemize}  

\item The appearance of a pseudogap regime below a critical doping ($p<p^*$). At low $T$ and high fields, quantum oscillations have revealed the existence of pocket Fermi surfaces \cite{DoironLeyraud2007,Proust2019}. These oscillations appear in a regime with long-range charge density wave order, but it is clear that a simple model of reconstruction of the large Fermi surface by the charge density wave order cannot explain the details of the quantum oscillations. At higher $T$, or at dopings 
$p_{\mathrm{CDW}}<p<p^*$ (where $p_{\mathrm{CDW}}$ is the doping below which there is charge density wave order), there is no known long-range order, and there is clear experimental evidence that the electronic spectrum cannot be described by the large Fermi surface. The observations include angle-dependent magnetoresistance \cite{Fang2020} and the `Fermi arcs' in angular resolved photoemission spectroscopy (ARPES)~\cite{zxs}.

\item Near $p^*$, several properties are evocative of quantum criticality, most notably: 
({\it i\/}) $T$-linear resistivity with a transport scattering time obeying Planckian behaviour 
$\tau \simeq \alpha \hbar/k_BT$ down to low temperatures~\cite{Hussey2008,Zaanen04,Homes2004,Bruin13,Legros19,
Grissonnanche2020,Varma2020}
({\it ii\/}) $\omega/T$ scaling observed in several spectroscopies, such as optical conductivity~\cite{Marel2003,Michon2022,Heumen22} 
and ARPES~\cite{Reber2019}. 
({\it iii\/}) 
A diverging specific-heat coefficient near $p^*$, with logarithmic dependence of $C/T$ upon $T$ at 
$p=p^*$~\cite{Michon2019}.
\end{itemize}

Seen in this perspective, the doped random exchange models discussed above offer a simple platform in 
which to study some of these phenomena. We have reviewed that they display a critical point upon doping at which 
quantum critical scaling is observed, and that the Luttinger theorem breaks down at this critical 
doping. We find clear evidence of the Luttinger breakdown in the value of the chemical potential at temperatures above the spin glass transition for $p<p_c$ in the Monte Carlo study \cite{Dumitrescu2021}, and at zero temperature within the metallic spin glass in the exact diagonalization study \cite{Shackleton2021}.
The precise nature of the Fermi surface 
reconstruction, and possible volume collapse, is still to be 
investigated in the low $T$ metallic spin-glass phase for $p<p_c$, and is one of the fascinating open questions 
in the field. 

Most notably, these doped random exchange and SYK models 
are among the few theoretical models in which Planckian 
behaviour of transport~\cite{Zaanen04} in the absence of coherent quasiparticles 
can be studied in a controlled manner 
(we note investigations of this issue \cite{Varma89,Varma2016,Varma2020} in the marginal Fermi liquid context). 
The randomness of the exchange constants helps introduce `frustration' and is, at the theoretical level, a simple way to 
account for the fact that the physics of short-range spin correlations is important in the 
pseudogap phase, but without true long-range order. 
One can also argue, as emphasized early on \cite{Parcollet1}, that randomness of the exchange constants can be motivated 
at a more microscopic level. In this respect, recent nuclear magnetic resonance and ultrasound
measurements have revealed that, remarkably, the spin-glass phase 
 extends up to $p=p^*$ for La$_{2-x}$Sr$_x$CuO$_4$ subject to a  high magnetic
field~\cite{Frachet2020}. 
The critical theory of the random exchange models is not particle-hole symmetric, and the possible relevance of the intrinsic particle-hole asymmetry of the $\omega/T$ scaling 
function associated with the scattering rate has been recently emphasized for the interpretation of Seebeck measurements 
on the cuprates \cite{Georges2021skewed,Gourgout2021}.

Another indication of Planckian behavior is the anomalous continuum observed in dynamic charge response measurements \cite{Abbamonte1,Abbamonte2}
on optimally doped Bi$_{2.1}$Sr$_{1.9}$Ca$_{1.0}$Cu$_{2.0}$O$_{8+x}$ (Bi-2212) using momentum-resolved electron energy-loss spectroscopy (M-EELS). This has been studied in a model with additional random density-density interactions \cite{Joshi20}.

%%%%%%%%%%%%%%%%%%%%%%%%%%%%%%%%%%%%%%%%%%%%
%%% KONDO HEISENBERG
%%%%%%%%%%%%%%%%%%%%%%%%%%%%%%%%%%%%%%%%%%%%

\section{Random exchange Kondo-Heisenberg model}
\label{sec:KH}

This section will combine the random matrix model of mobile electrons of Section~\ref{sec:matrix} with the random quantum magnet of Section~\ref{sec:rqm}, and couple them together with a {\it non-random}, antiferromagnetic, Kondo exchange coupling $J_K$.
So we have the Kondo-Heisenberg Hamiltonian
\bea
&& H_{KH} = \frac{1}{(N)^{1/2}} \sum_{i,j=1}^N  t_{ij} c_{i\alpha}^\dagger  c_{j\alpha} - \mu \sum_i c_{i \alpha}^\dagger c_{i \alpha} \label{kh1} \\
&&+ \frac{1}{\sqrt{N}}\sum_{1\leq i < j \leq N}  J_{ij} \vec{S}_i \cdot \vec{S}_j + \frac{J_K}{2} \sum_i \vec{S}_i \cdot \left(
c_{i \alpha}^\dagger \vec{\sigma}_{\alpha\beta} c_{i \beta} \right) \,, \nonumber
\eea
which has been used extensively as a theory of numerous rare-earth intermetallics (usually, in the absence of random exchange), the so-called heavy fermion compounds. This model exhibits a `heavy Fermi liquid' (HFL) ground state, which is a Fermi liquid with electron-like quasiparticle excitations with a large effective mass for models with non-random $t_{ij}$. Moreover, the Fermi energy is `large', because the occupied states count {\it both} 
the conduction electrons $c_{i \alpha}$, and the spins $\vec{S}_i$. The fully connected random model also has such a heavy Fermi liquid phase which obeys a Luttinger theorem with this large Fermi energy \cite{Burdin2002,Nikolaenko:2021vlw}, as we discuss further in Section~\ref{sec:Kondolut}.

Our interest here is in other possible phases of the Kondo-Heisenberg lattice model, and on the quantum critical points to these phases starting from the HFL. A possibility of particular interest is the `fractionalized Fermi liquid' (FL*) \cite{Burdin2002,FLSPRL,TSFL04,Paramekanti_2004} in which the Fermi surface is `small' and includes only the volume of the conduction electrons. The spins $\vec{S}_i$ form a spin liquid state with fractionalized excitations, and the fractionalized excitations are required to exist to allow deviation of the Fermi surface volume from the Luttinger value \cite{TSFL04,Paramekanti_2004,Bonderson16,ElseTS}; we also note other discussions of FL* and related states \cite{Coleman_2005,Norman07,Norman08,Norman13,Si2010,Si2014,PaschenSi21,Chowdhury_2018,AndreiColeman}. In the random fully connected model, the $\vec{S}_i$ spins form the SYK spin liquid of Section~\ref{sec:rqm} in the large $M$ limit, as we will describe in Section~\ref{sec:KondolargeM}. A number of recent experiments have reported the existence of a paramagnetic metallic phase with a Fermi surface volume that does not appear to include the local moment electrons in YbRh$_2$(Si$_{0:95}$Ge$_{0:05}$)$_2$ \cite{Paschen10,Paschen10a}, CePdAl \cite{Sun19}, and CeCoIn$_5$ \cite{Analytis20}, which shares resemblance with some aspects of the FL* phase.

A third possible phase of the Kondo-Heisenberg lattice model has broken spin rotation symmetry, and associated magnetic order. For the random fully connected model in Eq.~(\ref{kh1}) with SU(2) spin symmetry, this is likely realized as a spin glass phase. We will discuss a RG analysis of the SU(2) model in Section~\ref{sec:KHRG}, and this has a fixed point which is expected to describe the transition from the spin glass to the HFL. There have been a number of experimental studies of such a transition \cite{Cox91,Aronson95,Schroder98,Soldevilla00,Maple00,Maple01,Theumann04,Takagi12,Aronson18} in the heavy-fermion compounds.

We refer the reader to other reviews \cite{Stewart,SiRMP} for further discussion of the connections between the Kondo-Heisenberg model and experiments on the heavy-fermion compounds.

\subsection{Effective local action}

\label{sec:EDMFT_HKL}

We begin our analysis, as in Section~\ref{sec:EDMFT}, by averaging over disorder, and formulating the problem in terms of self-consistent single site problem.
We average over the $t_{ij}$ and $J_{ij}$ in Eq.~(\ref{kh1}), and in the large $N$ limit we obtain the following single site averaged partition function
\bea
&&\overline{\mathcal{Z}}_{KH} = \int \mathcal{D} c_{\alpha} (\tau) \mathcal{D} \vec{S} (\tau) \delta( \vec{S}^2 - 1) e^{-\mathcal{S}_B - \mathcal{S}_{KH}} \nonumber \\
&& \mathcal{S}_B = \frac{i}{2} \int_0^{1} du \int d \tau \vec{S} \cdot \left(\frac{\partial \vec{S}}{\partial \tau} \times \frac{\partial \vec{S}}{\partial u} \right) \nonumber \\
&&~~~~~~~~~~~~~~~~+ \int d \tau \left[~ c_{\alpha}^\dagger \frac{\partial c_{\alpha}}{\partial \tau} ~\right] \nonumber \\
&& \mathcal{S}_{KH} =  \int d\tau \left[\, -\mu\, c_\alpha^\dagger c_\alpha 
+ \frac{J_K}{2} \vec{S} \cdot \left( c_{\alpha}^\dagger \vec{\sigma}_{\alpha\beta} c_{\beta} \right)
\right] \nonumber\\ 
&&~~~- \frac{J^2}{2} \int d\tau d \tau' Q (\tau - \tau') \vec{S} (\tau) \cdot \vec{S} (\tau')  \nonumber \\
&&   - t^2 \int d\tau d \tau' R (\tau - \tau') c_\alpha^\dagger (\tau) c_\alpha (\tau') + \mbox{H.c.} \,. \label{kh2}
\end{eqnarray} 
From this action we determined the correlators
\begin{eqnarray}
\overline{R}(\tau - \tau') &=& - \frac{1}{2} \left\langle c^{}_{\alpha} (\tau) c^{\dagger}_\alpha (\tau') \right\rangle_{\mathcal{Z}_{KH}} \nonumber \\
\overline{Q} (\tau - \tau') &=&  \frac{1}{3} \left\langle \vec{S} (\tau) \cdot \vec{S} (\tau') \right\rangle_{\mathcal{Z}_{KH}} \label{kh3}
\end{eqnarray}
and finally impose the self-consistency conditions
\beq
R(\tau) = \overline{R}(\tau) \quad, \quad Q(\tau) = \overline{Q} (\tau). \label{kh4}
\eeq
As was the case for the $t$-$J$ model in Section~\ref{sec:dopedtJ}, closely related equations can also be obtained for the case of non-random $t_{ij}$, involving an electron dispersion $\epsilon_{\bf k}$.

It is interesting to note the difference between the single-site self-consistency problem for the $t$-$J$ model of Section~\ref{sec:dopedtJ}, and the present Kondo-Heisenberg model. The Berry phase term $\mathcal{S}_B$ reflects the different quantum degrees of freedom on the site: ({\it i\/}) for the $t$-$J$ model we have the 3 states of the SU($1|2$) superspin described above; ({\it ii\/}) for the Kondo-Heisenberg model we have the 2 state of the SU(2) spin 1/2 $\vec{S}$, and the 4 states of the electron $c_\alpha$, for a total of 8 states. Both models have very similar bosonic and fermionic baths, but do differ in the on-site Hamiltonian: the present model has a Kondo coupling $J_K$.

The self-consistent single-site quantum problem defined by Eqs.~(\ref{kh2},\ref{kh3},\ref{kh4}) cannot be solved exactly, and we will address it in the following subsections by the same methods used earlier for the random quantum magnet problem defined by Eqs.~(\ref{eq:BrayMoore},\ref{defQb},\ref{defQ}), and the Hubbard model problem defined by Eqs.~(\ref{eq:Seff_tUJ},\ref{eq:def_local_correlators},\ref{eq:EDMFT}).

\subsection{SU($M$) symmetry with $M$ large}
\label{sec:KondolargeM}

The large $M$ analysis of the fully connected Kondo-Heisenberg model \cite{Burdin2002} proceeds by generalizing the model in Eqs.~(\ref{kh2})-(\ref{kh4}) to SU($M$) symmetry just as in 
Section~\ref{sec:SY} for the random quantum magnet. We introduce fermionic spinons $f_{\alpha}$, $\alpha = 1 \ldots M$, treat the random $J_{ij}$ exchange as in Section~\ref{sec:SY}, and decouple the $J_K$ exchange by a bosonic field $P(\tau) \sim c_\alpha^\dagger (\tau) f_\alpha (\tau)$. 
Note that because the $J_K$ exchange is non-random, this decoupling variable is not bilocal in time. 

In this manner, Eqs.~(\ref{kh2})-(\ref{kh4}) reduce to the following equations for the fermion Green's functions, self energies, and time-independent saddle-point values $ i \lambda (\tau) = \overline{\lambda}$ and $P(\tau) = \overline{P}$. 
The Green's function acquires `band' indices associated with the $f$ and $c$ fermions, and so Dyson's equation has a matrix form
\bea
&& \left( 
\begin{array}{cc}
G_{f} (i \omega_n)    &  G_{fc}(i \omega_n) \\
G_{cf}(i \omega_n)     &  G_{c}(i \omega_n)
\end{array}\right)^{-1} = \label{kh20}  \\
&&~~~~\left( \begin{array}{cc}
i \omega_n - \overline{\lambda} - \Sigma_f (i \omega_n)  & - \overline{P} \\
-\overline{P} & i \omega_n + \mu - \Delta (i \omega_n) 
\end{array}
\right)\,. \nonumber
\eea
The $f$ fermion self energy $\Sigma_f$ is the same as that for the random magnet in Section~\ref{sec:SY}, 
and %$c$ fermion self energy is that 
the dynamical mean-field $\Delta$ is the same as that of the random matrix model in Eq.~(\ref{rm2}):
\bea
\Sigma_f (\tau) &=& - J^2 G_{f}^2 (\tau) G_{f} (-\tau) \nonumber \\
\Delta (\tau) &=& t^2 G_{c} (\tau)\,. \label{kh21}
\eea
Finally, the hybridization parameter, $\overline{P}$, is determined by the self-consistency equation
\beq
\overline{P} = J_K G_{fc} (\tau = 0^-)\,. \label{kh22}
\eeq
The equations can be obtained from a $G$-$\Sigma$ action analogous to Eqs.~(\ref{syk31a}) and (\ref{rqm6}) 
\begin{align}
& I[G, \Sigma, \Delta, \lambda, P] = \int_0^{\beta} \!\! d \tau \left[\frac{|P(\tau)|^2}{J_K} - \frac{i \lambda(\tau)}{2} \right] \nonumber \\
&-\ln \det \left[ \begin{array}{c}
(\partial_{\tau_1} + i \lambda (\tau_1)) \delta(\tau_1 - \tau_2)  + \Sigma_f (\tau_1, \tau_2)  \\ - P^\ast (\tau_1) \delta(\tau_1-\tau_2) \end{array} \right. \nonumber \\
&~~~~~~~~~~~~~~~~~~~~~~\left. \begin{array}{c} 
-P(\tau_1) \delta(\tau_1-\tau_2) \\ (\partial_{\tau_1}  -\mu) \delta(\tau_1 - \tau_2)  + \Delta (\tau_1, \tau_2) 
\end{array}
\right] \nonumber \\
&~~~~~~~~ - \mbox{Tr} \left( \Delta \cdot G_c \right) + \frac{t^2}{2} 
\mbox{Tr} \left( G_f \cdot G_f \right) \nonumber \\
&~~~~~~~~ - \mbox{Tr} \left( \Sigma_f \cdot G_f \right) - \frac{J^2}{4} 
\mbox{Tr} \left( G_f^2 \cdot G_f^2 \right)\,. \label{kh23}
\end{align}

\begin{figure}
\begin{center}
\includegraphics[width=3.25in]{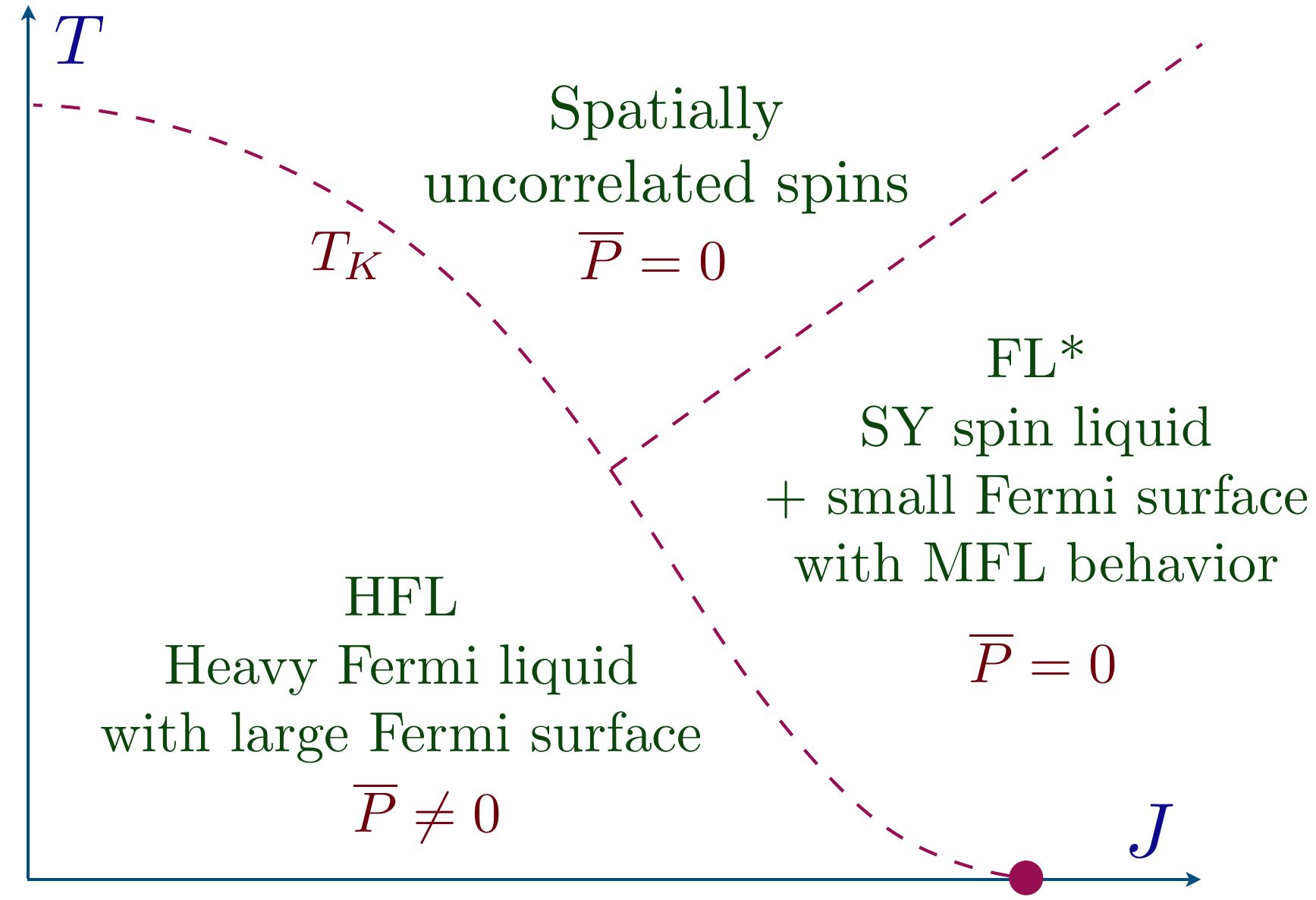}
\end{center}
\caption{Phase diagram of the large $M$ Kondo lattice \cite{Burdin2002} with random exchange. The dashed lines are crossovers, but the $T=0$ filled circle marks a quantum phase transition. The FL* phase and the quantum critical point exhibit linear-in-$T$ resistivity with the small carrier density of the conduction electrons. The HFL exhibits Fermi-liquid $T^2$ resistivity with a large carrier density of both the conduction electrons and the local moments. The critical theory of the HFL-FL* transition at $T=0$ 
has also been discussed for models with full translational symmetry \cite{TSFL04,FLSPRL}.
} 
\label{fig:BGG}
\end{figure}

A complete solution of Eqs.~(\ref{kh20}-\ref{kh22}) requires a numerical analysis, but much can be understood from a low frequency analysis similar to those in the preceding subsections \cite{Burdin2002}. The phase diagram as a function of $T$ and $J$ is shown in Fig.~\ref{fig:BGG}. 
A key determinant of the phase structure is the value of $\overline{P}$. We have $\overline{P} \neq 0$ below the line labeled $T_K$ in Fig.~\ref{fig:BGG}: this line approaches the single site Kondo temperature in the limit $J \rightarrow 0$. In this regime we have the HFL phase, in which both the spins and the electrons are part of the Fermi volume, as described in more detail in Section~\ref{sec:Kondolut}. The transition across the line where $\overline{P}$ vanishes is expected to turn into a smooth crossover once $1/M$ corrections have been included, as there is no underlying order in the HFL phase at $T>0$. However, the situation is different at $T=0$: $\overline{P}$ vanishes at the red circle in Fig.~\ref{fig:BGG}, which denotes a quantum critical point between the HFL and FL* phases: this point is expected to survive $1/M$ corrections because of the discontinuous change in the Fermi volume to be described in Section~\ref{sec:Kondolut}. Moreover, the critical theory has $\overline{P}=0$, and so the critical point has a `small' Fermi energy, in contrast to that for the $t$-$J$ model, as we will discuss further at the end of Section~\ref{sec:KHRG}.

Despite the absence of a sharp phase transition between them, there is a qualitative difference between the observable properties of the HFL and FL* phases at $T>0$. In the HFL phase, the non-zero $\overline{P}$ quenches the singular SYK behavior of the spins at low frequency, just as in Section~\ref{sec:tJPG}; consequently, we expect Fermi liquid-like behavior of the quasiparticles at non-zero $T$ around a large Fermi energy. In particular, the resistivity $\sim T^2$, and the associated carrier density will include both the conduction electrons and the spins. In contrast, while the FL* is also a metal, the carrier density is small, and includes only the conduction electrons. Moreover, in the present fully connected model, the singular SYK behavior of the spins survives. In the large $M$ limit, the spins are decoupled from the conduction electrons when $\overline{P}=0$, but there will be a coupling at higher orders in $1/M$. So although $\Sigma_c = 0$ at $M=\infty$, the leading correction to the imaginary part of the self energy \cite{Burdin2002}
\beq
\mbox{Im} \Sigma_c (\omega=0) \sim \left(\frac{J_K}{M} \right)^2  \int_0^\infty \frac{d \Omega}{\sinh (\beta \Omega)} \frac{\rho_Q (\Omega)}{t} \,,~~~~~
\eeq
where $\rho_Q (\Omega)$ is the SYK spin spectral density obtained in 
Eq.~(\ref{rqm7a}). This leads to marginal Fermi liquid behavior \cite{Varma89} for the small density of conduction electrons, with $\mbox{Im} \Sigma_c (\omega=0) \sim T$, and a linear-in-$T$ resistivity, using transport computations as defined in Section~\ref{sec:transport}. 

This mechanism for the linear-in-$T$ resistivity in the Kondo lattice model is distinct from that for the $t$-$J$ model in Section~\ref{sec:linearT}. Here the carrier density at the critical point is {\it small\/}, {\it i.e.\/} it does not involve the spins due to the breakdown of the Kondo effect. In contrast, the carrier density in Section~\ref{sec:linearT} was {\it large\/}, involving all the electrons. Moreover, here the linear-in-$T$ resistivity arises already in the leading scaling results for the SYK model, while those in Section~\ref{sec:linearT} required corrections to scaling.

\subsection{Luttinger theorem}
\label{sec:Kondolut}

The Luttinger theorem is normally applied to metallic phases of electrons, and we obtained an instance of this in Section~\ref{sec:tJPG} for the Fermi liquid phase of the $t$-$J$ model. But we also saw a modified Luttinger theorem in Section~\ref{sec:SY} for spins in an insulating Kondo magnet. The Kondo Hamiltonian Eq.~(\ref{kh1}) has both spins and mobile electrons, and there now are distinct realizations of the Luttinger theorem in the HFL and FL* phases \cite{FLSPRL,TSFL04}.

It is convenient to present the discussion in the large $M$ formulation of the theory in Section~\ref{sec:KondolargeM}, although all statements in the present subsection hold to all orders in $1/M$. When expressed in terms of the spinons $f_\alpha$, the theory in Eq.~(\ref{kh2}) has a U(1) gauge symmetry, along with global U(1) symmetries associated with the total charge of the $c_\alpha$ electrons, $(M/2) p$, and the total spin $\mathcal{S}_z$. In principle, all 3 U(1) symmetries will lead to their own and distinct Luttinger constraints, unless there are condensates of bosons carrying U(1) charges \cite{powell1,coleman1} (we review this connection between U(1) symmetries and the Luttinger constraints at the end of Section~\ref{sec:cfllutt}). In our discussion, the relevant boson is the hybridization $P \sim c_\alpha^\dagger f_\alpha$, and this is a Higgs boson because it carries a U(1) gauge charge. \\

\noindent
(A) {\it FL* phase\/}

In the FL* phase, there is no Higgs condensate $\langle P \rangle = 0$, so all 3 Luttinger constraints apply. An important property of this phase is that the off-diagonal Green's function vanishes at all frequencies $G_{fc} = 0$.
Consequently, the constraints arising from the gauge U(1) and $\mathcal{S}_z$ symmetries are essentially identical to those considered for insulating quantum magnets in Section~\ref{sec:SY}, which are in turn related to the discussion in Section~\ref{sec:syklutt}. So we need only consider the constraint associated with $c_\alpha$ fermion charge, which is
\beqn
G_{c} (\tau = 0^-) = \frac{p}{2}\,. \label{kondolut1}
\eeqn
We can write $G_{c}$ in the FL* phase in the general form 
\beqn
G_{c} (i \omega_n) = \frac{1}{i \omega_n + \mu - t^2 G_{c} (i \omega_n) - \Sigma_{c} (i\omega_n)} \label{kondolut2}
\eeqn
We have now included a self-energy $\Sigma_c (i \omega_n)$ which arises from $1/M$ corrections. This obeys
 $\mbox{Im} \Sigma_{c} (i0^+) = 0$ at $T=0$, and that is not the case for $\Delta (\omega)$ in Eq.~(\ref{kh21}). 
Another important point is that $1/M$ contributions to $\Sigma_{c} (i \omega_n)$
can be obtained from a Luttinger-Ward functional, and the Luttinger constraint will then follow straightforwardly.
First, we solve Eq.~(\ref{kondolut2}) to write
\beqn
G_{c} (i \omega_n) = \int_{-\infty}^{\infty} d \Omega \frac{ \rho(\Omega)}{i \omega_n + \mu - \Sigma_{c}(i \omega_n) - \Omega} \label{kondolut3}
\eeqn
where $\rho (\Omega)$ is the single particle density of states of the random matrix model in Eq.~(\ref{rm3a}). We now proceed with the analysis of the Luttinger constraint as in Section~\ref{sec:syklutt}: we expect that the contribution from the frequency derivative of the self energy vanishes, $I_2=0$, and then such an analysis shows that Eq.~(\ref{kondolut1}) implies
\beqn
\int_{-2t}^{E_F} d \Omega \, \rho(\Omega) = \frac{p}{2}\,, \label{kondolut4}
\eeqn
where the Fermi energy is 
\beqn
E_F = \mu - \Sigma_{c} (0)\,. \label{kondolut5}
\eeqn
We note that the analog of the analysis above applies also to the disordered Fermi liquid phase of Section~\ref{sec:tJPG} \cite{Parcollet1}.\\

\noindent
(B) {\it HFL phase\/}

In the HFL phase of the Kondo-Heisenberg lattice, we do have a Higgs condensate $\langle P \rangle \neq 0$, and so there is no separate Luttinger constraint from the U(1) gauge symmetry. The analysis of the Luttinger constraint \cite{Burdin_2000} with the conservation of the electron charge will now lead to a `large' Fermi energy of size $(1+p)/2$ per spin (for the particle-hole symmetric value $\kappa=1/2$ in Eq.~(\ref{rqm2}) for the SU($M$) spins) . 

We begin by writing Dyson's equation in Eq.~(\ref{kh20}) in a general form valid beyond the large $M$ limit. We define an auxiliary matrix Green's function by
\beqn
\left[\mathcal{G}(i \omega_n, \Omega)\right]^{-1} = 
\left( \begin{array}{cc} i \omega_n - \overline{\lambda} & 0 \\
0 & i \omega_n + \mu - \Omega \end{array} \right) - \Sigma (i \omega_n) \label{kondolut6}
\eeqn
where $\Sigma (i \omega_n)$ is the matrix self energy which obeys $\mbox{Im} \Sigma (i 0^+) = 0$ at $T=0$. As in Eq.~(\ref{kondolut3}), we have replaced the $t^2 G_c (i \omega_n)$ in Eqs.~(\ref{kh20}) and (\ref{kh21}) by $\Omega$. The presence of the Higgs condensate in the HFL phase requires that the off-diagonal matrix elements of $\Sigma (i \omega_n)$ are non-zero, and this is crucial for the Luttinger constraint here.

We now state a useful identity, which can be verified by explicit computation, for the trace of the matrix Green's function $G (i \omega)$
(which counts both the $f_\alpha$ and $c_\alpha$ fermions)
\bea
\mbox{Tr}\, G (i \omega) &=&  \int_{-\infty}^{\infty} d \Omega \rho (\Omega) \Biggl[  i \frac{d}{d \omega} \ln \det [\mathcal{G}(i \omega, \Omega)] \nn
&~&~~- i \mbox{Tr} \left( \mathcal{G}( i \omega, \Omega) \frac{d}{d \omega} \Sigma (i \omega) \right) \Biggr]  \,. \label{kondolut7}
\eea
Notice the similarity of Eq.~(\ref{kondolut7}) to the identity used for the SYK model in Eq.~(\ref{sykl1}). The subsequent analysis proceeds as there. In the present situation, the $I_2$ contribution of the second term in Eq.~(\ref{kondolut7}) vanishes from the usual Luttinger-Ward functional argument because we are in a Fermi liquid phase and there is no anomaly at $\omega=0$. For $p<1$, the first term in Eq.~(\ref{kondolut7}) yields the Luttinger constraint \cite{Burdin_2000,Nikolaenko:2021vlw}
\beqn
\int_{-2t}^{E_F} d \Omega\, \rho(\Omega) = \frac{1+p}{2}\,, \label{kondolut8}
\eeqn
which, unlike Eq.~(\ref{kondolut4}), counts both the $c_\alpha$ electrons and the spins. The expression for the Fermi energy in Eq.~(\ref{kondolut5}) is now replaced by 
\beqn
\det \left[\mathcal{G}(0, E_F)\right]^{-1} = 0\,. \label{kondolut9}
\eeqn

\subsection{RG analysis for SU(2) symmetry}
\label{sec:KHRG}

We will now analyze $\overline{\mathcal{Z}}_{KH}$ in Eq.~(\ref{kh2}) by combining the RG analysis of Section~\ref{sec:SYRG} with the `poor-man's scaling' RG of the Kondo problem.

This analysis will be carried out perturbatively in $J_K$, as in the `poor-man's scaling' \cite{hewson_book}.
Then at leading order, with $J_K =0$ but the mean-square hopping $t$ arbitrary, the equations for the $c_\alpha$ Green's function reduce precisely to those solved earlier in Section~\ref{sec:matrix} for the random matrix problem. These yield a fermion Green's function with a constant density of states at the Fermi level $\sim 1/t$, as in Eq.~(\ref{rm3a}). Note that this is a Fermi level only of the $c$ electrons, and so is a `small' Fermi `surface': so the present RG analysis is an expansion about the small Fermi surface. After a Fourier transform, the constant density of states implies $G(\tau) \sim 1/(t \tau) $ at large $|\tau|$. We therefore replace the fermion $c_\alpha$ with a `bath' fermion $\psi_{\alpha}$, which is the analog of the bosonic field $\phi_a $
that we introduced in Section~\ref{sec:SYRG} for the random quantum magnet. Similarly, we endow $\psi_\alpha$ with a momentum and a dispersion, with the dispersion chosen so that $\psi_\alpha (x=0)$ has the same temporal correlator as that of $c_\alpha$.
In this manner, we can express the problem in terms of an impurity Hamiltonian of a single $S=1/2$ spin coupled to fermionic and bosonic baths \cite{Sengupta2000}
\begin{eqnarray}
&& H_{\rm imp}  =  \gamma \vec{S} \cdot \vec{\phi} (0)  + \frac{J_K}{2} \vec{S} \cdot \left( 
\psi_\alpha^\dagger (0) \vec{\sigma}_{\alpha\beta} \psi_\beta (0) \right) \nonumber \\
&& + \int dk \, k \, \psi_{k\alpha}^\dagger \psi_{k \alpha} + \frac{1}{2} \int d^d x \left[ \pi_a^2 + (\partial_x \phi_a)^2 \right] \,.\label{kh24}
\end{eqnarray}
The bath correlators are 
\begin{equation}
Q(\tau) \sim \frac{1}{|\tau|^{d-1}} \quad, \quad R(\tau) \sim \frac{\mbox{sgn}(\tau)}{|\tau|}\,. \label{khrgqr}
\end{equation}
and the value of $d$ is to be determined by solving the self-consistency condition for $Q$ in Eq.~(\ref{kh4}). We have argued above that the self-consistency condition for $R$ is satisfied by a Fermi liquid constant density of states (of the small Fermi surface) at the Fermi level, and that dictated the $R(\tau)$ in Eq.~(\ref{khrgqr}). 

\begin{figure}
\centering
\includegraphics[width=3.2in]{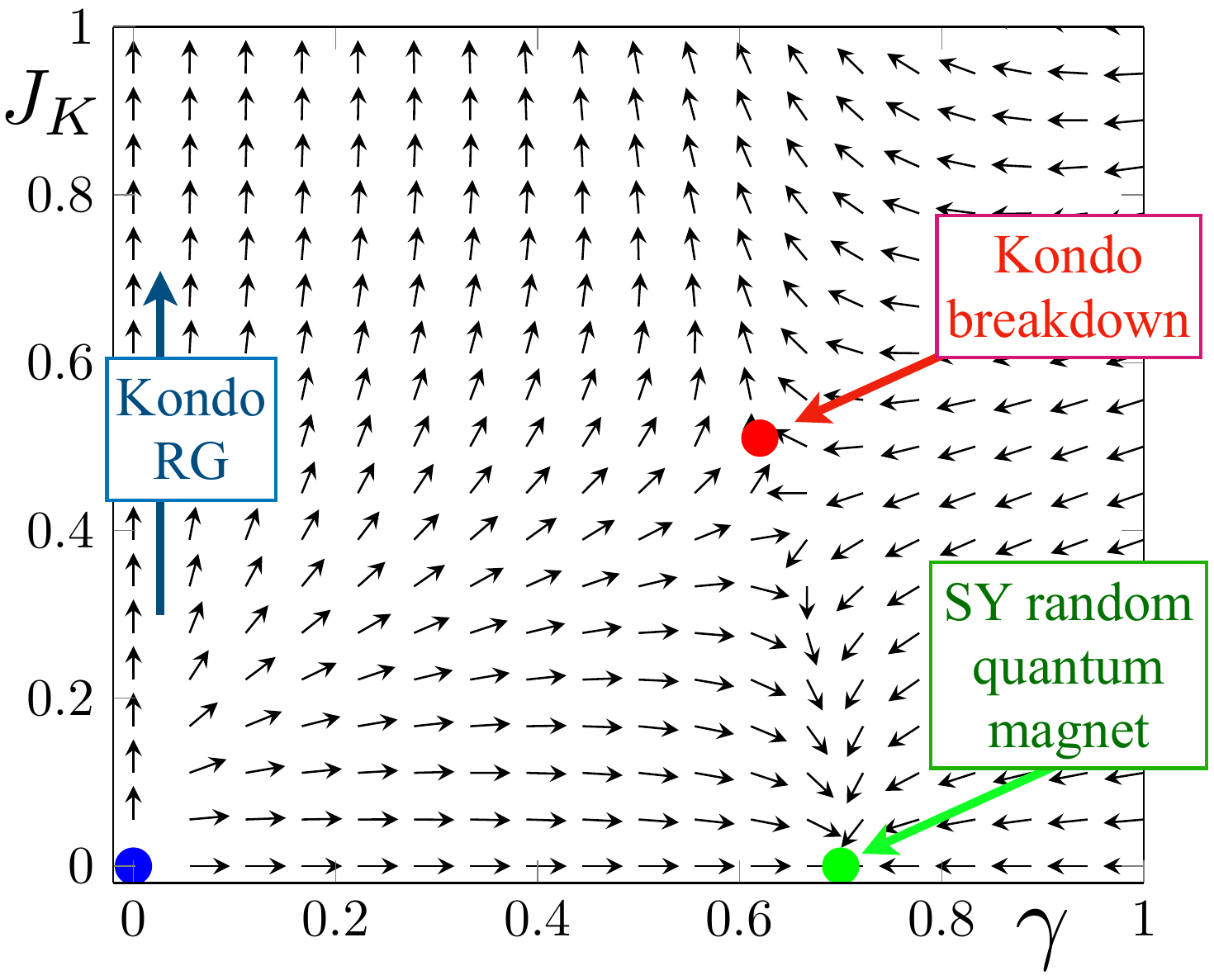}
\caption{RG flow \cite{Sengupta2000} of Eq.~(\ref{kondorg}) for $\epsilon = 1$. 
The $\gamma=0$ axis corresponds to the usual Kondo RG flow \cite{hewson_book}. The `Kondo breakdown' fixed point has one relevant direction, and describes a phase transition between a small Fermi surface phase (likely with magnetic order for SU(2)) associated with the random quantum magnet fixed point of Section~\ref{sec:SYRG}, and a large Fermi surface HFL at large $J_K$. Compare to Fig.~\ref{fig:tJrg} for the random $t$-$J$ model.}
\label{fig:kondorg}
\end{figure}
The impurity Hamiltonian in Eq.~(\ref{kh24}) has two couplings, $J_K$ and $\gamma$ and their RG flow equations can be computed by combining the analyses of the usual Kondo model \cite{hewson_book} and those for the random quantum magnet in Eq.~(\ref{betagamma}). This analysis is perturbative in $J_K$ and $\epsilon = 3-d$, and the one loop RG equations are  \cite{Sengupta2000,Si0,ZhuSi}
\begin{align}
\beta (\gamma) &= - \frac{\epsilon}{2} \gamma + \gamma^3  \nonumber\\
\beta( J_K) &=  \gamma^2 J_K - J_K^2 \,. \label{kondorg}
\end{align}
The resulting RG flow is plotted in Fig.~\ref{fig:kondorg}. The random quantum magnet fixed point of Section~\ref{sec:SYRG} is stable to turning on a small $J_K$, implying the stability of a small Fermi surface phase. For SU(2), this small Fermi surface phase has spin glass order; but more generally in models which are not fully connected, it could be a spin liquid, leading to a FL* state as in Section~\ref{sec:KondolargeM}. For larger $J_K$, there is an unstable fixed point beyond which the flow is towards $J_K \rightarrow \infty$, presumably to a large Fermi surface HFL. We have labeled this fixed point as `Kondo breakdown' \cite{Sengupta2000,Si1,Si2,Burdin2002,FLSPRL,TSFL04} because it separates the HFL phase with Kondo screening, from the small Fermi surface without Kondo screening.

It remains to solve the self-consistency equation in Eq.~(\ref{kh4}) to determine the value of $\epsilon$. As in Sections~\ref{sec:SYRG} and \ref{sec:tJRG}, the scaling dimension of the spin operator can be determined \cite{ZhuSi} to all orders at fixed point $\gamma^\ast \neq 0$, and we find the same result as in Eq.~(\ref{tJ30}). The self-consistent value is again
$\epsilon=1$, $d=2$, as for the $t$-$J$ model.

It is interesting to compare the RG flow diagram for the Kondo-Heisenberg model in Fig.~\ref{fig:kondorg} with that for the $t$-$J$ model in Fig.~\ref{fig:tJrg}. In both cases, we have a critical fixed point with one relevant direction, and similar critical correlators for the electron and spin operators: a Fermi liquid-like critical electron correlator, and a SYK-like critical spin correlator, as in Eq.~(\ref{tJ30}) for $d=2$ and $r=0$. However the density of electrons participating in the electron correlator in Eq.~(\ref{tJ30}) is different in the two cases: at the Kondo breakdown fixed point the density of electrons is {\it small\/}, and does not count the spins (as is clear from Section~\ref{sec:Kondolut} for $\overline{P} =0$, and from the large $M$ analysis in Section~\ref{sec:KondolargeM}), while at the $t$-$J$ model fixed point the density of electrons is {\it large\/} and counts all electrons. 

\subsection{Numerical analysis}
\label{sec:KHnum}

A complete numerical analysis of the single-site, self-consistent quantum problem defined by Eqs.~(\ref{kh2},\ref{kh3},\ref{kh4}) with SU(2) 
symmetry has not yet been carried out. However, there have been a number of studies of related models, motivated by an uncontrolled EDMFT analysis of low dimensional models with non-random exchange \cite{SiRMP,Si1,Si2}, and a self-consistency condition of the spin correlator which differs from that in Eq.~(\ref{kh4}). The self-consistency on spin correlators has only been systematically justified in models with random exchange, like those considered above, as we noted at the end of Section~\ref{sec:EDMFT}.
The numerical analyses were carried out for Ising spin symmetry \cite{SiIsing1,SiIsing2,SiIsing3,Ingersent1}, although recent works have also examined SU(2) spin symmetry \cite{SiSU21,SiSU22}. Aspects of these studies are similar to the RG results described in Section~\ref{sec:KHRG}, with a SYK-like spin spectral density ({\it i.e.\/} spin correlations similar to Eq.~(\ref{SStau})) at a critical point between a Fermi liquid phase and another phase which is presumed to break spin symmetry.

%%%%%%%%%%%%%%% Numerical methods

\section{Overview of numerical algorithms for fully connected SU(2) models} 
\label{sec:numerical_methods}

In the large-$M$ limit, the action in Eq.~(\ref{eq:Seff_tUJ}) is solved using a saddle point technique,
which reduces to non linear integral equations for the Green function $G$ as in Eq.~(\ref{eq:SYK}) and a
simple expression of higher correlators in terms of $G$ using Wick's theorem.  
The SU(2) case is more complex. The action in Eq.~(\ref{eq:Seff_tUJ}) is
still a (local) quantum many body problem (at ${\cal J} =0$, it is the
Anderson quantum impurity model) and more advanced algorithms are required to solve it.

In this section, we provide a brief introduction for non-experts to the algorithms used 
to solve the SU(2) models discussed above and discuss their strengths and limitations.
The goal is to solve the local action in Eq.~(\ref{eq:Seff_tUJ}), for fixed bath $\Delta$ and retarded spin-spin interaction ${\cal J}$.
The self-consistency condition on $\Delta$ and ${\cal J}$ is then solved with an iterative technique \cite{DMFT}.
Note however that the self-consistency can generate a non trivial frequency dependence for the bath $\Delta$ and the interaction ${\cal J}$, respectively, which complicates the solution.
For example, any technique based on a flat bath spectral function with a large high energy cut-off, e.g. integrability, is inoperable in this context.

A lot of progress has been made in the last two decades on numerical algorithms to solve such quantum impurity models with complex baths and interactions, in the context of DMFT and its extensions \cite{RevModPhys.83.349}.
Several classes of algorithms are available, in particular action based Quantum Monte Carlo (QMC) or Hamiltonian
based methods (exact diagonalization, NRG, DMRG, tensor network methods). 
The QMC are the methods of choice here, due to the retarded spin-spin interaction term in Eq.~(\ref{eq:Seff_tUJ}).

The SU(2) insulating case was studied first in the paramagnetic phase using an auxiliary field QMC \cite{GrempelRozenberg98}, 
with a sampling method of the auxiliary field in the Matsubara frequency space. 
Local moments solutions were obtained at low temperatures, as discussed in Section~\ref{sec:SYSU2} \cite{GrempelRozenberg98, Dumitrescu2021}.

Recent works however have used the ``Continuous Time'' QMC (CTQMC) family of algorithms for quantum impurity models.
The central idea is to perform an expansion of the partition function
$Z$ either in powers of the interaction $U$ and $\cal J$ around the non-interacting limit (CT-INT \cite{RubtsovCTQMC2005} or CT-AUX\cite{CTAUX2008}
algorithms), or in powers of the bath hybridisation $\Delta$  around the atomic
limit (CT-HYB algorithm \cite{cthyb2006}).

Let us consider first the CT-INT algorithm, used in Sections~\ref{sec:ChaHubbard} and \ref{sec:tJnumerics} \cite{Cha_2020,Dumitrescu2021}.
The partition function $Z$ 
\begin{equation}\label{eq:ZDefPathIntegral}
   Z \equiv \int {\cal D} c^\dagger_\sigma(\tau) {\cal D} c_\sigma(\tau) e^{- S_{tUJ}(c^\dagger_\sigma(\tau), c_\sigma(\tau))}
\end{equation}
is expanded in both $U$ and ${\cal J}$ to any order as 
%
%%%%%%%% CORRECTED FORMULA 9.2 %%%%%%%%%%%%%%%%%%%%%%
%
\begin{multline}
   \label{eq:ZexpansionCTINT}
   Z = \sum_{n\geq 0}\sum_{p\geq 0} 
    \dfrac{(-U)^n}{n! \, p! 2^p }
    \int_0^\beta 
    \prod_{i=1}^{n} d\tau_i
    \prod_{j=1}^{p} d\tau'_j d\tau''_j
   {\cal J} (\tau'_j - \tau''_j)
  \times \\
    \sum_{a_j = x,y,z} 
     \left \langle
	{\cal T}_\tau
	 \prod_{i=1}^{n} 
	 n_{\uparrow}(\tau_i) n_{\downarrow}(\tau_i)
	S^{a_j} (\tau'_j) S^{a_j}(\tau''_j) 
     \right \rangle_{0}.	
\end{multline}
%
%%%%%%%% END CORRECTED FORMULA 9.2 %%%%%%%%%%%%%%%%%%%%%%
%
The average is taken in the non-interacting model %$U= {\cal J} = 0$ 
and, via Wick's theorem, can be expressed as a determinant in terms 
of the non-interacting impurity Greens function.

The principle of the CTQMC is to sample  $Z$ stochastically with a Metropolis Monte Carlo algorithm, 
computing integrals of various dimensions simultaneously.
A Monte Carlo is defined by its configuration space and the elementary steps constituting the Markov chain in this space.
Discretizing the integrals with a Riemann sum on a regular grid of step $\delta \tau$, 
the configurations ${\cal C}$ are simply given by the orders $n$ and $p$ and the set of $\tau_i$, $\tau'_i, \tau''_j$.
Formally, $Z$ can then be written as 
\begin{equation}
   \label{eq:ZDiscrete}
   Z =  \sum_{n\geq 0}\sum_{p\geq 0} \sum_{{\cal C}_{np}} (\delta \tau)^{n + 2p} f_{{\cal C}_{np}}(\tau_i, \tau'_i, \tau''_j)
\end{equation}
where  $f_{{\cal C}_{np}}$ is given by the time-ordered correlator in Eq.~(\ref{eq:ZexpansionCTINT}).
The  weight of a configuration ${\cal C}_{np}$ is $ w_{{\cal C}_{np}} = (\delta \tau)^{n + 2p}  |f_{{\cal C}_{np}}|$.
The MC Markov chain consists of elementary steps in adding or removing one (or two) vertices at some randomly chosen times, 
sampling all the integrals simultaneously.
The various correlation functions are then computed from this Markov chain, as their expansions are very similar \cite{RevModPhys.83.349}.
The typical order of the expansion explored by the algorithm can be shown to be proportional to $\beta$, and in practice can go up to several hundreds.
In this model, CT-INT can develop a sign problem at low temperature in some parameter regimes due to the ${\cal J}$ term.
In practice however it can often be strongly reduced by using Gaussian counter-terms added to both the bare action and the $U$ interaction term \cite{RubtsovCTQMC2005, Dumitrescu2021}.

The CT-HYB algorithm is similar to CT-INT but is based on a double expansion around the atomic limit, 
{\it i.e.} in powers of $\Delta(\tau)$, and ${\cal J}_\perp$ where the retarded spin-spin interaction 
is rewritten $ {\cal J} \vec{S} (\tau) \cdot \vec{S} (\tau')  = {\cal J}_\parallel S^{z}(\tau) S^{z}(\tau') + {\cal J}_\perp \sum_{a= \pm}  S^{a}(\tau) S^{-a}(\tau')$.
Expanding Eq.~(\ref{eq:ZDefPathIntegral}) in $\Delta$ and ${\cal J}_\perp$, and using the antisymmetric property of time-ordered 
fermionic correlators, the partition function $Z$ reads
%
%%%%%%%% CORRECTED FORMULA 9.4 %%%%%%%%%%%%%%%%%%%%%%
%
\begin{multline}
   \label{eq:ZexpansionCTHYB}
Z =
 \sum_{n\geq 0}\sum_{p\geq 0} 
    \dfrac{1}{2^p p! n!^2}
    \int_0^\beta 
    \prod_{i=1}^n d\tau_i d\tau'_i
\prod_{j=1}^p d\bar\tau_j d\bar\tau'_j
  \times \\
\prod_{j=1}^p {\cal J}_\perp (\bar\tau_j- \bar \tau'_j)
 \sum_{\sigma_i = \uparrow, \downarrow} 
\sum_{a_j = \pm} 
\det_{1 \leq i,j\leq n} \left [ \Delta_{\sigma_i}(\tau_i - \tau'_j) \right ]
  \times \\
  \left \langle
   {\cal T}_\tau  \prod_{i=1}^n  c^\dagger_{\sigma_i}(\tau_i) c_{\sigma_i}(\tau'_i)
   	 \prod_{j=1}^{p}
	S^{a_j}(\bar\tau_j) S^{-a_j}(\bar \tau'_j) 
  \right \rangle_{\text{atomic}}	
\end{multline}
%%%%%%%% END CORRECTED FORMULA 9.4 %%%%%%%%%%%%%%%%%%%%%%
%
where the atomic correlators are taken in the isolated atom, {\it i.e.} $\Delta=0$ and ${\cal J}_\perp =0$.
The CT-HYB algorithm was introduced in the DMFT case ${\cal J}=0$ \cite{cthyb2006}, and later 
extended to the EDMFT case ${\cal J}\neq 0$ \cite{OtsukiCTHYBDoubleExpQMC, Otsuki2013}. 
The ${\cal J}_\parallel$ component of the retarded spin-spin interaction can be taken into account exactly in the atomic correlator.
The second expansion in ${\cal J}_\perp$ is however necessary,
since no efficient algorithm is known to compute the atomic correlators in the presence of a retarded non abelian spin-spin interaction term.

The CTQMC algorithms provide an exact solution in Matsubara time.
The main advantage over the previous generation of QMC impurity solvers \cite{HirschFye1986} is the ability to treat general atomic interactions, including 
retarded interactions, and the absence of time discretization as the algorithm can be performed directly in the continuous time 
limit $\delta \tau\rightarrow 0$ (hence their name). 
This last point can be illustrated easily, e.g. on a CT-INT.
Let us consider a Monte Carlo step from  a configuration $\cal C$ of order $(n,p)$ to a configuration $\cal C'$ of order
$(n+1,p)$. Their weights $w_{\cal C}$ and $w_{\cal C'}$ are proportional to $(\delta \tau)^{n + 2p}$ and
$(\delta \tau)^{n + 1 + 2p}$ respectively as seen in Eq.~(\ref{eq:ZDiscrete}). However, the Markov chain steps can be chosen so that the Metropolis ratio
\begin{equation}
R_{{\cal C}\rightarrow {\cal C'}} = \dfrac
{T_{{\cal C'}\rightarrow {\cal C}} w_{\cal C'} }
{T_{{\cal C}\rightarrow {\cal C'}} w_{\cal C} }
\end{equation}
(where $T_{{\cal C}\rightarrow {\cal C'}}$ is the proposition probability of the step), 
is {\it finite} for $\delta \tau\rightarrow 0$.
Indeed, $T_{{\cal C}\rightarrow {\cal C'}} = \delta\tau/\beta $ (the probability to randomly pick up one new time on the imaginary axis), 
and $T_{{\cal C'}\rightarrow {\cal C}} = 1/(n+1)$ (the probability to randomly select one time to remove from the configuration ${\cal C'}$).
As $R$ controls the Metropolis algorithm, its finite limit ensures the continuous time limit of the algorithm, 
even though the weights themselves vanishes at $\delta \tau\rightarrow 0$.
The absence of time grid extrapolation is a great advantage in practice at low temperatures \cite{RevModPhys.83.349}.

The main limitations of the CTQMC includes some sign problem (depending on the exact
algorithm and the parameter regime), a poor scaling with temperature (like
$\beta^3$), and most importantly their restriction to imaginary time.  Some
delicate analytical continuation are required to access real frequency
correlations.  Note that a third generation of QMC for impurity problems has
recently appeared that work directly in real time
\cite{InchwormCohenGullMillisRealTime2015, ProfumoRealTimeQMC, QQMCPRL}.  They
are based on diagrammatic computations of physical quantities rather than the
partition function.  It is an open question whether these new approaches, when
properly generalized to handle the retarded spin spin interaction, will allow
to solve some of the remaining challenge in these systems, including e.g. the low temperature behavior.

%%%%%%%%%%%%%%%%%%%%%%%%%%%%%%%%%%%%%%%%%%%

\section{Lattice models of SYK-atoms}
\label{sec:lattice}

This section returns to the SYK model of Section~\ref{sec:SYK}, and follows a different strategy towards connecting it to the physics of quantum matter. 
In Sections~\ref{sec:rqm}, \ref{sec:tJU}, and \ref{sec:KH} we imposed `Mottness' on the SYK model by adding an on-site repulsion on each site $i$; this approach then connected naturally to dynamical mean field theories of correlated materials. The present section will examine an alternative approach in which the SYK model is viewed as a multi-orbital atom, and $i$ labels the orbitals on such an atom. Then we will examine a lattice of such `SYK-atoms', and find that such models can also exhibit regimes of non-Fermi liquid behavior with linear-in-$T$ resistivity. For models with a single band of SYK-atoms, these non-Fermi liquids are invariably `bad metals' at temperatures higher than the renormalized bandwidth, in that the resistivity exceeds the MIR resistivity, where the quantum of resistance is redefined as $h/Ne^2$ for the $N-$orbital atom. This is in contrast to the non-Fermi liquids obtained using a two-band generalization of these models in Sec.~\ref{sec:MFL},  or, those introduced earlier in Section~\ref{sec:tJU}, which display resistivities smaller than the MIR resistivity. 

The models described in this section are interested in describing the anomalous transport properties of non-Fermi liquid metals with short-ranged interactions in crystalline settings. We seek to address the fate of the electronic Fermi surface in the regime of strong interactions when there are no long-lived low energy quasiparticles. Therefore it is natural to address the extent to which the models introduced thus far can serve as elementary building blocks for addressing these fundamental questions in a controlled setting. In the next few subsections, we discuss properties of a number of different variants of the SYK models.

\subsection{Breakdown of a heavy Fermi liquid}
\label{sec:hFL}

\begin{figure}[h]
\begin{center}
\includegraphics[scale=0.25]{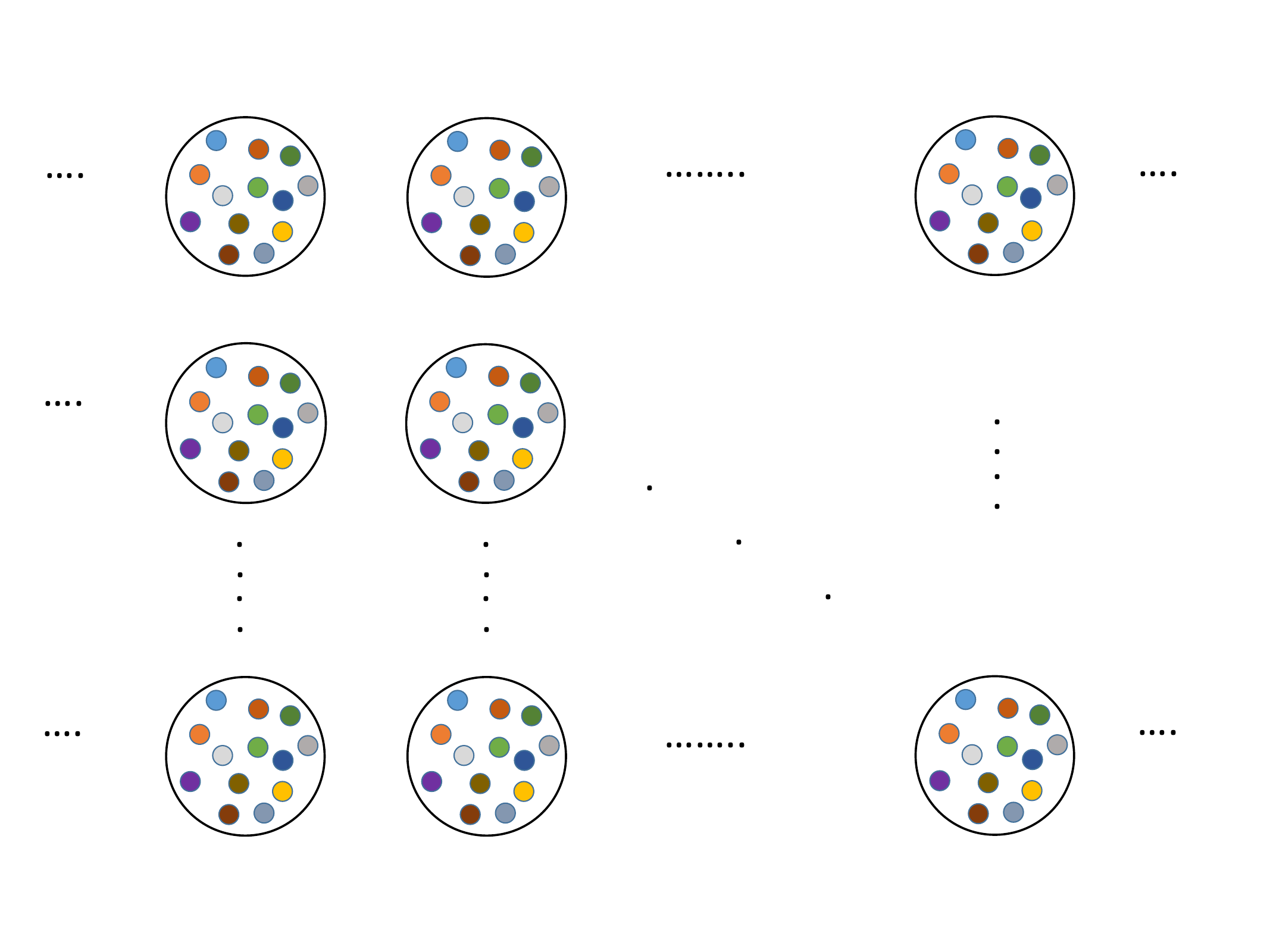}
\end{center}
\caption{A basic building block for studying translationally invariant lattice models constructed out of SYK-atoms with $N-$orbitals per site. The different sites are coupled together by single-electron hopping terms.}
\label{lattice}
\end{figure}

We begin by writing a model for electrons with orbital labels $i = 1, 2, .. , N$ and hopping on the sites, $\r$, of a $d-$dimensional hyper-cubic lattice (Fig.~\ref{lattice}). The Hamiltonian, $H_c = H_{\tn{kin}} + H_{\tn{int}}$, is given by
\begin{subequations}
\begin{align}
&H_{\tn{kin}} = -\sum_{\r,\r'} t_{\r\r'} c_{\r i}^\dagger c^{\vphantom \dagger}_{\r' i} -\mu \sum_{i} c_{\r i}^\dagger c_{\r i}^{\vphantom \dagger}  \label{lattice1a}\\
&H_{\tn{int}} = \frac{1}{(2 N)^{3/2}} \sum_{\r} \sum_{ijk\ell=1}^N U_{ij;k\ell} \, c_{\r i}^\dagger c_{\r j}^\dagger c_{\r k}^{\vphantom \dagger} c_{\r \ell}^{\vphantom \dagger} .
 \label{lattice1b}
\end{align}
\end{subequations}
The above Hamiltonian is thus a generalization of a completely (0+1)-dimensional model, $H_2 + H_4$, as introduced in Eqs. (\ref{rmt1}) and (\ref{syk1}), respectively. The simplest choice for the hopping and interaction parameters is to make them both random variables \cite{Balents}. Alternatively, the hopping parameters can be made translationally invariant such that they depend only on the spatial separation, $|\r-\r'|$ \cite{Zhang17,shenoy}. Of special interest is the situation where additionally the interaction terms, $U_{ij;k\ell}$, are also assumed to be {\it independent} of the site label $\r$ \cite{DCsyk}. Then, $H_{\tn{int}}$ is constructed as a repeated array of the $H_4$ term in Eq. (\ref{syk1}) for every site $\r$ and the $U_{ij;k\ell}$ are identical at every site, thereby preserving an exact (instead of statistical) translational invariance. The couplings are still chosen from a Gaussian random distribution with zero mean $\overline{U_{ij;k\ell}} = 0$ and variance $\overline{|U_{ij;k\ell}|^2} = U^2$. Appealing to the self-averaging properties of the SYK model in the large$-N$ limit, we can compute correlation functions of a typical translationally invariant realization (where crystalline momentum is a good quantum number) by averaging over the disorder realizations. The chemical potential, $\mu$, allows us to tune the electron density $\mathcal{Q}$. Variants of the one-band lattice model without any hopping terms (i.e. $t_{\r\r'}=0$) and with only four-fermion interactions that couple together different sites have also been studied \cite{Gu17,SS17}, with properties that are vastly different from what we discuss below. A different family of lattice SYK models defined in terms of Majorana fermions have been used to study insulating transitions out of a diffusive metal \cite{XuMIT17,Yao} and effects of longer ranged correlated couplings on diffusive transport \cite{DVK17}.

In the large$-N$ limit, once again only the `melon graphs' survive (Fig. \ref{melon}), but the Green's function now includes an additional contribution due to $H_{\tn{kin}}$ and takes a more non-trivial form compared to Eqs. (\ref{syk2a}) and (\ref{syk2b}):
\begin{subequations}
\begin{align}
G(i\omega_n,\k) & = \frac{1}{i \omega_n - \varepsilon_\k - \Sigma (i\omega_n,\k)} \label{lattice2a} \\ 
\Sigma (i\omega_n,\k) &= -  U^2 T\sum_{i\Omega_n}\int_{\k_1} G(i\Omega_n,\k_1)~\Pi(i\omega_n+i\Omega_n,\k+\k_1) 
\label{lattice2b} \\
\Pi(i\Omega_n,\q)  &= T\sum_{i\omega_n'}\int_{\k'} G(i\omega_n',\k')~G(i\omega_n'+i\Omega_n,\k'+\q), \nonumber
\end{align}
\end{subequations}
where $\int_\k\equiv \int d^d\k/(2\pi)^d$ and $\varepsilon_\k$ is the electron dispersion. These equations are reminiscent of the usual DMFT equations, but where the electron self-energy is allowed to be momentum dependent. As we discuss below, in the strong coupling limit, the momentum dependence becomes parametrically weaker compared to the frequency dependent renormalization, stemming from the local SYK physics \cite{DCsyk}. 

The above equations are difficult to solve analytically in general as a function of frequency and momenta; the full solution can be obtained numerically across the entire Brillouin zone. However, significant insights can be gained analytically by starting with a low energy guess for a self-consistent solution. Recall that in the limit where the sites are all decoupled, at energies $\omega\ll U$ for $H_4$ in Eq. (\ref{syk1}), the electron scaling dimension $\Delta=1/4$. By simple power counting arguments, $H_{\tn{kin}}$ is a relevant perturbation; as a result the power-law solution for the Green's function obtained earlier can not survive down to the lowest energies and there will be a crossover to a regime dominated by $H_{\tn{kin}}$ that can nevertheless be strongly renormalized due to interactions \cite{Parcollet1,Balents}.  

At the lowest energies, we assume the self-energy to take a Fermi liquid form,
\beqn
\Sigma(i\omega_n,\k) = -i (Z^{-1} - 1)\omega_n + \Delta\varepsilon_\k\,. \label{lattice3}
\eeqn
where $Z$ is the quasiparticle residue and $\Delta\varepsilon_\k$ is the renormalization associated with the dispersion, to be determined self-consistently. As a further simplification, zooming in on the near vicinity of the Fermi surface, we can parametrize $\Delta\varepsilon_\k= (\Delta v_F )k$ where $\Delta v_F$ is the Fermi velocity renormalization and $k$ is measured relative to the Fermi surface. The self-consistency condition then reduces to,
\begin{subequations}
\begin{eqnarray}
 Z^{-1} - 1 &=& \nu_0^2 U^2 Z \label{lattice4}\\
 \frac{\Delta v_F}{v_F} &\sim& \nu_0^2 U^2 Z^2, 
\end{eqnarray}
\end{subequations}
where $\nu_0$ is the bare density of states at the Fermi energy. In the strong-coupling limit $U\gg W$, where $W$ is the unrenormalized single-particle bandwidth, we immediately obtain $Z\sim 1/(\nu_0 U)$ and $(\Delta v_F)/v_F\sim O(1)$. Thus the dominant self-energy renormalization in Eq. (\ref{lattice3}) is frequency dependent, with a much weaker momentum dependence. As a result, we also immediately infer the effective mass renormalization, $(m^*/m)=1/Z$. The ground state is thus a {\it heavy} Fermi liquid with a sharp Fermi surface at any strength of interaction. 

The above picture of a Fermi liquid breaks down as a function of increasing energies. Naively, one would expect this to occur for energies comparable to $W$; this is incorrect and the crossover instead occurs at a much reduced scale of $W^*\sim W^2/U$ which also serves as the renormalized bandwidth of the heavy Fermi liquid. Consider the coherent part of the Green's function,
\begin{eqnarray}
 G(i\omega,\k) &=& \frac{Z}{i\omega - Z\overline\varepsilon_\k + i\alpha \nu_0^2 U |\omega|^2\ln\bigg(\frac{W^*}{|\omega|}\bigg)\tn{sgn}(\omega)}, \label{lattice5}\nn 
\end{eqnarray}
where $\overline\varepsilon_\k = \varepsilon_\k + \Delta\varepsilon_\k$ and $\alpha\sim O(1)$ constant; the $\tn{ln}(...)$ is specific to $d=2$. After analytically continuing to real frequencies, the imaginary part of the self-energy in Eq. (\ref{lattice5}) becomes $\Sigma''(\omega)\sim \omega^2/W^*$, such that $\Sigma''(W^*)\sim W^*$. Thus, at energies approaching $W^*$, the scattering rate of the quasiparticles becomes comparable to the renormalized bandwidth. This is a sign that the quasiparticle picture and the sharp Fermi surface associated with the low energy Fermi liquid is breaking down. 

We can instead approach the problem from higher energy scales. For $\omega\gg W^*$, it is appropriate to start from the solutions to Eqs. (\ref{lattice2a}) and (\ref{lattice2b}) in the decoupled limit and treat the hopping perturbatively (i.e. in powers of $\varepsilon_\k$). In this limit, we reproduce the completely local form of the electron Green's function obtained earlier in Eq. (\ref{syk3}). The leading momentum dependence can be obtained in the strong-coupling regime by expanding in powers of $\varepsilon_\k$, 
\beq
G(i\omega,\k) = \frac{i\tn{sgn}(\omega)}{\sqrt{U|\omega|}} - B(\omega) \frac{\varepsilon_\k}{U|\omega|},~~~ (W^*\ll |\omega| \ll U) \label{lattice6}\nn
\eeq
where $B(\omega)$ is a frequency independent constant whose value depends on the sign of $\omega$ and descends from the spectral asymmetry discussed earlier in Sec. (\ref{sec:SYK}). This is an incoherent regime where the electronic quasiparticles are not well defined. Note that the momentum-dependent correction becomes comparable to the local term at $\omega\sim W^*$. 

The above description leads to a simple picture for the properties of the model in Eq. (\ref{lattice1a}) and (\ref{lattice1b}). At the lowest energy scales, the system is a heavy Fermi liquid with a sharp Fermi surface satisfying Luttinger's theorem. All interaction induced corrections are predominantly frequency dependent, with a weak residual momentum dependence. The DMFT-like behavior is linked to the properties of the single SYK  cluster. As a function of increasing energies, the quasiparticle scattering rate increases until they are no longer well defined; at scales approaching the renormalized bandwidth, $W^*$, the Fermi surface and the quasiparticles are completely destroyed. Starting from higher energies, $W^*$ also marks the crossover where the completely local picture of the decoupled SYK dots with perturbative spatial corrections breaks down and is accompanied by the incipient formation of a Fermi surface. Going beyond the large$-N$ results discussed here, the fate of the low-temperature phase can be vastly different \cite{altland19}.   

We note that if the model in Eq. (\ref{lattice1a}) and (\ref{lattice1b}) is defined with a random $t_{\r\r'}$ and uncorrelated $U_{ij;k\ell}$ at different sites \cite{Balents}, the properties of the incoherent regime discussed above remain unchanged since the spatial correlations are completely local. The low energy {\it disordered} FL regime is similar in many aspects to the FL discussed above, but is notably different in the presence of the sharp Fermi surface. We return to some of the consequences of this subtle difference when we discuss transport in Sec. (\ref{sec:thermo}) below.

Finally, we note a model \cite{PatelPlanck} in which the random interactions are restricted to be `resonant': this has $W^\ast \rightarrow 0$, and the Planckian behavior holds down to zero temperature. The rationale for such a model is that the non-resonant interactions have already been absorbed in effective $t_{\r\r'}$ for the quasiparticles. The resonance condition can be interpreted in terms of a scalar field needed to impose the constraints, and this indicates that Planckian behavior should appear more readily and naturally in `Yukawa-SYK' models of fermions and bosons with random Yukawa couplings: we will consider such models in Section~\ref{sec:cfs}.

\subsection{Marginal Fermi liquid and critical Fermi surface from incoherent `flavor' fluctuations}
\label{sec:MFL}

Our theoretical discussion of the metallic non-Fermi liquids discussed in this review thus far have lacked any interesting spatial structure. Even for the lattice model considered in the previous section, the incoherent regime had no singular momentum-dependent features. However, it is possible to add additional electronic degrees of freedom to the model introduced in Eq. (\ref{lattice1a}), (\ref{lattice1b}) and engineer a {\it critical} Fermi surface --- a sharp electronic Fermi surface {\it without} any low-energy electronic quasiparticles --- over a wide range of energy scales. Interestingly, these additional electronic degrees of freedom realize a `marginal' Fermi liquid, where the single particle lifetime, $\Gamma_{\tn{sp}}\sim \tn{max}(\omega, T)$ \cite{SSmagneto,DCsyk}.  

Consider an additional band of electrons, $d_{\r i}$, defined on the sites of the same hyper-cubic lattice with orbital labels $i=1,.., N$, with a separately conserved density, $\mathcal{Q}_d$. We are interested in Hamiltonians of the form,
\begin{subequations}
\begin{align}
H &= H_c + H_d,\label{lattice7}\\
H_d &= \sum_{\k,i} \epsilon_\k d_{\k i}^\dagger d_{\k i}^{\vphantom \dagger} + \frac{1}{N^{3/2}}\sum_{\r}\sum_{ijk\ell=1}^N V_{ij;k\ell} c_{\r i}^\dagger d_{\r j}^\dagger d_{\r k}^{\vphantom \dagger} c_{\r \ell}^{\vphantom \dagger},
\end{align}
\end{subequations}
where $\epsilon_\k$ is the dispersion for $d-$electrons (including the respective chemical potential) and $H_c$ continues to be defined by Eq. (\ref{lattice1a}) and (\ref{lattice1b}). The $V_{ij;k\ell}$ are assumed to be identical at every site, thereby preserving translational symmetry, and chosen from a Gaussian random distribution with $\overline{V_{ij;k\ell}} = 0$ and variance $\overline{|V_{ij;k\ell}|^2} = V^2$. We are particularly interested in the regime where the bandwidth for $d-$electrons, $W_d$, far exceeds the $c-$electron bandwidth, $W$. The setup here is reminiscent of the periodic Anderson model for an itinerant `conduction' electron band coupled to a strongly interacting, narrow band \cite{hewson_book}, except that the interaction terms now are chosen to have a purely SYK form. A different variant of the two-band model involving an inter-band hybridization that conserves only the total density has also been analyzed \cite{mcgreevy}.  

\begin{figure}[h]
\begin{center}
\includegraphics[scale=0.14]{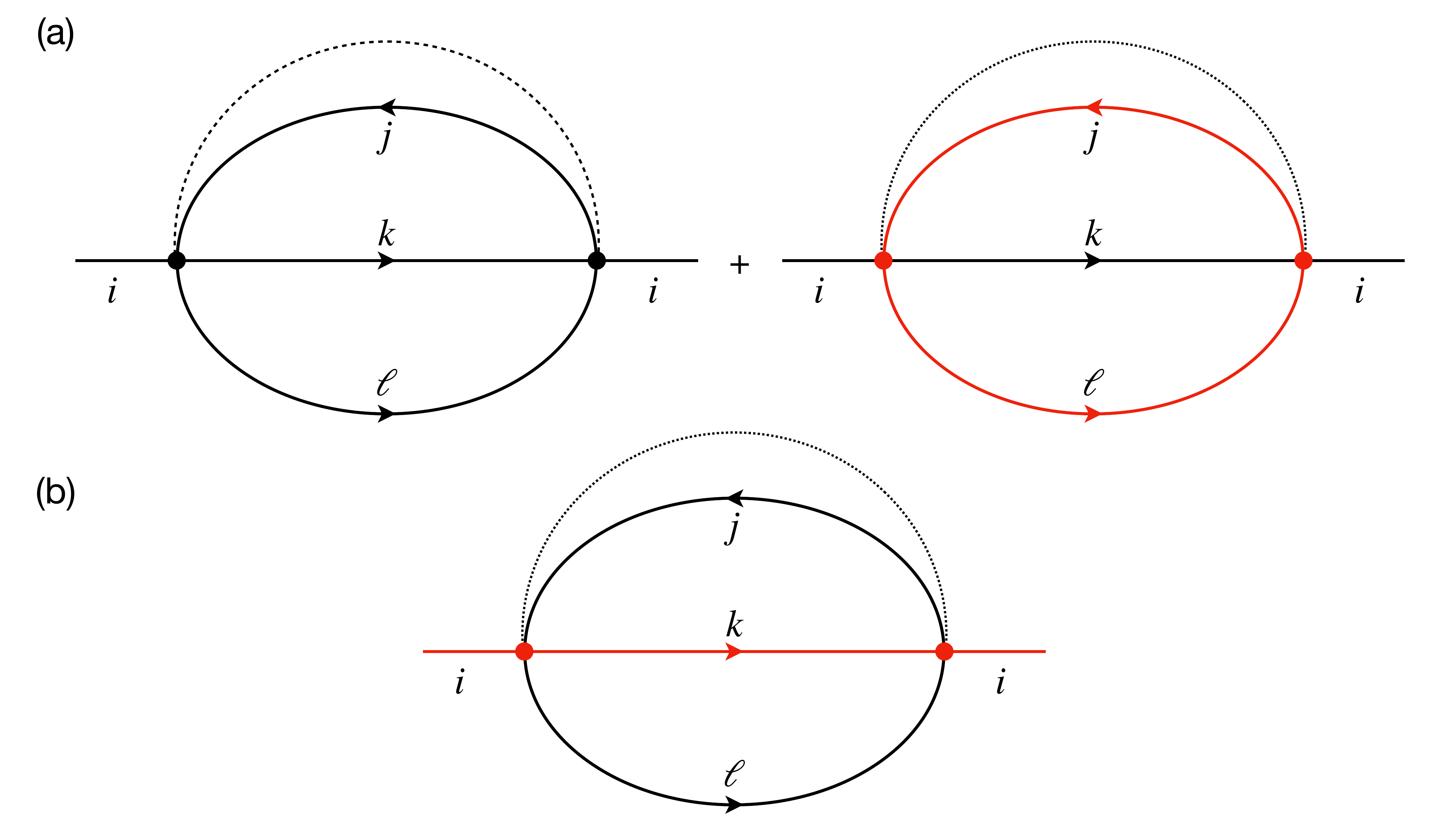}
\end{center}
\caption{The melon graphs for the model in Eq. (\ref{lattice7}) for the electron self-energies for (a) $c$, and (b) $d$ electrons, respectively. Solid black (red) lines denote fully dressed $c$ ($d$) Green's functions. The dashed (dotted) line represents the disorder averaging associated with the interaction vertices, $\overline{|U_{ij;k\ell}|^2}$ ($\overline{|V_{ij;k\ell}|^2}$), respectively. }
\label{melon2}
\end{figure}

In the large$-N$ limit, only a set of coupled melon graphs survive for the Green's function corresponding to both $c$ and $d$ electrons (Fig. \ref{melon2}),
\begin{subequations}
\begin{align}
G(i\omega_n,\k) & = \frac{1}{i \omega_n - \varepsilon_\k - \Sigma (i\omega_n,\k) - \Sigma_{d}' (i\omega_n,\k)} \label{lattice8a} \\ 
G_d(i\omega_n,\k) & = \frac{1}{i\omega_n - \epsilon_\k - \Sigma_{d}(i\omega_n,\k)}, \label{lattice8b} \\
\Sigma'_{d} (i\omega_n,\k) &\\
=-  V^2 T\sum_{i\Omega_n}&\int_{\k_1} G(i\Omega_n,\k_1)~\Pi_d(i\omega_n+i\Omega_n,\k+\k_1) \nonumber\\
\Sigma_{d} (i\omega_n,\k) &\\
= -  V^2 T\sum_{i\Omega_n}&\int_{\k_1} G_d(i\Omega_n,\k_1)~\Pi(i\omega_n+i\Omega_n,\k+\k_1) \nonumber \\
\Pi_d(i\Omega_n,\q)  &\\
= T\sum_{i\omega_n'} &\int_{\k'} G_d(i\omega_n',\k')~G_d(i\omega_n'+i\Omega_n,\k'+\q), \nonumber
\end{align}
\end{subequations}
where $\Sigma(i\omega_n,\k)$ and $\Pi(i\Omega_n,\q)$ are as defined earlier in Eq. (\ref{lattice2b}).

Over an energy window $W^* \ll \omega~ (\tn{or}~T) \ll \tn{min}(W_d, U)$, when the $d-$electrons scatter off the incoherent fluctuations associated with the  $c-$electrons, their self-energy is given by,
\beq
\Sigma_d(i\omega) \sim -i\omega\log\bigg(\frac{U}{|\omega|}\bigg),
\label{lattice9}
\eeq
which has the celebrated marginal Fermi liquid (MFL) form. It is worth emphasizing here that the MFL regime in the present setup is generated self-consistently---even after including its feedback on the $c-$electrons--- without having to postulate the existence of a featureless `bath' \cite{Varma89}. 

In the translationally invariant setting discussed here, the $d-$electrons have a sharp Fermi surface. To make this precise, we can take the limit of $W^*\rightarrow0$ at $T=0$ and identify the location of the Fermi surface from the solution to $G_d^{-1}(0,\k)=0$. The critical Fermi surface satisfies Luttinger's theorem, where its size is now determined solely by $\mathcal{Q}_d$, i.e. the density of $c-$electrons is not included in the size as anticipated, and can therefore be characterized as `small'. The proof of Luttinger's theorem for the critical Fermi surface follows the standard treatment in Fermi liquids \cite{AGD} and is based on the Luttinger-Ward functional. Interestingly, the two-particle correlators (e.g. in the density response) near the ``$2k_F$'' wavevector have a singular dependence as a function of energy. Note that the singular form of the self-energy in Eq. (\ref{lattice9}) is momentum independent and not tied to the near vicinity of the Fermi surface.

The above construction leads to a concrete realization of a `small' critical Fermi surface with marginally defined excitations. However, the critical Fermi surface obtained here is necessarily accompanied by a finite extensive entropy extrapolated to $T\rightarrow0$, which originates from the usual entropy associated with the incoherent regime of the local SYK islands of $c-$electrons. In Sec. \ref{sec:cfs} below, we will discuss a different class of models where the critical Fermi surface is `large' (i.e. the size is determined by the total electronic density) and can arise without an extensive entropy in the $T\rightarrow0$ limit.

\subsection{Thermodynamics and Transport}
\label{sec:thermo}

For the single-band model in Eq.~(\ref{lattice1a}) - (\ref{lattice1b}), the Fermi liquid at $T\ll W^*$ has an entropy density, $s\sim \gamma_{\tn{FL}} T$, where $\gamma_{\tn{FL}}\propto m^* \sim 1/W^*$. In the incoherent regime for $T\gg W^*$, the entropy density is given by that of a single SYK dot (Eq.~\ref{syk36c}) with weak perturbative corrections of order $(W/U)^2$; the extrapolated entropy in the limit of $T\rightarrow0$ from this regime is finite \cite{GPS2}, but the excess entropy is relieved at $T\sim W^*$ across the crossover into the Fermi liquid \cite{Balents}. At $T\gg W^*$, electrical transport occurs as a result of the (perturbative) electron hops between SYK dots. Starting from Kubo formula for the conductivity and given the completely local form of the single-electron Green's functions, the current-current correlation function reduces simply to a convolution of two spectral functions, much like standard computations of transport within DMFT. This leads to 
\beq
\sigma(\omega,T) = \frac{Ne^2}{h} \frac{W^*}{T} F\bigg(\frac{\omega}{T}\bigg),
\eeq
where $F(...)$ is a universal scaling function of $\omega/T$. This immediately leads to a bad metal $T-$linear resistivity (and scattering rate) with values that can far exceed $\rho_Q = h/Ne^2$ over a range of temperatures, $W^*\ll T\ll U$. In the Fermi liquid regime at $T\ll W^*$, the resistivity crosses over into a conventional regime with $\rho = BT^2$ as long as the Fermi surface is large enough and electron-electron umklapp scattering is allowed. Interestingly, the coefficient ($B$) of the $T^2$ term satisfies the `Kadowaki-Woods' scaling \cite{KW}, as can be verified simply by demanding that there is a smooth crossover at $T\sim W^*$ between the two different metallic regimes. We note that the `resonant' model \cite{PatelPlanck} has $W^\ast=0$, and it exhibits strange metal linear $T$ resistivity with values well below $\rho_Q$.

In the MFL regime of the two-band model introduced in Eq.~(\ref{lattice7}), the critical Fermi surface associated with the $d-$electrons gives rise to a singular specific heat, $C\sim T\ln(1/T)$, at low temperatures, in addition to the usual contribution from the SYK dot associated with the $c-$electrons. Once again, given the local form of the single-particle self-energy in the MFL regime, transport simplifies considerably leading to a $T-$linear resistivity associated with the $d-$electrons,
\beq
\rho_d(T) \sim \frac{h}{Ne^2} \bigg(\frac{V^2}{W_d^2 U}\bigg) T.
\eeq
In the translationally invariant setting of Eq.~(\ref{lattice7}), the finite resistivity arises as a result of momentum relaxation to the `bath' formed by the local $c$ electrons at every site \cite{DCsyk}. 

We end this section by noting that the extrapolated zero-temperature entropy from the strange metal regime of the cuprates vanishes \cite{loram}, unlike the residual extensive entropy in the limit of $T\rightarrow0$ associated with the models considered here displaying SYK-like critical correlations at large $N$. There are a number of other materials displaying nFL behavior over intermediate energy scales where the extrapolated entropy is also known to be extensive and finite but relieved below a certain low-temperature coherence scale \cite{allen,bruhwiler}.

\subsection{Superconductivity}
\label{sec:SC}

Conventional Fermi-liquid metals, even with purely repulsive interactions (i.e. in the absence of phonon-mediated attraction), are unstable to superconductivity at extremely low temperatures. This `Kohn-Luttinger' mechanism \cite{KL} relies on an effective attraction that is generated in a non-s-wave angular momentum channel at higher orders in the interaction strength. An analogous general statement can {\it not} be made about the non-Fermi liquid metals introduced in this article and their pairing instabilities, if any, have to be analyzed on an individual basis. 

The models introduced in this section so far do not have any pairing instabilities. By extending these models to include spinful fermions, a number of routes have been used to generate attraction via pair-hopping interactions \cite{Patel,CX18}, a random Yukawa interaction to a bosonic field (e.g., phonon) \cite{YW,JS,Classen} and introducing additional correlations between the interaction matrix-elements, $U_{ij;k\ell}$ \cite{ChowdhuryBerg}. At large$-N$, all of these models have a mean-field like transition to superconductivity where Eliashberg theory becomes asymptotically exact. However, the instability is not tied to the usual `Cooper-logarithm' \cite{AGD} associated with an underlying Fermi surface and the ratio of gap-magnitude to transition temperature is enhanced above the standard mean-field value. When supplemented by an on-site attractive Hubbard interaction, the above models display a fluctuation regime resembling a `pseudogap' \cite{kamenevSC} before the superconducting transition. Certain tensor models \cite{klebanov} and generalized SYK-type models \cite{Xu17,Xu19} defined in terms of real fermions have also been studied and found to exhibit spontaneous symmetry breaking analogous to pairing. 

\begin{figure}[h]
\begin{center}
\includegraphics[scale=0.18]{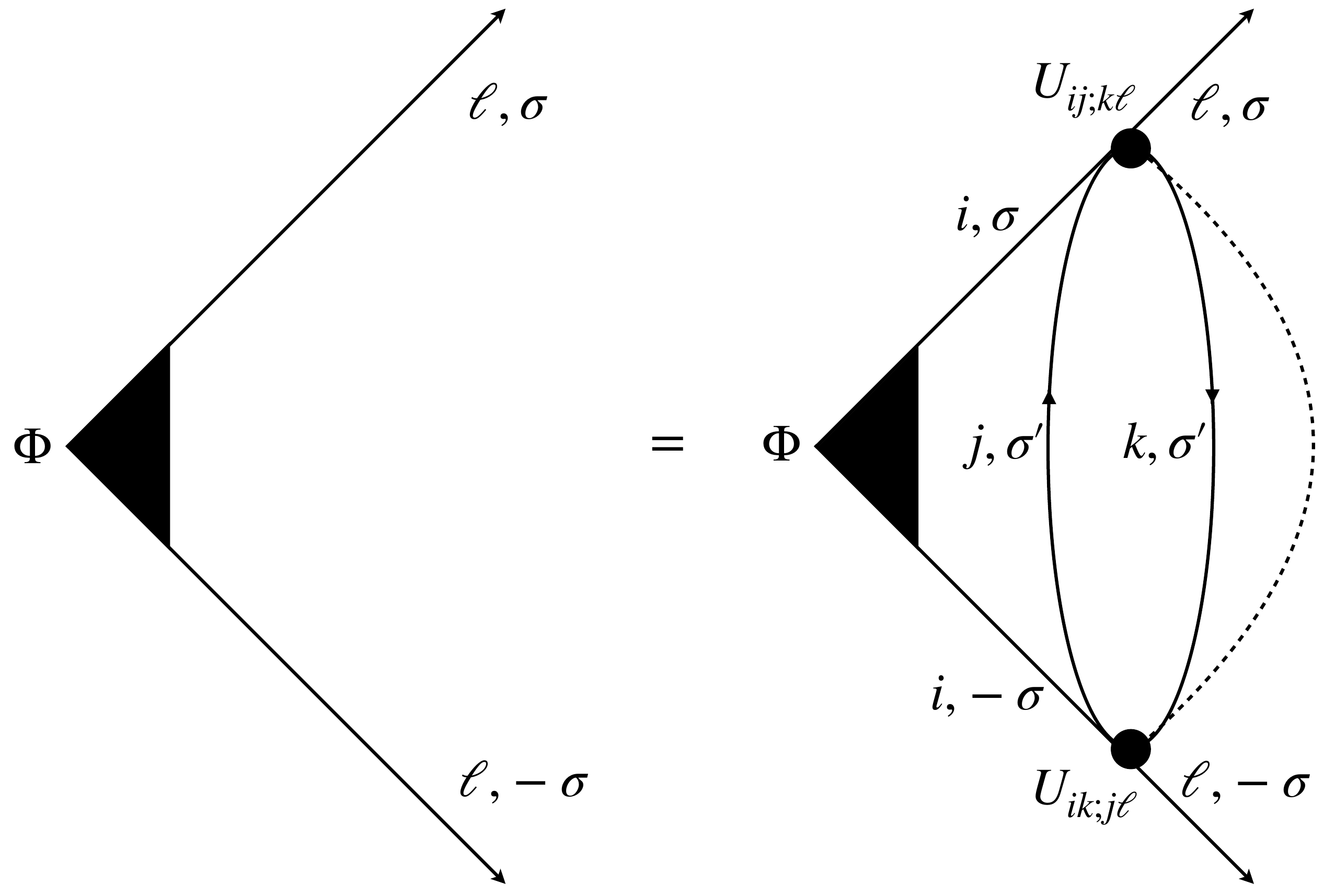}
\end{center}
\caption{The Bethe-Salpeter equation for the intra-orbital pairing vertex, $\Phi$, in the large$-N$ limit .}
\label{eliashberg}
\end{figure}

{\it Intrinsic} superconducting instabilities of the non-Fermi liquids introduced above and their analogy with the Kohn-Luttinger mechanism can be seen by introducing a spin-label, $\sigma=\uparrow,\downarrow$, modifying Eq.~\ref{lattice1b} to
\beq
H_{\tn{int}} \rightarrow \frac{1}{4N^{3/2}} \sum_{\r} \sum_{\sigma=\uparrow,\downarrow}\sum_{ijk\ell=1}^N U_{ij;k\ell} \, c_{\r i\sigma}^\dagger c_{\r j\sigma'}^\dagger c_{\r k\sigma'}^{\vphantom \dagger} c_{\r \ell\sigma}^{\vphantom \dagger}, \nonumber\\
\label{lattice_sc1}
\eeq
and including additional correlations between the interaction matrix elements as: $U_{ij;k\ell} = \pm U_{ik;j\ell}$. The physics is qualitatively different depending on the $\pm$ sign here, as can be seen most directly by writing down the Bethe-Salpeter equation (Fig.~\ref{eliashberg}) for the intra-orbital, spin-singlet vertex in the pairing channel: $\Phi_{\ell}(\r-\r') \equiv \epsilon_{\sigma\sigma'} c_{\r\ell \sigma} c_{\r'\ell \sigma'}$. 

At zero external center-of-mass momentum, the linearized equation for $\Phi_\ell$ becomes,
\beq
&&\Phi_{\ell}(\omega,\k) = \label{BSeqn}\\
&&\mp U^2T\sum_\Omega \int_\q \Phi_{i}(\Omega,\q) G(i\omega,\q)G(-i\omega,-\q) \Pi(i\omega-i\Omega,\k-\q), \nonumber
\eeq
where $G(i\omega,\q)$ and $\Pi(i\omega,\q)$ are as introduced earlier in Eqs.~\ref{lattice2a}, \ref{lattice2b}. Importantly, introducing the additional spin label and the matrix correlations, $U_{ij;k\ell} = \pm U_{ik;j\ell}$, does not change the asymptotic nature of the single-electron Green's functions, but can lead to preemptive instabilities to superconductivity depending on `$\pm$' sign  \cite{ChowdhuryBerg}. For the model with $U_{ij;k\ell} = U_{ik;j\ell}$, the eigenvalue problem in Eq.~\ref{BSeqn} has a non-trivial solution with a superconducting $T_c\sim U$. Importantly, superconductivity preempts the crossover into the heavy Fermi liquid and arises at the level of a single site due to effectively attractive interactions that are generated at $O(U^2)$; the superfluid stiffness is nevertheless finite and given by $NW^*\gg T_c$. On the other hand, for the model with $U_{ij;k\ell} = -U_{ik;j\ell}$, there is no instability at the level of a single-site and while the pairing susceptibility is enhanced approaching $T\sim W^*$ from above, the non-Fermi liquid is stable against pairing. However, across the crossover into the heavy Fermi liquid regime, the momentum dependence in $\Pi(\q)$ can drive a pairing transition, much like the Kohn-Luttinger mechanism, but where $T_c$ is now set by the only relevant scale in the problem, $W^*$. Similar generalizations can also be constructed for the two-band models in Sec.~\ref{sec:MFL}, to analyze the intrinsic pairing instabilities of the marginal Fermi liquid with a critical Fermi surface \cite{DCEB20b}. We end by noting that for a variety of non-Fermi liquids involving quantum critical degrees of freedom, the Eliashberg equations share a similar structure \cite{Chubukov20a,Chubukov20b}.  

\section{Fermi surfaces coupled to gapless bosons}
\label{sec:cfs}

This section will turn to a different, and extensively studied, approach to non-Fermi liquids in clean metals. We begin with a Fermi liquid with a well-defined Fermi surface and long-lived quasiparticles, and examine the breakdown of quasiparticles due to scattering from a gapless boson: this gapless boson can either be associated with an order parameter near a symmetry-breaking transition, or an emergent excitation associated with fractionalization. Note, however, that the Fermi surface remains sharp in momentum space, even though the quasiparticles are not well defined and the spectra are broad in energy space: this realizes a `critical Fermi surface', as discussed in Section~\ref{subsec:class}.

As we will describe below, there are difficulties \cite{sungsik1} in applying conventional large $N$ methods to the critical Fermi surface problem. However, progress has recently become possible \cite{Esterlis:2021eth} by incorporating insights from a class of `Yukawa-SYK' models describing fermions and bosons with a 3-body Yukawa coupling \cite{Fu:2016vas,Murugan:2017eto,Patel:2018zpy,Marcus:2018tsr,Wang:2019bpd,JS,Wang:2020dtj,Kim:2020jpz,Adalpe20,WangMeng21,Esterlis:2021eth}. These methods provide a systematic treatment of such critical Fermi surfaces, and also exposes similarities to SYK non-Fermi liquids. 
The new approach shows that required large $N$ limit can be obtained provided we allow {\it random\/} coupling constants, as in the Yukawa-SYK models. In the present situation, the couplings can be spatially uniform, so that translational invariance is maintained \cite{Esterlis:2021eth,Adalpe20}. Despite the presence of random couplings, many properties self-average in the large $N$ limit, just as in the Yukawa-SYK models. The central idea is that in a given finite $N$ system, with a fixed set of coupling constants, there is an RG flow to a common {\it universal\/} low energy theory. Assuming the existence of such a theory, we attempt to access the universal low energy physics simply by averaging over couplings. Upon carrying out this procedure, we find that only certain averages over the couplings matter, and the values of these averages {\it cancel out\/} in the low energy theory, thus supporting the existence of a universal theory.
We note that the idea of simplification realized by an average over similar strongly-coupled theories is also playing an important role in recent investigations in quantum gravity,
and averages over random matrices or conformal field theories yield systematic large $N$ holographic realizations of the path integral of simple theories of gravity \cite{Saad:2019lba,Stanford:2019vob,Afkhami-Jeddi:2020ezh,Maloney:2020nni,Perez:2020klz,Cotler:2020ugk,Engelhardt:2020qpv,Datta:2021ftn,Czech}.

%To begin, it is useful to recast some results of Fermi liquid theory using a `patch formalism', which is designed to simplify the treatment of long wavelength fluctuations near the Fermi surface. We summarize this treatment in Appendix \ref{sec:patch}. Such a formalism is not traditionally used in textbook treatments of Fermi liquid theory as one of their key features, namely the low energy excitations near the Fermi surface actually involve short wavelength fluctuations, associated with particle-hole excitations at large momenta but low energy. However, the breakdown of quasiparticles in the critical Fermi surface of interest to us here is dominated by long wavelength fluctuations; the structure of a `non-Fermi' liquid without quasiparticle excitations becomes clear in the patch theory.

We will consider a specific model of a critical Fermi surface --- fermions coupled to an emergent U(1) gauge field. As outlined in Sec. \ref{subsec:class}, such a theory arises in a number of different physical contexts, including spin liquid Mott insulators with a gapless Fermi surface of spinons \cite{PALee89,AIM,Polchinski:1993ii} and the compressible quantum Hall state in the half-filled Landau level with a gapless Fermi surface of composite fermions \cite{HLR}. The formalism is also easily extended to a number of other examples involving the onset of broken symmetries, identified by order parameters with vanishing lattice momentum, in a metal (e.g. Ising-nematic order in a Fermi liquid \cite{metlitski1}). 

\subsection{Fermi surface coupled to a dynamical U(1) gauge field}
\label{sec:cfs2}

Consider a non-zero density of fermions coupled to an emergent U(1) gauge field, $A_\mu$. 
In the presence of a Fermi surface, the longitudinal components of $A_\mu$ are screened just as in an ordinary metal with Coulomb interactions. However, there is no screening in the transverse sector, and so we shall focus only on the transverse spatial components $A_{x,y}$. We can schematically write the theory by generalizing the action for the Fermi liquid to 
\bea
\mathcal{S}_{cA} &=& \int d\tau \left[ \int \frac{d^d k}{( 2\pi)^d} c_{\k a}^\dagger \left( \frac{\partial}{\partial \tau} + \varepsilon(-i \nabla-A)
\right) c_{\k a} \right. \nonumber \\
&~&~~\left. + \frac{NK}{2} \int d^2 x ~(\nabla \times A)^2 \right] \,. \label{scA}
\eea
We have not included an explicit time derivative term for $A$ because it will turn to be subdominant to the frequency dependence induced by the Fermi surface. The co-efficient of the Maxwell term $(\nabla \times A)^2$ is determined by short distance physics, and we have included a prefactor of $N$ for future convenience; the gauge-coupling is denoted $K^{-1}$. We have restricted our considerations to spatial dimension $d=2$, where the frequency dependence for the self-energy will be most singular, and is also the dimension of most physical applications.

\begin{figure}
\centerline{\includegraphics[width=1.5in]{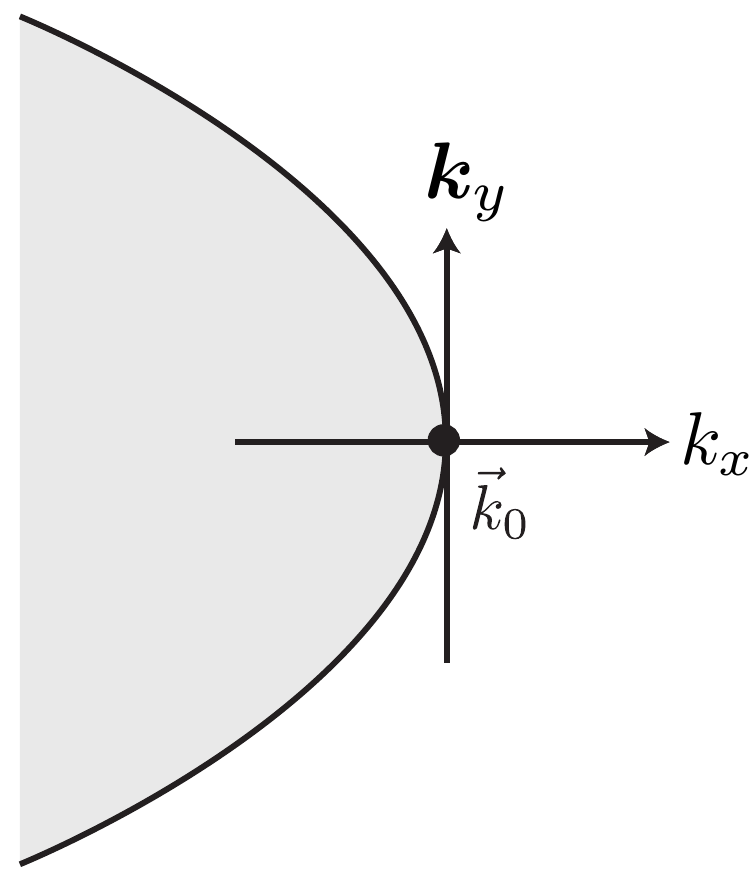}} 
\caption{We focus on an extended patch of the
Fermi surface, and expand in momenta about the point $\vec{k}_0$ on the Fermi surface. 
This yields a theory of $2$-dimensional fermions $\psi$ in (\ref{flt6A}).} \label{fig:sungsik}
\end{figure}

Let us now proceed with a perturbative, but self-consistent, analysis of $\mathcal{S}_{cA}$ in a `patch' theory: we focus on the vicinity of the point $\vec{k}_0$ on the Fermi surface, as in Fig.~\ref{fig:sungsik}. 
For the gauge field $A$, it turns out we need only include components of their momenta which are tangent to the Fermi surface, closely connected to the $1/|q_y|$ dependence of the fermion polarizibility that is obtained as in Fermi liquid theory
\begin{eqnarray}
\Pi (q, i\omega_n)
= -\frac{|\omega_n|}{4 \pi v_F \kappa |q_y|}\,. \label{flt182}
\end{eqnarray}
Recalling that we are focusing only on transverse gauge field fluctuations, we may replace the gauge field by a single scalar field $\phi = A_x$. In this manner, the patch theory limit of Eq.~(\ref{scA}) is 
\bea
\mathcal{S}_{\psi \phi} &=& \int \! d \tau  dx d y \Biggl[  \psi_a^\dagger \left( \frac{\partial}{\partial \tau} - i v_F \frac{\partial}{\partial x}
- \frac{\kappa}{2} \frac{\partial^2}{\partial y^2} \right) \psi_a \nonumber \\
&~&~~~~~~ + \frac{NK}{2} \left( \frac{\partial \phi}{\partial y} \right)^2 - v_F \phi \, \psi_a^\dagger \psi_a \Biggr]
\,, \label{flt6A}
\eea
where, for now, we are considering the case with $a=1 \ldots N$ fermion flavors.
This patch theory also applies to the other cases with order parameters, identified just before Section~\ref{sec:cfs2}.

The fermion polarizability will now appear as a self energy for the $\phi$ field, and so we can write the $\phi$ propagator, $D(q, i \Omega_n)$ as
\beq
D(q, i \Omega_n) = \frac{1}{N\left( K q_y^2 - v_F^2 \Pi (q, i\Omega_n) \right)},
\eeq
where $\Pi$ is given by (\ref{flt182}). The fermion Green's function is expressed in the usual way,
\begin{equation}
G (\vec{k}, i\omega_n) = \frac{1}{ i\omega_n - \varepsilon_k - \Sigma (k, i\omega)},
\label{nflG}
\end{equation}
where now
\begin{equation}
\varepsilon_k = v_F k_x + \kappa \frac{k_y^2}{2}\,.
\label{flt12a}
\end{equation}
The self energy, as a result of scattering off the fluctuations of $\phi$, can be evaluated as
\bea
&&\Sigma (k, i\omega_n) =  v_F^2 \int \frac{d^2 q }{(2 \pi)^2} T \sum_{\Omega_n \neq 0}  D (q, i\Omega_n) \nn
&&~~~~~~~~~~~~~~~~~~~~~~~~~~~~~~~~~~~~~\times G (k+q, i\Omega_n + i\omega_n)
\nonumber \\
&&~~~= -i \, \frac{v_F^2}{2N} \int \frac{d q_y}{2 \pi} T \sum_{\Omega_n \neq 0} \frac{ \mbox{sgn} (\omega_n + \Omega_n)}{\ds K q_y^2 + \frac{|\Omega_n|}{4 \pi v_F \kappa |q_y|}} \label{flt30}
 \\
&&~= - i \frac{v_F^2}{3 \sqrt{3} N K^{2/3}} (4 \pi v_F \kappa K)^{1/3} T \sum_{\Omega_n \neq 0} \frac{ \mbox{sgn} (\omega_n + \Omega_n)}{|\Omega_n|^{1/3}}. \nonumber
\eea
We have dropped the gauge fluctuations at $\Omega_n=0$ because they require a special treatment: this is likely an artifact of the fermion not being gauge invariant. The singularity at $\Omega_n=0$ in Eq.~(\ref{flt30}) will likely drop out of gauge-invariant observables. 
In any case, there are no issues at $T=0$, in which case we find the non-Fermi liquid self energy $\Sigma (\omega) \sim \omega^{2/3}$. At $T>0$, the result Eq.~(\ref{flt30}) obeys a scaling form similar to that for the SYK model in Eq.~(\ref{syk18a}) \cite{PALee89}
\beq
\Sigma (k, \omega, T) \propto T^{2/3} \Phi \left( \frac{\hbar\omega}{k_B T} \right)\,.
\label{flt30a}
\eeq
This is much larger than the bare $\omega$ term in the inverse Green's function, and leads to the absence of a quasiparticle pole at the Fermi surface, where the latter is defined as the location where $G^{-1}(k_F,\omega=0,T=0)=0$.

\subsubsection{Large $N$ limit}
\label{sec:cfsN}

As we have emphasized earlier, our apparent perturbative computations of the fermion Green's function are actually fully self-consistent in the self energies of both the gauge field and the fermion. In this sense, the equations have a structure very similar to that of the SYK models. So as in the Yukawa-SYK models, we ask if there is a systemic large $N$ approach in which these results can be obtained as the saddle point of an action? This will ensure that the solutions are locally stable against all perturbations, determine conditions under which superconducting or other instabilities could exist, and also allow a systematic treatment of corrections. 

Despite numerous attempts, a systematic and satisfactory treatment that relies only on a naive large-$N$ expansion has been lacking in the literature. The difficult is apparent from an examination of Eqs.~(\ref{nflG}) and (\ref{flt30}). In a model with $N$ fermion flavors, the singular self energy in Eq.~(\ref{flt30}) has a prefactor of $1/N$, and so is formally smaller than the bare dispersion, $v_F k_x + \kappa k_y^2/2$. However, the self energy has to be matched with the bare dispersion to obtain the physical excitations, and so a power of $N$ is unavoidable in the dispersion of the renormalized excitations. This implies that higher order Feynman graphs can be enhanced by powers of $N$ not associated with the symmetry factors of the graphs, leading to a breakdown of the $1/N$ expansion; this is indeed what happens \cite{sungsik1}. Various workarounds have been attempted \cite{Fitzpatrick:2013mja,Fitzpatrick:2013rfa,Torroba19}, but none have been entirely successful because they include $N$-dependent energy scales.

As we noted earlier, recent studies \cite{Adalpe20,Esterlis:2021eth} have shown that a systematic large-$N$ theory of the critical Fermi surface can be obtained in a theory with couplings which are random in flavor space, but are translationally invariant. We now show how such a theory leads to a $G$-$\Sigma$ formulation for the critical Fermi surface. We start with the theory Eq.~(\ref{flt6A}), promoting the scalar $\phi$ to now acquire $N$ indices, $\phi_a$, and introduce a set of couplings $g_{abc}$ which are random in flavor space, but spatially uniform; we also set $v_F=1$, $\kappa=2$. Then the required theory is \cite{Esterlis:2021eth}
\bea
\mathcal{S}_{\psi \phi} &=& \int \! d \tau  dx d y \Biggl[  \psi_a^\dagger \left( \frac{\partial}{\partial \tau} - i \frac{\partial}{\partial x}
-  \frac{\partial^2}{\partial y^2} \right) \psi_a \nonumber \\
&~&~~ + \frac{K}{2} \left( \frac{\partial \phi_a}{\partial y} \right)^2 - \frac{g_{abc}}{N} \phi_a \, \psi_b^\dagger \psi_c \Biggr]
. \label{flt60}
\eea
The key new feature is the set of space-independent random complex Yukawa couplings, $g_{abc}$, which have zero mean and variance $g^2$. 

We can now proceed just as in the Yukawa-SYK models: we obtain a theory for Green's functions which are bilocal in both space and time. Using the 
spacetime co-ordinate $X \equiv (\tau, x, y)$, we can write the averaged partition function
\bea
&& \overline{\mathcal{Z}}_{\psi \phi} = \int \mathcal{D} G (X_1, X_2) \mathcal{D} \Sigma (X_1, X_2) \mathcal{D} D(X_1, X_2)  \nonumber \\
&&~~~~~~~~\times \mathcal{D} \Pi (X_1, X_2) \exp\left[ - N I (G, \Sigma, D, \Pi) \right]\,.
\label{ZYSYK}
\eea
The $G$-$\Sigma$-$D$-$\Pi$ action is now
\bea
&& I (G, \Sigma, D, \Pi) = \frac{g^2}{2} \mbox{Tr} \left(G \cdot [G D] \right)\nn 
&&~~~~~~~~~~~~~- \mbox{Tr}(G \cdot \Sigma) + \frac{1}{2} \mbox{Tr}(D \cdot \Pi) \label{flt51} \\
&& -\ln \det \left[ \left(\partial_{\tau_1} - i \partial_{x_1} - \partial_{y_1}^2 \right) \delta(X_1-X_2)  + \Sigma(X_1,X_2) \right] \nonumber \\
&&~~~ + \frac{1}{2} \ln \det \left[ \left( - K \partial_{y_1}^2 \right)\delta(X_1-X_2)  - \Pi(X_1,X_2) \right] \,. \nonumber
\eea
where we have introduced notation analogous to Eq.~(\ref{syk31b}),
\beqn
\mbox{Tr} \left( f \cdot g \right) \equiv \int d X_1 d X_2 \,  f(X_2, X_1) g(X_1, X_2)\,. \label{flt52}
\eeqn
Note the crucial pre-factor of $N$ before $I$ in the path-integral.

The large $N$ saddle point equations of this action are precisely the self-consistent equations we have already solved above, apart from differences in factors of $N$. Assuming all saddle point Green's functions depend only upon spacetime differences, we can write them as 
\bea
G(k, i \omega_n) &=& \frac{1}{i \omega_n - k_x - k_y^2 - \Sigma (k, i \omega_n)} \nonumber \\
D(q, i \Omega_n) &=& \frac{1}{K q_y^2 - \Pi(q, \Omega_n)} \nonumber \\ 
\Sigma (X) &=& g^2 D (X) G(X) \nonumber \\
\Pi (X) &=& - g^2 G(X) G(-X) \,.
\eea
From the previous analysis, we can write down the solution to these equations as
\bea
\Pi(q, i \Omega_n) &=& - \frac{g^2}{8 \pi} \frac{|\Omega_n|}{|q_y|} \label{flt75} \\
\Sigma(k, i\omega_n) &=& -2i \frac{g^{4/3}\pi^{1/3}}{K^{1/3}} \frac{T}{3\sqrt{3}}\sum_{\Omega_n\neq0}\frac{\mathrm{sgn}(\omega_n+\Omega_n)}{|\Omega_n|^{1/3}} \,. \nonumber
\eea
Crucially, note that $N$ does not appear in these saddle point equations, unlike that in the self energy in Eq.~(\ref{flt30}).

\subsubsection{Luttinger's theorem}
\label{sec:cfllutt}

Despite the absence of a quasiparticle pole, Luttinger's theorem still applies to the critical Fermi surface with essentially no modifications. On general grounds we can expect that at $T=0$, $\mbox{Im} \, G^{-1} (k, i \eta) = 0$ at all $k$, where $\eta$ is a positive infinitesimal, and this is certainly obeyed by Eq.~(\ref{flt30}). Then, as in Fermi liquid theory, the Fermi surface is defined by $\mbox{Re} \, G^{-1} (k_F, i \eta) = 0$, with particle-like excitations for $\mbox{Re} \, G^{-1} (k_F, i \eta) < 0$, and hole-like excitations for $\mbox{Re} \, G^{-1} (k_F, i \eta) > 0$. Then, we proceed as in Section~\ref{sec:syklutt}, and decompose the expression for the charge density per flavor index $\mathcal{Q}$ into 2 terms:
\bea
\mathcal{Q}&=& \int_k \int_{-\infty}^{\infty} \frac{d \omega}{2 \pi} G(k,i \omega) e^{-i \omega 0^-} = I_1 + I_2 \nn
I_1 &=& i \int_k \int_{-\infty}^{\infty} \frac{d \omega}{2 \pi} \frac{d}{d \omega} \ln \left[ G(k,i \omega) \right] e^{-i \omega 0^-} \label{nfll1} \\
I_2 &=& -i \int_k \int_{-\infty}^{\infty} \frac{d \omega}{2 \pi} G(k,i \omega) \frac{d}{d \omega} \Sigma(k,i \omega)  e^{-i \omega 0^-}\,, \nonumber
\eea
where $\int_k \equiv \int d^d k/(2 \pi)^d$.
We evaluate $I_1$ as in Eq.~(\ref{sykl3}), and obtain
\bea
I_1 &=& i \lim_{\eta \rightarrow 0} \int_k \int_{-\infty}^{0} \frac{d \omega}{2 \pi} \frac{d}{d \omega} \ln \left[\frac{G^{-1}(k,\omega + i \eta)}{G^{-1} (k, \omega - i \eta)} \right] \label{nfll2} \\
&=&- \frac{1}{\pi} \lim_{\eta \rightarrow 0} \int_k \left[ 
\mbox{arg} \, G^{-1}(k, i \eta) -  \mbox{arg} \, G^{-1}(k,-\infty + i \eta) \right]\,. \nonumber
\eea
The momentum integrand evaluates to $-\pi$ for $\mbox{Re} \, G^{-1} (k_F, i \eta) > 0$, and 0 otherwise, and hence $I_1$ evaluates the momentum space volume enclosed by the Fermi surface, divided by $(2\pi)^d$.

It now remains to establish that $I_2=0$ for the critical Fermi surface case, unlike the SYK model results in Section~\ref{sec:syklutt}. The self energy of the critical Fermi surface in Eq.~(\ref{flt30a}) is singular at $\omega=0$, just like the self energy of the SYK model in Eq.~(\ref{syk18a}). So we might worry that there is an anomalous contribution to $I_2$ from the singularity at $\omega=0$, as there was in Section~\ref{sec:syklutt}. However, that is not the case here because the singularity of the Green's function is much weaker as a result of its momentum dependence; now the low-energy Green's function is
\beq
G^{-1} (k, \omega) = -v_F k_x - \frac{\kappa}{2} k_y^2 - \Sigma (\omega)\,, \label{nfll3}
\eeq
and this diverges at $\omega=0$ only on the Fermi surface $v_F k_x + \kappa k_y^2/2 = 0$.
Indeed, with this form, the local density of states is a constant at the Fermi level. Consequently, there is no anomaly at $T=0$, and $I_2 = 0$ from the Luttinger-Ward functional analysis. Incidentally, we note that the Luttinger-Ward functional in the large $N$ limit is just the first term in the action $I$ in Eq.~(\ref{flt51}), similar to the SYK model.

To complete this discussion, we add a few remarks on the structure of the Luttinger-Ward functional, and its connection to global U(1) symmetries \cite{powell1,coleman1}. Consider the general case where were are multiple Green's functions (of bosons or fermions) $G_\alpha (k_\alpha, 
\omega_\alpha)$. Let the $\alpha$'th particle have a charge $q_\alpha$ under a global U(1) symmetry. Then for each such U(1) symmetry, the Luttinger-Ward functional will obey the identity
\beq
\Phi_{LW} \left[ G_\alpha (k_\alpha, \omega_\alpha) \right] = \Phi_{LW} \left[ G_\alpha (k_\alpha, \omega_\alpha + q_\alpha \Omega) \right].~~ \label{nfll4}
\eeq
Here, we are regarding $\Phi_{LW}$ as functional of two distinct sets of functions $f_{1,2\alpha} (\omega_\alpha)$, with $f_{1\alpha} (\omega_\alpha) \equiv G_\alpha (k_\alpha, \omega_\alpha + q_\alpha \Omega)$ and  $f_{2\alpha} (\omega) \equiv G_\alpha (k_\alpha, \omega_\alpha) $, and $\Phi_{LW}$ evaluates to the same value for these two sets of functions.
Expanding Eq.~(\ref{nfll4}) to first order in $\Omega$, and integrating by parts, we establish the corresponding $I_2 = 0$. 

\subsubsection{Thermodynamics}
\label{sec:cfsthermo}

The grand potential can be computed by evaluating Eq.~(\ref{ZYSYK}) for the saddle point in Eq.~(\ref{flt75}). Such a computation \cite{HLR} shows that the entropy density
\beqn
s \sim T^{2/3}\,. \label{flt76}
\eeqn

It is useful to give a scaling interpretation of Eq.~(\ref{flt76}) \cite{Eberlein:2016jlt}. In a critical theory with dynamic critical exponent $z$ in spatial dimension $d$, we expect $s \sim T^{d/z}$. In our case, we have fermionic excitation which disperse as $\omega \sim k_x^{3/2}$, and so we identify $z=3/2$. In this case, Eq.~(\ref{flt76}) matches with the scaling expectations in $d=1$ dimension. Evidently, the free energy is similar to that of chiral fermions dispersing normal to the Fermi surface, and the integral along $k_y$ only determines the prefactor in Eq.~(\ref{flt76}) which is related to the area of the Fermi surface. In scaling terms, it is conventional to denote such a dimensional transmutation in terms of a violation of hyperscaling exponent $\theta$, so that the entropy density scales as $s \sim T^{(d-\theta)/z}$. Then Eq.~(\ref{flt76}) corresponds to $d=2$, $\theta =1$, $z=3/2$.

Let us now extend these scaling argument to a finite system volume $V$, and compare the behavior to that of the random matrix model in Section~\ref{sec:matrix2}, and of the SYK model in Section~\ref{sec:schwarzian}.
Following these earlier treatments, we deal with extensive quantities, such as the total entropy $S = s V$. We expect the scaling $V \sim T^{-d/z}$, and so $S \sim T^{-\theta/z}$. Similarly we, have for the energy density $e \sim T s \sim T^{(d+z-\theta)/z}$, and the total energy $E = e V \sim T^{(z-\theta)/z}$. Collecting these scaling forms, we express the total entropy $S$ as a function of the total energy $E$, and the volume $V$, as in Sections~\ref{sec:matrix2} and \ref{sec:schwarzian}
\beqn
S(E) = V^{\theta/d} \Phi_S \left( E V^{(z-\theta)/d} \right)\,, \label{flt77}
\eeqn
where $\Phi_S (y)$ is a scaling function. As $V \rightarrow \infty$, we expect the relationship to only involve intensive quantities, and so $S/V$ should only be a function of $E/V$. This is achieved if 
\beqn
\Phi_S (y \rightarrow \infty) \sim y^{(d-\theta)/(d-\theta+z)}\,. \label{flt78}
\eeqn 

The scaling results Eqs.~(\ref{flt77}) and (\ref{flt78}) are easily seen to be obeyed by both 
the random matrix and SYK models. For these models, we identify the system size $N$ with the volume $V$, but we cannot accord much meaning to the values of the exponents because there is no true sense of space. 
For the random matrix model, the result Eq.~(\ref{rm7}) is of the form Eq.~(\ref{flt78}) with the scaling function $\Phi_S (y) \sim \sqrt{y}$ and $\theta = d-z$.
For the SYK model, the result Eq.~(\ref{syk49}) corresponds to $\Phi_S (y) = c_1 + c_2 \sqrt{y}$, for some constants $c_{1,2}$, and the exponents $\theta=d$, $z=0$.

For the critical Fermi surface, the important open question is the behavior of $\Phi_S (y \rightarrow 0)$. A reasonable conjecture is that $\Phi_S ( y \rightarrow 0)$ is a non-zero constant. In this case, the total entropy in the $T \rightarrow 0$ or $E \rightarrow 0$ limit is $S \sim V^{\theta/d} = \sqrt{V}$. Note that this is different from the behavior of the entropy for the critical Fermi surface state obtained earlier in Sec.~\ref{sec:MFL}. 
In other words, the entropy of the critical Fermi surface here is {\it sub-extensive} at low energies, a behavior intermediate between the random matrix (which has $S(E\rightarrow 0) \sim V^0$) and SYK (which has $S(E\rightarrow 0) \sim V$) models. The many-body density of states would then behave as $\dos (E \rightarrow 0) \sim \exp({\sqrt{V}})$, although as in all systems $\dos (E) \sim \exp({V})$ when $E$ is extensive.

\subsubsection{Transport}
\label{sec:cfstransport}

We now couple the fermions on the critical Fermi surface to an external U(1) gauge field (distinct from $\vec{A}$ in Eq.~(\ref{scA})), and discuss the structure of the associated conductivity. The highly singular self energy in Eq.~(\ref{flt30a}) suggests that there will be strong scattering of charge carriers, and hence a low $T$ resistivity which is larger than the $\sim T^2$ resistivity of a Fermi liquid. Indeed, it was argued in an early work \cite{PALee89} that the resistivity $\sim T^{4/3}$; this is weaker than $\Sigma \sim T^{2/3}$, because of the $(1-\cos(\theta))$ factor in the transport scattering time, for scattering by an angle $\theta$, and the dominance of forward scattering. 

However, this argument ignores the strong constraints placed by momentum conservation \cite{Hartnoll:2007ih,Maslov2011,Hartnoll:2014gba,Eberlein:2016jlt,Hartnoll:2016apf} in a theory of critical fluctuations which is described by a translationally invariant continuum field theory, such as Eq.~(\ref{flt6A}). If we set up an initial state at $t=0$ with a non-zero current, such a state necessarily has a non-zero momentum, which will remain the same for $t>0$. The current will decay to a non-zero value which maximizes the entropy subject to the constraint of a non-zero momentum. This non-zero current as $t \rightarrow \infty$ implies that the d.c. conductivity is actually infinite.
These considerations are similar to those of `phonon drag' \cite{Peierls1930,Peierls1932} leading to the absence of resistivity from electron-phonon scattering. In practice, phonon drag is observed only in very clean samples \cite{APM12}, because otherwise the phonons rapidly lose their momentum to impurities. But the electron-phonon coupling is weak, allowing for phonon-impurity interactions before there are multiple electron-phonon interactions. In contrast, for the critical Fermi surface, the fermion-boson coupling is essentially infinite because it leads to the breakdown of electronic quasiparticles. So the critical Fermi surface must be studied in the limit of strong drag, with vanishing d.c. resistivity in the critical theory. 

Mechanisms extrinsic to the theory in Eq.~(\ref{flt6A}) are required to relax the current and obtain a finite d.c. conductivity. In a system with strong interactions, such processes are most conveniently addressed by a `memory matrix' approach that has been reviewed elsewhere \cite{Hartnoll:2016apf}; this approach also has close connections to holographic approaches \cite{Lucas:2015vna,Lucas:2015pxa}.
Various mechanisms have been considered \cite{Maslov2011,Hartnoll:2014gba,Patel:2014jfa,Berg19,Else20,Lee20} involving spatial disorder or umklapp processes, and these do lead to a singular resistivity at low $T$.

The behavior of the conductivity, $\sigma$ at non-zero frequency $\omega$ has been argued to be more universal, where the effects of total momentum conservation are not as singular. In a quantum-critical system, the naive scaling dimension is $d-2$, and so we expect $\sigma (\omega) \sim \omega^{(d-2)/z}$, which is frequency independent in $d=2$. However, we have noted violation of hyperscaling in the free energy in Section~\ref{sec:cfsthermo}, and a first guess would be that there is a similar violation of hyperscaling in the conductivity, with $\sigma (\omega) \sim \omega^{(d-2-\theta)/z}$. Using the values of $\theta$ and $z$, we can write the scaling form \cite{Eberlein:2016jlt}
\beqn
\mbox{Re}\, \sigma (\omega \neq 0, T) = \omega^{-2/3} \Phi_\sigma \left( \frac{\omega}{T} \right)
\eeqn
This scaling form is consistent with explicit computations of the frequency dependent conductivity \cite{YBK94,YBK95,Eberlein:2016jlt,ChubukovMaslov17}, but has been questioned in recent work analyses directly with a Fermi surface in $d=2$ \cite{Guo:2022zfl,Shi22}.

In a system with momentum conservation, we can sensibly define the shear viscosity, $\eta$, in the continuum field theory. This has been computed \cite{Patel:2016ymd}, and its hyperscaling violation however turns out to be different from that of the entropy and the conductivity. The ratio $\eta/s$, where $s$ is the entropy density, diverges as $T^{-2/z}$, a result that is consistent with the minimum viscosity conjecture \cite{KSS}.

\subsubsection{Pairing instability}
\label{sec:qcpairing}

As written in Eq.~(\ref{scA}), the gauge field mediates a repulsive interaction between antipodal points on the Fermi surface, and so does not lead to a Cooper pairing instability. However, we can consider closely related problems, either with critical order parameters or with fermions with multiple gauge charges, where the interactions between antipodal fermions is attractive \cite{MM15}. In the context of the large $N$ limit of Section~\ref{sec:cfsN}, the equations determining the pairing instability, remarkably, reduce \cite{Esterlis:2021eth} to precisely those associated with pairing instabilities of the SYK model \cite{klebanov,Klebanov:2020kck}. The pairing vertex $\Phi (i \Omega)$ obeys the integral equation \cite{Esterlis:2021eth}
\beq
E \Phi(i\Omega) = \frac{\mathcal{K}}{3}
\int\frac{d\omega}{2\pi}\frac{2\pi\Phi(i\omega)}{|\omega|^{2/3}|\omega-\Omega|^{1/3}}\,,
\label{Deltaclean}
\eeq
where $\omega, \Omega$ are imaginary frequencies, and $\mathcal{K}$ is a dimensionless number that can be determined from the structure of the critical Fermi surface problem being considered. Given the scale-invariant structure of Eq.~(\ref{Deltaclean}), we are search for solutions with
\beq
\Phi (i \Omega) = \frac{1}{|\Omega|^\alpha},
\eeq
and the physical solutions are those values of $\alpha$ for which the eigenvalue $E=1$. The pairing problem so defined appeared in the context of SYK models \cite{klebanov,Klebanov:2020kck}, but also in earlier studies of quantum critical pairing of Fermi surfaces \cite{MoonChubukov,Chubukov21}.
A solution with a real $0 < \alpha < 1/3$ implies that the critical Fermi surface state is stable, and the value of $\alpha$ determines the exponent of critical correlations of the pairing operator \cite{Esterlis:2021eth}. Otherwise, there are solutions with complex $\alpha$, and these imply a pairing instability. The critical temperature towards pairing is determined by solving a generalization Eq.~(\ref{Deltaclean}) at non-zero $T$, and examining the $T$ at which the complex solution first appears.

\subsection{Adding spatial disorder}
\label{sec:nFLdisorder}

Given the rather singular transport properties of the critical Fermi surface described in Section~\ref{sec:cfstransport}, it is valuable to have the corresponding large $N$ analysis of a model which includes the {\it self-consistent\/} influence of weak disorder on the critical Fermi surface, beyond the perturbative 
analysis provided by the memory function approach \cite{Hartnoll:2016apf}. The simplest spatial disorder we can add to Eq.~(\ref{flt60}) is potential disorder, similar in spirit to that in Section~\ref{sec:matrix}: this is a term $v_{ab} (x) \psi_a^\dagger (x) \psi_b (x)/\sqrt{N}$, in which $v_{ab}$ is a random matrix uncorrelated at different points in space, so that
\beq
\overline{v_{ab}^{\vphantom{\ast}} (x) v_{cd}^\ast (x')} = v^2 \delta_{ac} \delta_{bd} \delta^d (x-x')\,. \label{vvdelta}
\eeq
This potential leads to an additional term in the large $N$ action in Eq.~(\ref{flt51}). The solution of the saddle point equations in the theory with both $g$ and $v$ non-zero 
shows \cite{Guo:2022zfl} that the boson polarizibility in Eq.~(\ref{flt75}) is replaced by
\beqn
\Pi (q, i \Omega_n) \sim - \frac{g^2}{v^2} |\Omega_n|, \label{flt100}
\eeqn
which leads to $z=2$ behavior in the boson propagator. The corresponding fermion self energy has a familiar elastic impurity scattering contribution $\Sigma_v$, along with an inelastic term $\Sigma_{g}$ \cite{Patel:2022gdh} with the `marginal Fermi liquid' form \cite{Varma89}
\beqn
\Sigma_v (i \omega_n) \sim - i v^2 \mbox{sgn}(\omega_n), \quad 
\Sigma_{g} (i \omega_n) \sim - \frac{g^2}{v^2} \omega_n \ln (1/|\omega_n|)\,. \label{flt101}
\eeqn
Despite the promising singularity in $\Sigma_{g}$, Eq.~(\ref{flt101}) does not translate \cite{Patel:2022gdh} into interesting behavior in the transport: the scattering is mostly forward, and the resistivity is Fermi liquid-like with $\rho(T) = \rho(0) + AT^2$.

While the effect of potential scattering of fermions is weak, a related estimation of the effects of a spatially random $\phi^2$ term ({\it i.e.\/} a random `scalar mass' allowed when $\phi$ represents a symmetry breaking order parameter) turns out to be strong \cite{Patel:2014jfa}. It has been argued \cite{Patel:2022gdh} that such disorder should should be absorbed by transforming to eigenmodes of the quadratic $\phi$ action, at the price of introducing spatial randomness in the Yukawa coupling $g$. Remarkably, a theory with spatial randomness in the boson-fermion Yukawa coupling included at the outset leads to physical effects that are `just right' in the large $N$ limit.
We {\it add\/} to the spatially independent Yukawa couplings $g_{abc}$ in Eq.~(\ref{flt60}) a second coupling $g'_{abc} (x)$ which has both spatial and flavor randomness with vanishing first moment, and second moment
\beqn
\overline{ {g'}_{abc}^{\vphantom{\ast}} (x) {g'}_{a'b'c'}^\ast (x') } = g'^2 \delta^d (x-x') \delta_{aa'} \delta_{bb'} \delta_{cc'} \label{ggdelta}
\eeqn
Then, along with (\ref{flt100}),\ref{flt101}), we obtain additional contributions to the boson and fermion self energies \cite{Patel:2022gdh}
\beqn
\Pi_{g'} (q, i \Omega_n) \sim - g'^2 |\Omega_n| \,, \quad 
\Sigma_{g'} (i \omega_n) \sim -i g'^2 \omega_n \ln (1/|\omega_n|) \,.
\eeqn
Now the marginal Fermi liquid self energy does contribute significantly to transport \cite{Patel:2022gdh}, with a linear-$T$ resistivity $\sim g'^2 T$, while the residual resistivity is determined primarily by $v$. It is notable that it is the disorder in the interactions which determines the slope of the linear-$T$ resistivity, while it is the potential scattering disorder which determines the residual resistivity. 

It is also interesting to examine this theory for Planckian dissipation \cite{Legros19,Paglione,Cao20,efetov21,Grissonnanche2020,Paschen22}. This requires writing the conductivity in the form 
\beqn
\sigma = \frac{n e^2 \tau^\ast_{\rm tr}}{m^\ast} \label{mfl2}
\eeqn
where the effective mass $m^\ast$ is computed from the fermion self energy in a Fermi liquid state proximate to the critical theory. For $g' \gg g$, the transport scattering time is found to be \cite{Adalpe20,Esterlis:2021eth}
\beqn
\frac{1}{\tau^\ast_{\rm tr}} \approx \frac{\pi}{2} \frac{k_B T}{\hbar} \label{flt200}
\eeqn
along with factors which are slow logarithmic functions of temperature.
However, for smaller values of $g'/g$, there is a significant decrease from the value in (\ref{flt200}) \cite{Paschen22,Patel:2022gdh}.

\section{Connections to quantum gravity}
\label{sec:qg}

We saw in Section~\ref{sec:fluctuations} that the finite $N$ fluctuations of the SYK model were described by a path integral over time reparameterizations. This suggests a connection to a theory of quantum gravity. By the holographic principle \cite{tHooft:1999rgb}, we expect the gravity theory to acquire an emergent spatial direction. As the SYK path integral is over $0+1$ dimensions, we anticipate a connection to quantum gravity in 1+1 dimensions. However, Einstein gravity in 1+1 dimensions has no dynamical modes, and so cannot serve as a holographic partner to the SYK model. As we will see below in Section~\ref{sec:bh_fluc}, the appropriate theory is a class of 1+1 dimensional theories known as Jackiw-Teitelboim (JT) gravity, which has an additional scalar field $\Phi$. This gravity theory is most naturally obtained by dimensional reduction, from a charged black hole of Einstein gravity in $d+2$ spacetime dimensions ($d \geq 2$). Such a black hole has a AdS$_2 \times S^d$ near-horizon geometry; the JT gravity theory resides on AdS$_2$, and fluctuations of $\Phi$ represent the quantum fluctuations in the radius of $S^d$. The connection between the SYK model and charged black holes was first noted \cite{SS10,Sachdev:2010uj} by matching characteristics of the $N = \infty$ SYK theory and the classical gravity solution of charged black holes in Einstein gravity. It was pointed out later \cite{kitaev_talk} that the connection was stronger, and also held for a low energy sector of the fluctuations.

The AdS$_2$ near-horizon sector of charged black holes leads to a non-vanishing entropy as $T \rightarrow 0$, a key characteristic such black holes share with the SYK model \cite{SS10,Sachdev:2010uj}. Neutral black holes, such as the common Schwarzschild solution of Einstein gravity, do not have AdS$_2$ horizons, and have vanishing entropy as $T \rightarrow 0$. Such black holes display a Hawking-Page transition at a non-zero $T$, and have a distinct low $T$ behavior which we will not discuss further here \cite{Schlenker:2022dyo}.

We proceed by reviewing the quantum theory of charged black holes in $d+2$ spacetime dimensions. We will then discuss its low temperature limit, and show, by the dimensional reduction outlined above, that this yields a version of JT gravity, which is in turn equivalent to the Schwarzian theory of the SYK model in Section~\ref{sec:fluctuations}.

There is another, and closely related, connection between SYK models and black holes that we briefly mention now. Our discussion above (and below) focuses on the equilibrium thermodynamic properties. In dynamic properties, SYK models are characterized by Planckian time dynamics \cite{QPT}, as we discussed in Section~\ref{sec:SYKT}; other metallic systems also have a similar dynamics, in theory and experiment, as noted in Sections~\ref{sec:nFLdisorder} and \ref{subsec:planckian}. Remarkably, Einstein gravity also displays Planckian time dynamics for black holes responding to external perturbations. This is evident in computations of the damping rate of black hole quasi-normal modes \cite{Vishveshwara,Hod07}: this purely classical gravity rate is $\sim \hbar/(k_B T_H)$ where $T_H$ is the Hawking temperature of the black hole (the $\hbar$ in the Planckian time formula cancels with the $\hbar$ in $T_H$). A recent analysis of LIGO  data \cite{LIGO21} has confirmed this remarkable universality in black hole quasi-normal modes.

The quantum fluctuations of gravity and electromagnetism are formally defined by a path integral 
\beq
\mathcal{Z}_{EM} = \int \mathcal{D} g \mathcal{D} A \exp (- I_{EM} ) \label{em1}
\eeq
where $I_{EM}$ is the Einstein-Maxwell action (that we will write down below), and the path integral is over the metric $g$ of spacetime, and the electromagnetic vector potential $A$.
It is almost certainly true that the expression Eq.~(\ref{em1}) does not make sense as it stands, because of numerous ultraviolet divergencies and gauge-fixing issues. Nevertheless, it turns out to be possible to make sense of Eq.~(\ref{em1}) in certain limits. For black hole saddle-point solutions of $I_{EM}$, it was shown \cite{Gibbons_Hawking} that the evaluation of $I_{EM}$ at the saddle point in a Euclidean geometry, with a thermal circle of circumference $\hbar/(k_B T)$ along the temporal direction, gave a consistent description of the quantum thermodynamics of black holes. 
It is only via the $\hbar$ dependence of this circumference that Planck's constant appears in such computations: there is no $\hbar$ in $I_{EM}$, the classical Einstein-Maxwell action. We will set $\hbar=k_B = 1$ in the remainder of our discussion. 

We will review the Gibbons-Hawking description of a charged black hole \cite{Myers99} in Section~\ref{sec:bh_saddle}. Remarkably, there turn out to be precise quantitative connections to the thermodynamics of the SYK model \cite{SS10,Sachdev:2010uj,SS15}.

Fluctuation corrections to the Gibbons-Hawking thermodynamics were computed only recently in the low $T$ limit for charged holes. In principle, these corrections could have been computed decades ago, but the computations were undertaken only after the connection to the SYK model showed the route that was needed. These computations are reviewed in Section~\ref{sec:bh_fluc}, which shows that the low energy theory of charged black holes reduces to an effective theory which is identical to the theory in Eqs.~(\ref{syk32}) and (\ref{syk34}) obtained for the SYK model of complex fermions. 

Section~\ref{sec:wormholes} briefly surveys rapid recent developments on coupled SYK models in and out of equilibrium, which are holographically realized by solitons/instantons known as `wormholes'.

Section~\ref{sec:adscft} discusses approaches to the theory of strange metals using the AdS/CFT correspondence of supersymmetric Yang-Mills theory \cite{Hartnoll:2016apf}, and connects these to the SYK model by placing the Yang-Mills theory on a finite sphere. 

\subsection{Charged black holes: Einstein-Maxwell theory}
\label{sec:bh_saddle}

We consider the case of {\it spherical\/} black holes in $d+2$ spacetime dimensions; we assume $d \geq 2$ in all of the following discussions of quantum gravity.
The Einstein-Maxwell theory has the Euclidean action
\bea
I_{EM} & = & \int d^{d+2} x \sqrt{g} \Biggl[ -\frac{1}{2 \kappa^2} \left(\mathcal{R}_{d+2} + \frac{d(d+1)}{L^2} \right) \nonumber \\
&~&~~~~~~~~~~~~~~~~~~~~~~~~~ +  \frac{1}{4g_F^2} F^2 \Biggr], \label{EM}
\eea
where $\kappa^2 = 8 \pi G_N$ is the gravitational constant, $\mathcal{R}_{d+2}$ is the Ricci scalar, $F = dA$ is the electromagnetic flux, and $g_F$ is a U(1) gauge coupling constant. We have also included a negative cosmological constant term so that the spacetime at asymptotic infinity is AdS$_{d+2}$ with radius $L$; the limit of large $L$ can be taken at the end to obtain Minkowski spacetime at infinity. 

We now describe the spherical charged black hole saddle-point of $I_{EM}$. There is a 2 parameter family of such solutions, which we will specify by the temperature $T$, and the chemical potential $\mu$. All other properties of the black hole saddle point are determined by $T$, $\mu$, and the constants of nature in $I_{EM}$: this includes the spacetime metric, the U(1) gauge field, the radius of the horizon, $r_0$, the total charge in the black hole $\mathcal{Q}$, and the black hole entropy $S$. 

The classical Einstein-Maxwell equations yield the following expression for the metric expressed in terms of imaginary time $\tau$, radial co-ordinate $r$, and $d \Omega_d^2$ the metric of the $d$-sphere: \cite{Myers99}
\beqn
ds^2 =  V(r) d\tau^2 + r^2 d \Omega_d^2 + \frac{dr^2}{V(r)} \label{s1}
\eeqn
where 
\beqn
V(r) = 1 + \frac{r^2}{L^2} + \frac{\Theta^2}{r^{2d-2}} - \frac{M}{r^{d-1}}.
\eeqn
As $r \rightarrow \infty$, the metric in Eq.~(\ref{s1}) is AdS$_{d+2}$. The radius of the horizon is determined by $V(r_0)=0$, which we write as 
\beq
M = r_0^{d-1} \left( 1 + \frac{r_0^2}{L^2} + \frac{\Theta^2}{r_0^{2d-2}} \right). \label{a3}
\eeq
The gauge field solution has the form
\beq
A = i \mu \left( 1 - \frac{r_0^{d-1}}{r^{d-1}} \right) d \tau. \label{a1}
\eeq
The value of the gauge field at the AdS boundary defines the chemical potential $\mu$, provided $r_0$ is the horizon.
The Einstein-Maxwell equations applied to Eqs.~(\ref{s1}) and (\ref{a1}) also yield the condition
\beq
\Theta = \sqrt{\frac{(d-1)}{d}} \frac{\kappa r_0^{d-1}}{g_F} \mu\,.  \label{a2}
\eeq

So far, our analysis has been entirely classical. As stated above, quantum mechanics enters the picture only by the condition that the solution in Eq.~(\ref{s1}) yield a spacetime which is periodic as a function of $\tau$ with period $1/T$. We can impose periodicity as a function of $\tau$ by fiat, but have to ensure that there is no singularity at the horizon $r_0$ where $V(r_0)=0$. Let us change radial co-ordinates to $y$, where $r=r_0 + y^2$. Then near the horizon, the $(r, \tau)$ components of Eq.~(\ref{s1}) become
\beqn
ds^2 = \frac{4}{V'(r_0)} \left[ \frac{(V'(r_0))^2}{4} \, y^2 d \tau^2 + dy^2 \right].
\eeqn
Now notice that the expression in the square brackets is precisely the metric of the flat plane in polar co-ordinates, with radial co-ordinate $y$ and angular co-ordinate $\theta =  V'(r_0) \tau/2$. For there to be no real singularity at the origin of polar co-ordinates, only a co-ordinate singularity, we must have periodicity in $\theta$ with period $2 \pi$. Matching this with the period $1/T$ in $\tau$, we determine the Hawking temperature of the black hole
\beq
4 \pi T = V'(r_0)\,. \label{a4}
\eeq
 
The Eqs.~(\ref{a3},\ref{a2},\ref{a4}) determine all the parameters, $\Theta$, $M$, $r_0$ in terms of $\mu$ and $T$. So we have specified a black hole solution in terms of the independent thermodynamic parameters $\mu$ and $T$.

We now quote the free energy and entropy of this black hole, obtained by the evaluation of $I_{EM}$ at the saddle-point above. The action has to be supplemented by a Gibbons-Hawking boundary term which is required to obtain the classical Einstein-Maxwell equations as saddle-point equations of $I_{EM}$. Such an evaluation of $I_{EM}$ yields the 
grand potential \cite{Myers99}
\bea
\Omega (T, \mu) 
&=& \frac{s_d [r_0 (T, \mu)]^{d-1}}{2 \kappa^2} \left(1 - \frac{[r_0 (T, \mu)]^2}{L^2} \right) \nonumber \\
&~&~~~~ - \frac{s_d (d-1) \mu^2 [r_0 (T, \mu)]^{d-1}}{2d g_F^2}\,, \label{Omega}
\eea
where $s_d \equiv 2 \pi^{(d+1)/2} /\Gamma((d+1)/2)$ is the area of $S_d$ with unit radius.
We can evaluate the the total charge by taking the $\mu$ derivative of $\Omega$
\beqn
\mathcal{Q}(T, \mu) =\frac{s_d (d-1) \mu \, [r_0(T,\mu)]^{d-1}}{ g_F^2} \,, \label{s6}
\eeqn
and this expression can also be obtained from Gauss's law evaluated as $r \rightarrow \infty$. Similarly, the 
entropy by taking the temperature derivative of $\Omega$ to obtain 
\beqn
S(T, \mu) = \frac{2 \pi s_d}{\kappa^2} \, [r_0 (T,\mu)]^d, \label{s5}
\eeqn
which is precisely the expression expected from Hawking's celebrated result $\mathcal{A}/(4 G_N)$: $\mathcal{A} = s_d r_0^d$ is the area of the horizon.
The universality of the Hawking area result can be understood from the fact that the only explicit dependence of the action on $T$ arises from the identification in Eq.~(\ref{a4}) leading to a circumference $1/T$ on the time circle; then the $T$ derivative of $\Omega$ can be shown to arise only from the vicinity of the horizon at $r=r_0$, where the integral over the angular co-ordinates yields the area \cite{Ross:2005sc}.

\begin{figure}
\begin{center}
\includegraphics[width=3.5in]{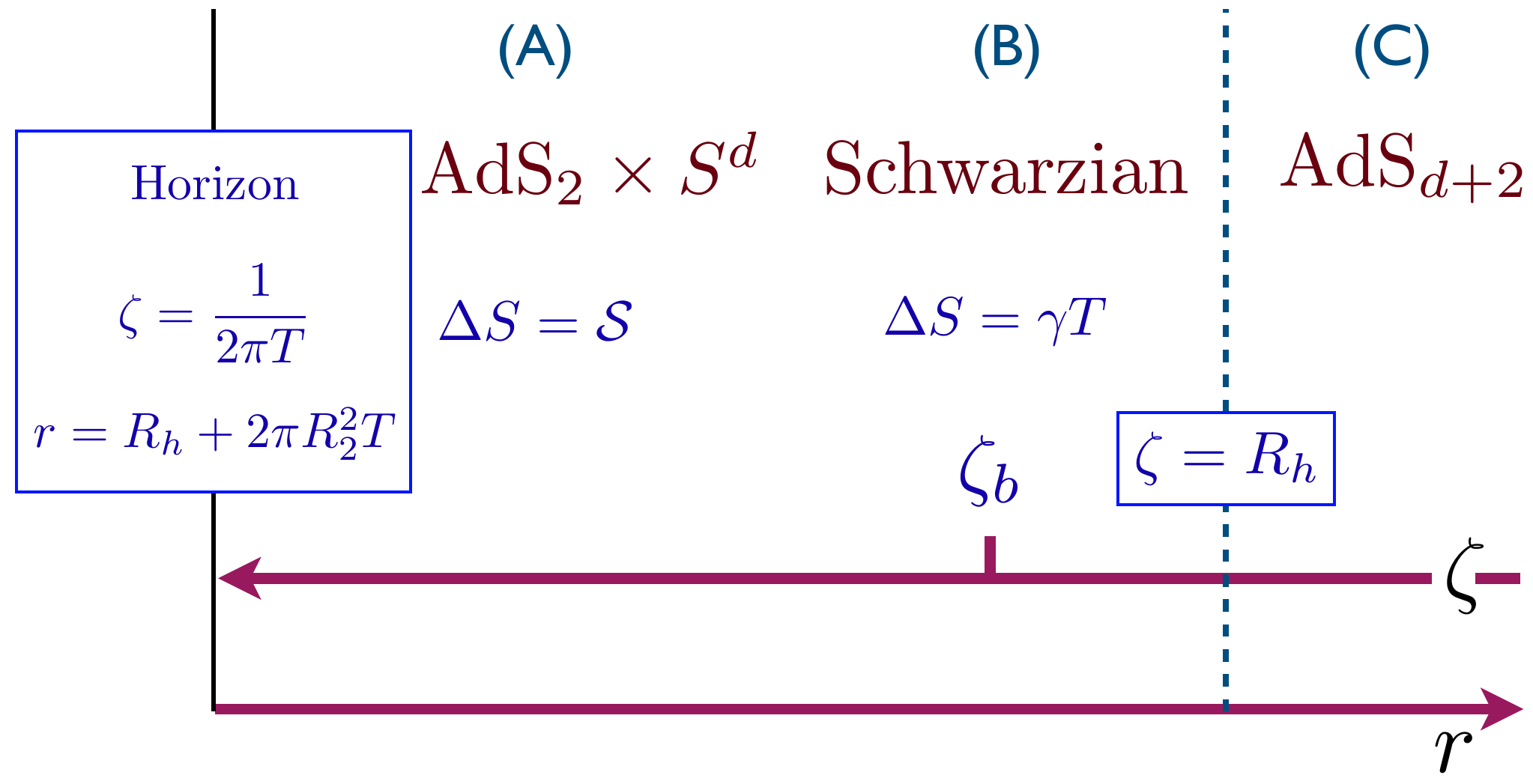}
\caption{Spatial crossover boundaries outside a black hole of charge $\mathcal{Q}$. The value of $R_h$ is determined from $\mathcal{Q}$ via Eq.~(\ref{Q1}), and we describe $T \ll 1/R_h$ at fixed $\mathcal{Q}$, and $R_2 \sim R_h$. We indicate contributions to the entropy $\Delta S$ from regions (A) and (B).} 
\label{fig:bh}
\end{center}
\end{figure}

We will now take the $T \rightarrow 0$ limit of all the results above, while keeping the charge $\mathcal{Q}$ fixed. Then the horizon radius $r_0 \rightarrow R_h$, where
\beqn
\mathcal{Q} = \frac{s_d R_h^{d-1} \sqrt{d \left[ (d+1)R_h^2  + (d-1)L^2  \right]}}{L \kappa g_F} \,.
\label{Q1}
\eeqn
We are interested in the structure of the metric near the horizon at $T=0$. For this purpose, 
we transform to near-horizon co-ordinates, by changing the radial co-ordinate from $r$ to the co-ordinate $\zeta$, where 
\beqn
r  = R_h + \frac{R_2^2}{\zeta}\,; \label{rzeta}
\eeqn
in these-co-ordinates, the $T=0$ horizon is at $\zeta=\infty$; see Fig.~\ref{fig:bh}.
We chose the length scale $R_2$ to be
\beqn
R_2  = \frac{L R_h}{\sqrt{d(d+1) R_h^2 + (d-1)^2 L^2}} \,. \label{valR2}
\eeqn
Then, as $T\rightarrow 0$, the metric Eq.~(\ref{s1}) for $\zeta \gg R_h$ (region (A) in Fig.~\ref{fig:bh}) becomes 
\beqn
ds^2 = \frac{R_2^2}{\zeta^2} \left[  d\tau^2 +d \zeta^2 \right] + R_h^2 d \Omega_d^2\,. \label{ads2}
\eeqn
The metric on the $(\tau, \zeta)$ spacetime is AdS$_2$, and the complete metric is AdS$_2 \times S^d$. In the same co-ordinate system, the U(1) gauge field becomes
\beqn
A = i \frac{\mathcal{E}}{\zeta} d \tau \,, \label{Azeta}
\eeqn
where the dimensionless prefactor
\beqn
\mathcal{E} = \frac{g_F R_h L \sqrt{d \left[ (d+1) R_h^2 + (d-1)L^2 \right]}}{\kappa \left[ d(d+1)R_h^2+  (d-1)^2 L^2 \right]}. \label{s4}
\eeqn
is a measure of the electric field on the horizon of the black hole.
We have chosen the same symbol $\mathcal{E}$ for this prefactor as that appearing in characterizing the particle-hole asymmetry of the SYK model in Eqs.~(\ref{syk7a}) and (\ref{syk18}). This is not arbitrary \cite{SS10,Sachdev:2010uj,SS15}: computations \cite{Liu2} of the Green's function of a fermion moving in the background specified by Eqs.~(\ref{ads2}) and (\ref{Azeta}) remarkably yields precisely the result in Eq.~(\ref{syk18}) . 

Let us now turn to a computation of the entropy, where we find remarkable connections to the SYK model. We write $S(T \rightarrow 0) = \mathcal{S}$, and then from Eq.~(\ref{s5})
\beqn
\mathcal{S} = \frac{2 \pi s_d}{\kappa^2} R_h^d \label{SBH}
\eeqn
So we obtain a non-vanishing entropy in the zero temperature limit, similar to the SYK model \cite{SS10,Sachdev:2010uj}.
Furthermore, by eliminating $R_h$ between Eqs.~(\ref{Q1}) and (\ref{SBH}), and using Eq.~(\ref{s4}), we find
\beqn
\left( \frac{ \partial \mathcal{S}}{\partial \mathcal{Q}} \right)_{T \rightarrow 0}
= 2 \pi \mathcal{E} \label{dSdQ}
\eeqn
which is exactly the relation Eq.~(\ref{syk23}) obtained for the SYK model \cite{SS15}.
We can also compute the low $T$ dependence of $\mu$, and verify that the Maxwell relation Eq.~(\ref{syk22a}) is satisfied. Furthermore, $T$
dependence of entropy computed from Eq.~(\ref{SBH}) is linear in $T$ at low $T$ and fixed $\mathcal{Q}$
\beqn
S(T \rightarrow 0, \mathcal{Q}) = \mathcal{S} + \gamma T\,,
\label{sgammabh}
\eeqn
where
\beqn
\gamma = \frac{4 \pi^2 d s_d R_2^2 R_h^{d-1}}{\kappa^2}\,. \label{gammares}
\eeqn
This is 
as in the SYK model in Eq.~(\ref{syk36c}), where the value of $\gamma$ was related to the co-efficient of a Schwarzian action, and we will do the same for the charged black hole in Section~\ref{sec:bh_fluc}. 

The appearance of the fundamental relation Eq.~(\ref{dSdQ}) of the SYK model in the theory of a charged black hole may appear like a co-incidence here, but it is not. In fact, Eq.~(\ref{dSdQ}) is a general property of black holes with AdS$_2$ horizons, and follows from careful consideration of their symmetries \cite{Sen05,Sen08}. These symmetries are similar to those described in Appendix~\ref{sec:symSYK} for the SYK model, which were exploited in 
Section~\ref{sec:SYKentropy} to obtain Eq.~(\ref{syk23}) \cite{SS15,SS17,GKST}. 

\subsubsection{Charged black branes}
\label{sec:brane}

This section briefly notes the limit of the above spherical solution when the black hole becomes an infinite, {\it flat\/}, charged `black brane', with a near-horizon geometry of AdS$_2 \times R^d$, in contrast to the near-horizon AdS$_2 \times S^d$ considered so far. These results will be helpful in Section~\ref{sec:adscft} where we discuss the connection to the AdS/CFT correspondence. This limit is obtained by taking $R_h \gg L$ in our results so far. 
We introduce the charge and entropy {\it densities}:
\beqn
\mathscr{Q} \equiv \frac{\mathcal{Q}}{s_d L^d} \quad, \quad \mathscr{S} \equiv \frac{S}{s_d L^d}\,. \label{Qdensity}
\eeqn
Then we have from Eq.~(\ref{Q1}) 
\beqn
\mathscr{Q} = \frac{ \sqrt{d(d+1)}}{L \kappa g_F} \left(\frac{R_h}{L} \right)^{d} \,.
\label{Q1a}
\eeqn
Similarly, the $T \rightarrow 0$ entropy in Eq.~(\ref{sgammabh}) becomes the Hawking entropy density
\beq
\mathscr{S} (T \rightarrow 0, \mathscr{Q}) = \frac{2 \pi}{\kappa^2} \left(\frac{R_h}{L} \right)^{d} \left[ 1 + \frac{2 \pi L^2 T}{(d+1) R_h} \right].~~~~~~
\eeq
These results for the densities correspond exactly to those obtained earlier \cite{Liu2} from direct solution of the flat black-brane geometry.

\subsection{Charged black holes: quantum fluctuations}
\label{sec:bh_fluc}

This section will examine quantum fluctuations about the saddle-point solution of Einstein-Maxwell theory described in Section~\ref{sec:bh_saddle}.
Remarkably, in the `extremal' limit $T \ll 1/R_h$, the theory of these fluctuations co-incides with those of the theory Eqs.~(\ref{syk32}) and (\ref{syk34}) obtained from the SYK model \cite{Nayak:2018qej,Moitra:2018jqs,Gaikwad:2018dfc,SachdevICMP,Heydeman:2020hhw,Iliesiu:2020qvm,Boruch:2022tno}. Below, we will outline how this theory may be obtained starting from Eq.~(\ref{em1}). 
A more detailed review of these fluctuation computations has been presented elsewhere \cite{SachdevICMP}, and here we shall highlight the key steps:
\begin{enumerate}
    \item Reduce the $d+2$ spacetime dimensional theory in $I_{EM}$ to a 1+1 dimensional theory $I_{EM,2}$ by taking all fields dependent only upon the radial co-ordinate $r$ and imaginary time $\tau$. 
    \item Take the low energy limit of $I_{EM,2}$ by mapping it to a near-horizon theory, $I_{JT}$, in a 1+1 dimensional spacetime with a boundary: so we ``integrate out'' region (C) in Fig.~\ref{fig:bh}, and obtain an effective theory in regions (A) and (B). In these regions, the near-horizon AdS$_2$ saddle point in Eqs.~(\ref{ads2}) and (\ref{Azeta}) is an exact saddle point of $I_{JT}$. Outside the boundary, there is a crossover to the full solution of $I_{EM}$ in Eqs.~(\ref{s1}) and (\ref{a1}) to region (C) where spacetime does not factorize into AdS$_2 \times S_d$.
    \item Compute fluctuations about the AdS$_2$ saddle point of $I_{JT}$. Einstein gravity in 1+1 dimensions has no graviton, and is `pure gauge'. In the JT-gravity theory with boundary, there is a remnant degree of freedom which is a boundary graviton. The action for this boundary graviton is precisely the Schwarzian theory in Eqs.~(\ref{syk32}) and (\ref{syk34}).
\end{enumerate}
We outline these steps for the gravity sector in the following subsections. The electromagnetic sector produces the action for the phase field $\phi$ in the Schwarzian theory, as is discussed elsewhere \cite{SachdevICMP}.

\subsubsection{Dimensional reduction from $d+2$ to $1+1$.}

We write the $(d+2)$-dimensional metric $g$ of $I_{EM}$ in Eq.~(\ref{EM}) in terms of a two-dimensional metric $h$ and a scalar field $\Phi$ \cite{SS17,Nayak:2018qej}:
\beq
ds^2 = \frac{ds_2^2}{\Phi^{d-1}} + \Phi^{2} \, d \Omega_d^2 \,. \label{dsds2}
\eeq
Both $h$ and $\Phi$, and the gauge field $A$, are allowed to be general functions of the two-dimensional co-ordinates $\zeta$ and $\tau$ (recall Eq.~(\ref{rzeta}) for the definition of the radial co-ordinate $\zeta$). Note that the scalar field $\Phi$ represents radial fluctuations in the size of the black hole.
Then Eq.~(\ref{EM})) and an associated Gibbons-Hawking boundary term reduce to ($x \equiv (\tau, \zeta)$)
\bea
I_{EM,2} &=& \int d^2 x \sqrt{h} \left[ - \frac{s_d}{2\kappa^2} \, \Phi^d \, \mathcal{R}_2 + U(\Phi) + \frac{Z(\Phi)}{4 g_F^2} F^2 \right]  \nn
I_{GH} &=& - \frac{s_d}{\kappa^2} \int_{\partial} dx \sqrt{h_b} \Phi^d \, \mathcal{K}_1 \label{2d1}
\eea
along with an additional term not displayed which cancels in $I_{EM,2} + I_{GH}$ \cite{Nayak:2018qej}.
The Gibbons-Hawking term is to be evaluated at the boundary at $\zeta \rightarrow 0$ or $r \rightarrow \infty$.
Here $\mathcal{R}_2$ is the two-dimensional Ricci scalar, the second integral is over a one-dimensional boundary with metric $h_b$ and extrinsic curvature $\mathcal{K}_1$. 
The explicit forms of the potentials $U(\Phi)$ and $Z(\Phi)$ are,
\bea
U(\Phi) &=& - \frac{s_d}{2 \kappa^2} \left( \frac{d(d-1)}{\Phi} + \frac{d(d+1) \Phi}{L^2} \right) \nonumber \\ 
Z(\Phi) &=& s_d \Phi^{2d-1}  \,. \label{2d2}
\eea
The 1+1 dimensional action in Eqs.~(\ref{2d1},\ref{2d2}) has exactly the same saddle point solution as that of the $d+2$ dimensional action in Eq.~(\ref{EM}). The 1+1 dimensional theory $I_{EM,2}$ now involves a metric $h$ and a scalar field $\Phi$, and in terms of the new variables, the solution is given by matching Eq.~(\ref{s1}) with the ansatz in Eq.~(\ref{2d1}). In this manner, it is easy to see that the exact solution for the scalar field is
\beq
\Phi (\zeta) = R_h + \frac{R_2^2}{\zeta}\,. \label{Phizeta}
\eeq

\subsubsection{JT gravity in the near-horizon limit}

Note that the $T=0$ horizon is obtained as $\zeta \rightarrow \infty$, and the factorization of the metric to AdS$_2 \times S^d$ fails for $\zeta \lesssim R_h$. So we reduce the theory to the near-horizon spatial region $\zeta > \zeta_b$, with 
\beqn
R_h \ll \zeta_b \ll \frac{1}{T} \, \label{zetab}
\eeqn
which applies in regions (A) and (B) of Fig.~\ref{fig:bh}.
The low energy limit  of the 1+1 dimensional theory of step 1 to $\zeta > \zeta_b$ was argued \cite{Almheiri15,JMDS16b} to be the JT gravity theory \cite{Teitelboim83,Jackiw85} of a metric $h$ and a scalar field $\Phi_1$ given by
\bea
I_{JT} &=& -\mathcal{S} + \int d^2 x \sqrt{h} \left[ - \frac{s_d}{2\kappa^2} \, \Phi_1 \left( \mathcal{R}_2 + \frac{2}{H_b} \right) \right]  \nn
I_{GH} &=& - \frac{s_d}{\kappa^2} \int_{\partial} dx \sqrt{h_b} \Phi_1 \, \mathcal{K}_1 \label{jt}
\eea
where $\mathcal{S}$ was defined in (\ref{SBH}). We also have the boundary conditions
\bea
h_{\tau\tau} (\zeta \searrow \zeta_b) &=& \frac{H_b}{\zeta^2} \nonumber \\
\Phi_1 (\zeta \searrow \zeta_b) &=&  \frac{\Phi_b}{\zeta} \,.\label{jtb}
\eea
This theory depends upon 2 constants, $H_b$ and $\Phi_b$, and we can obtain their values by matching to the solution for the two-dimensional metric $h$ and scalar field $\Phi$ obtained in Step 1, which was valid at all $\zeta$.
The boundary condition on $h_{\tau\tau}$ is obtained by comparing Eq.~(\ref{dsds2}) with 
with Eq.~(\ref{ads2}), and using the leading term in Eq.~(\ref{Phizeta}) for large $\zeta$ we obtain:
\beqn
H_b = R_2^2 R_h^{d-1}\,.
\eeqn
The subleading term in Eq.~(\ref{Phizeta}) contributes to the co-efficient of $\mathcal{R}_2$ in Eqs.~(\ref{2d1}) and (\ref{jt}),  which from Eq.~(\ref{Phizeta}) yields 
\beq
\lim_{\zeta \rightarrow \infty} [\Phi(\zeta)]^d  = R_h^d + \Phi_1 (\zeta) + \ldots \label{PhidPhi1}
\eeq
Then the boundary value of $\Phi_1$ in Eq.~(\ref{jtb}) determines
\beq
\Phi_b = d R_h^{d-1} R_2^2\,. \label{Phibval}
\eeq
Now, the saddle point solution of the $JT$ gravity theory in Eqs.~(\ref{jt}) and (\ref{jtb}) co-incides with the metric Eq.~(\ref{ads2}), which we now generalize to $T>0$:
\bea
ds_2^2 &=& \frac{R_2^2 R_h^{d-1}}{\zeta^2} \left[ (1 - 4 \pi^2 T^2 \zeta^2) d\tau^2 + \frac{d \zeta^2}{1 - 4 \pi^2 T^2 \zeta^2} \right] \nonumber \\
\Phi_1 (\zeta) &=&  \frac{\Phi_b}{\zeta}
\,. \label{ads2a}
\eea
Note that the boundary form of $\Phi_1$ in Eq.~(\ref{jtb}) holds for all $\zeta$ in the regime of validity of the JT theory, a result also evident from Eq.~(\ref{PhidPhi1}). The horizon is at $\zeta = 1/(2 \pi T)$, and it can be verified that the analog of Eq.~(\ref{a4}) for the Hawking temperature is satisfied here.

\subsubsection{From JT gravity to the Schwarzian}
\label{sec:JTtoSchwarzian}

We address fluctuations about the saddle-point solution Eq.~(\ref{ads2a}) of the JT gravity theory defined by Eqs.~(\ref{jt}) and (\ref{jtb}). The effective theory now has a simple enough form that these fluctuations can be evaluated reliably \cite{JMDS16b}. The integral over $\Phi_1$ in Eq.~(\ref{jt}) can be evaluated exactly, and yields a constraint on the bulk metric and the only dynamical degree of freedom in JT gravity is a time reparameterization 
along the boundary $\tau \rightarrow f(\tau)$. 
To ensure that the bulk metric obeys the boundary condition Eq.~(\ref{jtb}), we also have to make the spatial co-ordinate $\zeta$ a function of $\tau$, so we map
$(\tau, \zeta) \rightarrow (f(\tau), \zeta(\tau))$. Then 
the boundary metric induced by Eq.~(\ref{ads2a}) equals the value in Eq.~(\ref{jtb}) 
provided $\zeta (\tau)$ is related to $f(\tau)$ by
\bea
\zeta (\tau) &=& \zeta_b f' (\tau) + \zeta_b^3 \left( \frac{\left[f'' (\tau) \right]^2}{2 f'(\tau)} - 2 \pi^2 T^2  \left[f' (\tau) \right]^3 \right)\nonumber \\
&~&~~~+ \mathcal{O}(\zeta_b^4 ) \,. \label{zetacurve}
\eea
Finally, we evaluate $I_{GH}$ in Eq.~(\ref{jt}) along this boundary curve (the bulk contribution $I_{JT}$ vanishes from equation of motion of $\Phi_1$, which is $\mathcal{R}_2 + 2/H_b = 0$).  In this manner we obtain the action \cite{JMDS16b,SachdevICMP} $I_{1,{\rm eff}} = - \mathcal{S} + I_{\rm eff}$ with
\bea
I_{{\rm eff}} [f] &=& -  \frac{s_d \Phi_b}{\kappa^2} \int_0^{1/T} d \tau \, \left( \{ f(\tau), \tau\}
+ 2 \pi^2 T^2 \left[f' (\tau) \right]^2 \right)
\nonumber \\
&=&   -  \frac{s_d \Phi_b}{\kappa^2} \int_0^{1/T} d \tau \, \{ \tan (\pi T f(\tau)), \tau\}\,. \label{Seff1}
\eea
Notice that the arbitrary value of $\zeta_b$ has cancelled out, and this is an important consistency check on our steps. Remarkably, we have obtained the Schwarzain action, found earlier for the SYK model. Here, its presence is a consequence of the SL(2,R) symmetry of pure AdS$_2$ discussed in Appendix~\ref{sec:symAdS}, which require that the action vanish for $f(\tau)$ which are isometries of AdS$_2$. The action for other $f(\tau)$ appears from the `boundary graviton' \cite{JMDS16b} obtained by embedding of AdS$_2$ in the $d+2$ dimensional geometry of a charged black hole. 

Comparing the action Eq.~(\ref{Seff1}) with the action for the SYK model in Eq.~(\ref{syk34}), we obtain from the co-efficient of the Schwarzian (ignoring the $N$ prefactor in Eq.~(\ref{syk34})) 
\beq
\gamma = \frac{4 \pi^2 s_d \Phi_b}{\kappa^2}\,. \label{gammares2}
\eeq
After using the value of $\Phi_b$ in Eq.~(\ref{Phibval}), we find that this value of $\gamma$ is in precise agreement with the value in Eq.~(\ref{gammares}), which was computed the $T$ dependence of the entropy in Eq.~(\ref{s5}) for the full $d+2$ dimensional theory. 
So the $\gamma$ co-efficients of both charged black holes and the SYK model (in Eq.~(\ref{syk36c})) are given by the co-efficient of the Schwarzian effective action. 

Finally, we can combine the Schwarzian fluctuation contribution to the entropy in Eq.~(\ref{syk45}) with leading Bekenstein-Hawking entropy in Eqs.~(\ref{s5},\ref{valR2},\ref{sgammabh},\ref{gammares}) to obtain the universal, leading, low $T$ form of the entropy of charged black holes when the AdS$_{d+2}$ radius is much larger than the size of the black hole ($L \gg R_h$) \cite{SachdevICMP,Iliesiu:2020qvm}
\beqn
S(T) = \frac{1}{G_N} \left[ \frac{\mathcal{A}_0}{4} + \frac{\pi d \mathcal{A}_0^{(d+1)/d}}{2(d-1)^2 s_d^{1/d}}\, T \right] - \frac{3}{2} \ln \left( \frac{U}{T} \right)\,, \label{blackholelog}
\eeqn
where $\mathcal{A}_0 = s_d R_h^d$ is the horizon area at $T=0$, and the factor is square brackets accounts for the change in the horizon area with increasing $T$ at fixed $\mathcal{Q}$.
The non-universal energy scale $U$ is now presumably a Planck scale energy, but the $3/2$ co-efficient of the logarithm is independent of the nature of the high energy cutoff. The Schwarzian fluctuation correction to the entropy becomes of order the Bekenstein-Hawking term only at an exponentially low temperature $T \sim U \exp(-\mathcal{A}_0/(6G_N))$, when the theory breaks down, and the discrete level spacing of the black hole has to be accounted for: the path integral over the Einstein-Maxwell theory Eq.~(\ref{EM}) only has information on the density of states coarse-grained over the exponentially small level spacing, and determining the precise energy levels requires embedding Eq.~(\ref{EM}) in a higher energy theory like string theory.  As in Section~\ref{sec:schwarzian}, the logarithmic correction to the entropy in Eq.~(\ref{blackholelog}) translates to the coarse-grained density of many-body states in Eq.~(\ref{syk47}); for a charged black hole in $d+2$ dimensional Minkowski spacetime with $L \gg R_h$ the density of states takes the form \cite{Sachdev:2022qnu}
\begin{align}
D(E) \sim & \exp\left( \frac{\mathcal{A}_0 c^3}{4 \hbar G_N} \right) \nonumber \\ &~~\times  \sinh \left( \left[\frac{ \pi d \mathcal{A}_0^{(d+1)/d} }{(d-1)^2 s_d^{1/d}} \frac{c^3}{\hbar G} \frac{E}{\hbar c} \right]^{1/2} \right)  \label{DEF}
\end{align}
after restoring factors of $\hbar$ and $c$. Eq.~(\ref{DEF}) is a rare formula which combines Planck's constant $\hbar$ with Newton's gravitational constant $G_N$: the exponential prefactor was obtained by Hawking, the sinh follows from developments ensuing from the solution of the SYK model. Both terms depend only upon the $T=0$ area of the black hole horizon, $\mathcal{A}_0$, and fundamental constants of nature. Note also that there is no dependence upon the electromagnetic coupling, $g_F$.

We note that the black hole density of states obtained above is very different from that obtained in supersymmetric SYK models and black hole solutions of string theory \cite{Fu:2016vas,Heydeman:2020hhw,Boruch:2022tno}: the latter have an exponentially large {\it exact\/} degeneracy of ground states with multiplicity $\sim \exp (\mathcal{A}_0/(4G_N))$, and a gap $\sim 1/\mathcal{A}_0^{1/d}$ to the first excited state.
Contrast this with the generic non-supersymmetric situation with an exponentially small level spacing down to the ground state illustrated in Fig.~\ref{fig:specq4}. Indeed, it was the determination of the density of states of the SYK model which led to the understanding 
that black holes with AdS$_2$ horizons and no low-energy supersymmetry do not have ground states with an exponentially large degeneracy.

\subsection{Wormholes}
\label{sec:wormholes}

We have so far considered a single SYK model in thermal equilibrium, and argued that it is equivalent to a charged black hole, also in thermal equilibrium. 
The past few years have seen very rapid developments on the theory of more complex configurations of SYK models and black holes, including remarkable progress in resolving Hawking's quantum information paradox on evaporating black holes. A common thread in these developments have been `wormholes', which are the analogs of solitons or instanton tunneling events in quantum gravity.

Consider a pair of identical coupled SYK models, {\it i.e.\/} a `homonuclear diatomic SYK-molecule', with Hamiltonian\cite{Sahoo:2020unu}
\bea
H &=& \sum_{ij;k\ell} U_{ij;k\ell} \sum_{a=1,2} c_{ia}^\dagger c_{ja}^\dagger c_{k a} c_{\ell a} 
- \mu \sum_{i,a} c_{ia}^\dagger c_{ia} \nonumber \\
&+& \sum_i \kappa \left( c_{i 1}^\dagger c_{i 2} + c_{i 2}^\dagger c_{i 1} \right)\,.
\eea
Here $a = 1,2$ labels the two SYK-atoms, and $\kappa$ is the tunneling amplitude between them. Notice that the random interactions
$U_{ij,k\ell}$ are the same on both SYK-atoms. This 2-atom model is clearly similar to the lattices of SYK-atoms considered in Section~\ref{sec:hFL}.
At half-filling, this model can acquire a gapped ground state, when the fermions occupy only the lower energy `bonding' orbitals which are eigenstates of the $\kappa$ term. Holographically, this gapped state corresponds to an eternal wormhole between 2 black holes with AdS$_2$ horizons, as has been discussed in many recent works \cite{Maldacena:2018lmt,Gao:2019nyj,Plugge:2020wgc,Sahoo:2020unu,Zhou:2020kxb,Zhou:2020wgh,Zhang:2020szi,Garcia-Garcia:2019poj,Nikolaenko:2021vlw,Zhang:2022yaw}.

Next consider a single Majorana $q=4$ SYK model of $N$ sites (as in Section~\ref{sec:SYK}) coupled to a Majorana $q=2$ random matrix model of $M$ sites (as in Section~\ref{sec:matrix}), with $M \gg N$; this is a `heteronuclear diatomic SYK-molecule' with one atom much larger than the other, and is
described by the Hamiltonian \cite{Su:2020quk,Zhang:2022yaw}
\bea
   H&=&   \sum_{i<j<k<\ell}U^S_{ijkl}\psi_i \psi_j \psi_k \psi_\ell + i\sum_{a<b}U^\text{E}_{ab}\chi_a \chi_b \nonumber \\
   &~&~~~~~+ i\sum_{i,a}V_{ia}\psi_i \chi_a\,.
\eea        
Here $i,j,k,\ell = 1 \ldots N$ and $a,b=1 \ldots M$. (The same considerations apply to models of complex fermions, but the authors chose Majorana fermions for simplicity.) The SYK-atom of $\psi$ fermions models a black hole, and we consider a situation in which it is in some pure excited state with energy $E$ at time $t=0$. The $\chi$ free fermions represent the environment into which the black hole is going to radiate its energy, and so this setup models an evaporating black hole. At the initial time, the black hole is presumed to be decoupled from the environment, and so the entanglement entropy between the black hole and the environment vanishes. In the early stages of the evaporation, the energy $E$ will radiate out into the environment, and so the entanglement entropy will increase with time. However, we can also see that as $t \rightarrow \infty$, the energy $E$ will be essentially all absorbed by the environment (because $M \gg N$), and so the SYK model will be in a low energy state, with small entanglement with the environment. This time evolution of the entanglement is a model of the black hole Page curve \cite{Su:2020quk,Zhang:2022yaw}. In the holographic representation, the computation of such a Page curve involves spacetime wormholes \cite{Chen:2020wiq,Penington:2019kki,Almheiri:2020cfm,Almheiri:2019qdq,Saad:2019lba,Czech}. These works have led to the realization \cite{Bousso:2022ntt} that, upon including wormhole contributions, the path integrals over Einstein-Maxwell theories like (\ref{EM}) are also able to properly compute the time evolution of entanglement entropy in black hole evaporation, along with the density of states noted at end of Section~\ref{sec:JTtoSchwarzian}, despite their lack of knowledge of the precise black hole energy levels.

\subsection{AdS/CFT correspondence}
\label{sec:adscft}

An alternative route to a connection between strange metals and quantum gravity uses the AdS/CFT correspondence of string theory. This is a correspondence between a conformal field theory (CFT) in flat $d$-dimensional space, and gravity on a AdS$_{d+2}$ spacetime \cite{Maldacena:1997re,Witten:1998qj}. The canonical example in spatial dimension $d=3$ is SU($N_{YM})$ Yang-Mills gauge theory with $\mathcal{N}=4$ supersymmetry \cite{Maldacena:1997re}, and in spatial dimension $d=2$ is SU($N_{YM})$ Yang-Mills gauge theory with $\mathcal{N}=8$ supersymmetry \cite{Aharony:2008ug}. Both theories are conformally invariant, and map to {\it neutral} $\mathcal{Q}=0$ black hole solutions of the action Eq.~(\ref{EM}), with coupling constants 
\beqn
\kappa = \bar{\kappa} \, N_{YM}^{-a} \, L^{d/2} \quad, \quad 
g_F = \bar{g}_F \, N_{YM}^{-a}\, L^{(d-2)/2}\,,
\eeqn
where $\bar{\kappa}$ and $\bar{g}_F$ are dimensionless constants of order unity, and $a=1$ for $d=3$,
and $a=3/4$ for $d=2$.

To obtain a connection to strange metals, we have to `dope' these CFTs {\it i.e.\/} we have to place them in a chemical potential coupling to a global U(1) symmetry, which induces a conjugate charge density $N_{YM}^{2a} \mathscr{Q}_{YM}$ \cite{Hartnoll:2007ih}. In the gravity theory, this doped CFT maps to the same charged black hole solutions we have considered for the SYK model, with the crucial difference that the relevant solutions are the flat {\it black-brane\/} solutions in Section~\ref{sec:brane}, which describe
the strange metals produced by doped supersymmetric Yang-Mills theory in infinite $d$-dimensional space in the limit of large $N_{YM}$. We note that the doping breaks the supersymmetry, so the low energy theory has no supersymmetry.
The non-zero charge density in the supersymmetric Yang-Mills theory
introduces a length scale of order $[\mathscr{Q}_{YM}]^{-1/d}$, and we are interested in physics at longer length scales. At these length scales, black brane solutions described in Section~\ref{sec:brane} have a AdS$_2 \times R^d$ geometry \cite{Liu2}. 
The doped Yang-Mills theories are described by continuum Lagrangians similar to the disorder-free models of non-Fermi liquids we considered in Section~\ref{sec:cfs} \cite{Huijse:2011ef,Huijse:2011hp}. The holographic flow of the doped Yang-Mills theory to a AdS$_2$ geometry is therefore evidence that models in the class of Section~\ref{sec:cfs} could have an intermediate energy range over which their physics is described by the SYK-like local criticality. While the SYK-critical state of Section~\ref{sec:tJU} is unstable to spin glass order at the lowest temperatures, there could be a crossover from local criticality to the momentum-dependent Fermi surface physics at the lowest energies for the models of Section~\ref{sec:cfs}. This is in contrast to the supersymmetric doped Yang-Mills theories, for which the AdS$_2$ geometry is stable down to zero temperature in the large $N_{YM}$ limit. We note another discussion \cite{Iqbal:2011in,Iqbal:2011ae} with a related point of view.

Some studies of the  AdS$_2 \times R^d$ black brane solutions have focused on their response to additional {\it probe} fermions \cite{Liu1,Liu2,Liu3,Cubrovic:2009ye,Cubrovic:2010bf}.
In particular, it was shown that probe fermions in such a geometry acquired a Fermi surface and a self energy with some similarities to the critical Fermi surface described in Section~\ref{sec:cfs2}, with a self-energy which obeyed a scaling form similar to Eq.~(\ref{flt30a}). But there were also significant differences from the microscopic critical Fermi surface theory of Section~\ref{sec:cfs2}: ({\it i\/}) the self energy of the probe fermions had an exponent which varied with momentum across the Fermi surface; ({\it ii\/}) the size of the Fermi surface of the probe fermions was determined by the density of the probe fermions, and did not include the large density $N_{YM}^{2a} \mathscr{Q}_{YM}$ of the Yang-Mills theory itself. There is expected to be a separate Fermi surface of the latter background fermions upon including finite $N_{YM}$ corrections \cite{Faulkner:2012gt,Sachdev:2012tj}.
These features imply that the probe fermion black-brane strange metal is really a description of a {\it spectator\/} band of fermions \cite{SS10,Huijse:2011ef,Huijse:2011hp} scattering off a background which has a large density of low energy excitations, and the source of the breakdown of the quasiparticles does not arise from interactions between the putative quasiparticles themselves.

\subsubsection{Connection to the SYK model}
\label{sec:YMtoSYK}

The SYK model mapping of Section~\ref{sec:bh_fluc} appeared for a {\it spherical black hole\/} horizon of radius $R_h$, which at temperatures $T \ll 1/R_h$ mapped on to the SYK model at $T \ll U$. We can also place the supersymmetric Yang-Mills theory on a sphere of radius $R_{YM}$, and then this supersymmetric Yang-Mills theories is connected to the Schwarzian path integral in Eqs.~(\ref{syk34}) and (\ref{Seff1}), as we now discuss.

The Yang-Mills theory is characterized by 2 length scales, $R_{YM}$ and $[\mathscr{Q}_{YM}]^{-1/d}$, and the charged black hole solution of Sections~\ref{sec:bh_saddle} and \ref{sec:bh_fluc}, with a near-horizon AdS$_2 \times S^d$ geometry, provides a complete holographic description as $1/T$ is varied across these length scales. To make this correspondence precise, we have to relate $R_{YM}$ and $[\mathscr{Q}_{YM}]^{-1/d}$ to the length scales in the black hole solution, which are $R_h$, $L$, and $R_2$. The connection between the total charge and charge density in Eq.~(\ref{Qdensity}) immediately implies
\beqn
L = R_{YM}\,, \label{LRYM}
\eeqn
while the total charge of the black hole solution in Eq.~(\ref{Q1}) leads to 
\beqn
\mathscr{Q}_{YM} = \frac{R_h^{d-1} \sqrt{d \left[ (d+1)R_h^2  + (d-1)L^2  \right]}}{L^{2d} \bar{\kappa} \bar{g}_F} \,.
\label{Q1s}
\eeqn
The value of $R_2$ remains connected to $R_h$ and $L$ as in Eq.~(\ref{valR2}). 

Finally, we connect to the low energy Schwarzian approximation of the charged black hole. The charge density breaks the supersymmetry of the Yang-Mills theory, so we don't need to consider the super-Schwarzian theories that are needed for supersymmetric SYK models and supersymmetric black holes \cite{Fu:2016vas,Stanford:2017thb,Heydeman:2020hhw,Boruch:2022tno}.
If we are at low temperatures so that the thermal length is larger than charge length $T \ll [\mathscr{Q}_{YM}]^{1/d}$, and also so that fluctuations of non-constant horizon modes can be neglected $T \ll 1/R_h$, we can map these values $L$, $R_h$, and $R_2$ to obtain the dimensionless coupling constant $g_{\rm Sch}$ \cite{Stanford:2017thb} of the low energy Schwarzian theory from Eq.~(\ref{gammares2})
\bea
\frac{\pi}{g_{\rm Sch}^2} &=& \gamma T \nn
&=& \frac{4 \pi^2d s_d N_{YM}^{2a}}{\bar{\kappa}^2} \frac{R_2^2 R_h^{d-1} T}{L^d}\,. \label{gYM}
\eea
The ratio of length scales $R_2^2 R_h^{d-1} /L^d$ is to be determined as a function of the length scales $R_{YM}$ and $[\mathscr{Q}_{YM}]^{-1/d}$ by solving Eqs.~(\ref{LRYM}), (\ref{Q1s}) and (\ref{valR2}). Thus Eq.~(\ref{gYM}) is the main result of this subsection, determining the Schwarzian coupling $g_{\rm Sch}$ as a function of the parameteres of the Yang-Mills theory, which are the 
temperature $T$, the radius of the sphere $R_{YM}$, and the charge density $N_{YM}^{2a} \mathscr{Q}_{YM}$.
Note that the coupling $g$ becomes small in the limit of large $N_{YM}$.

Let us examine the value of $g_{\rm Sch}$ in the limiting regime when the size of the sphere of the Yang-Mills theory is much larger than the size set by the charge density, $R_{YM} \gg [\mathscr{Q}_{YM}]^{-1/d}$. Then we find $R_h \gg L$ with
\beq
R_h \sim R_{YM} [\mathscr{Q}_{YM} R_{YM}^d]^{1/d} , \quad R_2 \sim R_{YM}
\eeq
so that
\beqn
\frac{1}{g_{\rm Sch}^2} \sim N_{YM}^{2a} \left[ \mathscr{Q}_{YM} R_{YM}^d \right]^{(d-1)/d} R_{YM} T\,.
\eeqn
We observe that $g_{\rm Sch}^2 \sim \left[R_{YM} \right]^{-d}$ so the coupling becomes weak in the limit of a large sphere.
Of course, as always, we have to maintain $T \ll 1/R_h$ to apply the Schwarian theory, so the minimum possible value of the Schwarzian coupling is
\beqn
g_{\rm Sch,min}^2 \sim N_{YM}^{-2a} \left[ \mathscr{Q}_{YM} R_{YM}^d\right]^{(2-d)/d} \,.
\eeqn

\subsection{Out-of-time-order correlations}
\label{sec:otoc}

The connections to quantum gravity have also introduced a new diagnostic---the out-of-time-order correlator--- for detecting how quickly local perturbations become entangled with a macroscopic number of degrees of freedom in quantum many-body systems evolving under their own unitary dynamics. Out-of-time-order correlations (OTOCs) were studied a long time ago \cite{LO69} as an approach to diagnosing the semiclassical consequences of classical chaos in a quantum system. The modern incarnation of OTOCs appeared \cite{Shenker:2013pqa} in the study of shock waves in black holes \cite{Dray85}, where they were proposed as a signature of intrinsically quantum chaos in a strongly interacting many-body system. Shenker and Stanford argued that any strongly interacting quantum system, which is holographically dual to a black hole described by a theory containing Einstein gravity, has an OTOC of local operators $V$, $W$ which has an exponential growth at early times:
\beqn
\left\langle W(t) V(0) W(t) V(0) \right\rangle \sim \exp (\lambda_L t)\,,
\label{otoc1}
\eeqn
and the `Lyapunov growth rate exponent' is given by
\beqn
\lambda_L = 2 \pi T\,.
\eeqn
This value of $\lambda_L$ is a direct consequence of Einstein gravity, and the circumference of the Euclidean temporal circle being equal to $\hbar/k_B T$.
This exponential growth was argued to be related to a rapid loss of memory of the initial perturbations with time, a characteristic also expected from the absence of quasiparticle excitations. It was subsequently argued \cite{Maldacena2016}, without using any holographic connection, that the inequality $\lambda_L \leq 2 \pi T$ must apply to all strongly interacting quantum systems. The bound has also been shown to follow directly from the structure of generic operators that satisfy the eigenstate thermalization hypothesis \cite{murthy19}. A complementary bound has also been proposed on a closely related quantity that diagnoses operator growth \cite{parker19}. However, none of these statements suggest that generic quantum many body systems necessarily display an exponential growth of the OTOC.

The OTOC ideas have found a precise realization in the SYK model. For the model of Section~\ref{sec:SYK}, we define the OTOC by
\bea
&& {\rm OTOC} (t_1, t_3, ; t_2,  t_4) = \nonumber \\
&&~~~ \frac{1}{N^2} \sum_{i,j} \left\langle c_i^\dagger (t_1) c_j (t_3) c_i (t_2) c_j^\dagger (t_4) \right\rangle_{\rm conn.}
\eea
We examine the real time regime with $t_1 \approx t_2 \gg 1/T$, and $t_3 \approx t_4$, and define the `center of mass' time separation
\beq
t = \frac{1}{2}( t_1 + t_2 - t_3 - t_4 )
\eeq
The OTOC of the SYK model can be computed by generalizing the expression Eq.~(\ref{syk33}) for Schwarzian fluctuations corrections from 2 point to 4 point correlators. In imaginary time, we have the 4-point correlator
\bea
&& \mathcal{F} (\tau_1, \tau_3; \tau_2, \tau_4) = 
\Bigl\langle  [f'(\tau_1) f'(\tau_2)f'(\tau_3) f' (\tau_4) ]^{1/4}  \nonumber \\
&&~~ \times G_c (f(\tau_1) - f(\tau_2)) \, G_c (f(\tau_3)- f(\tau_4) \Bigr\rangle_{\overline{\mathcal{Z}}} \,, \label{otoc3}
\eea
where the average is over the Schwarzian path integral in Eq.~(\ref{syk32}) (we have omitted the unimportant fluctuations of $\phi$), and the conformal saddle-point Green's function $G_c (\tau)$  is given by Eq.~(\ref{syk17}). After careful analytic continuation of this correlator to real times, it was found that in the time range $1 \lesssim T t \ll \ln N$ there is an exponential growth of the OTOC \cite{kitaev_talk,Maldacena_syk,kitaevsuh}
\beq
{\rm OTOC} (t_1, t_3, ; t_2,  t_4)  \propto \frac{1}{N}\exp (2 \pi T t)
\eeq
So the chaos inequality \cite{Maldacena2016} is saturated by the SYK model, which, not surprisingly, has the same chaos growth rate as systems holographically dual to Einstein gravity.

The spatial structure associated with the OTOC is equally interesting and directly diagnoses operator growth. In Eq.~(\ref{otoc1}), if the operators are spatially separated, $W(t,\r),~V(0,0)$, the OTOC exhibits a ballistic wavefront associated with the growing operators as a function of $(t-|\r|/v_B)$. The `butterfly-velocity', $v_B$, is an intrinsic speed associated with the quantum many-body state and can, in principle, be parametrically smaller than the microscopic scales associated with the Hamiltonian \cite{BSDC17}. 

OTOCs have been studied in a variety of models, including the critical Fermi surface model of Section~\ref{sec:cfs2} \cite{Patel:2016wdy,Tikhanovskaya:2022zqq}, the lattice models related to those of Section~\ref{sec:lattice} \cite{Gu:2017ohj,Gu:2018jsv,Guo:2019csw}, disordered metals \cite{Patel:2017vfp}, and conformal field theories \cite{Stanford:2015owe,Chowdhury:2017jzb,Steinberg:2019uqb,Grozdanov:2018atb,Kim:2020jpz}, and all find a regime of exponential growth with a $\lambda_L$ that obeys the chaos bound, accompanied by a sharp ballistic wavefront. All of these settings involve a large$-N$ or a weak-coupling semiclassical limit. Direct numerical studies of realistic lattice models in one dimension \cite{Knap16,Luitz,BS20} have revealed a ballistic growth of operators but no indication of a well defined (i.e. position and velocity independent) Lyapunov exponent and a sharp front.

There have also been studies involving random unitary circuits with a finite dimensional local Hilbert space and no semi-classical limit, which observed a behavior of the OTOC that is qualitatively distinct from the above models  \cite{AN18,FP18b,Khemani:2018sdn,Xu:2018dfp}; importantly the growth is not identified by a well-defined $\lambda_L$ and the ballistic wavefront is not sharp. However, these models do not have a conserved energy and an associated notion of temperature, thereby making a direct comparison to the chaos bound far from being clear. A recent study \cite{Keselman:2020fmo} has demonstrated a way to access a regime of exponential growth of the OTOC even in random unitary circuits by effectively tuning $v_B\gg\lambda_L\times($microscopic length scale$)$, thereby presenting evidence that a finite Hilbert space can have an exponential growth of the OTOC.

The relevance of $\lambda_L$ and $v_B$ for measurable transport quantities has been scrutinized in a number of works. Bounds on transport quantities, such as the viscosity \cite{KSS} and charge diffusion coefficient \cite{Hartnoll2014}, have been suggested to hold for strongly interacting phases without quasiparticle excitations. Both of these bounds can be interpreted in terms of a bound on the diffusion coefficient, $D\sim \hbar v^2/k_BT$, where $v$ is a characteristic (but unknown) velocity scale in the problem. The statement of the bound was sharpened with the bold proposition \cite{Blake16} that the relevant velocity scale is set by $v=v_B$. While there are a number of holographic examples where these bounds have been shown to apply and even be saturated \cite{Gu17}, there are explicit counterexamples where the proposed bounds are violated \cite{Lucas1,Gu:2017ohj}. Stepping away from concrete models, a `hydrodynamic' understanding of some aspects of operator growth and chaos has also been developed in situations where the exponential regime exists \cite{Blake18}. 

In general, the relation between diffusive spreading of conserved charges and ballistic growth of non-conserved operators is complicated. For a class of generic random unitary circuits with conserved charges, it was shown that a spreading operator consists of a conserved part spreading diffusively, which acts as a source of nonconserved operators and leads to dissipation at a rate set by the local diffusion current \cite{Khemani_prx}. The nonconserved operators spread ballistically at a butterfly speed, becoming increasingly entangled with a macroscopic number of degrees of freedom in the system, acting as a dissipative `bath'. So in this random unitary circuit approach, the diffusion coefficient need not be related to any of the metrics associated with the OTOCs.

However a close relationship has been found between the OTOC $\lambda_L$ and the {\it thermal\/} diffusivity in computations for the critical Fermi surface \cite{Patel:2016wdy}, and in a wide class of holographic models \cite{Blake:2017qgd}. The relationship between the thermal diffusivity and $v_B^2/\lambda_L$ has also been analyzed in a family of strongly interacting bosonic variants of the SYK model \cite{EB21}, which are more closely related to the quantum spherical $p-$spin-glass model \cite{LC01}, inspired by the observation of Planckian diffusivities in a class of complex insulators \cite{AK19,SH20}. A simplified interpretation is that both quantum chaos and thermal diffusivity are related to loss of phase coherence. The time derivative of a local phase is the local energy density, and the fluctuation-dissipation theorem relates energy fluctuations to thermal transport. 

 The exact relation between OTOCs and universal aspects of transport remains unclear. Inspired by the universality of scattering rates across distinct materials displaying non-Fermi liquid properties, it has been conjectured  \cite{DCsyk} that there is an emergent length scale, $\ell\gg a (\equiv$ lattice spacing$)$, which is characterized by maximal chaos with a Lyapunov exponent $\lambda_L = 2\pi T$ at low temperature (i.e. either as $T \rightarrow 0$, or, $T>W^*$ but still small compared to microscopic energy scales) and effectively reaches local thermal equilibrium in a time of order $1/T$ \cite{QPT}. The universal coarse-grained description for the non-Fermi liquids can then possibly be built by coupling the islands of typical size $\ell$. This does not imply that the system is necessarily maximally chaotic at the scale of the system size.  In contrast, in a system with quasiparticles that does not display any non-Fermi liquid behavior, we expect that $\lambda_L \ll T$ as $T \rightarrow 0$. 

We end by noting that a different diagnostic of quantum chaos, which measures the correlations between energy levels and diagnoses the spectral `rigidity', is the spectral form factor (SFF). The SFF has been analyzed in the past in the context of mesoscopic physics and random matrix theory \cite{AS86}. The celebrated `ramp-plateau' form of the SFF beyond the Thouless time, signifies the onset of chaotic random matrix like behavior and has been analyzed for the SYK model using a variety of different methods \cite{Cotler:2016fpe,ShenkerRMT,Saad18,Garcia-Garcia:2017pzl,altland_npb,Winer20,sonner21,Liao:2020lac}.

\section{Outlook}
\label{sec:outlook}

Finding models of interacting electrons that can be solved reliably in the regime of strong interactions and at finite temperatures, without making uncontrolled approximations, remains a key challenge in quantum many-body physics. The family of models studied in this review offer a remarkably useful starting point for describing compressible metallic phases without any `Landau’ quasiparticles at strong interactions. Furthermore, they naturally lead to nFL regimes exhibiting electronic interaction induced $T-$linear resistivity and Planckian behavior over a wide range of energy scales and are accompanied by $(\omega/T)-$scaling. The theoretical results reviewed here are consistent with much of the universal experimental nFL phenomenology across numerous distinct microscopic materials. Therefore, it is natural to consider the possibility that a large class of strongly interacting microscopic models describing real materials {\it flow} (in a RG sense) to the different families of models considered in this article, over a significant intermediate energy range. Proving this remains an outstanding challenge. 

A notable recent result in the study of non-Fermi liquids is the phase diagram of Fig.~\ref{fig:tJU_phasediagram} \cite{Dumitrescu2021,Shackleton2021}. This presents the results of a numerical study of the doped random exchange $t$-$U$-$J$ Hubbard model. Many features of the phase diagram are reminiscent of the observations in the hole-doped cuprates, as we discussed in Section~\ref{sec:exprel}.  These include a doping induced transition from a disordered Fermi liquid satisfying Luttinger's theorem for $p>p_c$ to a low temperature metallic spin glass for $p<p_c$. At higher temperatures, the latter has a small carrier density and violates Luttinger's theorem. The quantum critical metal near $p=p_c$ exhibits a single-particle lifetime that has a Planckian form (see Fig.~\ref{fig:tJU_tau_star}) with an $O(1)$ coefficient; in the low-temperature limit the inferred resistivity is significantly below the MIR value. The quantum critical spin correlations are given by those of the SY spin liquid. 

At first sight, this concordance is remarkable and puzzling: the theory relies 
on a random exchange coupling with zero mean, which is far from the physical situation in the cuprates. We can take the concordance as an indication that AdS$_2$/SYK `local' criticality has a robustness, and can be present in models over a significant intermediate energy range. We note the renormalization group arguments \cite{PatelPlanck} that enhancement of resonant scattering can lead to the emergence of local SYK criticality. We also discussed holographic evidence of such a crossover \cite{Iqbal:2011in,Iqbal:2011ae,Liu:2020rrn} in disorder-free non-Fermi liquids of Fermi surfaces coupled to gauge fields in Section~\ref{sec:adscft}. See also other thoughts  \cite{Khveshchenko:2018nod,Khveshchenko:2022gcd} on the emergence of SYK local criticality.

The universality of the models studied is also 
encoded in their remarkable maximal many-body chaos, as diagnosed using the OTOC. Whether this aspect also indirectly controls the universality of Planckian transport scattering rates across distinct nFL materials is an important and non-trivial theoretical question.  A recent work has highlighted some of the fundamental differences between the growth of operators in maximally chaotic vs. non maximally chaotic quantum systems \cite{Blake21}, which could be of some relevance to understanding transport in nFL without quasiparticles.

For the disordered models considered in Section~\ref{sec:tJU}, the SY spin liquid behavior \cite{Joshi:2019csz} cannot extend down to $T=0$ because of the divergence of the spin glass susceptibility \cite{GPS1,GPS2} (although this instability is not visible over the accessible temperature range in the Planckian behavior in Fig.~\ref{fig:tJU_tau_star}). So we expect the eventual appearance of a metallic spin glass or a disordered Fermi liquid in which the zero temperature entropy is quenched, with the SY spin liquid surviving at $T=0$ only at the `spinodal' critical point 
where the Fermi liquid solution disappears. 
In more realistic models with weak disorder, we can expect the pseudogap to acquire the topological order of a fractionalized Fermi liquid, or have spin or charge density wave order. The critical theory asymptotically close to the pseudogap critical point will also be different: for the models of Section~\ref{sec:tJU}, we can expect a transition from a disordered Fermi liquid to a metallic spin glass, as described in theories without fractionalization \cite{SRO95,SG95}. Another possibility, present in the non-Fermi liquid large $M$ limit of Section~\ref{sec:JoshiM}, is that the entire overdoped regime is a critical metal with linear-$T$ resistivity \cite{Christos:toappear}, and then the critical point is also a deconfined theory. The connections between the transition outlined above and `deconfined' metallic criticality \cite{YZSS20,LZDC20} associated with abrupt Fermi surface changing transitions in clean systems, and in the absence of fractionalization in the adjacent phase, remains an interesting open problem.

A consequence of models with $J_{ij}$ having zero mean is that there is no superconductivity. Adding a non-zero mean $J_{ij}$, or other attractive interaction should lead to superconductivity \cite{JS,Patel,kamenevSC,YW,ChowdhuryBerg,Hauck20}, and 
a theory is needed for the onset of superconductivity from the Planckian metal phase of Figs.~\ref{fig:tJU_phasediagram} and \ref{fig:tJU_tau_star}. 

We discussed theories of non-Fermi liquids with critical Fermi surfaces in Section~\ref{sec:cfs}. Without disorder, such theories have zero resistivity in the absence of exponentially weak umklapp scattering, and so cannot produce linear-in-$T$ resistivity at low $T$. Adding potential scattering disorder to such a critical Fermi surface does produce a non-zero residual resistivity, but the temperature dependence of the resistivity is Fermi liquid-like, even though there is marginal Fermi liquid behavior in the fermion self energy \cite{Patel:2022gdh}. An interesting recent observation \cite{Patel:2022gdh} is that spatial disorder in the interaction strength does indeed produce a linear-$T$ resistivity (along with a $T \ln (1/T)$ specific heat). Two different types of disorder are therefore responsible for the residual resistivity and the slope of the linear-$T$ resistivity: the former arises from potential disorder, and the latter from interaction disorder. This feature has promise in explaining observations, and a better understanding is needed of the strengths of these disorders in the context of microscopic models.

An emerging application of the SYK model is to mesoscopic systems, and this has not been covered in our review. In this context, the behavior of the SYK model at finite $N$ is important, and we have to reverse the orders of limit of $N \rightarrow \infty$ (which we generally have taken first) and long time $t \rightarrow \infty$. The SYK model has a new emergent criticality for $t > N/U$, some aspects of which were covered in Section~\ref{sec:fluctuations}. Note that even for finite but large $N$, we do not immediately have a crossover to a regime where the discreteness of the energy spectrum is important; in a many-body system, the energy level spacing $\sim \exp (- \alpha N)$, and so even for $t > N/U$ we deal with an effectively continuous spectrum.
We refer the reader to the literature~\cite{Franz2018_review} 
for studies of applications to quantum dots 
and graphene flakes~\cite{Franz17,Beenakker18,micklitz19,Kruchkov:2019idx,Altland:2019czw,altland19,Kobrin:2020xms,Kiselev21,Khveshchenko19}, lattices of quantum dots \cite{altland19,Khveshchenko:2020rai}, Majorana fermions \cite{Alicea17,Franz2018}, 
ultracold atoms \cite{Danshita:2016xbo,wei_2021}, and quantum simulation \cite{Garcia-Alvarez:2016wem,Xu19}. The latter keep $N$ finite, and so differ from the models in Section~\ref{sec:lattice}, which take the $N \rightarrow \infty$ first. In Sections~\ref{sec:rqm}, \ref{sec:tJU}, and \ref{sec:KH}, our interest was in dynamical mean field theories of lattice systems in the thermodynamic limit, and so it was appropriate to take the $N \rightarrow \infty$ limit first.

\subsection*{Acknowledgements}

We are grateful to the many colleagues with whom we collaborated on projects 
related to the topic of this review, who generously shared their insights.
We thank Darshan Joshi for help with Figs.~\ref{fig:tJrg} and \ref{fig:kondorg}, and Grigory Tarnopolsky for help with Figs.~\ref{fig:specq2} and \ref{fig:specq4}. D.C. was supported by a faculty startup grant at Cornell University. S.S. was supported by the National Science Foundation under Grant No.~DMR-2002850 and by the U.S.~Department of Energy under Grant $\mbox{DE-SC0019030}$.  This work was also supported by the Simons Collaboration on Ultra-Quantum Matter, which is a grant from the Simons Foundation (651440, S.S.). 
The Flatiron Institute is a division of the Simons Foundation. 

\appendix

\section{Time reparametrization and gauge symmetries of the SYK model}
\label{sec:timerepar}

In this appendix, we will elaborate on the origin of Eq. (\ref{syk12}) from a more fundamental basis, and generalize it to the particle-hole asymmetric case. We return to the original equations Eq. (\ref{syk2a}) and Eq. (\ref{syk2b}), and simplify them in the low energy limit. As we saw in Eq. (\ref{syk10}), 
at frequencies $\ll U$, the $i \omega + \mu$ can be dropped, because $\mu - \Sigma(0) = 0$ and the $i \omega_n$ term is smaller than the singular frequency dependence in $\Sigma (i \omega_n)$. After Fourier transforming to the time domain, we can rewrite the original saddle-point equations as
\begin{subequations}
\begin{align}
& \int_0^\beta d\tau_2 \, \Sigma_{\rm sing} (\tau_1, \tau_2) G(\tau_2, \tau_3) = - \delta(\tau_1 - \tau_3) \label{syk14a} \\
&~~~~~~~~~~~\Sigma_{\rm sing} (\tau_1, \tau_2) = - U^2 G^2(\tau_1, \tau_2) G(\tau_2, \tau_1)\,, \label{syk14b}
\end{align}
\end{subequations}
where $\Sigma_{\rm sing}$ is the singular part of $\Sigma$. Also the saddle point Green's functions and self energies are functions only of time differences, like $\tau_1 - \tau_2$. Nevertheless, we have written them as a function of two independent times, because the fluctuations about the saddle point will involve the bilocal fields, as we will see. Moreover, the symmetries are more transparent in the bilocal formulation.

It is now not difficult to verify that Eq. (\ref{syk14a}) and Eq. (\ref{syk14b}) are invariant under the following transformation
\begin{subequations}
\begin{align}
\tau &= f (\sigma) \label{param} \\
G(\tau_1 , \tau_2) &= \left[ f' (\sigma_1) f' (\sigma_2) \right]^{-1/4} \frac{ g (\sigma_1)}{g (\sigma_2)} \, \widetilde{G} (\sigma_1, \sigma_2)  \\
{\Sigma} (\tau_1 , \tau_2) &= \left[ f' (\sigma_1) f' (\sigma_2) \right]^{-3/4} \frac{ g (\sigma_1)}{g (\sigma_2)} \, \widetilde{\Sigma} (\sigma_1, \sigma_2) \label{syk15}
\end{align}
\end{subequations}
where $f(\sigma)$ and $g(\sigma)$ are arbitrary functions. Here $f(\sigma)$ is a time reparametrization, and $g(\sigma)$ is a U(1) gauge transformation in imaginary time. These are emergent symmetries because the form of the equations obeyed by $\widetilde{G} (\sigma_1, \sigma_2)$ and $\widetilde{\Sigma}(\sigma_1, \sigma_2) $ is the same as Eq. (\ref{syk14a}) and Eq. (\ref{syk14b}) obeyed by $G(\tau_1, \tau_2)$, and $\Sigma(\tau_1, \tau_2)$. 

We obtain the non-zero temperature solution by choosing the time reparametrization in Eq. (\ref{param}) as the conformal map
\beqn
\tau = \frac{1}{\pi T} \tan ( \pi T \sigma )
\eeqn
where $\sigma$ is the periodic imaginary time co-ordinate with period $1/T$.
Applying this map to Eq.~(\ref{syk7}) we obtain
\beqn
G(\pm \sigma) = \mp C g(\pm \sigma) \sin (\pi/4 + \theta) \left( \frac{T}{\sin (\pi T \sigma)} \right)^{1/2} \,,
\eeqn
for $0 < \pm \sigma < 1/T$.
The function $g (\sigma)$ is so far undetermined apart from a normalization choice $g(0)=1$.
We can now determine $g (\sigma) $ by imposing the KMS condition
\beqn
G (\sigma + 1/T) = - G(\sigma) \label{KMS}
\eeqn
which implies
\beqn
g(\sigma) = \tan(\pi/4 + \theta) g(\sigma + 1/T).
\eeqn
The solution is clearly 
\beqn
g(\sigma) = e^{-2 \pi \mathcal{E} T \sigma} 
\eeqn
where the new parameter $\mathcal{E}$ and the angle $\theta$ are related as in Eq. (\ref{syk7b}).
This yields the final expression for $G(\sigma)$ in (\ref{syk17}). 

\section{Symmetries of the SYK saddle point}
\label{sec:symSYK}

We showed in Appendix~\ref{sec:timerepar} that the low energy limit of the saddle point equations in Eq. (\ref{syk14a}) and Eq. (\ref{syk14b}) have a very large set of symmetries, when expressed in terms of bilocal correlators of 2 times. However, the actual solution of the saddle point equations in Eq. (\ref{syk17}) is a function only of time differences. Here we ask a somewhat different question: what subgroup of the symmetries in Appendix~\ref{sec:timerepar} apply to the thermal solution in Eq. (\ref{syk17}). In other words, how are the emergent low energy time reparametrization and gauge symmetries broken by the low $T$ thermal state?

First, let us consider the simplest case with particle-hole symmetry at $T=0$, when we can schematically represent the large $N$ solutions in Section~\ref{sec:sykT0} as
\begin{eqnarray}
G_c (\tau_1 - \tau_2) &\sim& (\tau_1-\tau_2)^{-1/2} \nonumber \\ 
\Sigma_c (\tau_1 - \tau_2) &\sim& (\tau_1 - \tau_2)^{-3/2}. \nonumber
\end{eqnarray}
The saddle point will be invariant under a reparameterization $f(\tau)$ when
choosing $G(\tau_1, \tau_2) = G_c (\tau_1 - \tau_2)$ leads to a transformed
$\widetilde{G}(\sigma_1 , \sigma_2) = G_c (\sigma_1 - \sigma_2)$ (and similarly for $\Sigma$). It turns out this
is true only for the SL(2, R) transformations under which 
\begin{equation}
f(\tau) = \frac{a \tau + b}{c \tau + d} \quad, \quad ad -bc =1. \label{sl2r1}
\end{equation}
So the (approximate) reparametrization symmetry is spontaneously broken
down to SL(2, R) by the saddle point.

Now let us consider the most general case with $T>0$ and no particle-hole symmetry. We write Eq. (\ref{syk15}) as
\bea
&& G(\tau_1, \tau_2) = [f'(\tau_1) f'(\tau_2)]^{1/4} \nonumber \\
&&~~~~~~~~~~\times G_c (f(\tau_1) - f(\tau_2)) e^{i \phi (\tau_1) - i \phi (\tau_2)}\,, \label{syk19}
\eea
where $G_c (\tau)$ is the conformal saddle point solution given in Eq. (\ref{syk17}). 
Here, we have parameterized $g(\tau) = e^{-i \phi (\tau)}$ in terms of a phase field $\phi$; we will soon see that the derivative of $\phi$ is conjugate to density fluctuations.

It can now be checked that the $G(\tau_1, \tau_2)$ obtained from Eq. (\ref{syk19}) equals $G_c (\tau_1 - \tau_2)$ only if the transformations $f(\tau)$ and $\phi(\tau)$ satisfy
\begin{align}
\frac{\tan (\pi T f(\tau))}{\pi T}  &=  \frac{a \displaystyle \frac{\tan(\pi T \tau)}{\pi T} + b }{c \displaystyle \frac{\tan(\pi T \tau)}{\pi T} + d } \quad, \quad ad - bc =1, \nonumber \\
-i \phi (\tau) &= - i \phi_0 + 2 \pi \mathcal{E} T ( \tau - f(\tau)) \label{syk20}
\end{align}
The transformation of $f(\tau)$ looks rather mysterious, but we can simplify it as follows: we define
\beqn
z = e^{2 \pi i T \tau} \quad, \quad z_f = e^{2 \pi i T f(\tau)}
\eeqn
and then the transformation in Eq. (\ref{syk20}) is between unimodular complex numbers representing the thermal circle
\beqn
z_f = \frac{ w_1 \, z + w_2 }{w_2^\ast \, z + w_1^\ast} \quad, \quad |w_1|^2 - |w_2|^2 = 1, \label{syk21}
\eeqn
where $w_{1,2}$ are complex numbers. In this form, we have a SU(1,1) transformation, a group which is isomorphic to SL(2,R).

The symmetries in Eq.~(\ref{syk20}) and (\ref{syk21}) are crucial in determining the structure of the low energy action for fluctuations.

\section{Symmetries of AdS$_2$}
\label{sec:symAdS}

This Appendix notes that the AdS$_2$ metric
\beqn
ds^2 = \frac{d \tau^2 + d \zeta^2}{\zeta^2} \label{appads2}
\eeqn
is invariant under isometries which are SL(2,R) transformations, as in Eq.~(\ref{sl2r1}). It is easy to verify that the co-ordinate change
\beqn
\tau' + i \zeta' = \frac{a (\tau + i \zeta) + b}{c (\tau + i \zeta) + d}\,, \quad ad-bc =1\,,
\eeqn
with $a$,$b$,$c$,$d$ real, leaves the metric (\ref{appads2}) invariant.

\section{Schwarzian determinant}
\label{app:schwarzian}

This appendix will evaluate quadratic fluctuation correction to the free energy of the SYK model in Eq.~(\ref{syk44a}) arising from the time reparameterization mode in Eq.~(\ref{syk44b}). The formal expression for this correction is
\beqn
\mathcal{I} = \frac{1}{2} \sum_{n\neq 0, \pm 1} \ln \left[ 2 \pi^2 N \gamma T n^2 (n^2 - 1) \right]\,.
\label{appsch1}
\eeqn
This expression is clearly divergent, and we have to regulate it by finding the proper measure over the path integral of the $\epsilon_n$ in Eq.~(\ref{syk44b}) \cite{JMDS16b,Stanford:2017thb}.
For simplicity, we will only consider the particle-hole symmetric case $\mu=0$ in our discussion below, but the final result is more general.

We will regulate the divergence in Eq.~(\ref{appsch1}) by returning to the original $G$-$\Sigma$ path integral in Eq.~(\ref{syk31a}) to which the Schwarzian path integral in Eq.~(\ref{syk43}) is a low energy approximation. The saddle-point equations of Eq.~(\ref{syk31a}) are simply the original SYK equations Eqs.~(\ref{syk2a},\ref{syk2b}). Denoting the exact saddle point solution of the latter as $\overline{G}$ and $\overline{\Sigma}$, we can write the fluctuations as
\beqn
G = \overline{G} + \delta G\,, \quad \Sigma = \overline{\Sigma} + \delta \Sigma\,. \label{appsch2}
\eeqn
Then we expand the action in Eq.~(\ref{syk31a}) to quadratic order, and find that the needed eigenmodes of the quadratic fluctuations are eigenmodes of the kernels \cite{GKST,Tikhanovskaya:2020elb} which generalize that in Eq.~(\ref{syks10})
\begin{eqnarray}
&& K_{\textrm{A}/\textrm{S}}(\tau_{1},\tau_{2};\tau_{3},\tau_{4}) =  \label{appsch3} \\
&&~~ -\Big(\frac{q}{2}\pm \Big(\frac{q}{2}-1\Big)\Big)U^{2} \overline{G}(\tau_{13})\overline{G}(\tau_{24}) \overline{G}(\tau_{34})^{q-2}\,. \nonumber
\end{eqnarray}
We are considering the general case of SYK model with $q$ fermion terms, and $\tau_{ij} \equiv \tau_i - \tau_j$. The eigenmodes are defined by 
the equations (which generalize Eq.~(\ref{eq:DS3ptfa}))
\begin{eqnarray}
 \label{eq:DS3ptf}
   && k_{\textrm{A}/\textrm{S}} (h) v^{\textrm{A}/\textrm{S}}_{h}(\tau_{1},\tau_{2},\tau_{0})=  \\ 
   &&~~ \int d\tau_{3}d\tau_{4}K_{\textrm{A}/\textrm{S}}(\tau_{1},\tau_{2};\tau_{3},\tau_{4})v^{\textrm{A}/\textrm{S}}_{h}(\tau_{3},\tau_{4},\tau_{0})\,, \nonumber
  \end{eqnarray}
with dimensionless eigenvalue $k_{\textrm{A}/\textrm{S}}(h)$. For $k_{\textrm{A}/\textrm{S}}(h)=1$ we obtain the scaling dimension $h$ of composite operators associated with the fermion bilinears in the conformal limit theory.
Our overall task is to expand $\delta G$ and $\delta \Sigma$ is terms of the eigenmodes of $K_{A/S}$, each of which will also be eigenmodes of the quadratic fluctuation of the action in Eq.~(\ref{syk31a}).

The Schwarzian fluctuation focuses on a specific eigenmode, $v^A_2$, which is associated with time reparameterization symmetry. The infinitesimal version of the time reparameterization in Eq.~(\ref{syk19}), using Eq.~(\ref{syk36a}), is
\bea
\delta G (\tau_1, \tau_2) &=& \left[ \Delta \epsilon' (\tau_1) + \Delta \epsilon' (\tau_2) \right. \label{appsch4} \\
&~&~~\left. +\epsilon (\tau_1) \partial_{\tau_1} + \epsilon(\tau_2) \partial_{\tau_2} \right] \overline{G} (\tau_1 - \tau_2)\,. \nonumber
\eea
For the conformal limit result, $\overline{G} = G_c$ in Eq.~(\ref{syk17}), and also for conformal Green's functions in Eq.~(\ref{appsch3}), $\delta G$ in Eq.~(\ref{appsch4}) is indeed an eigenmode $v^A_2$ of Eq.~(\ref{eq:DS3ptf}) with $k_A(2)=1$, as can be verified by explicit evaluation. 

We now have all the ingredients necessary to expand the time reparameterization eigenmode of $K_A$ in terms of the eigenmodes $\epsilon_n$ in Eq.~(\ref{syk44b}). One needed technical step is that we multiply the $K_A$ eigenmode by $G_c^{(q-2)/2}$ to make the kernel in Eq.~(\ref{appsch3}) a Hermitian operator. Then we write
\bea
&& [G_c (\tau_1, \tau_2)]^{(q-2)/2} \delta G (\tau_1, \tau_2) = \nn 
&&~~~~~~\sum_n \epsilon_n \,f_n \left(2 \pi T [\tau_1 - \tau_2 ]\right) e^{-i \pi n T (\tau_1 + \tau_2)}\,. \label{appsch6}
\eea
We can easily obtain the explicit form of the co-efficients $f_n (\theta)$ in this expansion by using Eqs.~(\ref{syk17}), (\ref{syk44b}) and (\ref{appsch4}):
\beqn
f_n (\theta) =  \frac{\sin(n \theta/2)\cos(\theta/2)}{\sin^2(\theta/2)} - n \frac{\cos(n\theta/2)}{\sin(\theta/2)}\,.
\label{appsch7}
\eeqn
Recall we are working at $\mu=\mathcal{E}=0$, and we have dropped an unimportant $n$-independent prefactor in Eq.~(\ref{appsch7}). The functions $f_n (\theta)$ are analogs for SL(2,R) of the Legendre polynomials for SO(3). As expected, they vanish identically for $n=0,\pm 1$ because $G_c$ is invariant under SL(2,R) transformations. The property we need here is the $n$-dependence of their normalization
\beqn
\int_0^{2 \pi} \frac{d \theta}{2 \pi} [f_n (\theta)]^2 = \frac{|n| (n^2 - 1)}{3}\,. \label{appsch8}
\eeqn

Using the eigenmodes of Eq.~(\ref{appsch3}), the Gaussian fluctuation contribution to the free energy from the $G$-$\Sigma$ path integral in Eq.~(\ref{syk31a}) can be written schematically as \cite{JMDS16b}
\beqn
\mathcal{I}_{G{\rm -}\Sigma} = \frac{1}{2} \sum \ln \left( \frac{1}{k_{A/S}(h)} - 1 \right)\,. \label{appsch9}
\eeqn
We now compare this $G$-$\Sigma$ form of the fluctuation contribution, with the $\epsilon_n$ fluctuation contribution in Eq.~(\ref{appsch1}). Given the transformation between the eigenmodes in Eq.~(\ref{appsch6}), and the normalization in Eq.~(\ref{appsch8}), we conclude that the $n^2 (n^2-1)$ factor in Eq.~(\ref{appsch1}) should be identified with the product of a $|n|(n^2-1)$ factor from Eq.~(\ref{appsch8}), and a $(1 - k_A (2)) \sim T|n|/U$ factor. The deviation of $k_A(2)$ from unity arises from conformal corrections to the saddle point $\overline{G}-G_c$, and arguments have been given \cite{JMDS16b} for their $|n|$ dependence. 

With this corrected measure for $\epsilon_n$ fluctuations, we conclude  that the properly regulated form of Eq.~(\ref{appsch1}) is that deduced from Eq.~(\ref{appsch9}) \cite{JMDS16b}
\beqn
\widetilde{\mathcal{I}} = \frac{1}{2} \sum_{|n|\neq 0, \pm 1}^{|n| < c_1 U/T} \ln \left( \frac{T |n|}{U} \right)\,,
\label{appsch10}
\eeqn
where we have dropped $T$-independent constants, and $c_1$ is a non-universal number determining the high energy cutoff. 
We can now apply the $\zeta$ function theory result
\beqn
\sum_{n=1}^{m} \ln \left(a n \right) =  m\ln (a m) - m + \frac{\ln(2 \pi m)}{2} + \mathcal{O}\left(\frac{1}{m}\right)
\label{appsch11}
\eeqn
to Eq.~(\ref{appsch10}), and obtain Eq.~(\ref{syk45}). Note the $3/2$ co-efficient of the logarithm in Eq.~(\ref{syk45}) is independent of $c_1$; it is the sum of the $1/2$ co-efficient in Eq.~(\ref{appsch11}), and the omitted $n=\pm 1$ contributions in Eq.~(\ref{appsch10}).

\section{Generalization to SYK$_q$ model}
\label{app:SYKq}
Much of our discussion of the SYK model has focused on the physically motivated problem with four-fermion interactions. However, the model can be readily generalized to $q\geq 4$ fermion interactions \cite{Gross:2016kjj}, otherwise referred to as the SYK$_q$ model. We review here the low-energy properties of a local SYK$_q$ model and the effect of perturbing it by a quadratic (hopping) term. The interaction Hamiltonian for electrons occupying orbitals labeled $i_\ell = 1,..., N$ is given by,
\beq
H_q &=& \frac{\left(q/2\right)!}{N^{\frac{q-1}{2}}}\sum_{\{i_\ell\}}U_{i_1i_2...i_q} \bigg[c^\dagger_{i_1}c^\dagger_{i_2}...c^\dagger_{i_{q/2}}c{\vphantom \dagger}_{i_{q/2+1}}...c{\vphantom \dagger}_{i_{q-1}}c_{i_q} \bigg] \nonumber \\
&& -\mu \sum_{i_\ell} c^\dagger_{i_\ell}c{\vphantom \dagger}_{i_\ell},  \label{sykq_1}
\eeq
where as before we choose the couplings $U_{i_1i_2...i_q}$ to be independent random variables with  $\overline{U_{i_1i_2...i_q}}= 0$, and $\overline{(U_{i_1i_2...i_q})^2}=U^2$. The density, $\mathcal{Q}$, can be tuned by an external chemical potential, $\mu$.

In the large $N$ limit, once again only the melon graphs survive, but the number of internal legs is now $(q-1)$. The on-site Green's function reduces to the solution of the equations
\begin{subequations}
\begin{align}
G(i\omega_n) & = \frac{1}{i \omega_n + \mu - \Sigma (i\omega_n)} \label{sykq_2a} \\ 
\Sigma (\tau) &= -  U^2 [G(\tau)]^{q/2} [G(-\tau)]^{q/2-1} \label{sykq_2b} \\
G(\tau = 0^-) & = \mathcal{Q}. \label{sykq_2c}
\end{align}
\end{subequations}
Following the analysis in Sec.~\ref{sec:sykT0}, we can obtain the low energy solution at $T=0$ for the electron Green's function. Importantly, the power-law singularity at low frequencies is now determined by the dimension, $\Delta=1/q$, such that the Green's function has the form,
\begin{subequations}
\begin{align}
G(\tau) &\sim \frac{\tn{sgn}(\tau)}{(U|\tau|)^{2/q}}, ~~~ |\tau|\gg 1/U \\
G(i\omega) &\sim \frac{i\tn{sgn}(\omega)}{U^{2/q} |\omega|^{1-2/q}}, ~~~ |\omega|\ll U.
\end{align}
\end{subequations}
For the sake of simplicity, we chose the density to be at half-filling where the spectral asymmetry vanishes. In spite of the different scaling dimension, the finite compressibility and residual entropy (including the $T-$linear correction) have the same qualitative behavior as the model with $q=4$. Generalizations to two-band models involving distinct $q-$body interactions have also been studied \cite{shenoy2}.

Let us now consider a lattice generalization of the model as in Sec. \ref{sec:lattice}, where the local interaction at every site is given by $H_q$, and the sites are coupled together via uniform translationally invariant hopping terms, $H_{\tn{kin}}$ (see Eq. \ref{lattice1a}). The hopping term is a relevant perturbation and the gapless scale invariant solutions can not survive down to the lowest energies. Starting from the decoupled limit, one finds that the coherence scale is given by
\beq
W_q^* \sim t \bigg(\frac{t}{U}\bigg)^{\frac{2}{(q-2)}},
\eeq
below which the hopping terms can no longer be treated perturbatively and the ground state is a Fermi liquid. In spite of the similarities in the thermodynamic properties with the $q=4$ model, charge transport is dramatically different for $T\gg W_q^*$. The electrical resistivity in the incoherent regime is now given by,
\beq
\rho_{\tn{dc}}\sim \frac{h}{Ne^2} \bigg(\frac{T}{W_q^*}\bigg)^{2-4/q}.
\eeq
Interestingly, for $q\neq 4$ the resistivity scales faster than $T$ (but slower than $T^2$) with increasing temperature. Importantly, the $T-$linearity of the resistivity is tied to the electron scaling dimension of $\Delta=1/4$ for $q=4$. 

\bibliography{refs}

%apsrmp4-2.bst 2018-12-27 (MD) hand-edited version of apsrmp4-1.bst
%Control: key (0)
%Control: author (3) reversed first dotless
%Control: editor formatted (0) differently from author
%Control: production of article title (0) allowed
%Control: page (1) range
%Control: year (0) verbatim
%Control: production of eprint (0) enabled
\begin{thebibliography}{459}%
\makeatletter
\providecommand \@ifxundefined [1]{%
 \@ifx{#1\undefined}
}%
\providecommand \@ifnum [1]{%
 \ifnum #1\expandafter \@firstoftwo
 \else \expandafter \@secondoftwo
 \fi
}%
\providecommand \@ifx [1]{%
 \ifx #1\expandafter \@firstoftwo
 \else \expandafter \@secondoftwo
 \fi
}%
\providecommand \natexlab [1]{#1}%
\providecommand \enquote  [1]{``#1''}%
\providecommand \bibnamefont  [1]{#1}%
\providecommand \bibfnamefont [1]{#1}%
\providecommand \citenamefont [1]{#1}%
\providecommand \href@noop [0]{\@secondoftwo}%
\providecommand \href [0]{\begingroup \@sanitize@url \@href}%
\providecommand \@href[1]{\@@startlink{#1}\@@href}%
\providecommand \@@href[1]{\endgroup#1\@@endlink}%
\providecommand \@sanitize@url [0]{\catcode `\\12\catcode `\$12\catcode
  `\&12\catcode `\#12\catcode `\^12\catcode `\_12\catcode `\%12\relax}%
\providecommand \@@startlink[1]{}%
\providecommand \@@endlink[0]{}%
\providecommand \url  [0]{\begingroup\@sanitize@url \@url }%
\providecommand \@url [1]{\endgroup\@href {#1}{\urlprefix }}%
\providecommand \urlprefix  [0]{URL }%
\providecommand \Eprint [0]{\href }%
\providecommand \doibase [0]{https://doi.org/}%
\providecommand \selectlanguage [0]{\@gobble}%
\providecommand \bibinfo  [0]{\@secondoftwo}%
\providecommand \bibfield  [0]{\@secondoftwo}%
\providecommand \translation [1]{[#1]}%
\providecommand \BibitemOpen [0]{}%
\providecommand \bibitemStop [0]{}%
\providecommand \bibitemNoStop [0]{.\EOS\space}%
\providecommand \EOS [0]{\spacefactor3000\relax}%
\providecommand \BibitemShut  [1]{\csname bibitem#1\endcsname}%
\let\auto@bib@innerbib\@empty
%</preamble>
\bibitem [{\citenamefont {{Abanov}}\ and\ \citenamefont
  {{Chubukov}}(2020)}]{Chubukov20a}%
  \BibitemOpen
  \bibfield  {author} {\bibinfo {author} {\bibnamefont {{Abanov}},
  \bibfnamefont {Artem}}, and\ \bibinfo {author} {\bibfnamefont {Andrey~V.}\
  \bibnamefont {{Chubukov}}}} (\bibinfo {year} {2020}),\ \bibfield  {title}
  {\enquote {\bibinfo {title} {{Interplay between superconductivity and
  non-Fermi liquid at a quantum critical point in a metal. I. The $\gamma$
  model and its phase diagram at $T =0$ : The case $0 <\gamma <1$}},}\ }\href
  {https://doi.org/10.1103/PhysRevB.102.024524} {\bibfield  {journal} {\bibinfo
   {journal} {Phys. Rev. B}\ }\textbf {\bibinfo {volume} {102}}~(\bibinfo
  {number} {2}),\ \bibinfo {eid} {024524}},\ \Eprint
  {https://arxiv.org/abs/2004.13220} {arXiv:2004.13220 [cond-mat.str-el]}
  \BibitemShut {NoStop}%
\bibitem [{\citenamefont {Abrahams}\ \emph {et~al.}(1981)\citenamefont
  {Abrahams}, \citenamefont {Anderson}, \citenamefont {Lee},\ and\
  \citenamefont {Ramakrishnan}}]{Abrahams81}%
  \BibitemOpen
  \bibfield  {author} {\bibinfo {author} {\bibnamefont {Abrahams},
  \bibfnamefont {Elihu}}, \bibinfo {author} {\bibfnamefont {P.~W.}\
  \bibnamefont {Anderson}}, \bibinfo {author} {\bibfnamefont {P.~A.}\
  \bibnamefont {Lee}}, and\ \bibinfo {author} {\bibfnamefont {T.~V.}\
  \bibnamefont {Ramakrishnan}}} (\bibinfo {year} {1981}),\ \bibfield  {title}
  {\enquote {\bibinfo {title} {Quasiparticle lifetime in disordered
  two-dimensional metals},}\ }\href {https://doi.org/10.1103/PhysRevB.24.6783}
  {\bibfield  {journal} {\bibinfo  {journal} {Phys. Rev. B}\ }\textbf {\bibinfo
  {volume} {24}},\ \bibinfo {pages} {6783--6789}}\BibitemShut {NoStop}%
\bibitem [{\citenamefont {Abrikosov}\ \emph {et~al.}(1963)\citenamefont
  {Abrikosov}, \citenamefont {Gor’kov},\ and\ \citenamefont
  {Dzyaloshinskii}}]{AGD}%
  \BibitemOpen
  \bibfield  {author} {\bibinfo {author} {\bibnamefont {Abrikosov},
  \bibfnamefont {AA}}, \bibinfo {author} {\bibfnamefont {LP}~\bibnamefont
  {Gor’kov}}, and\ \bibinfo {author} {\bibfnamefont {IE}~\bibnamefont
  {Dzyaloshinskii}}} (\bibinfo {year} {1963}),\ \href@noop {} {\emph {\bibinfo
  {title} {Methods of Quantum Field Theory in Statistical Physics}}}\ (\bibinfo
   {publisher} {Prentice-Hall, Englewood Cliffs, NJ})\BibitemShut {NoStop}%
\bibitem [{\citenamefont {Afkhami-Jeddi}\ \emph {et~al.}(2020)\citenamefont
  {Afkhami-Jeddi}, \citenamefont {Cohn}, \citenamefont {Hartman},\ and\
  \citenamefont {Tajdini}}]{Afkhami-Jeddi:2020ezh}%
  \BibitemOpen
  \bibfield  {author} {\bibinfo {author} {\bibnamefont {Afkhami-Jeddi},
  \bibfnamefont {Nima}}, \bibinfo {author} {\bibfnamefont {Henry}\ \bibnamefont
  {Cohn}}, \bibinfo {author} {\bibfnamefont {Thomas}\ \bibnamefont {Hartman}},
  and\ \bibinfo {author} {\bibfnamefont {Amirhossein}\ \bibnamefont {Tajdini}}}
  (\bibinfo {year} {2020}),\ \bibfield  {title} {\enquote {\bibinfo {title}
  {{Free partition functions and an averaged holographic duality}},}\
  }\href@noop {} {\ }\Eprint {https://arxiv.org/abs/2006.04839}
  {arXiv:2006.04839 [hep-th]} \BibitemShut {NoStop}%
\bibitem [{\citenamefont {Aharony}\ \emph {et~al.}(2008)\citenamefont
  {Aharony}, \citenamefont {Bergman}, \citenamefont {Jafferis},\ and\
  \citenamefont {Maldacena}}]{Aharony:2008ug}%
  \BibitemOpen
  \bibfield  {author} {\bibinfo {author} {\bibnamefont {Aharony}, \bibfnamefont
  {Ofer}}, \bibinfo {author} {\bibfnamefont {Oren}\ \bibnamefont {Bergman}},
  \bibinfo {author} {\bibfnamefont {Daniel~Louis}\ \bibnamefont {Jafferis}},
  and\ \bibinfo {author} {\bibfnamefont {Juan}\ \bibnamefont {Maldacena}}}
  (\bibinfo {year} {2008}),\ \bibfield  {title} {\enquote {\bibinfo {title}
  {{$N=6$ superconformal Chern-Simons-matter theories, M2-branes and their
  gravity duals}},}\ }\href {https://doi.org/10.1088/1126-6708/2008/10/091}
  {\bibfield  {journal} {\bibinfo  {journal} {JHEP}\ }\textbf {\bibinfo
  {volume} {10}},\ \bibinfo {pages} {091}},\ \Eprint
  {https://arxiv.org/abs/0806.1218} {arXiv:0806.1218 [hep-th]} \BibitemShut
  {NoStop}%
\bibitem [{\citenamefont {{Aldape}}\ \emph {et~al.}(2020)\citenamefont
  {{Aldape}}, \citenamefont {{Cookmeyer}}, \citenamefont {{Patel}},\ and\
  \citenamefont {{Altman}}}]{Adalpe20}%
  \BibitemOpen
  \bibfield  {author} {\bibinfo {author} {\bibnamefont {{Aldape}},
  \bibfnamefont {Erik~E}}, \bibinfo {author} {\bibfnamefont {Taylor}\
  \bibnamefont {{Cookmeyer}}}, \bibinfo {author} {\bibfnamefont {Aavishkar~A.}\
  \bibnamefont {{Patel}}}, and\ \bibinfo {author} {\bibfnamefont {Ehud}\
  \bibnamefont {{Altman}}}} (\bibinfo {year} {2020}),\ \bibfield  {title}
  {\enquote {\bibinfo {title} {{Solvable Theory of a Strange Metal at the
  Breakdown of a Heavy Fermi Liquid}},}\ }\href@noop {} {\ }\Eprint
  {https://arxiv.org/abs/2012.00763} {arXiv:2012.00763 [cond-mat.str-el]}
  \BibitemShut {NoStop}%
\bibitem [{\citenamefont {Alhassid}(2000)}]{alhassid}%
  \BibitemOpen
  \bibfield  {author} {\bibinfo {author} {\bibnamefont {Alhassid},
  \bibfnamefont {Y}}} (\bibinfo {year} {2000}),\ \bibfield  {title} {\enquote
  {\bibinfo {title} {The statistical theory of quantum dots},}\ }\href
  {https://doi.org/10.1103/RevModPhys.72.895} {\bibfield  {journal} {\bibinfo
  {journal} {Rev. Mod. Phys.}\ }\textbf {\bibinfo {volume} {72}},\ \bibinfo
  {pages} {895--968}}\BibitemShut {NoStop}%
\bibitem [{\citenamefont {Allen}\ \emph {et~al.}(1996)\citenamefont {Allen},
  \citenamefont {Berger}, \citenamefont {Chauvet}, \citenamefont {Forro},
  \citenamefont {Jarlborg}, \citenamefont {Junod}, \citenamefont {Revaz},\ and\
  \citenamefont {Santi}}]{allen}%
  \BibitemOpen
  \bibfield  {author} {\bibinfo {author} {\bibnamefont {Allen}, \bibfnamefont
  {P~B}}, \bibinfo {author} {\bibfnamefont {H.}~\bibnamefont {Berger}},
  \bibinfo {author} {\bibfnamefont {O.}~\bibnamefont {Chauvet}}, \bibinfo
  {author} {\bibfnamefont {L.}~\bibnamefont {Forro}}, \bibinfo {author}
  {\bibfnamefont {T.}~\bibnamefont {Jarlborg}}, \bibinfo {author}
  {\bibfnamefont {A.}~\bibnamefont {Junod}}, \bibinfo {author} {\bibfnamefont
  {B.}~\bibnamefont {Revaz}}, and\ \bibinfo {author} {\bibfnamefont
  {G.}~\bibnamefont {Santi}}} (\bibinfo {year} {1996}),\ \bibfield  {title}
  {\enquote {\bibinfo {title} {{Transport properties, thermodynamic properties,
  and electronic structure of SrRuO$_3$}},}\ }\href
  {https://doi.org/10.1103/PhysRevB.53.4393} {\bibfield  {journal} {\bibinfo
  {journal} {Phys. Rev. B}\ }\textbf {\bibinfo {volume} {53}},\ \bibinfo
  {pages} {4393--4398}}\BibitemShut {NoStop}%
\bibitem [{\citenamefont {Almheiri}\ \emph {et~al.}(2020)\citenamefont
  {Almheiri}, \citenamefont {Hartman}, \citenamefont {Maldacena}, \citenamefont
  {Shaghoulian},\ and\ \citenamefont {Tajdini}}]{Almheiri:2019qdq}%
  \BibitemOpen
  \bibfield  {author} {\bibinfo {author} {\bibnamefont {Almheiri},
  \bibfnamefont {Ahmed}}, \bibinfo {author} {\bibfnamefont {Thomas}\
  \bibnamefont {Hartman}}, \bibinfo {author} {\bibfnamefont {Juan}\
  \bibnamefont {Maldacena}}, \bibinfo {author} {\bibfnamefont {Edgar}\
  \bibnamefont {Shaghoulian}}, and\ \bibinfo {author} {\bibfnamefont
  {Amirhossein}\ \bibnamefont {Tajdini}}} (\bibinfo {year} {2020}),\ \bibfield
  {title} {\enquote {\bibinfo {title} {{Replica Wormholes and the Entropy of
  Hawking Radiation}},}\ }\href {https://doi.org/10.1007/JHEP05(2020)013}
  {\bibfield  {journal} {\bibinfo  {journal} {JHEP}\ }\textbf {\bibinfo
  {volume} {05}},\ \bibinfo {pages} {013}},\ \Eprint
  {https://arxiv.org/abs/1911.12333} {arXiv:1911.12333 [hep-th]} \BibitemShut
  {NoStop}%
\bibitem [{\citenamefont {Almheiri}\ \emph {et~al.}(2021)\citenamefont
  {Almheiri}, \citenamefont {Hartman}, \citenamefont {Maldacena}, \citenamefont
  {Shaghoulian},\ and\ \citenamefont {Tajdini}}]{Almheiri:2020cfm}%
  \BibitemOpen
  \bibfield  {author} {\bibinfo {author} {\bibnamefont {Almheiri},
  \bibfnamefont {Ahmed}}, \bibinfo {author} {\bibfnamefont {Thomas}\
  \bibnamefont {Hartman}}, \bibinfo {author} {\bibfnamefont {Juan}\
  \bibnamefont {Maldacena}}, \bibinfo {author} {\bibfnamefont {Edgar}\
  \bibnamefont {Shaghoulian}}, and\ \bibinfo {author} {\bibfnamefont
  {Amirhossein}\ \bibnamefont {Tajdini}}} (\bibinfo {year} {2021}),\ \bibfield
  {title} {\enquote {\bibinfo {title} {{The entropy of Hawking radiation}},}\
  }\href {https://doi.org/10.1103/RevModPhys.93.035002} {\bibfield  {journal}
  {\bibinfo  {journal} {Rev. Mod. Phys.}\ }\textbf {\bibinfo {volume}
  {93}}~(\bibinfo {number} {3}),\ \bibinfo {pages} {035002}},\ \Eprint
  {https://arxiv.org/abs/2006.06872} {arXiv:2006.06872 [hep-th]} \BibitemShut
  {NoStop}%
\bibitem [{\citenamefont {Almheiri}\ \emph {et~al.}(2019)\citenamefont
  {Almheiri}, \citenamefont {Milekhin},\ and\ \citenamefont
  {Swingle}}]{Almheiri:2019jqq}%
  \BibitemOpen
  \bibfield  {author} {\bibinfo {author} {\bibnamefont {Almheiri},
  \bibfnamefont {Ahmed}}, \bibinfo {author} {\bibfnamefont {Alexey}\
  \bibnamefont {Milekhin}}, and\ \bibinfo {author} {\bibfnamefont {Brian}\
  \bibnamefont {Swingle}}} (\bibinfo {year} {2019}),\ \bibfield  {title}
  {\enquote {\bibinfo {title} {{Universal Constraints on Energy Flow and SYK
  Thermalization}},}\ }\href@noop {} {\ }\Eprint
  {https://arxiv.org/abs/1912.04912} {arXiv:1912.04912 [hep-th]} \BibitemShut
  {NoStop}%
\bibitem [{\citenamefont {Almheiri}\ and\ \citenamefont
  {Polchinski}(2015)}]{Almheiri15}%
  \BibitemOpen
  \bibfield  {author} {\bibinfo {author} {\bibnamefont {Almheiri},
  \bibfnamefont {Ahmed}}, and\ \bibinfo {author} {\bibfnamefont {Joseph}\
  \bibnamefont {Polchinski}}} (\bibinfo {year} {2015}),\ \bibfield  {title}
  {\enquote {\bibinfo {title} {{Models of AdS$_{2}$ backreaction and
  holography}},}\ }\href {https://doi.org/10.1007/JHEP11(2015)014} {\bibfield
  {journal} {\bibinfo  {journal} {JHEP}\ }\textbf {\bibinfo {volume} {11}},\
  \bibinfo {pages} {014}},\ \Eprint {https://arxiv.org/abs/1402.6334}
  {arXiv:1402.6334 [hep-th]} \BibitemShut {NoStop}%
\bibitem [{\citenamefont {Altland}\ and\ \citenamefont
  {Bagrets}(2018)}]{altland_npb}%
  \BibitemOpen
  \bibfield  {author} {\bibinfo {author} {\bibnamefont {Altland}, \bibfnamefont
  {Alexander}}, and\ \bibinfo {author} {\bibfnamefont {Dmitry}\ \bibnamefont
  {Bagrets}}} (\bibinfo {year} {2018}),\ \bibfield  {title} {\enquote {\bibinfo
  {title} {{Quantum ergodicity in the SYK model}},}\ }\href
  {https://doi.org/https://doi.org/10.1016/j.nuclphysb.2018.02.015} {\bibfield
  {journal} {\bibinfo  {journal} {Nuclear Physics B}\ }\textbf {\bibinfo
  {volume} {930}},\ \bibinfo {pages} {45--68}}\BibitemShut {NoStop}%
\bibitem [{\citenamefont {Altland}\ \emph
  {et~al.}(2019{\natexlab{a}})\citenamefont {Altland}, \citenamefont
  {Bagrets},\ and\ \citenamefont {Kamenev}}]{altland19}%
  \BibitemOpen
  \bibfield  {author} {\bibinfo {author} {\bibnamefont {Altland}, \bibfnamefont
  {Alexander}}, \bibinfo {author} {\bibfnamefont {Dmitry}\ \bibnamefont
  {Bagrets}}, and\ \bibinfo {author} {\bibfnamefont {Alex}\ \bibnamefont
  {Kamenev}}} (\bibinfo {year} {2019}{\natexlab{a}}),\ \bibfield  {title}
  {\enquote {\bibinfo {title} {{Quantum Criticality of Granular
  Sachdev-Ye-Kitaev Matter}},}\ }\href
  {https://doi.org/10.1103/PhysRevLett.123.106601} {\bibfield  {journal}
  {\bibinfo  {journal} {Phys. Rev. Lett.}\ }\textbf {\bibinfo {volume} {123}},\
  \bibinfo {pages} {106601}}\BibitemShut {NoStop}%
\bibitem [{\citenamefont {Altland}\ \emph
  {et~al.}(2019{\natexlab{b}})\citenamefont {Altland}, \citenamefont
  {Bagrets},\ and\ \citenamefont {Kamenev}}]{Altland:2019czw}%
  \BibitemOpen
  \bibfield  {author} {\bibinfo {author} {\bibnamefont {Altland}, \bibfnamefont
  {Alexander}}, \bibinfo {author} {\bibfnamefont {Dmitry}\ \bibnamefont
  {Bagrets}}, and\ \bibinfo {author} {\bibfnamefont {Alex}\ \bibnamefont
  {Kamenev}}} (\bibinfo {year} {2019}{\natexlab{b}}),\ \bibfield  {title}
  {\enquote {\bibinfo {title} {{Sachdev-Ye-Kitaev Non-Fermi-Liquid Correlations
  in Nanoscopic Quantum Transport}},}\ }\href
  {https://doi.org/10.1103/PhysRevLett.123.226801} {\bibfield  {journal}
  {\bibinfo  {journal} {Phys. Rev. Lett.}\ }\textbf {\bibinfo {volume}
  {123}}~(\bibinfo {number} {22}),\ \bibinfo {pages} {226801}},\ \Eprint
  {https://arxiv.org/abs/1908.11351} {arXiv:1908.11351 [cond-mat.str-el]}
  \BibitemShut {NoStop}%
\bibitem [{\citenamefont {Altland}\ \emph {et~al.}(2021)\citenamefont
  {Altland}, \citenamefont {Bagrets}, \citenamefont {Nayak}, \citenamefont
  {Sonner},\ and\ \citenamefont {Vielma}}]{sonner21}%
  \BibitemOpen
  \bibfield  {author} {\bibinfo {author} {\bibnamefont {Altland}, \bibfnamefont
  {Alexander}}, \bibinfo {author} {\bibfnamefont {Dmitry}\ \bibnamefont
  {Bagrets}}, \bibinfo {author} {\bibfnamefont {Pranjal}\ \bibnamefont
  {Nayak}}, \bibinfo {author} {\bibfnamefont {Julian}\ \bibnamefont {Sonner}},
  and\ \bibinfo {author} {\bibfnamefont {Manuel}\ \bibnamefont {Vielma}}}
  (\bibinfo {year} {2021}),\ \bibfield  {title} {\enquote {\bibinfo {title}
  {{From operator statistics to wormholes}},}\ }\href
  {https://doi.org/10.1103/PhysRevResearch.3.033259} {\bibfield  {journal}
  {\bibinfo  {journal} {Phys. Rev. Res.}\ }\textbf {\bibinfo {volume}
  {3}}~(\bibinfo {number} {3}),\ \bibinfo {pages} {033259}},\ \Eprint
  {https://arxiv.org/abs/2105.12129} {arXiv:2105.12129 [hep-th]} \BibitemShut
  {NoStop}%
\bibitem [{\citenamefont {Altshuler}\ \emph {et~al.}(1994)\citenamefont
  {Altshuler}, \citenamefont {Ioffe},\ and\ \citenamefont {Millis}}]{AIM}%
  \BibitemOpen
  \bibfield  {author} {\bibinfo {author} {\bibnamefont {Altshuler},
  \bibfnamefont {B~L}}, \bibinfo {author} {\bibfnamefont {L.~B.}\ \bibnamefont
  {Ioffe}}, and\ \bibinfo {author} {\bibfnamefont {A.~J.}\ \bibnamefont
  {Millis}}} (\bibinfo {year} {1994}),\ \bibfield  {title} {\enquote {\bibinfo
  {title} {Low-energy properties of fermions with singular interactions},}\
  }\href {https://doi.org/10.1103/PhysRevB.50.14048} {\bibfield  {journal}
  {\bibinfo  {journal} {Phys. Rev. B}\ }\textbf {\bibinfo {volume} {50}},\
  \bibinfo {pages} {14048--14064}}\BibitemShut {NoStop}%
\bibitem [{\citenamefont {Altshuler}\ and\ \citenamefont
  {Shklovskii}(1986)}]{AS86}%
  \BibitemOpen
  \bibfield  {author} {\bibinfo {author} {\bibnamefont {Altshuler},
  \bibfnamefont {BL}}, and\ \bibinfo {author} {\bibfnamefont {BI}~\bibnamefont
  {Shklovskii}}} (\bibinfo {year} {1986}),\ \bibfield  {title} {\enquote
  {\bibinfo {title} {Repulsion of energy levels and conductivity of small metal
  samples},}\ }\href@noop {} {\bibfield  {journal} {\bibinfo  {journal} {Sov.
  Phys. JETP}\ }\textbf {\bibinfo {volume} {64}}~(\bibinfo {number} {1}),\
  \bibinfo {pages} {127--135}}\BibitemShut {NoStop}%
\bibitem [{\citenamefont {Anantharaman}\ and\ \citenamefont
  {Macia}(2011)}]{anantharaman2011}%
  \BibitemOpen
  \bibfield  {author} {\bibinfo {author} {\bibnamefont {Anantharaman},
  \bibfnamefont {Nalini}}, and\ \bibinfo {author} {\bibfnamefont {Fabricio}\
  \bibnamefont {Macia}}} (\bibinfo {year} {2011}),\ \href@noop {} {\enquote
  {\bibinfo {title} {The dynamics of the schr\"odinger flow from the point of
  view of semiclassical measures},}\ }\Eprint {https://arxiv.org/abs/1102.0970}
  {arXiv:1102.0970 [math.AP]} \BibitemShut {NoStop}%
\bibitem [{\citenamefont {Anderson}\ \emph {et~al.}(2019)\citenamefont
  {Anderson}, \citenamefont {Wang}, \citenamefont {Xu}, \citenamefont {Venu},
  \citenamefont {Trotzky}, \citenamefont {Chevy},\ and\ \citenamefont
  {Thywissen}}]{Anderson2019}%
  \BibitemOpen
  \bibfield  {author} {\bibinfo {author} {\bibnamefont {Anderson},
  \bibfnamefont {Rhys}}, \bibinfo {author} {\bibfnamefont {Fudong}\
  \bibnamefont {Wang}}, \bibinfo {author} {\bibfnamefont {Peihang}\
  \bibnamefont {Xu}}, \bibinfo {author} {\bibfnamefont {Vijin}\ \bibnamefont
  {Venu}}, \bibinfo {author} {\bibfnamefont {Stefan}\ \bibnamefont {Trotzky}},
  \bibinfo {author} {\bibfnamefont {Fr\'ed\'eric}\ \bibnamefont {Chevy}}, and\
  \bibinfo {author} {\bibfnamefont {Joseph~H.}\ \bibnamefont {Thywissen}}}
  (\bibinfo {year} {2019}),\ \bibfield  {title} {\enquote {\bibinfo {title}
  {Conductivity spectrum of ultracold atoms in an optical lattice},}\ }\href
  {https://doi.org/10.1103/PhysRevLett.122.153602} {\bibfield  {journal}
  {\bibinfo  {journal} {Phys. Rev. Lett.}\ }\textbf {\bibinfo {volume} {122}},\
  \bibinfo {pages} {153602}}\BibitemShut {NoStop}%
\bibitem [{\citenamefont {Andrei}\ and\ \citenamefont
  {Coleman}(1989)}]{AndreiColeman}%
  \BibitemOpen
  \bibfield  {author} {\bibinfo {author} {\bibnamefont {Andrei}, \bibfnamefont
  {N}}, and\ \bibinfo {author} {\bibfnamefont {P.}~\bibnamefont {Coleman}}}
  (\bibinfo {year} {1989}),\ \bibfield  {title} {\enquote {\bibinfo {title}
  {Cooper instability in the presence of a spin liquid},}\ }\href
  {https://doi.org/10.1103/PhysRevLett.62.595} {\bibfield  {journal} {\bibinfo
  {journal} {Phys. Rev. Lett.}\ }\textbf {\bibinfo {volume} {62}},\ \bibinfo
  {pages} {595--598}}\BibitemShut {NoStop}%
\bibitem [{\citenamefont {Aronson}\ \emph {et~al.}(1995)\citenamefont
  {Aronson}, \citenamefont {Osborn}, \citenamefont {Robinson}, \citenamefont
  {Lynn}, \citenamefont {Chau}, \citenamefont {Seaman},\ and\ \citenamefont
  {Maple}}]{Aronson95}%
  \BibitemOpen
  \bibfield  {author} {\bibinfo {author} {\bibnamefont {Aronson}, \bibfnamefont
  {M~C}}, \bibinfo {author} {\bibfnamefont {R.}~\bibnamefont {Osborn}},
  \bibinfo {author} {\bibfnamefont {R.~A.}\ \bibnamefont {Robinson}}, \bibinfo
  {author} {\bibfnamefont {J.~W.}\ \bibnamefont {Lynn}}, \bibinfo {author}
  {\bibfnamefont {R.}~\bibnamefont {Chau}}, \bibinfo {author} {\bibfnamefont
  {C.~L.}\ \bibnamefont {Seaman}}, and\ \bibinfo {author} {\bibfnamefont
  {M.~B.}\ \bibnamefont {Maple}}} (\bibinfo {year} {1995}),\ \bibfield  {title}
  {\enquote {\bibinfo {title} {{Non-Fermi-Liquid Scaling of the Magnetic
  Response in UCu$_{5-x}$Pd$_x$ ($x=1,1.5$)}},}\ }\href
  {https://doi.org/10.1103/PhysRevLett.75.725} {\bibfield  {journal} {\bibinfo
  {journal} {Phys. Rev. Lett.}\ }\textbf {\bibinfo {volume} {75}},\ \bibinfo
  {pages} {725--728}}\BibitemShut {NoStop}%
\bibitem [{\citenamefont {Arrachea}\ and\ \citenamefont
  {Rozenberg}(2002)}]{ArracheaRozenbergSG2002}%
  \BibitemOpen
  \bibfield  {author} {\bibinfo {author} {\bibnamefont {Arrachea},
  \bibfnamefont {Liliana}}, and\ \bibinfo {author} {\bibfnamefont {Marcelo~J.}\
  \bibnamefont {Rozenberg}}} (\bibinfo {year} {2002}),\ \bibfield  {title}
  {\enquote {\bibinfo {title} {{Infinite-range quantum random Heisenberg
  magnet}},}\ }\href {https://doi.org/10.1103/PhysRevB.65.224430} {\bibfield
  {journal} {\bibinfo  {journal} {Phys. Rev. B}\ }\textbf {\bibinfo {volume}
  {65}},\ \bibinfo {pages} {224430}}\BibitemShut {NoStop}%
\bibitem [{\citenamefont {Azeyanagi}\ \emph {et~al.}(2018)\citenamefont
  {Azeyanagi}, \citenamefont {Ferrari},\ and\ \citenamefont
  {Schaposnik~Massolo}}]{Azeyanagi:2017drg}%
  \BibitemOpen
  \bibfield  {author} {\bibinfo {author} {\bibnamefont {Azeyanagi},
  \bibfnamefont {Tatsuo}}, \bibinfo {author} {\bibfnamefont {Frank}\
  \bibnamefont {Ferrari}}, and\ \bibinfo {author} {\bibfnamefont {Fidel~I.}\
  \bibnamefont {Schaposnik~Massolo}}} (\bibinfo {year} {2018}),\ \bibfield
  {title} {\enquote {\bibinfo {title} {{Phase Diagram of Planar Matrix Quantum
  Mechanics, Tensor, and Sachdev-Ye-Kitaev Models}},}\ }\href
  {https://doi.org/10.1103/PhysRevLett.120.061602} {\bibfield  {journal}
  {\bibinfo  {journal} {Phys. Rev. Lett.}\ }\textbf {\bibinfo {volume}
  {120}}~(\bibinfo {number} {6}),\ \bibinfo {pages} {061602}},\ \Eprint
  {https://arxiv.org/abs/1707.03431} {arXiv:1707.03431 [hep-th]} \BibitemShut
  {NoStop}%
\bibitem [{\citenamefont {Bagrets}\ \emph {et~al.}(2016)\citenamefont
  {Bagrets}, \citenamefont {Altland},\ and\ \citenamefont
  {Kamenev}}]{Bagrets:2016cdf}%
  \BibitemOpen
  \bibfield  {author} {\bibinfo {author} {\bibnamefont {Bagrets}, \bibfnamefont
  {Dmitry}}, \bibinfo {author} {\bibfnamefont {Alexander}\ \bibnamefont
  {Altland}}, and\ \bibinfo {author} {\bibfnamefont {Alex}\ \bibnamefont
  {Kamenev}}} (\bibinfo {year} {2016}),\ \bibfield  {title} {\enquote {\bibinfo
  {title} {{Sachdev\textendash{}Ye\textendash{}Kitaev model as Liouville
  quantum mechanics}},}\ }\href
  {https://doi.org/10.1016/j.nuclphysb.2016.08.002} {\bibfield  {journal}
  {\bibinfo  {journal} {Nucl. Phys. B}\ }\textbf {\bibinfo {volume} {911}},\
  \bibinfo {pages} {191--205}},\ \Eprint {https://arxiv.org/abs/1607.00694}
  {arXiv:1607.00694 [cond-mat.str-el]} \BibitemShut {NoStop}%
\bibitem [{\citenamefont {Bagrets}\ \emph {et~al.}(2017)\citenamefont
  {Bagrets}, \citenamefont {Altland},\ and\ \citenamefont
  {Kamenev}}]{Bagrets:2017pwq}%
  \BibitemOpen
  \bibfield  {author} {\bibinfo {author} {\bibnamefont {Bagrets}, \bibfnamefont
  {Dmitry}}, \bibinfo {author} {\bibfnamefont {Alexander}\ \bibnamefont
  {Altland}}, and\ \bibinfo {author} {\bibfnamefont {Alex}\ \bibnamefont
  {Kamenev}}} (\bibinfo {year} {2017}),\ \bibfield  {title} {\enquote {\bibinfo
  {title} {{Power-law out of time order correlation functions in the SYK
  model}},}\ }\href {https://doi.org/10.1016/j.nuclphysb.2017.06.012}
  {\bibfield  {journal} {\bibinfo  {journal} {Nucl. Phys. B}\ }\textbf
  {\bibinfo {volume} {921}},\ \bibinfo {pages} {727--752}},\ \Eprint
  {https://arxiv.org/abs/1702.08902} {arXiv:1702.08902 [cond-mat.str-el]}
  \BibitemShut {NoStop}%
\bibitem [{\citenamefont {Bandyopadhyay}\ \emph {et~al.}(2021)\citenamefont
  {Bandyopadhyay}, \citenamefont {Uhrich}, \citenamefont {Paviglianiti},\ and\
  \citenamefont {Hauke}}]{Bandyopadhyay:2021wpy}%
  \BibitemOpen
  \bibfield  {author} {\bibinfo {author} {\bibnamefont {Bandyopadhyay},
  \bibfnamefont {Soumik}}, \bibinfo {author} {\bibfnamefont {Philipp}\
  \bibnamefont {Uhrich}}, \bibinfo {author} {\bibfnamefont {Alessio}\
  \bibnamefont {Paviglianiti}}, and\ \bibinfo {author} {\bibfnamefont
  {Philipp}\ \bibnamefont {Hauke}}} (\bibinfo {year} {2021}),\ \bibfield
  {title} {\enquote {\bibinfo {title} {{Universal equilibration dynamics of the
  Sachdev-Ye-Kitaev model}},}\ }\href@noop {} {\ }\Eprint
  {https://arxiv.org/abs/2108.01718} {arXiv:2108.01718 [cond-mat.str-el]}
  \BibitemShut {NoStop}%
\bibitem [{\citenamefont {{Baraduc}}\ \emph {et~al.}(1996)\citenamefont
  {{Baraduc}}, \citenamefont {{Azrak}},\ and\ \citenamefont
  {{Bontemps}}}]{Baraduc_1996}%
  \BibitemOpen
  \bibfield  {author} {\bibinfo {author} {\bibnamefont {{Baraduc}},
  \bibfnamefont {Claire}}, \bibinfo {author} {\bibfnamefont {Abdellatif}\
  \bibnamefont {{Azrak}}}, and\ \bibinfo {author} {\bibfnamefont {Nicole}\
  \bibnamefont {{Bontemps}}}} (\bibinfo {year} {1996}),\ \bibfield  {title}
  {\enquote {\bibinfo {title} {{Infrared conductivity in the normal state of
  cuprate thin films}},}\ }\href {https://doi.org/10.1007/BF00728415}
  {\bibfield  {journal} {\bibinfo  {journal} {Journal of Superconductivity}\
  }\textbf {\bibinfo {volume} {9}}~(\bibinfo {number} {1}),\ \bibinfo {pages}
  {3--6}}\BibitemShut {NoStop}%
\bibitem [{\citenamefont {Basov}\ \emph {et~al.}(2011)\citenamefont {Basov},
  \citenamefont {Averitt}, \citenamefont {van~der Marel}, \citenamefont
  {Dressel},\ and\ \citenamefont {Haule}}]{Basov_RMP}%
  \BibitemOpen
  \bibfield  {author} {\bibinfo {author} {\bibnamefont {Basov}, \bibfnamefont
  {D~N}}, \bibinfo {author} {\bibfnamefont {Richard~D.}\ \bibnamefont
  {Averitt}}, \bibinfo {author} {\bibfnamefont {Dirk}\ \bibnamefont {van~der
  Marel}}, \bibinfo {author} {\bibfnamefont {Martin}\ \bibnamefont {Dressel}},
  and\ \bibinfo {author} {\bibfnamefont {Kristjan}\ \bibnamefont {Haule}}}
  (\bibinfo {year} {2011}),\ \bibfield  {title} {\enquote {\bibinfo {title}
  {Electrodynamics of correlated electron materials},}\ }\href
  {https://doi.org/10.1103/RevModPhys.83.471} {\bibfield  {journal} {\bibinfo
  {journal} {Rev. Mod. Phys.}\ }\textbf {\bibinfo {volume} {83}},\ \bibinfo
  {pages} {471--541}}\BibitemShut {NoStop}%
\bibitem [{\citenamefont {Beccaria}\ \emph {et~al.}(2022)\citenamefont
  {Beccaria}, \citenamefont {Giombi},\ and\ \citenamefont
  {Tseytlin}}]{Beccaria:2022bcr}%
  \BibitemOpen
  \bibfield  {author} {\bibinfo {author} {\bibnamefont {Beccaria},
  \bibfnamefont {Matteo}}, \bibinfo {author} {\bibfnamefont {Simone}\
  \bibnamefont {Giombi}}, and\ \bibinfo {author} {\bibfnamefont {Arkady~A}\
  \bibnamefont {Tseytlin}}} (\bibinfo {year} {2022}),\ \bibfield  {title}
  {\enquote {\bibinfo {title} {{Wilson loop in general representation and RG
  flow in 1d defect QFT}},}\ }\href@noop {} {\ }\Eprint
  {https://arxiv.org/abs/2202.00028} {arXiv:2202.00028 [hep-th]} \BibitemShut
  {NoStop}%
\bibitem [{\citenamefont {Ben-Zion}\ and\ \citenamefont
  {McGreevy}(2018)}]{mcgreevy}%
  \BibitemOpen
  \bibfield  {author} {\bibinfo {author} {\bibnamefont {Ben-Zion},
  \bibfnamefont {Daniel}}, and\ \bibinfo {author} {\bibfnamefont {John}\
  \bibnamefont {McGreevy}}} (\bibinfo {year} {2018}),\ \bibfield  {title}
  {\enquote {\bibinfo {title} {Strange metal from local quantum chaos},}\
  }\href {https://doi.org/10.1103/PhysRevB.97.155117} {\bibfield  {journal}
  {\bibinfo  {journal} {Phys. Rev. B}\ }\textbf {\bibinfo {volume} {97}},\
  \bibinfo {pages} {155117}}\BibitemShut {NoStop}%
\bibitem [{\citenamefont {Berg}\ \emph {et~al.}(2019)\citenamefont {Berg},
  \citenamefont {Lederer}, \citenamefont {Schattner},\ and\ \citenamefont
  {Trebst}}]{BergAR19}%
  \BibitemOpen
  \bibfield  {author} {\bibinfo {author} {\bibnamefont {Berg}, \bibfnamefont
  {Erez}}, \bibinfo {author} {\bibfnamefont {Samuel}\ \bibnamefont {Lederer}},
  \bibinfo {author} {\bibfnamefont {Yoni}\ \bibnamefont {Schattner}}, and\
  \bibinfo {author} {\bibfnamefont {Simon}\ \bibnamefont {Trebst}}} (\bibinfo
  {year} {2019}),\ \bibfield  {title} {\enquote {\bibinfo {title} {{Monte Carlo
  Studies of Quantum Critical Metals}},}\ }\href
  {https://doi.org/10.1146/annurev-conmatphys-031218-013339} {\bibfield
  {journal} {\bibinfo  {journal} {Annual Review of Condensed Matter Physics}\
  }\textbf {\bibinfo {volume} {10}}~(\bibinfo {number} {1}),\ \bibinfo {pages}
  {63--84}}\BibitemShut {NoStop}%
\bibitem [{\citenamefont {Bhattacharya}\ \emph {et~al.}(2019)\citenamefont
  {Bhattacharya}, \citenamefont {Jatkar},\ and\ \citenamefont
  {Sorokhaibam}}]{Jatkar}%
  \BibitemOpen
  \bibfield  {author} {\bibinfo {author} {\bibnamefont {Bhattacharya},
  \bibfnamefont {Ritabrata}}, \bibinfo {author} {\bibfnamefont {Dileep~P.}\
  \bibnamefont {Jatkar}}, and\ \bibinfo {author} {\bibfnamefont {Nilakash}\
  \bibnamefont {Sorokhaibam}}} (\bibinfo {year} {2019}),\ \bibfield  {title}
  {\enquote {\bibinfo {title} {{Quantum quenches and thermalization in SYK
  models}},}\ }\href {https://doi.org/10.1007/JHEP07(2019)066} {\bibfield
  {journal} {\bibinfo  {journal} {Journal of High Energy Physics}\ }\textbf
  {\bibinfo {volume} {2019}}~(\bibinfo {number} {7}),\ \bibinfo {pages}
  {66}}\BibitemShut {NoStop}%
\bibitem [{\citenamefont {Bi}\ \emph {et~al.}(2017)\citenamefont {Bi},
  \citenamefont {Jian}, \citenamefont {You}, \citenamefont {Pawlak},\ and\
  \citenamefont {Xu}}]{Xu17}%
  \BibitemOpen
  \bibfield  {author} {\bibinfo {author} {\bibnamefont {Bi}, \bibfnamefont
  {Zhen}}, \bibinfo {author} {\bibfnamefont {Chao-Ming}\ \bibnamefont {Jian}},
  \bibinfo {author} {\bibfnamefont {Yi-Zhuang}\ \bibnamefont {You}}, \bibinfo
  {author} {\bibfnamefont {Kelly~Ann}\ \bibnamefont {Pawlak}}, and\ \bibinfo
  {author} {\bibfnamefont {Cenke}\ \bibnamefont {Xu}}} (\bibinfo {year}
  {2017}),\ \bibfield  {title} {\enquote {\bibinfo {title} {{Instability of the
  non-Fermi-liquid state of the Sachdev-Ye-Kitaev model}},}\ }\href
  {https://doi.org/10.1103/PhysRevB.95.205105} {\bibfield  {journal} {\bibinfo
  {journal} {Phys. Rev. B}\ }\textbf {\bibinfo {volume} {95}},\ \bibinfo
  {pages} {205105}}\BibitemShut {NoStop}%
\bibitem [{\citenamefont {Biroli}\ and\ \citenamefont
  {Parcollet}(2002)}]{Biroli_OP2002}%
  \BibitemOpen
  \bibfield  {author} {\bibinfo {author} {\bibnamefont {Biroli}, \bibfnamefont
  {Giulio}}, and\ \bibinfo {author} {\bibfnamefont {Olivier}\ \bibnamefont
  {Parcollet}}} (\bibinfo {year} {2002}),\ \bibfield  {title} {\enquote
  {\bibinfo {title} {{Out-of-equilibrium dynamics of a quantum Heisenberg spin
  glass}},}\ }\href {https://doi.org/10.1103/PhysRevB.65.094414} {\bibfield
  {journal} {\bibinfo  {journal} {Phys. Rev. B}\ }\textbf {\bibinfo {volume}
  {65}},\ \bibinfo {pages} {094414}}\BibitemShut {NoStop}%
\bibitem [{\citenamefont {Blake}(2016)}]{Blake16}%
  \BibitemOpen
  \bibfield  {author} {\bibinfo {author} {\bibnamefont {Blake}, \bibfnamefont
  {Mike}}} (\bibinfo {year} {2016}),\ \bibfield  {title} {\enquote {\bibinfo
  {title} {{Universal Charge Diffusion and the Butterfly Effect in Holographic
  Theories}},}\ }\href {https://doi.org/10.1103/PhysRevLett.117.091601}
  {\bibfield  {journal} {\bibinfo  {journal} {Phys. Rev. Lett.}\ }\textbf
  {\bibinfo {volume} {117}},\ \bibinfo {pages} {091601}}\BibitemShut {NoStop}%
\bibitem [{\citenamefont {Blake}\ \emph {et~al.}(2017)\citenamefont {Blake},
  \citenamefont {Davison},\ and\ \citenamefont {Sachdev}}]{Blake:2017qgd}%
  \BibitemOpen
  \bibfield  {author} {\bibinfo {author} {\bibnamefont {Blake}, \bibfnamefont
  {Mike}}, \bibinfo {author} {\bibfnamefont {Richard~A.}\ \bibnamefont
  {Davison}}, and\ \bibinfo {author} {\bibfnamefont {Subir}\ \bibnamefont
  {Sachdev}}} (\bibinfo {year} {2017}),\ \bibfield  {title} {\enquote {\bibinfo
  {title} {{Thermal diffusivity and chaos in metals without quasiparticles}},}\
  }\href {https://doi.org/10.1103/PhysRevD.96.106008} {\bibfield  {journal}
  {\bibinfo  {journal} {Phys. Rev. D}\ }\textbf {\bibinfo {volume}
  {96}}~(\bibinfo {number} {10}),\ \bibinfo {pages} {106008}},\ \Eprint
  {https://arxiv.org/abs/1705.07896} {arXiv:1705.07896 [hep-th]} \BibitemShut
  {NoStop}%
\bibitem [{\citenamefont {Blake}\ \emph {et~al.}(2018)\citenamefont {Blake},
  \citenamefont {Lee},\ and\ \citenamefont {Liu}}]{Blake18}%
  \BibitemOpen
  \bibfield  {author} {\bibinfo {author} {\bibnamefont {Blake}, \bibfnamefont
  {Mike}}, \bibinfo {author} {\bibfnamefont {Hyunseok}\ \bibnamefont {Lee}},
  and\ \bibinfo {author} {\bibfnamefont {Hong}\ \bibnamefont {Liu}}} (\bibinfo
  {year} {2018}),\ \bibfield  {title} {\enquote {\bibinfo {title} {A quantum
  hydrodynamical description for scrambling and many-body chaos},}\ }\href
  {https://doi.org/10.1007/JHEP10(2018)127} {\bibfield  {journal} {\bibinfo
  {journal} {Journal of High Energy Physics}\ }\textbf {\bibinfo {volume}
  {2018}}~(\bibinfo {number} {10}),\ \bibinfo {pages} {127}}\BibitemShut
  {NoStop}%
\bibitem [{\citenamefont {{Blake}}\ and\ \citenamefont
  {{Liu}}(2021)}]{Blake21}%
  \BibitemOpen
  \bibfield  {author} {\bibinfo {author} {\bibnamefont {{Blake}}, \bibfnamefont
  {Mike}}, and\ \bibinfo {author} {\bibfnamefont {Hong}\ \bibnamefont {{Liu}}}}
  (\bibinfo {year} {2021}),\ \bibfield  {title} {\enquote {\bibinfo {title}
  {{On systems of maximal quantum chaos}},}\ }\href
  {https://doi.org/10.1007/JHEP05(2021)229} {\bibfield  {journal} {\bibinfo
  {journal} {Journal of High Energy Physics}\ }\textbf {\bibinfo {volume}
  {2021}}~(\bibinfo {number} {5}),\ \bibinfo {eid} {229}},\ \Eprint
  {https://arxiv.org/abs/2102.11294} {arXiv:2102.11294 [hep-th]} \BibitemShut
  {NoStop}%
\bibitem [{\citenamefont {Bloch}(1929)}]{Bloch29}%
  \BibitemOpen
  \bibfield  {author} {\bibinfo {author} {\bibnamefont {Bloch}, \bibfnamefont
  {Felix}}} (\bibinfo {year} {1929}),\ \bibfield  {title} {\enquote {\bibinfo
  {title} {{\"U}ber die quantenmechanik der elektronen in kristallgittern},}\
  }\href {https://doi.org/10.1007/BF01339455} {\bibfield  {journal} {\bibinfo
  {journal} {Zeitschrift f{\"u}r Physik}\ }\textbf {\bibinfo {volume}
  {52}}~(\bibinfo {number} {7}),\ \bibinfo {pages} {555--600}}\BibitemShut
  {NoStop}%
\bibitem [{\citenamefont {Bohigas}\ and\ \citenamefont
  {Flores}(1971)}]{Bohigas71}%
  \BibitemOpen
  \bibfield  {author} {\bibinfo {author} {\bibnamefont {Bohigas}, \bibfnamefont
  {O}}, and\ \bibinfo {author} {\bibfnamefont {J.}~\bibnamefont {Flores}}}
  (\bibinfo {year} {1971}),\ \bibfield  {title} {\enquote {\bibinfo {title}
  {{Two-body random Hamiltonian and level density}},}\ }\href
  {https://doi.org/https://doi.org/10.1016/0370-2693(71)90598-3} {\bibfield
  {journal} {\bibinfo  {journal} {Physics Letters B}\ }\textbf {\bibinfo
  {volume} {34}}~(\bibinfo {number} {4}),\ \bibinfo {pages}
  {261--263}}\BibitemShut {NoStop}%
\bibitem [{\citenamefont {Bohigas}\ \emph {et~al.}(1984)\citenamefont
  {Bohigas}, \citenamefont {Giannoni},\ and\ \citenamefont {Schmit}}]{BGS}%
  \BibitemOpen
  \bibfield  {author} {\bibinfo {author} {\bibnamefont {Bohigas}, \bibfnamefont
  {O}}, \bibinfo {author} {\bibfnamefont {M.~J.}\ \bibnamefont {Giannoni}},
  and\ \bibinfo {author} {\bibfnamefont {C.}~\bibnamefont {Schmit}}} (\bibinfo
  {year} {1984}),\ \bibfield  {title} {\enquote {\bibinfo {title}
  {Characterization of chaotic quantum spectra and universality of level
  fluctuation laws},}\ }\href {https://doi.org/10.1103/PhysRevLett.52.1}
  {\bibfield  {journal} {\bibinfo  {journal} {Phys. Rev. Lett.}\ }\textbf
  {\bibinfo {volume} {52}},\ \bibinfo {pages} {1--4}}\BibitemShut {NoStop}%
\bibitem [{\citenamefont {Bohrdt}\ \emph {et~al.}(2017)\citenamefont {Bohrdt},
  \citenamefont {Mendl}, \citenamefont {Endres},\ and\ \citenamefont
  {Knap}}]{Knap16}%
  \BibitemOpen
  \bibfield  {author} {\bibinfo {author} {\bibnamefont {Bohrdt}, \bibfnamefont
  {A}}, \bibinfo {author} {\bibfnamefont {C~B}\ \bibnamefont {Mendl}}, \bibinfo
  {author} {\bibfnamefont {M}~\bibnamefont {Endres}}, and\ \bibinfo {author}
  {\bibfnamefont {M}~\bibnamefont {Knap}}} (\bibinfo {year} {2017}),\ \bibfield
   {title} {\enquote {\bibinfo {title} {Scrambling and thermalization in a
  diffusive quantum many-body system},}\ }\href
  {http://stacks.iop.org/1367-2630/19/i=6/a=063001} {\bibfield  {journal}
  {\bibinfo  {journal} {New Journal of Physics}\ }\textbf {\bibinfo {volume}
  {19}}~(\bibinfo {number} {6}),\ \bibinfo {pages} {063001}}\BibitemShut
  {NoStop}%
\bibitem [{\citenamefont {Boltzmann}(1872)}]{boltzmann}%
  \BibitemOpen
  \bibfield  {author} {\bibinfo {author} {\bibnamefont {Boltzmann},
  \bibfnamefont {Ludwig}}} (\bibinfo {year} {1872}),\ \bibfield  {title}
  {\enquote {\bibinfo {title} {Weitere studien {\"u}ber das
  w{\"a}rmegleichgewicht unter gasmolek{\"u}len},}\ }\href@noop {} {\bibfield
  {journal} {\bibinfo  {journal} {Wiener Berichte}\ }\textbf {\bibinfo {volume}
  {66}},\ \bibinfo {pages} {275–370}}\BibitemShut {NoStop}%
\bibitem [{\citenamefont {{Bonderson}}\ \emph {et~al.}(2016)\citenamefont
  {{Bonderson}}, \citenamefont {{Cheng}}, \citenamefont {{Patel}},\ and\
  \citenamefont {{Plamadeala}}}]{Bonderson16}%
  \BibitemOpen
  \bibfield  {author} {\bibinfo {author} {\bibnamefont {{Bonderson}},
  \bibfnamefont {Parsa}}, \bibinfo {author} {\bibfnamefont {Meng}\ \bibnamefont
  {{Cheng}}}, \bibinfo {author} {\bibfnamefont {Kaushal}\ \bibnamefont
  {{Patel}}}, and\ \bibinfo {author} {\bibfnamefont {Eugeniu}\ \bibnamefont
  {{Plamadeala}}}} (\bibinfo {year} {2016}),\ \bibfield  {title} {\enquote
  {\bibinfo {title} {{Topological Enrichment of Luttinger's Theorem}},}\
  }\href@noop {} {\ }\Eprint {https://arxiv.org/abs/1601.07902}
  {arXiv:1601.07902 [cond-mat.str-el]} \BibitemShut {NoStop}%
\bibitem [{\citenamefont {Boruch}\ \emph {et~al.}(2022)\citenamefont {Boruch},
  \citenamefont {Heydeman}, \citenamefont {Iliesiu},\ and\ \citenamefont
  {Turiaci}}]{Boruch:2022tno}%
  \BibitemOpen
  \bibfield  {author} {\bibinfo {author} {\bibnamefont {Boruch}, \bibfnamefont
  {Jan}}, \bibinfo {author} {\bibfnamefont {Matthew~T.}\ \bibnamefont
  {Heydeman}}, \bibinfo {author} {\bibfnamefont {Luca~V.}\ \bibnamefont
  {Iliesiu}}, and\ \bibinfo {author} {\bibfnamefont {Gustavo~J.}\ \bibnamefont
  {Turiaci}}} (\bibinfo {year} {2022}),\ \bibfield  {title} {\enquote {\bibinfo
  {title} {{BPS and near-BPS black holes in $AdS_5$ and their spectrum in
  $\mathcal{N}=4$ SYM}},}\ }\href@noop {} {\ }\Eprint
  {https://arxiv.org/abs/2203.01331} {arXiv:2203.01331 [hep-th]} \BibitemShut
  {NoStop}%
\bibitem [{\citenamefont {Bousso}\ \emph {et~al.}(2022)\citenamefont {Bousso},
  \citenamefont {Dong}, \citenamefont {Engelhardt}, \citenamefont {Faulkner},
  \citenamefont {Hartman}, \citenamefont {Shenker},\ and\ \citenamefont
  {Stanford}}]{Bousso:2022ntt}%
  \BibitemOpen
  \bibfield  {author} {\bibinfo {author} {\bibnamefont {Bousso}, \bibfnamefont
  {Raphael}}, \bibinfo {author} {\bibfnamefont {Xi}~\bibnamefont {Dong}},
  \bibinfo {author} {\bibfnamefont {Netta}\ \bibnamefont {Engelhardt}},
  \bibinfo {author} {\bibfnamefont {Thomas}\ \bibnamefont {Faulkner}}, \bibinfo
  {author} {\bibfnamefont {Thomas}\ \bibnamefont {Hartman}}, \bibinfo {author}
  {\bibfnamefont {Stephen~H.}\ \bibnamefont {Shenker}}, and\ \bibinfo {author}
  {\bibfnamefont {Douglas}\ \bibnamefont {Stanford}}} (\bibinfo {year}
  {2022}),\ \bibfield  {title} {\enquote {\bibinfo {title} {{Snowmass White
  Paper: Quantum Aspects of Black Holes and the Emergence of Spacetime}},}\
  }\href@noop {} {\ }\Eprint {https://arxiv.org/abs/2201.03096}
  {arXiv:2201.03096 [hep-th]} \BibitemShut {NoStop}%
\bibitem [{\citenamefont {Bozovic}\ \emph {et~al.}(1987)\citenamefont
  {Bozovic}, \citenamefont {Kirillov}, \citenamefont {Kapitulnik},
  \citenamefont {Char}, \citenamefont {Hahn}, \citenamefont {Beasley},
  \citenamefont {Geballe}, \citenamefont {Kim},\ and\ \citenamefont
  {Heeger}}]{Bozovic87}%
  \BibitemOpen
  \bibfield  {author} {\bibinfo {author} {\bibnamefont {Bozovic}, \bibfnamefont
  {I}}, \bibinfo {author} {\bibfnamefont {D.}~\bibnamefont {Kirillov}},
  \bibinfo {author} {\bibfnamefont {A.}~\bibnamefont {Kapitulnik}}, \bibinfo
  {author} {\bibfnamefont {K.}~\bibnamefont {Char}}, \bibinfo {author}
  {\bibfnamefont {M.~R.}\ \bibnamefont {Hahn}}, \bibinfo {author}
  {\bibfnamefont {M.~R.}\ \bibnamefont {Beasley}}, \bibinfo {author}
  {\bibfnamefont {T.~H.}\ \bibnamefont {Geballe}}, \bibinfo {author}
  {\bibfnamefont {Y.~H.}\ \bibnamefont {Kim}}, and\ \bibinfo {author}
  {\bibfnamefont {A.~J.}\ \bibnamefont {Heeger}}} (\bibinfo {year} {1987}),\
  \bibfield  {title} {\enquote {\bibinfo {title} {{Optical measurements on
  oriented thin
  ${\mathrm{YBa}}_{2}$${\mathrm{Cu}}_{3}$${\mathrm{O}}_{7\mathrm{\ensuremath{-}}\mathrm{\ensuremath{\delta}}}$
  films: Lack of evidence for excitonic superconductivity}},}\ }\href
  {https://doi.org/10.1103/PhysRevLett.59.2219} {\bibfield  {journal} {\bibinfo
   {journal} {Phys. Rev. Lett.}\ }\textbf {\bibinfo {volume} {59}},\ \bibinfo
  {pages} {2219--2221}}\BibitemShut {NoStop}%
\bibitem [{\citenamefont {Bray}\ and\ \citenamefont {Moore}(1980)}]{BrayMoore}%
  \BibitemOpen
  \bibfield  {author} {\bibinfo {author} {\bibnamefont {Bray}, \bibfnamefont
  {A~J}}, and\ \bibinfo {author} {\bibfnamefont {M~A}\ \bibnamefont {Moore}}}
  (\bibinfo {year} {1980}),\ \bibfield  {title} {\enquote {\bibinfo {title}
  {Replica theory of quantum spin glasses},}\ }\href
  {https://doi.org/10.1088/0022-3719/13/24/005} {\bibfield  {journal} {\bibinfo
   {journal} {Journal of Physics C: Solid State Physics}\ }\textbf {\bibinfo
  {volume} {13}}~(\bibinfo {number} {24}),\ \bibinfo {pages}
  {L655--L660}}\BibitemShut {NoStop}%
\bibitem [{\citenamefont {Brinkman}\ and\ \citenamefont
  {Rice}(1970)}]{BrinkmanRice}%
  \BibitemOpen
  \bibfield  {author} {\bibinfo {author} {\bibnamefont {Brinkman},
  \bibfnamefont {W~F}}, and\ \bibinfo {author} {\bibfnamefont {T.~M.}\
  \bibnamefont {Rice}}} (\bibinfo {year} {1970}),\ \bibfield  {title} {\enquote
  {\bibinfo {title} {{Application of Gutzwiller's Variational Method to the
  Metal-Insulator Transition}},}\ }\href
  {https://doi.org/10.1103/PhysRevB.2.4302} {\bibfield  {journal} {\bibinfo
  {journal} {Phys. Rev. B}\ }\textbf {\bibinfo {volume} {2}},\ \bibinfo {pages}
  {4302--4304}}\BibitemShut {NoStop}%
\bibitem [{\citenamefont {Brody}\ \emph {et~al.}(1981)\citenamefont {Brody},
  \citenamefont {Flores}, \citenamefont {French}, \citenamefont {Mello},
  \citenamefont {Pandey},\ and\ \citenamefont {Wong}}]{French81}%
  \BibitemOpen
  \bibfield  {author} {\bibinfo {author} {\bibnamefont {Brody}, \bibfnamefont
  {T~A}}, \bibinfo {author} {\bibfnamefont {J.}~\bibnamefont {Flores}},
  \bibinfo {author} {\bibfnamefont {J.~B.}\ \bibnamefont {French}}, \bibinfo
  {author} {\bibfnamefont {P.~A.}\ \bibnamefont {Mello}}, \bibinfo {author}
  {\bibfnamefont {A.}~\bibnamefont {Pandey}}, and\ \bibinfo {author}
  {\bibfnamefont {S.~S.~M.}\ \bibnamefont {Wong}}} (\bibinfo {year} {1981}),\
  \bibfield  {title} {\enquote {\bibinfo {title} {Random-matrix physics:
  spectrum and strength fluctuations},}\ }\href
  {https://doi.org/10.1103/RevModPhys.53.385} {\bibfield  {journal} {\bibinfo
  {journal} {Rev. Mod. Phys.}\ }\textbf {\bibinfo {volume} {53}},\ \bibinfo
  {pages} {385--479}}\BibitemShut {NoStop}%
\bibitem [{\citenamefont {Broholm}\ \emph {et~al.}(2020)\citenamefont
  {Broholm}, \citenamefont {Cava}, \citenamefont {Kivelson}, \citenamefont
  {Nocera}, \citenamefont {Norman},\ and\ \citenamefont {Senthil}}]{QSL}%
  \BibitemOpen
  \bibfield  {author} {\bibinfo {author} {\bibnamefont {Broholm}, \bibfnamefont
  {C}}, \bibinfo {author} {\bibfnamefont {R.~J.}\ \bibnamefont {Cava}},
  \bibinfo {author} {\bibfnamefont {S.~A.}\ \bibnamefont {Kivelson}}, \bibinfo
  {author} {\bibfnamefont {D.~G.}\ \bibnamefont {Nocera}}, \bibinfo {author}
  {\bibfnamefont {M.~R.}\ \bibnamefont {Norman}}, and\ \bibinfo {author}
  {\bibfnamefont {T.}~\bibnamefont {Senthil}}} (\bibinfo {year} {2020}),\
  \bibfield  {title} {\enquote {\bibinfo {title} {Quantum spin liquids},}\
  }\href {https://doi.org/10.1126/science.aay0668} {\bibfield  {journal}
  {\bibinfo  {journal} {Science}\ }\textbf {\bibinfo {volume} {367}}~(\bibinfo
  {number} {6475}),\ 10.1126/science.aay0668}\BibitemShut {NoStop}%
\bibitem [{\citenamefont {Brown}\ \emph {et~al.}(2019)\citenamefont {Brown},
  \citenamefont {Mitra}, \citenamefont {Guardado-Sanchez}, \citenamefont
  {Nourafkan}, \citenamefont {Reymbaut}, \citenamefont {H{\'e}bert},
  \citenamefont {Bergeron}, \citenamefont {Tremblay}, \citenamefont {Kokalj},
  \citenamefont {Huse}, \citenamefont {Schau{\ss}},\ and\ \citenamefont
  {Bakr}}]{Bakr}%
  \BibitemOpen
  \bibfield  {author} {\bibinfo {author} {\bibnamefont {Brown}, \bibfnamefont
  {Peter~T}}, \bibinfo {author} {\bibfnamefont {Debayan}\ \bibnamefont
  {Mitra}}, \bibinfo {author} {\bibfnamefont {Elmer}\ \bibnamefont
  {Guardado-Sanchez}}, \bibinfo {author} {\bibfnamefont {Reza}\ \bibnamefont
  {Nourafkan}}, \bibinfo {author} {\bibfnamefont {Alexis}\ \bibnamefont
  {Reymbaut}}, \bibinfo {author} {\bibfnamefont {Charles-David}\ \bibnamefont
  {H{\'e}bert}}, \bibinfo {author} {\bibfnamefont {Simon}\ \bibnamefont
  {Bergeron}}, \bibinfo {author} {\bibfnamefont {A.-M.~S.}\ \bibnamefont
  {Tremblay}}, \bibinfo {author} {\bibfnamefont {Jure}\ \bibnamefont {Kokalj}},
  \bibinfo {author} {\bibfnamefont {David~A.}\ \bibnamefont {Huse}}, \bibinfo
  {author} {\bibfnamefont {Peter}\ \bibnamefont {Schau{\ss}}}, and\ \bibinfo
  {author} {\bibfnamefont {Waseem~S.}\ \bibnamefont {Bakr}}} (\bibinfo {year}
  {2019}),\ \bibfield  {title} {\enquote {\bibinfo {title} {{Bad metallic
  transport in a cold atom Fermi-Hubbard system}},}\ }\href
  {https://doi.org/10.1126/science.aat4134} {\bibfield  {journal} {\bibinfo
  {journal} {Science}\ }\textbf {\bibinfo {volume} {363}}~(\bibinfo {number}
  {6425}),\ \bibinfo {pages} {379--382}}\BibitemShut {NoStop}%
\bibitem [{\citenamefont {Bruin}\ \emph {et~al.}(2013)\citenamefont {Bruin},
  \citenamefont {Sakai}, \citenamefont {Perry},\ and\ \citenamefont
  {Mackenzie}}]{Bruin13}%
  \BibitemOpen
  \bibfield  {author} {\bibinfo {author} {\bibnamefont {Bruin}, \bibfnamefont
  {J~A~N}}, \bibinfo {author} {\bibfnamefont {H.}~\bibnamefont {Sakai}},
  \bibinfo {author} {\bibfnamefont {R.~S.}\ \bibnamefont {Perry}}, and\
  \bibinfo {author} {\bibfnamefont {A.~P.}\ \bibnamefont {Mackenzie}}}
  (\bibinfo {year} {2013}),\ \href {https://doi.org/10.1126/science.1227612}
  {\enquote {\bibinfo {title} {Similarity of scattering rates in metals showing
  t-linear resistivity},}\ }\BibitemShut {NoStop}%
\bibitem [{\citenamefont {Brühwiler}\ \emph {et~al.}(2006)\citenamefont
  {Brühwiler}, \citenamefont {Batlogg}, \citenamefont {Kazakov}, \citenamefont
  {Niedermayer},\ and\ \citenamefont {Karpinski}}]{bruhwiler}%
  \BibitemOpen
  \bibfield  {author} {\bibinfo {author} {\bibnamefont {Brühwiler},
  \bibfnamefont {M}}, \bibinfo {author} {\bibfnamefont {B.}~\bibnamefont
  {Batlogg}}, \bibinfo {author} {\bibfnamefont {S.M.}\ \bibnamefont {Kazakov}},
  \bibinfo {author} {\bibfnamefont {Ch.}\ \bibnamefont {Niedermayer}}, and\
  \bibinfo {author} {\bibfnamefont {J.}~\bibnamefont {Karpinski}}} (\bibinfo
  {year} {2006}),\ \bibfield  {title} {\enquote {\bibinfo {title}
  {{Na$_x$CoO$_2$: Enhanced low-energy excitations of electrons on a 2d
  triangular lattice}},}\ }\href
  {https://doi.org/https://doi.org/10.1016/j.physb.2006.01.422} {\bibfield
  {journal} {\bibinfo  {journal} {Physica B: Condensed Matter}\ }\textbf
  {\bibinfo {volume} {378-380}},\ \bibinfo {pages} {630 -- 631}}\BibitemShut
  {NoStop}%
\bibitem [{\citenamefont {{Burdin}}\ \emph {et~al.}(2000)\citenamefont
  {{Burdin}}, \citenamefont {{Georges}},\ and\ \citenamefont
  {{Grempel}}}]{Burdin_2000}%
  \BibitemOpen
  \bibfield  {author} {\bibinfo {author} {\bibnamefont {{Burdin}},
  \bibfnamefont {S}}, \bibinfo {author} {\bibfnamefont {A.}~\bibnamefont
  {{Georges}}}, and\ \bibinfo {author} {\bibfnamefont {D.~R.}\ \bibnamefont
  {{Grempel}}}} (\bibinfo {year} {2000}),\ \bibfield  {title} {\enquote
  {\bibinfo {title} {{Coherence Scale of the Kondo Lattice}},}\ }\href
  {https://doi.org/10.1103/PhysRevLett.85.1048} {\bibfield  {journal} {\bibinfo
   {journal} {Phys. Rev. Lett.}\ }\textbf {\bibinfo {volume} {85}}~(\bibinfo
  {number} {5}),\ \bibinfo {pages} {1048--1051}},\ \Eprint
  {https://arxiv.org/abs/cond-mat/0004043} {arXiv:cond-mat/0004043
  [cond-mat.str-el]} \BibitemShut {NoStop}%
\bibitem [{\citenamefont {Burdin}\ \emph {et~al.}(2002)\citenamefont {Burdin},
  \citenamefont {Grempel},\ and\ \citenamefont {Georges}}]{Burdin2002}%
  \BibitemOpen
  \bibfield  {author} {\bibinfo {author} {\bibnamefont {Burdin}, \bibfnamefont
  {S}}, \bibinfo {author} {\bibfnamefont {D.~R.}\ \bibnamefont {Grempel}}, and\
  \bibinfo {author} {\bibfnamefont {A.}~\bibnamefont {Georges}}} (\bibinfo
  {year} {2002}),\ \bibfield  {title} {\enquote {\bibinfo {title}
  {{Heavy-fermion and spin-liquid behavior in a Kondo lattice with magnetic
  frustration}},}\ }\href {https://doi.org/10.1103/PhysRevB.66.045111}
  {\bibfield  {journal} {\bibinfo  {journal} {Phys. Rev. B}\ }\textbf {\bibinfo
  {volume} {66}},\ \bibinfo {pages} {045111}}\BibitemShut {NoStop}%
\bibitem [{\citenamefont {Cai}\ \emph {et~al.}(2020)\citenamefont {Cai},
  \citenamefont {Yu}, \citenamefont {Hu}, \citenamefont {Kirchner},\ and\
  \citenamefont {Si}}]{SiSU21}%
  \BibitemOpen
  \bibfield  {author} {\bibinfo {author} {\bibnamefont {Cai}, \bibfnamefont
  {Ang}}, \bibinfo {author} {\bibfnamefont {Zuodong}\ \bibnamefont {Yu}},
  \bibinfo {author} {\bibfnamefont {Haoyu}\ \bibnamefont {Hu}}, \bibinfo
  {author} {\bibfnamefont {Stefan}\ \bibnamefont {Kirchner}}, and\ \bibinfo
  {author} {\bibfnamefont {Qimiao}\ \bibnamefont {Si}}} (\bibinfo {year}
  {2020}),\ \bibfield  {title} {\enquote {\bibinfo {title} {{Dynamical Scaling
  of Charge and Spin Responses at a Kondo Destruction Quantum Critical
  Point}},}\ }\href {https://doi.org/10.1103/PhysRevLett.124.027205} {\bibfield
   {journal} {\bibinfo  {journal} {Phys. Rev. Lett.}\ }\textbf {\bibinfo
  {volume} {124}},\ \bibinfo {pages} {027205}}\BibitemShut {NoStop}%
\bibitem [{\citenamefont {Cao}\ \emph {et~al.}(2020)\citenamefont {Cao},
  \citenamefont {Chowdhury}, \citenamefont {Rodan-Legrain}, \citenamefont
  {Rubies-Bigorda}, \citenamefont {Watanabe}, \citenamefont {Taniguchi},
  \citenamefont {Senthil},\ and\ \citenamefont {Jarillo-Herrero}}]{Cao20}%
  \BibitemOpen
  \bibfield  {author} {\bibinfo {author} {\bibnamefont {Cao}, \bibfnamefont
  {Yuan}}, \bibinfo {author} {\bibfnamefont {Debanjan}\ \bibnamefont
  {Chowdhury}}, \bibinfo {author} {\bibfnamefont {Daniel}\ \bibnamefont
  {Rodan-Legrain}}, \bibinfo {author} {\bibfnamefont {Oriol}\ \bibnamefont
  {Rubies-Bigorda}}, \bibinfo {author} {\bibfnamefont {Kenji}\ \bibnamefont
  {Watanabe}}, \bibinfo {author} {\bibfnamefont {Takashi}\ \bibnamefont
  {Taniguchi}}, \bibinfo {author} {\bibfnamefont {T.}~\bibnamefont {Senthil}},
  and\ \bibinfo {author} {\bibfnamefont {Pablo}\ \bibnamefont
  {Jarillo-Herrero}}} (\bibinfo {year} {2020}),\ \bibfield  {title} {\enquote
  {\bibinfo {title} {{Strange Metal in Magic-Angle Graphene with near Planckian
  Dissipation}},}\ }\href {https://doi.org/10.1103/PhysRevLett.124.076801}
  {\bibfield  {journal} {\bibinfo  {journal} {Phys. Rev. Lett.}\ }\textbf
  {\bibinfo {volume} {124}},\ \bibinfo {pages} {076801}},\ \Eprint
  {https://arxiv.org/abs/1901.03710} {arXiv:1901.03710 [cond-mat.str-el]}
  \BibitemShut {NoStop}%
\bibitem [{\citenamefont {Carullo}\ \emph {et~al.}(2021)\citenamefont
  {Carullo}, \citenamefont {Laghi}, \citenamefont {Veitch},\ and\ \citenamefont
  {Del~Pozzo}}]{LIGO21}%
  \BibitemOpen
  \bibfield  {author} {\bibinfo {author} {\bibnamefont {Carullo}, \bibfnamefont
  {Gregorio}}, \bibinfo {author} {\bibfnamefont {Danny}\ \bibnamefont {Laghi}},
  \bibinfo {author} {\bibfnamefont {John}\ \bibnamefont {Veitch}}, and\
  \bibinfo {author} {\bibfnamefont {Walter}\ \bibnamefont {Del~Pozzo}}}
  (\bibinfo {year} {2021}),\ \bibfield  {title} {\enquote {\bibinfo {title}
  {{Bekenstein-Hod Universal Bound on Information Emission Rate Is Obeyed by
  LIGO-Virgo Binary Black Hole Remnants}},}\ }\href
  {https://doi.org/10.1103/PhysRevLett.126.161102} {\bibfield  {journal}
  {\bibinfo  {journal} {Phys. Rev. Lett.}\ }\textbf {\bibinfo {volume} {126}},\
  \bibinfo {pages} {161102}}\BibitemShut {NoStop}%
\bibitem [{\citenamefont {Vrani\ifmmode~\acute{c}\else \'{c}\fi{}}\ \emph
  {et~al.}(2020)\citenamefont {Vrani\ifmmode~\acute{c}\else \'{c}\fi{}},
  \citenamefont {Vu\ifmmode \check{c}\else \v{c}\fi{}i\ifmmode \check{c}\else
  \v{c}\fi{}evi\ifmmode~\acute{c}\else \'{c}\fi{}}, \citenamefont {Kokalj},
  \citenamefont {Skolimowski}, \citenamefont {\ifmmode~\check{Z}\else
  \v{Z}\fi{}itko}, \citenamefont {Mravlje},\ and\ \citenamefont
  {Tanaskovi\ifmmode~\acute{c}\else \'{c}\fi{}}}]{Vranic2020}%
  \BibitemOpen
  \bibfield  {author} {\bibinfo {author} {\bibnamefont
  {Vrani\ifmmode~\acute{c}\else \'{c}\fi{}}, \bibfnamefont {A}}, \bibinfo
  {author} {\bibfnamefont {J.}~\bibnamefont {Vu\ifmmode \check{c}\else
  \v{c}\fi{}i\ifmmode \check{c}\else \v{c}\fi{}evi\ifmmode~\acute{c}\else
  \'{c}\fi{}}}, \bibinfo {author} {\bibfnamefont {J.}~\bibnamefont {Kokalj}},
  \bibinfo {author} {\bibfnamefont {J.}~\bibnamefont {Skolimowski}}, \bibinfo
  {author} {\bibfnamefont {R.}~\bibnamefont {\ifmmode~\check{Z}\else
  \v{Z}\fi{}itko}}, \bibinfo {author} {\bibfnamefont {J.}~\bibnamefont
  {Mravlje}}, and\ \bibinfo {author} {\bibfnamefont {D.}~\bibnamefont
  {Tanaskovi\ifmmode~\acute{c}\else \'{c}\fi{}}}} (\bibinfo {year} {2020}),\
  \bibfield  {title} {\enquote {\bibinfo {title} {{Charge transport in the
  Hubbard model at high temperatures: Triangular versus square lattice}},}\
  }\href {https://doi.org/10.1103/PhysRevB.102.115142} {\bibfield  {journal}
  {\bibinfo  {journal} {Phys. Rev. B}\ }\textbf {\bibinfo {volume} {102}},\
  \bibinfo {pages} {115142}}\BibitemShut {NoStop}%
\bibitem [{\citenamefont {Vu\ifmmode \check{c}\else \v{c}\fi{}i\ifmmode
  \check{c}\else \v{c}\fi{}evi\ifmmode~\acute{c}\else \'{c}\fi{}}\ \emph
  {et~al.}(2019)\citenamefont {Vu\ifmmode \check{c}\else \v{c}\fi{}i\ifmmode
  \check{c}\else \v{c}\fi{}evi\ifmmode~\acute{c}\else \'{c}\fi{}},
  \citenamefont {Kokalj}, \citenamefont {\ifmmode~\check{Z}\else
  \v{Z}\fi{}itko}, \citenamefont {Wentzell}, \citenamefont
  {Tanaskovi\ifmmode~\acute{c}\else \'{c}\fi{}},\ and\ \citenamefont
  {Mravlje}}]{Vucicevic2019}%
  \BibitemOpen
  \bibfield  {author} {\bibinfo {author} {\bibnamefont {Vu\ifmmode
  \check{c}\else \v{c}\fi{}i\ifmmode \check{c}\else
  \v{c}\fi{}evi\ifmmode~\acute{c}\else \'{c}\fi{}}, \bibfnamefont {J}},
  \bibinfo {author} {\bibfnamefont {J.}~\bibnamefont {Kokalj}}, \bibinfo
  {author} {\bibfnamefont {R.}~\bibnamefont {\ifmmode~\check{Z}\else
  \v{Z}\fi{}itko}}, \bibinfo {author} {\bibfnamefont {N.}~\bibnamefont
  {Wentzell}}, \bibinfo {author} {\bibfnamefont {D.}~\bibnamefont
  {Tanaskovi\ifmmode~\acute{c}\else \'{c}\fi{}}}, and\ \bibinfo {author}
  {\bibfnamefont {J.}~\bibnamefont {Mravlje}}} (\bibinfo {year} {2019}),\
  \bibfield  {title} {\enquote {\bibinfo {title} {{Conductivity in the Square
  Lattice Hubbard Model at High Temperatures: Importance of Vertex
  Corrections}},}\ }\href {https://doi.org/10.1103/PhysRevLett.123.036601}
  {\bibfield  {journal} {\bibinfo  {journal} {Phys. Rev. Lett.}\ }\textbf
  {\bibinfo {volume} {123}},\ \bibinfo {pages} {036601}}\BibitemShut {NoStop}%
\bibitem [{\citenamefont {Dobrosavljevi\ifmmode~\acute{c}\else \'{c}\fi{}}\
  and\ \citenamefont {Kotliar}(1997)}]{Dobro1997}%
  \BibitemOpen
  \bibfield  {author} {\bibinfo {author} {\bibnamefont
  {Dobrosavljevi\ifmmode~\acute{c}\else \'{c}\fi{}}, \bibfnamefont {V}}, and\
  \bibinfo {author} {\bibfnamefont {G.}~\bibnamefont {Kotliar}}} (\bibinfo
  {year} {1997}),\ \bibfield  {title} {\enquote {\bibinfo {title} {{Mean Field
  Theory of the Mott-Anderson Transition}},}\ }\href
  {https://doi.org/10.1103/PhysRevLett.78.3943} {\bibfield  {journal} {\bibinfo
   {journal} {Phys. Rev. Lett.}\ }\textbf {\bibinfo {volume} {78}},\ \bibinfo
  {pages} {3943--3946}}\BibitemShut {NoStop}%
\bibitem [{\citenamefont {Cha}\ \emph {et~al.}(2020{\natexlab{a}})\citenamefont
  {Cha}, \citenamefont {Patel}, \citenamefont {Gull},\ and\ \citenamefont
  {Kim}}]{Cha_Patel_2020}%
  \BibitemOpen
  \bibfield  {author} {\bibinfo {author} {\bibnamefont {Cha}, \bibfnamefont
  {Peter}}, \bibinfo {author} {\bibfnamefont {Aavishkar~A.}\ \bibnamefont
  {Patel}}, \bibinfo {author} {\bibfnamefont {Emanuel}\ \bibnamefont {Gull}},
  and\ \bibinfo {author} {\bibfnamefont {Eun-Ah}\ \bibnamefont {Kim}}}
  (\bibinfo {year} {2020}{\natexlab{a}}),\ \bibfield  {title} {\enquote
  {\bibinfo {title} {{Slope invariant $T$-linear resistivity from local
  self-energy}},}\ }\href {https://doi.org/10.1103/PhysRevResearch.2.033434}
  {\bibfield  {journal} {\bibinfo  {journal} {Phys. Rev. Research}\ }\textbf
  {\bibinfo {volume} {2}},\ \bibinfo {pages} {033434}}\BibitemShut {NoStop}%
\bibitem [{\citenamefont {Cha}\ \emph {et~al.}(2020{\natexlab{b}})\citenamefont
  {Cha}, \citenamefont {Wentzell}, \citenamefont {Parcollet}, \citenamefont
  {Georges},\ and\ \citenamefont {Kim}}]{Cha_2020}%
  \BibitemOpen
  \bibfield  {author} {\bibinfo {author} {\bibnamefont {Cha}, \bibfnamefont
  {Peter}}, \bibinfo {author} {\bibfnamefont {Nils}\ \bibnamefont {Wentzell}},
  \bibinfo {author} {\bibfnamefont {Olivier}\ \bibnamefont {Parcollet}},
  \bibinfo {author} {\bibfnamefont {Antoine}\ \bibnamefont {Georges}}, and\
  \bibinfo {author} {\bibfnamefont {Eun-Ah}\ \bibnamefont {Kim}}} (\bibinfo
  {year} {2020}{\natexlab{b}}),\ \bibfield  {title} {\enquote {\bibinfo {title}
  {{Linear resistivity and Sachdev-Ye-Kitaev (SYK) spin liquid behavior in a
  quantum critical metal with spin-1/2 fermions}},}\ }\href
  {https://doi.org/10.1073/pnas.2003179117} {\bibfield  {journal} {\bibinfo
  {journal} {Proceedings of the National Academy of Sciences}\ }\textbf
  {\bibinfo {volume} {117}}~(\bibinfo {number} {31}),\ \bibinfo {pages}
  {18341–18346}}\BibitemShut {NoStop}%
\bibitem [{\citenamefont {Chamblin}\ \emph {et~al.}(1999)\citenamefont
  {Chamblin}, \citenamefont {Emparan}, \citenamefont {Johnson},\ and\
  \citenamefont {Myers}}]{Myers99}%
  \BibitemOpen
  \bibfield  {author} {\bibinfo {author} {\bibnamefont {Chamblin},
  \bibfnamefont {Andrew}}, \bibinfo {author} {\bibfnamefont {Roberto}\
  \bibnamefont {Emparan}}, \bibinfo {author} {\bibfnamefont {Clifford~V.}\
  \bibnamefont {Johnson}}, and\ \bibinfo {author} {\bibfnamefont {Robert~C.}\
  \bibnamefont {Myers}}} (\bibinfo {year} {1999}),\ \bibfield  {title}
  {\enquote {\bibinfo {title} {{Charged AdS black holes and catastrophic
  holography}},}\ }\href {https://doi.org/10.1103/PhysRevD.60.064018}
  {\bibfield  {journal} {\bibinfo  {journal} {Phys. Rev. D}\ }\textbf {\bibinfo
  {volume} {60}},\ \bibinfo {pages} {064018}},\ \Eprint
  {https://arxiv.org/abs/hep-th/9902170} {arXiv:hep-th/9902170 [hep-th]}
  \BibitemShut {NoStop}%
%%CITATION = HEP-TH/9902170;%%
\bibitem [{\citenamefont {Cheipesh}\ \emph {et~al.}(2021)\citenamefont
  {Cheipesh}, \citenamefont {Pavlov}, \citenamefont {Ohanesjan}, \citenamefont
  {Schalm},\ and\ \citenamefont {Gnezdilov}}]{Cheipesh20}%
  \BibitemOpen
  \bibfield  {author} {\bibinfo {author} {\bibnamefont {Cheipesh},
  \bibfnamefont {Y}}, \bibinfo {author} {\bibfnamefont {A.~I.}\ \bibnamefont
  {Pavlov}}, \bibinfo {author} {\bibfnamefont {V.}~\bibnamefont {Ohanesjan}},
  \bibinfo {author} {\bibfnamefont {K.}~\bibnamefont {Schalm}}, and\ \bibinfo
  {author} {\bibfnamefont {N.~V.}\ \bibnamefont {Gnezdilov}}} (\bibinfo {year}
  {2021}),\ \bibfield  {title} {\enquote {\bibinfo {title} {Quantum tunneling
  dynamics in a complex-valued sachdev-ye-kitaev model quench-coupled to a cool
  bath},}\ }\href {https://doi.org/10.1103/PhysRevB.104.115134} {\bibfield
  {journal} {\bibinfo  {journal} {Phys. Rev. B}\ }\textbf {\bibinfo {volume}
  {104}},\ \bibinfo {pages} {115134}}\BibitemShut {NoStop}%
\bibitem [{\citenamefont {Chen}\ \emph {et~al.}(2018)\citenamefont {Chen},
  \citenamefont {Ilan}, \citenamefont {de~Juan}, \citenamefont {Pikulin},\ and\
  \citenamefont {Franz}}]{Franz2018}%
  \BibitemOpen
  \bibfield  {author} {\bibinfo {author} {\bibnamefont {Chen}, \bibfnamefont
  {Anffany}}, \bibinfo {author} {\bibfnamefont {R.}~\bibnamefont {Ilan}},
  \bibinfo {author} {\bibfnamefont {F.}~\bibnamefont {de~Juan}}, \bibinfo
  {author} {\bibfnamefont {D.~I.}\ \bibnamefont {Pikulin}}, and\ \bibinfo
  {author} {\bibfnamefont {M.}~\bibnamefont {Franz}}} (\bibinfo {year}
  {2018}),\ \bibfield  {title} {\enquote {\bibinfo {title} {Quantum holography
  in a graphene flake with an irregular boundary},}\ }\href
  {https://doi.org/10.1103/PhysRevLett.121.036403} {\bibfield  {journal}
  {\bibinfo  {journal} {Phys. Rev. Lett.}\ }\textbf {\bibinfo {volume} {121}},\
  \bibinfo {pages} {036403}}\BibitemShut {NoStop}%
\bibitem [{\citenamefont {Chen}\ \emph {et~al.}(2021)\citenamefont {Chen},
  \citenamefont {Czech},\ and\ \citenamefont {Wang}}]{Czech}%
  \BibitemOpen
  \bibfield  {author} {\bibinfo {author} {\bibnamefont {Chen}, \bibfnamefont
  {Bowen}}, \bibinfo {author} {\bibfnamefont {Bartlomiej}\ \bibnamefont
  {Czech}}, and\ \bibinfo {author} {\bibfnamefont {Zi-zhi}\ \bibnamefont
  {Wang}}} (\bibinfo {year} {2021}),\ \bibfield  {title} {\enquote {\bibinfo
  {title} {{Quantum Information in Holographic Duality}},}\ }\href@noop {} {\
  }\Eprint {https://arxiv.org/abs/2108.09188} {arXiv:2108.09188 [hep-th]}
  \BibitemShut {NoStop}%
\bibitem [{\citenamefont {Chen}\ \emph {et~al.}(2020)\citenamefont {Chen},
  \citenamefont {Qi},\ and\ \citenamefont {Zhang}}]{Chen:2020wiq}%
  \BibitemOpen
  \bibfield  {author} {\bibinfo {author} {\bibnamefont {Chen}, \bibfnamefont
  {Yiming}}, \bibinfo {author} {\bibfnamefont {Xiao-Liang}\ \bibnamefont {Qi}},
  and\ \bibinfo {author} {\bibfnamefont {Pengfei}\ \bibnamefont {Zhang}}}
  (\bibinfo {year} {2020}),\ \bibfield  {title} {\enquote {\bibinfo {title}
  {{Replica wormhole and information retrieval in the SYK model coupled to
  Majorana chains}},}\ }\href {https://doi.org/10.1007/JHEP06(2020)121}
  {\bibfield  {journal} {\bibinfo  {journal} {JHEP}\ }\textbf {\bibinfo
  {volume} {06}},\ \bibinfo {pages} {121}},\ \Eprint
  {https://arxiv.org/abs/2003.13147} {arXiv:2003.13147 [hep-th]} \BibitemShut
  {NoStop}%
\bibitem [{\citenamefont {{Chew}}\ \emph {et~al.}(2017)\citenamefont {{Chew}},
  \citenamefont {{Essin}},\ and\ \citenamefont {{Alicea}}}]{Alicea17}%
  \BibitemOpen
  \bibfield  {author} {\bibinfo {author} {\bibnamefont {{Chew}}, \bibfnamefont
  {Aaron}}, \bibinfo {author} {\bibfnamefont {Andrew}\ \bibnamefont {{Essin}}},
  and\ \bibinfo {author} {\bibfnamefont {Jason}\ \bibnamefont {{Alicea}}}}
  (\bibinfo {year} {2017}),\ \bibfield  {title} {\enquote {\bibinfo {title}
  {{Approximating the Sachdev-Ye-Kitaev model with Majorana wires}},}\ }\href
  {https://doi.org/10.1103/PhysRevB.96.121119} {\bibfield  {journal} {\bibinfo
  {journal} {Phys. Rev. B}\ }\textbf {\bibinfo {volume} {96}}~(\bibinfo
  {number} {12}),\ \bibinfo {eid} {121119}},\ \Eprint
  {https://arxiv.org/abs/1703.06890} {arXiv:1703.06890 [cond-mat.dis-nn]}
  \BibitemShut {NoStop}%
\bibitem [{\citenamefont {Chitra}\ and\ \citenamefont
  {Kotliar}(2000)}]{Chitra2000}%
  \BibitemOpen
  \bibfield  {author} {\bibinfo {author} {\bibnamefont {Chitra}, \bibfnamefont
  {R}}, and\ \bibinfo {author} {\bibfnamefont {G.}~\bibnamefont {Kotliar}}}
  (\bibinfo {year} {2000}),\ \bibfield  {title} {\enquote {\bibinfo {title}
  {{Effect of Long Range Coulomb Interactions on the Mott Transition}},}\
  }\href {https://doi.org/10.1103/PhysRevLett.84.3678} {\bibfield  {journal}
  {\bibinfo  {journal} {Phys. Rev. Lett.}\ }\textbf {\bibinfo {volume} {84}},\
  \bibinfo {pages} {3678--3681}}\BibitemShut {NoStop}%
\bibitem [{\citenamefont {{Chowdhury}}\ and\ \citenamefont
  {{Berg}}(2020)}]{ChowdhuryBerg}%
  \BibitemOpen
  \bibfield  {author} {\bibinfo {author} {\bibnamefont {{Chowdhury}},
  \bibfnamefont {Debanjan}}, and\ \bibinfo {author} {\bibfnamefont {Erez}\
  \bibnamefont {{Berg}}}} (\bibinfo {year} {2020}),\ \bibfield  {title}
  {\enquote {\bibinfo {title} {{Intrinsic superconducting instabilities of a
  solvable model for an incoherent metal}},}\ }\href
  {https://doi.org/10.1103/PhysRevResearch.2.013301} {\bibfield  {journal}
  {\bibinfo  {journal} {Physical Review Research}\ }\textbf {\bibinfo {volume}
  {2}}~(\bibinfo {number} {1}),\ \bibinfo {eid} {013301}},\ \Eprint
  {https://arxiv.org/abs/1908.02757} {arXiv:1908.02757 [cond-mat.str-el]}
  \BibitemShut {NoStop}%
\bibitem [{\citenamefont {Chowdhury}\ and\ \citenamefont
  {Berg}(2020)}]{DCEB20b}%
  \BibitemOpen
  \bibfield  {author} {\bibinfo {author} {\bibnamefont {Chowdhury},
  \bibfnamefont {Debanjan}}, and\ \bibinfo {author} {\bibfnamefont {Erez}\
  \bibnamefont {Berg}}} (\bibinfo {year} {2020}),\ \bibfield  {title} {\enquote
  {\bibinfo {title} {{The unreasonable effectiveness of Eliashberg theory for
  pairing of non-Fermi liquids}},}\ }\href
  {https://doi.org/https://doi.org/10.1016/j.aop.2020.168125} {\bibfield
  {journal} {\bibinfo  {journal} {Annals of Physics}\ }\textbf {\bibinfo
  {volume} {417}},\ \bibinfo {pages} {168125}},\ \bibinfo {note} {{Eliashberg
  theory at 60: Strong-coupling superconductivity and beyond}}\BibitemShut
  {NoStop}%
\bibitem [{\citenamefont {{Chowdhury}}\ \emph {et~al.}(2018)\citenamefont
  {{Chowdhury}}, \citenamefont {{Sodemann}},\ and\ \citenamefont
  {{Senthil}}}]{Chowdhury_2018}%
  \BibitemOpen
  \bibfield  {author} {\bibinfo {author} {\bibnamefont {{Chowdhury}},
  \bibfnamefont {Debanjan}}, \bibinfo {author} {\bibfnamefont {Inti}\
  \bibnamefont {{Sodemann}}}, and\ \bibinfo {author} {\bibfnamefont
  {T.}~\bibnamefont {{Senthil}}}} (\bibinfo {year} {2018}),\ \bibfield  {title}
  {\enquote {\bibinfo {title} {{Mixed-valence insulators with neutral Fermi
  surfaces}},}\ }\href {https://doi.org/10.1038/s41467-018-04163-2} {\bibfield
  {journal} {\bibinfo  {journal} {Nature Communications}\ }\textbf {\bibinfo
  {volume} {9}},\ \bibinfo {eid} {1766}},\ \Eprint
  {https://arxiv.org/abs/1706.00418} {arXiv:1706.00418 [cond-mat.str-el]}
  \BibitemShut {NoStop}%
\bibitem [{\citenamefont {Chowdhury}\ and\ \citenamefont
  {Swingle}(2017)}]{Chowdhury:2017jzb}%
  \BibitemOpen
  \bibfield  {author} {\bibinfo {author} {\bibnamefont {Chowdhury},
  \bibfnamefont {Debanjan}}, and\ \bibinfo {author} {\bibfnamefont {Brian}\
  \bibnamefont {Swingle}}} (\bibinfo {year} {2017}),\ \bibfield  {title}
  {\enquote {\bibinfo {title} {{Onset of many-body chaos in the O$(N)$
  model}},}\ }\href {https://doi.org/10.1103/PhysRevD.96.065005} {\bibfield
  {journal} {\bibinfo  {journal} {Phys. Rev. D}\ }\textbf {\bibinfo {volume}
  {96}}~(\bibinfo {number} {6}),\ \bibinfo {pages} {065005}},\ \Eprint
  {https://arxiv.org/abs/1703.02545} {arXiv:1703.02545 [cond-mat.str-el]}
  \BibitemShut {NoStop}%
\bibitem [{\citenamefont {Chowdhury}\ \emph {et~al.}(2018)\citenamefont
  {Chowdhury}, \citenamefont {Werman}, \citenamefont {Berg},\ and\
  \citenamefont {Senthil}}]{DCsyk}%
  \BibitemOpen
  \bibfield  {author} {\bibinfo {author} {\bibnamefont {Chowdhury},
  \bibfnamefont {Debanjan}}, \bibinfo {author} {\bibfnamefont {Yochai}\
  \bibnamefont {Werman}}, \bibinfo {author} {\bibfnamefont {Erez}\ \bibnamefont
  {Berg}}, and\ \bibinfo {author} {\bibfnamefont {T.}~\bibnamefont {Senthil}}}
  (\bibinfo {year} {2018}),\ \bibfield  {title} {\enquote {\bibinfo {title}
  {{Translationally Invariant Non-Fermi-Liquid Metals with Critical Fermi
  Surfaces: Solvable Models}},}\ }\href
  {https://doi.org/10.1103/PhysRevX.8.031024} {\bibfield  {journal} {\bibinfo
  {journal} {Phys. Rev. X}\ }\textbf {\bibinfo {volume} {8}},\ \bibinfo {pages}
  {031024}}\BibitemShut {NoStop}%
\bibitem [{\citenamefont {Christos}\ \emph
  {et~al.}(2022{\natexlab{a}})\citenamefont {Christos}, \citenamefont {Haehl},\
  and\ \citenamefont {Sachdev}}]{Christos:2021wno}%
  \BibitemOpen
  \bibfield  {author} {\bibinfo {author} {\bibnamefont {Christos},
  \bibfnamefont {Maine}}, \bibinfo {author} {\bibfnamefont {Felix~M.}\
  \bibnamefont {Haehl}}, and\ \bibinfo {author} {\bibfnamefont {Subir}\
  \bibnamefont {Sachdev}}} (\bibinfo {year} {2022}{\natexlab{a}}),\ \bibfield
  {title} {\enquote {\bibinfo {title} {{Spin liquid to spin glass crossover in
  the random quantum Heisenberg magnet}},}\ }\href
  {https://doi.org/10.1103/PhysRevB.105.085120} {\bibfield  {journal} {\bibinfo
   {journal} {Phys. Rev. B}\ }\textbf {\bibinfo {volume} {105}},\ \bibinfo
  {pages} {085120}},\ \Eprint {https://arxiv.org/abs/2110.00007}
  {arXiv:2110.00007 [cond-mat.str-el]} \BibitemShut {NoStop}%
\bibitem [{\citenamefont {Christos}\ \emph
  {et~al.}(2022{\natexlab{b}})\citenamefont {Christos}, \citenamefont {Joshi},
  \citenamefont {Sachdev},\ and\ \citenamefont
  {Tikhanovskaya}}]{Christos:toappear}%
  \BibitemOpen
  \bibfield  {author} {\bibinfo {author} {\bibnamefont {Christos},
  \bibfnamefont {Maine}}, \bibinfo {author} {\bibfnamefont {Darshan~G.}\
  \bibnamefont {Joshi}}, \bibinfo {author} {\bibfnamefont {Subir}\ \bibnamefont
  {Sachdev}}, and\ \bibinfo {author} {\bibfnamefont {Maria}\ \bibnamefont
  {Tikhanovskaya}}} (\bibinfo {year} {2022}{\natexlab{b}}),\ \bibfield  {title}
  {\enquote {\bibinfo {title} {{Critical metallic phase in the overdoped random
  $t$-$J$ model}},}\ }\href {https://doi.org/10.1073/pnas.2206921119}
  {\bibfield  {journal} {\bibinfo  {journal} {Proc. Nat. Acad. Sci.}\ }\textbf
  {\bibinfo {volume} {119}},\ \bibinfo {pages} {e2206921119}},\ \Eprint
  {https://arxiv.org/abs/2203.16548} {arXiv:2203.16548 [cond-mat.str-el]}
  \BibitemShut {NoStop}%
\bibitem [{\citenamefont {{Chubukov}}\ and\ \citenamefont
  {{Abanov}}(2021)}]{Chubukov21}%
  \BibitemOpen
  \bibfield  {author} {\bibinfo {author} {\bibnamefont {{Chubukov}},
  \bibfnamefont {A~V}}, and\ \bibinfo {author} {\bibfnamefont {A.}~\bibnamefont
  {{Abanov}}}} (\bibinfo {year} {2021}),\ \bibfield  {title} {\enquote
  {\bibinfo {title} {{Pairing by a Dynamical Interaction in a Metal}},}\ }\href
  {https://doi.org/10.1134/S1063776121040051} {\bibfield  {journal} {\bibinfo
  {journal} {Soviet Journal of Experimental and Theoretical Physics}\ }\textbf
  {\bibinfo {volume} {132}}~(\bibinfo {number} {4}),\ \bibinfo {pages}
  {606--617}},\ \Eprint {https://arxiv.org/abs/2012.11777} {arXiv:2012.11777
  [cond-mat.supr-con]} \BibitemShut {NoStop}%
\bibitem [{\citenamefont {{Chubukov}}\ and\ \citenamefont
  {{Maslov}}(2017)}]{ChubukovMaslov17}%
  \BibitemOpen
  \bibfield  {author} {\bibinfo {author} {\bibnamefont {{Chubukov}},
  \bibfnamefont {Andrey~V}}, and\ \bibinfo {author} {\bibfnamefont
  {Dmitrii~L.}\ \bibnamefont {{Maslov}}}} (\bibinfo {year} {2017}),\ \bibfield
  {title} {\enquote {\bibinfo {title} {{Optical conductivity of a
  two-dimensional metal near a quantum critical point: The status of the
  extended Drude formula}},}\ }\href
  {https://doi.org/10.1103/PhysRevB.96.205136} {\bibfield  {journal} {\bibinfo
  {journal} {Phys. Rev. B}\ }\textbf {\bibinfo {volume} {96}}~(\bibinfo
  {number} {20}),\ \bibinfo {eid} {205136}},\ \Eprint
  {https://arxiv.org/abs/1707.07352} {arXiv:1707.07352 [cond-mat.str-el]}
  \BibitemShut {NoStop}%
\bibitem [{\citenamefont {Classen}\ and\ \citenamefont
  {Chubukov}(2021)}]{Classen}%
  \BibitemOpen
  \bibfield  {author} {\bibinfo {author} {\bibnamefont {Classen}, \bibfnamefont
  {Laura}}, and\ \bibinfo {author} {\bibfnamefont {Andrey}\ \bibnamefont
  {Chubukov}}} (\bibinfo {year} {2021}),\ \bibfield  {title} {\enquote
  {\bibinfo {title} {{Superconductivity of incoherent electrons in the Yukawa
  Sachdev-Ye-Kitaev model}},}\ }\href
  {https://doi.org/10.1103/PhysRevB.104.125120} {\bibfield  {journal} {\bibinfo
   {journal} {Phys. Rev. B}\ }\textbf {\bibinfo {volume} {104}},\ \bibinfo
  {pages} {125120}}\BibitemShut {NoStop}%
\bibitem [{\citenamefont {Cohen}\ \emph {et~al.}(2015)\citenamefont {Cohen},
  \citenamefont {Gull}, \citenamefont {Reichman},\ and\ \citenamefont
  {Millis}}]{InchwormCohenGullMillisRealTime2015}%
  \BibitemOpen
  \bibfield  {author} {\bibinfo {author} {\bibnamefont {Cohen}, \bibfnamefont
  {Guy}}, \bibinfo {author} {\bibfnamefont {Emanuel}\ \bibnamefont {Gull}},
  \bibinfo {author} {\bibfnamefont {David~R.}\ \bibnamefont {Reichman}}, and\
  \bibinfo {author} {\bibfnamefont {Andrew~J.}\ \bibnamefont {Millis}}}
  (\bibinfo {year} {2015}),\ \bibfield  {title} {\enquote {\bibinfo {title}
  {Taming the dynamical sign problem in real-time evolution of quantum
  many-body problems},}\ }\href
  {https://doi.org/10.1103/PhysRevLett.115.266802} {\bibfield  {journal}
  {\bibinfo  {journal} {Phys. Rev. Lett.}\ }\textbf {\bibinfo {volume} {115}},\
  \bibinfo {pages} {266802}}\BibitemShut {NoStop}%
\bibitem [{\citenamefont {{Coleman}}\ \emph
  {et~al.}(2005{\natexlab{a}})\citenamefont {{Coleman}}, \citenamefont
  {{Marston}},\ and\ \citenamefont {{Schofield}}}]{Coleman_2005}%
  \BibitemOpen
  \bibfield  {author} {\bibinfo {author} {\bibnamefont {{Coleman}},
  \bibfnamefont {P}}, \bibinfo {author} {\bibfnamefont {J.~B.}\ \bibnamefont
  {{Marston}}}, and\ \bibinfo {author} {\bibfnamefont {A.~J.}\ \bibnamefont
  {{Schofield}}}} (\bibinfo {year} {2005}{\natexlab{a}}),\ \bibfield  {title}
  {\enquote {\bibinfo {title} {{Transport anomalies in a simplified model for a
  heavy-electron quantum critical point}},}\ }\href
  {https://doi.org/10.1103/PhysRevB.72.245111} {\bibfield  {journal} {\bibinfo
  {journal} {Phys. Rev. B}\ }\textbf {\bibinfo {volume} {72}}~(\bibinfo
  {number} {24}),\ \bibinfo {eid} {245111}},\ \Eprint
  {https://arxiv.org/abs/cond-mat/0507003} {arXiv:cond-mat/0507003
  [cond-mat.str-el]} \BibitemShut {NoStop}%
\bibitem [{\citenamefont {{Coleman}}\ \emph
  {et~al.}(2005{\natexlab{b}})\citenamefont {{Coleman}}, \citenamefont
  {{Paul}},\ and\ \citenamefont {{Rech}}}]{coleman1}%
  \BibitemOpen
  \bibfield  {author} {\bibinfo {author} {\bibnamefont {{Coleman}},
  \bibfnamefont {P}}, \bibinfo {author} {\bibfnamefont {I.}~\bibnamefont
  {{Paul}}}, and\ \bibinfo {author} {\bibfnamefont {J.}~\bibnamefont {{Rech}}}}
  (\bibinfo {year} {2005}{\natexlab{b}}),\ \bibfield  {title} {\enquote
  {\bibinfo {title} {{Sum rules and Ward identities in the Kondo lattice}},}\
  }\href {https://doi.org/10.1103/PhysRevB.72.094430} {\bibfield  {journal}
  {\bibinfo  {journal} {Phys. Rev. B}\ }\textbf {\bibinfo {volume}
  {72}}~(\bibinfo {number} {9}),\ \bibinfo {eid} {094430}},\ \Eprint
  {https://arxiv.org/abs/cond-mat/0503001} {cond-mat/0503001} \BibitemShut
  {NoStop}%
\bibitem [{\citenamefont {Coleman}\ \emph {et~al.}(2001)\citenamefont
  {Coleman}, \citenamefont {Pépin}, \citenamefont {Si},\ and\ \citenamefont
  {Ramazashvili}}]{Colemanetal}%
  \BibitemOpen
  \bibfield  {author} {\bibinfo {author} {\bibnamefont {Coleman}, \bibfnamefont
  {P}}, \bibinfo {author} {\bibfnamefont {C}~\bibnamefont {Pépin}}, \bibinfo
  {author} {\bibfnamefont {Qimiao}\ \bibnamefont {Si}}, and\ \bibinfo {author}
  {\bibfnamefont {R}~\bibnamefont {Ramazashvili}}} (\bibinfo {year} {2001}),\
  \bibfield  {title} {\enquote {\bibinfo {title} {{How do Fermi liquids get
  heavy and die?}}}\ }\href {http://stacks.iop.org/0953-8984/13/i=35/a=202}
  {\bibfield  {journal} {\bibinfo  {journal} {Journal of Physics: Condensed
  Matter}\ }\textbf {\bibinfo {volume} {13}}~(\bibinfo {number} {35}),\
  \bibinfo {pages} {R723}}\BibitemShut {NoStop}%
\bibitem [{\citenamefont {Collignon}\ \emph {et~al.}(2020)\citenamefont
  {Collignon}, \citenamefont {Bourges}, \citenamefont {Fauqu\'e},\ and\
  \citenamefont {Behnia}}]{BehniaPRX}%
  \BibitemOpen
  \bibfield  {author} {\bibinfo {author} {\bibnamefont {Collignon},
  \bibfnamefont {Cl\'ement}}, \bibinfo {author} {\bibfnamefont {Phillipe}\
  \bibnamefont {Bourges}}, \bibinfo {author} {\bibfnamefont {Beno\^{\i}t}\
  \bibnamefont {Fauqu\'e}}, and\ \bibinfo {author} {\bibfnamefont {Kamran}\
  \bibnamefont {Behnia}}} (\bibinfo {year} {2020}),\ \bibfield  {title}
  {\enquote {\bibinfo {title} {Heavy nondegenerate electrons in doped strontium
  titanate},}\ }\href {https://doi.org/10.1103/PhysRevX.10.031025} {\bibfield
  {journal} {\bibinfo  {journal} {Phys. Rev. X}\ }\textbf {\bibinfo {volume}
  {10}},\ \bibinfo {pages} {031025}}\BibitemShut {NoStop}%
\bibitem [{\citenamefont {Collignon}\ \emph {et~al.}(2019)\citenamefont
  {Collignon}, \citenamefont {Lin}, \citenamefont {Rischau}, \citenamefont
  {Fauqué},\ and\ \citenamefont {Behnia}}]{behnia_review}%
  \BibitemOpen
  \bibfield  {author} {\bibinfo {author} {\bibnamefont {Collignon},
  \bibfnamefont {Clément}}, \bibinfo {author} {\bibfnamefont {Xiao}\
  \bibnamefont {Lin}}, \bibinfo {author} {\bibfnamefont {Carl~Willem}\
  \bibnamefont {Rischau}}, \bibinfo {author} {\bibfnamefont {Benoît}\
  \bibnamefont {Fauqué}}, and\ \bibinfo {author} {\bibfnamefont {Kamran}\
  \bibnamefont {Behnia}}} (\bibinfo {year} {2019}),\ \bibfield  {title}
  {\enquote {\bibinfo {title} {Metallicity and superconductivity in doped
  strontium titanate},}\ }\href
  {https://doi.org/10.1146/annurev-conmatphys-031218-013144} {\bibfield
  {journal} {\bibinfo  {journal} {Annual Review of Condensed Matter Physics}\
  }\textbf {\bibinfo {volume} {10}}~(\bibinfo {number} {1}),\ \bibinfo {pages}
  {25--44}}\BibitemShut {NoStop}%
\bibitem [{\citenamefont {Cotler}\ and\ \citenamefont
  {Jensen}(2021)}]{Cotler:2020ugk}%
  \BibitemOpen
  \bibfield  {author} {\bibinfo {author} {\bibnamefont {Cotler}, \bibfnamefont
  {Jordan}}, and\ \bibinfo {author} {\bibfnamefont {Kristan}\ \bibnamefont
  {Jensen}}} (\bibinfo {year} {2021}),\ \bibfield  {title} {\enquote {\bibinfo
  {title} {Ads3 gravity and random cft},}\ }\href
  {https://doi.org/10.1007/JHEP04(2021)033} {\bibfield  {journal} {\bibinfo
  {journal} {Journal of High Energy Physics}\ }\textbf {\bibinfo {volume}
  {2021}}~(\bibinfo {number} {4}),\ \bibinfo {pages} {33}}\BibitemShut
  {NoStop}%
\bibitem [{\citenamefont {Cotler}\ \emph {et~al.}(2017)\citenamefont {Cotler},
  \citenamefont {Gur-Ari}, \citenamefont {Hanada}, \citenamefont {Polchinski},
  \citenamefont {Saad}, \citenamefont {Shenker}, \citenamefont {Stanford},
  \citenamefont {Streicher},\ and\ \citenamefont {Tezuka}}]{Cotler:2016fpe}%
  \BibitemOpen
  \bibfield  {author} {\bibinfo {author} {\bibnamefont {Cotler}, \bibfnamefont
  {Jordan~S}}, \bibinfo {author} {\bibfnamefont {Guy}\ \bibnamefont {Gur-Ari}},
  \bibinfo {author} {\bibfnamefont {Masanori}\ \bibnamefont {Hanada}}, \bibinfo
  {author} {\bibfnamefont {Joseph}\ \bibnamefont {Polchinski}}, \bibinfo
  {author} {\bibfnamefont {Phil}\ \bibnamefont {Saad}}, \bibinfo {author}
  {\bibfnamefont {Stephen~H.}\ \bibnamefont {Shenker}}, \bibinfo {author}
  {\bibfnamefont {Douglas}\ \bibnamefont {Stanford}}, \bibinfo {author}
  {\bibfnamefont {Alexandre}\ \bibnamefont {Streicher}}, and\ \bibinfo {author}
  {\bibfnamefont {Masaki}\ \bibnamefont {Tezuka}}} (\bibinfo {year} {2017}),\
  \bibfield  {title} {\enquote {\bibinfo {title} {{Black Holes and Random
  Matrices}},}\ }\href {https://doi.org/10.1007/JHEP05(2017)118} {\bibfield
  {journal} {\bibinfo  {journal} {JHEP}\ }\textbf {\bibinfo {volume} {05}},\
  \bibinfo {pages} {118}},\ \bibinfo {note} {[Erratum: JHEP 09, 002 (2018)]},\
  \Eprint {https://arxiv.org/abs/1611.04650} {arXiv:1611.04650 [hep-th]}
  \BibitemShut {NoStop}%
\bibitem [{\citenamefont {Cubrovic}\ \emph {et~al.}(2009)\citenamefont
  {Cubrovic}, \citenamefont {Zaanen},\ and\ \citenamefont
  {Schalm}}]{Cubrovic:2009ye}%
  \BibitemOpen
  \bibfield  {author} {\bibinfo {author} {\bibnamefont {Cubrovic},
  \bibfnamefont {Mihailo}}, \bibinfo {author} {\bibfnamefont {Jan}\
  \bibnamefont {Zaanen}}, and\ \bibinfo {author} {\bibfnamefont {Koenraad}\
  \bibnamefont {Schalm}}} (\bibinfo {year} {2009}),\ \bibfield  {title}
  {\enquote {\bibinfo {title} {{String Theory, Quantum Phase Transitions and
  the Emergent Fermi-Liquid}},}\ }\href
  {https://doi.org/10.1126/science.1174962} {\bibfield  {journal} {\bibinfo
  {journal} {Science}\ }\textbf {\bibinfo {volume} {325}},\ \bibinfo {pages}
  {439--444}},\ \Eprint {https://arxiv.org/abs/0904.1993} {arXiv:0904.1993
  [hep-th]} \BibitemShut {NoStop}%
\bibitem [{\citenamefont {Cubrovic}\ \emph {et~al.}(2011)\citenamefont
  {Cubrovic}, \citenamefont {Zaanen},\ and\ \citenamefont
  {Schalm}}]{Cubrovic:2010bf}%
  \BibitemOpen
  \bibfield  {author} {\bibinfo {author} {\bibnamefont {Cubrovic},
  \bibfnamefont {Mihailo}}, \bibinfo {author} {\bibfnamefont {Jan}\
  \bibnamefont {Zaanen}}, and\ \bibinfo {author} {\bibfnamefont {Koenraad}\
  \bibnamefont {Schalm}}} (\bibinfo {year} {2011}),\ \bibfield  {title}
  {\enquote {\bibinfo {title} {{Constructing the AdS Dual of a Fermi Liquid:
  AdS Black Holes with Dirac Hair}},}\ }\href
  {https://doi.org/10.1007/JHEP10(2011)017} {\bibfield  {journal} {\bibinfo
  {journal} {JHEP}\ }\textbf {\bibinfo {volume} {10}},\ \bibinfo {pages}
  {017}},\ \Eprint {https://arxiv.org/abs/1012.5681} {arXiv:1012.5681 [hep-th]}
  \BibitemShut {NoStop}%
\bibitem [{\citenamefont {Cugliandolo}\ \emph {et~al.}(2001)\citenamefont
  {Cugliandolo}, \citenamefont {Grempel},\ and\ \citenamefont
  {da~Silva~Santos}}]{LC01}%
  \BibitemOpen
  \bibfield  {author} {\bibinfo {author} {\bibnamefont {Cugliandolo},
  \bibfnamefont {Leticia~F}}, \bibinfo {author} {\bibfnamefont {D.~R.}\
  \bibnamefont {Grempel}}, and\ \bibinfo {author} {\bibfnamefont
  {Constantino~A.}\ \bibnamefont {da~Silva~Santos}}} (\bibinfo {year} {2001}),\
  \bibfield  {title} {\enquote {\bibinfo {title} {Imaginary-time replica
  formalism study of a quantum spherical p-spin-glass model},}\ }\href
  {https://doi.org/10.1103/PhysRevB.64.014403} {\bibfield  {journal} {\bibinfo
  {journal} {Phys. Rev. B}\ }\textbf {\bibinfo {volume} {64}},\ \bibinfo
  {pages} {014403}}\BibitemShut {NoStop}%
\bibitem [{\citenamefont {Cuomo}\ \emph {et~al.}(2022)\citenamefont {Cuomo},
  \citenamefont {Komargodski}, \citenamefont {Mezei},\ and\ \citenamefont
  {Raviv-Moshe}}]{Cuomo:2022xgw}%
  \BibitemOpen
  \bibfield  {author} {\bibinfo {author} {\bibnamefont {Cuomo}, \bibfnamefont
  {Gabriel}}, \bibinfo {author} {\bibfnamefont {Zohar}\ \bibnamefont
  {Komargodski}}, \bibinfo {author} {\bibfnamefont {M\'ark}\ \bibnamefont
  {Mezei}}, and\ \bibinfo {author} {\bibfnamefont {Avia}\ \bibnamefont
  {Raviv-Moshe}}} (\bibinfo {year} {2022}),\ \bibfield  {title} {\enquote
  {\bibinfo {title} {{Spin Impurities, Wilson Lines and Semiclassics}},}\
  }\href@noop {} {\ }\Eprint {https://arxiv.org/abs/2202.00040}
  {arXiv:2202.00040 [hep-th]} \BibitemShut {NoStop}%
\bibitem [{\citenamefont {{Custers}}\ \emph {et~al.}(2010)\citenamefont
  {{Custers}}, \citenamefont {{Gegenwart}}, \citenamefont {{Geibel}},
  \citenamefont {{Steglich}}, \citenamefont {{Coleman}},\ and\ \citenamefont
  {{Paschen}}}]{Paschen10}%
  \BibitemOpen
  \bibfield  {author} {\bibinfo {author} {\bibnamefont {{Custers}},
  \bibfnamefont {J}}, \bibinfo {author} {\bibfnamefont {P.}~\bibnamefont
  {{Gegenwart}}}, \bibinfo {author} {\bibfnamefont {C.}~\bibnamefont
  {{Geibel}}}, \bibinfo {author} {\bibfnamefont {F.}~\bibnamefont
  {{Steglich}}}, \bibinfo {author} {\bibfnamefont {P.}~\bibnamefont
  {{Coleman}}}, and\ \bibinfo {author} {\bibfnamefont {S.}~\bibnamefont
  {{Paschen}}}} (\bibinfo {year} {2010}),\ \bibfield  {title} {\enquote
  {\bibinfo {title} {{Evidence for a Non-Fermi-Liquid Phase in Ge-Substituted
  YbRh$_{2}$Si$_{2}$}},}\ }\href
  {https://doi.org/10.1103/PhysRevLett.104.186402} {\bibfield  {journal}
  {\bibinfo  {journal} {Phys. Rev. Lett.}\ }\textbf {\bibinfo {volume}
  {104}}~(\bibinfo {number} {18}),\ \bibinfo {eid} {186402}},\ \Eprint
  {https://arxiv.org/abs/1004.0107} {arXiv:1004.0107 [cond-mat.str-el]}
  \BibitemShut {NoStop}%
\bibitem [{\citenamefont {Custers}\ \emph {et~al.}(2003)\citenamefont
  {Custers}, \citenamefont {Gegenwart}, \citenamefont {Wilhelm}, \citenamefont
  {Neumaier}, \citenamefont {Tokiwa}, \citenamefont {Trovarelli}, \citenamefont
  {Geibel}, \citenamefont {Steglich}, \citenamefont {P{\'e}pin},\ and\
  \citenamefont {Coleman}}]{Paschen10a}%
  \BibitemOpen
  \bibfield  {author} {\bibinfo {author} {\bibnamefont {Custers}, \bibfnamefont
  {J}}, \bibinfo {author} {\bibfnamefont {P.}~\bibnamefont {Gegenwart}},
  \bibinfo {author} {\bibfnamefont {H.}~\bibnamefont {Wilhelm}}, \bibinfo
  {author} {\bibfnamefont {K.}~\bibnamefont {Neumaier}}, \bibinfo {author}
  {\bibfnamefont {Y.}~\bibnamefont {Tokiwa}}, \bibinfo {author} {\bibfnamefont
  {O.}~\bibnamefont {Trovarelli}}, \bibinfo {author} {\bibfnamefont
  {C.}~\bibnamefont {Geibel}}, \bibinfo {author} {\bibfnamefont
  {F.}~\bibnamefont {Steglich}}, \bibinfo {author} {\bibfnamefont
  {C.}~\bibnamefont {P{\'e}pin}}, and\ \bibinfo {author} {\bibfnamefont
  {P.}~\bibnamefont {Coleman}}} (\bibinfo {year} {2003}),\ \bibfield  {title}
  {\enquote {\bibinfo {title} {The break-up of heavy electrons at a quantum
  critical point},}\ }\href {https://doi.org/10.1038/nature01774} {\bibfield
  {journal} {\bibinfo  {journal} {Nature}\ }\textbf {\bibinfo {volume}
  {424}}~(\bibinfo {number} {6948}),\ \bibinfo {pages} {524--527}}\BibitemShut
  {NoStop}%
\bibitem [{\citenamefont {Damascelli}\ \emph {et~al.}(2003)\citenamefont
  {Damascelli}, \citenamefont {Hussain},\ and\ \citenamefont {Shen}}]{zxs}%
  \BibitemOpen
  \bibfield  {author} {\bibinfo {author} {\bibnamefont {Damascelli},
  \bibfnamefont {Andrea}}, \bibinfo {author} {\bibfnamefont {Zahid}\
  \bibnamefont {Hussain}}, and\ \bibinfo {author} {\bibfnamefont {Zhi-Xun}\
  \bibnamefont {Shen}}} (\bibinfo {year} {2003}),\ \bibfield  {title} {\enquote
  {\bibinfo {title} {Angle-resolved photoemission studies of the cuprate
  superconductors},}\ }\href {https://doi.org/10.1103/RevModPhys.75.473}
  {\bibfield  {journal} {\bibinfo  {journal} {Rev. Mod. Phys.}\ }\textbf
  {\bibinfo {volume} {75}},\ \bibinfo {pages} {473--541}}\BibitemShut {NoStop}%
\bibitem [{\citenamefont {Damia}\ \emph {et~al.}(2019)\citenamefont {Damia},
  \citenamefont {Kachru}, \citenamefont {Raghu},\ and\ \citenamefont
  {Torroba}}]{Torroba19}%
  \BibitemOpen
  \bibfield  {author} {\bibinfo {author} {\bibnamefont {Damia}, \bibfnamefont
  {Jeremias~Aguilera}}, \bibinfo {author} {\bibfnamefont {Shamit}\ \bibnamefont
  {Kachru}}, \bibinfo {author} {\bibfnamefont {Srinivas}\ \bibnamefont
  {Raghu}}, and\ \bibinfo {author} {\bibfnamefont {Gonzalo}\ \bibnamefont
  {Torroba}}} (\bibinfo {year} {2019}),\ \bibfield  {title} {\enquote {\bibinfo
  {title} {{Two-Dimensional Non-Fermi-Liquid Metals: A Solvable Large-$N$
  Limit}},}\ }\href {https://doi.org/10.1103/PhysRevLett.123.096402} {\bibfield
   {journal} {\bibinfo  {journal} {Phys. Rev. Lett.}\ }\textbf {\bibinfo
  {volume} {123}},\ \bibinfo {pages} {096402}}\BibitemShut {NoStop}%
\bibitem [{\citenamefont {Danshita}\ \emph {et~al.}(2017)\citenamefont
  {Danshita}, \citenamefont {Hanada},\ and\ \citenamefont
  {Tezuka}}]{Danshita:2016xbo}%
  \BibitemOpen
  \bibfield  {author} {\bibinfo {author} {\bibnamefont {Danshita},
  \bibfnamefont {Ippei}}, \bibinfo {author} {\bibfnamefont {Masanori}\
  \bibnamefont {Hanada}}, and\ \bibinfo {author} {\bibfnamefont {Masaki}\
  \bibnamefont {Tezuka}}} (\bibinfo {year} {2017}),\ \bibfield  {title}
  {\enquote {\bibinfo {title} {{Creating and probing the Sachdev-Ye-Kitaev
  model with ultracold gases: Towards experimental studies of quantum
  gravity}},}\ }\href {https://doi.org/10.1093/ptep/ptx108} {\bibfield
  {journal} {\bibinfo  {journal} {PTEP}\ }\textbf {\bibinfo {volume}
  {2017}}~(\bibinfo {number} {8}),\ \bibinfo {pages} {083I01}},\ \Eprint
  {https://arxiv.org/abs/1606.02454} {arXiv:1606.02454 [cond-mat.quant-gas]}
  \BibitemShut {NoStop}%
\bibitem [{\citenamefont {{Darius Shi}}\ \emph {et~al.}(2022)\citenamefont
  {{Darius Shi}}, \citenamefont {{Goldman}}, \citenamefont {{Else}},\ and\
  \citenamefont {{Senthil}}}]{Shi22}%
  \BibitemOpen
  \bibfield  {author} {\bibinfo {author} {\bibnamefont {{Darius Shi}},
  \bibfnamefont {Zhengyan}}, \bibinfo {author} {\bibfnamefont {Hart}\
  \bibnamefont {{Goldman}}}, \bibinfo {author} {\bibfnamefont {Dominic~V.}\
  \bibnamefont {{Else}}}, and\ \bibinfo {author} {\bibfnamefont
  {T.}~\bibnamefont {{Senthil}}}} (\bibinfo {year} {2022}),\ \bibfield  {title}
  {\enquote {\bibinfo {title} {{Gifts from anomalies: Exact results for Landau
  phase transitions in metals}},}\ }\href@noop {} {\ }\Eprint
  {https://arxiv.org/abs/2204.07585} {arXiv:2204.07585 [cond-mat.str-el]}
  \BibitemShut {NoStop}%
\bibitem [{\citenamefont {{Datta}}\ \emph {et~al.}(2022)\citenamefont
  {{Datta}}, \citenamefont {{Duary}}, \citenamefont {{Kraus}}, \citenamefont
  {{Maity}},\ and\ \citenamefont {{Maloney}}}]{Datta:2021ftn}%
  \BibitemOpen
  \bibfield  {author} {\bibinfo {author} {\bibnamefont {{Datta}}, \bibfnamefont
  {Shouvik}}, \bibinfo {author} {\bibfnamefont {Sarthak}\ \bibnamefont
  {{Duary}}}, \bibinfo {author} {\bibfnamefont {Per}\ \bibnamefont {{Kraus}}},
  \bibinfo {author} {\bibfnamefont {Pronobesh}\ \bibnamefont {{Maity}}}, and\
  \bibinfo {author} {\bibfnamefont {Alexander}\ \bibnamefont {{Maloney}}}}
  (\bibinfo {year} {2022}),\ \bibfield  {title} {\enquote {\bibinfo {title}
  {{Adding flavor to the Narain ensemble}},}\ }\href
  {https://doi.org/10.1007/JHEP05(2022)090} {\bibfield  {journal} {\bibinfo
  {journal} {Journal of High Energy Physics}\ }\textbf {\bibinfo {volume}
  {2022}}~(\bibinfo {number} {5}),\ \bibinfo {eid} {90}},\ \Eprint
  {https://arxiv.org/abs/2102.12509} {arXiv:2102.12509 [hep-th]} \BibitemShut
  {NoStop}%
\bibitem [{\citenamefont {Davison}\ \emph {et~al.}(2017)\citenamefont
  {Davison}, \citenamefont {Fu}, \citenamefont {Georges}, \citenamefont {Gu},
  \citenamefont {Jensen},\ and\ \citenamefont {Sachdev}}]{SS17}%
  \BibitemOpen
  \bibfield  {author} {\bibinfo {author} {\bibnamefont {Davison}, \bibfnamefont
  {Richard~A}}, \bibinfo {author} {\bibfnamefont {Wenbo}\ \bibnamefont {Fu}},
  \bibinfo {author} {\bibfnamefont {Antoine}\ \bibnamefont {Georges}}, \bibinfo
  {author} {\bibfnamefont {Yingfei}\ \bibnamefont {Gu}}, \bibinfo {author}
  {\bibfnamefont {Kristan}\ \bibnamefont {Jensen}}, and\ \bibinfo {author}
  {\bibfnamefont {Subir}\ \bibnamefont {Sachdev}}} (\bibinfo {year} {2017}),\
  \bibfield  {title} {\enquote {\bibinfo {title} {{Thermoelectric transport in
  disordered metals without quasiparticles: The Sachdev-Ye-Kitaev models and
  holography}},}\ }\href {https://doi.org/10.1103/PhysRevB.95.155131}
  {\bibfield  {journal} {\bibinfo  {journal} {Phys. Rev. B}\ }\textbf {\bibinfo
  {volume} {95}},\ \bibinfo {pages} {155131}}\BibitemShut {NoStop}%
\bibitem [{\citenamefont {Deng}\ \emph {et~al.}(2013)\citenamefont {Deng},
  \citenamefont {Mravlje}, \citenamefont {\ifmmode~\check{Z}\else
  \v{Z}\fi{}itko}, \citenamefont {Ferrero}, \citenamefont {Kotliar},\ and\
  \citenamefont {Georges}}]{Deng2013}%
  \BibitemOpen
  \bibfield  {author} {\bibinfo {author} {\bibnamefont {Deng}, \bibfnamefont
  {Xiaoyu}}, \bibinfo {author} {\bibfnamefont {Jernej}\ \bibnamefont
  {Mravlje}}, \bibinfo {author} {\bibfnamefont {Rok}\ \bibnamefont
  {\ifmmode~\check{Z}\else \v{Z}\fi{}itko}}, \bibinfo {author} {\bibfnamefont
  {Michel}\ \bibnamefont {Ferrero}}, \bibinfo {author} {\bibfnamefont
  {Gabriel}\ \bibnamefont {Kotliar}}, and\ \bibinfo {author} {\bibfnamefont
  {Antoine}\ \bibnamefont {Georges}}} (\bibinfo {year} {2013}),\ \bibfield
  {title} {\enquote {\bibinfo {title} {{How Bad Metals Turn Good: Spectroscopic
  Signatures of Resilient Quasiparticles}},}\ }\href
  {https://doi.org/10.1103/PhysRevLett.110.086401} {\bibfield  {journal}
  {\bibinfo  {journal} {Phys. Rev. Lett.}\ }\textbf {\bibinfo {volume} {110}},\
  \bibinfo {pages} {086401}}\BibitemShut {NoStop}%
\bibitem [{\citenamefont {Deng}\ \emph {et~al.}(2014)\citenamefont {Deng},
  \citenamefont {Sternbach}, \citenamefont {Haule}, \citenamefont {Basov},\
  and\ \citenamefont {Kotliar}}]{Deng2014}%
  \BibitemOpen
  \bibfield  {author} {\bibinfo {author} {\bibnamefont {Deng}, \bibfnamefont
  {Xiaoyu}}, \bibinfo {author} {\bibfnamefont {Aaron}\ \bibnamefont
  {Sternbach}}, \bibinfo {author} {\bibfnamefont {Kristjan}\ \bibnamefont
  {Haule}}, \bibinfo {author} {\bibfnamefont {D.~N.}\ \bibnamefont {Basov}},
  and\ \bibinfo {author} {\bibfnamefont {Gabriel}\ \bibnamefont {Kotliar}}}
  (\bibinfo {year} {2014}),\ \bibfield  {title} {\enquote {\bibinfo {title}
  {{Shining Light on Transition-Metal Oxides: Unveiling the Hidden Fermi
  Liquid}},}\ }\href {https://doi.org/10.1103/PhysRevLett.113.246404}
  {\bibfield  {journal} {\bibinfo  {journal} {Phys. Rev. Lett.}\ }\textbf
  {\bibinfo {volume} {113}},\ \bibinfo {pages} {246404}}\BibitemShut {NoStop}%
\bibitem [{\citenamefont {Dhar}\ \emph {et~al.}(2019)\citenamefont {Dhar},
  \citenamefont {Gaikwad}, \citenamefont {Joshi}, \citenamefont {Mandal},\ and\
  \citenamefont {Wadia}}]{Dhar:2018pii}%
  \BibitemOpen
  \bibfield  {author} {\bibinfo {author} {\bibnamefont {Dhar}, \bibfnamefont
  {Avinash}}, \bibinfo {author} {\bibfnamefont {Adwait}\ \bibnamefont
  {Gaikwad}}, \bibinfo {author} {\bibfnamefont {Lata~Kh.}\ \bibnamefont
  {Joshi}}, \bibinfo {author} {\bibfnamefont {Gautam}\ \bibnamefont {Mandal}},
  and\ \bibinfo {author} {\bibfnamefont {Spenta~R.}\ \bibnamefont {Wadia}}}
  (\bibinfo {year} {2019}),\ \bibfield  {title} {\enquote {\bibinfo {title}
  {{Gravitational collapse in SYK models and Choptuik-like phenomenon}},}\
  }\href {https://doi.org/10.1007/JHEP11(2019)067} {\bibfield  {journal}
  {\bibinfo  {journal} {JHEP}\ }\textbf {\bibinfo {volume} {11}},\ \bibinfo
  {pages} {067}},\ \Eprint {https://arxiv.org/abs/1812.03979} {arXiv:1812.03979
  [hep-th]} \BibitemShut {NoStop}%
\bibitem [{\citenamefont {Dodge}\ \emph {et~al.}(2000)\citenamefont {Dodge},
  \citenamefont {Weber}, \citenamefont {Corson}, \citenamefont {Orenstein},
  \citenamefont {Schlesinger}, \citenamefont {Reiner},\ and\ \citenamefont
  {Beasley}}]{Dodge_2000}%
  \BibitemOpen
  \bibfield  {author} {\bibinfo {author} {\bibnamefont {Dodge}, \bibfnamefont
  {J~S}}, \bibinfo {author} {\bibfnamefont {C.~P.}\ \bibnamefont {Weber}},
  \bibinfo {author} {\bibfnamefont {J.}~\bibnamefont {Corson}}, \bibinfo
  {author} {\bibfnamefont {J.}~\bibnamefont {Orenstein}}, \bibinfo {author}
  {\bibfnamefont {Z.}~\bibnamefont {Schlesinger}}, \bibinfo {author}
  {\bibfnamefont {J.~W.}\ \bibnamefont {Reiner}}, and\ \bibinfo {author}
  {\bibfnamefont {M.~R.}\ \bibnamefont {Beasley}}} (\bibinfo {year} {2000}),\
  \bibfield  {title} {\enquote {\bibinfo {title} {Low-frequency crossover of
  the fractional power-law conductivity in ${\mathrm{srruo}}_{3}$},}\ }\href
  {https://doi.org/10.1103/PhysRevLett.85.4932} {\bibfield  {journal} {\bibinfo
   {journal} {Phys. Rev. Lett.}\ }\textbf {\bibinfo {volume} {85}},\ \bibinfo
  {pages} {4932--4935}}\BibitemShut {NoStop}%
\bibitem [{\citenamefont {Doiron-Leyraud}\ \emph {et~al.}(2007)\citenamefont
  {Doiron-Leyraud}, \citenamefont {Proust}, \citenamefont {LeBoeuf},
  \citenamefont {Levallois}, \citenamefont {Bonnemaison}, \citenamefont
  {Liang}, \citenamefont {Bonn}, \citenamefont {Hardy},\ and\ \citenamefont
  {Taillefer}}]{DoironLeyraud2007}%
  \BibitemOpen
  \bibfield  {author} {\bibinfo {author} {\bibnamefont {Doiron-Leyraud},
  \bibfnamefont {Nicolas}}, \bibinfo {author} {\bibfnamefont {Cyril}\
  \bibnamefont {Proust}}, \bibinfo {author} {\bibfnamefont {David}\
  \bibnamefont {LeBoeuf}}, \bibinfo {author} {\bibfnamefont {Julien}\
  \bibnamefont {Levallois}}, \bibinfo {author} {\bibfnamefont {Jean-Baptiste}\
  \bibnamefont {Bonnemaison}}, \bibinfo {author} {\bibfnamefont {Ruixing}\
  \bibnamefont {Liang}}, \bibinfo {author} {\bibfnamefont {D.~A.}\ \bibnamefont
  {Bonn}}, \bibinfo {author} {\bibfnamefont {W.~N.}\ \bibnamefont {Hardy}},
  and\ \bibinfo {author} {\bibfnamefont {Louis}\ \bibnamefont {Taillefer}}}
  (\bibinfo {year} {2007}),\ \bibfield  {title} {\enquote {\bibinfo {title}
  {{Quantum oscillations and the Fermi surface in an underdoped high-$T_c$
  superconductor}},}\ }\href {https://doi.org/10.1038/nature05872} {\bibfield
  {journal} {\bibinfo  {journal} {Nature}\ }\textbf {\bibinfo {volume}
  {447}}~(\bibinfo {number} {7144}),\ \bibinfo {pages} {565--568}}\BibitemShut
  {NoStop}%
\bibitem [{\citenamefont {Dray}\ and\ \citenamefont {{'t
  Hooft}}(1985)}]{Dray85}%
  \BibitemOpen
  \bibfield  {author} {\bibinfo {author} {\bibnamefont {Dray}, \bibfnamefont
  {Tevian}}, and\ \bibinfo {author} {\bibfnamefont {Gerard}\ \bibnamefont {{'t
  Hooft}}}} (\bibinfo {year} {1985}),\ \bibfield  {title} {\enquote {\bibinfo
  {title} {The gravitational shock wave of a massless particle},}\ }\href
  {https://doi.org/https://doi.org/10.1016/0550-3213(85)90525-5} {\bibfield
  {journal} {\bibinfo  {journal} {Nuclear Physics B}\ }\textbf {\bibinfo
  {volume} {253}},\ \bibinfo {pages} {173--188}}\BibitemShut {NoStop}%
\bibitem [{\citenamefont {{Dumitrescu}}\ \emph {et~al.}(2022)\citenamefont
  {{Dumitrescu}}, \citenamefont {{Wentzell}}, \citenamefont {{Georges}},\ and\
  \citenamefont {{Parcollet}}}]{Dumitrescu2021}%
  \BibitemOpen
  \bibfield  {author} {\bibinfo {author} {\bibnamefont {{Dumitrescu}},
  \bibfnamefont {Philipp~T}}, \bibinfo {author} {\bibfnamefont {Nils}\
  \bibnamefont {{Wentzell}}}, \bibinfo {author} {\bibfnamefont {Antoine}\
  \bibnamefont {{Georges}}}, and\ \bibinfo {author} {\bibfnamefont {Olivier}\
  \bibnamefont {{Parcollet}}}} (\bibinfo {year} {2022}),\ \bibfield  {title}
  {\enquote {\bibinfo {title} {{Planckian metal at a doping-induced quantum
  critical point}},}\ }\href {https://doi.org/10.1103/PhysRevB.105.L180404}
  {\bibfield  {journal} {\bibinfo  {journal} {Phys. Rev. B}\ }\textbf {\bibinfo
  {volume} {105}}~(\bibinfo {number} {18}),\ \bibinfo {eid} {L180404}},\
  \Eprint {https://arxiv.org/abs/2103.08607} {arXiv:2103.08607
  [cond-mat.str-el]} \BibitemShut {NoStop}%
\bibitem [{\citenamefont {Eberlein}\ \emph
  {et~al.}(2017{\natexlab{a}})\citenamefont {Eberlein}, \citenamefont {Kasper},
  \citenamefont {Sachdev},\ and\ \citenamefont {Steinberg}}]{Eberlein:2017wah}%
  \BibitemOpen
  \bibfield  {author} {\bibinfo {author} {\bibnamefont {Eberlein},
  \bibfnamefont {Andreas}}, \bibinfo {author} {\bibfnamefont {Valentin}\
  \bibnamefont {Kasper}}, \bibinfo {author} {\bibfnamefont {Subir}\
  \bibnamefont {Sachdev}}, and\ \bibinfo {author} {\bibfnamefont {Julia}\
  \bibnamefont {Steinberg}}} (\bibinfo {year} {2017}{\natexlab{a}}),\ \bibfield
   {title} {\enquote {\bibinfo {title} {{Quantum quench of the
  Sachdev-Ye-Kitaev Model}},}\ }\href
  {https://doi.org/10.1103/PhysRevB.96.205123} {\bibfield  {journal} {\bibinfo
  {journal} {Phys. Rev. B}\ }\textbf {\bibinfo {volume} {96}}~(\bibinfo
  {number} {20}),\ \bibinfo {pages} {205123}},\ \Eprint
  {https://arxiv.org/abs/1706.07803} {arXiv:1706.07803 [cond-mat.str-el]}
  \BibitemShut {NoStop}%
\bibitem [{\citenamefont {Eberlein}\ \emph {et~al.}(2016)\citenamefont
  {Eberlein}, \citenamefont {Mandal},\ and\ \citenamefont
  {Sachdev}}]{Eberlein:2016jlt}%
  \BibitemOpen
  \bibfield  {author} {\bibinfo {author} {\bibnamefont {Eberlein},
  \bibfnamefont {Andreas}}, \bibinfo {author} {\bibfnamefont {Ipsita}\
  \bibnamefont {Mandal}}, and\ \bibinfo {author} {\bibfnamefont {Subir}\
  \bibnamefont {Sachdev}}} (\bibinfo {year} {2016}),\ \bibfield  {title}
  {\enquote {\bibinfo {title} {{Hyperscaling violation at the Ising-nematic
  quantum critical point in two dimensional metals}},}\ }\href
  {https://doi.org/10.1103/PhysRevB.94.045133} {\bibfield  {journal} {\bibinfo
  {journal} {Phys. Rev. B}\ }\textbf {\bibinfo {volume} {94}}~(\bibinfo
  {number} {4}),\ \bibinfo {pages} {045133}},\ \Eprint
  {https://arxiv.org/abs/1605.00657} {arXiv:1605.00657 [cond-mat.str-el]}
  \BibitemShut {NoStop}%
\bibitem [{\citenamefont {Eberlein}\ \emph
  {et~al.}(2017{\natexlab{b}})\citenamefont {Eberlein}, \citenamefont {Patel},\
  and\ \citenamefont {Sachdev}}]{Patel:2016ymd}%
  \BibitemOpen
  \bibfield  {author} {\bibinfo {author} {\bibnamefont {Eberlein},
  \bibfnamefont {Andreas}}, \bibinfo {author} {\bibfnamefont {Aavishkar~A.}\
  \bibnamefont {Patel}}, and\ \bibinfo {author} {\bibfnamefont {Subir}\
  \bibnamefont {Sachdev}}} (\bibinfo {year} {2017}{\natexlab{b}}),\ \bibfield
  {title} {\enquote {\bibinfo {title} {{Shear viscosity at the Ising-nematic
  quantum critical point in two dimensional metals}},}\ }\href
  {https://doi.org/10.1103/PhysRevB.95.075127} {\bibfield  {journal} {\bibinfo
  {journal} {Phys. Rev. B}\ }\textbf {\bibinfo {volume} {95}}~(\bibinfo
  {number} {7}),\ \bibinfo {pages} {075127}},\ \Eprint
  {https://arxiv.org/abs/1607.03894} {arXiv:1607.03894 [cond-mat.str-el]}
  \BibitemShut {NoStop}%
\bibitem [{\citenamefont {El~Azrak}\ \emph {et~al.}(1994)\citenamefont
  {El~Azrak}, \citenamefont {Nahoum}, \citenamefont {Bontemps}, \citenamefont
  {Guilloux-Viry}, \citenamefont {Thivet}, \citenamefont {Perrin},
  \citenamefont {Labdi}, \citenamefont {Li},\ and\ \citenamefont
  {Raffy}}]{ElAzrak_1994}%
  \BibitemOpen
  \bibfield  {author} {\bibinfo {author} {\bibnamefont {El~Azrak},
  \bibfnamefont {A}}, \bibinfo {author} {\bibfnamefont {R.}~\bibnamefont
  {Nahoum}}, \bibinfo {author} {\bibfnamefont {N.}~\bibnamefont {Bontemps}},
  \bibinfo {author} {\bibfnamefont {M.}~\bibnamefont {Guilloux-Viry}}, \bibinfo
  {author} {\bibfnamefont {C.}~\bibnamefont {Thivet}}, \bibinfo {author}
  {\bibfnamefont {A.}~\bibnamefont {Perrin}}, \bibinfo {author} {\bibfnamefont
  {S.}~\bibnamefont {Labdi}}, \bibinfo {author} {\bibfnamefont {Z.~Z.}\
  \bibnamefont {Li}}, and\ \bibinfo {author} {\bibfnamefont {H.}~\bibnamefont
  {Raffy}}} (\bibinfo {year} {1994}),\ \bibfield  {title} {\enquote {\bibinfo
  {title} {{Infrared properties of
  ${\mathrm{YBa}}_{2}$${\mathrm{Cu}}_{3}$${\mathrm{O}}_{7}$ and
  ${\mathrm{Bi}}_{2}$${\mathrm{Sr}}_{2}$${\mathrm{Ca}}_{\mathit{n}\mathrm{\ensuremath{-}}1}$${\mathrm{Cu}}_{\mathit{n}}$${\mathrm{O}}_{2\mathit{n}+4}$
  thin films}},}\ }\href {https://doi.org/10.1103/PhysRevB.49.9846} {\bibfield
  {journal} {\bibinfo  {journal} {Phys. Rev. B}\ }\textbf {\bibinfo {volume}
  {49}},\ \bibinfo {pages} {9846--9856}}\BibitemShut {NoStop}%
\bibitem [{\citenamefont {{Else}}\ and\ \citenamefont
  {{Senthil}}(2021)}]{Else20}%
  \BibitemOpen
  \bibfield  {author} {\bibinfo {author} {\bibnamefont {{Else}}, \bibfnamefont
  {Dominic~V}}, and\ \bibinfo {author} {\bibfnamefont {T.}~\bibnamefont
  {{Senthil}}}} (\bibinfo {year} {2021}),\ \bibfield  {title} {\enquote
  {\bibinfo {title} {{Strange Metals as Ersatz Fermi Liquids}},}\ }\href
  {https://doi.org/10.1103/PhysRevLett.127.086601} {\bibfield  {journal}
  {\bibinfo  {journal} {Phys. Rev. Lett.}\ }\textbf {\bibinfo {volume}
  {127}}~(\bibinfo {number} {8}),\ \bibinfo {eid} {086601}},\ \Eprint
  {https://arxiv.org/abs/2010.10523} {arXiv:2010.10523 [cond-mat.str-el]}
  \BibitemShut {NoStop}%
\bibitem [{\citenamefont {{Else}}\ \emph {et~al.}(2021)\citenamefont {{Else}},
  \citenamefont {{Thorngren}},\ and\ \citenamefont {{Senthil}}}]{ElseTS}%
  \BibitemOpen
  \bibfield  {author} {\bibinfo {author} {\bibnamefont {{Else}}, \bibfnamefont
  {Dominic~V}}, \bibinfo {author} {\bibfnamefont {Ryan}\ \bibnamefont
  {{Thorngren}}}, and\ \bibinfo {author} {\bibfnamefont {T.}~\bibnamefont
  {{Senthil}}}} (\bibinfo {year} {2021}),\ \bibfield  {title} {\enquote
  {\bibinfo {title} {{Non-Fermi Liquids as Ersatz Fermi Liquids: General
  Constraints on Compressible Metals}},}\ }\href
  {https://doi.org/10.1103/PhysRevX.11.021005} {\bibfield  {journal} {\bibinfo
  {journal} {Physical Review X}\ }\textbf {\bibinfo {volume} {11}}~(\bibinfo
  {number} {2}),\ \bibinfo {eid} {021005}},\ \Eprint
  {https://arxiv.org/abs/2007.07896} {arXiv:2007.07896 [cond-mat.str-el]}
  \BibitemShut {NoStop}%
\bibitem [{\citenamefont {Emery}\ and\ \citenamefont
  {Kivelson}(1995)}]{emery_kivelson_prl_1995}%
  \BibitemOpen
  \bibfield  {author} {\bibinfo {author} {\bibnamefont {Emery}, \bibfnamefont
  {V~J}}, and\ \bibinfo {author} {\bibfnamefont {S.~A.}\ \bibnamefont
  {Kivelson}}} (\bibinfo {year} {1995}),\ \bibfield  {title} {\enquote
  {\bibinfo {title} {Superconductivity in bad metals},}\ }\href
  {https://doi.org/10.1103/PhysRevLett.74.3253} {\bibfield  {journal} {\bibinfo
   {journal} {Phys. Rev. Lett.}\ }\textbf {\bibinfo {volume} {74}},\ \bibinfo
  {pages} {3253--3256}}\BibitemShut {NoStop}%
\bibitem [{\citenamefont {{Engelhardt}}\ \emph {et~al.}(2021)\citenamefont
  {{Engelhardt}}, \citenamefont {{Fischetti}},\ and\ \citenamefont
  {{Maloney}}}]{Engelhardt:2020qpv}%
  \BibitemOpen
  \bibfield  {author} {\bibinfo {author} {\bibnamefont {{Engelhardt}},
  \bibfnamefont {Netta}}, \bibinfo {author} {\bibfnamefont {Sebastian}\
  \bibnamefont {{Fischetti}}}, and\ \bibinfo {author} {\bibfnamefont
  {Alexander}\ \bibnamefont {{Maloney}}}} (\bibinfo {year} {2021}),\ \bibfield
  {title} {\enquote {\bibinfo {title} {{Free energy from replica wormholes}},}\
  }\href {https://doi.org/10.1103/PhysRevD.103.046021} {\bibfield  {journal}
  {\bibinfo  {journal} {\prd}\ }\textbf {\bibinfo {volume} {103}}~(\bibinfo
  {number} {4}),\ \bibinfo {eid} {046021}},\ \Eprint
  {https://arxiv.org/abs/2007.07444} {arXiv:2007.07444 [hep-th]} \BibitemShut
  {NoStop}%
\bibitem [{\citenamefont {Esterlis}\ \emph {et~al.}(2021)\citenamefont
  {Esterlis}, \citenamefont {Guo}, \citenamefont {Patel},\ and\ \citenamefont
  {Sachdev}}]{Esterlis:2021eth}%
  \BibitemOpen
  \bibfield  {author} {\bibinfo {author} {\bibnamefont {Esterlis},
  \bibfnamefont {Ilya}}, \bibinfo {author} {\bibfnamefont {Haoyu}\ \bibnamefont
  {Guo}}, \bibinfo {author} {\bibfnamefont {Aavishkar~A.}\ \bibnamefont
  {Patel}}, and\ \bibinfo {author} {\bibfnamefont {Subir}\ \bibnamefont
  {Sachdev}}} (\bibinfo {year} {2021}),\ \bibfield  {title} {\enquote {\bibinfo
  {title} {{Large $N$ theory of critical Fermi surfaces}},}\ }\href
  {https://doi.org/10.1103/PhysRevB.103.235129} {\bibfield  {journal} {\bibinfo
   {journal} {Phys. Rev. B}\ }\textbf {\bibinfo {volume} {103}}~(\bibinfo
  {number} {23}),\ \bibinfo {pages} {235129}},\ \Eprint
  {https://arxiv.org/abs/2103.08615} {arXiv:2103.08615 [cond-mat.str-el]}
  \BibitemShut {NoStop}%
\bibitem [{\citenamefont {Esterlis}\ and\ \citenamefont
  {Schmalian}(2019)}]{JS}%
  \BibitemOpen
  \bibfield  {author} {\bibinfo {author} {\bibnamefont {Esterlis},
  \bibfnamefont {Ilya}}, and\ \bibinfo {author} {\bibfnamefont {J\"org}\
  \bibnamefont {Schmalian}}} (\bibinfo {year} {2019}),\ \bibfield  {title}
  {\enquote {\bibinfo {title} {{Cooper pairing of incoherent electrons: An
  electron-phonon version of the Sachdev-Ye-Kitaev model}},}\ }\href
  {https://doi.org/10.1103/PhysRevB.100.115132} {\bibfield  {journal} {\bibinfo
   {journal} {Phys. Rev. B}\ }\textbf {\bibinfo {volume} {100}},\ \bibinfo
  {pages} {115132}}\BibitemShut {NoStop}%
\bibitem [{\citenamefont {Fang}\ \emph {et~al.}(2022)\citenamefont {Fang},
  \citenamefont {Grissonnanche}, \citenamefont {Legros}, \citenamefont
  {Verret}, \citenamefont {Laliberte}, \citenamefont {Collignon}, \citenamefont
  {Ataei}, \citenamefont {Dion}, \citenamefont {Zhou}, \citenamefont {Graf},
  \citenamefont {Lawler}, \citenamefont {Goddard}, \citenamefont {Taillefer},\
  and\ \citenamefont {Ramshaw}}]{Fang2020}%
  \BibitemOpen
  \bibfield  {author} {\bibinfo {author} {\bibnamefont {Fang}, \bibfnamefont
  {Yawen}}, \bibinfo {author} {\bibfnamefont {Gael}\ \bibnamefont
  {Grissonnanche}}, \bibinfo {author} {\bibfnamefont {Anaelle}\ \bibnamefont
  {Legros}}, \bibinfo {author} {\bibfnamefont {Simon}\ \bibnamefont {Verret}},
  \bibinfo {author} {\bibfnamefont {Francis}\ \bibnamefont {Laliberte}},
  \bibinfo {author} {\bibfnamefont {Clement}\ \bibnamefont {Collignon}},
  \bibinfo {author} {\bibfnamefont {Amirreza}\ \bibnamefont {Ataei}}, \bibinfo
  {author} {\bibfnamefont {Maxime}\ \bibnamefont {Dion}}, \bibinfo {author}
  {\bibfnamefont {Jianshi}\ \bibnamefont {Zhou}}, \bibinfo {author}
  {\bibfnamefont {David}\ \bibnamefont {Graf}}, \bibinfo {author}
  {\bibfnamefont {M.~J.}\ \bibnamefont {Lawler}}, \bibinfo {author}
  {\bibfnamefont {Paul}\ \bibnamefont {Goddard}}, \bibinfo {author}
  {\bibfnamefont {Louis}\ \bibnamefont {Taillefer}}, and\ \bibinfo {author}
  {\bibfnamefont {B.~J.}\ \bibnamefont {Ramshaw}}} (\bibinfo {year} {2022}),\
  \bibfield  {title} {\enquote {\bibinfo {title} {Fermi surface transformation
  at the pseudogap critical point of a cuprate superconductor},}\ }\href
  {https://doi.org/10.1038/s41567-022-01514-1} {\bibfield  {journal} {\bibinfo
  {journal} {Nature Physics}\ }10.1038/s41567-022-01514-1},\ \Eprint
  {https://arxiv.org/abs/2004.01725} {arXiv:2004.01725 [cond-mat.str-el]}
  \BibitemShut {NoStop}%
\bibitem [{\citenamefont {Faulkner}\ and\ \citenamefont
  {Iqbal}(2013)}]{Faulkner:2012gt}%
  \BibitemOpen
  \bibfield  {author} {\bibinfo {author} {\bibnamefont {Faulkner},
  \bibfnamefont {Thomas}}, and\ \bibinfo {author} {\bibfnamefont {Nabil}\
  \bibnamefont {Iqbal}}} (\bibinfo {year} {2013}),\ \bibfield  {title}
  {\enquote {\bibinfo {title} {{Friedel oscillations and horizon charge in 1D
  holographic liquids}},}\ }\href {https://doi.org/10.1007/JHEP07(2013)060}
  {\bibfield  {journal} {\bibinfo  {journal} {JHEP}\ }\textbf {\bibinfo
  {volume} {07}},\ \bibinfo {pages} {060}},\ \Eprint
  {https://arxiv.org/abs/1207.4208} {arXiv:1207.4208 [hep-th]} \BibitemShut
  {NoStop}%
\bibitem [{\citenamefont {Faulkner}\ \emph
  {et~al.}(2011{\natexlab{a}})\citenamefont {Faulkner}, \citenamefont {Iqbal},
  \citenamefont {Liu}, \citenamefont {McGreevy},\ and\ \citenamefont
  {Vegh}}]{Liu3}%
  \BibitemOpen
  \bibfield  {author} {\bibinfo {author} {\bibnamefont {Faulkner},
  \bibfnamefont {Thomas}}, \bibinfo {author} {\bibfnamefont {Nabil}\
  \bibnamefont {Iqbal}}, \bibinfo {author} {\bibfnamefont {Hong}\ \bibnamefont
  {Liu}}, \bibinfo {author} {\bibfnamefont {John}\ \bibnamefont {McGreevy}},
  and\ \bibinfo {author} {\bibfnamefont {David}\ \bibnamefont {Vegh}}}
  (\bibinfo {year} {2011}{\natexlab{a}}),\ \bibfield  {title} {\enquote
  {\bibinfo {title} {{Holographic non-Fermi liquid fixed points}},}\ }\href
  {https://doi.org/10.1098/rsta.2010.0354} {\bibfield  {journal} {\bibinfo
  {journal} {Phil. Trans. Roy. Soc.}\ }\textbf {\bibinfo {volume} {A 369}},\
  \bibinfo {pages} {1640}},\ \Eprint {https://arxiv.org/abs/1101.0597}
  {arXiv:1101.0597 [hep-th]} \BibitemShut {NoStop}%
\bibitem [{\citenamefont {Faulkner}\ \emph
  {et~al.}(2011{\natexlab{b}})\citenamefont {Faulkner}, \citenamefont {Liu},
  \citenamefont {McGreevy},\ and\ \citenamefont {Vegh}}]{Liu2}%
  \BibitemOpen
  \bibfield  {author} {\bibinfo {author} {\bibnamefont {Faulkner},
  \bibfnamefont {Thomas}}, \bibinfo {author} {\bibfnamefont {Hong}\
  \bibnamefont {Liu}}, \bibinfo {author} {\bibfnamefont {John}\ \bibnamefont
  {McGreevy}}, and\ \bibinfo {author} {\bibfnamefont {David}\ \bibnamefont
  {Vegh}}} (\bibinfo {year} {2011}{\natexlab{b}}),\ \bibfield  {title}
  {\enquote {\bibinfo {title} {{Emergent quantum criticality, Fermi surfaces,
  and AdS(2)}},}\ }\href {https://doi.org/10.1103/PhysRevD.83.125002}
  {\bibfield  {journal} {\bibinfo  {journal} {Phys. Rev. D}\ }\textbf {\bibinfo
  {volume} {83}},\ \bibinfo {pages} {125002}},\ \Eprint
  {https://arxiv.org/abs/0907.2694} {arXiv:0907.2694 [hep-th]} \BibitemShut
  {NoStop}%
\bibitem [{\citenamefont {Fiete}(2007)}]{sill}%
  \BibitemOpen
  \bibfield  {author} {\bibinfo {author} {\bibnamefont {Fiete}, \bibfnamefont
  {Gregory~A}}} (\bibinfo {year} {2007}),\ \bibfield  {title} {\enquote
  {\bibinfo {title} {Colloquium: The spin-incoherent luttinger liquid},}\
  }\href {https://doi.org/10.1103/RevModPhys.79.801} {\bibfield  {journal}
  {\bibinfo  {journal} {Rev. Mod. Phys.}\ }\textbf {\bibinfo {volume} {79}},\
  \bibinfo {pages} {801--820}}\BibitemShut {NoStop}%
\bibitem [{\citenamefont {Fisk}\ and\ \citenamefont {Webb}(1976)}]{Fisk76}%
  \BibitemOpen
  \bibfield  {author} {\bibinfo {author} {\bibnamefont {Fisk}, \bibfnamefont
  {Z}}, and\ \bibinfo {author} {\bibfnamefont {G.~W.}\ \bibnamefont {Webb}}}
  (\bibinfo {year} {1976}),\ \bibfield  {title} {\enquote {\bibinfo {title}
  {Saturation of the high-temperature normal-state electrical resistivity of
  superconductors},}\ }\href {https://doi.org/10.1103/PhysRevLett.36.1084}
  {\bibfield  {journal} {\bibinfo  {journal} {Phys. Rev. Lett.}\ }\textbf
  {\bibinfo {volume} {36}},\ \bibinfo {pages} {1084--1086}}\BibitemShut
  {NoStop}%
\bibitem [{\citenamefont {Fitzpatrick}\ \emph {et~al.}(2013)\citenamefont
  {Fitzpatrick}, \citenamefont {Kachru}, \citenamefont {Kaplan},\ and\
  \citenamefont {Raghu}}]{Fitzpatrick:2013mja}%
  \BibitemOpen
  \bibfield  {author} {\bibinfo {author} {\bibnamefont {Fitzpatrick},
  \bibfnamefont {A~Liam}}, \bibinfo {author} {\bibfnamefont {Shamit}\
  \bibnamefont {Kachru}}, \bibinfo {author} {\bibfnamefont {Jared}\
  \bibnamefont {Kaplan}}, and\ \bibinfo {author} {\bibfnamefont
  {S.}~\bibnamefont {Raghu}}} (\bibinfo {year} {2013}),\ \bibfield  {title}
  {\enquote {\bibinfo {title} {{Non-Fermi liquid fixed point in a Wilsonian
  theory of quantum critical metals}},}\ }\href
  {https://doi.org/10.1103/PhysRevB.88.125116} {\bibfield  {journal} {\bibinfo
  {journal} {Phys. Rev. B}\ }\textbf {\bibinfo {volume} {88}},\ \bibinfo
  {pages} {125116}},\ \Eprint {https://arxiv.org/abs/1307.0004}
  {arXiv:1307.0004 [cond-mat.str-el]} \BibitemShut {NoStop}%
\bibitem [{\citenamefont {Fitzpatrick}\ \emph {et~al.}(2014)\citenamefont
  {Fitzpatrick}, \citenamefont {Kachru}, \citenamefont {Kaplan},\ and\
  \citenamefont {Raghu}}]{Fitzpatrick:2013rfa}%
  \BibitemOpen
  \bibfield  {author} {\bibinfo {author} {\bibnamefont {Fitzpatrick},
  \bibfnamefont {A~Liam}}, \bibinfo {author} {\bibfnamefont {Shamit}\
  \bibnamefont {Kachru}}, \bibinfo {author} {\bibfnamefont {Jared}\
  \bibnamefont {Kaplan}}, and\ \bibinfo {author} {\bibfnamefont
  {S.}~\bibnamefont {Raghu}}} (\bibinfo {year} {2014}),\ \bibfield  {title}
  {\enquote {\bibinfo {title} {{Non-Fermi-liquid behavior of large-$N_B$
  quantum critical metals}},}\ }\href
  {https://doi.org/10.1103/PhysRevB.89.165114} {\bibfield  {journal} {\bibinfo
  {journal} {Phys. Rev. B}\ }\textbf {\bibinfo {volume} {89}}~(\bibinfo
  {number} {16}),\ \bibinfo {pages} {165114}},\ \Eprint
  {https://arxiv.org/abs/1312.3321} {arXiv:1312.3321 [cond-mat.str-el]}
  \BibitemShut {NoStop}%
\bibitem [{\citenamefont {Florens}\ and\ \citenamefont
  {Georges}(2004)}]{Florens2004}%
  \BibitemOpen
  \bibfield  {author} {\bibinfo {author} {\bibnamefont {Florens}, \bibfnamefont
  {Serge}}, and\ \bibinfo {author} {\bibfnamefont {Antoine}\ \bibnamefont
  {Georges}}} (\bibinfo {year} {2004}),\ \bibfield  {title} {\enquote {\bibinfo
  {title} {Slave-rotor mean-field theories of strongly correlated systems and
  the mott transition in finite dimensions},}\ }\href
  {https://doi.org/10.1103/PhysRevB.70.035114} {\bibfield  {journal} {\bibinfo
  {journal} {Phys. Rev. B}\ }\textbf {\bibinfo {volume} {70}},\ \bibinfo
  {pages} {035114}}\BibitemShut {NoStop}%
\bibitem [{\citenamefont {Florens}\ \emph {et~al.}(2013)\citenamefont
  {Florens}, \citenamefont {Mohan}, \citenamefont {Janani}, \citenamefont
  {Gupta},\ and\ \citenamefont {Narayanan}}]{Florens2013}%
  \BibitemOpen
  \bibfield  {author} {\bibinfo {author} {\bibnamefont {Florens}, \bibfnamefont
  {Serge}}, \bibinfo {author} {\bibfnamefont {Priyanka}\ \bibnamefont {Mohan}},
  \bibinfo {author} {\bibfnamefont {C.}~\bibnamefont {Janani}}, \bibinfo
  {author} {\bibfnamefont {T.}~\bibnamefont {Gupta}}, and\ \bibinfo {author}
  {\bibfnamefont {R.}~\bibnamefont {Narayanan}}} (\bibinfo {year} {2013}),\
  \bibfield  {title} {\enquote {\bibinfo {title} {{Magnetic fluctuations near
  the Mott transition towards a spin liquid state}},}\ }\href
  {https://doi.org/10.1209/0295-5075/103/17002} {\bibfield  {journal} {\bibinfo
   {journal} {Europhysics Letters}\ }\textbf {\bibinfo {volume}
  {103}}~(\bibinfo {number} {1}),\ \bibinfo {pages} {17002}}\BibitemShut
  {NoStop}%
\bibitem [{\citenamefont {Frachet}\ \emph {et~al.}(2020)\citenamefont
  {Frachet}, \citenamefont {Vinograd}, \citenamefont {Zhou}, \citenamefont
  {Benhabib}, \citenamefont {Wu}, \citenamefont {Mayaffre}, \citenamefont
  {Kr{\"a}mer}, \citenamefont {Ramakrishna}, \citenamefont {Reyes},
  \citenamefont {Debray}, \citenamefont {Kurosawa}, \citenamefont {Momono},
  \citenamefont {Oda}, \citenamefont {Komiya}, \citenamefont {Ono},
  \citenamefont {Horio}, \citenamefont {Chang}, \citenamefont {Proust},
  \citenamefont {LeBoeuf},\ and\ \citenamefont {Julien}}]{Frachet2020}%
  \BibitemOpen
  \bibfield  {author} {\bibinfo {author} {\bibnamefont {Frachet}, \bibfnamefont
  {Mehdi}}, \bibinfo {author} {\bibfnamefont {Igor}\ \bibnamefont {Vinograd}},
  \bibinfo {author} {\bibfnamefont {Rui}\ \bibnamefont {Zhou}}, \bibinfo
  {author} {\bibfnamefont {Siham}\ \bibnamefont {Benhabib}}, \bibinfo {author}
  {\bibfnamefont {Shangfei}\ \bibnamefont {Wu}}, \bibinfo {author}
  {\bibfnamefont {Hadrien}\ \bibnamefont {Mayaffre}}, \bibinfo {author}
  {\bibfnamefont {Steffen}\ \bibnamefont {Kr{\"a}mer}}, \bibinfo {author}
  {\bibfnamefont {Sanath~K.}\ \bibnamefont {Ramakrishna}}, \bibinfo {author}
  {\bibfnamefont {Arneil~P.}\ \bibnamefont {Reyes}}, \bibinfo {author}
  {\bibfnamefont {J{\'e}r{\^o}me}\ \bibnamefont {Debray}}, \bibinfo {author}
  {\bibfnamefont {Tohru}\ \bibnamefont {Kurosawa}}, \bibinfo {author}
  {\bibfnamefont {Naoki}\ \bibnamefont {Momono}}, \bibinfo {author}
  {\bibfnamefont {Migaku}\ \bibnamefont {Oda}}, \bibinfo {author}
  {\bibfnamefont {Seiki}\ \bibnamefont {Komiya}}, \bibinfo {author}
  {\bibfnamefont {Shimpei}\ \bibnamefont {Ono}}, \bibinfo {author}
  {\bibfnamefont {Masafumi}\ \bibnamefont {Horio}}, \bibinfo {author}
  {\bibfnamefont {Johan}\ \bibnamefont {Chang}}, \bibinfo {author}
  {\bibfnamefont {Cyril}\ \bibnamefont {Proust}}, \bibinfo {author}
  {\bibfnamefont {David}\ \bibnamefont {LeBoeuf}}, and\ \bibinfo {author}
  {\bibfnamefont {Marc-Henri}\ \bibnamefont {Julien}}} (\bibinfo {year}
  {2020}),\ \bibfield  {title} {\enquote {\bibinfo {title} {Hidden magnetism at
  the pseudogap critical point of a cuprate superconductor},}\ }\href
  {https://doi.org/10.1038/s41567-020-0950-5} {\bibfield  {journal} {\bibinfo
  {journal} {Nature Physics}\ }\textbf {\bibinfo {volume} {16}}~(\bibinfo
  {number} {10}),\ \bibinfo {pages} {1064--1068}}\BibitemShut {NoStop}%
\bibitem [{\citenamefont {Franz}\ and\ \citenamefont
  {Rozali}(2018)}]{Franz2018_review}%
  \BibitemOpen
  \bibfield  {author} {\bibinfo {author} {\bibnamefont {Franz}, \bibfnamefont
  {Marcel}}, and\ \bibinfo {author} {\bibfnamefont {Moshe}\ \bibnamefont
  {Rozali}}} (\bibinfo {year} {2018}),\ \bibfield  {title} {\enquote {\bibinfo
  {title} {Mimicking black hole event horizons in atomic and solid-state
  systems},}\ }\href {https://doi.org/10.1038/s41578-018-0058-z} {\bibfield
  {journal} {\bibinfo  {journal} {Nature Reviews Materials}\ }\textbf {\bibinfo
  {volume} {3}}~(\bibinfo {number} {12}),\ \bibinfo {pages}
  {491--501}}\BibitemShut {NoStop}%
\bibitem [{\citenamefont {{Fritz}}\ and\ \citenamefont
  {{Vojta}}(2004)}]{FritzVojta2004}%
  \BibitemOpen
  \bibfield  {author} {\bibinfo {author} {\bibnamefont {{Fritz}}, \bibfnamefont
  {Lars}}, and\ \bibinfo {author} {\bibfnamefont {Matthias}\ \bibnamefont
  {{Vojta}}}} (\bibinfo {year} {2004}),\ \bibfield  {title} {\enquote {\bibinfo
  {title} {{Phase transitions in the pseudogap Anderson and Kondo
  models:{\quad} Critical dimensions, renormalization group, and local-moment
  criticality}},}\ }\href {https://doi.org/10.1103/PhysRevB.70.214427}
  {\bibfield  {journal} {\bibinfo  {journal} {Phys. Rev. B}\ }\textbf {\bibinfo
  {volume} {70}}~(\bibinfo {number} {21}),\ \bibinfo {eid} {214427}},\ \Eprint
  {https://arxiv.org/abs/cond-mat/0408543} {arXiv:cond-mat/0408543
  [cond-mat.str-el]} \BibitemShut {NoStop}%
\bibitem [{\citenamefont {Fu}\ \emph {et~al.}(2017)\citenamefont {Fu},
  \citenamefont {Gaiotto}, \citenamefont {Maldacena},\ and\ \citenamefont
  {Sachdev}}]{Fu:2016vas}%
  \BibitemOpen
  \bibfield  {author} {\bibinfo {author} {\bibnamefont {Fu}, \bibfnamefont
  {Wenbo}}, \bibinfo {author} {\bibfnamefont {Davide}\ \bibnamefont {Gaiotto}},
  \bibinfo {author} {\bibfnamefont {Juan}\ \bibnamefont {Maldacena}}, and\
  \bibinfo {author} {\bibfnamefont {Subir}\ \bibnamefont {Sachdev}}} (\bibinfo
  {year} {2017}),\ \bibfield  {title} {\enquote {\bibinfo {title}
  {{Supersymmetric Sachdev-Ye-Kitaev models}},}\ }\href
  {https://doi.org/10.1103/PhysRevD.95.026009} {\bibfield  {journal} {\bibinfo
  {journal} {Phys. Rev. D}\ }\textbf {\bibinfo {volume} {95}}~(\bibinfo
  {number} {2}),\ \bibinfo {pages} {026009}},\ \bibinfo {note} {[Addendum:
  Phys.Rev.D 95, 069904 (2017)]},\ \Eprint {https://arxiv.org/abs/1610.08917}
  {arXiv:1610.08917 [hep-th]} \BibitemShut {NoStop}%
\bibitem [{\citenamefont {Fu}\ and\ \citenamefont {Sachdev}(2016)}]{FuSS}%
  \BibitemOpen
  \bibfield  {author} {\bibinfo {author} {\bibnamefont {Fu}, \bibfnamefont
  {Wenbo}}, and\ \bibinfo {author} {\bibfnamefont {Subir}\ \bibnamefont
  {Sachdev}}} (\bibinfo {year} {2016}),\ \bibfield  {title} {\enquote {\bibinfo
  {title} {Numerical study of fermion and boson models with infinite-range
  random interactions},}\ }\href {https://doi.org/10.1103/PhysRevB.94.035135}
  {\bibfield  {journal} {\bibinfo  {journal} {Phys. Rev. B}\ }\textbf {\bibinfo
  {volume} {94}},\ \bibinfo {pages} {035135}}\BibitemShut {NoStop}%
\bibitem [{\citenamefont {Gaikwad}\ \emph {et~al.}(2020)\citenamefont
  {Gaikwad}, \citenamefont {Joshi}, \citenamefont {Mandal},\ and\ \citenamefont
  {Wadia}}]{Gaikwad:2018dfc}%
  \BibitemOpen
  \bibfield  {author} {\bibinfo {author} {\bibnamefont {Gaikwad}, \bibfnamefont
  {Adwait}}, \bibinfo {author} {\bibfnamefont {Lata~Kh}\ \bibnamefont {Joshi}},
  \bibinfo {author} {\bibfnamefont {Gautam}\ \bibnamefont {Mandal}}, and\
  \bibinfo {author} {\bibfnamefont {Spenta~R.}\ \bibnamefont {Wadia}}}
  (\bibinfo {year} {2020}),\ \bibfield  {title} {\enquote {\bibinfo {title}
  {{Holographic dual to charged SYK from 3D Gravity and Chern-Simons}},}\
  }\href {https://doi.org/10.1007/JHEP02(2020)033} {\bibfield  {journal}
  {\bibinfo  {journal} {JHEP}\ }\textbf {\bibinfo {volume} {02}},\ \bibinfo
  {pages} {033}},\ \Eprint {https://arxiv.org/abs/1802.07746} {arXiv:1802.07746
  [hep-th]} \BibitemShut {NoStop}%
\bibitem [{\citenamefont {{Gannon}}\ \emph {et~al.}(2018)\citenamefont
  {{Gannon}}, \citenamefont {{Wu}}, \citenamefont {{Zaliznyak}}, \citenamefont
  {{Xu}}, \citenamefont {{Tsvelik}}, \citenamefont {{Qiu}}, \citenamefont
  {{Rodriguez-Rivera}},\ and\ \citenamefont {{Aronson}}}]{Aronson18}%
  \BibitemOpen
  \bibfield  {author} {\bibinfo {author} {\bibnamefont {{Gannon}},
  \bibfnamefont {W~J}}, \bibinfo {author} {\bibfnamefont {L.~S.}\ \bibnamefont
  {{Wu}}}, \bibinfo {author} {\bibfnamefont {I.~A.}\ \bibnamefont
  {{Zaliznyak}}}, \bibinfo {author} {\bibfnamefont {W.~H.}\ \bibnamefont
  {{Xu}}}, \bibinfo {author} {\bibfnamefont {A.~M.}\ \bibnamefont {{Tsvelik}}},
  \bibinfo {author} {\bibfnamefont {Y.}~\bibnamefont {{Qiu}}}, \bibinfo
  {author} {\bibfnamefont {J.~A.}\ \bibnamefont {{Rodriguez-Rivera}}}, and\
  \bibinfo {author} {\bibfnamefont {M.~C.}\ \bibnamefont {{Aronson}}}}
  (\bibinfo {year} {2018}),\ \bibfield  {title} {\enquote {\bibinfo {title}
  {{Local quantum phase transition in YFe$_2$Al$_{10}$}},}\ }\href
  {https://doi.org/10.1073/pnas.1721493115} {\bibfield  {journal} {\bibinfo
  {journal} {Proceedings of the National Academy of Science}\ }\textbf
  {\bibinfo {volume} {115}}~(\bibinfo {number} {27}),\ \bibinfo {pages}
  {6995--6999}},\ \Eprint {https://arxiv.org/abs/1712.04033} {arXiv:1712.04033
  [cond-mat.str-el]} \BibitemShut {NoStop}%
\bibitem [{\citenamefont {Gao}\ and\ \citenamefont
  {Jafferis}(2021)}]{Gao:2019nyj}%
  \BibitemOpen
  \bibfield  {author} {\bibinfo {author} {\bibnamefont {Gao}, \bibfnamefont
  {Ping}}, and\ \bibinfo {author} {\bibfnamefont {Daniel~Louis}\ \bibnamefont
  {Jafferis}}} (\bibinfo {year} {2021}),\ \bibfield  {title} {\enquote
  {\bibinfo {title} {{A traversable wormhole teleportation protocol in the SYK
  model}},}\ }\href {https://doi.org/10.1007/JHEP07(2021)097} {\bibfield
  {journal} {\bibinfo  {journal} {JHEP}\ }\textbf {\bibinfo {volume} {07}},\
  \bibinfo {pages} {097}},\ \Eprint {https://arxiv.org/abs/1911.07416}
  {arXiv:1911.07416 [hep-th]} \BibitemShut {NoStop}%
\bibitem [{\citenamefont {Garc\'\i{}a-\'Alvarez}\ \emph
  {et~al.}(2017)\citenamefont {Garc\'\i{}a-\'Alvarez}, \citenamefont
  {Egusquiza}, \citenamefont {Lamata}, \citenamefont {del Campo}, \citenamefont
  {Sonner},\ and\ \citenamefont {Solano}}]{Garcia-Alvarez:2016wem}%
  \BibitemOpen
  \bibfield  {author} {\bibinfo {author} {\bibnamefont {Garc\'\i{}a-\'Alvarez},
  \bibfnamefont {L}}, \bibinfo {author} {\bibfnamefont {I.~L.}\ \bibnamefont
  {Egusquiza}}, \bibinfo {author} {\bibfnamefont {L.}~\bibnamefont {Lamata}},
  \bibinfo {author} {\bibfnamefont {A.}~\bibnamefont {del Campo}}, \bibinfo
  {author} {\bibfnamefont {J.}~\bibnamefont {Sonner}}, and\ \bibinfo {author}
  {\bibfnamefont {E.}~\bibnamefont {Solano}}} (\bibinfo {year} {2017}),\
  \bibfield  {title} {\enquote {\bibinfo {title} {{Digital Quantum Simulation
  of Minimal AdS/CFT}},}\ }\href
  {https://doi.org/10.1103/PhysRevLett.119.040501} {\bibfield  {journal}
  {\bibinfo  {journal} {Phys. Rev. Lett.}\ }\textbf {\bibinfo {volume}
  {119}}~(\bibinfo {number} {4}),\ \bibinfo {pages} {040501}},\ \Eprint
  {https://arxiv.org/abs/1607.08560} {arXiv:1607.08560 [quant-ph]} \BibitemShut
  {NoStop}%
\bibitem [{\citenamefont {Garc\'\i{}a-Garc\'\i{}a}\ \emph
  {et~al.}(2019)\citenamefont {Garc\'\i{}a-Garc\'\i{}a}, \citenamefont
  {Nosaka}, \citenamefont {Rosa},\ and\ \citenamefont
  {Verbaarschot}}]{Garcia-Garcia:2019poj}%
  \BibitemOpen
  \bibfield  {author} {\bibinfo {author} {\bibnamefont
  {Garc\'\i{}a-Garc\'\i{}a}, \bibfnamefont {Antonio~M}}, \bibinfo {author}
  {\bibfnamefont {Tomoki}\ \bibnamefont {Nosaka}}, \bibinfo {author}
  {\bibfnamefont {Dario}\ \bibnamefont {Rosa}}, and\ \bibinfo {author}
  {\bibfnamefont {Jacobus J.~M.}\ \bibnamefont {Verbaarschot}}} (\bibinfo
  {year} {2019}),\ \bibfield  {title} {\enquote {\bibinfo {title} {{Quantum
  chaos transition in a two-site Sachdev-Ye-Kitaev model dual to an eternal
  traversable wormhole}},}\ }\href
  {https://doi.org/10.1103/PhysRevD.100.026002} {\bibfield  {journal} {\bibinfo
   {journal} {Phys. Rev. D}\ }\textbf {\bibinfo {volume} {100}},\ \bibinfo
  {pages} {026002}},\ \Eprint {https://arxiv.org/abs/1901.06031}
  {arXiv:1901.06031 [hep-th]} \BibitemShut {NoStop}%
\bibitem [{\citenamefont {Garc\'\i{}a-Garc\'\i{}a}\ and\ \citenamefont
  {Verbaarschot}(2017)}]{Garcia-Garcia:2017pzl}%
  \BibitemOpen
  \bibfield  {author} {\bibinfo {author} {\bibnamefont
  {Garc\'\i{}a-Garc\'\i{}a}, \bibfnamefont {Antonio~M}}, and\ \bibinfo {author}
  {\bibfnamefont {Jacobus J.~M.}\ \bibnamefont {Verbaarschot}}} (\bibinfo
  {year} {2017}),\ \bibfield  {title} {\enquote {\bibinfo {title} {{Analytical
  Spectral Density of the Sachdev-Ye-Kitaev Model at finite $N$}},}\ }\href
  {https://doi.org/10.1103/PhysRevD.96.066012} {\bibfield  {journal} {\bibinfo
  {journal} {Phys. Rev. D}\ }\textbf {\bibinfo {volume} {96}},\ \bibinfo
  {pages} {066012}},\ \Eprint {https://arxiv.org/abs/1701.06593}
  {arXiv:1701.06593 [hep-th]} \BibitemShut {NoStop}%
\bibitem [{\citenamefont {Georges}\ \emph {et~al.}(1996)\citenamefont
  {Georges}, \citenamefont {Kotliar}, \citenamefont {Krauth},\ and\
  \citenamefont {Rozenberg}}]{DMFT}%
  \BibitemOpen
  \bibfield  {author} {\bibinfo {author} {\bibnamefont {Georges}, \bibfnamefont
  {Antoine}}, \bibinfo {author} {\bibfnamefont {Gabriel}\ \bibnamefont
  {Kotliar}}, \bibinfo {author} {\bibfnamefont {Werner}\ \bibnamefont
  {Krauth}}, and\ \bibinfo {author} {\bibfnamefont {Marcelo~J.}\ \bibnamefont
  {Rozenberg}}} (\bibinfo {year} {1996}),\ \bibfield  {title} {\enquote
  {\bibinfo {title} {Dynamical mean-field theory of strongly correlated fermion
  systems and the limit of infinite dimensions},}\ }\href
  {https://doi.org/10.1103/RevModPhys.68.13} {\bibfield  {journal} {\bibinfo
  {journal} {Rev. Mod. Phys.}\ }\textbf {\bibinfo {volume} {68}},\ \bibinfo
  {pages} {13--125}}\BibitemShut {NoStop}%
\bibitem [{\citenamefont {Georges}\ and\ \citenamefont
  {Mravlje}(2021)}]{Georges2021skewed}%
  \BibitemOpen
  \bibfield  {author} {\bibinfo {author} {\bibnamefont {Georges}, \bibfnamefont
  {Antoine}}, and\ \bibinfo {author} {\bibfnamefont {Jernej}\ \bibnamefont
  {Mravlje}}} (\bibinfo {year} {2021}),\ \href@noop {} {\enquote {\bibinfo
  {title} {{Skewed Non-Fermi Liquids and the Seebeck Effect}},}\ }\Eprint
  {https://arxiv.org/abs/2102.13224} {arXiv:2102.13224 [cond-mat.str-el]}
  \BibitemShut {NoStop}%
\bibitem [{\citenamefont {Georges}\ \emph {et~al.}(2000)\citenamefont
  {Georges}, \citenamefont {Parcollet},\ and\ \citenamefont {Sachdev}}]{GPS1}%
  \BibitemOpen
  \bibfield  {author} {\bibinfo {author} {\bibnamefont {Georges}, \bibfnamefont
  {Antoine}}, \bibinfo {author} {\bibfnamefont {Olivier}\ \bibnamefont
  {Parcollet}}, and\ \bibinfo {author} {\bibfnamefont {Subir}\ \bibnamefont
  {Sachdev}}} (\bibinfo {year} {2000}),\ \bibfield  {title} {\enquote {\bibinfo
  {title} {{Mean Field Theory of a Quantum Heisenberg Spin Glass}},}\ }\href
  {https://doi.org/10.1103/PhysRevLett.85.840} {\bibfield  {journal} {\bibinfo
  {journal} {Phys. Rev. Lett.}\ }\textbf {\bibinfo {volume} {85}},\ \bibinfo
  {pages} {840--843}}\BibitemShut {NoStop}%
\bibitem [{\citenamefont {Georges}\ \emph {et~al.}(2001)\citenamefont
  {Georges}, \citenamefont {Parcollet},\ and\ \citenamefont {Sachdev}}]{GPS2}%
  \BibitemOpen
  \bibfield  {author} {\bibinfo {author} {\bibnamefont {Georges}, \bibfnamefont
  {Antoine}}, \bibinfo {author} {\bibfnamefont {Olivier}\ \bibnamefont
  {Parcollet}}, and\ \bibinfo {author} {\bibfnamefont {Subir}\ \bibnamefont
  {Sachdev}}} (\bibinfo {year} {2001}),\ \bibfield  {title} {\enquote {\bibinfo
  {title} {{Quantum fluctuations of a nearly critical Heisenberg spin
  glass}},}\ }\href {https://doi.org/10.1103/PhysRevB.63.134406} {\bibfield
  {journal} {\bibinfo  {journal} {Phys. Rev. B}\ }\textbf {\bibinfo {volume}
  {63}},\ \bibinfo {pages} {134406}}\BibitemShut {NoStop}%
\bibitem [{\citenamefont {Gharibyan}\ \emph {et~al.}(2018)\citenamefont
  {Gharibyan}, \citenamefont {Hanada}, \citenamefont {Shenker},\ and\
  \citenamefont {Tezuka}}]{ShenkerRMT}%
  \BibitemOpen
  \bibfield  {author} {\bibinfo {author} {\bibnamefont {Gharibyan},
  \bibfnamefont {Hrant}}, \bibinfo {author} {\bibfnamefont {Masanori}\
  \bibnamefont {Hanada}}, \bibinfo {author} {\bibfnamefont {Stephen~H.}\
  \bibnamefont {Shenker}}, and\ \bibinfo {author} {\bibfnamefont {Masaki}\
  \bibnamefont {Tezuka}}} (\bibinfo {year} {2018}),\ \bibfield  {title}
  {\enquote {\bibinfo {title} {Onset of random matrix behavior in scrambling
  systems},}\ }\href {https://doi.org/10.1007/JHEP07(2018)124} {\bibfield
  {journal} {\bibinfo  {journal} {Journal of High Energy Physics}\ }\textbf
  {\bibinfo {volume} {2018}}~(\bibinfo {number} {7}),\ \bibinfo {pages}
  {124}}\BibitemShut {NoStop}%
\bibitem [{\citenamefont {{Ghiotto}}\ \emph {et~al.}(2021)\citenamefont
  {{Ghiotto}}, \citenamefont {{Shih}}, \citenamefont {{Pereira}}, \citenamefont
  {{Rhodes}}, \citenamefont {{Kim}}, \citenamefont {{Zang}}, \citenamefont
  {{Millis}}, \citenamefont {{Watanabe}}, \citenamefont {{Taniguchi}},
  \citenamefont {{Hone}}, \citenamefont {{Wang}}, \citenamefont {{Dean}},\ and\
  \citenamefont {{Pasupathy}}}]{AP21}%
  \BibitemOpen
  \bibfield  {author} {\bibinfo {author} {\bibnamefont {{Ghiotto}},
  \bibfnamefont {Augusto}}, \bibinfo {author} {\bibfnamefont {En-Min}\
  \bibnamefont {{Shih}}}, \bibinfo {author} {\bibfnamefont {Giancarlo
  S.~S.~G.}\ \bibnamefont {{Pereira}}}, \bibinfo {author} {\bibfnamefont
  {Daniel~A.}\ \bibnamefont {{Rhodes}}}, \bibinfo {author} {\bibfnamefont
  {Bumho}\ \bibnamefont {{Kim}}}, \bibinfo {author} {\bibfnamefont {Jiawei}\
  \bibnamefont {{Zang}}}, \bibinfo {author} {\bibfnamefont {Andrew~J.}\
  \bibnamefont {{Millis}}}, \bibinfo {author} {\bibfnamefont {Kenji}\
  \bibnamefont {{Watanabe}}}, \bibinfo {author} {\bibfnamefont {Takashi}\
  \bibnamefont {{Taniguchi}}}, \bibinfo {author} {\bibfnamefont {James~C.}\
  \bibnamefont {{Hone}}}, \bibinfo {author} {\bibfnamefont {Lei}\ \bibnamefont
  {{Wang}}}, \bibinfo {author} {\bibfnamefont {Cory~R.}\ \bibnamefont
  {{Dean}}}, and\ \bibinfo {author} {\bibfnamefont {Abhay~N.}\ \bibnamefont
  {{Pasupathy}}}} (\bibinfo {year} {2021}),\ \bibfield  {title} {\enquote
  {\bibinfo {title} {{Quantum criticality in twisted transition metal
  dichalcogenides}},}\ }\href {https://doi.org/10.1038/s41586-021-03815-6}
  {\bibfield  {journal} {\bibinfo  {journal} {Nature}\ }\textbf {\bibinfo
  {volume} {597}}~(\bibinfo {number} {7876}),\ \bibinfo {pages} {345--349}},\
  \Eprint {https://arxiv.org/abs/2103.09796} {arXiv:2103.09796
  [cond-mat.mes-hall]} \BibitemShut {NoStop}%
\bibitem [{\citenamefont {Giamarchi}(2003)}]{giamarchi}%
  \BibitemOpen
  \bibfield  {author} {\bibinfo {author} {\bibnamefont {Giamarchi},
  \bibfnamefont {Thierry}}} (\bibinfo {year} {2003}),\ \href@noop {} {\emph
  {\bibinfo {title} {Quantum physics in one dimension}}},\ Vol.\ \bibinfo
  {volume} {121}\ (\bibinfo  {publisher} {Clarendon press})\BibitemShut
  {NoStop}%
\bibitem [{\citenamefont {Gibbons}\ and\ \citenamefont
  {Hawking}(1977)}]{Gibbons_Hawking}%
  \BibitemOpen
  \bibfield  {author} {\bibinfo {author} {\bibnamefont {Gibbons}, \bibfnamefont
  {G~W}}, and\ \bibinfo {author} {\bibfnamefont {S.~W.}\ \bibnamefont
  {Hawking}}} (\bibinfo {year} {1977}),\ \bibfield  {title} {\enquote {\bibinfo
  {title} {Action integrals and partition functions in quantum gravity},}\
  }\href {https://doi.org/10.1103/PhysRevD.15.2752} {\bibfield  {journal}
  {\bibinfo  {journal} {Phys. Rev. D}\ }\textbf {\bibinfo {volume} {15}},\
  \bibinfo {pages} {2752--2756}}\BibitemShut {NoStop}%
\bibitem [{\citenamefont {Giraldo-Gallo}\ \emph {et~al.}(2018)\citenamefont
  {Giraldo-Gallo}, \citenamefont {Galvis}, \citenamefont {Stegen},
  \citenamefont {Modic}, \citenamefont {Balakirev}, \citenamefont {Betts},
  \citenamefont {Lian}, \citenamefont {Moir}, \citenamefont {Riggs},
  \citenamefont {Wu}, \citenamefont {Bollinger}, \citenamefont {He},
  \citenamefont {Bo{\v z}ovi{\'c}}, \citenamefont {Ramshaw}, \citenamefont
  {McDonald}, \citenamefont {Boebinger},\ and\ \citenamefont
  {Shekhter}}]{boebinger}%
  \BibitemOpen
  \bibfield  {author} {\bibinfo {author} {\bibnamefont {Giraldo-Gallo},
  \bibfnamefont {P}}, \bibinfo {author} {\bibfnamefont {J.~A.}\ \bibnamefont
  {Galvis}}, \bibinfo {author} {\bibfnamefont {Z.}~\bibnamefont {Stegen}},
  \bibinfo {author} {\bibfnamefont {K.~A.}\ \bibnamefont {Modic}}, \bibinfo
  {author} {\bibfnamefont {F.~F.}\ \bibnamefont {Balakirev}}, \bibinfo {author}
  {\bibfnamefont {J.~B.}\ \bibnamefont {Betts}}, \bibinfo {author}
  {\bibfnamefont {X.}~\bibnamefont {Lian}}, \bibinfo {author} {\bibfnamefont
  {C.}~\bibnamefont {Moir}}, \bibinfo {author} {\bibfnamefont {S.~C.}\
  \bibnamefont {Riggs}}, \bibinfo {author} {\bibfnamefont {J.}~\bibnamefont
  {Wu}}, \bibinfo {author} {\bibfnamefont {A.~T.}\ \bibnamefont {Bollinger}},
  \bibinfo {author} {\bibfnamefont {X.}~\bibnamefont {He}}, \bibinfo {author}
  {\bibfnamefont {I.}~\bibnamefont {Bo{\v z}ovi{\'c}}}, \bibinfo {author}
  {\bibfnamefont {B.~J.}\ \bibnamefont {Ramshaw}}, \bibinfo {author}
  {\bibfnamefont {R.~D.}\ \bibnamefont {McDonald}}, \bibinfo {author}
  {\bibfnamefont {G.~S.}\ \bibnamefont {Boebinger}}, and\ \bibinfo {author}
  {\bibfnamefont {A.}~\bibnamefont {Shekhter}}} (\bibinfo {year} {2018}),\
  \bibfield  {title} {\enquote {\bibinfo {title} {Scale-invariant
  magnetoresistance in a cuprate superconductor},}\ }\href
  {https://doi.org/10.1126/science.aan3178} {\bibfield  {journal} {\bibinfo
  {journal} {Science}\ }\textbf {\bibinfo {volume} {361}}~(\bibinfo {number}
  {6401}),\ \bibinfo {pages} {479--481}}\BibitemShut {NoStop}%
\bibitem [{\citenamefont {Glossop}\ and\ \citenamefont
  {Ingersent}(2007)}]{Ingersent1}%
  \BibitemOpen
  \bibfield  {author} {\bibinfo {author} {\bibnamefont {Glossop}, \bibfnamefont
  {Matthew~T}}, and\ \bibinfo {author} {\bibfnamefont {Kevin}\ \bibnamefont
  {Ingersent}}} (\bibinfo {year} {2007}),\ \bibfield  {title} {\enquote
  {\bibinfo {title} {{Magnetic Quantum Phase Transition in an Anisotropic Kondo
  Lattice}},}\ }\href {https://doi.org/10.1103/PhysRevLett.99.227203}
  {\bibfield  {journal} {\bibinfo  {journal} {Phys. Rev. Lett.}\ }\textbf
  {\bibinfo {volume} {99}},\ \bibinfo {pages} {227203}}\BibitemShut {NoStop}%
\bibitem [{\citenamefont {Gnezdilov}\ \emph {et~al.}(2018)\citenamefont
  {Gnezdilov}, \citenamefont {Hutasoit},\ and\ \citenamefont
  {Beenakker}}]{Beenakker18}%
  \BibitemOpen
  \bibfield  {author} {\bibinfo {author} {\bibnamefont {Gnezdilov},
  \bibfnamefont {N~V}}, \bibinfo {author} {\bibfnamefont {J.~A.}\ \bibnamefont
  {Hutasoit}}, and\ \bibinfo {author} {\bibfnamefont {C.~W.~J.}\ \bibnamefont
  {Beenakker}}} (\bibinfo {year} {2018}),\ \bibfield  {title} {\enquote
  {\bibinfo {title} {{Low-high voltage duality in tunneling spectroscopy of the
  Sachdev-Ye-Kitaev model}},}\ }\href
  {https://doi.org/10.1103/PhysRevB.98.081413} {\bibfield  {journal} {\bibinfo
  {journal} {Phys. Rev. B}\ }\textbf {\bibinfo {volume} {98}},\ \bibinfo
  {pages} {081413}}\BibitemShut {NoStop}%
\bibitem [{\citenamefont {Gourgout}\ \emph {et~al.}(2021)\citenamefont
  {Gourgout}, \citenamefont {Grissonnanche}, \citenamefont {Laliberté},
  \citenamefont {Ataei}, \citenamefont {Chen}, \citenamefont {Verret},
  \citenamefont {Zhou}, \citenamefont {Mravlje}, \citenamefont {Georges},
  \citenamefont {Doiron-Leyraud},\ and\ \citenamefont
  {Taillefer}}]{Gourgout2021}%
  \BibitemOpen
  \bibfield  {author} {\bibinfo {author} {\bibnamefont {Gourgout},
  \bibfnamefont {A}}, \bibinfo {author} {\bibfnamefont {G.}~\bibnamefont
  {Grissonnanche}}, \bibinfo {author} {\bibfnamefont {F.}~\bibnamefont
  {Laliberté}}, \bibinfo {author} {\bibfnamefont {A.}~\bibnamefont {Ataei}},
  \bibinfo {author} {\bibfnamefont {L.}~\bibnamefont {Chen}}, \bibinfo {author}
  {\bibfnamefont {S.}~\bibnamefont {Verret}}, \bibinfo {author} {\bibfnamefont
  {J.~S.}\ \bibnamefont {Zhou}}, \bibinfo {author} {\bibfnamefont
  {J.}~\bibnamefont {Mravlje}}, \bibinfo {author} {\bibfnamefont
  {A.}~\bibnamefont {Georges}}, \bibinfo {author} {\bibfnamefont
  {N.}~\bibnamefont {Doiron-Leyraud}}, and\ \bibinfo {author} {\bibfnamefont
  {L.}~\bibnamefont {Taillefer}}} (\bibinfo {year} {2021}),\ \href@noop {}
  {\enquote {\bibinfo {title} {{Out-of-plane Seebeck coefficient of the cuprate
  La$_{1.6-x}$Nd$_{0.4}$Sr$_{x}$CuO$_4$ across the pseudogap critical point:
  particle-hole asymmetry and Fermi surface transformation}},}\ }\Eprint
  {https://arxiv.org/abs/2106.05959} {arXiv:2106.05959 [cond-mat.str-el]}
  \BibitemShut {NoStop}%
\bibitem [{\citenamefont {Grempel}\ and\ \citenamefont
  {Rozenberg}(1998)}]{GrempelRozenberg98}%
  \BibitemOpen
  \bibfield  {author} {\bibinfo {author} {\bibnamefont {Grempel}, \bibfnamefont
  {D~R}}, and\ \bibinfo {author} {\bibfnamefont {M.~J.}\ \bibnamefont
  {Rozenberg}}} (\bibinfo {year} {1998}),\ \bibfield  {title} {\enquote
  {\bibinfo {title} {{Fluctuations in a Quantum Random Heisenberg
  Paramagnet}},}\ }\href {https://doi.org/10.1103/PhysRevLett.80.389}
  {\bibfield  {journal} {\bibinfo  {journal} {Phys. Rev. Lett.}\ }\textbf
  {\bibinfo {volume} {80}},\ \bibinfo {pages} {389--392}}\BibitemShut {NoStop}%
\bibitem [{\citenamefont {Grempel}\ and\ \citenamefont {Si}(2003)}]{SiIsing1}%
  \BibitemOpen
  \bibfield  {author} {\bibinfo {author} {\bibnamefont {Grempel}, \bibfnamefont
  {D~R}}, and\ \bibinfo {author} {\bibfnamefont {Qimiao}\ \bibnamefont {Si}}}
  (\bibinfo {year} {2003}),\ \bibfield  {title} {\enquote {\bibinfo {title}
  {{Locally Critical Point in an Anisotropic Kondo Lattice}},}\ }\href
  {https://doi.org/10.1103/PhysRevLett.91.026401} {\bibfield  {journal}
  {\bibinfo  {journal} {Phys. Rev. Lett.}\ }\textbf {\bibinfo {volume} {91}},\
  \bibinfo {pages} {026401}}\BibitemShut {NoStop}%
\bibitem [{\citenamefont {{Grissonnanche}}\ \emph {et~al.}(2021)\citenamefont
  {{Grissonnanche}}, \citenamefont {{Fang}}, \citenamefont {{Legros}},
  \citenamefont {{Verret}}, \citenamefont {{Lalibert{\'e}}}, \citenamefont
  {{Collignon}}, \citenamefont {{Zhou}}, \citenamefont {{Graf}}, \citenamefont
  {{Goddard}}, \citenamefont {{Taillefer}},\ and\ \citenamefont
  {{Ramshaw}}}]{Grissonnanche2020}%
  \BibitemOpen
  \bibfield  {author} {\bibinfo {author} {\bibnamefont {{Grissonnanche}},
  \bibfnamefont {G}}, \bibinfo {author} {\bibfnamefont {Y.}~\bibnamefont
  {{Fang}}}, \bibinfo {author} {\bibfnamefont {A.}~\bibnamefont {{Legros}}},
  \bibinfo {author} {\bibfnamefont {S.}~\bibnamefont {{Verret}}}, \bibinfo
  {author} {\bibfnamefont {F.}~\bibnamefont {{Lalibert{\'e}}}}, \bibinfo
  {author} {\bibfnamefont {C.}~\bibnamefont {{Collignon}}}, \bibinfo {author}
  {\bibfnamefont {J.}~\bibnamefont {{Zhou}}}, \bibinfo {author} {\bibfnamefont
  {D.}~\bibnamefont {{Graf}}}, \bibinfo {author} {\bibfnamefont
  {P.}~\bibnamefont {{Goddard}}}, \bibinfo {author} {\bibfnamefont
  {L.}~\bibnamefont {{Taillefer}}}, and\ \bibinfo {author} {\bibfnamefont
  {B.~J.}\ \bibnamefont {{Ramshaw}}}} (\bibinfo {year} {2021}),\ \bibfield
  {title} {\enquote {\bibinfo {title} {{Linear-in temperature resistivity from
  an isotropic Planckian scattering rate}},}\ }\href
  {https://doi.org/10.1038/s41586-021-03697-8} {\bibfield  {journal} {\bibinfo
  {journal} {Nature}\ }\textbf {\bibinfo {volume} {595}},\ \bibinfo {pages}
  {667}},\ \Eprint {https://arxiv.org/abs/2011.13054} {arXiv:2011.13054
  [cond-mat.str-el]} \BibitemShut {NoStop}%
\bibitem [{\citenamefont {Gross}\ and\ \citenamefont
  {Rosenhaus}(2017)}]{Gross:2016kjj}%
  \BibitemOpen
  \bibfield  {author} {\bibinfo {author} {\bibnamefont {Gross}, \bibfnamefont
  {David~J}}, and\ \bibinfo {author} {\bibfnamefont {Vladimir}\ \bibnamefont
  {Rosenhaus}}} (\bibinfo {year} {2017}),\ \bibfield  {title} {\enquote
  {\bibinfo {title} {{A Generalization of Sachdev-Ye-Kitaev}},}\ }\href
  {https://doi.org/10.1007/JHEP02(2017)093} {\bibfield  {journal} {\bibinfo
  {journal} {JHEP}\ }\textbf {\bibinfo {volume} {02}},\ \bibinfo {pages}
  {093}},\ \Eprint {https://arxiv.org/abs/1610.01569} {arXiv:1610.01569
  [hep-th]} \BibitemShut {NoStop}%
\bibitem [{\citenamefont {Grozdanov}\ \emph {et~al.}(2019)\citenamefont
  {Grozdanov}, \citenamefont {Schalm},\ and\ \citenamefont
  {Scopelliti}}]{Grozdanov:2018atb}%
  \BibitemOpen
  \bibfield  {author} {\bibinfo {author} {\bibnamefont {Grozdanov},
  \bibfnamefont {Sa\v{s}o}}, \bibinfo {author} {\bibfnamefont {Koenraad}\
  \bibnamefont {Schalm}}, and\ \bibinfo {author} {\bibfnamefont {Vincenzo}\
  \bibnamefont {Scopelliti}}} (\bibinfo {year} {2019}),\ \bibfield  {title}
  {\enquote {\bibinfo {title} {{Kinetic theory for classical and quantum
  many-body chaos}},}\ }\href {https://doi.org/10.1103/PhysRevE.99.012206}
  {\bibfield  {journal} {\bibinfo  {journal} {Phys. Rev. E}\ }\textbf {\bibinfo
  {volume} {99}}~(\bibinfo {number} {1}),\ \bibinfo {pages} {012206}},\ \Eprint
  {https://arxiv.org/abs/1804.09182} {arXiv:1804.09182 [hep-th]} \BibitemShut
  {NoStop}%
\bibitem [{\citenamefont {Gu}\ and\ \citenamefont {Kitaev}(2019)}]{Gu:2018jsv}%
  \BibitemOpen
  \bibfield  {author} {\bibinfo {author} {\bibnamefont {Gu}, \bibfnamefont
  {Yingfei}}, and\ \bibinfo {author} {\bibfnamefont {Alexei}\ \bibnamefont
  {Kitaev}}} (\bibinfo {year} {2019}),\ \bibfield  {title} {\enquote {\bibinfo
  {title} {{On the relation between the magnitude and exponent of OTOCs}},}\
  }\href {https://doi.org/10.1007/JHEP02(2019)075} {\bibfield  {journal}
  {\bibinfo  {journal} {JHEP}\ }\textbf {\bibinfo {volume} {02}},\ \bibinfo
  {pages} {075}},\ \Eprint {https://arxiv.org/abs/1812.00120} {arXiv:1812.00120
  [hep-th]} \BibitemShut {NoStop}%
\bibitem [{\citenamefont {Gu}\ \emph {et~al.}(2020)\citenamefont {Gu},
  \citenamefont {Kitaev}, \citenamefont {Sachdev},\ and\ \citenamefont
  {Tarnopolsky}}]{GKST}%
  \BibitemOpen
  \bibfield  {author} {\bibinfo {author} {\bibnamefont {Gu}, \bibfnamefont
  {Yingfei}}, \bibinfo {author} {\bibfnamefont {Alexei}\ \bibnamefont
  {Kitaev}}, \bibinfo {author} {\bibfnamefont {Subir}\ \bibnamefont {Sachdev}},
  and\ \bibinfo {author} {\bibfnamefont {Grigory}\ \bibnamefont {Tarnopolsky}}}
  (\bibinfo {year} {2020}),\ \bibfield  {title} {\enquote {\bibinfo {title}
  {{Notes on the complex Sachdev-Ye-Kitaev model}},}\ }\href
  {https://doi.org/10.1007/JHEP02(2020)157} {\bibfield  {journal} {\bibinfo
  {journal} {JHEP}\ }\textbf {\bibinfo {volume} {02}},\ \bibinfo {pages}
  {157}},\ \Eprint {https://arxiv.org/abs/1910.14099} {arXiv:1910.14099
  [hep-th]} \BibitemShut {NoStop}%
\bibitem [{\citenamefont {Gu}\ \emph {et~al.}(2017{\natexlab{a}})\citenamefont
  {Gu}, \citenamefont {Lucas},\ and\ \citenamefont {Qi}}]{Gu:2017ohj}%
  \BibitemOpen
  \bibfield  {author} {\bibinfo {author} {\bibnamefont {Gu}, \bibfnamefont
  {Yingfei}}, \bibinfo {author} {\bibfnamefont {Andrew}\ \bibnamefont {Lucas}},
  and\ \bibinfo {author} {\bibfnamefont {Xiao-Liang}\ \bibnamefont {Qi}}}
  (\bibinfo {year} {2017}{\natexlab{a}}),\ \bibfield  {title} {\enquote
  {\bibinfo {title} {{Energy diffusion and the butterfly effect in
  inhomogeneous Sachdev-Ye-Kitaev chains}},}\ }\href
  {https://doi.org/10.21468/SciPostPhys.2.3.018} {\bibfield  {journal}
  {\bibinfo  {journal} {SciPost Phys.}\ }\textbf {\bibinfo {volume}
  {2}}~(\bibinfo {number} {3}),\ \bibinfo {pages} {018}},\ \Eprint
  {https://arxiv.org/abs/1702.08462} {arXiv:1702.08462 [hep-th]} \BibitemShut
  {NoStop}%
\bibitem [{\citenamefont {Gu}\ \emph {et~al.}(2017{\natexlab{b}})\citenamefont
  {Gu}, \citenamefont {Qi},\ and\ \citenamefont {Stanford}}]{Gu17}%
  \BibitemOpen
  \bibfield  {author} {\bibinfo {author} {\bibnamefont {Gu}, \bibfnamefont
  {Yingfei}}, \bibinfo {author} {\bibfnamefont {Xiao-Liang}\ \bibnamefont
  {Qi}}, and\ \bibinfo {author} {\bibfnamefont {Douglas}\ \bibnamefont
  {Stanford}}} (\bibinfo {year} {2017}{\natexlab{b}}),\ \bibfield  {title}
  {\enquote {\bibinfo {title} {{Local criticality, diffusion and chaos in
  generalized Sachdev-Ye-Kitaev models}},}\ }\href
  {https://doi.org/10.1007/JHEP05(2017)125} {\bibfield  {journal} {\bibinfo
  {journal} {Journal of High Energy Physics}\ }\textbf {\bibinfo {volume}
  {2017}}~(\bibinfo {number} {5}),\ \bibinfo {pages} {125}}\BibitemShut
  {NoStop}%
\bibitem [{\citenamefont {Gull}\ \emph {et~al.}(2008)\citenamefont {Gull},
  \citenamefont {Werner}, \citenamefont {Parcollet},\ and\ \citenamefont
  {Troyer}}]{CTAUX2008}%
  \BibitemOpen
  \bibfield  {author} {\bibinfo {author} {\bibnamefont {Gull}, \bibfnamefont
  {E}}, \bibinfo {author} {\bibfnamefont {P.}~\bibnamefont {Werner}}, \bibinfo
  {author} {\bibfnamefont {O.}~\bibnamefont {Parcollet}}, and\ \bibinfo
  {author} {\bibfnamefont {M.}~\bibnamefont {Troyer}}} (\bibinfo {year}
  {2008}),\ \bibfield  {title} {\enquote {\bibinfo {title} {{Continuous-time
  auxiliary-field Monte Carlo for quantum impurity models}},}\ }\href
  {https://doi.org/10.1209/0295-5075/82/57003} {\bibfield  {journal} {\bibinfo
  {journal} {Europhysics Letters}\ }\textbf {\bibinfo {volume} {82}}~(\bibinfo
  {number} {5}),\ \bibinfo {pages} {57003}}\BibitemShut {NoStop}%
\bibitem [{\citenamefont {Gull}\ \emph {et~al.}(2011)\citenamefont {Gull},
  \citenamefont {Millis}, \citenamefont {Lichtenstein}, \citenamefont
  {Rubtsov}, \citenamefont {Troyer},\ and\ \citenamefont
  {Werner}}]{RevModPhys.83.349}%
  \BibitemOpen
  \bibfield  {author} {\bibinfo {author} {\bibnamefont {Gull}, \bibfnamefont
  {Emanuel}}, \bibinfo {author} {\bibfnamefont {Andrew~J.}\ \bibnamefont
  {Millis}}, \bibinfo {author} {\bibfnamefont {Alexander~I.}\ \bibnamefont
  {Lichtenstein}}, \bibinfo {author} {\bibfnamefont {Alexey~N.}\ \bibnamefont
  {Rubtsov}}, \bibinfo {author} {\bibfnamefont {Matthias}\ \bibnamefont
  {Troyer}}, and\ \bibinfo {author} {\bibfnamefont {Philipp}\ \bibnamefont
  {Werner}}} (\bibinfo {year} {2011}),\ \bibfield  {title} {\enquote {\bibinfo
  {title} {{Continuous-time Monte Carlo methods for quantum impurity
  models}},}\ }\href {https://doi.org/10.1103/RevModPhys.83.349} {\bibfield
  {journal} {\bibinfo  {journal} {Rev. Mod. Phys.}\ }\textbf {\bibinfo {volume}
  {83}},\ \bibinfo {pages} {349--404}}\BibitemShut {NoStop}%
\bibitem [{\citenamefont {Gunnarsson}\ \emph {et~al.}(2003)\citenamefont
  {Gunnarsson}, \citenamefont {Calandra},\ and\ \citenamefont
  {Han}}]{GunnarssonRMP}%
  \BibitemOpen
  \bibfield  {author} {\bibinfo {author} {\bibnamefont {Gunnarsson},
  \bibfnamefont {O}}, \bibinfo {author} {\bibfnamefont {M.}~\bibnamefont
  {Calandra}}, and\ \bibinfo {author} {\bibfnamefont {J.~E.}\ \bibnamefont
  {Han}}} (\bibinfo {year} {2003}),\ \bibfield  {title} {\enquote {\bibinfo
  {title} {Colloquium: Saturation of electrical resistivity},}\ }\href
  {https://doi.org/10.1103/RevModPhys.75.1085} {\bibfield  {journal} {\bibinfo
  {journal} {Rev. Mod. Phys.}\ }\textbf {\bibinfo {volume} {75}},\ \bibinfo
  {pages} {1085--1099}}\BibitemShut {NoStop}%
\bibitem [{\citenamefont {Guo}\ \emph {et~al.}(2019)\citenamefont {Guo},
  \citenamefont {Gu},\ and\ \citenamefont {Sachdev}}]{Guo:2019csw}%
  \BibitemOpen
  \bibfield  {author} {\bibinfo {author} {\bibnamefont {Guo}, \bibfnamefont
  {Haoyu}}, \bibinfo {author} {\bibfnamefont {Yingfei}\ \bibnamefont {Gu}},
  and\ \bibinfo {author} {\bibfnamefont {Subir}\ \bibnamefont {Sachdev}}}
  (\bibinfo {year} {2019}),\ \bibfield  {title} {\enquote {\bibinfo {title}
  {{Transport and chaos in lattice Sachdev-Ye-Kitaev models}},}\ }\href
  {https://doi.org/10.1103/PhysRevB.100.045140} {\bibfield  {journal} {\bibinfo
   {journal} {Phys. Rev. B}\ }\textbf {\bibinfo {volume} {100}}~(\bibinfo
  {number} {4}),\ \bibinfo {pages} {045140}},\ \Eprint
  {https://arxiv.org/abs/1904.02174} {arXiv:1904.02174 [cond-mat.str-el]}
  \BibitemShut {NoStop}%
\bibitem [{\citenamefont {Guo}\ \emph {et~al.}(2020)\citenamefont {Guo},
  \citenamefont {Gu},\ and\ \citenamefont {Sachdev}}]{Guo:2020aog}%
  \BibitemOpen
  \bibfield  {author} {\bibinfo {author} {\bibnamefont {Guo}, \bibfnamefont
  {Haoyu}}, \bibinfo {author} {\bibfnamefont {Yingfei}\ \bibnamefont {Gu}},
  and\ \bibinfo {author} {\bibfnamefont {Subir}\ \bibnamefont {Sachdev}}}
  (\bibinfo {year} {2020}),\ \bibfield  {title} {\enquote {\bibinfo {title}
  {{Linear in temperature resistivity in the limit of zero temperature from the
  time reparameterization soft mode}},}\ }\href
  {https://doi.org/10.1016/j.aop.2020.168202} {\bibfield  {journal} {\bibinfo
  {journal} {Annals Phys.}\ }\textbf {\bibinfo {volume} {418}},\ \bibinfo
  {pages} {168202}},\ \Eprint {https://arxiv.org/abs/2004.05182}
  {arXiv:2004.05182 [cond-mat.str-el]} \BibitemShut {NoStop}%
\bibitem [{\citenamefont {Guo}\ \emph {et~al.}(2022)\citenamefont {Guo},
  \citenamefont {Patel}, \citenamefont {Esterlis},\ and\ \citenamefont
  {Sachdev}}]{Guo:2022zfl}%
  \BibitemOpen
  \bibfield  {author} {\bibinfo {author} {\bibnamefont {Guo}, \bibfnamefont
  {Haoyu}}, \bibinfo {author} {\bibfnamefont {Aavishkar~A.}\ \bibnamefont
  {Patel}}, \bibinfo {author} {\bibfnamefont {Ilya}\ \bibnamefont {Esterlis}},
  and\ \bibinfo {author} {\bibfnamefont {Subir}\ \bibnamefont {Sachdev}}}
  (\bibinfo {year} {2022}),\ \bibfield  {title} {\enquote {\bibinfo {title}
  {{Large $N$ theory of critical Fermi surfaces II: conductivity}},}\
  }\href@noop {} {\ }\Eprint {https://arxiv.org/abs/2207.08841}
  {arXiv:2207.08841 [cond-mat.str-el]} \BibitemShut {NoStop}%
\bibitem [{\citenamefont {Haldar}\ \emph {et~al.}(2018)\citenamefont {Haldar},
  \citenamefont {Banerjee},\ and\ \citenamefont {Shenoy}}]{shenoy}%
  \BibitemOpen
  \bibfield  {author} {\bibinfo {author} {\bibnamefont {Haldar}, \bibfnamefont
  {Arijit}}, \bibinfo {author} {\bibfnamefont {Sumilan}\ \bibnamefont
  {Banerjee}}, and\ \bibinfo {author} {\bibfnamefont {Vijay~B.}\ \bibnamefont
  {Shenoy}}} (\bibinfo {year} {2018}),\ \bibfield  {title} {\enquote {\bibinfo
  {title} {{Higher-dimensional Sachdev-Ye-Kitaev non-Fermi liquids at Lifshitz
  transitions}},}\ }\href {https://doi.org/10.1103/PhysRevB.97.241106}
  {\bibfield  {journal} {\bibinfo  {journal} {Phys. Rev. B}\ }\textbf {\bibinfo
  {volume} {97}},\ \bibinfo {pages} {241106}}\BibitemShut {NoStop}%
\bibitem [{\citenamefont {Haldar}\ \emph {et~al.}(2020)\citenamefont {Haldar},
  \citenamefont {Haldar}, \citenamefont {Bera}, \citenamefont {Mandal},\ and\
  \citenamefont {Banerjee}}]{Banerjee20}%
  \BibitemOpen
  \bibfield  {author} {\bibinfo {author} {\bibnamefont {Haldar}, \bibfnamefont
  {Arijit}}, \bibinfo {author} {\bibfnamefont {Prosenjit}\ \bibnamefont
  {Haldar}}, \bibinfo {author} {\bibfnamefont {Surajit}\ \bibnamefont {Bera}},
  \bibinfo {author} {\bibfnamefont {Ipsita}\ \bibnamefont {Mandal}}, and\
  \bibinfo {author} {\bibfnamefont {Sumilan}\ \bibnamefont {Banerjee}}}
  (\bibinfo {year} {2020}),\ \bibfield  {title} {\enquote {\bibinfo {title}
  {{Quench, thermalization, and residual entropy across a non-Fermi liquid to
  Fermi liquid transition}},}\ }\href
  {https://doi.org/10.1103/PhysRevResearch.2.013307} {\bibfield  {journal}
  {\bibinfo  {journal} {Phys. Rev. Research}\ }\textbf {\bibinfo {volume}
  {2}},\ \bibinfo {pages} {013307}}\BibitemShut {NoStop}%
\bibitem [{\citenamefont {Haldar}\ and\ \citenamefont
  {Shenoy}(2018)}]{shenoy2}%
  \BibitemOpen
  \bibfield  {author} {\bibinfo {author} {\bibnamefont {Haldar}, \bibfnamefont
  {Arijit}}, and\ \bibinfo {author} {\bibfnamefont {Vijay~B.}\ \bibnamefont
  {Shenoy}}} (\bibinfo {year} {2018}),\ \bibfield  {title} {\enquote {\bibinfo
  {title} {{Strange half-metals and Mott insulators in Sachdev-Ye-Kitaev
  models}},}\ }\href {https://doi.org/10.1103/PhysRevB.98.165135} {\bibfield
  {journal} {\bibinfo  {journal} {Phys. Rev. B}\ }\textbf {\bibinfo {volume}
  {98}},\ \bibinfo {pages} {165135}}\BibitemShut {NoStop}%
\bibitem [{\citenamefont {Halperin}\ \emph {et~al.}(1993)\citenamefont
  {Halperin}, \citenamefont {Lee},\ and\ \citenamefont {Read}}]{HLR}%
  \BibitemOpen
  \bibfield  {author} {\bibinfo {author} {\bibnamefont {Halperin},
  \bibfnamefont {B~I}}, \bibinfo {author} {\bibfnamefont {Patrick~A.}\
  \bibnamefont {Lee}}, and\ \bibinfo {author} {\bibfnamefont {Nicholas}\
  \bibnamefont {Read}}} (\bibinfo {year} {1993}),\ \bibfield  {title} {\enquote
  {\bibinfo {title} {{Theory of the half-filled Landau level}},}\ }\href
  {https://doi.org/10.1103/PhysRevB.47.7312} {\bibfield  {journal} {\bibinfo
  {journal} {Phys. Rev. B}\ }\textbf {\bibinfo {volume} {47}},\ \bibinfo
  {pages} {7312--7343}}\BibitemShut {NoStop}%
\bibitem [{\citenamefont {Haque}\ and\ \citenamefont
  {McClarty}(2019)}]{Haque19}%
  \BibitemOpen
  \bibfield  {author} {\bibinfo {author} {\bibnamefont {Haque}, \bibfnamefont
  {Masudul}}, and\ \bibinfo {author} {\bibfnamefont {P.~A.}\ \bibnamefont
  {McClarty}}} (\bibinfo {year} {2019}),\ \bibfield  {title} {\enquote
  {\bibinfo {title} {{Eigenstate thermalization scaling in Majorana clusters:
  From chaotic to integrable Sachdev-Ye-Kitaev models}},}\ }\href
  {https://doi.org/10.1103/PhysRevB.100.115122} {\bibfield  {journal} {\bibinfo
   {journal} {Phys. Rev. B}\ }\textbf {\bibinfo {volume} {100}},\ \bibinfo
  {pages} {115122}}\BibitemShut {NoStop}%
\bibitem [{\citenamefont {Hartnoll}(2014)}]{Hartnoll2014}%
  \BibitemOpen
  \bibfield  {author} {\bibinfo {author} {\bibnamefont {Hartnoll},
  \bibfnamefont {Sean~A}}} (\bibinfo {year} {2014}),\ \bibfield  {title}
  {\enquote {\bibinfo {title} {Theory of universal incoherent metallic
  transport},}\ }\href {https://doi.org/10.1038/nphys3174} {\bibfield
  {journal} {\bibinfo  {journal} {Nature Physics}\ }\textbf {\bibinfo {volume}
  {11}}~(\bibinfo {number} {1}),\ \bibinfo {pages} {54--61}}\BibitemShut
  {NoStop}%
\bibitem [{\citenamefont {Hartnoll}\ \emph {et~al.}(2007)\citenamefont
  {Hartnoll}, \citenamefont {Kovtun}, \citenamefont {Muller},\ and\
  \citenamefont {Sachdev}}]{Hartnoll:2007ih}%
  \BibitemOpen
  \bibfield  {author} {\bibinfo {author} {\bibnamefont {Hartnoll},
  \bibfnamefont {Sean~A}}, \bibinfo {author} {\bibfnamefont {Pavel~K.}\
  \bibnamefont {Kovtun}}, \bibinfo {author} {\bibfnamefont {Markus}\
  \bibnamefont {Muller}}, and\ \bibinfo {author} {\bibfnamefont {Subir}\
  \bibnamefont {Sachdev}}} (\bibinfo {year} {2007}),\ \bibfield  {title}
  {\enquote {\bibinfo {title} {{Theory of the Nernst effect near quantum phase
  transitions in condensed matter, and in dyonic black holes}},}\ }\href
  {https://doi.org/10.1103/PhysRevB.76.144502} {\bibfield  {journal} {\bibinfo
  {journal} {Phys. Rev. B}\ }\textbf {\bibinfo {volume} {76}},\ \bibinfo
  {pages} {144502}},\ \Eprint {https://arxiv.org/abs/0706.3215}
  {arXiv:0706.3215 [cond-mat.str-el]} \BibitemShut {NoStop}%
\bibitem [{\citenamefont {Hartnoll}\ \emph {et~al.}(2016)\citenamefont
  {Hartnoll}, \citenamefont {Lucas},\ and\ \citenamefont
  {Sachdev}}]{Hartnoll:2016apf}%
  \BibitemOpen
  \bibfield  {author} {\bibinfo {author} {\bibnamefont {Hartnoll},
  \bibfnamefont {Sean~A}}, \bibinfo {author} {\bibfnamefont {Andrew}\
  \bibnamefont {Lucas}}, and\ \bibinfo {author} {\bibfnamefont {Subir}\
  \bibnamefont {Sachdev}}} (\bibinfo {year} {2016}),\ \bibfield  {title}
  {\enquote {\bibinfo {title} {{Holographic quantum matter}},}\ }\href
  {https://mitpress.mit.edu/books/holographic-quantum-matter} {\bibfield
  {journal} {\bibinfo  {journal} {MIT Press, Cambridge MA}\ }}\Eprint
  {https://arxiv.org/abs/1612.07324} {arXiv:1612.07324 [hep-th]} \BibitemShut
  {NoStop}%
%%CITATION = ARXIV:1612.07324;%%
\bibitem [{\citenamefont {{Hartnoll}}\ and\ \citenamefont
  {{MacKenzie}}(2021)}]{Mackenzie_rev}%
  \BibitemOpen
  \bibfield  {author} {\bibinfo {author} {\bibnamefont {{Hartnoll}},
  \bibfnamefont {Sean~A}}, and\ \bibinfo {author} {\bibfnamefont {Andrew~P.}\
  \bibnamefont {{MacKenzie}}}} (\bibinfo {year} {2021}),\ \bibfield  {title}
  {\enquote {\bibinfo {title} {{Planckian Dissipation in Metals}},}\
  }\href@noop {} {\ }\Eprint {https://arxiv.org/abs/2107.07802}
  {arXiv:2107.07802 [cond-mat.str-el]} \BibitemShut {NoStop}%
\bibitem [{\citenamefont {Hartnoll}\ \emph {et~al.}(2014)\citenamefont
  {Hartnoll}, \citenamefont {Mahajan}, \citenamefont {Punk},\ and\
  \citenamefont {Sachdev}}]{Hartnoll:2014gba}%
  \BibitemOpen
  \bibfield  {author} {\bibinfo {author} {\bibnamefont {Hartnoll},
  \bibfnamefont {Sean~A}}, \bibinfo {author} {\bibfnamefont {Raghu}\
  \bibnamefont {Mahajan}}, \bibinfo {author} {\bibfnamefont {Matthias}\
  \bibnamefont {Punk}}, and\ \bibinfo {author} {\bibfnamefont {Subir}\
  \bibnamefont {Sachdev}}} (\bibinfo {year} {2014}),\ \bibfield  {title}
  {\enquote {\bibinfo {title} {{Transport near the Ising-nematic quantum
  critical point of metals in two dimensions}},}\ }\href
  {https://doi.org/10.1103/PhysRevB.89.155130} {\bibfield  {journal} {\bibinfo
  {journal} {Phys. Rev. B}\ }\textbf {\bibinfo {volume} {89}}~(\bibinfo
  {number} {15}),\ \bibinfo {pages} {155130}},\ \Eprint
  {https://arxiv.org/abs/1401.7012} {arXiv:1401.7012 [cond-mat.str-el]}
  \BibitemShut {NoStop}%
\bibitem [{\citenamefont {{Hauck}}\ \emph {et~al.}(2020)\citenamefont
  {{Hauck}}, \citenamefont {{Klug}}, \citenamefont {{Esterlis}},\ and\
  \citenamefont {{Schmalian}}}]{Hauck20}%
  \BibitemOpen
  \bibfield  {author} {\bibinfo {author} {\bibnamefont {{Hauck}}, \bibfnamefont
  {Daniel}}, \bibinfo {author} {\bibfnamefont {Markus~J.}\ \bibnamefont
  {{Klug}}}, \bibinfo {author} {\bibfnamefont {Ilya}\ \bibnamefont
  {{Esterlis}}}, and\ \bibinfo {author} {\bibfnamefont {J{\"o}rg}\ \bibnamefont
  {{Schmalian}}}} (\bibinfo {year} {2020}),\ \bibfield  {title} {\enquote
  {\bibinfo {title} {{Eliashberg equations for an electron-phonon version of
  the Sachdev-Ye-Kitaev model: Pair breaking in non-Fermi liquid
  superconductors}},}\ }\href {https://doi.org/10.1016/j.aop.2020.168120}
  {\bibfield  {journal} {\bibinfo  {journal} {Annals of Physics}\ }\textbf
  {\bibinfo {volume} {417}},\ \bibinfo {eid} {168120}},\ \Eprint
  {https://arxiv.org/abs/1911.04328} {arXiv:1911.04328 [cond-mat.str-el]}
  \BibitemShut {NoStop}%
\bibitem [{\citenamefont {Haule}\ \emph {et~al.}(2002)\citenamefont {Haule},
  \citenamefont {Rosch}, \citenamefont {Kroha},\ and\ \citenamefont
  {W\"olfle}}]{Haule2002}%
  \BibitemOpen
  \bibfield  {author} {\bibinfo {author} {\bibnamefont {Haule}, \bibfnamefont
  {K}}, \bibinfo {author} {\bibfnamefont {A.}~\bibnamefont {Rosch}}, \bibinfo
  {author} {\bibfnamefont {J.}~\bibnamefont {Kroha}}, and\ \bibinfo {author}
  {\bibfnamefont {P.}~\bibnamefont {W\"olfle}}} (\bibinfo {year} {2002}),\
  \bibfield  {title} {\enquote {\bibinfo {title} {Pseudogaps in an incoherent
  metal},}\ }\href {https://doi.org/10.1103/PhysRevLett.89.236402} {\bibfield
  {journal} {\bibinfo  {journal} {Phys. Rev. Lett.}\ }\textbf {\bibinfo
  {volume} {89}},\ \bibinfo {pages} {236402}}\BibitemShut {NoStop}%
\bibitem [{\citenamefont {Held}\ \emph {et~al.}(2013)\citenamefont {Held},
  \citenamefont {Peters},\ and\ \citenamefont {Toschi}}]{Held2013}%
  \BibitemOpen
  \bibfield  {author} {\bibinfo {author} {\bibnamefont {Held}, \bibfnamefont
  {K}}, \bibinfo {author} {\bibfnamefont {R.}~\bibnamefont {Peters}}, and\
  \bibinfo {author} {\bibfnamefont {A.}~\bibnamefont {Toschi}}} (\bibinfo
  {year} {2013}),\ \bibfield  {title} {\enquote {\bibinfo {title} {Poor man's
  understanding of kinks originating from strong electronic correlations},}\
  }\href {https://doi.org/10.1103/PhysRevLett.110.246402} {\bibfield  {journal}
  {\bibinfo  {journal} {Phys. Rev. Lett.}\ }\textbf {\bibinfo {volume} {110}},\
  \bibinfo {pages} {246402}}\BibitemShut {NoStop}%
\bibitem [{\citenamefont {Hewson}(1997)}]{hewson_book}%
  \BibitemOpen
  \bibfield  {author} {\bibinfo {author} {\bibnamefont {Hewson}, \bibfnamefont
  {A~C}}} (\bibinfo {year} {1997}),\ \href
  {https://books.google.com/books?id=fPzgHneNFDAC} {\emph {\bibinfo {title}
  {{The Kondo Problem to Heavy Fermions}}}}\ (\bibinfo  {publisher} {Cambridge
  University Press})\BibitemShut {NoStop}%
\bibitem [{\citenamefont {Heydeman}\ \emph {et~al.}(2020)\citenamefont
  {Heydeman}, \citenamefont {Iliesiu}, \citenamefont {Turiaci},\ and\
  \citenamefont {Zhao}}]{Heydeman:2020hhw}%
  \BibitemOpen
  \bibfield  {author} {\bibinfo {author} {\bibnamefont {Heydeman},
  \bibfnamefont {Matthew}}, \bibinfo {author} {\bibfnamefont {Luca~V.}\
  \bibnamefont {Iliesiu}}, \bibinfo {author} {\bibfnamefont {Gustavo~J.}\
  \bibnamefont {Turiaci}}, and\ \bibinfo {author} {\bibfnamefont {Wenli}\
  \bibnamefont {Zhao}}} (\bibinfo {year} {2020}),\ \bibfield  {title} {\enquote
  {\bibinfo {title} {{The statistical mechanics of near-BPS black holes}},}\
  }\href@noop {} {\ }\Eprint {https://arxiv.org/abs/2011.01953}
  {arXiv:2011.01953 [hep-th]} \BibitemShut {NoStop}%
\bibitem [{\citenamefont {{Hicks}}\ \emph {et~al.}(2012)\citenamefont
  {{Hicks}}, \citenamefont {{Gibbs}}, \citenamefont {{Mackenzie}},
  \citenamefont {{Takatsu}}, \citenamefont {{Maeno}},\ and\ \citenamefont
  {{Yelland}}}]{APM12}%
  \BibitemOpen
  \bibfield  {author} {\bibinfo {author} {\bibnamefont {{Hicks}}, \bibfnamefont
  {Clifford~W}}, \bibinfo {author} {\bibfnamefont {Alexandra~S.}\ \bibnamefont
  {{Gibbs}}}, \bibinfo {author} {\bibfnamefont {Andrew~P.}\ \bibnamefont
  {{Mackenzie}}}, \bibinfo {author} {\bibfnamefont {Hiroshi}\ \bibnamefont
  {{Takatsu}}}, \bibinfo {author} {\bibfnamefont {Yoshiteru}\ \bibnamefont
  {{Maeno}}}, and\ \bibinfo {author} {\bibfnamefont {Edward~A.}\ \bibnamefont
  {{Yelland}}}} (\bibinfo {year} {2012}),\ \bibfield  {title} {\enquote
  {\bibinfo {title} {{Quantum Oscillations and High Carrier Mobility in the
  Delafossite PdCoO$_{2}$}},}\ }\href
  {https://doi.org/10.1103/PhysRevLett.109.116401} {\bibfield  {journal}
  {\bibinfo  {journal} {Phys. Rev. Lett.}\ }\textbf {\bibinfo {volume}
  {109}}~(\bibinfo {number} {11}),\ \bibinfo {eid} {116401}},\ \Eprint
  {https://arxiv.org/abs/1207.5402} {arXiv:1207.5402 [cond-mat.str-el]}
  \BibitemShut {NoStop}%
\bibitem [{\citenamefont {Hirsch}\ and\ \citenamefont
  {Fye}(1986)}]{HirschFye1986}%
  \BibitemOpen
  \bibfield  {author} {\bibinfo {author} {\bibnamefont {Hirsch}, \bibfnamefont
  {J~E}}, and\ \bibinfo {author} {\bibfnamefont {R.~M.}\ \bibnamefont {Fye}}}
  (\bibinfo {year} {1986}),\ \bibfield  {title} {\enquote {\bibinfo {title}
  {{Monte Carlo Method for Magnetic Impurities in Metals}},}\ }\href
  {https://doi.org/10.1103/PhysRevLett.56.2521} {\bibfield  {journal} {\bibinfo
   {journal} {Phys. Rev. Lett.}\ }\textbf {\bibinfo {volume} {56}},\ \bibinfo
  {pages} {2521--2524}}\BibitemShut {NoStop}%
\bibitem [{\citenamefont {Hod}(2007)}]{Hod07}%
  \BibitemOpen
  \bibfield  {author} {\bibinfo {author} {\bibnamefont {Hod}, \bibfnamefont
  {Shahar}}} (\bibinfo {year} {2007}),\ \bibfield  {title} {\enquote {\bibinfo
  {title} {Universal bound on dynamical relaxation times and black-hole
  quasinormal ringing},}\ }\href {https://doi.org/10.1103/PhysRevD.75.064013}
  {\bibfield  {journal} {\bibinfo  {journal} {Phys. Rev. D}\ }\textbf {\bibinfo
  {volume} {75}},\ \bibinfo {pages} {064013}}\BibitemShut {NoStop}%
\bibitem [{\citenamefont {Homes}\ \emph {et~al.}(2004)\citenamefont {Homes},
  \citenamefont {Dordevic}, \citenamefont {Strongin}, \citenamefont {Bonn},
  \citenamefont {Liang}, \citenamefont {Hardy}, \citenamefont {Komiya},
  \citenamefont {Ando}, \citenamefont {Yu}, \citenamefont {Kaneko},
  \citenamefont {Zhao}, \citenamefont {Greven}, \citenamefont {Basov},\ and\
  \citenamefont {Timusk}}]{Homes2004}%
  \BibitemOpen
  \bibfield  {author} {\bibinfo {author} {\bibnamefont {Homes}, \bibfnamefont
  {C~C}}, \bibinfo {author} {\bibfnamefont {S.~V.}\ \bibnamefont {Dordevic}},
  \bibinfo {author} {\bibfnamefont {M.}~\bibnamefont {Strongin}}, \bibinfo
  {author} {\bibfnamefont {D.~A.}\ \bibnamefont {Bonn}}, \bibinfo {author}
  {\bibfnamefont {Ruixing}\ \bibnamefont {Liang}}, \bibinfo {author}
  {\bibfnamefont {W.~N.}\ \bibnamefont {Hardy}}, \bibinfo {author}
  {\bibfnamefont {Seiki}\ \bibnamefont {Komiya}}, \bibinfo {author}
  {\bibfnamefont {Yoichi}\ \bibnamefont {Ando}}, \bibinfo {author}
  {\bibfnamefont {G.}~\bibnamefont {Yu}}, \bibinfo {author} {\bibfnamefont
  {N.}~\bibnamefont {Kaneko}}, \bibinfo {author} {\bibfnamefont
  {X.}~\bibnamefont {Zhao}}, \bibinfo {author} {\bibfnamefont {M.}~\bibnamefont
  {Greven}}, \bibinfo {author} {\bibfnamefont {D.~N.}\ \bibnamefont {Basov}},
  and\ \bibinfo {author} {\bibfnamefont {T.}~\bibnamefont {Timusk}}} (\bibinfo
  {year} {2004}),\ \bibfield  {title} {\enquote {\bibinfo {title} {A universal
  scaling relation in high-temperature superconductors},}\ }\href
  {https://doi.org/10.1038/nature02673} {\bibfield  {journal} {\bibinfo
  {journal} {Nature}\ }\textbf {\bibinfo {volume} {430}}~(\bibinfo {number}
  {6999}),\ \bibinfo {pages} {539--541}}\BibitemShut {NoStop}%
\bibitem [{\citenamefont {'t~Hooft}(2001)}]{tHooft:1999rgb}%
  \BibitemOpen
  \bibfield  {author} {\bibinfo {author} {\bibnamefont {'t~Hooft},
  \bibfnamefont {Gerard}}} (\bibinfo {year} {2001}),\ \bibfield  {title}
  {\enquote {\bibinfo {title} {{The Holographic principle: Opening lecture}},}\
  }\href {https://doi.org/10.1142/9789812811585_0005} {\bibfield  {journal}
  {\bibinfo  {journal} {Subnucl. Ser.}\ }\textbf {\bibinfo {volume} {37}},\
  \bibinfo {pages} {72--100}},\ \Eprint {https://arxiv.org/abs/hep-th/0003004}
  {arXiv:hep-th/0003004} \BibitemShut {NoStop}%
\bibitem [{\citenamefont {{Hu}}\ \emph {et~al.}(2020)\citenamefont {{Hu}},
  \citenamefont {{Cai}},\ and\ \citenamefont {{Si}}}]{SiSU22}%
  \BibitemOpen
  \bibfield  {author} {\bibinfo {author} {\bibnamefont {{Hu}}, \bibfnamefont
  {H}}, \bibinfo {author} {\bibfnamefont {A.}~\bibnamefont {{Cai}}}, and\
  \bibinfo {author} {\bibfnamefont {Q.}~\bibnamefont {{Si}}}} (\bibinfo {year}
  {2020}),\ \bibfield  {title} {\enquote {\bibinfo {title} {{Quantum
  Criticality and Dynamical Kondo Effect in an SU(2) Anderson Lattice
  Model}},}\ }\href@noop {} {\ }\Eprint {https://arxiv.org/abs/2004.04679}
  {arXiv:2004.04679 [cond-mat.str-el]} \BibitemShut {NoStop}%
\bibitem [{\citenamefont {Huang}\ \emph {et~al.}(2019)\citenamefont {Huang},
  \citenamefont {Sheppard}, \citenamefont {Moritz},\ and\ \citenamefont
  {Devereaux}}]{Huang19}%
  \BibitemOpen
  \bibfield  {author} {\bibinfo {author} {\bibnamefont {Huang}, \bibfnamefont
  {Edwin~W}}, \bibinfo {author} {\bibfnamefont {Ryan}\ \bibnamefont
  {Sheppard}}, \bibinfo {author} {\bibfnamefont {Brian}\ \bibnamefont
  {Moritz}}, and\ \bibinfo {author} {\bibfnamefont {Thomas~P.}\ \bibnamefont
  {Devereaux}}} (\bibinfo {year} {2019}),\ \bibfield  {title} {\enquote
  {\bibinfo {title} {{Strange metallicity in the doped Hubbard model}},}\
  }\href {https://doi.org/10.1126/science.aau7063} {\bibfield  {journal}
  {\bibinfo  {journal} {Science}\ }\textbf {\bibinfo {volume} {366}}~(\bibinfo
  {number} {6468}),\ \bibinfo {pages} {987--990}}\BibitemShut {NoStop}%
\bibitem [{\citenamefont {Huijse}\ and\ \citenamefont
  {Sachdev}(2011)}]{Huijse:2011hp}%
  \BibitemOpen
  \bibfield  {author} {\bibinfo {author} {\bibnamefont {Huijse}, \bibfnamefont
  {Liza}}, and\ \bibinfo {author} {\bibfnamefont {Subir}\ \bibnamefont
  {Sachdev}}} (\bibinfo {year} {2011}),\ \bibfield  {title} {\enquote {\bibinfo
  {title} {{Fermi surfaces and gauge-gravity duality}},}\ }\href
  {https://doi.org/10.1103/PhysRevD.84.026001} {\bibfield  {journal} {\bibinfo
  {journal} {Phys. Rev. D}\ }\textbf {\bibinfo {volume} {84}},\ \bibinfo
  {pages} {026001}},\ \Eprint {https://arxiv.org/abs/1104.5022}
  {arXiv:1104.5022 [hep-th]} \BibitemShut {NoStop}%
\bibitem [{\citenamefont {Huijse}\ \emph {et~al.}(2012)\citenamefont {Huijse},
  \citenamefont {Sachdev},\ and\ \citenamefont {Swingle}}]{Huijse:2011ef}%
  \BibitemOpen
  \bibfield  {author} {\bibinfo {author} {\bibnamefont {Huijse}, \bibfnamefont
  {Liza}}, \bibinfo {author} {\bibfnamefont {Subir}\ \bibnamefont {Sachdev}},
  and\ \bibinfo {author} {\bibfnamefont {Brian}\ \bibnamefont {Swingle}}}
  (\bibinfo {year} {2012}),\ \bibfield  {title} {\enquote {\bibinfo {title}
  {{Hidden Fermi surfaces in compressible states of gauge-gravity duality}},}\
  }\href {https://doi.org/10.1103/PhysRevB.85.035121} {\bibfield  {journal}
  {\bibinfo  {journal} {Phys. Rev. B}\ }\textbf {\bibinfo {volume} {85}},\
  \bibinfo {pages} {035121}},\ \Eprint {https://arxiv.org/abs/1112.0573}
  {arXiv:1112.0573 [cond-mat.str-el]} \BibitemShut {NoStop}%
\bibitem [{\citenamefont {{Husain}}\ \emph {et~al.}(2020)\citenamefont
  {{Husain}}, \citenamefont {{Huang}}, \citenamefont {{Mitrano}}, \citenamefont
  {{Rak}}, \citenamefont {{Rubeck}}, \citenamefont {{Guo}}, \citenamefont
  {{Yang}}, \citenamefont {{Sow}}, \citenamefont {{Maeno}}, \citenamefont
  {{Uchoa}}, \citenamefont {{Chiang}}, \citenamefont {{Batson}}, \citenamefont
  {{Phillips}},\ and\ \citenamefont {{Abbamonte}}}]{Abbamonte3}%
  \BibitemOpen
  \bibfield  {author} {\bibinfo {author} {\bibnamefont {{Husain}},
  \bibfnamefont {A~A}}, \bibinfo {author} {\bibfnamefont {E.~W.}\ \bibnamefont
  {{Huang}}}, \bibinfo {author} {\bibfnamefont {M.}~\bibnamefont {{Mitrano}}},
  \bibinfo {author} {\bibfnamefont {M.~S.}\ \bibnamefont {{Rak}}}, \bibinfo
  {author} {\bibfnamefont {S.~I.}\ \bibnamefont {{Rubeck}}}, \bibinfo {author}
  {\bibfnamefont {X.}~\bibnamefont {{Guo}}}, \bibinfo {author} {\bibfnamefont
  {H.}~\bibnamefont {{Yang}}}, \bibinfo {author} {\bibfnamefont
  {C.}~\bibnamefont {{Sow}}}, \bibinfo {author} {\bibfnamefont
  {Y.}~\bibnamefont {{Maeno}}}, \bibinfo {author} {\bibfnamefont
  {B.}~\bibnamefont {{Uchoa}}}, \bibinfo {author} {\bibfnamefont {T.~C.}\
  \bibnamefont {{Chiang}}}, \bibinfo {author} {\bibfnamefont {P.~E.}\
  \bibnamefont {{Batson}}}, \bibinfo {author} {\bibfnamefont {P.~W.}\
  \bibnamefont {{Phillips}}}, and\ \bibinfo {author} {\bibfnamefont
  {P.}~\bibnamefont {{Abbamonte}}}} (\bibinfo {year} {2020}),\ \bibfield
  {title} {\enquote {\bibinfo {title} {{Observation of Pines' Demon in
  Sr$_2$RuO$_4$}},}\ }\href@noop {} {\bibfield  {journal} {\bibinfo  {journal}
  {arXiv e-prints}\ ,\ \bibinfo {eid} {arXiv:2007.06670}}}\Eprint
  {https://arxiv.org/abs/2007.06670} {arXiv:2007.06670 [cond-mat.str-el]}
  \BibitemShut {NoStop}%
\bibitem [{\citenamefont {Husain}\ \emph {et~al.}(2019)\citenamefont {Husain},
  \citenamefont {Mitrano}, \citenamefont {Rak}, \citenamefont {Rubeck},
  \citenamefont {Uchoa}, \citenamefont {March}, \citenamefont {Dwyer},
  \citenamefont {Schneeloch}, \citenamefont {Zhong}, \citenamefont {Gu},\ and\
  \citenamefont {Abbamonte}}]{Abbamonte2}%
  \BibitemOpen
  \bibfield  {author} {\bibinfo {author} {\bibnamefont {Husain}, \bibfnamefont
  {Ali~A}}, \bibinfo {author} {\bibfnamefont {Matteo}\ \bibnamefont {Mitrano}},
  \bibinfo {author} {\bibfnamefont {Melinda~S.}\ \bibnamefont {Rak}}, \bibinfo
  {author} {\bibfnamefont {Samantha}\ \bibnamefont {Rubeck}}, \bibinfo {author}
  {\bibfnamefont {Bruno}\ \bibnamefont {Uchoa}}, \bibinfo {author}
  {\bibfnamefont {Katia}\ \bibnamefont {March}}, \bibinfo {author}
  {\bibfnamefont {Christian}\ \bibnamefont {Dwyer}}, \bibinfo {author}
  {\bibfnamefont {John}\ \bibnamefont {Schneeloch}}, \bibinfo {author}
  {\bibfnamefont {Ruidan}\ \bibnamefont {Zhong}}, \bibinfo {author}
  {\bibfnamefont {G.~D.}\ \bibnamefont {Gu}}, and\ \bibinfo {author}
  {\bibfnamefont {Peter}\ \bibnamefont {Abbamonte}}} (\bibinfo {year} {2019}),\
  \bibfield  {title} {\enquote {\bibinfo {title} {Crossover of charge
  fluctuations across the strange metal phase diagram},}\ }\href
  {https://doi.org/10.1103/PhysRevX.9.041062} {\bibfield  {journal} {\bibinfo
  {journal} {Phys. Rev. X}\ }\textbf {\bibinfo {volume} {9}},\ \bibinfo {pages}
  {041062}}\BibitemShut {NoStop}%
\bibitem [{\citenamefont {Hussey}(2008)}]{Hussey2008}%
  \BibitemOpen
  \bibfield  {author} {\bibinfo {author} {\bibnamefont {Hussey}, \bibfnamefont
  {N~E}}} (\bibinfo {year} {2008}),\ \bibfield  {title} {\enquote {\bibinfo
  {title} {{Phenomenology of the normal state in-plane transport properties of
  high-$T_c$ cuprates}},}\ }\href
  {https://doi.org/10.1088/0953-8984/20/12/123201} {\bibfield  {journal}
  {\bibinfo  {journal} {Journal of Physics: Condensed Matter}\ }\textbf
  {\bibinfo {volume} {20}}~(\bibinfo {number} {12}),\ \bibinfo {pages}
  {123201}}\BibitemShut {NoStop}%
\bibitem [{\citenamefont {Hussey}\ \emph {et~al.}(2004)\citenamefont {Hussey},
  \citenamefont {Takenaka},\ and\ \citenamefont {Takagi}}]{Hussey04}%
  \BibitemOpen
  \bibfield  {author} {\bibinfo {author} {\bibnamefont {Hussey}, \bibfnamefont
  {N~E}}, \bibinfo {author} {\bibfnamefont {K.}~\bibnamefont {Takenaka}}, and\
  \bibinfo {author} {\bibfnamefont {H.}~\bibnamefont {Takagi}}} (\bibinfo
  {year} {2004}),\ \bibfield  {title} {\enquote {\bibinfo {title}
  {{Universality of the Mott Ioffe Regel limit in metals}},}\ }\href
  {https://doi.org/10.1080/14786430410001716944} {\bibfield  {journal}
  {\bibinfo  {journal} {Philosophical Magazine}\ }\textbf {\bibinfo {volume}
  {84}}~(\bibinfo {number} {27}),\ \bibinfo {pages} {2847--2864}}\BibitemShut
  {NoStop}%
\bibitem [{\citenamefont {Hwang}\ \emph {et~al.}(2007)\citenamefont {Hwang},
  \citenamefont {Timusk},\ and\ \citenamefont {Gu}}]{Hwang_2007}%
  \BibitemOpen
  \bibfield  {author} {\bibinfo {author} {\bibnamefont {Hwang}, \bibfnamefont
  {J}}, \bibinfo {author} {\bibfnamefont {T}~\bibnamefont {Timusk}}, and\
  \bibinfo {author} {\bibfnamefont {G~D}\ \bibnamefont {Gu}}} (\bibinfo {year}
  {2007}),\ \bibfield  {title} {\enquote {\bibinfo {title} {{Doping dependent
  optical properties of Bi$_2$Sr$_2$CaCu$_2$O$_{8+\delta}$}},}\ }\href
  {https://doi.org/10.1088/0953-8984/19/12/125208} {\bibfield  {journal}
  {\bibinfo  {journal} {Journal of Physics: Condensed Matter}\ }\textbf
  {\bibinfo {volume} {19}}~(\bibinfo {number} {12}),\ \bibinfo {pages}
  {125208}}\BibitemShut {NoStop}%
\bibitem [{\citenamefont {Iliesiu}\ and\ \citenamefont
  {Turiaci}(2020)}]{Iliesiu:2020qvm}%
  \BibitemOpen
  \bibfield  {author} {\bibinfo {author} {\bibnamefont {Iliesiu}, \bibfnamefont
  {Luca~V}}, and\ \bibinfo {author} {\bibfnamefont {Gustavo~J.}\ \bibnamefont
  {Turiaci}}} (\bibinfo {year} {2020}),\ \bibfield  {title} {\enquote {\bibinfo
  {title} {{The statistical mechanics of near-extremal black holes}},}\
  }\href@noop {} {\ }\Eprint {https://arxiv.org/abs/2003.02860}
  {arXiv:2003.02860 [hep-th]} \BibitemShut {NoStop}%
\bibitem [{\citenamefont {Ioffe}\ and\ \citenamefont {Regel}(1960)}]{IR}%
  \BibitemOpen
  \bibfield  {author} {\bibinfo {author} {\bibnamefont {Ioffe}, \bibfnamefont
  {A~F}}, and\ \bibinfo {author} {\bibfnamefont {A.~R.}\ \bibnamefont {Regel}}}
  (\bibinfo {year} {1960}),\ \bibfield  {title} {\enquote {\bibinfo {title}
  {Non-crystalline, amorphous and liquid electronic semiconductors},}\
  }\href@noop {} {\bibfield  {journal} {\bibinfo  {journal} {Prog. Semicond.}\
  }\textbf {\bibinfo {volume} {4}},\ \bibinfo {pages} {237--291}}\BibitemShut
  {NoStop}%
\bibitem [{\citenamefont {Iqbal}\ \emph {et~al.}(2011)\citenamefont {Iqbal},
  \citenamefont {Liu},\ and\ \citenamefont {Mezei}}]{Iqbal:2011ae}%
  \BibitemOpen
  \bibfield  {author} {\bibinfo {author} {\bibnamefont {Iqbal}, \bibfnamefont
  {Nabil}}, \bibinfo {author} {\bibfnamefont {Hong}\ \bibnamefont {Liu}}, and\
  \bibinfo {author} {\bibfnamefont {Mark}\ \bibnamefont {Mezei}}} (\bibinfo
  {year} {2011}),\ \bibfield  {title} {\enquote {\bibinfo {title} {{Lectures on
  holographic non-Fermi liquids and quantum phase transitions}},}\ }in\ \href
  {https://doi.org/10.1142/9789814350525_0013} {\emph {\bibinfo {booktitle}
  {{Theoretical Advanced Study Institute in Elementary Particle Physics}:
  {String theory and its Applications: From meV to the Planck Scale}}}},\ pp.\
  \bibinfo {pages} {707--816},\ \Eprint {https://arxiv.org/abs/1110.3814}
  {arXiv:1110.3814 [hep-th]} \BibitemShut {NoStop}%
\bibitem [{\citenamefont {Iqbal}\ \emph {et~al.}(2012)\citenamefont {Iqbal},
  \citenamefont {Liu},\ and\ \citenamefont {Mezei}}]{Iqbal:2011in}%
  \BibitemOpen
  \bibfield  {author} {\bibinfo {author} {\bibnamefont {Iqbal}, \bibfnamefont
  {Nabil}}, \bibinfo {author} {\bibfnamefont {Hong}\ \bibnamefont {Liu}}, and\
  \bibinfo {author} {\bibfnamefont {Mark}\ \bibnamefont {Mezei}}} (\bibinfo
  {year} {2012}),\ \bibfield  {title} {\enquote {\bibinfo {title} {{Semi-local
  quantum liquids}},}\ }\href {https://doi.org/10.1007/JHEP04(2012)086}
  {\bibfield  {journal} {\bibinfo  {journal} {JHEP}\ }\textbf {\bibinfo
  {volume} {04}},\ \bibinfo {pages} {086}},\ \Eprint
  {https://arxiv.org/abs/1105.4621} {arXiv:1105.4621 [hep-th]} \BibitemShut
  {NoStop}%
\bibitem [{\citenamefont {Jackiw}(1985)}]{Jackiw85}%
  \BibitemOpen
  \bibfield  {author} {\bibinfo {author} {\bibnamefont {Jackiw}, \bibfnamefont
  {Roman}}} (\bibinfo {year} {1985}),\ \bibfield  {title} {\enquote {\bibinfo
  {title} {Lower dimensional gravity},}\ }\href
  {https://doi.org/https://doi.org/10.1016/0550-3213(85)90448-1} {\bibfield
  {journal} {\bibinfo  {journal} {Nuclear Physics B}\ }\textbf {\bibinfo
  {volume} {252}},\ \bibinfo {pages} {343--356}}\BibitemShut {NoStop}%
\bibitem [{\citenamefont {Jain}(2007)}]{JJCF}%
  \BibitemOpen
  \bibfield  {author} {\bibinfo {author} {\bibnamefont {Jain}, \bibfnamefont
  {Jainendra~K}}} (\bibinfo {year} {2007}),\ \href@noop {} {\emph {\bibinfo
  {title} {Composite fermions}}}\ (\bibinfo  {publisher} {Cambridge University
  Press})\BibitemShut {NoStop}%
\bibitem [{\citenamefont {Jaoui}\ \emph {et~al.}(2022)\citenamefont {Jaoui},
  \citenamefont {Das}, \citenamefont {Di~Battista}, \citenamefont
  {D{\'i}ez-M{\'e}rida}, \citenamefont {Lu}, \citenamefont {Watanabe},
  \citenamefont {Taniguchi}, \citenamefont {Ishizuka}, \citenamefont
  {Levitov},\ and\ \citenamefont {Efetov}}]{efetov21}%
  \BibitemOpen
  \bibfield  {author} {\bibinfo {author} {\bibnamefont {Jaoui}, \bibfnamefont
  {Alexandre}}, \bibinfo {author} {\bibfnamefont {Ipsita}\ \bibnamefont {Das}},
  \bibinfo {author} {\bibfnamefont {Giorgio}\ \bibnamefont {Di~Battista}},
  \bibinfo {author} {\bibfnamefont {Jaime}\ \bibnamefont
  {D{\'i}ez-M{\'e}rida}}, \bibinfo {author} {\bibfnamefont {Xiaobo}\
  \bibnamefont {Lu}}, \bibinfo {author} {\bibfnamefont {Kenji}\ \bibnamefont
  {Watanabe}}, \bibinfo {author} {\bibfnamefont {Takashi}\ \bibnamefont
  {Taniguchi}}, \bibinfo {author} {\bibfnamefont {Hiroaki}\ \bibnamefont
  {Ishizuka}}, \bibinfo {author} {\bibfnamefont {Leonid}\ \bibnamefont
  {Levitov}}, and\ \bibinfo {author} {\bibfnamefont {Dmitri~K.}\ \bibnamefont
  {Efetov}}} (\bibinfo {year} {2022}),\ \bibfield  {title} {\enquote {\bibinfo
  {title} {Quantum critical behaviour in magic-angle twisted bilayer
  graphene},}\ }\href {https://doi.org/10.1038/s41567-022-01556-5} {\bibfield
  {journal} {\bibinfo  {journal} {Nature Physics}\
  }10.1038/s41567-022-01556-5}\BibitemShut {NoStop}%
\bibitem [{\citenamefont {Jian}\ \emph {et~al.}(2017)\citenamefont {Jian},
  \citenamefont {Bi},\ and\ \citenamefont {Xu}}]{XuMIT17}%
  \BibitemOpen
  \bibfield  {author} {\bibinfo {author} {\bibnamefont {Jian}, \bibfnamefont
  {Chao-Ming}}, \bibinfo {author} {\bibfnamefont {Zhen}\ \bibnamefont {Bi}},
  and\ \bibinfo {author} {\bibfnamefont {Cenke}\ \bibnamefont {Xu}}} (\bibinfo
  {year} {2017}),\ \bibfield  {title} {\enquote {\bibinfo {title} {Model for
  continuous thermal metal to insulator transition},}\ }\href
  {https://doi.org/10.1103/PhysRevB.96.115122} {\bibfield  {journal} {\bibinfo
  {journal} {Phys. Rev. B}\ }\textbf {\bibinfo {volume} {96}},\ \bibinfo
  {pages} {115122}}\BibitemShut {NoStop}%
\bibitem [{\citenamefont {Jian}\ and\ \citenamefont {Yao}(2017)}]{Yao}%
  \BibitemOpen
  \bibfield  {author} {\bibinfo {author} {\bibnamefont {Jian}, \bibfnamefont
  {Shao-Kai}}, and\ \bibinfo {author} {\bibfnamefont {Hong}\ \bibnamefont
  {Yao}}} (\bibinfo {year} {2017}),\ \bibfield  {title} {\enquote {\bibinfo
  {title} {{Solvable Sachdev-Ye-Kitaev Models in Higher Dimensions: From
  Diffusion to Many-Body Localization}},}\ }\href
  {https://doi.org/10.1103/PhysRevLett.119.206602} {\bibfield  {journal}
  {\bibinfo  {journal} {Phys. Rev. Lett.}\ }\textbf {\bibinfo {volume} {119}},\
  \bibinfo {pages} {206602}}\BibitemShut {NoStop}%
\bibitem [{\citenamefont {Jiang}\ \emph {et~al.}(2012)\citenamefont {Jiang},
  \citenamefont {Block}, \citenamefont {Mishmash}, \citenamefont {Garrison},
  \citenamefont {Sheng}, \citenamefont {Motrunich},\ and\ \citenamefont
  {Fisher}}]{mishmash}%
  \BibitemOpen
  \bibfield  {author} {\bibinfo {author} {\bibnamefont {Jiang}, \bibfnamefont
  {Hong-Chen}}, \bibinfo {author} {\bibfnamefont {Matthew~S.}\ \bibnamefont
  {Block}}, \bibinfo {author} {\bibfnamefont {Ryan~V.}\ \bibnamefont
  {Mishmash}}, \bibinfo {author} {\bibfnamefont {James~R.}\ \bibnamefont
  {Garrison}}, \bibinfo {author} {\bibfnamefont {D.~N.}\ \bibnamefont {Sheng}},
  \bibinfo {author} {\bibfnamefont {Olexei~I.}\ \bibnamefont {Motrunich}}, and\
  \bibinfo {author} {\bibfnamefont {Matthew P.~A.}\ \bibnamefont {Fisher}}}
  (\bibinfo {year} {2012}),\ \bibfield  {title} {\enquote {\bibinfo {title}
  {{Non-Fermi-liquid $d$-wave metal phase of strongly interacting
  electrons}},}\ }\href {http://dx.doi.org/10.1038/nature11732} {\bibfield
  {journal} {\bibinfo  {journal} {Nature}\ }\textbf {\bibinfo {volume} {493}},\
  \bibinfo {pages} {39}}\BibitemShut {NoStop}%
\bibitem [{\citenamefont {Joshi}\ \emph {et~al.}(2020)\citenamefont {Joshi},
  \citenamefont {Li}, \citenamefont {Tarnopolsky}, \citenamefont {Georges},\
  and\ \citenamefont {Sachdev}}]{Joshi:2019csz}%
  \BibitemOpen
  \bibfield  {author} {\bibinfo {author} {\bibnamefont {Joshi}, \bibfnamefont
  {Darshan~G}}, \bibinfo {author} {\bibfnamefont {Chenyuan}\ \bibnamefont
  {Li}}, \bibinfo {author} {\bibfnamefont {Grigory}\ \bibnamefont
  {Tarnopolsky}}, \bibinfo {author} {\bibfnamefont {Antoine}\ \bibnamefont
  {Georges}}, and\ \bibinfo {author} {\bibfnamefont {Subir}\ \bibnamefont
  {Sachdev}}} (\bibinfo {year} {2020}),\ \bibfield  {title} {\enquote {\bibinfo
  {title} {{Deconfined critical point in a doped random quantum Heisenberg
  magnet}},}\ }\href {https://doi.org/10.1103/PhysRevX.10.021033} {\bibfield
  {journal} {\bibinfo  {journal} {Phys. Rev. X}\ }\textbf {\bibinfo {volume}
  {10}}~(\bibinfo {number} {2}),\ \bibinfo {pages} {021033}},\ \Eprint
  {https://arxiv.org/abs/1912.08822} {arXiv:1912.08822 [cond-mat.str-el]}
  \BibitemShut {NoStop}%
\bibitem [{\citenamefont {{Joshi}}\ and\ \citenamefont
  {{Sachdev}}(2020)}]{Joshi20}%
  \BibitemOpen
  \bibfield  {author} {\bibinfo {author} {\bibnamefont {{Joshi}}, \bibfnamefont
  {Darshan~G}}, and\ \bibinfo {author} {\bibfnamefont {Subir}\ \bibnamefont
  {{Sachdev}}}} (\bibinfo {year} {2020}),\ \bibfield  {title} {\enquote
  {\bibinfo {title} {{Anomalous density fluctuations in a random $t$-$J$
  model}},}\ }\href {https://doi.org/10.1103/PhysRevB.102.165146} {\bibfield
  {journal} {\bibinfo  {journal} {Phys. Rev. B}\ }\textbf {\bibinfo {volume}
  {102}}~(\bibinfo {number} {16}),\ \bibinfo {eid} {165146}},\ \Eprint
  {https://arxiv.org/abs/2006.13947} {arXiv:2006.13947 [cond-mat.str-el]}
  \BibitemShut {NoStop}%
\bibitem [{\citenamefont {Kadowaki}\ and\ \citenamefont {Woods}(1986)}]{KW}%
  \BibitemOpen
  \bibfield  {author} {\bibinfo {author} {\bibnamefont {Kadowaki},
  \bibfnamefont {K}}, and\ \bibinfo {author} {\bibfnamefont {S.B.}\
  \bibnamefont {Woods}}} (\bibinfo {year} {1986}),\ \bibfield  {title}
  {\enquote {\bibinfo {title} {Universal relationship of the resistivity and
  specific heat in heavy-fermion compounds},}\ }\href
  {https://doi.org/https://doi.org/10.1016/0038-1098(86)90785-4} {\bibfield
  {journal} {\bibinfo  {journal} {Solid State Communications}\ }\textbf
  {\bibinfo {volume} {58}}~(\bibinfo {number} {8}),\ \bibinfo {pages} {507 --
  509}}\BibitemShut {NoStop}%
\bibitem [{\citenamefont {Keselman}\ \emph {et~al.}(2021)\citenamefont
  {Keselman}, \citenamefont {Nie},\ and\ \citenamefont
  {Berg}}]{Keselman:2020fmo}%
  \BibitemOpen
  \bibfield  {author} {\bibinfo {author} {\bibnamefont {Keselman},
  \bibfnamefont {Anna}}, \bibinfo {author} {\bibfnamefont {Laimei}\
  \bibnamefont {Nie}}, and\ \bibinfo {author} {\bibfnamefont {Erez}\
  \bibnamefont {Berg}}} (\bibinfo {year} {2021}),\ \bibfield  {title} {\enquote
  {\bibinfo {title} {{Scrambling and Lyapunov exponent in spatially extended
  systems}},}\ }\href {https://doi.org/10.1103/PhysRevB.103.L121111} {\bibfield
   {journal} {\bibinfo  {journal} {Phys. Rev. B}\ }\textbf {\bibinfo {volume}
  {103}}~(\bibinfo {number} {12}),\ \bibinfo {pages} {L121111}},\ \Eprint
  {https://arxiv.org/abs/2009.10104} {arXiv:2009.10104 [cond-mat.str-el]}
  \BibitemShut {NoStop}%
\bibitem [{\citenamefont {von Keyserlingk}\ \emph {et~al.}(2018)\citenamefont
  {von Keyserlingk}, \citenamefont {Rakovszky}, \citenamefont {Pollmann},\ and\
  \citenamefont {Sondhi}}]{FP18b}%
  \BibitemOpen
  \bibfield  {author} {\bibinfo {author} {\bibnamefont {von Keyserlingk},
  \bibfnamefont {C~W}}, \bibinfo {author} {\bibfnamefont {Tibor}\ \bibnamefont
  {Rakovszky}}, \bibinfo {author} {\bibfnamefont {Frank}\ \bibnamefont
  {Pollmann}}, and\ \bibinfo {author} {\bibfnamefont {S.~L.}\ \bibnamefont
  {Sondhi}}} (\bibinfo {year} {2018}),\ \bibfield  {title} {\enquote {\bibinfo
  {title} {{Operator Hydrodynamics, OTOCs, and Entanglement Growth in Systems
  without Conservation Laws}},}\ }\href
  {https://doi.org/10.1103/PhysRevX.8.021013} {\bibfield  {journal} {\bibinfo
  {journal} {Phys. Rev. X}\ }\textbf {\bibinfo {volume} {8}},\ \bibinfo {pages}
  {021013}}\BibitemShut {NoStop}%
\bibitem [{\citenamefont {Khemani}\ \emph
  {et~al.}(2018{\natexlab{a}})\citenamefont {Khemani}, \citenamefont {Huse},\
  and\ \citenamefont {Nahum}}]{Khemani:2018sdn}%
  \BibitemOpen
  \bibfield  {author} {\bibinfo {author} {\bibnamefont {Khemani}, \bibfnamefont
  {Vedika}}, \bibinfo {author} {\bibfnamefont {David~A.}\ \bibnamefont {Huse}},
  and\ \bibinfo {author} {\bibfnamefont {Adam}\ \bibnamefont {Nahum}}}
  (\bibinfo {year} {2018}{\natexlab{a}}),\ \bibfield  {title} {\enquote
  {\bibinfo {title} {{Velocity-dependent Lyapunov exponents in many-body
  quantum, semiclassical, and classical chaos}},}\ }\href
  {https://doi.org/10.1103/PhysRevB.98.144304} {\bibfield  {journal} {\bibinfo
  {journal} {Phys. Rev. B}\ }\textbf {\bibinfo {volume} {98}}~(\bibinfo
  {number} {14}),\ \bibinfo {pages} {144304}},\ \Eprint
  {https://arxiv.org/abs/1803.05902} {arXiv:1803.05902 [cond-mat.stat-mech]}
  \BibitemShut {NoStop}%
\bibitem [{\citenamefont {Khemani}\ \emph
  {et~al.}(2018{\natexlab{b}})\citenamefont {Khemani}, \citenamefont
  {Vishwanath},\ and\ \citenamefont {Huse}}]{Khemani_prx}%
  \BibitemOpen
  \bibfield  {author} {\bibinfo {author} {\bibnamefont {Khemani}, \bibfnamefont
  {Vedika}}, \bibinfo {author} {\bibfnamefont {Ashvin}\ \bibnamefont
  {Vishwanath}}, and\ \bibinfo {author} {\bibfnamefont {David~A.}\ \bibnamefont
  {Huse}}} (\bibinfo {year} {2018}{\natexlab{b}}),\ \bibfield  {title}
  {\enquote {\bibinfo {title} {Operator spreading and the emergence of
  dissipative hydrodynamics under unitary evolution with conservation laws},}\
  }\href {https://doi.org/10.1103/PhysRevX.8.031057} {\bibfield  {journal}
  {\bibinfo  {journal} {Phys. Rev. X}\ }\textbf {\bibinfo {volume} {8}},\
  \bibinfo {pages} {031057}}\BibitemShut {NoStop}%
\bibitem [{\citenamefont {Khurana}(1990)}]{Khurana1990}%
  \BibitemOpen
  \bibfield  {author} {\bibinfo {author} {\bibnamefont {Khurana}, \bibfnamefont
  {Anil}}} (\bibinfo {year} {1990}),\ \bibfield  {title} {\enquote {\bibinfo
  {title} {{Electrical conductivity in the infinite-dimensional Hubbard
  model}},}\ }\href {https://doi.org/10.1103/PhysRevLett.64.1990} {\bibfield
  {journal} {\bibinfo  {journal} {Phys. Rev. Lett.}\ }\textbf {\bibinfo
  {volume} {64}},\ \bibinfo {pages} {1990--1990}}\BibitemShut {NoStop}%
\bibitem [{\citenamefont
  {Khveshchenko}(2018{\natexlab{a}})}]{Khveshchenko:2018nod}%
  \BibitemOpen
  \bibfield  {author} {\bibinfo {author} {\bibnamefont {Khveshchenko},
  \bibfnamefont {D~V}}} (\bibinfo {year} {2018}{\natexlab{a}}),\ \bibfield
  {title} {\enquote {\bibinfo {title} {{Seeking to develop global SYK-ness}},}\
  }\href {https://doi.org/10.3390/condmat3040040} {\bibfield  {journal}
  {\bibinfo  {journal} {Condens. Mat.}\ }\textbf {\bibinfo {volume}
  {3}}~(\bibinfo {number} {4}),\ \bibinfo {pages} {40}},\ \Eprint
  {https://arxiv.org/abs/1805.00870} {arXiv:1805.00870 [cond-mat.str-el]}
  \BibitemShut {NoStop}%
\bibitem [{\citenamefont {Khveshchenko}(2018{\natexlab{b}})}]{DVK17}%
  \BibitemOpen
  \bibfield  {author} {\bibinfo {author} {\bibnamefont {Khveshchenko},
  \bibfnamefont {D~V}}} (\bibinfo {year} {2018}{\natexlab{b}}),\ \bibfield
  {title} {\enquote {\bibinfo {title} {{Thickening and sickening the SYK
  model}},}\ }\href {https://doi.org/10.21468/SciPostPhys.5.1.012} {\bibfield
  {journal} {\bibinfo  {journal} {SciPost Phys.}\ }\textbf {\bibinfo {volume}
  {5}},\ \bibinfo {pages} {12}}\BibitemShut {NoStop}%
\bibitem [{\citenamefont {{Khveshchenko}}(2019)}]{Khveshchenko19}%
  \BibitemOpen
  \bibfield  {author} {\bibinfo {author} {\bibnamefont {{Khveshchenko}},
  \bibfnamefont {D~V}}} (\bibinfo {year} {2019}),\ \bibfield  {title} {\enquote
  {\bibinfo {title} {{One SYK SET}},}\ }\href
  {https://doi.org/10.3952/physics.v60i3.4305} {\bibfield  {journal} {\bibinfo
  {journal} {Lithuanian Journal of Physics}\ }\textbf {\bibinfo {volume}
  {60}},\ \bibinfo {pages} {185}},\ \Eprint {https://arxiv.org/abs/1912.05691}
  {arXiv:1912.05691 [cond-mat.str-el]} \BibitemShut {NoStop}%
\bibitem [{\citenamefont {Khveshchenko}(2020)}]{Khveshchenko:2020rai}%
  \BibitemOpen
  \bibfield  {author} {\bibinfo {author} {\bibnamefont {Khveshchenko},
  \bibfnamefont {D~V}}} (\bibinfo {year} {2020}),\ \bibfield  {title} {\enquote
  {\bibinfo {title} {{Connecting the SYK Dots}},}\ }\href
  {https://doi.org/10.3390/condmat5020037} {\bibfield  {journal} {\bibinfo
  {journal} {Condens. Mat.}\ }\textbf {\bibinfo {volume} {5}}~(\bibinfo
  {number} {2}),\ \bibinfo {pages} {37}},\ \Eprint
  {https://arxiv.org/abs/2004.06646} {arXiv:2004.06646 [cond-mat.str-el]}
  \BibitemShut {NoStop}%
\bibitem [{\citenamefont {Khveshchenko}(2022)}]{Khveshchenko:2022gcd}%
  \BibitemOpen
  \bibfield  {author} {\bibinfo {author} {\bibnamefont {Khveshchenko},
  \bibfnamefont {D~V}}} (\bibinfo {year} {2022}),\ \bibfield  {title} {\enquote
  {\bibinfo {title} {{SYK does not Transit Gloria Mundi just yet}},}\
  }\href@noop {} {\ }\Eprint {https://arxiv.org/abs/2205.11478}
  {arXiv:2205.11478 [cond-mat.str-el]} \BibitemShut {NoStop}%
\bibitem [{\citenamefont {Kim}\ \emph {et~al.}(2021)\citenamefont {Kim},
  \citenamefont {Altman},\ and\ \citenamefont {Cao}}]{Kim:2020jpz}%
  \BibitemOpen
  \bibfield  {author} {\bibinfo {author} {\bibnamefont {Kim}, \bibfnamefont
  {Jaewon}}, \bibinfo {author} {\bibfnamefont {Ehud}\ \bibnamefont {Altman}},
  and\ \bibinfo {author} {\bibfnamefont {Xiangyu}\ \bibnamefont {Cao}}}
  (\bibinfo {year} {2021}),\ \bibfield  {title} {\enquote {\bibinfo {title}
  {{Dirac Fast Scramblers}},}\ }\href
  {https://doi.org/10.1103/PhysRevB.103.L081113} {\bibfield  {journal}
  {\bibinfo  {journal} {Phys. Rev. B}\ }\textbf {\bibinfo {volume}
  {103}}~(\bibinfo {number} {8}),\ \bibinfo {pages} {081113}},\ \Eprint
  {https://arxiv.org/abs/2010.10545} {arXiv:2010.10545 [cond-mat.str-el]}
  \BibitemShut {NoStop}%
\bibitem [{\citenamefont {Kim}\ \emph {et~al.}(2019)\citenamefont {Kim},
  \citenamefont {Klebanov}, \citenamefont {Tarnopolsky},\ and\ \citenamefont
  {Zhao}}]{klebanov}%
  \BibitemOpen
  \bibfield  {author} {\bibinfo {author} {\bibnamefont {Kim}, \bibfnamefont
  {Jaewon}}, \bibinfo {author} {\bibfnamefont {Igor~R.}\ \bibnamefont
  {Klebanov}}, \bibinfo {author} {\bibfnamefont {Grigory}\ \bibnamefont
  {Tarnopolsky}}, and\ \bibinfo {author} {\bibfnamefont {Wenli}\ \bibnamefont
  {Zhao}}} (\bibinfo {year} {2019}),\ \bibfield  {title} {\enquote {\bibinfo
  {title} {{Symmetry Breaking in Coupled SYK or Tensor Models}},}\ }\href
  {https://doi.org/10.1103/PhysRevX.9.021043} {\bibfield  {journal} {\bibinfo
  {journal} {Phys. Rev. X}\ }\textbf {\bibinfo {volume} {9}}~(\bibinfo {number}
  {2}),\ \bibinfo {pages} {021043}},\ \Eprint
  {https://arxiv.org/abs/1902.02287} {arXiv:1902.02287 [hep-th]} \BibitemShut
  {NoStop}%
\bibitem [{\citenamefont {{Kim}}\ \emph {et~al.}(1994)\citenamefont {{Kim}},
  \citenamefont {{Furusaki}}, \citenamefont {{Wen}},\ and\ \citenamefont
  {{Lee}}}]{YBK94}%
  \BibitemOpen
  \bibfield  {author} {\bibinfo {author} {\bibnamefont {{Kim}}, \bibfnamefont
  {Yong~Baek}}, \bibinfo {author} {\bibfnamefont {Akira}\ \bibnamefont
  {{Furusaki}}}, \bibinfo {author} {\bibfnamefont {Xiao-Gang}\ \bibnamefont
  {{Wen}}}, and\ \bibinfo {author} {\bibfnamefont {Patrick~A.}\ \bibnamefont
  {{Lee}}}} (\bibinfo {year} {1994}),\ \bibfield  {title} {\enquote {\bibinfo
  {title} {{Gauge-invariant response functions of fermions coupled to a gauge
  field}},}\ }\href {https://doi.org/10.1103/PhysRevB.50.17917} {\bibfield
  {journal} {\bibinfo  {journal} {Phys. Rev. B}\ }\textbf {\bibinfo {volume}
  {50}}~(\bibinfo {number} {24}),\ \bibinfo {pages} {17917--17932}},\ \Eprint
  {https://arxiv.org/abs/cond-mat/9405083} {arXiv:cond-mat/9405083 [cond-mat]}
  \BibitemShut {NoStop}%
\bibitem [{\citenamefont {Kim}\ \emph {et~al.}(1995)\citenamefont {Kim},
  \citenamefont {Lee},\ and\ \citenamefont {Wen}}]{YBK95}%
  \BibitemOpen
  \bibfield  {author} {\bibinfo {author} {\bibnamefont {Kim}, \bibfnamefont
  {Yong~Baek}}, \bibinfo {author} {\bibfnamefont {Patrick~A.}\ \bibnamefont
  {Lee}}, and\ \bibinfo {author} {\bibfnamefont {Xiao-Gang}\ \bibnamefont
  {Wen}}} (\bibinfo {year} {1995}),\ \bibfield  {title} {\enquote {\bibinfo
  {title} {{Quantum Boltzmann equation of composite fermions interacting with a
  gauge field}},}\ }\href {https://doi.org/10.1103/PhysRevB.52.17275}
  {\bibfield  {journal} {\bibinfo  {journal} {Phys. Rev. B}\ }\textbf {\bibinfo
  {volume} {52}},\ \bibinfo {pages} {17275--17292}}\BibitemShut {NoStop}%
\bibitem [{\citenamefont {Kirchner}\ \emph {et~al.}(2020)\citenamefont
  {Kirchner}, \citenamefont {Paschen}, \citenamefont {Chen}, \citenamefont
  {Wirth}, \citenamefont {Feng}, \citenamefont {Thompson},\ and\ \citenamefont
  {Si}}]{SiRMP}%
  \BibitemOpen
  \bibfield  {author} {\bibinfo {author} {\bibnamefont {Kirchner},
  \bibfnamefont {Stefan}}, \bibinfo {author} {\bibfnamefont {Silke}\
  \bibnamefont {Paschen}}, \bibinfo {author} {\bibfnamefont {Qiuyun}\
  \bibnamefont {Chen}}, \bibinfo {author} {\bibfnamefont {Steffen}\
  \bibnamefont {Wirth}}, \bibinfo {author} {\bibfnamefont {Donglai}\
  \bibnamefont {Feng}}, \bibinfo {author} {\bibfnamefont {Joe~D.}\ \bibnamefont
  {Thompson}}, and\ \bibinfo {author} {\bibfnamefont {Qimiao}\ \bibnamefont
  {Si}}} (\bibinfo {year} {2020}),\ \bibfield  {title} {\enquote {\bibinfo
  {title} {{Colloquium: Heavy-electron quantum criticality and single-particle
  spectroscopy}},}\ }\href {https://doi.org/10.1103/RevModPhys.92.011002}
  {\bibfield  {journal} {\bibinfo  {journal} {Rev. Mod. Phys.}\ }\textbf
  {\bibinfo {volume} {92}},\ \bibinfo {pages} {011002}},\ \Eprint
  {https://arxiv.org/abs/1810.13293} {arXiv:1810.13293 [cond-mat.str-el]}
  \BibitemShut {NoStop}%
\bibitem [{\citenamefont {Kitaev}(2015)}]{kitaev_talk}%
  \BibitemOpen
  \bibfield  {author} {\bibinfo {author} {\bibnamefont {Kitaev}, \bibfnamefont
  {Alexei}}} (\bibinfo {year} {2015}),\ \bibfield  {title} {\enquote {\bibinfo
  {title} {A simple model of quantum holography, talk given at kitp program:
  entanglement in strongly-correlated quantum matter},}\ }\href@noop {}
  {\bibinfo  {journal} {USA April 2015}\ }\BibitemShut {NoStop}%
\bibitem [{\citenamefont {Kitaev}\ and\ \citenamefont {Suh}(2018)}]{kitaevsuh}%
  \BibitemOpen
\bibfield  {journal} {  }\bibfield  {author} {\bibinfo {author} {\bibnamefont
  {Kitaev}, \bibfnamefont {Alexei}}, and\ \bibinfo {author} {\bibfnamefont
  {S.~Josephine}\ \bibnamefont {Suh}}} (\bibinfo {year} {2018}),\ \bibfield
  {title} {\enquote {\bibinfo {title} {{The soft mode in the Sachdev-Ye-Kitaev
  model and its gravity dual}},}\ }\href
  {https://doi.org/10.1007/JHEP05(2018)183} {\bibfield  {journal} {\bibinfo
  {journal} {JHEP}\ }\textbf {\bibinfo {volume} {05}},\ \bibinfo {pages}
  {183}},\ \Eprint {https://arxiv.org/abs/1711.08467} {arXiv:1711.08467
  [hep-th]} \BibitemShut {NoStop}%
\bibitem [{\citenamefont {Klebanov}\ \emph {et~al.}(2020)\citenamefont
  {Klebanov}, \citenamefont {Milekhin}, \citenamefont {Tarnopolsky},\ and\
  \citenamefont {Zhao}}]{Klebanov:2020kck}%
  \BibitemOpen
  \bibfield  {author} {\bibinfo {author} {\bibnamefont {Klebanov},
  \bibfnamefont {Igor~R}}, \bibinfo {author} {\bibfnamefont {Alexey}\
  \bibnamefont {Milekhin}}, \bibinfo {author} {\bibfnamefont {Grigory}\
  \bibnamefont {Tarnopolsky}}, and\ \bibinfo {author} {\bibfnamefont {Wenli}\
  \bibnamefont {Zhao}}} (\bibinfo {year} {2020}),\ \bibfield  {title} {\enquote
  {\bibinfo {title} {{Spontaneous Breaking of $U(1)$ Symmetry in Coupled
  Complex SYK Models}},}\ }\href {https://doi.org/10.1007/JHEP11(2020)162}
  {\bibfield  {journal} {\bibinfo  {journal} {JHEP}\ }\textbf {\bibinfo
  {volume} {11}},\ \bibinfo {pages} {162}},\ \Eprint
  {https://arxiv.org/abs/2006.07317} {arXiv:2006.07317 [hep-th]} \BibitemShut
  {NoStop}%
\bibitem [{\citenamefont {Klebanov}\ \emph {et~al.}(2018)\citenamefont
  {Klebanov}, \citenamefont {Popov},\ and\ \citenamefont
  {Tarnopolsky}}]{Klebanov:2018fzb}%
  \BibitemOpen
  \bibfield  {author} {\bibinfo {author} {\bibnamefont {Klebanov},
  \bibfnamefont {Igor~R}}, \bibinfo {author} {\bibfnamefont {Fedor}\
  \bibnamefont {Popov}}, and\ \bibinfo {author} {\bibfnamefont {Grigory}\
  \bibnamefont {Tarnopolsky}}} (\bibinfo {year} {2018}),\ \bibfield  {title}
  {\enquote {\bibinfo {title} {{TASI Lectures on Large $N$ Tensor Models}},}\
  }\href {https://doi.org/10.22323/1.305.0004} {\bibfield  {journal} {\bibinfo
  {journal} {PoS}\ }\textbf {\bibinfo {volume} {TASI2017}},\ \bibinfo {pages}
  {004}},\ \Eprint {https://arxiv.org/abs/1808.09434} {arXiv:1808.09434
  [hep-th]} \BibitemShut {NoStop}%
\bibitem [{\citenamefont {Klebanov}\ and\ \citenamefont
  {Tarnopolsky}(2017)}]{Klebanov:2016xxf}%
  \BibitemOpen
  \bibfield  {author} {\bibinfo {author} {\bibnamefont {Klebanov},
  \bibfnamefont {Igor~R}}, and\ \bibinfo {author} {\bibfnamefont {Grigory}\
  \bibnamefont {Tarnopolsky}}} (\bibinfo {year} {2017}),\ \bibfield  {title}
  {\enquote {\bibinfo {title} {{Uncolored random tensors, melon diagrams, and
  the Sachdev-Ye-Kitaev models}},}\ }\href
  {https://doi.org/10.1103/PhysRevD.95.046004} {\bibfield  {journal} {\bibinfo
  {journal} {Phys. Rev. D}\ }\textbf {\bibinfo {volume} {95}}~(\bibinfo
  {number} {4}),\ \bibinfo {pages} {046004}},\ \Eprint
  {https://arxiv.org/abs/1611.08915} {arXiv:1611.08915 [hep-th]} \BibitemShut
  {NoStop}%
\bibitem [{\citenamefont {Kobrin}\ \emph {et~al.}(2021)\citenamefont {Kobrin},
  \citenamefont {Yang}, \citenamefont {Kahanamoku-Meyer}, \citenamefont
  {Olund}, \citenamefont {Moore}, \citenamefont {Stanford},\ and\ \citenamefont
  {Yao}}]{Kobrin:2020xms}%
  \BibitemOpen
  \bibfield  {author} {\bibinfo {author} {\bibnamefont {Kobrin}, \bibfnamefont
  {Bryce}}, \bibinfo {author} {\bibfnamefont {Zhenbin}\ \bibnamefont {Yang}},
  \bibinfo {author} {\bibfnamefont {Gregory~D.}\ \bibnamefont
  {Kahanamoku-Meyer}}, \bibinfo {author} {\bibfnamefont {Christopher~T.}\
  \bibnamefont {Olund}}, \bibinfo {author} {\bibfnamefont {Joel~E.}\
  \bibnamefont {Moore}}, \bibinfo {author} {\bibfnamefont {Douglas}\
  \bibnamefont {Stanford}}, and\ \bibinfo {author} {\bibfnamefont {Norman~Y.}\
  \bibnamefont {Yao}}} (\bibinfo {year} {2021}),\ \bibfield  {title} {\enquote
  {\bibinfo {title} {{Many-Body Chaos in the Sachdev-Ye-Kitaev Model}},}\
  }\href {https://doi.org/10.1103/PhysRevLett.126.030602} {\bibfield  {journal}
  {\bibinfo  {journal} {Phys. Rev. Lett.}\ }\textbf {\bibinfo {volume}
  {126}}~(\bibinfo {number} {3}),\ \bibinfo {pages} {030602}},\ \Eprint
  {https://arxiv.org/abs/2002.05725} {arXiv:2002.05725 [hep-th]} \BibitemShut
  {NoStop}%
\bibitem [{\citenamefont {Kohn}\ and\ \citenamefont {Luttinger}(1965)}]{KL}%
  \BibitemOpen
  \bibfield  {author} {\bibinfo {author} {\bibnamefont {Kohn}, \bibfnamefont
  {W}}, and\ \bibinfo {author} {\bibfnamefont {J.~M.}\ \bibnamefont
  {Luttinger}}} (\bibinfo {year} {1965}),\ \bibfield  {title} {\enquote
  {\bibinfo {title} {New mechanism for superconductivity},}\ }\href
  {https://doi.org/10.1103/PhysRevLett.15.524} {\bibfield  {journal} {\bibinfo
  {journal} {Phys. Rev. Lett.}\ }\textbf {\bibinfo {volume} {15}},\ \bibinfo
  {pages} {524--526}}\BibitemShut {NoStop}%
\bibitem [{\citenamefont {Kostic}\ \emph {et~al.}(1998)\citenamefont {Kostic},
  \citenamefont {Okada}, \citenamefont {Collins}, \citenamefont {Schlesinger},
  \citenamefont {Reiner}, \citenamefont {Klein}, \citenamefont {Kapitulnik},
  \citenamefont {Geballe},\ and\ \citenamefont {Beasley}}]{Kostic_1998}%
  \BibitemOpen
  \bibfield  {author} {\bibinfo {author} {\bibnamefont {Kostic}, \bibfnamefont
  {P}}, \bibinfo {author} {\bibfnamefont {Y.}~\bibnamefont {Okada}}, \bibinfo
  {author} {\bibfnamefont {N.~C.}\ \bibnamefont {Collins}}, \bibinfo {author}
  {\bibfnamefont {Z.}~\bibnamefont {Schlesinger}}, \bibinfo {author}
  {\bibfnamefont {J.~W.}\ \bibnamefont {Reiner}}, \bibinfo {author}
  {\bibfnamefont {L.}~\bibnamefont {Klein}}, \bibinfo {author} {\bibfnamefont
  {A.}~\bibnamefont {Kapitulnik}}, \bibinfo {author} {\bibfnamefont {T.~H.}\
  \bibnamefont {Geballe}}, and\ \bibinfo {author} {\bibfnamefont {M.~R.}\
  \bibnamefont {Beasley}}} (\bibinfo {year} {1998}),\ \bibfield  {title}
  {\enquote {\bibinfo {title} {{Non-Fermi-Liquid Behavior of
  $\mathrm{SrRuO}{}_{3}$: Evidence from Infrared Conductivity}},}\ }\href
  {https://doi.org/10.1103/PhysRevLett.81.2498} {\bibfield  {journal} {\bibinfo
   {journal} {Phys. Rev. Lett.}\ }\textbf {\bibinfo {volume} {81}},\ \bibinfo
  {pages} {2498--2501}}\BibitemShut {NoStop}%
\bibitem [{\citenamefont {Kotliar}(1995)}]{kotliar_largeN}%
  \BibitemOpen
  \bibfield  {author} {\bibinfo {author} {\bibnamefont {Kotliar}, \bibfnamefont
  {G}}} (\bibinfo {year} {1995}),\ \bibfield  {title} {\enquote {\bibinfo
  {title} {{The Large $N$ Expansion and the Strong Correlation Problem}},}\
  }in\ \href@noop {} {\emph {\bibinfo {booktitle} {Strongly Interacting
  Fermions and High-T$_c$ superconductivity}}},\ \bibinfo {series and number}
  {Les Houches, Session LVI},\ \bibinfo {editor} {edited by\ \bibinfo {editor}
  {\bibfnamefont {B.}~\bibnamefont {Doucot}}\ and\ \bibinfo {editor}
  {\bibfnamefont {J.}~\bibnamefont {Zinn-Justin}}}\ (\bibinfo  {publisher}
  {Elsevier})\ p.\ \bibinfo {pages} {197}\BibitemShut {NoStop}%
\bibitem [{\citenamefont {Kourkoulou}\ and\ \citenamefont
  {Maldacena}(2017)}]{Kourkoulou:2017zaj}%
  \BibitemOpen
  \bibfield  {author} {\bibinfo {author} {\bibnamefont {Kourkoulou},
  \bibfnamefont {Ioanna}}, and\ \bibinfo {author} {\bibfnamefont {Juan}\
  \bibnamefont {Maldacena}}} (\bibinfo {year} {2017}),\ \bibfield  {title}
  {\enquote {\bibinfo {title} {{Pure states in the SYK model and nearly-$AdS_2$
  gravity}},}\ }\href@noop {} {\ }\Eprint {https://arxiv.org/abs/1707.02325}
  {arXiv:1707.02325 [hep-th]} \BibitemShut {NoStop}%
\bibitem [{\citenamefont {Kovtun}\ \emph {et~al.}(2005)\citenamefont {Kovtun},
  \citenamefont {Son},\ and\ \citenamefont {Starinets}}]{KSS}%
  \BibitemOpen
  \bibfield  {author} {\bibinfo {author} {\bibnamefont {Kovtun}, \bibfnamefont
  {P}}, \bibinfo {author} {\bibfnamefont {Dan~T.}\ \bibnamefont {Son}}, and\
  \bibinfo {author} {\bibfnamefont {Andrei~O.}\ \bibnamefont {Starinets}}}
  (\bibinfo {year} {2005}),\ \bibfield  {title} {\enquote {\bibinfo {title}
  {{Viscosity in strongly interacting quantum field theories from black hole
  physics}},}\ }\href {https://doi.org/10.1103/PhysRevLett.94.111601}
  {\bibfield  {journal} {\bibinfo  {journal} {Phys. Rev. Lett.}\ }\textbf
  {\bibinfo {volume} {94}},\ \bibinfo {pages} {111601}},\ \Eprint
  {https://arxiv.org/abs/hep-th/0405231} {arXiv:hep-th/0405231} \BibitemShut
  {NoStop}%
\bibitem [{\citenamefont {Kruchkov}\ \emph {et~al.}(2020)\citenamefont
  {Kruchkov}, \citenamefont {Patel}, \citenamefont {Kim},\ and\ \citenamefont
  {Sachdev}}]{Kruchkov:2019idx}%
  \BibitemOpen
  \bibfield  {author} {\bibinfo {author} {\bibnamefont {Kruchkov},
  \bibfnamefont {Alexander}}, \bibinfo {author} {\bibfnamefont {Aavishkar~A.}\
  \bibnamefont {Patel}}, \bibinfo {author} {\bibfnamefont {Philip}\
  \bibnamefont {Kim}}, and\ \bibinfo {author} {\bibfnamefont {Subir}\
  \bibnamefont {Sachdev}}} (\bibinfo {year} {2020}),\ \bibfield  {title}
  {\enquote {\bibinfo {title} {{Thermoelectric power of Sachdev-Ye-Kitaev
  islands: Probing Bekenstein-Hawking entropy in quantum matter
  experiments}},}\ }\href {https://doi.org/10.1103/PhysRevB.101.205148}
  {\bibfield  {journal} {\bibinfo  {journal} {Phys. Rev. B}\ }\textbf {\bibinfo
  {volume} {101}},\ \bibinfo {pages} {205148}}\BibitemShut {NoStop}%
\bibitem [{\citenamefont {Kuhlenkamp}\ and\ \citenamefont
  {Knap}(2020)}]{Knap20}%
  \BibitemOpen
  \bibfield  {author} {\bibinfo {author} {\bibnamefont {Kuhlenkamp},
  \bibfnamefont {Clemens}}, and\ \bibinfo {author} {\bibfnamefont {Michael}\
  \bibnamefont {Knap}}} (\bibinfo {year} {2020}),\ \bibfield  {title} {\enquote
  {\bibinfo {title} {{Periodically Driven Sachdev-Ye-Kitaev Models}},}\ }\href
  {https://doi.org/10.1103/PhysRevLett.124.106401} {\bibfield  {journal}
  {\bibinfo  {journal} {Phys. Rev. Lett.}\ }\textbf {\bibinfo {volume} {124}},\
  \bibinfo {pages} {106401}}\BibitemShut {NoStop}%
\bibitem [{\citenamefont {{Kumar}}\ \emph {et~al.}(2021)\citenamefont
  {{Kumar}}, \citenamefont {{Sachdev}},\ and\ \citenamefont
  {{Tripathi}}}]{Kumar21}%
  \BibitemOpen
  \bibfield  {author} {\bibinfo {author} {\bibnamefont {{Kumar}}, \bibfnamefont
  {Aman}}, \bibinfo {author} {\bibfnamefont {Subir}\ \bibnamefont {{Sachdev}}},
  and\ \bibinfo {author} {\bibfnamefont {Vikram}\ \bibnamefont {{Tripathi}}}}
  (\bibinfo {year} {2021}),\ \bibfield  {title} {\enquote {\bibinfo {title}
  {{Quasiparticle metamorphosis in the random $t$-$J$ model}},}\ }\href@noop {}
  {\ }\Eprint {https://arxiv.org/abs/2112.01760} {arXiv:2112.01760
  [cond-mat.str-el]} \BibitemShut {NoStop}%
\bibitem [{\citenamefont {Landau}(1957)}]{landau}%
  \BibitemOpen
  \bibfield  {author} {\bibinfo {author} {\bibnamefont {Landau}, \bibfnamefont
  {Lev~Davidovich}}} (\bibinfo {year} {1957}),\ \bibfield  {title} {\enquote
  {\bibinfo {title} {{The theory of a Fermi liquid}},}\ }\href
  {http://www.jetp.ras.ru/cgi-bin/e/index/e/3/6/p920?a=list} {\bibfield
  {journal} {\bibinfo  {journal} {Soviet Physics JETP}\ }\textbf {\bibinfo
  {volume} {3}}~(\bibinfo {number} {6}),\ \bibinfo {pages}
  {920--925}}\BibitemShut {NoStop}%
\bibitem [{\citenamefont {Larkin}\ and\ \citenamefont
  {Ovchinnikov}(1969)}]{LO69}%
  \BibitemOpen
  \bibfield  {author} {\bibinfo {author} {\bibnamefont {Larkin}, \bibfnamefont
  {AI}}, and\ \bibinfo {author} {\bibfnamefont {Yu~N}\ \bibnamefont
  {Ovchinnikov}}} (\bibinfo {year} {1969}),\ \bibfield  {title} {\enquote
  {\bibinfo {title} {Quasiclassical method in the theory of
  superconductivity},}\ }\href@noop {} {\bibfield  {journal} {\bibinfo
  {journal} {Soviet Journal of Experimental and Theoretical Physics}\ }\textbf
  {\bibinfo {volume} {28}},\ \bibinfo {pages} {1200}}\BibitemShut {NoStop}%
\bibitem [{\citenamefont {{Larzul}}\ and\ \citenamefont
  {{Schir{\'o}}}(2021)}]{Schiro21}%
  \BibitemOpen
  \bibfield  {author} {\bibinfo {author} {\bibnamefont {{Larzul}},
  \bibfnamefont {Ancel}}, and\ \bibinfo {author} {\bibfnamefont {Marco}\
  \bibnamefont {{Schir{\'o}}}}} (\bibinfo {year} {2021}),\ \bibfield  {title}
  {\enquote {\bibinfo {title} {{Quenches and (Pre)Thermalisation in a mixed
  Sachdev-Ye-Kitaev Model}},}\ }\href@noop {} {\ }\Eprint
  {https://arxiv.org/abs/2107.07781} {arXiv:2107.07781 [cond-mat.str-el]}
  \BibitemShut {NoStop}%
\bibitem [{\citenamefont {Lee}(1989)}]{PALee89}%
  \BibitemOpen
  \bibfield  {author} {\bibinfo {author} {\bibnamefont {Lee}, \bibfnamefont
  {Patrick~A}}} (\bibinfo {year} {1989}),\ \bibfield  {title} {\enquote
  {\bibinfo {title} {{Gauge field, Aharonov-Bohm flux, and high-${T}_{c}$
  superconductivity}},}\ }\href {https://doi.org/10.1103/PhysRevLett.63.680}
  {\bibfield  {journal} {\bibinfo  {journal} {Phys. Rev. Lett.}\ }\textbf
  {\bibinfo {volume} {63}},\ \bibinfo {pages} {680--683}}\BibitemShut {NoStop}%
\bibitem [{\citenamefont {{Lee}}(2021)}]{Lee20}%
  \BibitemOpen
  \bibfield  {author} {\bibinfo {author} {\bibnamefont {{Lee}}, \bibfnamefont
  {Patrick~A}}} (\bibinfo {year} {2021}),\ \bibfield  {title} {\enquote
  {\bibinfo {title} {{Low-temperature $T$-linear resistivity due to umklapp
  scattering from a critical mode}},}\ }\href
  {https://doi.org/10.1103/PhysRevB.104.035140} {\bibfield  {journal} {\bibinfo
   {journal} {Phys. Rev. B}\ }\textbf {\bibinfo {volume} {104}}~(\bibinfo
  {number} {3}),\ \bibinfo {eid} {035140}},\ \Eprint
  {https://arxiv.org/abs/2012.09339} {arXiv:2012.09339 [cond-mat.str-el]}
  \BibitemShut {NoStop}%
\bibitem [{\citenamefont {Lee}\ \emph {et~al.}(2006)\citenamefont {Lee},
  \citenamefont {Nagaosa},\ and\ \citenamefont {Wen}}]{LeeRMP}%
  \BibitemOpen
  \bibfield  {author} {\bibinfo {author} {\bibnamefont {Lee}, \bibfnamefont
  {Patrick~A}}, \bibinfo {author} {\bibfnamefont {Naoto}\ \bibnamefont
  {Nagaosa}}, and\ \bibinfo {author} {\bibfnamefont {Xiao-Gang}\ \bibnamefont
  {Wen}}} (\bibinfo {year} {2006}),\ \bibfield  {title} {\enquote {\bibinfo
  {title} {{Doping a Mott insulator: Physics of high-temperature
  superconductivity}},}\ }\href {https://doi.org/10.1103/RevModPhys.78.17}
  {\bibfield  {journal} {\bibinfo  {journal} {Rev. Mod. Phys.}\ }\textbf
  {\bibinfo {volume} {78}},\ \bibinfo {pages} {17--85}}\BibitemShut {NoStop}%
\bibitem [{\citenamefont {Lee}\ and\ \citenamefont
  {Ramakrishnan}(1985)}]{LeeTVRRMP}%
  \BibitemOpen
  \bibfield  {author} {\bibinfo {author} {\bibnamefont {Lee}, \bibfnamefont
  {Patrick~A}}, and\ \bibinfo {author} {\bibfnamefont {T.~V.}\ \bibnamefont
  {Ramakrishnan}}} (\bibinfo {year} {1985}),\ \bibfield  {title} {\enquote
  {\bibinfo {title} {Disordered electronic systems},}\ }\href
  {https://doi.org/10.1103/RevModPhys.57.287} {\bibfield  {journal} {\bibinfo
  {journal} {Rev. Mod. Phys.}\ }\textbf {\bibinfo {volume} {57}},\ \bibinfo
  {pages} {287--337}}\BibitemShut {NoStop}%
\bibitem [{\citenamefont {{Lee}}(2009)}]{sungsik1}%
  \BibitemOpen
  \bibfield  {author} {\bibinfo {author} {\bibnamefont {{Lee}}, \bibfnamefont
  {Sung-Sik}}} (\bibinfo {year} {2009}),\ \bibfield  {title} {\enquote
  {\bibinfo {title} {{Low-energy effective theory of Fermi surface coupled with
  U(1) gauge field in 2+1 dimensions}},}\ }\href
  {https://doi.org/10.1103/PhysRevB.80.165102} {\bibfield  {journal} {\bibinfo
  {journal} {Phys. Rev. B}\ }\textbf {\bibinfo {volume} {80}}~(\bibinfo
  {number} {16}),\ \bibinfo {eid} {165102}},\ \Eprint
  {https://arxiv.org/abs/0905.4532} {arXiv:0905.4532 [cond-mat.str-el]}
  \BibitemShut {NoStop}%
\bibitem [{\citenamefont {{Lee}}(2018)}]{lee_review}%
  \BibitemOpen
  \bibfield  {author} {\bibinfo {author} {\bibnamefont {{Lee}}, \bibfnamefont
  {Sung-Sik}}} (\bibinfo {year} {2018}),\ \bibfield  {title} {\enquote
  {\bibinfo {title} {{Recent Developments in Non-Fermi Liquid Theory}},}\
  }\href {https://doi.org/10.1146/annurev-conmatphys-031016-025531} {\bibfield
  {journal} {\bibinfo  {journal} {Annual Review of Condensed Matter Physics}\
  }\textbf {\bibinfo {volume} {9}},\ \bibinfo {pages} {227--244}},\ \Eprint
  {https://arxiv.org/abs/1703.08172} {arXiv:1703.08172 [cond-mat.str-el]}
  \BibitemShut {NoStop}%
\bibitem [{\citenamefont {Lee}\ \emph {et~al.}(2002)\citenamefont {Lee},
  \citenamefont {Yu}, \citenamefont {Lee}, \citenamefont {Noh}, \citenamefont
  {Gimm}, \citenamefont {Choi},\ and\ \citenamefont {Eom}}]{Eom02}%
  \BibitemOpen
  \bibfield  {author} {\bibinfo {author} {\bibnamefont {Lee}, \bibfnamefont
  {Y~S}}, \bibinfo {author} {\bibfnamefont {Jaejun}\ \bibnamefont {Yu}},
  \bibinfo {author} {\bibfnamefont {J.~S.}\ \bibnamefont {Lee}}, \bibinfo
  {author} {\bibfnamefont {T.~W.}\ \bibnamefont {Noh}}, \bibinfo {author}
  {\bibfnamefont {T.-H.}\ \bibnamefont {Gimm}}, \bibinfo {author}
  {\bibfnamefont {Han-Yong}\ \bibnamefont {Choi}}, and\ \bibinfo {author}
  {\bibfnamefont {C.~B.}\ \bibnamefont {Eom}}} (\bibinfo {year} {2002}),\
  \bibfield  {title} {\enquote {\bibinfo {title} {{Non-Fermi liquid behavior
  and scaling of the low-frequency suppression in the optical conductivity
  spectra of CaRuO$_3$}},}\ }\href {https://doi.org/10.1103/PhysRevB.66.041104}
  {\bibfield  {journal} {\bibinfo  {journal} {Phys. Rev. B}\ }\textbf {\bibinfo
  {volume} {66}},\ \bibinfo {pages} {041104}}\BibitemShut {NoStop}%
\bibitem [{\citenamefont {Legros}\ \emph {et~al.}(2019)\citenamefont {Legros},
  \citenamefont {Benhabib}, \citenamefont {Tabis}, \citenamefont {Laliberte},
  \citenamefont {Dion}, \citenamefont {Lizaire}, \citenamefont {Vignolle},
  \citenamefont {Vignolles}, \citenamefont {Raffy}, \citenamefont {Li},
  \citenamefont {Auban-Senzier}, \citenamefont {Doiron-Leyraud}, \citenamefont
  {Fournier}, \citenamefont {Colson}, \citenamefont {Taillefer},\ and\
  \citenamefont {Proust}}]{Legros19}%
  \BibitemOpen
  \bibfield  {author} {\bibinfo {author} {\bibnamefont {Legros}, \bibfnamefont
  {A}}, \bibinfo {author} {\bibfnamefont {S.}~\bibnamefont {Benhabib}},
  \bibinfo {author} {\bibfnamefont {W.}~\bibnamefont {Tabis}}, \bibinfo
  {author} {\bibfnamefont {F.}~\bibnamefont {Laliberte}}, \bibinfo {author}
  {\bibfnamefont {M.}~\bibnamefont {Dion}}, \bibinfo {author} {\bibfnamefont
  {M.}~\bibnamefont {Lizaire}}, \bibinfo {author} {\bibfnamefont
  {B.}~\bibnamefont {Vignolle}}, \bibinfo {author} {\bibfnamefont
  {D.}~\bibnamefont {Vignolles}}, \bibinfo {author} {\bibfnamefont
  {H.}~\bibnamefont {Raffy}}, \bibinfo {author} {\bibfnamefont {Z.~Z.}\
  \bibnamefont {Li}}, \bibinfo {author} {\bibfnamefont {P.}~\bibnamefont
  {Auban-Senzier}}, \bibinfo {author} {\bibfnamefont {N.}~\bibnamefont
  {Doiron-Leyraud}}, \bibinfo {author} {\bibfnamefont {P.}~\bibnamefont
  {Fournier}}, \bibinfo {author} {\bibfnamefont {D.}~\bibnamefont {Colson}},
  \bibinfo {author} {\bibfnamefont {L.}~\bibnamefont {Taillefer}}, and\
  \bibinfo {author} {\bibfnamefont {C.}~\bibnamefont {Proust}}} (\bibinfo
  {year} {2019}),\ \bibfield  {title} {\enquote {\bibinfo {title} {{Universal
  $T$-linear resistivity and Planckian dissipation in overdoped cuprates}},}\
  }\href {https://doi.org/10.1038/s41567-018-0334-2} {\bibfield  {journal}
  {\bibinfo  {journal} {Nature Physics}\ }\textbf {\bibinfo {volume}
  {15}}~(\bibinfo {number} {2}),\ \bibinfo {pages} {142--147}}\BibitemShut
  {NoStop}%
\bibitem [{\citenamefont {Lensky}\ and\ \citenamefont
  {Qi}(2021)}]{Lensky:2020fqf}%
  \BibitemOpen
  \bibfield  {author} {\bibinfo {author} {\bibnamefont {Lensky}, \bibfnamefont
  {Yuri~D}}, and\ \bibinfo {author} {\bibfnamefont {Xiao-Liang}\ \bibnamefont
  {Qi}}} (\bibinfo {year} {2021}),\ \bibfield  {title} {\enquote {\bibinfo
  {title} {{Rescuing a black hole in the large-$q$ coupled SYK model}},}\
  }\href {https://doi.org/10.1007/JHEP04(2021)116} {\bibfield  {journal}
  {\bibinfo  {journal} {JHEP}\ }\textbf {\bibinfo {volume} {04}},\ \bibinfo
  {pages} {116}},\ \Eprint {https://arxiv.org/abs/2012.15798} {arXiv:2012.15798
  [hep-th]} \BibitemShut {NoStop}%
\bibitem [{\citenamefont {Liao}\ and\ \citenamefont
  {Galitski}(2021)}]{Liao:2021ofk}%
  \BibitemOpen
  \bibfield  {author} {\bibinfo {author} {\bibnamefont {Liao}, \bibfnamefont
  {Yunxiang}}, and\ \bibinfo {author} {\bibfnamefont {Victor}\ \bibnamefont
  {Galitski}}} (\bibinfo {year} {2021}),\ \bibfield  {title} {\enquote
  {\bibinfo {title} {{Emergence of many-body quantum chaos via spontaneous
  breaking of unitarity}},}\ }\href@noop {} {\ }\Eprint
  {https://arxiv.org/abs/2104.05721} {arXiv:2104.05721 [cond-mat.stat-mech]}
  \BibitemShut {NoStop}%
\bibitem [{\citenamefont {Liao}\ \emph {et~al.}(2020)\citenamefont {Liao},
  \citenamefont {Vikram},\ and\ \citenamefont {Galitski}}]{Liao:2020lac}%
  \BibitemOpen
  \bibfield  {author} {\bibinfo {author} {\bibnamefont {Liao}, \bibfnamefont
  {Yunxiang}}, \bibinfo {author} {\bibfnamefont {Amit}\ \bibnamefont {Vikram}},
  and\ \bibinfo {author} {\bibfnamefont {Victor}\ \bibnamefont {Galitski}}}
  (\bibinfo {year} {2020}),\ \bibfield  {title} {\enquote {\bibinfo {title}
  {{Many-body level statistics of single-particle quantum chaos}},}\ }\href
  {https://doi.org/10.1103/PhysRevLett.125.250601} {\bibfield  {journal}
  {\bibinfo  {journal} {Phys. Rev. Lett.}\ }\textbf {\bibinfo {volume} {125}},\
  \bibinfo {pages} {250601}},\ \Eprint {https://arxiv.org/abs/2005.08991}
  {arXiv:2005.08991 [cond-mat.stat-mech]} \BibitemShut {NoStop}%
\bibitem [{\citenamefont {Limelette}\ \emph {et~al.}(2013)\citenamefont
  {Limelette}, \citenamefont {Ta~Phuoc}, \citenamefont {Gervais},\ and\
  \citenamefont {Fr\'esard}}]{Limelette_2013}%
  \BibitemOpen
  \bibfield  {author} {\bibinfo {author} {\bibnamefont {Limelette},
  \bibfnamefont {P}}, \bibinfo {author} {\bibfnamefont {V.}~\bibnamefont
  {Ta~Phuoc}}, \bibinfo {author} {\bibfnamefont {F.}~\bibnamefont {Gervais}},
  and\ \bibinfo {author} {\bibfnamefont {R.}~\bibnamefont {Fr\'esard}}}
  (\bibinfo {year} {2013}),\ \bibfield  {title} {\enquote {\bibinfo {title}
  {{$\ensuremath{\omega}/T$ scaling of the optical conductivity in strongly
  correlated layered cobalt oxide}},}\ }\href
  {https://doi.org/10.1103/PhysRevB.87.035102} {\bibfield  {journal} {\bibinfo
  {journal} {Phys. Rev. B}\ }\textbf {\bibinfo {volume} {87}},\ \bibinfo
  {pages} {035102}}\BibitemShut {NoStop}%
\bibitem [{\citenamefont {Lindner}\ and\ \citenamefont
  {Auerbach}(2010)}]{lindner}%
  \BibitemOpen
  \bibfield  {author} {\bibinfo {author} {\bibnamefont {Lindner}, \bibfnamefont
  {Netanel~H}}, and\ \bibinfo {author} {\bibfnamefont {Assa}\ \bibnamefont
  {Auerbach}}} (\bibinfo {year} {2010}),\ \bibfield  {title} {\enquote
  {\bibinfo {title} {Conductivity of hard core bosons: A paradigm of a bad
  metal},}\ }\href {https://doi.org/10.1103/PhysRevB.81.054512} {\bibfield
  {journal} {\bibinfo  {journal} {Phys. Rev. B}\ }\textbf {\bibinfo {volume}
  {81}},\ \bibinfo {pages} {054512}}\BibitemShut {NoStop}%
\bibitem [{\citenamefont {Liu}\ \emph {et~al.}(2011)\citenamefont {Liu},
  \citenamefont {McGreevy},\ and\ \citenamefont {Vegh}}]{Liu1}%
  \BibitemOpen
  \bibfield  {author} {\bibinfo {author} {\bibnamefont {Liu}, \bibfnamefont
  {Hong}}, \bibinfo {author} {\bibfnamefont {John}\ \bibnamefont {McGreevy}},
  and\ \bibinfo {author} {\bibfnamefont {David}\ \bibnamefont {Vegh}}}
  (\bibinfo {year} {2011}),\ \bibfield  {title} {\enquote {\bibinfo {title}
  {{Non-Fermi liquids from holography}},}\ }\href
  {https://doi.org/10.1103/PhysRevD.83.065029} {\bibfield  {journal} {\bibinfo
  {journal} {Phys. Rev. D}\ }\textbf {\bibinfo {volume} {83}},\ \bibinfo
  {pages} {065029}},\ \Eprint {https://arxiv.org/abs/0903.2477}
  {arXiv:0903.2477 [hep-th]} \BibitemShut {NoStop}%
\bibitem [{\citenamefont {Liu}\ and\ \citenamefont
  {Sonner}(2020)}]{Liu:2020rrn}%
  \BibitemOpen
  \bibfield  {author} {\bibinfo {author} {\bibnamefont {Liu}, \bibfnamefont
  {Hong}}, and\ \bibinfo {author} {\bibfnamefont {Julian}\ \bibnamefont
  {Sonner}}} (\bibinfo {year} {2020}),\ \bibfield  {title} {\enquote {\bibinfo
  {title} {{Quantum many-body physics from a gravitational lens}},}\ }\href
  {https://doi.org/10.1038/s42254-020-0225-1} {\bibfield  {journal} {\bibinfo
  {journal} {Nature Rev. Phys.}\ }\textbf {\bibinfo {volume} {2}}~(\bibinfo
  {number} {11}),\ \bibinfo {pages} {615--633}},\ \Eprint
  {https://arxiv.org/abs/2004.06159} {arXiv:2004.06159 [hep-th]} \BibitemShut
  {NoStop}%
\bibitem [{\citenamefont {L\"ohneysen}\ \emph {et~al.}(2007)\citenamefont
  {L\"ohneysen}, \citenamefont {Rosch}, \citenamefont {Vojta},\ and\
  \citenamefont {W\"olfle}}]{rmpqcp}%
  \BibitemOpen
  \bibfield  {author} {\bibinfo {author} {\bibnamefont {L\"ohneysen},
  \bibfnamefont {Hilbert~v}}, \bibinfo {author} {\bibfnamefont {Achim}\
  \bibnamefont {Rosch}}, \bibinfo {author} {\bibfnamefont {Matthias}\
  \bibnamefont {Vojta}}, and\ \bibinfo {author} {\bibfnamefont {Peter}\
  \bibnamefont {W\"olfle}}} (\bibinfo {year} {2007}),\ \bibfield  {title}
  {\enquote {\bibinfo {title} {Fermi-liquid instabilities at magnetic quantum
  phase transitions},}\ }\href {https://doi.org/10.1103/RevModPhys.79.1015}
  {\bibfield  {journal} {\bibinfo  {journal} {Rev. Mod. Phys.}\ }\textbf
  {\bibinfo {volume} {79}},\ \bibinfo {pages} {1015--1075}}\BibitemShut
  {NoStop}%
\bibitem [{\citenamefont {Loram}\ \emph {et~al.}(1994)\citenamefont {Loram},
  \citenamefont {Mirza}, \citenamefont {Wade}, \citenamefont {Cooper},\ and\
  \citenamefont {Liang}}]{loram}%
  \BibitemOpen
  \bibfield  {author} {\bibinfo {author} {\bibnamefont {Loram}, \bibfnamefont
  {JW}}, \bibinfo {author} {\bibfnamefont {K.A.}\ \bibnamefont {Mirza}},
  \bibinfo {author} {\bibfnamefont {J.M.}\ \bibnamefont {Wade}}, \bibinfo
  {author} {\bibfnamefont {J.R.}\ \bibnamefont {Cooper}}, and\ \bibinfo
  {author} {\bibfnamefont {W.Y.}\ \bibnamefont {Liang}}} (\bibinfo {year}
  {1994}),\ \bibfield  {title} {\enquote {\bibinfo {title} {The electronic
  specific heat of cuprate superconductors},}\ }\href
  {https://doi.org/https://doi.org/10.1016/0921-4534(94)91331-5} {\bibfield
  {journal} {\bibinfo  {journal} {Physica C: Superconductivity}\ }\textbf
  {\bibinfo {volume} {235-240}},\ \bibinfo {pages} {134 -- 137}}\BibitemShut
  {NoStop}%
\bibitem [{\citenamefont {Lucas}(2015)}]{Lucas:2015vna}%
  \BibitemOpen
  \bibfield  {author} {\bibinfo {author} {\bibnamefont {Lucas}, \bibfnamefont
  {Andrew}}} (\bibinfo {year} {2015}),\ \bibfield  {title} {\enquote {\bibinfo
  {title} {{Conductivity of a strange metal: from holography to memory
  functions}},}\ }\href {https://doi.org/10.1007/JHEP03(2015)071} {\bibfield
  {journal} {\bibinfo  {journal} {JHEP}\ }\textbf {\bibinfo {volume} {03}},\
  \bibinfo {pages} {071}},\ \Eprint {https://arxiv.org/abs/1501.05656}
  {arXiv:1501.05656 [hep-th]} \BibitemShut {NoStop}%
\bibitem [{\citenamefont {Lucas}\ and\ \citenamefont
  {Sachdev}(2015)}]{Lucas:2015pxa}%
  \BibitemOpen
  \bibfield  {author} {\bibinfo {author} {\bibnamefont {Lucas}, \bibfnamefont
  {Andrew}}, and\ \bibinfo {author} {\bibfnamefont {Subir}\ \bibnamefont
  {Sachdev}}} (\bibinfo {year} {2015}),\ \bibfield  {title} {\enquote {\bibinfo
  {title} {{Memory matrix theory of magnetotransport in strange metals}},}\
  }\href {https://doi.org/10.1103/PhysRevB.91.195122} {\bibfield  {journal}
  {\bibinfo  {journal} {Phys. Rev. B}\ }\textbf {\bibinfo {volume}
  {91}}~(\bibinfo {number} {19}),\ \bibinfo {pages} {195122}},\ \Eprint
  {https://arxiv.org/abs/1502.04704} {arXiv:1502.04704 [cond-mat.str-el]}
  \BibitemShut {NoStop}%
\bibitem [{\citenamefont {Lucas}\ and\ \citenamefont
  {Steinberg}(2016)}]{Lucas1}%
  \BibitemOpen
  \bibfield  {author} {\bibinfo {author} {\bibnamefont {Lucas}, \bibfnamefont
  {Andrew}}, and\ \bibinfo {author} {\bibfnamefont {Julia}\ \bibnamefont
  {Steinberg}}} (\bibinfo {year} {2016}),\ \bibfield  {title} {\enquote
  {\bibinfo {title} {Charge diffusion and the butterfly effect in striped
  holographic matter},}\ }\href {https://doi.org/10.1007/JHEP10(2016)143}
  {\bibfield  {journal} {\bibinfo  {journal} {Journal of High Energy Physics}\
  }\textbf {\bibinfo {volume} {2016}}~(\bibinfo {number} {10}),\ \bibinfo
  {pages} {143}}\BibitemShut {NoStop}%
\bibitem [{\citenamefont {Luitz}\ and\ \citenamefont {Bar~Lev}(2017)}]{Luitz}%
  \BibitemOpen
  \bibfield  {author} {\bibinfo {author} {\bibnamefont {Luitz}, \bibfnamefont
  {David~J}}, and\ \bibinfo {author} {\bibfnamefont {Yevgeny}\ \bibnamefont
  {Bar~Lev}}} (\bibinfo {year} {2017}),\ \bibfield  {title} {\enquote {\bibinfo
  {title} {Information propagation in isolated quantum systems},}\ }\href
  {https://doi.org/10.1103/PhysRevB.96.020406} {\bibfield  {journal} {\bibinfo
  {journal} {Phys. Rev. B}\ }\textbf {\bibinfo {volume} {96}},\ \bibinfo
  {pages} {020406}}\BibitemShut {NoStop}%
\bibitem [{\citenamefont {{Luo}}\ \emph {et~al.}(2019)\citenamefont {{Luo}},
  \citenamefont {{You}}, \citenamefont {{Li}}, \citenamefont {{Jian}},
  \citenamefont {{Lu}}, \citenamefont {{Xu}}, \citenamefont {{Zeng}},\ and\
  \citenamefont {{Laflamme}}}]{Xu19}%
  \BibitemOpen
  \bibfield  {author} {\bibinfo {author} {\bibnamefont {{Luo}}, \bibfnamefont
  {Zhihuang}}, \bibinfo {author} {\bibfnamefont {Yi-Zhuang}\ \bibnamefont
  {{You}}}, \bibinfo {author} {\bibfnamefont {Jun}\ \bibnamefont {{Li}}},
  \bibinfo {author} {\bibfnamefont {Chao-Ming}\ \bibnamefont {{Jian}}},
  \bibinfo {author} {\bibfnamefont {Dawei}\ \bibnamefont {{Lu}}}, \bibinfo
  {author} {\bibfnamefont {Cenke}\ \bibnamefont {{Xu}}}, \bibinfo {author}
  {\bibfnamefont {Bei}\ \bibnamefont {{Zeng}}}, and\ \bibinfo {author}
  {\bibfnamefont {Raymond}\ \bibnamefont {{Laflamme}}}} (\bibinfo {year}
  {2019}),\ \bibfield  {title} {\enquote {\bibinfo {title} {{Quantum simulation
  of the non-Fermi-liquid state of Sachdev-Ye-Kitaev model}},}\ }\href
  {https://doi.org/10.1038/s41534-019-0166-7} {\bibfield  {journal} {\bibinfo
  {journal} {npj Quantum Information}\ }\textbf {\bibinfo {volume} {5}},\
  \bibinfo {eid} {53}},\ \Eprint {https://arxiv.org/abs/1712.06458}
  {arXiv:1712.06458 [quant-ph]} \BibitemShut {NoStop}%
\bibitem [{\citenamefont {Luttinger}\ and\ \citenamefont
  {Ward}(1960)}]{Luttinger_Ward_1960}%
  \BibitemOpen
  \bibfield  {author} {\bibinfo {author} {\bibnamefont {Luttinger},
  \bibfnamefont {J~M}}, and\ \bibinfo {author} {\bibfnamefont {J.~C.}\
  \bibnamefont {Ward}}} (\bibinfo {year} {1960}),\ \bibfield  {title} {\enquote
  {\bibinfo {title} {Ground-state energy of a many-fermion system. ii},}\
  }\href {https://doi.org/10.1103/PhysRev.118.1417} {\bibfield  {journal}
  {\bibinfo  {journal} {Phys. Rev.}\ }\textbf {\bibinfo {volume} {118}},\
  \bibinfo {pages} {1417--1427}}\BibitemShut {NoStop}%
\bibitem [{\citenamefont {Maksimovic}\ \emph {et~al.}(2022)\citenamefont
  {Maksimovic}, \citenamefont {Eilbott}, \citenamefont {Cookmeyer},
  \citenamefont {Wan}, \citenamefont {Rusz}, \citenamefont {Nagarajan},
  \citenamefont {Haley}, \citenamefont {Maniv}, \citenamefont {Gong},
  \citenamefont {Faubel}, \citenamefont {Hayes}, \citenamefont {Bangura},
  \citenamefont {Singleton}, \citenamefont {Palmstrom}, \citenamefont {Winter},
  \citenamefont {McDonald}, \citenamefont {Jang}, \citenamefont {Ai},
  \citenamefont {Lin}, \citenamefont {Ciocys}, \citenamefont {Gobbo},
  \citenamefont {Werman}, \citenamefont {Oppeneer}, \citenamefont {Altman},
  \citenamefont {Lanzara},\ and\ \citenamefont {Analytis}}]{Analytis20}%
  \BibitemOpen
  \bibfield  {author} {\bibinfo {author} {\bibnamefont {Maksimovic},
  \bibfnamefont {Nikola}}, \bibinfo {author} {\bibfnamefont {Daniel~H.}\
  \bibnamefont {Eilbott}}, \bibinfo {author} {\bibfnamefont {Tessa}\
  \bibnamefont {Cookmeyer}}, \bibinfo {author} {\bibfnamefont {Fanghui}\
  \bibnamefont {Wan}}, \bibinfo {author} {\bibfnamefont {Jan}\ \bibnamefont
  {Rusz}}, \bibinfo {author} {\bibfnamefont {Vikram}\ \bibnamefont
  {Nagarajan}}, \bibinfo {author} {\bibfnamefont {Shannon~C.}\ \bibnamefont
  {Haley}}, \bibinfo {author} {\bibfnamefont {Eran}\ \bibnamefont {Maniv}},
  \bibinfo {author} {\bibfnamefont {Amanda}\ \bibnamefont {Gong}}, \bibinfo
  {author} {\bibfnamefont {Stefano}\ \bibnamefont {Faubel}}, \bibinfo {author}
  {\bibfnamefont {Ian~M.}\ \bibnamefont {Hayes}}, \bibinfo {author}
  {\bibfnamefont {Ali}\ \bibnamefont {Bangura}}, \bibinfo {author}
  {\bibfnamefont {John}\ \bibnamefont {Singleton}}, \bibinfo {author}
  {\bibfnamefont {Johanna~C.}\ \bibnamefont {Palmstrom}}, \bibinfo {author}
  {\bibfnamefont {Laurel}\ \bibnamefont {Winter}}, \bibinfo {author}
  {\bibfnamefont {Ross}\ \bibnamefont {McDonald}}, \bibinfo {author}
  {\bibfnamefont {Sooyoung}\ \bibnamefont {Jang}}, \bibinfo {author}
  {\bibfnamefont {Ping}\ \bibnamefont {Ai}}, \bibinfo {author} {\bibfnamefont
  {Yi}~\bibnamefont {Lin}}, \bibinfo {author} {\bibfnamefont {Samuel}\
  \bibnamefont {Ciocys}}, \bibinfo {author} {\bibfnamefont {Jacob}\
  \bibnamefont {Gobbo}}, \bibinfo {author} {\bibfnamefont {Yochai}\
  \bibnamefont {Werman}}, \bibinfo {author} {\bibfnamefont {Peter~M.}\
  \bibnamefont {Oppeneer}}, \bibinfo {author} {\bibfnamefont {Ehud}\
  \bibnamefont {Altman}}, \bibinfo {author} {\bibfnamefont {Alessandra}\
  \bibnamefont {Lanzara}}, and\ \bibinfo {author} {\bibfnamefont {James~G.}\
  \bibnamefont {Analytis}}} (\bibinfo {year} {2022}),\ \bibfield  {title}
  {\enquote {\bibinfo {title} {{Evidence for a delocalization quantum phase
  transition without symmetry breaking in CeCoIn$_5$}},}\ }\href
  {https://doi.org/10.1126/science.aaz4566} {\bibfield  {journal} {\bibinfo
  {journal} {Science}\ }\textbf {\bibinfo {volume} {375}}~(\bibinfo {number}
  {6576}),\ \bibinfo {pages} {76--81}}\BibitemShut {NoStop}%
\bibitem [{\citenamefont {Maldacena}\ and\ \citenamefont
  {Qi}(2018)}]{Maldacena:2018lmt}%
  \BibitemOpen
  \bibfield  {author} {\bibinfo {author} {\bibnamefont {Maldacena},
  \bibfnamefont {Juan}}, and\ \bibinfo {author} {\bibfnamefont {Xiao-Liang}\
  \bibnamefont {Qi}}} (\bibinfo {year} {2018}),\ \bibfield  {title} {\enquote
  {\bibinfo {title} {{Eternal traversable wormhole}},}\ }\href@noop {} {\
  }\Eprint {https://arxiv.org/abs/1804.00491} {arXiv:1804.00491 [hep-th]}
  \BibitemShut {NoStop}%
\bibitem [{\citenamefont {Maldacena}\ \emph
  {et~al.}(2016{\natexlab{a}})\citenamefont {Maldacena}, \citenamefont
  {Shenker},\ and\ \citenamefont {Stanford}}]{Maldacena2016}%
  \BibitemOpen
  \bibfield  {author} {\bibinfo {author} {\bibnamefont {Maldacena},
  \bibfnamefont {Juan}}, \bibinfo {author} {\bibfnamefont {Stephen~H.}\
  \bibnamefont {Shenker}}, and\ \bibinfo {author} {\bibfnamefont {Douglas}\
  \bibnamefont {Stanford}}} (\bibinfo {year} {2016}{\natexlab{a}}),\ \bibfield
  {title} {\enquote {\bibinfo {title} {A bound on chaos},}\ }\href
  {https://doi.org/10.1007/JHEP08(2016)106} {\bibfield  {journal} {\bibinfo
  {journal} {Journal of High Energy Physics}\ }\textbf {\bibinfo {volume}
  {2016}}~(\bibinfo {number} {8}),\ \bibinfo {pages} {106}}\BibitemShut
  {NoStop}%
\bibitem [{\citenamefont {Maldacena}\ and\ \citenamefont
  {Stanford}(2016)}]{Maldacena_syk}%
  \BibitemOpen
  \bibfield  {author} {\bibinfo {author} {\bibnamefont {Maldacena},
  \bibfnamefont {Juan}}, and\ \bibinfo {author} {\bibfnamefont {Douglas}\
  \bibnamefont {Stanford}}} (\bibinfo {year} {2016}),\ \bibfield  {title}
  {\enquote {\bibinfo {title} {{Remarks on the Sachdev-Ye-Kitaev model}},}\
  }\href {https://doi.org/10.1103/PhysRevD.94.106002} {\bibfield  {journal}
  {\bibinfo  {journal} {Phys. Rev. D}\ }\textbf {\bibinfo {volume} {94}},\
  \bibinfo {pages} {106002}}\BibitemShut {NoStop}%
\bibitem [{\citenamefont {Maldacena}\ \emph
  {et~al.}(2016{\natexlab{b}})\citenamefont {Maldacena}, \citenamefont
  {Stanford},\ and\ \citenamefont {Yang}}]{JMDS16b}%
  \BibitemOpen
  \bibfield  {author} {\bibinfo {author} {\bibnamefont {Maldacena},
  \bibfnamefont {Juan}}, \bibinfo {author} {\bibfnamefont {Douglas}\
  \bibnamefont {Stanford}}, and\ \bibinfo {author} {\bibfnamefont {Zhenbin}\
  \bibnamefont {Yang}}} (\bibinfo {year} {2016}{\natexlab{b}}),\ \bibfield
  {title} {\enquote {\bibinfo {title} {{Conformal symmetry and its breaking in
  two dimensional Nearly Anti-de-Sitter space}},}\ }\href
  {https://doi.org/10.1093/ptep/ptw124} {\bibfield  {journal} {\bibinfo
  {journal} {Prog. Theor. Exp. Phys.}\ }\textbf {\bibinfo {volume}
  {2016}}~(\bibinfo {number} {12}),\ \bibinfo {pages} {12C104}},\ \Eprint
  {https://arxiv.org/abs/1606.01857} {arXiv:1606.01857 [hep-th]} \BibitemShut
  {NoStop}%
%%CITATION = ARXIV:1606.01857;%%
\bibitem [{\citenamefont {Maldacena}(1998)}]{Maldacena:1997re}%
  \BibitemOpen
  \bibfield  {author} {\bibinfo {author} {\bibnamefont {Maldacena},
  \bibfnamefont {Juan~Martin}}} (\bibinfo {year} {1998}),\ \bibfield  {title}
  {\enquote {\bibinfo {title} {{The Large $N$ limit of superconformal field
  theories and supergravity}},}\ }\href
  {https://doi.org/10.1023/A:1026654312961} {\bibfield  {journal} {\bibinfo
  {journal} {Adv. Theor. Math. Phys.}\ }\textbf {\bibinfo {volume} {2}},\
  \bibinfo {pages} {231--252}},\ \Eprint {https://arxiv.org/abs/hep-th/9711200}
  {arXiv:hep-th/9711200} \BibitemShut {NoStop}%
\bibitem [{\citenamefont {Maloney}\ and\ \citenamefont
  {Witten}(2020)}]{Maloney:2020nni}%
  \BibitemOpen
  \bibfield  {author} {\bibinfo {author} {\bibnamefont {Maloney}, \bibfnamefont
  {Alexander}}, and\ \bibinfo {author} {\bibfnamefont {Edward}\ \bibnamefont
  {Witten}}} (\bibinfo {year} {2020}),\ \bibfield  {title} {\enquote {\bibinfo
  {title} {{Averaging over Narain moduli space}},}\ }\href
  {https://doi.org/10.1007/JHEP10(2020)187} {\bibfield  {journal} {\bibinfo
  {journal} {JHEP}\ }\textbf {\bibinfo {volume} {10}},\ \bibinfo {pages}
  {187}},\ \Eprint {https://arxiv.org/abs/2006.04855} {arXiv:2006.04855
  [hep-th]} \BibitemShut {NoStop}%
\bibitem [{\citenamefont {Marcus}\ and\ \citenamefont
  {Vandoren}(2019)}]{Marcus:2018tsr}%
  \BibitemOpen
  \bibfield  {author} {\bibinfo {author} {\bibnamefont {Marcus}, \bibfnamefont
  {Eric}}, and\ \bibinfo {author} {\bibfnamefont {Stefan}\ \bibnamefont
  {Vandoren}}} (\bibinfo {year} {2019}),\ \bibfield  {title} {\enquote
  {\bibinfo {title} {{A new class of SYK-like models with maximal chaos}},}\
  }\href {https://doi.org/10.1007/JHEP01(2019)166} {\bibfield  {journal}
  {\bibinfo  {journal} {JHEP}\ }\textbf {\bibinfo {volume} {01}},\ \bibinfo
  {pages} {166}},\ \Eprint {https://arxiv.org/abs/1808.01190} {arXiv:1808.01190
  [hep-th]} \BibitemShut {NoStop}%
\bibitem [{\citenamefont {van~der Marel}\ \emph {et~al.}(2003)\citenamefont
  {van~der Marel}, \citenamefont {Molegraaf}, \citenamefont {Zaanen},
  \citenamefont {Nussinov}, \citenamefont {Carbone}, \citenamefont
  {Damascelli}, \citenamefont {Eisaki}, \citenamefont {Greven}, \citenamefont
  {Kes},\ and\ \citenamefont {Li}}]{Marel2003}%
  \BibitemOpen
  \bibfield  {author} {\bibinfo {author} {\bibnamefont {van~der Marel},
  \bibfnamefont {D}}, \bibinfo {author} {\bibfnamefont {H.~J.~A.}\ \bibnamefont
  {Molegraaf}}, \bibinfo {author} {\bibfnamefont {J.}~\bibnamefont {Zaanen}},
  \bibinfo {author} {\bibfnamefont {Z.}~\bibnamefont {Nussinov}}, \bibinfo
  {author} {\bibfnamefont {F.}~\bibnamefont {Carbone}}, \bibinfo {author}
  {\bibfnamefont {A.}~\bibnamefont {Damascelli}}, \bibinfo {author}
  {\bibfnamefont {H.}~\bibnamefont {Eisaki}}, \bibinfo {author} {\bibfnamefont
  {M.}~\bibnamefont {Greven}}, \bibinfo {author} {\bibfnamefont {P.~H.}\
  \bibnamefont {Kes}}, and\ \bibinfo {author} {\bibfnamefont {M.}~\bibnamefont
  {Li}}} (\bibinfo {year} {2003}),\ \bibfield  {title} {\enquote {\bibinfo
  {title} {{Quantum critical behaviour in a high-$T_c$ superconductor}},}\
  }\href {https://doi.org/10.1038/nature01978} {\bibfield  {journal} {\bibinfo
  {journal} {Nature}\ }\textbf {\bibinfo {volume} {425}},\ \bibinfo {pages}
  {271}}\BibitemShut {NoStop}%
\bibitem [{\citenamefont {{Maslov}}\ \emph {et~al.}(2011)\citenamefont
  {{Maslov}}, \citenamefont {{Yudson}},\ and\ \citenamefont
  {{Chubukov}}}]{Maslov2011}%
  \BibitemOpen
  \bibfield  {author} {\bibinfo {author} {\bibnamefont {{Maslov}},
  \bibfnamefont {Dmitrii~L}}, \bibinfo {author} {\bibfnamefont {Vladimir~I.}\
  \bibnamefont {{Yudson}}}, and\ \bibinfo {author} {\bibfnamefont {Andrey~V.}\
  \bibnamefont {{Chubukov}}}} (\bibinfo {year} {2011}),\ \bibfield  {title}
  {\enquote {\bibinfo {title} {{Resistivity of a Non-Galilean-Invariant Fermi
  Liquid near Pomeranchuk Quantum Criticality}},}\ }\href
  {https://doi.org/10.1103/PhysRevLett.106.106403} {\bibfield  {journal}
  {\bibinfo  {journal} {Phys. Rev. Lett.}\ }\textbf {\bibinfo {volume}
  {106}}~(\bibinfo {number} {10}),\ \bibinfo {eid} {106403}},\ \Eprint
  {https://arxiv.org/abs/1012.0069} {arXiv:1012.0069 [cond-mat.str-el]}
  \BibitemShut {NoStop}%
\bibitem [{\citenamefont {Mehta}(2004)}]{Mehta}%
  \BibitemOpen
  \bibfield  {author} {\bibinfo {author} {\bibnamefont {Mehta}, \bibfnamefont
  {Madan~Lal}}} (\bibinfo {year} {2004}),\ \href@noop {} {\emph {\bibinfo
  {title} {Random matrices}}}\ (\bibinfo  {publisher} {Elsevier})\BibitemShut
  {NoStop}%
\bibitem [{\citenamefont {Mena}\ \emph {et~al.}(2003)\citenamefont {Mena},
  \citenamefont {van~der Marel}, \citenamefont {Damascelli}, \citenamefont
  {F\"ath}, \citenamefont {Menovsky},\ and\ \citenamefont
  {Mydosh}}]{Mena_2003}%
  \BibitemOpen
  \bibfield  {author} {\bibinfo {author} {\bibnamefont {Mena}, \bibfnamefont
  {F~P}}, \bibinfo {author} {\bibfnamefont {D.}~\bibnamefont {van~der Marel}},
  \bibinfo {author} {\bibfnamefont {A.}~\bibnamefont {Damascelli}}, \bibinfo
  {author} {\bibfnamefont {M.}~\bibnamefont {F\"ath}}, \bibinfo {author}
  {\bibfnamefont {A.~A.}\ \bibnamefont {Menovsky}}, and\ \bibinfo {author}
  {\bibfnamefont {J.~A.}\ \bibnamefont {Mydosh}}} (\bibinfo {year} {2003}),\
  \bibfield  {title} {\enquote {\bibinfo {title} {Heavy carriers and non-drude
  optical conductivity in mnsi},}\ }\href
  {https://doi.org/10.1103/PhysRevB.67.241101} {\bibfield  {journal} {\bibinfo
  {journal} {Phys. Rev. B}\ }\textbf {\bibinfo {volume} {67}},\ \bibinfo
  {pages} {241101}}\BibitemShut {NoStop}%
\bibitem [{\citenamefont {{Metlitski}}\ and\ \citenamefont
  {{Sachdev}}(2010)}]{metlitski1}%
  \BibitemOpen
  \bibfield  {author} {\bibinfo {author} {\bibnamefont {{Metlitski}},
  \bibfnamefont {M~A}}, and\ \bibinfo {author} {\bibfnamefont {S.}~\bibnamefont
  {{Sachdev}}}} (\bibinfo {year} {2010}),\ \bibfield  {title} {\enquote
  {\bibinfo {title} {{Quantum phase transitions of metals in two spatial
  dimensions. I. Ising-nematic order}},}\ }\href
  {https://doi.org/10.1103/PhysRevB.82.075127} {\bibfield  {journal} {\bibinfo
  {journal} {Phys. Rev. B}\ }\textbf {\bibinfo {volume} {82}}~(\bibinfo
  {number} {7}),\ \bibinfo {eid} {075127}},\ \Eprint
  {https://arxiv.org/abs/1001.1153} {arXiv:1001.1153 [cond-mat.str-el]}
  \BibitemShut {NoStop}%
\bibitem [{\citenamefont {Metlitski}\ \emph {et~al.}(2015)\citenamefont
  {Metlitski}, \citenamefont {Mross}, \citenamefont {Sachdev},\ and\
  \citenamefont {Senthil}}]{MM15}%
  \BibitemOpen
  \bibfield  {author} {\bibinfo {author} {\bibnamefont {Metlitski},
  \bibfnamefont {Max~A}}, \bibinfo {author} {\bibfnamefont {David~F.}\
  \bibnamefont {Mross}}, \bibinfo {author} {\bibfnamefont {Subir}\ \bibnamefont
  {Sachdev}}, and\ \bibinfo {author} {\bibfnamefont {T.}~\bibnamefont
  {Senthil}}} (\bibinfo {year} {2015}),\ \bibfield  {title} {\enquote {\bibinfo
  {title} {{Cooper pairing in non-Fermi liquids}},}\ }\href
  {https://doi.org/10.1103/PhysRevB.91.115111} {\bibfield  {journal} {\bibinfo
  {journal} {Phys. Rev. B}\ }\textbf {\bibinfo {volume} {91}},\ \bibinfo
  {pages} {115111}}\BibitemShut {NoStop}%
\bibitem [{\citenamefont {Mezard}\ \emph {et~al.}(1987)\citenamefont {Mezard},
  \citenamefont {Parisi},\ and\ \citenamefont {Virasoro}}]{mezard1987spin}%
  \BibitemOpen
  \bibfield  {author} {\bibinfo {author} {\bibnamefont {Mezard}, \bibfnamefont
  {M}}, \bibinfo {author} {\bibfnamefont {G.}~\bibnamefont {Parisi}}, and\
  \bibinfo {author} {\bibfnamefont {M.A.}\ \bibnamefont {Virasoro}}} (\bibinfo
  {year} {1987}),\ \href {https://books.google.com/books?id=DwY8DQAAQBAJ}
  {\emph {\bibinfo {title} {Spin Glass Theory And Beyond: An Introduction To
  The Replica Method And Its Applications}}},\ World Scientific Lecture Notes
  In Physics\ (\bibinfo  {publisher} {World Scientific Publishing
  Company})\BibitemShut {NoStop}%
\bibitem [{\citenamefont {{Michon}}\ \emph {et~al.}(2022)\citenamefont
  {{Michon}}, \citenamefont {{Berthod}}, \citenamefont {{Rischau}},
  \citenamefont {{Ataei}}, \citenamefont {{Chen}}, \citenamefont {{Komiya}},
  \citenamefont {{Ono}}, \citenamefont {{Taillefer}}, \citenamefont {{van der
  Marel}},\ and\ \citenamefont {{Georges}}}]{Michon2022}%
  \BibitemOpen
  \bibfield  {author} {\bibinfo {author} {\bibnamefont {{Michon}},
  \bibfnamefont {B}}, \bibinfo {author} {\bibfnamefont {C.}~\bibnamefont
  {{Berthod}}}, \bibinfo {author} {\bibfnamefont {C.~W.}\ \bibnamefont
  {{Rischau}}}, \bibinfo {author} {\bibfnamefont {A.}~\bibnamefont {{Ataei}}},
  \bibinfo {author} {\bibfnamefont {L.}~\bibnamefont {{Chen}}}, \bibinfo
  {author} {\bibfnamefont {S.}~\bibnamefont {{Komiya}}}, \bibinfo {author}
  {\bibfnamefont {S.}~\bibnamefont {{Ono}}}, \bibinfo {author} {\bibfnamefont
  {L.}~\bibnamefont {{Taillefer}}}, \bibinfo {author} {\bibfnamefont
  {D.}~\bibnamefont {{van der Marel}}}, and\ \bibinfo {author} {\bibfnamefont
  {A.}~\bibnamefont {{Georges}}}} (\bibinfo {year} {2022}),\ \bibfield  {title}
  {\enquote {\bibinfo {title} {{Planckian behavior of cuprate superconductors:
  Reconciling the scaling of optical conductivity with resistivity and specific
  heat}},}\ }\href@noop {} {\ }\Eprint {https://arxiv.org/abs/2205.04030}
  {arXiv:2205.04030 [cond-mat.str-el]} \BibitemShut {NoStop}%
\bibitem [{\citenamefont {Michon}\ \emph {et~al.}(2019)\citenamefont {Michon},
  \citenamefont {Girod}, \citenamefont {Badoux}, \citenamefont
  {Ka{\v{c}}mar{\v{c}}{\'{\i}}k}, \citenamefont {Ma}, \citenamefont {Dragomir},
  \citenamefont {Dabkowska}, \citenamefont {Gaulin}, \citenamefont {Zhou},
  \citenamefont {Pyon}, \citenamefont {Takayama}, \citenamefont {Takagi},
  \citenamefont {Verret}, \citenamefont {Doiron-Leyraud}, \citenamefont
  {Marcenat}, \citenamefont {Taillefer},\ and\ \citenamefont
  {Klein}}]{Michon2019}%
  \BibitemOpen
  \bibfield  {author} {\bibinfo {author} {\bibnamefont {Michon}, \bibfnamefont
  {B}}, \bibinfo {author} {\bibfnamefont {C.}~\bibnamefont {Girod}}, \bibinfo
  {author} {\bibfnamefont {S.}~\bibnamefont {Badoux}}, \bibinfo {author}
  {\bibfnamefont {J.}~\bibnamefont {Ka{\v{c}}mar{\v{c}}{\'{\i}}k}}, \bibinfo
  {author} {\bibfnamefont {Q.}~\bibnamefont {Ma}}, \bibinfo {author}
  {\bibfnamefont {M.}~\bibnamefont {Dragomir}}, \bibinfo {author}
  {\bibfnamefont {H.~A.}\ \bibnamefont {Dabkowska}}, \bibinfo {author}
  {\bibfnamefont {B.~D.}\ \bibnamefont {Gaulin}}, \bibinfo {author}
  {\bibfnamefont {J.-S.}\ \bibnamefont {Zhou}}, \bibinfo {author}
  {\bibfnamefont {S.}~\bibnamefont {Pyon}}, \bibinfo {author} {\bibfnamefont
  {T.}~\bibnamefont {Takayama}}, \bibinfo {author} {\bibfnamefont
  {H.}~\bibnamefont {Takagi}}, \bibinfo {author} {\bibfnamefont
  {S.}~\bibnamefont {Verret}}, \bibinfo {author} {\bibfnamefont
  {N.}~\bibnamefont {Doiron-Leyraud}}, \bibinfo {author} {\bibfnamefont
  {C.}~\bibnamefont {Marcenat}}, \bibinfo {author} {\bibfnamefont
  {L.}~\bibnamefont {Taillefer}}, and\ \bibinfo {author} {\bibfnamefont
  {T.}~\bibnamefont {Klein}}} (\bibinfo {year} {2019}),\ \bibfield  {title}
  {\enquote {\bibinfo {title} {Thermodynamic signatures of quantum criticality
  in cuprate superconductors},}\ }\href
  {https://doi.org/10.1038/s41586-019-0932-x} {\bibfield  {journal} {\bibinfo
  {journal} {Nature}\ }\textbf {\bibinfo {volume} {567}}~(\bibinfo {number}
  {7747}),\ \bibinfo {pages} {218--222}}\BibitemShut {NoStop}%
\bibitem [{\citenamefont {Micklitz}\ \emph {et~al.}(2019)\citenamefont
  {Micklitz}, \citenamefont {Monteiro},\ and\ \citenamefont
  {Altland}}]{micklitz19}%
  \BibitemOpen
  \bibfield  {author} {\bibinfo {author} {\bibnamefont {Micklitz},
  \bibfnamefont {T}}, \bibinfo {author} {\bibfnamefont {Felipe}\ \bibnamefont
  {Monteiro}}, and\ \bibinfo {author} {\bibfnamefont {Alexander}\ \bibnamefont
  {Altland}}} (\bibinfo {year} {2019}),\ \bibfield  {title} {\enquote {\bibinfo
  {title} {{Nonergodic Extended States in the Sachdev-Ye-Kitaev Model}},}\
  }\href {https://doi.org/10.1103/PhysRevLett.123.125701} {\bibfield  {journal}
  {\bibinfo  {journal} {Phys. Rev. Lett.}\ }\textbf {\bibinfo {volume} {123}},\
  \bibinfo {pages} {125701}}\BibitemShut {NoStop}%
\bibitem [{\citenamefont {Milekhin}(2021)}]{Milekhin:2021cou}%
  \BibitemOpen
  \bibfield  {author} {\bibinfo {author} {\bibnamefont {Milekhin},
  \bibfnamefont {Alexey}}} (\bibinfo {year} {2021}),\ \bibfield  {title}
  {\enquote {\bibinfo {title} {{Non-local reparametrization action in coupled
  Sachdev-Ye-Kitaev models}},}\ }\href
  {https://doi.org/10.1007/JHEP12(2021)114} {\bibfield  {journal} {\bibinfo
  {journal} {JHEP}\ }\textbf {\bibinfo {volume} {12}},\ \bibinfo {pages}
  {114}},\ \Eprint {https://arxiv.org/abs/2102.06647} {arXiv:2102.06647
  [hep-th]} \BibitemShut {NoStop}%
\bibitem [{\citenamefont {Millis}(1993)}]{Millis1993}%
  \BibitemOpen
  \bibfield  {author} {\bibinfo {author} {\bibnamefont {Millis}, \bibfnamefont
  {A~J}}} (\bibinfo {year} {1993}),\ \bibfield  {title} {\enquote {\bibinfo
  {title} {Effect of a nonzero temperature on quantum critical points in
  itinerant fermion systems},}\ }\href
  {https://doi.org/10.1103/PhysRevB.48.7183} {\bibfield  {journal} {\bibinfo
  {journal} {Phys. Rev. B}\ }\textbf {\bibinfo {volume} {48}},\ \bibinfo
  {pages} {7183--7196}}\BibitemShut {NoStop}%
\bibitem [{\citenamefont {Mitrano}\ \emph {et~al.}(2018)\citenamefont
  {Mitrano}, \citenamefont {Husain}, \citenamefont {Vig}, \citenamefont
  {Kogar}, \citenamefont {Rak}, \citenamefont {Rubeck}, \citenamefont
  {Schmalian}, \citenamefont {Uchoa}, \citenamefont {Schneeloch}, \citenamefont
  {Zhong}, \citenamefont {Gu},\ and\ \citenamefont {Abbamonte}}]{Abbamonte1}%
  \BibitemOpen
  \bibfield  {author} {\bibinfo {author} {\bibnamefont {Mitrano}, \bibfnamefont
  {M}}, \bibinfo {author} {\bibfnamefont {A.~A.}\ \bibnamefont {Husain}},
  \bibinfo {author} {\bibfnamefont {S.}~\bibnamefont {Vig}}, \bibinfo {author}
  {\bibfnamefont {A.}~\bibnamefont {Kogar}}, \bibinfo {author} {\bibfnamefont
  {M.~S.}\ \bibnamefont {Rak}}, \bibinfo {author} {\bibfnamefont {S.~I.}\
  \bibnamefont {Rubeck}}, \bibinfo {author} {\bibfnamefont {J.}~\bibnamefont
  {Schmalian}}, \bibinfo {author} {\bibfnamefont {B.}~\bibnamefont {Uchoa}},
  \bibinfo {author} {\bibfnamefont {J.}~\bibnamefont {Schneeloch}}, \bibinfo
  {author} {\bibfnamefont {R.}~\bibnamefont {Zhong}}, \bibinfo {author}
  {\bibfnamefont {G.~D.}\ \bibnamefont {Gu}}, and\ \bibinfo {author}
  {\bibfnamefont {P.}~\bibnamefont {Abbamonte}}} (\bibinfo {year} {2018}),\
  \bibfield  {title} {\enquote {\bibinfo {title} {Anomalous density
  fluctuations in a strange metal},}\ }\href
  {https://doi.org/10.1073/pnas.1721495115} {\bibfield  {journal} {\bibinfo
  {journal} {Proceedings of the National Academy of Sciences}\ }\textbf
  {\bibinfo {volume} {115}}~(\bibinfo {number} {21}),\ \bibinfo {pages}
  {5392--5396}}\BibitemShut {NoStop}%
\bibitem [{\citenamefont {Moitra}\ \emph {et~al.}(2019)\citenamefont {Moitra},
  \citenamefont {Trivedi},\ and\ \citenamefont {Vishal}}]{Moitra:2018jqs}%
  \BibitemOpen
  \bibfield  {author} {\bibinfo {author} {\bibnamefont {Moitra}, \bibfnamefont
  {Upamanyu}}, \bibinfo {author} {\bibfnamefont {Sandip~P.}\ \bibnamefont
  {Trivedi}}, and\ \bibinfo {author} {\bibfnamefont {V.}~\bibnamefont
  {Vishal}}} (\bibinfo {year} {2019}),\ \bibfield  {title} {\enquote {\bibinfo
  {title} {{Extremal and near-extremal black holes and near-CFT$_{1}$}},}\
  }\href {https://doi.org/10.1007/JHEP07(2019)055} {\bibfield  {journal}
  {\bibinfo  {journal} {JHEP}\ }\textbf {\bibinfo {volume} {07}},\ \bibinfo
  {pages} {055}},\ \Eprint {https://arxiv.org/abs/1808.08239} {arXiv:1808.08239
  [hep-th]} \BibitemShut {NoStop}%
\bibitem [{\citenamefont {{Moon}}\ and\ \citenamefont
  {{Chubukov}}(2010)}]{MoonChubukov}%
  \BibitemOpen
  \bibfield  {author} {\bibinfo {author} {\bibnamefont {{Moon}}, \bibfnamefont
  {Eun-Gook}}, and\ \bibinfo {author} {\bibfnamefont {Andrey}\ \bibnamefont
  {{Chubukov}}}} (\bibinfo {year} {2010}),\ \bibfield  {title} {\enquote
  {\bibinfo {title} {{Quantum-critical Pairing with Varying Exponents}},}\
  }\href {https://doi.org/10.1007/s10909-010-0199-y} {\bibfield  {journal}
  {\bibinfo  {journal} {Journal of Low Temperature Physics}\ }\textbf {\bibinfo
  {volume} {161}}~(\bibinfo {number} {1-2}),\ \bibinfo {pages} {263--281}},\
  \Eprint {https://arxiv.org/abs/1005.0356} {arXiv:1005.0356
  [cond-mat.supr-con]} \BibitemShut {NoStop}%
\bibitem [{\citenamefont {Moriya}(1985)}]{Moriya1985}%
  \BibitemOpen
  \bibfield  {author} {\bibinfo {author} {\bibnamefont {Moriya}, \bibfnamefont
  {T}}} (\bibinfo {year} {1985}),\ \href@noop {} {\emph {\bibinfo {title} {Spin
  Fluctuations in Itinerant Electron Magnetism}}},\ edited by\ \bibinfo
  {editor} {\bibfnamefont {M.}~\bibnamefont {Cardona}}, \bibinfo {editor}
  {\bibfnamefont {P.}~\bibnamefont {Fulde}}, \ and\ \bibinfo {editor}
  {\bibfnamefont {H.-J.}\ \bibnamefont {Queisser}}\ (\bibinfo  {publisher}
  {Springer Verlag, Berlin, Heidelberg})\BibitemShut {NoStop}%
\bibitem [{\citenamefont {Mott}(1974)}]{mott}%
  \BibitemOpen
  \bibfield  {author} {\bibinfo {author} {\bibnamefont {Mott}, \bibfnamefont
  {N}}} (\bibinfo {year} {1974}),\ \href@noop {} {\emph {\bibinfo {title}
  {Metal-Insulator Transitions}}}\ (\bibinfo  {publisher} {Taylor and Francis,
  London})\BibitemShut {NoStop}%
\bibitem [{\citenamefont {Mousatov}\ and\ \citenamefont
  {Hartnoll}(2020)}]{SH20}%
  \BibitemOpen
  \bibfield  {author} {\bibinfo {author} {\bibnamefont {Mousatov},
  \bibfnamefont {Connie~H}}, and\ \bibinfo {author} {\bibfnamefont {Sean~A.}\
  \bibnamefont {Hartnoll}}} (\bibinfo {year} {2020}),\ \bibfield  {title}
  {\enquote {\bibinfo {title} {On the planckian bound for heat diffusion in
  insulators},}\ }\href {https://doi.org/10.1038/s41567-020-0828-6} {\bibfield
  {journal} {\bibinfo  {journal} {Nature Physics}\ }\textbf {\bibinfo {volume}
  {16}}~(\bibinfo {number} {5}),\ \bibinfo {pages} {579--584}}\BibitemShut
  {NoStop}%
\bibitem [{\citenamefont {Mross}\ \emph {et~al.}(2010)\citenamefont {Mross},
  \citenamefont {McGreevy}, \citenamefont {Liu},\ and\ \citenamefont
  {Senthil}}]{mross}%
  \BibitemOpen
  \bibfield  {author} {\bibinfo {author} {\bibnamefont {Mross}, \bibfnamefont
  {David~F}}, \bibinfo {author} {\bibfnamefont {John}\ \bibnamefont
  {McGreevy}}, \bibinfo {author} {\bibfnamefont {Hong}\ \bibnamefont {Liu}},
  and\ \bibinfo {author} {\bibfnamefont {T.}~\bibnamefont {Senthil}}} (\bibinfo
  {year} {2010}),\ \bibfield  {title} {\enquote {\bibinfo {title} {{Controlled
  expansion for certain non-Fermi-liquid metals}},}\ }\href
  {https://doi.org/10.1103/PhysRevB.82.045121} {\bibfield  {journal} {\bibinfo
  {journal} {Phys. Rev. B}\ }\textbf {\bibinfo {volume} {82}},\ \bibinfo
  {pages} {045121}}\BibitemShut {NoStop}%
\bibitem [{\citenamefont {Mukerjee}\ \emph {et~al.}(2006)\citenamefont
  {Mukerjee}, \citenamefont {Oganesyan},\ and\ \citenamefont
  {Huse}}]{Oganesyan}%
  \BibitemOpen
  \bibfield  {author} {\bibinfo {author} {\bibnamefont {Mukerjee},
  \bibfnamefont {Subroto}}, \bibinfo {author} {\bibfnamefont {Vadim}\
  \bibnamefont {Oganesyan}}, and\ \bibinfo {author} {\bibfnamefont {David}\
  \bibnamefont {Huse}}} (\bibinfo {year} {2006}),\ \bibfield  {title} {\enquote
  {\bibinfo {title} {Statistical theory of transport by strongly interacting
  lattice fermions},}\ }\href {https://doi.org/10.1103/PhysRevB.73.035113}
  {\bibfield  {journal} {\bibinfo  {journal} {Phys. Rev. B}\ }\textbf {\bibinfo
  {volume} {73}},\ \bibinfo {pages} {035113}}\BibitemShut {NoStop}%
\bibitem [{\citenamefont {{M{\"u}ller}}\ \emph {et~al.}(2009)\citenamefont
  {{M{\"u}ller}}, \citenamefont {{Heusler}}, \citenamefont {{Altland}},
  \citenamefont {{Braun}},\ and\ \citenamefont {{Haake}}}]{Altland09}%
  \BibitemOpen
  \bibfield  {author} {\bibinfo {author} {\bibnamefont {{M{\"u}ller}},
  \bibfnamefont {Sebastian}}, \bibinfo {author} {\bibfnamefont {Stefan}\
  \bibnamefont {{Heusler}}}, \bibinfo {author} {\bibfnamefont {Alexander}\
  \bibnamefont {{Altland}}}, \bibinfo {author} {\bibfnamefont {Petr}\
  \bibnamefont {{Braun}}}, and\ \bibinfo {author} {\bibfnamefont {Fritz}\
  \bibnamefont {{Haake}}}} (\bibinfo {year} {2009}),\ \bibfield  {title}
  {\enquote {\bibinfo {title} {{Periodic-orbit theory of universal level
  correlations in quantum chaos}},}\ }\href
  {https://doi.org/10.1088/1367-2630/11/10/103025} {\bibfield  {journal}
  {\bibinfo  {journal} {New Journal of Physics}\ }\textbf {\bibinfo {volume}
  {11}}~(\bibinfo {number} {10}),\ \bibinfo {eid} {103025}},\ \Eprint
  {https://arxiv.org/abs/0906.1960} {arXiv:0906.1960 [nlin.CD]} \BibitemShut
  {NoStop}%
\bibitem [{\citenamefont {Murthy}\ and\ \citenamefont
  {Srednicki}(2019)}]{murthy19}%
  \BibitemOpen
  \bibfield  {author} {\bibinfo {author} {\bibnamefont {Murthy}, \bibfnamefont
  {Chaitanya}}, and\ \bibinfo {author} {\bibfnamefont {Mark}\ \bibnamefont
  {Srednicki}}} (\bibinfo {year} {2019}),\ \bibfield  {title} {\enquote
  {\bibinfo {title} {Bounds on chaos from the eigenstate thermalization
  hypothesis},}\ }\href {https://doi.org/10.1103/PhysRevLett.123.230606}
  {\bibfield  {journal} {\bibinfo  {journal} {Phys. Rev. Lett.}\ }\textbf
  {\bibinfo {volume} {123}},\ \bibinfo {pages} {230606}}\BibitemShut {NoStop}%
\bibitem [{\citenamefont {Murugan}\ \emph {et~al.}(2017)\citenamefont
  {Murugan}, \citenamefont {Stanford},\ and\ \citenamefont
  {Witten}}]{Murugan:2017eto}%
  \BibitemOpen
  \bibfield  {author} {\bibinfo {author} {\bibnamefont {Murugan}, \bibfnamefont
  {Jeff}}, \bibinfo {author} {\bibfnamefont {Douglas}\ \bibnamefont
  {Stanford}}, and\ \bibinfo {author} {\bibfnamefont {Edward}\ \bibnamefont
  {Witten}}} (\bibinfo {year} {2017}),\ \bibfield  {title} {\enquote {\bibinfo
  {title} {{More on Supersymmetric and 2d Analogs of the SYK Model}},}\ }\href
  {https://doi.org/10.1007/JHEP08(2017)146} {\bibfield  {journal} {\bibinfo
  {journal} {JHEP}\ }\textbf {\bibinfo {volume} {08}},\ \bibinfo {pages}
  {146}},\ \Eprint {https://arxiv.org/abs/1706.05362} {arXiv:1706.05362
  [hep-th]} \BibitemShut {NoStop}%
\bibitem [{\citenamefont {Nahum}(2022)}]{Nahum:2022fqw}%
  \BibitemOpen
  \bibfield  {author} {\bibinfo {author} {\bibnamefont {Nahum}, \bibfnamefont
  {Adam}}} (\bibinfo {year} {2022}),\ \bibfield  {title} {\enquote {\bibinfo
  {title} {{Fixed point annihilation for a spin in a fluctuating field}},}\
  }\href@noop {} {\ }\Eprint {https://arxiv.org/abs/2202.08431}
  {arXiv:2202.08431 [cond-mat.str-el]} \BibitemShut {NoStop}%
\bibitem [{\citenamefont {Nahum}\ \emph {et~al.}(2018)\citenamefont {Nahum},
  \citenamefont {Vijay},\ and\ \citenamefont {Haah}}]{AN18}%
  \BibitemOpen
  \bibfield  {author} {\bibinfo {author} {\bibnamefont {Nahum}, \bibfnamefont
  {Adam}}, \bibinfo {author} {\bibfnamefont {Sagar}\ \bibnamefont {Vijay}},
  and\ \bibinfo {author} {\bibfnamefont {Jeongwan}\ \bibnamefont {Haah}}}
  (\bibinfo {year} {2018}),\ \bibfield  {title} {\enquote {\bibinfo {title}
  {Operator spreading in random unitary circuits},}\ }\href
  {https://doi.org/10.1103/PhysRevX.8.021014} {\bibfield  {journal} {\bibinfo
  {journal} {Phys. Rev. X}\ }\textbf {\bibinfo {volume} {8}},\ \bibinfo {pages}
  {021014}}\BibitemShut {NoStop}%
\bibitem [{\citenamefont {Nakajima}\ \emph {et~al.}(2020)\citenamefont
  {Nakajima}, \citenamefont {Metz}, \citenamefont {Eckberg}, \citenamefont
  {Kirshenbaum}, \citenamefont {Hughes}, \citenamefont {Wang}, \citenamefont
  {Wang}, \citenamefont {Saha}, \citenamefont {Liu}, \citenamefont {Butch},
  \citenamefont {Campbell}, \citenamefont {Eo}, \citenamefont {Graf},
  \citenamefont {Liu}, \citenamefont {Borisenko}, \citenamefont {Zavalij},\
  and\ \citenamefont {Paglione}}]{Paglione}%
  \BibitemOpen
  \bibfield  {author} {\bibinfo {author} {\bibnamefont {Nakajima},
  \bibfnamefont {Yasuyuki}}, \bibinfo {author} {\bibfnamefont {Tristin}\
  \bibnamefont {Metz}}, \bibinfo {author} {\bibfnamefont {Christopher}\
  \bibnamefont {Eckberg}}, \bibinfo {author} {\bibfnamefont {Kevin}\
  \bibnamefont {Kirshenbaum}}, \bibinfo {author} {\bibfnamefont {Alex}\
  \bibnamefont {Hughes}}, \bibinfo {author} {\bibfnamefont {Renxiong}\
  \bibnamefont {Wang}}, \bibinfo {author} {\bibfnamefont {Limin}\ \bibnamefont
  {Wang}}, \bibinfo {author} {\bibfnamefont {Shanta~R.}\ \bibnamefont {Saha}},
  \bibinfo {author} {\bibfnamefont {I-Lin}\ \bibnamefont {Liu}}, \bibinfo
  {author} {\bibfnamefont {Nicholas~P.}\ \bibnamefont {Butch}}, \bibinfo
  {author} {\bibfnamefont {Daniel}\ \bibnamefont {Campbell}}, \bibinfo {author}
  {\bibfnamefont {Yun~Suk}\ \bibnamefont {Eo}}, \bibinfo {author}
  {\bibfnamefont {David}\ \bibnamefont {Graf}}, \bibinfo {author}
  {\bibfnamefont {Zhonghao}\ \bibnamefont {Liu}}, \bibinfo {author}
  {\bibfnamefont {Sergey~V.}\ \bibnamefont {Borisenko}}, \bibinfo {author}
  {\bibfnamefont {Peter~Y.}\ \bibnamefont {Zavalij}}, and\ \bibinfo {author}
  {\bibfnamefont {Johnpierre}\ \bibnamefont {Paglione}}} (\bibinfo {year}
  {2020}),\ \bibfield  {title} {\enquote {\bibinfo {title} {Quantum-critical
  scale invariance in a transition metal alloy},}\ }\href
  {https://doi.org/10.1038/s42005-020-00448-5} {\bibfield  {journal} {\bibinfo
  {journal} {Communications Physics}\ }\textbf {\bibinfo {volume}
  {3}}~(\bibinfo {number} {1}),\ \bibinfo {pages} {181}}\BibitemShut {NoStop}%
\bibitem [{\citenamefont {Nayak}\ \emph {et~al.}(2018)\citenamefont {Nayak},
  \citenamefont {Shukla}, \citenamefont {Soni}, \citenamefont {Trivedi},\ and\
  \citenamefont {Vishal}}]{Nayak:2018qej}%
  \BibitemOpen
  \bibfield  {author} {\bibinfo {author} {\bibnamefont {Nayak}, \bibfnamefont
  {Pranjal}}, \bibinfo {author} {\bibfnamefont {Ashish}\ \bibnamefont
  {Shukla}}, \bibinfo {author} {\bibfnamefont {Ronak~M.}\ \bibnamefont {Soni}},
  \bibinfo {author} {\bibfnamefont {Sandip~P.}\ \bibnamefont {Trivedi}}, and\
  \bibinfo {author} {\bibfnamefont {V.}~\bibnamefont {Vishal}}} (\bibinfo
  {year} {2018}),\ \bibfield  {title} {\enquote {\bibinfo {title} {{On the
  Dynamics of Near-Extremal Black Holes}},}\ }\href
  {https://doi.org/10.1007/JHEP09(2018)048} {\bibfield  {journal} {\bibinfo
  {journal} {JHEP}\ }\textbf {\bibinfo {volume} {09}},\ \bibinfo {pages}
  {048}},\ \Eprint {https://arxiv.org/abs/1802.09547} {arXiv:1802.09547
  [hep-th]} \BibitemShut {NoStop}%
%%CITATION = ARXIV:1802.09547;%%
\bibitem [{\citenamefont {Nikolaenko}\ \emph {et~al.}(2021)\citenamefont
  {Nikolaenko}, \citenamefont {Tikhanovskaya}, \citenamefont {Sachdev},\ and\
  \citenamefont {Zhang}}]{Nikolaenko:2021vlw}%
  \BibitemOpen
  \bibfield  {author} {\bibinfo {author} {\bibnamefont {Nikolaenko},
  \bibfnamefont {Alexander}}, \bibinfo {author} {\bibfnamefont {Maria}\
  \bibnamefont {Tikhanovskaya}}, \bibinfo {author} {\bibfnamefont {Subir}\
  \bibnamefont {Sachdev}}, and\ \bibinfo {author} {\bibfnamefont {Ya-Hui}\
  \bibnamefont {Zhang}}} (\bibinfo {year} {2021}),\ \bibfield  {title}
  {\enquote {\bibinfo {title} {{Small to large Fermi surface transition in a
  single band model, using randomly coupled ancillas}},}\ }\href
  {https://doi.org/10.1103/PhysRevB.103.235138} {\bibfield  {journal} {\bibinfo
   {journal} {Phys. Rev. B}\ }\textbf {\bibinfo {volume} {103}}~(\bibinfo
  {number} {23}),\ \bibinfo {pages} {235138}},\ \Eprint
  {https://arxiv.org/abs/2103.05009} {arXiv:2103.05009 [cond-mat.str-el]}
  \BibitemShut {NoStop}%
\bibitem [{\citenamefont {Otsuki}(2013)}]{OtsukiCTHYBDoubleExpQMC}%
  \BibitemOpen
  \bibfield  {author} {\bibinfo {author} {\bibnamefont {Otsuki}, \bibfnamefont
  {Junya}}} (\bibinfo {year} {2013}),\ \bibfield  {title} {\enquote {\bibinfo
  {title} {Spin-boson coupling in continuous-time quantum monte carlo},}\
  }\href {https://doi.org/10.1103/PhysRevB.87.125102} {\bibfield  {journal}
  {\bibinfo  {journal} {Phys. Rev. B}\ }\textbf {\bibinfo {volume} {87}},\
  \bibinfo {pages} {125102}}\BibitemShut {NoStop}%
\bibitem [{\citenamefont {Otsuki}\ and\ \citenamefont
  {Vollhardt}(2013)}]{Otsuki2013}%
  \BibitemOpen
  \bibfield  {author} {\bibinfo {author} {\bibnamefont {Otsuki}, \bibfnamefont
  {Junya}}, and\ \bibinfo {author} {\bibfnamefont {Dieter}\ \bibnamefont
  {Vollhardt}}} (\bibinfo {year} {2013}),\ \bibfield  {title} {\enquote
  {\bibinfo {title} {{Numerical Solution of the $t$-$J$ Model with Random
  Exchange Couplings in $d=\infty$ Dimensions}},}\ }\href
  {https://doi.org/10.1103/PhysRevLett.110.196407} {\bibfield  {journal}
  {\bibinfo  {journal} {Phys. Rev. Lett.}\ }\textbf {\bibinfo {volume} {110}},\
  \bibinfo {pages} {196407}}\BibitemShut {NoStop}%
\bibitem [{\citenamefont {P\'alsson}\ and\ \citenamefont
  {Kotliar}(1998)}]{Palsson_1998}%
  \BibitemOpen
  \bibfield  {author} {\bibinfo {author} {\bibnamefont {P\'alsson},
  \bibfnamefont {Gunnar}}, and\ \bibinfo {author} {\bibfnamefont {Gabriel}\
  \bibnamefont {Kotliar}}} (\bibinfo {year} {1998}),\ \bibfield  {title}
  {\enquote {\bibinfo {title} {{Thermoelectric Response Near the Density Driven
  Mott Transition}},}\ }\href {https://doi.org/10.1103/PhysRevLett.80.4775}
  {\bibfield  {journal} {\bibinfo  {journal} {Phys. Rev. Lett.}\ }\textbf
  {\bibinfo {volume} {80}},\ \bibinfo {pages} {4775--4778}}\BibitemShut
  {NoStop}%
\bibitem [{\citenamefont {Pankov}\ \emph {et~al.}(2002)\citenamefont {Pankov},
  \citenamefont {Kotliar},\ and\ \citenamefont {Motome}}]{Pankov2002}%
  \BibitemOpen
  \bibfield  {author} {\bibinfo {author} {\bibnamefont {Pankov}, \bibfnamefont
  {Sergey}}, \bibinfo {author} {\bibfnamefont {Gabriel}\ \bibnamefont
  {Kotliar}}, and\ \bibinfo {author} {\bibfnamefont {Yukitoshi}\ \bibnamefont
  {Motome}}} (\bibinfo {year} {2002}),\ \bibfield  {title} {\enquote {\bibinfo
  {title} {Semiclassical analysis of extended dynamical mean-field
  equations},}\ }\href {https://doi.org/10.1103/PhysRevB.66.045117} {\bibfield
  {journal} {\bibinfo  {journal} {Phys. Rev. B}\ }\textbf {\bibinfo {volume}
  {66}},\ \bibinfo {pages} {045117}}\BibitemShut {NoStop}%
\bibitem [{\citenamefont {{Paramekanti}}\ and\ \citenamefont
  {{Vishwanath}}(2004)}]{Paramekanti_2004}%
  \BibitemOpen
  \bibfield  {author} {\bibinfo {author} {\bibnamefont {{Paramekanti}},
  \bibfnamefont {Arun}}, and\ \bibinfo {author} {\bibfnamefont {Ashvin}\
  \bibnamefont {{Vishwanath}}}} (\bibinfo {year} {2004}),\ \bibfield  {title}
  {\enquote {\bibinfo {title} {{Extending Luttinger's theorem to ${Z}_{2}$
  fractionalized phases of matter}},}\ }\href
  {https://doi.org/10.1103/PhysRevB.70.245118} {\bibfield  {journal} {\bibinfo
  {journal} {Phys. Rev. B}\ }\textbf {\bibinfo {volume} {70}}~(\bibinfo
  {number} {24}),\ \bibinfo {eid} {245118}},\ \Eprint
  {https://arxiv.org/abs/cond-mat/0406619} {arXiv:cond-mat/0406619
  [cond-mat.str-el]} \BibitemShut {NoStop}%
\bibitem [{\citenamefont {Parcollet}\ and\ \citenamefont
  {Georges}(1999)}]{Parcollet1}%
  \BibitemOpen
  \bibfield  {author} {\bibinfo {author} {\bibnamefont {Parcollet},
  \bibfnamefont {Olivier}}, and\ \bibinfo {author} {\bibfnamefont {Antoine}\
  \bibnamefont {Georges}}} (\bibinfo {year} {1999}),\ \bibfield  {title}
  {\enquote {\bibinfo {title} {{Non-Fermi-liquid regime of a doped Mott
  insulator}},}\ }\href {https://doi.org/10.1103/PhysRevB.59.5341} {\bibfield
  {journal} {\bibinfo  {journal} {Phys. Rev. B}\ }\textbf {\bibinfo {volume}
  {59}},\ \bibinfo {pages} {5341--5360}}\BibitemShut {NoStop}%
\bibitem [{\citenamefont {Parcollet}\ \emph {et~al.}(1998)\citenamefont
  {Parcollet}, \citenamefont {Georges}, \citenamefont {Kotliar},\ and\
  \citenamefont {Sengupta}}]{ParcolletKondo}%
  \BibitemOpen
  \bibfield  {author} {\bibinfo {author} {\bibnamefont {Parcollet},
  \bibfnamefont {Olivier}}, \bibinfo {author} {\bibfnamefont {Antoine}\
  \bibnamefont {Georges}}, \bibinfo {author} {\bibfnamefont {Gabriel}\
  \bibnamefont {Kotliar}}, and\ \bibinfo {author} {\bibfnamefont {Anirvan}\
  \bibnamefont {Sengupta}}} (\bibinfo {year} {1998}),\ \bibfield  {title}
  {\enquote {\bibinfo {title} {{Overscreened multichannel SU$(N)$ Kondo model:
  Large-$N$ solution and conformal field theory}},}\ }\href
  {https://doi.org/10.1103/PhysRevB.58.3794} {\bibfield  {journal} {\bibinfo
  {journal} {Phys. Rev. B}\ }\textbf {\bibinfo {volume} {58}},\ \bibinfo
  {pages} {3794--3813}}\BibitemShut {NoStop}%
\bibitem [{\citenamefont {Park}\ \emph {et~al.}(2021)\citenamefont {Park},
  \citenamefont {Cao}, \citenamefont {Watanabe}, \citenamefont {Taniguchi},\
  and\ \citenamefont {Jarillo-Herrero}}]{ParkDiffusivity2020}%
  \BibitemOpen
  \bibfield  {author} {\bibinfo {author} {\bibnamefont {Park}, \bibfnamefont
  {Jeong~Min}}, \bibinfo {author} {\bibfnamefont {Yuan}\ \bibnamefont {Cao}},
  \bibinfo {author} {\bibfnamefont {Kenji}\ \bibnamefont {Watanabe}}, \bibinfo
  {author} {\bibfnamefont {Takashi}\ \bibnamefont {Taniguchi}}, and\ \bibinfo
  {author} {\bibfnamefont {Pablo}\ \bibnamefont {Jarillo-Herrero}}} (\bibinfo
  {year} {2021}),\ \bibfield  {title} {\enquote {\bibinfo {title} {{Flavour
  Hund's coupling, Chern gaps and charge diffusivity in moir{\'e} graphene}},}\
  }\href {https://doi.org/10.1038/s41586-021-03366-w} {\bibfield  {journal}
  {\bibinfo  {journal} {Nature}\ }\textbf {\bibinfo {volume} {592}}~(\bibinfo
  {number} {7852}),\ \bibinfo {pages} {43--48}}\BibitemShut {NoStop}%
\bibitem [{\citenamefont {Parker}\ \emph {et~al.}(2019)\citenamefont {Parker},
  \citenamefont {Cao}, \citenamefont {Avdoshkin}, \citenamefont {Scaffidi},\
  and\ \citenamefont {Altman}}]{parker19}%
  \BibitemOpen
  \bibfield  {author} {\bibinfo {author} {\bibnamefont {Parker}, \bibfnamefont
  {Daniel~E}}, \bibinfo {author} {\bibfnamefont {Xiangyu}\ \bibnamefont {Cao}},
  \bibinfo {author} {\bibfnamefont {Alexander}\ \bibnamefont {Avdoshkin}},
  \bibinfo {author} {\bibfnamefont {Thomas}\ \bibnamefont {Scaffidi}}, and\
  \bibinfo {author} {\bibfnamefont {Ehud}\ \bibnamefont {Altman}}} (\bibinfo
  {year} {2019}),\ \bibfield  {title} {\enquote {\bibinfo {title} {A universal
  operator growth hypothesis},}\ }\href
  {https://doi.org/10.1103/PhysRevX.9.041017} {\bibfield  {journal} {\bibinfo
  {journal} {Phys. Rev. X}\ }\textbf {\bibinfo {volume} {9}},\ \bibinfo {pages}
  {041017}}\BibitemShut {NoStop}%
\bibitem [{\citenamefont {{Paschen}}\ and\ \citenamefont
  {{Si}}(2021)}]{PaschenSi21}%
  \BibitemOpen
  \bibfield  {author} {\bibinfo {author} {\bibnamefont {{Paschen}},
  \bibfnamefont {Silke}}, and\ \bibinfo {author} {\bibfnamefont {Qimiao}\
  \bibnamefont {{Si}}}} (\bibinfo {year} {2021}),\ \bibfield  {title} {\enquote
  {\bibinfo {title} {{Quantum phases driven by strong correlations}},}\ }\href
  {https://doi.org/10.1038/s42254-020-00262-6} {\bibfield  {journal} {\bibinfo
  {journal} {Nature Reviews Physics}\ }\textbf {\bibinfo {volume}
  {3}}~(\bibinfo {number} {1}),\ \bibinfo {pages} {9--26}},\ \Eprint
  {https://arxiv.org/abs/2009.03602} {arXiv:2009.03602 [cond-mat.str-el]}
  \BibitemShut {NoStop}%
\bibitem [{\citenamefont {Patel}\ and\ \citenamefont
  {Changlani}(2022)}]{Changlani}%
  \BibitemOpen
  \bibfield  {author} {\bibinfo {author} {\bibnamefont {Patel}, \bibfnamefont
  {Aavishkar~A}}, and\ \bibinfo {author} {\bibfnamefont {Hitesh~J.}\
  \bibnamefont {Changlani}}} (\bibinfo {year} {2022}),\ \bibfield  {title}
  {\enquote {\bibinfo {title} {Many-body energy invariant for $t$-linear
  resistivity},}\ }\href {https://doi.org/10.1103/PhysRevB.105.L201108}
  {\bibfield  {journal} {\bibinfo  {journal} {Phys. Rev. B}\ }\textbf {\bibinfo
  {volume} {105}},\ \bibinfo {pages} {L201108}}\BibitemShut {NoStop}%
\bibitem [{\citenamefont {Patel}\ \emph {et~al.}(2017)\citenamefont {Patel},
  \citenamefont {Chowdhury}, \citenamefont {Sachdev},\ and\ \citenamefont
  {Swingle}}]{Patel:2017vfp}%
  \BibitemOpen
  \bibfield  {author} {\bibinfo {author} {\bibnamefont {Patel}, \bibfnamefont
  {Aavishkar~A}}, \bibinfo {author} {\bibfnamefont {Debanjan}\ \bibnamefont
  {Chowdhury}}, \bibinfo {author} {\bibfnamefont {Subir}\ \bibnamefont
  {Sachdev}}, and\ \bibinfo {author} {\bibfnamefont {Brian}\ \bibnamefont
  {Swingle}}} (\bibinfo {year} {2017}),\ \bibfield  {title} {\enquote {\bibinfo
  {title} {{Quantum butterfly effect in weakly interacting diffusive
  metals}},}\ }\href {https://doi.org/10.1103/PhysRevX.7.031047} {\bibfield
  {journal} {\bibinfo  {journal} {Phys. Rev. X}\ }\textbf {\bibinfo {volume}
  {7}}~(\bibinfo {number} {3}),\ \bibinfo {pages} {031047}},\ \Eprint
  {https://arxiv.org/abs/1703.07353} {arXiv:1703.07353 [cond-mat.str-el]}
  \BibitemShut {NoStop}%
\bibitem [{\citenamefont {Patel}\ \emph {et~al.}(2022)\citenamefont {Patel},
  \citenamefont {Guo}, \citenamefont {Esterlis},\ and\ \citenamefont
  {Sachdev}}]{Patel:2022gdh}%
  \BibitemOpen
  \bibfield  {author} {\bibinfo {author} {\bibnamefont {Patel}, \bibfnamefont
  {Aavishkar~A}}, \bibinfo {author} {\bibfnamefont {Haoyu}\ \bibnamefont
  {Guo}}, \bibinfo {author} {\bibfnamefont {Ilya}\ \bibnamefont {Esterlis}},
  and\ \bibinfo {author} {\bibfnamefont {Subir}\ \bibnamefont {Sachdev}}}
  (\bibinfo {year} {2022}),\ \bibfield  {title} {\enquote {\bibinfo {title}
  {{Universal, low temperature, $T$-linear resistivity in two-dimensional
  quantum-critical metals from spatially random interactions}},}\ }\href@noop
  {} {\ }\Eprint {https://arxiv.org/abs/2203.04990} {arXiv:2203.04990
  [cond-mat.str-el]} \BibitemShut {NoStop}%
\bibitem [{\citenamefont {Patel}\ \emph
  {et~al.}(2018{\natexlab{a}})\citenamefont {Patel}, \citenamefont {Lawler},\
  and\ \citenamefont {Kim}}]{Patel}%
  \BibitemOpen
  \bibfield  {author} {\bibinfo {author} {\bibnamefont {Patel}, \bibfnamefont
  {Aavishkar~A}}, \bibinfo {author} {\bibfnamefont {Michael~J.}\ \bibnamefont
  {Lawler}}, and\ \bibinfo {author} {\bibfnamefont {Eun-Ah}\ \bibnamefont
  {Kim}}} (\bibinfo {year} {2018}{\natexlab{a}}),\ \bibfield  {title} {\enquote
  {\bibinfo {title} {Coherent superconductivity with a large gap ratio from
  incoherent metals},}\ }\href {https://doi.org/10.1103/PhysRevLett.121.187001}
  {\bibfield  {journal} {\bibinfo  {journal} {Phys. Rev. Lett.}\ }\textbf
  {\bibinfo {volume} {121}},\ \bibinfo {pages} {187001}}\BibitemShut {NoStop}%
\bibitem [{\citenamefont {Patel}\ \emph
  {et~al.}(2018{\natexlab{b}})\citenamefont {Patel}, \citenamefont {McGreevy},
  \citenamefont {Arovas},\ and\ \citenamefont {Sachdev}}]{SSmagneto}%
  \BibitemOpen
  \bibfield  {author} {\bibinfo {author} {\bibnamefont {Patel}, \bibfnamefont
  {Aavishkar~A}}, \bibinfo {author} {\bibfnamefont {John}\ \bibnamefont
  {McGreevy}}, \bibinfo {author} {\bibfnamefont {Daniel~P.}\ \bibnamefont
  {Arovas}}, and\ \bibinfo {author} {\bibfnamefont {Subir}\ \bibnamefont
  {Sachdev}}} (\bibinfo {year} {2018}{\natexlab{b}}),\ \bibfield  {title}
  {\enquote {\bibinfo {title} {Magnetotransport in a model of a disordered
  strange metal},}\ }\href {https://doi.org/10.1103/PhysRevX.8.021049}
  {\bibfield  {journal} {\bibinfo  {journal} {Phys. Rev. X}\ }\textbf {\bibinfo
  {volume} {8}},\ \bibinfo {pages} {021049}}\BibitemShut {NoStop}%
\bibitem [{\citenamefont {Patel}\ and\ \citenamefont
  {Sachdev}(2014)}]{Patel:2014jfa}%
  \BibitemOpen
  \bibfield  {author} {\bibinfo {author} {\bibnamefont {Patel}, \bibfnamefont
  {Aavishkar~A}}, and\ \bibinfo {author} {\bibfnamefont {Subir}\ \bibnamefont
  {Sachdev}}} (\bibinfo {year} {2014}),\ \bibfield  {title} {\enquote {\bibinfo
  {title} {{DC resistivity at the onset of spin density wave order in
  two-dimensional metals}},}\ }\href
  {https://doi.org/10.1103/PhysRevB.90.165146} {\bibfield  {journal} {\bibinfo
  {journal} {Phys. Rev. B}\ }\textbf {\bibinfo {volume} {90}}~(\bibinfo
  {number} {16}),\ \bibinfo {pages} {165146}},\ \Eprint
  {https://arxiv.org/abs/1408.6549} {arXiv:1408.6549 [cond-mat.str-el]}
  \BibitemShut {NoStop}%
\bibitem [{\citenamefont {Patel}\ and\ \citenamefont
  {Sachdev}(2017)}]{Patel:2016wdy}%
  \BibitemOpen
  \bibfield  {author} {\bibinfo {author} {\bibnamefont {Patel}, \bibfnamefont
  {Aavishkar~A}}, and\ \bibinfo {author} {\bibfnamefont {Subir}\ \bibnamefont
  {Sachdev}}} (\bibinfo {year} {2017}),\ \bibfield  {title} {\enquote {\bibinfo
  {title} {{Quantum chaos on a critical Fermi surface}},}\ }\href
  {https://doi.org/10.1073/pnas.1618185114} {\bibfield  {journal} {\bibinfo
  {journal} {Proc. Nat. Acad. Sci.}\ }\textbf {\bibinfo {volume} {114}},\
  \bibinfo {pages} {1844--1849}},\ \Eprint {https://arxiv.org/abs/1611.00003}
  {arXiv:1611.00003 [cond-mat.str-el]} \BibitemShut {NoStop}%
\bibitem [{\citenamefont {Patel}\ and\ \citenamefont
  {Sachdev}(2018)}]{Patel:2018zpy}%
  \BibitemOpen
  \bibfield  {author} {\bibinfo {author} {\bibnamefont {Patel}, \bibfnamefont
  {Aavishkar~A}}, and\ \bibinfo {author} {\bibfnamefont {Subir}\ \bibnamefont
  {Sachdev}}} (\bibinfo {year} {2018}),\ \bibfield  {title} {\enquote {\bibinfo
  {title} {{Critical strange metal from fluctuating gauge fields in a solvable
  random model}},}\ }\href {https://doi.org/10.1103/PhysRevB.98.125134}
  {\bibfield  {journal} {\bibinfo  {journal} {Phys. Rev. B}\ }\textbf {\bibinfo
  {volume} {98}}~(\bibinfo {number} {12}),\ \bibinfo {pages} {125134}},\
  \Eprint {https://arxiv.org/abs/1807.04754} {arXiv:1807.04754
  [cond-mat.str-el]} \BibitemShut {NoStop}%
\bibitem [{\citenamefont {Patel}\ and\ \citenamefont
  {Sachdev}(2019)}]{PatelPlanck}%
  \BibitemOpen
  \bibfield  {author} {\bibinfo {author} {\bibnamefont {Patel}, \bibfnamefont
  {Aavishkar~A}}, and\ \bibinfo {author} {\bibfnamefont {Subir}\ \bibnamefont
  {Sachdev}}} (\bibinfo {year} {2019}),\ \bibfield  {title} {\enquote {\bibinfo
  {title} {{Theory of a Planckian Metal}},}\ }\href
  {https://doi.org/10.1103/PhysRevLett.123.066601} {\bibfield  {journal}
  {\bibinfo  {journal} {Phys. Rev. Lett.}\ }\textbf {\bibinfo {volume} {123}},\
  \bibinfo {pages} {066601}}\BibitemShut {NoStop}%
\bibitem [{\citenamefont {{Paul}}\ \emph {et~al.}(2007)\citenamefont {{Paul}},
  \citenamefont {{P{\'e}pin}},\ and\ \citenamefont {{Norman}}}]{Norman07}%
  \BibitemOpen
  \bibfield  {author} {\bibinfo {author} {\bibnamefont {{Paul}}, \bibfnamefont
  {I}}, \bibinfo {author} {\bibfnamefont {C.}~\bibnamefont {{P{\'e}pin}}}, and\
  \bibinfo {author} {\bibfnamefont {M.~R.}\ \bibnamefont {{Norman}}}} (\bibinfo
  {year} {2007}),\ \bibfield  {title} {\enquote {\bibinfo {title} {{Kondo
  Breakdown and Hybridization Fluctuations in the Kondo-Heisenberg Lattice}},}\
  }\href {https://doi.org/10.1103/PhysRevLett.98.026402} {\bibfield  {journal}
  {\bibinfo  {journal} {Phys. Rev. Lett.}\ }\textbf {\bibinfo {volume}
  {98}}~(\bibinfo {number} {2}),\ \bibinfo {eid} {026402}},\ \Eprint
  {https://arxiv.org/abs/cond-mat/0605152} {arXiv:cond-mat/0605152
  [cond-mat.str-el]} \BibitemShut {NoStop}%
\bibitem [{\citenamefont {{Paul}}\ \emph {et~al.}(2008)\citenamefont {{Paul}},
  \citenamefont {{P{\'e}pin}},\ and\ \citenamefont {{Norman}}}]{Norman08}%
  \BibitemOpen
  \bibfield  {author} {\bibinfo {author} {\bibnamefont {{Paul}}, \bibfnamefont
  {I}}, \bibinfo {author} {\bibfnamefont {C.}~\bibnamefont {{P{\'e}pin}}}, and\
  \bibinfo {author} {\bibfnamefont {M.~R.}\ \bibnamefont {{Norman}}}} (\bibinfo
  {year} {2008}),\ \bibfield  {title} {\enquote {\bibinfo {title} {{Multiscale
  fluctuations near a Kondo breakdown quantum critical point}},}\ }\href
  {https://doi.org/10.1103/PhysRevB.78.035109} {\bibfield  {journal} {\bibinfo
  {journal} {Phys. Rev. B}\ }\textbf {\bibinfo {volume} {78}}~(\bibinfo
  {number} {3}),\ \bibinfo {eid} {035109}},\ \Eprint
  {https://arxiv.org/abs/0804.1808} {arXiv:0804.1808 [cond-mat.str-el]}
  \BibitemShut {NoStop}%
\bibitem [{\citenamefont {{Paul}}\ \emph {et~al.}(2013)\citenamefont {{Paul}},
  \citenamefont {{P{\'e}pin}},\ and\ \citenamefont {{Norman}}}]{Norman13}%
  \BibitemOpen
  \bibfield  {author} {\bibinfo {author} {\bibnamefont {{Paul}}, \bibfnamefont
  {I}}, \bibinfo {author} {\bibfnamefont {C.}~\bibnamefont {{P{\'e}pin}}}, and\
  \bibinfo {author} {\bibfnamefont {M.~R.}\ \bibnamefont {{Norman}}}} (\bibinfo
  {year} {2013}),\ \bibfield  {title} {\enquote {\bibinfo {title} {{Equivalence
  of Single-Particle and Transport Lifetimes from Hybridization
  Fluctuations}},}\ }\href {https://doi.org/10.1103/PhysRevLett.110.066402}
  {\bibfield  {journal} {\bibinfo  {journal} {Phys. Rev. Lett.}\ }\textbf
  {\bibinfo {volume} {110}}~(\bibinfo {number} {6}),\ \bibinfo {eid}
  {066402}},\ \Eprint {https://arxiv.org/abs/1211.5809} {arXiv:1211.5809
  [cond-mat.str-el]} \BibitemShut {NoStop}%
\bibitem [{\citenamefont {Pavlov}\ and\ \citenamefont
  {Kiselev}(2021)}]{Kiselev21}%
  \BibitemOpen
  \bibfield  {author} {\bibinfo {author} {\bibnamefont {Pavlov}, \bibfnamefont
  {Andrei~I}}, and\ \bibinfo {author} {\bibfnamefont {Mikhail~N.}\ \bibnamefont
  {Kiselev}}} (\bibinfo {year} {2021}),\ \bibfield  {title} {\enquote {\bibinfo
  {title} {{Quantum thermal transport in the charged Sachdev-Ye-Kitaev model:
  Thermoelectric Coulomb blockade}},}\ }\href
  {https://doi.org/10.1103/PhysRevB.103.L201107} {\bibfield  {journal}
  {\bibinfo  {journal} {Phys. Rev. B}\ }\textbf {\bibinfo {volume} {103}},\
  \bibinfo {pages} {L201107}},\ \Eprint {https://arxiv.org/abs/2010.13592}
  {arXiv:2010.13592 [cond-mat.str-el]} \BibitemShut {NoStop}%
\bibitem [{\citenamefont {Peierls}(1930)}]{Peierls1930}%
  \BibitemOpen
  \bibfield  {author} {\bibinfo {author} {\bibnamefont {Peierls}, \bibfnamefont
  {R}}} (\bibinfo {year} {1930}),\ \bibfield  {title} {\enquote {\bibinfo
  {title} {{Zur Theorie der elektrischen und thermischen Leitf\"ahigkeit von
  Metallen}},}\ }\href
  {https://doi.org/https://doi.org/10.1002/andp.19303960202} {\bibfield
  {journal} {\bibinfo  {journal} {Annalen der Physik}\ }\textbf {\bibinfo
  {volume} {396}}~(\bibinfo {number} {2}),\ \bibinfo {pages}
  {121--148}}\BibitemShut {NoStop}%
\bibitem [{\citenamefont {Peierls}(1932)}]{Peierls1932}%
  \BibitemOpen
  \bibfield  {author} {\bibinfo {author} {\bibnamefont {Peierls}, \bibfnamefont
  {R}}} (\bibinfo {year} {1932}),\ \bibfield  {title} {\enquote {\bibinfo
  {title} {{Zur Frage des elektrischen Widerstandsgesetzes f\"ur tiefe
  Temperaturen}},}\ }\href
  {https://doi.org/https://doi.org/10.1002/andp.19324040203} {\bibfield
  {journal} {\bibinfo  {journal} {Annalen der Physik}\ }\textbf {\bibinfo
  {volume} {404}}~(\bibinfo {number} {2}),\ \bibinfo {pages}
  {154--168}}\BibitemShut {NoStop}%
\bibitem [{\citenamefont {Penington}\ \emph {et~al.}(2019)\citenamefont
  {Penington}, \citenamefont {Shenker}, \citenamefont {Stanford},\ and\
  \citenamefont {Yang}}]{Penington:2019kki}%
  \BibitemOpen
  \bibfield  {author} {\bibinfo {author} {\bibnamefont {Penington},
  \bibfnamefont {Geoff}}, \bibinfo {author} {\bibfnamefont {Stephen~H.}\
  \bibnamefont {Shenker}}, \bibinfo {author} {\bibfnamefont {Douglas}\
  \bibnamefont {Stanford}}, and\ \bibinfo {author} {\bibfnamefont {Zhenbin}\
  \bibnamefont {Yang}}} (\bibinfo {year} {2019}),\ \bibfield  {title} {\enquote
  {\bibinfo {title} {{Replica wormholes and the black hole interior}},}\
  }\href@noop {} {\ }\Eprint {https://arxiv.org/abs/1911.11977}
  {arXiv:1911.11977 [hep-th]} \BibitemShut {NoStop}%
\bibitem [{\citenamefont {Perepelitsky}\ \emph {et~al.}(2016)\citenamefont
  {Perepelitsky}, \citenamefont {Galatas}, \citenamefont {Mravlje},
  \citenamefont {\ifmmode~\check{Z}\else \v{Z}\fi{}itko}, \citenamefont
  {Khatami}, \citenamefont {Shastry},\ and\ \citenamefont
  {Georges}}]{Perepelitsky2016}%
  \BibitemOpen
  \bibfield  {author} {\bibinfo {author} {\bibnamefont {Perepelitsky},
  \bibfnamefont {Edward}}, \bibinfo {author} {\bibfnamefont {Andrew}\
  \bibnamefont {Galatas}}, \bibinfo {author} {\bibfnamefont {Jernej}\
  \bibnamefont {Mravlje}}, \bibinfo {author} {\bibfnamefont {Rok}\ \bibnamefont
  {\ifmmode~\check{Z}\else \v{Z}\fi{}itko}}, \bibinfo {author} {\bibfnamefont
  {Ehsan}\ \bibnamefont {Khatami}}, \bibinfo {author} {\bibfnamefont
  {B.~Sriram}\ \bibnamefont {Shastry}}, and\ \bibinfo {author} {\bibfnamefont
  {Antoine}\ \bibnamefont {Georges}}} (\bibinfo {year} {2016}),\ \bibfield
  {title} {\enquote {\bibinfo {title} {{Transport and optical conductivity in
  the Hubbard model: A high-temperature expansion perspective}},}\ }\href
  {https://doi.org/10.1103/PhysRevB.94.235115} {\bibfield  {journal} {\bibinfo
  {journal} {Phys. Rev. B}\ }\textbf {\bibinfo {volume} {94}},\ \bibinfo
  {pages} {235115}}\BibitemShut {NoStop}%
\bibitem [{\citenamefont {P\'erez}\ and\ \citenamefont
  {Troncoso}(2020)}]{Perez:2020klz}%
  \BibitemOpen
  \bibfield  {author} {\bibinfo {author} {\bibnamefont {P\'erez}, \bibfnamefont
  {Alfredo}}, and\ \bibinfo {author} {\bibfnamefont {Ricardo}\ \bibnamefont
  {Troncoso}}} (\bibinfo {year} {2020}),\ \bibfield  {title} {\enquote
  {\bibinfo {title} {{Gravitational dual of averaged free CFT's over the Narain
  lattice}},}\ }\href {https://doi.org/10.1007/JHEP11(2020)015} {\
  10.1007/JHEP11(2020)015},\ \Eprint {https://arxiv.org/abs/2006.08216}
  {arXiv:2006.08216 [hep-th]} \BibitemShut {NoStop}%
\bibitem [{\citenamefont {Phanindra}\ \emph {et~al.}(2018)\citenamefont
  {Phanindra}, \citenamefont {Agarwal},\ and\ \citenamefont
  {Rana}}]{Phanindra_2018}%
  \BibitemOpen
  \bibfield  {author} {\bibinfo {author} {\bibnamefont {Phanindra},
  \bibfnamefont {V~Eswara}}, \bibinfo {author} {\bibfnamefont {Piyush}\
  \bibnamefont {Agarwal}}, and\ \bibinfo {author} {\bibfnamefont {D.~S.}\
  \bibnamefont {Rana}}} (\bibinfo {year} {2018}),\ \bibfield  {title} {\enquote
  {\bibinfo {title} {Terahertz spectroscopic evidence of non-fermi-liquid-like
  behavior in structurally modulated $\mathrm{PrNi}{\mathrm{o}}_{3}$ thin
  films},}\ }\href {https://doi.org/10.1103/PhysRevMaterials.2.015001}
  {\bibfield  {journal} {\bibinfo  {journal} {Phys. Rev. Materials}\ }\textbf
  {\bibinfo {volume} {2}},\ \bibinfo {pages} {015001}}\BibitemShut {NoStop}%
\bibitem [{\citenamefont {Pikulin}\ and\ \citenamefont
  {Franz}(2017)}]{Franz17}%
  \BibitemOpen
  \bibfield  {author} {\bibinfo {author} {\bibnamefont {Pikulin}, \bibfnamefont
  {D~I}}, and\ \bibinfo {author} {\bibfnamefont {M.}~\bibnamefont {Franz}}}
  (\bibinfo {year} {2017}),\ \bibfield  {title} {\enquote {\bibinfo {title}
  {{Black Hole on a Chip: Proposal for a Physical Realization of the
  Sachdev-Ye-Kitaev model in a Solid-State System}},}\ }\href
  {https://doi.org/10.1103/PhysRevX.7.031006} {\bibfield  {journal} {\bibinfo
  {journal} {Phys. Rev. X}\ }\textbf {\bibinfo {volume} {7}},\ \bibinfo {pages}
  {031006}}\BibitemShut {NoStop}%
\bibitem [{\citenamefont {{Pixley}}\ \emph {et~al.}(2014)\citenamefont
  {{Pixley}}, \citenamefont {{Yu}},\ and\ \citenamefont {{Si}}}]{Si2014}%
  \BibitemOpen
  \bibfield  {author} {\bibinfo {author} {\bibnamefont {{Pixley}},
  \bibfnamefont {J~H}}, \bibinfo {author} {\bibfnamefont {Rong}\ \bibnamefont
  {{Yu}}}, and\ \bibinfo {author} {\bibfnamefont {Qimiao}\ \bibnamefont
  {{Si}}}} (\bibinfo {year} {2014}),\ \bibfield  {title} {\enquote {\bibinfo
  {title} {{Quantum Phases of the Shastry-Sutherland Kondo Lattice:
  Implications for the Global Phase Diagram of Heavy-Fermion Metals}},}\ }\href
  {https://doi.org/10.1103/PhysRevLett.113.176402} {\bibfield  {journal}
  {\bibinfo  {journal} {Phys. Rev. Lett.}\ }\textbf {\bibinfo {volume}
  {113}}~(\bibinfo {number} {17}),\ \bibinfo {eid} {176402}},\ \Eprint
  {https://arxiv.org/abs/1309.0581} {arXiv:1309.0581 [cond-mat.str-el]}
  \BibitemShut {NoStop}%
\bibitem [{\citenamefont {Plugge}\ \emph {et~al.}(2020)\citenamefont {Plugge},
  \citenamefont {Lantagne-Hurtubise},\ and\ \citenamefont
  {Franz}}]{Plugge:2020wgc}%
  \BibitemOpen
  \bibfield  {author} {\bibinfo {author} {\bibnamefont {Plugge}, \bibfnamefont
  {Stephan}}, \bibinfo {author} {\bibfnamefont {\'Etienne}\ \bibnamefont
  {Lantagne-Hurtubise}}, and\ \bibinfo {author} {\bibfnamefont {Marcel}\
  \bibnamefont {Franz}}} (\bibinfo {year} {2020}),\ \bibfield  {title}
  {\enquote {\bibinfo {title} {{Revival Dynamics in a Traversable Wormhole}},}\
  }\href {https://doi.org/10.1103/PhysRevLett.124.221601} {\bibfield  {journal}
  {\bibinfo  {journal} {Phys. Rev. Lett.}\ }\textbf {\bibinfo {volume}
  {124}}~(\bibinfo {number} {22}),\ \bibinfo {pages} {221601}},\ \Eprint
  {https://arxiv.org/abs/2003.03914} {arXiv:2003.03914 [cond-mat.str-el]}
  \BibitemShut {NoStop}%
\bibitem [{\citenamefont {Polchinski}(1994)}]{Polchinski:1993ii}%
  \BibitemOpen
  \bibfield  {author} {\bibinfo {author} {\bibnamefont {Polchinski},
  \bibfnamefont {Joseph}}} (\bibinfo {year} {1994}),\ \bibfield  {title}
  {\enquote {\bibinfo {title} {{Low-energy dynamics of the spinon gauge
  system}},}\ }\href {https://doi.org/10.1016/0550-3213(94)90449-9} {\bibfield
  {journal} {\bibinfo  {journal} {Nucl. Phys. B}\ }\textbf {\bibinfo {volume}
  {422}},\ \bibinfo {pages} {617--633}},\ \Eprint
  {https://arxiv.org/abs/cond-mat/9303037} {arXiv:cond-mat/9303037}
  \BibitemShut {NoStop}%
\bibitem [{\citenamefont {Polshyn}\ \emph {et~al.}(2019)\citenamefont
  {Polshyn}, \citenamefont {Yankowitz}, \citenamefont {Chen}, \citenamefont
  {Zhang}, \citenamefont {Watanabe}, \citenamefont {Taniguchi}, \citenamefont
  {Dean},\ and\ \citenamefont {Young}}]{Young19}%
  \BibitemOpen
  \bibfield  {author} {\bibinfo {author} {\bibnamefont {Polshyn}, \bibfnamefont
  {Hryhoriy}}, \bibinfo {author} {\bibfnamefont {Matthew}\ \bibnamefont
  {Yankowitz}}, \bibinfo {author} {\bibfnamefont {Shaowen}\ \bibnamefont
  {Chen}}, \bibinfo {author} {\bibfnamefont {Yuxuan}\ \bibnamefont {Zhang}},
  \bibinfo {author} {\bibfnamefont {K.}~\bibnamefont {Watanabe}}, \bibinfo
  {author} {\bibfnamefont {T.}~\bibnamefont {Taniguchi}}, \bibinfo {author}
  {\bibfnamefont {Cory~R.}\ \bibnamefont {Dean}}, and\ \bibinfo {author}
  {\bibfnamefont {Andrea~F.}\ \bibnamefont {Young}}} (\bibinfo {year} {2019}),\
  \bibfield  {title} {\enquote {\bibinfo {title} {Large linear-in-temperature
  resistivity in twisted bilayer graphene},}\ }\href
  {https://doi.org/10.1038/s41567-019-0596-3} {\bibfield  {journal} {\bibinfo
  {journal} {Nature Physics}\ }\textbf {\bibinfo {volume} {15}}~(\bibinfo
  {number} {10}),\ \bibinfo {pages} {1011--1016}}\BibitemShut {NoStop}%
\bibitem [{\citenamefont {{Powell}}\ \emph {et~al.}(2005)\citenamefont
  {{Powell}}, \citenamefont {{Sachdev}},\ and\ \citenamefont
  {{B{\"u}chler}}}]{powell1}%
  \BibitemOpen
  \bibfield  {author} {\bibinfo {author} {\bibnamefont {{Powell}},
  \bibfnamefont {S}}, \bibinfo {author} {\bibfnamefont {S.}~\bibnamefont
  {{Sachdev}}}, and\ \bibinfo {author} {\bibfnamefont {H.~P.}\ \bibnamefont
  {{B{\"u}chler}}}} (\bibinfo {year} {2005}),\ \bibfield  {title} {\enquote
  {\bibinfo {title} {{Depletion of the Bose-Einstein condensate in Bose-Fermi
  mixtures}},}\ }\href {https://doi.org/10.1103/PhysRevB.72.024534} {\bibfield
  {journal} {\bibinfo  {journal} {Phys. Rev. B}\ }\textbf {\bibinfo {volume}
  {72}}~(\bibinfo {number} {2}),\ \bibinfo {eid} {024534}},\ \Eprint
  {https://arxiv.org/abs/cond-mat/0502299} {cond-mat/0502299} \BibitemShut
  {NoStop}%
\bibitem [{\citenamefont {Prange}\ and\ \citenamefont
  {Kadanoff}(1964)}]{Prange}%
  \BibitemOpen
  \bibfield  {author} {\bibinfo {author} {\bibnamefont {Prange}, \bibfnamefont
  {Richard~E}}, and\ \bibinfo {author} {\bibfnamefont {Leo~P.}\ \bibnamefont
  {Kadanoff}}} (\bibinfo {year} {1964}),\ \bibfield  {title} {\enquote
  {\bibinfo {title} {Transport theory for electron-phonon interactions in
  metals},}\ }\href {https://doi.org/10.1103/PhysRev.134.A566} {\bibfield
  {journal} {\bibinfo  {journal} {Phys. Rev.}\ }\textbf {\bibinfo {volume}
  {134}},\ \bibinfo {pages} {A566--A580}}\BibitemShut {NoStop}%
\bibitem [{\citenamefont {Profumo}\ \emph {et~al.}(2015)\citenamefont
  {Profumo}, \citenamefont {Groth}, \citenamefont {Messio}, \citenamefont
  {Parcollet},\ and\ \citenamefont {Waintal}}]{ProfumoRealTimeQMC}%
  \BibitemOpen
  \bibfield  {author} {\bibinfo {author} {\bibnamefont {Profumo}, \bibfnamefont
  {Rosario E~V}}, \bibinfo {author} {\bibfnamefont {Christoph}\ \bibnamefont
  {Groth}}, \bibinfo {author} {\bibfnamefont {Laura}\ \bibnamefont {Messio}},
  \bibinfo {author} {\bibfnamefont {Olivier}\ \bibnamefont {Parcollet}}, and\
  \bibinfo {author} {\bibfnamefont {Xavier}\ \bibnamefont {Waintal}}} (\bibinfo
  {year} {2015}),\ \bibfield  {title} {\enquote {\bibinfo {title} {Quantum
  monte carlo for correlated out-of-equilibrium nanoelectronic devices},}\
  }\href {https://doi.org/10.1103/PhysRevB.91.245154} {\bibfield  {journal}
  {\bibinfo  {journal} {Phys. Rev. B}\ }\textbf {\bibinfo {volume} {91}},\
  \bibinfo {pages} {245154}}\BibitemShut {NoStop}%
\bibitem [{\citenamefont {Proust}\ and\ \citenamefont
  {Taillefer}(2019)}]{Proust2019}%
  \BibitemOpen
  \bibfield  {author} {\bibinfo {author} {\bibnamefont {Proust}, \bibfnamefont
  {Cyril}}, and\ \bibinfo {author} {\bibfnamefont {Louis}\ \bibnamefont
  {Taillefer}}} (\bibinfo {year} {2019}),\ \bibfield  {title} {\enquote
  {\bibinfo {title} {The remarkable underlying ground states of cuprate
  superconductors},}\ }\href
  {https://doi.org/10.1146/annurev-conmatphys-031218-013210} {\bibfield
  {journal} {\bibinfo  {journal} {Annual Review of Condensed Matter Physics}\
  }\textbf {\bibinfo {volume} {10}}~(\bibinfo {number} {1}),\ \bibinfo {pages}
  {409--429}}\BibitemShut {NoStop}%
\bibitem [{\citenamefont {{Read}}\ \emph {et~al.}(1995)\citenamefont {{Read}},
  \citenamefont {{Sachdev}},\ and\ \citenamefont {{Ye}}}]{RSY95}%
  \BibitemOpen
  \bibfield  {author} {\bibinfo {author} {\bibnamefont {{Read}}, \bibfnamefont
  {N}}, \bibinfo {author} {\bibfnamefont {Subir}\ \bibnamefont {{Sachdev}}},
  and\ \bibinfo {author} {\bibfnamefont {J.}~\bibnamefont {{Ye}}}} (\bibinfo
  {year} {1995}),\ \bibfield  {title} {\enquote {\bibinfo {title} {{Landau
  theory of quantum spin glasses of rotors and Ising spins}},}\ }\href
  {https://doi.org/10.1103/PhysRevB.52.384} {\bibfield  {journal} {\bibinfo
  {journal} {Phys. Rev. B}\ }\textbf {\bibinfo {volume} {52}}~(\bibinfo
  {number} {1}),\ \bibinfo {pages} {384--410}},\ \Eprint
  {https://arxiv.org/abs/cond-mat/9412032} {arXiv:cond-mat/9412032 [cond-mat]}
  \BibitemShut {NoStop}%
\bibitem [{\citenamefont {Reber}\ \emph {et~al.}(2019)\citenamefont {Reber},
  \citenamefont {Zhou}, \citenamefont {Plumb}, \citenamefont {Parham},
  \citenamefont {Waugh}, \citenamefont {Cao}, \citenamefont {Sun},
  \citenamefont {Li}, \citenamefont {Wang}, \citenamefont {Wen}, \citenamefont
  {Xu}, \citenamefont {Gu}, \citenamefont {Yoshida}, \citenamefont {Eisaki},
  \citenamefont {Arnold},\ and\ \citenamefont {Dessau}}]{Reber2019}%
  \BibitemOpen
  \bibfield  {author} {\bibinfo {author} {\bibnamefont {Reber}, \bibfnamefont
  {T~J}}, \bibinfo {author} {\bibfnamefont {X.}~\bibnamefont {Zhou}}, \bibinfo
  {author} {\bibfnamefont {N.~C.}\ \bibnamefont {Plumb}}, \bibinfo {author}
  {\bibfnamefont {S.}~\bibnamefont {Parham}}, \bibinfo {author} {\bibfnamefont
  {J.~A.}\ \bibnamefont {Waugh}}, \bibinfo {author} {\bibfnamefont
  {Y.}~\bibnamefont {Cao}}, \bibinfo {author} {\bibfnamefont {Z.}~\bibnamefont
  {Sun}}, \bibinfo {author} {\bibfnamefont {H.}~\bibnamefont {Li}}, \bibinfo
  {author} {\bibfnamefont {Q.}~\bibnamefont {Wang}}, \bibinfo {author}
  {\bibfnamefont {J.~S.}\ \bibnamefont {Wen}}, \bibinfo {author} {\bibfnamefont
  {Z.~J.}\ \bibnamefont {Xu}}, \bibinfo {author} {\bibfnamefont
  {G.}~\bibnamefont {Gu}}, \bibinfo {author} {\bibfnamefont {Y.}~\bibnamefont
  {Yoshida}}, \bibinfo {author} {\bibfnamefont {H.}~\bibnamefont {Eisaki}},
  \bibinfo {author} {\bibfnamefont {G.~B.}\ \bibnamefont {Arnold}}, and\
  \bibinfo {author} {\bibfnamefont {D.~S.}\ \bibnamefont {Dessau}}} (\bibinfo
  {year} {2019}),\ \bibfield  {title} {\enquote {\bibinfo {title} {{A unified
  form of low-energy nodal electronic interactions in hole-doped cuprate
  superconductors}},}\ }\href {https://doi.org/10.1038/s41467-019-13497-4}
  {\bibfield  {journal} {\bibinfo  {journal} {Nature Communications}\ }\textbf
  {\bibinfo {volume} {10}}~(\bibinfo {number} {1}),\ \bibinfo {pages}
  {5737}}\BibitemShut {NoStop}%
\bibitem [{\citenamefont {Ross}(2005)}]{Ross:2005sc}%
  \BibitemOpen
  \bibfield  {author} {\bibinfo {author} {\bibnamefont {Ross}, \bibfnamefont
  {Simon~F}}} (\bibinfo {year} {2005}),\ \bibfield  {title} {\enquote {\bibinfo
  {title} {{Black hole thermodynamics}},}\ }\href@noop {} {\ }\Eprint
  {https://arxiv.org/abs/hep-th/0502195} {arXiv:hep-th/0502195} \BibitemShut
  {NoStop}%
\bibitem [{\citenamefont {Rossini}\ \emph {et~al.}(2020)\citenamefont
  {Rossini}, \citenamefont {Andolina}, \citenamefont {Rosa}, \citenamefont
  {Carrega},\ and\ \citenamefont {Polini}}]{Rossini:2019nfu}%
  \BibitemOpen
  \bibfield  {author} {\bibinfo {author} {\bibnamefont {Rossini}, \bibfnamefont
  {Davide}}, \bibinfo {author} {\bibfnamefont {Gian~Marcello}\ \bibnamefont
  {Andolina}}, \bibinfo {author} {\bibfnamefont {Dario}\ \bibnamefont {Rosa}},
  \bibinfo {author} {\bibfnamefont {Matteo}\ \bibnamefont {Carrega}}, and\
  \bibinfo {author} {\bibfnamefont {Marco}\ \bibnamefont {Polini}}} (\bibinfo
  {year} {2020}),\ \bibfield  {title} {\enquote {\bibinfo {title} {{Quantum
  advantage in the charging process of Sachdev-Ye-Kitaev batteries}},}\ }\href
  {https://doi.org/10.1103/PhysRevLett.125.236402} {\bibfield  {journal}
  {\bibinfo  {journal} {Phys. Rev. Lett.}\ }\textbf {\bibinfo {volume}
  {125}}~(\bibinfo {number} {23}),\ \bibinfo {pages} {236402}},\ \Eprint
  {https://arxiv.org/abs/1912.07234} {arXiv:1912.07234 [cond-mat.str-el]}
  \BibitemShut {NoStop}%
\bibitem [{\citenamefont {Rozenberg}\ \emph {et~al.}(1996)\citenamefont
  {Rozenberg}, \citenamefont {Kotliar},\ and\ \citenamefont
  {Kajueter}}]{Rozenberg_1996}%
  \BibitemOpen
  \bibfield  {author} {\bibinfo {author} {\bibnamefont {Rozenberg},
  \bibfnamefont {M~J}}, \bibinfo {author} {\bibfnamefont {G.}~\bibnamefont
  {Kotliar}}, and\ \bibinfo {author} {\bibfnamefont {H.}~\bibnamefont
  {Kajueter}}} (\bibinfo {year} {1996}),\ \bibfield  {title} {\enquote
  {\bibinfo {title} {Transfer of spectral weight in spectroscopies of
  correlated electron systems},}\ }\href
  {https://doi.org/10.1103/PhysRevB.54.8452} {\bibfield  {journal} {\bibinfo
  {journal} {Phys. Rev. B}\ }\textbf {\bibinfo {volume} {54}},\ \bibinfo
  {pages} {8452--8468}}\BibitemShut {NoStop}%
\bibitem [{\citenamefont {Rubtsov}\ \emph {et~al.}(2005)\citenamefont
  {Rubtsov}, \citenamefont {Savkin},\ and\ \citenamefont
  {Lichtenstein}}]{RubtsovCTQMC2005}%
  \BibitemOpen
  \bibfield  {author} {\bibinfo {author} {\bibnamefont {Rubtsov}, \bibfnamefont
  {A~N}}, \bibinfo {author} {\bibfnamefont {V.~V.}\ \bibnamefont {Savkin}},
  and\ \bibinfo {author} {\bibfnamefont {A.~I.}\ \bibnamefont {Lichtenstein}}}
  (\bibinfo {year} {2005}),\ \bibfield  {title} {\enquote {\bibinfo {title}
  {{Continuous-time quantum Monte Carlo method for fermions}},}\ }\href
  {https://doi.org/10.1103/PhysRevB.72.035122} {\bibfield  {journal} {\bibinfo
  {journal} {Phys. Rev. B}\ }\textbf {\bibinfo {volume} {72}},\ \bibinfo
  {pages} {035122}}\BibitemShut {NoStop}%
\bibitem [{\citenamefont {{Saad}}\ \emph {et~al.}(2018)\citenamefont {{Saad}},
  \citenamefont {{Shenker}},\ and\ \citenamefont {{Stanford}}}]{Saad18}%
  \BibitemOpen
  \bibfield  {author} {\bibinfo {author} {\bibnamefont {{Saad}}, \bibfnamefont
  {Phil}}, \bibinfo {author} {\bibfnamefont {Stephen~H.}\ \bibnamefont
  {{Shenker}}}, and\ \bibinfo {author} {\bibfnamefont {Douglas}\ \bibnamefont
  {{Stanford}}}} (\bibinfo {year} {2018}),\ \bibfield  {title} {\enquote
  {\bibinfo {title} {{A semiclassical ramp in SYK and in gravity}},}\
  }\href@noop {} {\ }\Eprint {https://arxiv.org/abs/1806.06840}
  {arXiv:1806.06840 [hep-th]} \BibitemShut {NoStop}%
\bibitem [{\citenamefont {Saad}\ \emph {et~al.}(2019)\citenamefont {Saad},
  \citenamefont {Shenker},\ and\ \citenamefont {Stanford}}]{Saad:2019lba}%
  \BibitemOpen
  \bibfield  {author} {\bibinfo {author} {\bibnamefont {Saad}, \bibfnamefont
  {Phil}}, \bibinfo {author} {\bibfnamefont {Stephen~H.}\ \bibnamefont
  {Shenker}}, and\ \bibinfo {author} {\bibfnamefont {Douglas}\ \bibnamefont
  {Stanford}}} (\bibinfo {year} {2019}),\ \bibfield  {title} {\enquote
  {\bibinfo {title} {{JT gravity as a matrix integral}},}\ }\href@noop {} {\
  }\Eprint {https://arxiv.org/abs/1903.11115} {arXiv:1903.11115 [hep-th]}
  \BibitemShut {NoStop}%
\bibitem [{\citenamefont {Sachdev}(1999)}]{QPT}%
  \BibitemOpen
  \bibfield  {author} {\bibinfo {author} {\bibnamefont {Sachdev}, \bibfnamefont
  {Subir}}} (\bibinfo {year} {1999}),\ \href@noop {} {\emph {\bibinfo {title}
  {Quantum phase transitions}}}\ (\bibinfo  {publisher} {Cambridge University
  Press})\BibitemShut {NoStop}%
\bibitem [{\citenamefont {Sachdev}(2001)}]{Sachdev2001}%
  \BibitemOpen
  \bibfield  {author} {\bibinfo {author} {\bibnamefont {Sachdev}, \bibfnamefont
  {Subir}}} (\bibinfo {year} {2001}),\ \bibfield  {title} {\enquote {\bibinfo
  {title} {Static hole in a critical antiferromagnet: field-theoretic
  renormalization group},}\ }\href
  {https://doi.org/10.1016/s0921-4534(01)00198-8} {\bibfield  {journal}
  {\bibinfo  {journal} {Physica C: Superconductivity}\ }\textbf {\bibinfo
  {volume} {357-360}},\ \bibinfo {pages} {78--81}}\BibitemShut {NoStop}%
\bibitem [{\citenamefont {Sachdev}(2010{\natexlab{a}})}]{SS10}%
  \BibitemOpen
  \bibfield  {author} {\bibinfo {author} {\bibnamefont {Sachdev}, \bibfnamefont
  {Subir}}} (\bibinfo {year} {2010}{\natexlab{a}}),\ \bibfield  {title}
  {\enquote {\bibinfo {title} {{Holographic metals and the fractionalized Fermi
  liquid}},}\ }\href {https://doi.org/10.1103/PhysRevLett.105.151602}
  {\bibfield  {journal} {\bibinfo  {journal} {Phys. Rev. Lett.}\ }\textbf
  {\bibinfo {volume} {105}},\ \bibinfo {pages} {151602}},\ \Eprint
  {https://arxiv.org/abs/1006.3794} {arXiv:1006.3794 [hep-th]} \BibitemShut
  {NoStop}%
%%CITATION = ARXIV:1006.3794;%%
\bibitem [{\citenamefont {Sachdev}(2010{\natexlab{b}})}]{Sachdev:2010uj}%
  \BibitemOpen
  \bibfield  {author} {\bibinfo {author} {\bibnamefont {Sachdev}, \bibfnamefont
  {Subir}}} (\bibinfo {year} {2010}{\natexlab{b}}),\ \bibfield  {title}
  {\enquote {\bibinfo {title} {{Strange metals and the AdS/CFT
  correspondence}},}\ }\href {https://doi.org/10.1088/1742-5468/2010/11/P11022}
  {\bibfield  {journal} {\bibinfo  {journal} {J. Stat. Mech.}\ }\textbf
  {\bibinfo {volume} {1011}},\ \bibinfo {pages} {P11022}},\ \Eprint
  {https://arxiv.org/abs/1010.0682} {arXiv:1010.0682 [cond-mat.str-el]}
  \BibitemShut {NoStop}%
\bibitem [{\citenamefont {Sachdev}(2012)}]{Sachdev:2012tj}%
  \BibitemOpen
  \bibfield  {author} {\bibinfo {author} {\bibnamefont {Sachdev}, \bibfnamefont
  {Subir}}} (\bibinfo {year} {2012}),\ \bibfield  {title} {\enquote {\bibinfo
  {title} {{Compressible quantum phases from conformal field theories in 2+1
  dimensions}},}\ }\href {https://doi.org/10.1103/PhysRevD.86.126003}
  {\bibfield  {journal} {\bibinfo  {journal} {Phys. Rev. D}\ }\textbf {\bibinfo
  {volume} {86}},\ \bibinfo {pages} {126003}},\ \Eprint
  {https://arxiv.org/abs/1209.1637} {arXiv:1209.1637 [hep-th]} \BibitemShut
  {NoStop}%
\bibitem [{\citenamefont {Sachdev}(2015)}]{SS15}%
  \BibitemOpen
  \bibfield  {author} {\bibinfo {author} {\bibnamefont {Sachdev}, \bibfnamefont
  {Subir}}} (\bibinfo {year} {2015}),\ \bibfield  {title} {\enquote {\bibinfo
  {title} {{Bekenstein-Hawking Entropy and Strange Metals}},}\ }\href
  {https://doi.org/10.1103/PhysRevX.5.041025} {\bibfield  {journal} {\bibinfo
  {journal} {Phys. Rev. X}\ }\textbf {\bibinfo {volume} {5}},\ \bibinfo {pages}
  {041025}}\BibitemShut {NoStop}%
\bibitem [{\citenamefont {Sachdev}(2019)}]{SachdevICMP}%
  \BibitemOpen
  \bibfield  {author} {\bibinfo {author} {\bibnamefont {Sachdev}, \bibfnamefont
  {Subir}}} (\bibinfo {year} {2019}),\ \bibfield  {title} {\enquote {\bibinfo
  {title} {{Universal low temperature theory of charged black holes with
  AdS$_2$ horizons}},}\ }\href {https://doi.org/10.1063/1.5092726} {\bibfield
  {journal} {\bibinfo  {journal} {J. Math. Phys.}\ }\textbf {\bibinfo {volume}
  {60}}~(\bibinfo {number} {5}),\ \bibinfo {pages} {052303}},\ \Eprint
  {https://arxiv.org/abs/1902.04078} {arXiv:1902.04078 [hep-th]} \BibitemShut
  {NoStop}%
\bibitem [{\citenamefont {Sachdev}(2022)}]{Sachdev:2022qnu}%
  \BibitemOpen
  \bibfield  {author} {\bibinfo {author} {\bibnamefont {Sachdev}, \bibfnamefont
  {Subir}}} (\bibinfo {year} {2022}),\ \bibfield  {title} {\enquote {\bibinfo
  {title} {{Statistical mechanics of strange metals and black holes}},}\
  }\href@noop {} {\ }\Eprint {https://arxiv.org/abs/2205.02285}
  {arXiv:2205.02285 [hep-th]} \BibitemShut {NoStop}%
\bibitem [{\citenamefont {Sachdev}\ \emph {et~al.}(1999)\citenamefont
  {Sachdev}, \citenamefont {Buragohain},\ and\ \citenamefont
  {Vojta}}]{SBV1999}%
  \BibitemOpen
  \bibfield  {author} {\bibinfo {author} {\bibnamefont {Sachdev}, \bibfnamefont
  {Subir}}, \bibinfo {author} {\bibfnamefont {Chiranjeeb}\ \bibnamefont
  {Buragohain}}, and\ \bibinfo {author} {\bibfnamefont {Matthias}\ \bibnamefont
  {Vojta}}} (\bibinfo {year} {1999}),\ \bibfield  {title} {\enquote {\bibinfo
  {title} {{Quantum Impurity in a Nearly Critical Two-Dimensional
  Antiferromagnet}},}\ }\href {https://doi.org/10.1126/science.286.5449.2479}
  {\bibfield  {journal} {\bibinfo  {journal} {Science}\ }\textbf {\bibinfo
  {volume} {286}}~(\bibinfo {number} {5449}),\ \bibinfo {pages} {2479--2482}},\
  \Eprint {https://arxiv.org/abs/cond-mat/0004156} {arXiv:cond-mat/0004156
  [cond-mat.str-el]} \BibitemShut {NoStop}%
\bibitem [{\citenamefont {{Sachdev}}\ \emph {et~al.}(1995)\citenamefont
  {{Sachdev}}, \citenamefont {{Read}},\ and\ \citenamefont
  {{Oppermann}}}]{SRO95}%
  \BibitemOpen
  \bibfield  {author} {\bibinfo {author} {\bibnamefont {{Sachdev}},
  \bibfnamefont {Subir}}, \bibinfo {author} {\bibfnamefont {N.}~\bibnamefont
  {{Read}}}, and\ \bibinfo {author} {\bibfnamefont {R.}~\bibnamefont
  {{Oppermann}}}} (\bibinfo {year} {1995}),\ \bibfield  {title} {\enquote
  {\bibinfo {title} {{Quantum field theory of metallic spin glasses}},}\ }\href
  {https://doi.org/10.1103/PhysRevB.52.10286} {\bibfield  {journal} {\bibinfo
  {journal} {Physical Review B}\ }\textbf {\bibinfo {volume} {52}}~(\bibinfo
  {number} {14}),\ \bibinfo {pages} {10286--10294}},\ \Eprint
  {https://arxiv.org/abs/cond-mat/9504036} {arXiv:cond-mat/9504036 [cond-mat]}
  \BibitemShut {NoStop}%
\bibitem [{\citenamefont {Sachdev}\ and\ \citenamefont {Ye}(1993)}]{SY}%
  \BibitemOpen
  \bibfield  {author} {\bibinfo {author} {\bibnamefont {Sachdev}, \bibfnamefont
  {Subir}}, and\ \bibinfo {author} {\bibfnamefont {Jinwu}\ \bibnamefont {Ye}}}
  (\bibinfo {year} {1993}),\ \bibfield  {title} {\enquote {\bibinfo {title}
  {{Gapless spin-fluid ground state in a random quantum Heisenberg magnet}},}\
  }\href {https://doi.org/10.1103/PhysRevLett.70.3339} {\bibfield  {journal}
  {\bibinfo  {journal} {Phys. Rev. Lett.}\ }\textbf {\bibinfo {volume} {70}},\
  \bibinfo {pages} {3339--3342}}\BibitemShut {NoStop}%
\bibitem [{\citenamefont {{Sadovskii}}(2020)}]{Sadovskii2020}%
  \BibitemOpen
  \bibfield  {author} {\bibinfo {author} {\bibnamefont {{Sadovskii}},
  \bibfnamefont {M~V}}} (\bibinfo {year} {2020}),\ \bibfield  {title} {\enquote
  {\bibinfo {title} {{On the Planckian Limit for Inelastic Relaxation in
  Metals}},}\ }\href {https://doi.org/10.1134/S0021364020030029} {\bibfield
  {journal} {\bibinfo  {journal} {Soviet Journal of Experimental and
  Theoretical Physics Letters}\ }\textbf {\bibinfo {volume} {111}}~(\bibinfo
  {number} {3}),\ \bibinfo {pages} {188--192}},\ \Eprint
  {https://arxiv.org/abs/2001.01393} {arXiv:2001.01393 [cond-mat.supr-con]}
  \BibitemShut {NoStop}%
\bibitem [{\citenamefont {{Sadovskii}}(2021)}]{Sadovskii2021}%
  \BibitemOpen
  \bibfield  {author} {\bibinfo {author} {\bibnamefont {{Sadovskii}},
  \bibfnamefont {M~V}}} (\bibinfo {year} {2021}),\ \bibfield  {title} {\enquote
  {\bibinfo {title} {{Planckian relaxation delusion in metals}},}\ }\href
  {https://doi.org/10.3367/UFNe.2020.08.038821} {\bibfield  {journal} {\bibinfo
   {journal} {Physics Uspekhi}\ }\textbf {\bibinfo {volume} {64}}~(\bibinfo
  {number} {2}),\ \bibinfo {pages} {175--190}},\ \Eprint
  {https://arxiv.org/abs/2008.09300} {arXiv:2008.09300 [cond-mat.supr-con]}
  \BibitemShut {NoStop}%
\bibitem [{\citenamefont {Sahoo}\ \emph {et~al.}(2020)\citenamefont {Sahoo},
  \citenamefont {Lantagne-Hurtubise}, \citenamefont {Plugge},\ and\
  \citenamefont {Franz}}]{Sahoo:2020unu}%
  \BibitemOpen
  \bibfield  {author} {\bibinfo {author} {\bibnamefont {Sahoo}, \bibfnamefont
  {Sharmistha}}, \bibinfo {author} {\bibfnamefont {\'Etienne}\ \bibnamefont
  {Lantagne-Hurtubise}}, \bibinfo {author} {\bibfnamefont {Stephan}\
  \bibnamefont {Plugge}}, and\ \bibinfo {author} {\bibfnamefont {Marcel}\
  \bibnamefont {Franz}}} (\bibinfo {year} {2020}),\ \bibfield  {title}
  {\enquote {\bibinfo {title} {{Traversable wormhole and Hawking-Page
  transition in coupled complex SYK models}},}\ }\href
  {https://doi.org/10.1103/PhysRevResearch.2.043049} {\bibfield  {journal}
  {\bibinfo  {journal} {Phys. Rev. Res.}\ }\textbf {\bibinfo {volume}
  {2}}~(\bibinfo {number} {4}),\ \bibinfo {pages} {043049}},\ \Eprint
  {https://arxiv.org/abs/2006.06019} {arXiv:2006.06019 [cond-mat.str-el]}
  \BibitemShut {NoStop}%
\bibitem [{\citenamefont {Samui}\ and\ \citenamefont
  {Sorokhaibam}(2021)}]{Samui:2020jli}%
  \BibitemOpen
  \bibfield  {author} {\bibinfo {author} {\bibnamefont {Samui}, \bibfnamefont
  {Tousik}}, and\ \bibinfo {author} {\bibfnamefont {Nilakash}\ \bibnamefont
  {Sorokhaibam}}} (\bibinfo {year} {2021}),\ \bibfield  {title} {\enquote
  {\bibinfo {title} {{Thermalization in different phases of charged SYK
  model}},}\ }\href {https://doi.org/10.1007/JHEP04(2021)157} {\bibfield
  {journal} {\bibinfo  {journal} {JHEP}\ }\textbf {\bibinfo {volume} {04}},\
  \bibinfo {pages} {157}},\ \Eprint {https://arxiv.org/abs/2004.14376}
  {arXiv:2004.14376 [hep-th]} \BibitemShut {NoStop}%
\bibitem [{\citenamefont {Schlenker}\ and\ \citenamefont
  {Witten}(2022)}]{Schlenker:2022dyo}%
  \BibitemOpen
  \bibfield  {author} {\bibinfo {author} {\bibnamefont {Schlenker},
  \bibfnamefont {Jean-Marc}}, and\ \bibinfo {author} {\bibfnamefont {Edward}\
  \bibnamefont {Witten}}} (\bibinfo {year} {2022}),\ \bibfield  {title}
  {\enquote {\bibinfo {title} {{No Ensemble Averaging Below the Black Hole
  Threshold}},}\ }\href@noop {} {\ }\Eprint {https://arxiv.org/abs/2202.01372}
  {arXiv:2202.01372 [hep-th]} \BibitemShut {NoStop}%
\bibitem [{\citenamefont {Schlesinger}\ \emph {et~al.}(1990)\citenamefont
  {Schlesinger}, \citenamefont {Collins}, \citenamefont {Holtzberg},
  \citenamefont {Feild}, \citenamefont {Blanton}, \citenamefont {Welp},
  \citenamefont {Crabtree}, \citenamefont {Fang},\ and\ \citenamefont
  {Liu}}]{Schlesinger_1990}%
  \BibitemOpen
  \bibfield  {author} {\bibinfo {author} {\bibnamefont {Schlesinger},
  \bibfnamefont {Z}}, \bibinfo {author} {\bibfnamefont {R.~T.}\ \bibnamefont
  {Collins}}, \bibinfo {author} {\bibfnamefont {F.}~\bibnamefont {Holtzberg}},
  \bibinfo {author} {\bibfnamefont {C.}~\bibnamefont {Feild}}, \bibinfo
  {author} {\bibfnamefont {S.~H.}\ \bibnamefont {Blanton}}, \bibinfo {author}
  {\bibfnamefont {U.}~\bibnamefont {Welp}}, \bibinfo {author} {\bibfnamefont
  {G.~W.}\ \bibnamefont {Crabtree}}, \bibinfo {author} {\bibfnamefont
  {Y.}~\bibnamefont {Fang}}, and\ \bibinfo {author} {\bibfnamefont {J.~Z.}\
  \bibnamefont {Liu}}} (\bibinfo {year} {1990}),\ \bibfield  {title} {\enquote
  {\bibinfo {title} {Superconducting energy gap and normal-state conductivity
  of a single-domain ${\mathrm{yba}}_{2}$${\mathrm{cu}}_{3}$${\mathrm{o}}_{7}$
  crystal},}\ }\href {https://doi.org/10.1103/PhysRevLett.65.801} {\bibfield
  {journal} {\bibinfo  {journal} {Phys. Rev. Lett.}\ }\textbf {\bibinfo
  {volume} {65}},\ \bibinfo {pages} {801--804}}\BibitemShut {NoStop}%
\bibitem [{\citenamefont {{Schr{\"o}der}}\ \emph {et~al.}(1998)\citenamefont
  {{Schr{\"o}der}}, \citenamefont {{Aeppli}}, \citenamefont {{Bucher}},
  \citenamefont {{Ramazashvili}},\ and\ \citenamefont
  {{Coleman}}}]{Schroder98}%
  \BibitemOpen
  \bibfield  {author} {\bibinfo {author} {\bibnamefont {{Schr{\"o}der}},
  \bibfnamefont {A}}, \bibinfo {author} {\bibfnamefont {G.}~\bibnamefont
  {{Aeppli}}}, \bibinfo {author} {\bibfnamefont {E.}~\bibnamefont {{Bucher}}},
  \bibinfo {author} {\bibfnamefont {R.}~\bibnamefont {{Ramazashvili}}}, and\
  \bibinfo {author} {\bibfnamefont {P.}~\bibnamefont {{Coleman}}}} (\bibinfo
  {year} {1998}),\ \bibfield  {title} {\enquote {\bibinfo {title} {{Scaling of
  Magnetic Fluctuations near a Quantum Phase Transition}},}\ }\href
  {https://doi.org/10.1103/PhysRevLett.80.5623} {\bibfield  {journal} {\bibinfo
   {journal} {Phys. Rev. Lett.}\ }\textbf {\bibinfo {volume} {80}}~(\bibinfo
  {number} {25}),\ \bibinfo {pages} {5623--5626}},\ \Eprint
  {https://arxiv.org/abs/cond-mat/9803004} {arXiv:cond-mat/9803004
  [cond-mat.str-el]} \BibitemShut {NoStop}%
\bibitem [{\citenamefont {Schr{\"o}der}\ \emph {et~al.}(2000)\citenamefont
  {Schr{\"o}der}, \citenamefont {Aeppli}, \citenamefont {Coldea}, \citenamefont
  {Adams}, \citenamefont {Stockert}, \citenamefont {L{\"o}hneysen},
  \citenamefont {Bucher}, \citenamefont {Ramazashvili},\ and\ \citenamefont
  {Coleman}}]{Coleman2000}%
  \BibitemOpen
  \bibfield  {author} {\bibinfo {author} {\bibnamefont {Schr{\"o}der},
  \bibfnamefont {A}}, \bibinfo {author} {\bibfnamefont {G.}~\bibnamefont
  {Aeppli}}, \bibinfo {author} {\bibfnamefont {R.}~\bibnamefont {Coldea}},
  \bibinfo {author} {\bibfnamefont {M.}~\bibnamefont {Adams}}, \bibinfo
  {author} {\bibfnamefont {O.}~\bibnamefont {Stockert}}, \bibinfo {author}
  {\bibfnamefont {H.~v.}\ \bibnamefont {L{\"o}hneysen}}, \bibinfo {author}
  {\bibfnamefont {E.}~\bibnamefont {Bucher}}, \bibinfo {author} {\bibfnamefont
  {R.}~\bibnamefont {Ramazashvili}}, and\ \bibinfo {author} {\bibfnamefont
  {P.}~\bibnamefont {Coleman}}} (\bibinfo {year} {2000}),\ \bibfield  {title}
  {\enquote {\bibinfo {title} {Onset of antiferromagnetism in heavy-fermion
  metals},}\ }\href {http://dx.doi.org/10.1038/35030039} {\bibfield  {journal}
  {\bibinfo  {journal} {Nature}\ }\textbf {\bibinfo {volume} {407}},\ \bibinfo
  {pages} {351}}\BibitemShut {NoStop}%
\bibitem [{\citenamefont {Schwartz}\ \emph {et~al.}(1998)\citenamefont
  {Schwartz}, \citenamefont {Dressel}, \citenamefont {Gr\"uner}, \citenamefont
  {Vescoli}, \citenamefont {Degiorgi},\ and\ \citenamefont
  {Giamarchi}}]{Schwartz_1998}%
  \BibitemOpen
  \bibfield  {author} {\bibinfo {author} {\bibnamefont {Schwartz},
  \bibfnamefont {A}}, \bibinfo {author} {\bibfnamefont {M.}~\bibnamefont
  {Dressel}}, \bibinfo {author} {\bibfnamefont {G.}~\bibnamefont {Gr\"uner}},
  \bibinfo {author} {\bibfnamefont {V.}~\bibnamefont {Vescoli}}, \bibinfo
  {author} {\bibfnamefont {L.}~\bibnamefont {Degiorgi}}, and\ \bibinfo {author}
  {\bibfnamefont {T.}~\bibnamefont {Giamarchi}}} (\bibinfo {year} {1998}),\
  \bibfield  {title} {\enquote {\bibinfo {title} {On-chain electrodynamics of
  metallic $(\mathrm{TMTSF}{)}_{2}x$ salts: Observation of tomonaga-luttinger
  liquid response},}\ }\href {https://doi.org/10.1103/PhysRevB.58.1261}
  {\bibfield  {journal} {\bibinfo  {journal} {Phys. Rev. B}\ }\textbf {\bibinfo
  {volume} {58}},\ \bibinfo {pages} {1261--1271}}\BibitemShut {NoStop}%
\bibitem [{\citenamefont {Seaman}\ \emph {et~al.}(1991)\citenamefont {Seaman},
  \citenamefont {Maple}, \citenamefont {Lee}, \citenamefont {Ghamaty},
  \citenamefont {Torikachvili}, \citenamefont {Kang}, \citenamefont {Liu},
  \citenamefont {Allen},\ and\ \citenamefont {Cox}}]{Cox91}%
  \BibitemOpen
  \bibfield  {author} {\bibinfo {author} {\bibnamefont {Seaman}, \bibfnamefont
  {C~L}}, \bibinfo {author} {\bibfnamefont {M.~B.}\ \bibnamefont {Maple}},
  \bibinfo {author} {\bibfnamefont {B.~W.}\ \bibnamefont {Lee}}, \bibinfo
  {author} {\bibfnamefont {S.}~\bibnamefont {Ghamaty}}, \bibinfo {author}
  {\bibfnamefont {M.~S.}\ \bibnamefont {Torikachvili}}, \bibinfo {author}
  {\bibfnamefont {J.-S.}\ \bibnamefont {Kang}}, \bibinfo {author}
  {\bibfnamefont {L.~Z.}\ \bibnamefont {Liu}}, \bibinfo {author} {\bibfnamefont
  {J.~W.}\ \bibnamefont {Allen}}, and\ \bibinfo {author} {\bibfnamefont
  {D.~L.}\ \bibnamefont {Cox}}} (\bibinfo {year} {1991}),\ \bibfield  {title}
  {\enquote {\bibinfo {title} {{Evidence for non-Fermi liquid behavior in the
  Kondo alloy Y$_{1-x}$U$_x$Pd$_3$}},}\ }\href
  {https://doi.org/10.1103/PhysRevLett.67.2882} {\bibfield  {journal} {\bibinfo
   {journal} {Phys. Rev. Lett.}\ }\textbf {\bibinfo {volume} {67}},\ \bibinfo
  {pages} {2882--2885}}\BibitemShut {NoStop}%
\bibitem [{\citenamefont {Sen}(2005)}]{Sen05}%
  \BibitemOpen
  \bibfield  {author} {\bibinfo {author} {\bibnamefont {Sen}, \bibfnamefont
  {Ashoke}}} (\bibinfo {year} {2005}),\ \bibfield  {title} {\enquote {\bibinfo
  {title} {{Black hole entropy function and the attractor mechanism in higher
  derivative gravity}},}\ }\href
  {https://doi.org/10.1088/1126-6708/2005/09/038} {\bibfield  {journal}
  {\bibinfo  {journal} {JHEP}\ }\textbf {\bibinfo {volume} {0509}},\ \bibinfo
  {pages} {038}},\ \Eprint {https://arxiv.org/abs/hep-th/0506177}
  {arXiv:hep-th/0506177 [hep-th]} \BibitemShut {NoStop}%
%%CITATION = HEP-TH/0506177;%%
\bibitem [{\citenamefont {Sen}(2008)}]{Sen08}%
  \BibitemOpen
  \bibfield  {author} {\bibinfo {author} {\bibnamefont {Sen}, \bibfnamefont
  {Ashoke}}} (\bibinfo {year} {2008}),\ \bibfield  {title} {\enquote {\bibinfo
  {title} {{Entropy Function and AdS(2) / CFT(1) Correspondence}},}\ }\href
  {https://doi.org/10.1088/1126-6708/2008/11/075} {\bibfield  {journal}
  {\bibinfo  {journal} {JHEP}\ }\textbf {\bibinfo {volume} {0811}},\ \bibinfo
  {pages} {075}},\ \Eprint {https://arxiv.org/abs/0805.0095} {arXiv:0805.0095
  [hep-th]} \BibitemShut {NoStop}%
%%CITATION = ARXIV:0805.0095;%%
\bibitem [{\citenamefont {Sengupta}(2000)}]{Sengupta2000}%
  \BibitemOpen
  \bibfield  {author} {\bibinfo {author} {\bibnamefont {Sengupta},
  \bibfnamefont {Anirvan~M}}} (\bibinfo {year} {2000}),\ \bibfield  {title}
  {\enquote {\bibinfo {title} {{Spin in a fluctuating field: The Bose(+Fermi)
  Kondo models}},}\ }\href {https://doi.org/10.1103/PhysRevB.61.4041}
  {\bibfield  {journal} {\bibinfo  {journal} {Phys. Rev. B}\ }\textbf {\bibinfo
  {volume} {61}},\ \bibinfo {pages} {4041--4043}},\ \Eprint
  {https://arxiv.org/abs/cond-mat/9707316} {arXiv:cond-mat/9707316
  [cond-mat.str-el]} \BibitemShut {NoStop}%
\bibitem [{\citenamefont {{Sengupta}}\ and\ \citenamefont
  {{Georges}}(1995)}]{SG95}%
  \BibitemOpen
  \bibfield  {author} {\bibinfo {author} {\bibnamefont {{Sengupta}},
  \bibfnamefont {Anirvan~M}}, and\ \bibinfo {author} {\bibfnamefont {Antoine}\
  \bibnamefont {{Georges}}}} (\bibinfo {year} {1995}),\ \bibfield  {title}
  {\enquote {\bibinfo {title} {{Non-Fermi-liquid behavior near a T=0 spin-glass
  transition}},}\ }\href {https://doi.org/10.1103/PhysRevB.52.10295} {\bibfield
   {journal} {\bibinfo  {journal} {Phys. Rev. B}\ }\textbf {\bibinfo {volume}
  {52}}~(\bibinfo {number} {14}),\ \bibinfo {pages} {10295--10302}},\ \Eprint
  {https://arxiv.org/abs/cond-mat/9504120} {arXiv:cond-mat/9504120 [cond-mat]}
  \BibitemShut {NoStop}%
\bibitem [{\citenamefont {Senthil}(2008{\natexlab{a}})}]{TS08}%
  \BibitemOpen
  \bibfield  {author} {\bibinfo {author} {\bibnamefont {Senthil}, \bibfnamefont
  {T}}} (\bibinfo {year} {2008}{\natexlab{a}}),\ \bibfield  {title} {\enquote
  {\bibinfo {title} {{Critical Fermi surfaces and non-Fermi liquid metals}},}\
  }\href {https://doi.org/10.1103/PhysRevB.78.035103} {\bibfield  {journal}
  {\bibinfo  {journal} {Phys. Rev. B}\ }\textbf {\bibinfo {volume} {78}},\
  \bibinfo {pages} {035103}}\BibitemShut {NoStop}%
\bibitem [{\citenamefont {Senthil}(2008{\natexlab{b}})}]{TSmott}%
  \BibitemOpen
  \bibfield  {author} {\bibinfo {author} {\bibnamefont {Senthil}, \bibfnamefont
  {T}}} (\bibinfo {year} {2008}{\natexlab{b}}),\ \bibfield  {title} {\enquote
  {\bibinfo {title} {{Theory of a continuous Mott transition in two
  dimensions}},}\ }\href {https://doi.org/10.1103/PhysRevB.78.045109}
  {\bibfield  {journal} {\bibinfo  {journal} {Phys. Rev. B}\ }\textbf {\bibinfo
  {volume} {78}},\ \bibinfo {pages} {045109}}\BibitemShut {NoStop}%
\bibitem [{\citenamefont {{Senthil}}\ \emph {et~al.}(2003)\citenamefont
  {{Senthil}}, \citenamefont {{Sachdev}},\ and\ \citenamefont
  {{Vojta}}}]{FLSPRL}%
  \BibitemOpen
  \bibfield  {author} {\bibinfo {author} {\bibnamefont {{Senthil}},
  \bibfnamefont {T}}, \bibinfo {author} {\bibfnamefont {Subir}\ \bibnamefont
  {{Sachdev}}}, and\ \bibinfo {author} {\bibfnamefont {Matthias}\ \bibnamefont
  {{Vojta}}}} (\bibinfo {year} {2003}),\ \bibfield  {title} {\enquote {\bibinfo
  {title} {{Fractionalized Fermi Liquids}},}\ }\href
  {https://doi.org/10.1103/PhysRevLett.90.216403} {\bibfield  {journal}
  {\bibinfo  {journal} {Phys. Rev. Lett.}\ }\textbf {\bibinfo {volume}
  {90}}~(\bibinfo {number} {21}),\ \bibinfo {eid} {216403}},\ \Eprint
  {https://arxiv.org/abs/cond-mat/0209144} {arXiv:cond-mat/0209144
  [cond-mat.str-el]} \BibitemShut {NoStop}%
\bibitem [{\citenamefont {{Senthil}}\ \emph {et~al.}(2004)\citenamefont
  {{Senthil}}, \citenamefont {{Vojta}},\ and\ \citenamefont
  {{Sachdev}}}]{TSFL04}%
  \BibitemOpen
  \bibfield  {author} {\bibinfo {author} {\bibnamefont {{Senthil}},
  \bibfnamefont {T}}, \bibinfo {author} {\bibfnamefont {Matthias}\ \bibnamefont
  {{Vojta}}}, and\ \bibinfo {author} {\bibfnamefont {Subir}\ \bibnamefont
  {{Sachdev}}}} (\bibinfo {year} {2004}),\ \bibfield  {title} {\enquote
  {\bibinfo {title} {{Weak magnetism and non-Fermi liquids near heavy-fermion
  critical points}},}\ }\href {https://doi.org/10.1103/PhysRevB.69.035111}
  {\bibfield  {journal} {\bibinfo  {journal} {Phys. Rev. B}\ }\textbf {\bibinfo
  {volume} {69}}~(\bibinfo {number} {3}),\ \bibinfo {eid} {035111}},\ \Eprint
  {https://arxiv.org/abs/cond-mat/0305193} {arXiv:cond-mat/0305193
  [cond-mat.str-el]} \BibitemShut {NoStop}%
\bibitem [{\citenamefont {Setty}(2020)}]{Setty20}%
  \BibitemOpen
  \bibfield  {author} {\bibinfo {author} {\bibnamefont {Setty}, \bibfnamefont
  {Chandan}}} (\bibinfo {year} {2020}),\ \bibfield  {title} {\enquote {\bibinfo
  {title} {{Pairing instability on a Luttinger surface: A non-Fermi liquid to
  superconductor transition and its Sachdev-Ye-Kitaev dual}},}\ }\href
  {https://doi.org/10.1103/PhysRevB.101.184506} {\bibfield  {journal} {\bibinfo
   {journal} {Phys. Rev. B}\ }\textbf {\bibinfo {volume} {101}},\ \bibinfo
  {pages} {184506}}\BibitemShut {NoStop}%
\bibitem [{\citenamefont {Setty}(2021)}]{Setty21}%
  \BibitemOpen
  \bibfield  {author} {\bibinfo {author} {\bibnamefont {Setty}, \bibfnamefont
  {Chandan}}} (\bibinfo {year} {2021}),\ \bibfield  {title} {\enquote {\bibinfo
  {title} {{Superconductivity from Luttinger surfaces: Emergent
  Sachdev-Ye-Kitaev physics with infinite-body interactions}},}\ }\href
  {https://doi.org/10.1103/PhysRevB.103.014501} {\bibfield  {journal} {\bibinfo
   {journal} {Phys. Rev. B}\ }\textbf {\bibinfo {volume} {103}},\ \bibinfo
  {pages} {014501}}\BibitemShut {NoStop}%
\bibitem [{\citenamefont {Shackleton}\ \emph {et~al.}(2021)\citenamefont
  {Shackleton}, \citenamefont {Wietek}, \citenamefont {Georges},\ and\
  \citenamefont {Sachdev}}]{Shackleton2021}%
  \BibitemOpen
  \bibfield  {author} {\bibinfo {author} {\bibnamefont {Shackleton},
  \bibfnamefont {Henry}}, \bibinfo {author} {\bibfnamefont {Alexander}\
  \bibnamefont {Wietek}}, \bibinfo {author} {\bibfnamefont {Antoine}\
  \bibnamefont {Georges}}, and\ \bibinfo {author} {\bibfnamefont {Subir}\
  \bibnamefont {Sachdev}}} (\bibinfo {year} {2021}),\ \bibfield  {title}
  {\enquote {\bibinfo {title} {{Quantum Phase Transition at Nonzero Doping in a
  Random $t\text{\ensuremath{-}}J$ Model}},}\ }\href
  {https://doi.org/10.1103/PhysRevLett.126.136602} {\bibfield  {journal}
  {\bibinfo  {journal} {Phys. Rev. Lett.}\ }\textbf {\bibinfo {volume} {126}},\
  \bibinfo {pages} {136602}}\BibitemShut {NoStop}%
\bibitem [{\citenamefont {Shenker}\ and\ \citenamefont
  {Stanford}(2014)}]{Shenker:2013pqa}%
  \BibitemOpen
  \bibfield  {author} {\bibinfo {author} {\bibnamefont {Shenker}, \bibfnamefont
  {Stephen~H}}, and\ \bibinfo {author} {\bibfnamefont {Douglas}\ \bibnamefont
  {Stanford}}} (\bibinfo {year} {2014}),\ \bibfield  {title} {\enquote
  {\bibinfo {title} {{Black holes and the butterfly effect}},}\ }\href
  {https://doi.org/10.1007/JHEP03(2014)067} {\bibfield  {journal} {\bibinfo
  {journal} {JHEP}\ }\textbf {\bibinfo {volume} {03}},\ \bibinfo {pages}
  {067}},\ \Eprint {https://arxiv.org/abs/1306.0622} {arXiv:1306.0622 [hep-th]}
  \BibitemShut {NoStop}%
\bibitem [{\citenamefont {Shimizu}\ \emph {et~al.}(2012)\citenamefont
  {Shimizu}, \citenamefont {Takeda}, \citenamefont {Tanaka}, \citenamefont
  {Itoh}, \citenamefont {Niitaka},\ and\ \citenamefont {Takagi}}]{Takagi12}%
  \BibitemOpen
  \bibfield  {author} {\bibinfo {author} {\bibnamefont {Shimizu}, \bibfnamefont
  {Yasuhiro}}, \bibinfo {author} {\bibfnamefont {Hikaru}\ \bibnamefont
  {Takeda}}, \bibinfo {author} {\bibfnamefont {Moe}\ \bibnamefont {Tanaka}},
  \bibinfo {author} {\bibfnamefont {Masayuki}\ \bibnamefont {Itoh}}, \bibinfo
  {author} {\bibfnamefont {Seiji}\ \bibnamefont {Niitaka}}, and\ \bibinfo
  {author} {\bibfnamefont {Hidenori}\ \bibnamefont {Takagi}}} (\bibinfo {year}
  {2012}),\ \bibfield  {title} {\enquote {\bibinfo {title} {{An
  orbital-selective spin liquid in a frustrated heavy fermion spinel
  LiV$_2$O$_4$}},}\ }\href {https://doi.org/10.1038/ncomms1979} {\bibfield
  {journal} {\bibinfo  {journal} {Nature Communications}\ }\textbf {\bibinfo
  {volume} {3}}~(\bibinfo {number} {1}),\ \bibinfo {pages} {981}}\BibitemShut
  {NoStop}%
\bibitem [{\citenamefont {{Si}}(2010)}]{Si2010}%
  \BibitemOpen
  \bibfield  {author} {\bibinfo {author} {\bibnamefont {{Si}}, \bibfnamefont
  {Qimiao}}} (\bibinfo {year} {2010}),\ \bibfield  {title} {\enquote {\bibinfo
  {title} {{Quantum criticality and global phase diagram of magnetic heavy
  fermions}},}\ }\href {https://doi.org/10.1002/pssb.200983082} {\bibfield
  {journal} {\bibinfo  {journal} {Physica Status Solidi B Basic Research}\
  }\textbf {\bibinfo {volume} {247}}~(\bibinfo {number} {3}),\ \bibinfo {pages}
  {476--484}},\ \Eprint {https://arxiv.org/abs/0912.0040} {arXiv:0912.0040
  [cond-mat.str-el]} \BibitemShut {NoStop}%
\bibitem [{\citenamefont {{Si}}\ \emph {et~al.}(2001)\citenamefont {{Si}},
  \citenamefont {{Rabello}}, \citenamefont {{Ingersent}},\ and\ \citenamefont
  {{Smith}}}]{Si1}%
  \BibitemOpen
  \bibfield  {author} {\bibinfo {author} {\bibnamefont {{Si}}, \bibfnamefont
  {Qimiao}}, \bibinfo {author} {\bibfnamefont {Silvio}\ \bibnamefont
  {{Rabello}}}, \bibinfo {author} {\bibfnamefont {Kevin}\ \bibnamefont
  {{Ingersent}}}, and\ \bibinfo {author} {\bibfnamefont {J.~Lleweilun}\
  \bibnamefont {{Smith}}}} (\bibinfo {year} {2001}),\ \bibfield  {title}
  {\enquote {\bibinfo {title} {{Locally critical quantum phase transitions in
  strongly correlated metals}},}\ }\href {https://doi.org/10.1038/35101507}
  {\bibfield  {journal} {\bibinfo  {journal} {Nature}\ }\textbf {\bibinfo
  {volume} {413}}~(\bibinfo {number} {6858}),\ \bibinfo {pages} {804--808}},\
  \Eprint {https://arxiv.org/abs/cond-mat/0011477} {arXiv:cond-mat/0011477
  [cond-mat.str-el]} \BibitemShut {NoStop}%
\bibitem [{\citenamefont {{Si}}\ \emph {et~al.}(2003)\citenamefont {{Si}},
  \citenamefont {{Rabello}}, \citenamefont {{Ingersent}},\ and\ \citenamefont
  {{Smith}}}]{Si2}%
  \BibitemOpen
  \bibfield  {author} {\bibinfo {author} {\bibnamefont {{Si}}, \bibfnamefont
  {Qimiao}}, \bibinfo {author} {\bibfnamefont {Silvio}\ \bibnamefont
  {{Rabello}}}, \bibinfo {author} {\bibfnamefont {Kevin}\ \bibnamefont
  {{Ingersent}}}, and\ \bibinfo {author} {\bibfnamefont {J.~Lleweilun}\
  \bibnamefont {{Smith}}}} (\bibinfo {year} {2003}),\ \bibfield  {title}
  {\enquote {\bibinfo {title} {{Local fluctuations in quantum critical
  metals}},}\ }\href {https://doi.org/10.1103/PhysRevB.68.115103} {\bibfield
  {journal} {\bibinfo  {journal} {Phys. Rev. B}\ }\textbf {\bibinfo {volume}
  {68}}~(\bibinfo {number} {11}),\ \bibinfo {eid} {115103}},\ \Eprint
  {https://arxiv.org/abs/cond-mat/0202414} {arXiv:cond-mat/0202414
  [cond-mat.str-el]} \BibitemShut {NoStop}%
\bibitem [{\citenamefont {Si}\ and\ \citenamefont {Smith}(1996)}]{Si_1996}%
  \BibitemOpen
  \bibfield  {author} {\bibinfo {author} {\bibnamefont {Si}, \bibfnamefont
  {Qimiao}}, and\ \bibinfo {author} {\bibfnamefont {J.~Lleweilun}\ \bibnamefont
  {Smith}}} (\bibinfo {year} {1996}),\ \bibfield  {title} {\enquote {\bibinfo
  {title} {{Kosterlitz-Thouless Transition and Short Range Spatial Correlations
  in an Extended Hubbard Model}},}\ }\href
  {https://doi.org/10.1103/PhysRevLett.77.3391} {\bibfield  {journal} {\bibinfo
   {journal} {Phys. Rev. Lett.}\ }\textbf {\bibinfo {volume} {77}},\ \bibinfo
  {pages} {3391--3394}}\BibitemShut {NoStop}%
\bibitem [{\citenamefont {Slakey}\ \emph {et~al.}(1991)\citenamefont {Slakey},
  \citenamefont {Klein}, \citenamefont {Rice},\ and\ \citenamefont
  {Ginsberg}}]{Ginsberg91}%
  \BibitemOpen
  \bibfield  {author} {\bibinfo {author} {\bibnamefont {Slakey}, \bibfnamefont
  {F}}, \bibinfo {author} {\bibfnamefont {M.~V.}\ \bibnamefont {Klein}},
  \bibinfo {author} {\bibfnamefont {J.~P.}\ \bibnamefont {Rice}}, and\ \bibinfo
  {author} {\bibfnamefont {D.~M.}\ \bibnamefont {Ginsberg}}} (\bibinfo {year}
  {1991}),\ \bibfield  {title} {\enquote {\bibinfo {title} {Raman investigation
  of the ${\mathrm{yba}}_{2}$${\mathrm{cu}}_{3}$${\mathrm{o}}_{7}$ imaginary
  response function},}\ }\href {https://doi.org/10.1103/PhysRevB.43.3764}
  {\bibfield  {journal} {\bibinfo  {journal} {Phys. Rev. B}\ }\textbf {\bibinfo
  {volume} {43}},\ \bibinfo {pages} {3764--3767}}\BibitemShut {NoStop}%
\bibitem [{\citenamefont {{Smith}}\ and\ \citenamefont {{Si}}(1999)}]{Si0}%
  \BibitemOpen
  \bibfield  {author} {\bibinfo {author} {\bibnamefont {{Smith}}, \bibfnamefont
  {J~L}}, and\ \bibinfo {author} {\bibfnamefont {Q.}~\bibnamefont {{Si}}}}
  (\bibinfo {year} {1999}),\ \bibfield  {title} {\enquote {\bibinfo {title}
  {{Non-Fermi liquids in the two-band extended Hubbard model}},}\ }\href
  {https://doi.org/10.1209/epl/i1999-00151-4} {\bibfield  {journal} {\bibinfo
  {journal} {Europhys. Lett.}\ }\textbf {\bibinfo {volume} {45}}~(\bibinfo
  {number} {2}),\ \bibinfo {pages} {228--234}}\BibitemShut {NoStop}%
\bibitem [{\citenamefont {Smith}\ and\ \citenamefont
  {Si}(2000)}]{smith_si_prb_2000}%
  \BibitemOpen
  \bibfield  {author} {\bibinfo {author} {\bibnamefont {Smith}, \bibfnamefont
  {J~Lleweilun}}, and\ \bibinfo {author} {\bibfnamefont {Qimiao}\ \bibnamefont
  {Si}}} (\bibinfo {year} {2000}),\ \bibfield  {title} {\enquote {\bibinfo
  {title} {{Spatial correlations in dynamical mean-field theory}},}\ }\href
  {https://doi.org/10.1103/PhysRevB.61.5184} {\bibfield  {journal} {\bibinfo
  {journal} {Phys. Rev. B}\ }\textbf {\bibinfo {volume} {61}},\ \bibinfo
  {pages} {5184--5193}}\BibitemShut {NoStop}%
\bibitem [{\citenamefont {Soldevilla}\ \emph {et~al.}(2000)\citenamefont
  {Soldevilla}, \citenamefont {Sal}, \citenamefont {Blanco}, \citenamefont
  {Espeso},\ and\ \citenamefont {Fern\'andez}}]{Soldevilla00}%
  \BibitemOpen
  \bibfield  {author} {\bibinfo {author} {\bibnamefont {Soldevilla},
  \bibfnamefont {J~Garc\'{\i}a}}, \bibinfo {author} {\bibfnamefont
  {J.~C.~G\'omez}\ \bibnamefont {Sal}}, \bibinfo {author} {\bibfnamefont
  {J.~A.}\ \bibnamefont {Blanco}}, \bibinfo {author} {\bibfnamefont {J.~I.}\
  \bibnamefont {Espeso}}, and\ \bibinfo {author} {\bibfnamefont
  {J.~Rodr\'{\i}guez}\ \bibnamefont {Fern\'andez}}} (\bibinfo {year} {2000}),\
  \bibfield  {title} {\enquote {\bibinfo {title} {{Phase diagram of the
  ${\mathrm{CeNi}}_{1\ensuremath{-}x}{\mathrm{Cu}}_{x}$ Kondo system with
  spin-glass-like behavior favored by hybridization}},}\ }\href
  {https://doi.org/10.1103/PhysRevB.61.6821} {\bibfield  {journal} {\bibinfo
  {journal} {Phys. Rev. B}\ }\textbf {\bibinfo {volume} {61}},\ \bibinfo
  {pages} {6821--6825}}\BibitemShut {NoStop}%
\bibitem [{\citenamefont {Son}(2015)}]{Son15}%
  \BibitemOpen
  \bibfield  {author} {\bibinfo {author} {\bibnamefont {Son}, \bibfnamefont
  {Dam~Thanh}}} (\bibinfo {year} {2015}),\ \bibfield  {title} {\enquote
  {\bibinfo {title} {{Is the Composite Fermion a Dirac Particle?}}}\ }\href
  {https://doi.org/10.1103/PhysRevX.5.031027} {\bibfield  {journal} {\bibinfo
  {journal} {Phys. Rev. X}\ }\textbf {\bibinfo {volume} {5}},\ \bibinfo {pages}
  {031027}}\BibitemShut {NoStop}%
\bibitem [{\citenamefont {Song}\ \emph {et~al.}(2017)\citenamefont {Song},
  \citenamefont {Jian},\ and\ \citenamefont {Balents}}]{Balents}%
  \BibitemOpen
  \bibfield  {author} {\bibinfo {author} {\bibnamefont {Song}, \bibfnamefont
  {Xue-Yang}}, \bibinfo {author} {\bibfnamefont {Chao-Ming}\ \bibnamefont
  {Jian}}, and\ \bibinfo {author} {\bibfnamefont {Leon}\ \bibnamefont
  {Balents}}} (\bibinfo {year} {2017}),\ \bibfield  {title} {\enquote {\bibinfo
  {title} {{Strongly Correlated Metal Built from Sachdev-Ye-Kitaev Models}},}\
  }\href {https://doi.org/10.1103/PhysRevLett.119.216601} {\bibfield  {journal}
  {\bibinfo  {journal} {Phys. Rev. Lett.}\ }\textbf {\bibinfo {volume} {119}},\
  \bibinfo {pages} {216601}}\BibitemShut {NoStop}%
\bibitem [{\citenamefont {Sonner}\ and\ \citenamefont
  {Vielma}(2017)}]{Sonner2017}%
  \BibitemOpen
  \bibfield  {author} {\bibinfo {author} {\bibnamefont {Sonner}, \bibfnamefont
  {Julian}}, and\ \bibinfo {author} {\bibfnamefont {Manuel}\ \bibnamefont
  {Vielma}}} (\bibinfo {year} {2017}),\ \bibfield  {title} {\enquote {\bibinfo
  {title} {{Eigenstate thermalization in the Sachdev-Ye-Kitaev model}},}\
  }\href {https://doi.org/10.1007/JHEP11(2017)149} {\bibfield  {journal}
  {\bibinfo  {journal} {Journal of High Energy Physics}\ }\textbf {\bibinfo
  {volume} {2017}}~(\bibinfo {number} {11}),\ \bibinfo {pages}
  {149}}\BibitemShut {NoStop}%
\bibitem [{\citenamefont {Stanford}(2016)}]{Stanford:2015owe}%
  \BibitemOpen
  \bibfield  {author} {\bibinfo {author} {\bibnamefont {Stanford},
  \bibfnamefont {Douglas}}} (\bibinfo {year} {2016}),\ \bibfield  {title}
  {\enquote {\bibinfo {title} {{Many-body chaos at weak coupling}},}\ }\href
  {https://doi.org/10.1007/JHEP10(2016)009} {\bibfield  {journal} {\bibinfo
  {journal} {JHEP}\ }\textbf {\bibinfo {volume} {10}},\ \bibinfo {pages}
  {009}},\ \Eprint {https://arxiv.org/abs/1512.07687} {arXiv:1512.07687
  [hep-th]} \BibitemShut {NoStop}%
\bibitem [{\citenamefont {Stanford}\ and\ \citenamefont
  {Witten}(2017)}]{Stanford:2017thb}%
  \BibitemOpen
  \bibfield  {author} {\bibinfo {author} {\bibnamefont {Stanford},
  \bibfnamefont {Douglas}}, and\ \bibinfo {author} {\bibfnamefont {Edward}\
  \bibnamefont {Witten}}} (\bibinfo {year} {2017}),\ \bibfield  {title}
  {\enquote {\bibinfo {title} {{Fermionic Localization of the Schwarzian
  Theory}},}\ }\href {https://doi.org/10.1007/JHEP10(2017)008} {\bibfield
  {journal} {\bibinfo  {journal} {JHEP}\ }\textbf {\bibinfo {volume} {10}},\
  \bibinfo {pages} {008}},\ \Eprint {https://arxiv.org/abs/1703.04612}
  {arXiv:1703.04612 [hep-th]} \BibitemShut {NoStop}%
\bibitem [{\citenamefont {Stanford}\ and\ \citenamefont
  {Witten}(2019)}]{Stanford:2019vob}%
  \BibitemOpen
  \bibfield  {author} {\bibinfo {author} {\bibnamefont {Stanford},
  \bibfnamefont {Douglas}}, and\ \bibinfo {author} {\bibfnamefont {Edward}\
  \bibnamefont {Witten}}} (\bibinfo {year} {2019}),\ \bibfield  {title}
  {\enquote {\bibinfo {title} {{JT Gravity and the Ensembles of Random Matrix
  Theory}},}\ }\href@noop {} {\ }\Eprint {https://arxiv.org/abs/1907.03363}
  {arXiv:1907.03363 [hep-th]} \BibitemShut {NoStop}%
\bibitem [{\citenamefont {Steinberg}\ and\ \citenamefont
  {Swingle}(2019)}]{Steinberg:2019uqb}%
  \BibitemOpen
  \bibfield  {author} {\bibinfo {author} {\bibnamefont {Steinberg},
  \bibfnamefont {Julia}}, and\ \bibinfo {author} {\bibfnamefont {Brian}\
  \bibnamefont {Swingle}}} (\bibinfo {year} {2019}),\ \bibfield  {title}
  {\enquote {\bibinfo {title} {{Thermalization and chaos in QED$_{3}$}},}\
  }\href {https://doi.org/10.1103/PhysRevD.99.076007} {\bibfield  {journal}
  {\bibinfo  {journal} {Phys. Rev. D}\ }\textbf {\bibinfo {volume}
  {99}}~(\bibinfo {number} {7}),\ \bibinfo {pages} {076007}},\ \Eprint
  {https://arxiv.org/abs/1901.04984} {arXiv:1901.04984 [cond-mat.str-el]}
  \BibitemShut {NoStop}%
\bibitem [{\citenamefont {Stewart}(2001)}]{Stewart}%
  \BibitemOpen
  \bibfield  {author} {\bibinfo {author} {\bibnamefont {Stewart}, \bibfnamefont
  {G~R}}} (\bibinfo {year} {2001}),\ \bibfield  {title} {\enquote {\bibinfo
  {title} {{Non-Fermi-liquid behavior in $d$- and $f$-electron metals}},}\
  }\href {https://doi.org/10.1103/RevModPhys.73.797} {\bibfield  {journal}
  {\bibinfo  {journal} {Rev. Mod. Phys.}\ }\textbf {\bibinfo {volume} {73}},\
  \bibinfo {pages} {797--855}}\BibitemShut {NoStop}%
\bibitem [{\citenamefont {Su}\ \emph {et~al.}(2020)\citenamefont {Su},
  \citenamefont {Zhang},\ and\ \citenamefont {Zhai}}]{Su:2020quk}%
  \BibitemOpen
  \bibfield  {author} {\bibinfo {author} {\bibnamefont {Su}, \bibfnamefont
  {Kaixiang}}, \bibinfo {author} {\bibfnamefont {Pengfei}\ \bibnamefont
  {Zhang}}, and\ \bibinfo {author} {\bibfnamefont {Hui}\ \bibnamefont {Zhai}}}
  (\bibinfo {year} {2020}),\ \bibfield  {title} {\enquote {\bibinfo {title}
  {{Page curve from non-Markovianity}},}\ }\href
  {https://doi.org/10.1007/JHEP06(2021)156} {\bibfield  {journal} {\bibinfo
  {journal} {JHEP}\ }\textbf {\bibinfo {volume} {21}},\ \bibinfo {pages}
  {156}},\ \Eprint {https://arxiv.org/abs/2101.11238} {arXiv:2101.11238
  [cond-mat.str-el]} \BibitemShut {NoStop}%
\bibitem [{\citenamefont {Swingle}\ and\ \citenamefont
  {Chowdhury}(2017)}]{BSDC17}%
  \BibitemOpen
  \bibfield  {author} {\bibinfo {author} {\bibnamefont {Swingle}, \bibfnamefont
  {Brian}}, and\ \bibinfo {author} {\bibfnamefont {Debanjan}\ \bibnamefont
  {Chowdhury}}} (\bibinfo {year} {2017}),\ \bibfield  {title} {\enquote
  {\bibinfo {title} {Slow scrambling in disordered quantum systems},}\ }\href
  {https://doi.org/10.1103/PhysRevB.95.060201} {\bibfield  {journal} {\bibinfo
  {journal} {Phys. Rev. B}\ }\textbf {\bibinfo {volume} {95}},\ \bibinfo
  {pages} {060201}}\BibitemShut {NoStop}%
\bibitem [{\citenamefont {Tarnopolsky}\ \emph {et~al.}(2020)\citenamefont
  {Tarnopolsky}, \citenamefont {Li}, \citenamefont {Joshi},\ and\ \citenamefont
  {Sachdev}}]{Tarnopolsky:2020spd}%
  \BibitemOpen
  \bibfield  {author} {\bibinfo {author} {\bibnamefont {Tarnopolsky},
  \bibfnamefont {Grigory}}, \bibinfo {author} {\bibfnamefont {Chenyuan}\
  \bibnamefont {Li}}, \bibinfo {author} {\bibfnamefont {Darshan~G.}\
  \bibnamefont {Joshi}}, and\ \bibinfo {author} {\bibfnamefont {Subir}\
  \bibnamefont {Sachdev}}} (\bibinfo {year} {2020}),\ \bibfield  {title}
  {\enquote {\bibinfo {title} {{Metal-insulator transition in a random Hubbard
  model}},}\ }\href {https://doi.org/10.1103/PhysRevB.101.205106} {\bibfield
  {journal} {\bibinfo  {journal} {Phys. Rev. B}\ }\textbf {\bibinfo {volume}
  {101}}~(\bibinfo {number} {20}),\ \bibinfo {pages} {205106}},\ \Eprint
  {https://arxiv.org/abs/2002.12381} {arXiv:2002.12381 [cond-mat.str-el]}
  \BibitemShut {NoStop}%
\bibitem [{\citenamefont {{Taupin}}\ and\ \citenamefont
  {{Paschen}}(2022)}]{Paschen22}%
  \BibitemOpen
  \bibfield  {author} {\bibinfo {author} {\bibnamefont {{Taupin}},
  \bibfnamefont {Mathieu}}, and\ \bibinfo {author} {\bibfnamefont {Silke}\
  \bibnamefont {{Paschen}}}} (\bibinfo {year} {2022}),\ \bibfield  {title}
  {\enquote {\bibinfo {title} {{Are Heavy Fermion Strange Metals Planckian?}}}\
  }\href@noop {} {\bibfield  {journal} {\bibinfo  {journal} {arXiv e-prints}\
  ,\ \bibinfo {eid} {arXiv:2201.02820}}}\Eprint
  {https://arxiv.org/abs/2201.02820} {arXiv:2201.02820 [cond-mat.str-el]}
  \BibitemShut {NoStop}%
\bibitem [{\citenamefont {Teitelboim}(1983)}]{Teitelboim83}%
  \BibitemOpen
  \bibfield  {author} {\bibinfo {author} {\bibnamefont {Teitelboim},
  \bibfnamefont {Claudio}}} (\bibinfo {year} {1983}),\ \bibfield  {title}
  {\enquote {\bibinfo {title} {{Gravitation and Hamiltonian structure in two
  spacetime dimensions}},}\ }\href
  {https://doi.org/https://doi.org/10.1016/0370-2693(83)90012-6} {\bibfield
  {journal} {\bibinfo  {journal} {Physics Letters B}\ }\textbf {\bibinfo
  {volume} {126}}~(\bibinfo {number} {1}),\ \bibinfo {pages}
  {41--45}}\BibitemShut {NoStop}%
\bibitem [{\citenamefont {{Theumann}}\ and\ \citenamefont
  {{Coqblin}}(2004)}]{Theumann04}%
  \BibitemOpen
  \bibfield  {author} {\bibinfo {author} {\bibnamefont {{Theumann}},
  \bibfnamefont {Alba}}, and\ \bibinfo {author} {\bibfnamefont
  {B.}~\bibnamefont {{Coqblin}}}} (\bibinfo {year} {2004}),\ \bibfield  {title}
  {\enquote {\bibinfo {title} {{Quantum critical point in the spin glass Kondo
  transition in heavy-fermion systems}},}\ }\href
  {https://doi.org/10.1103/PhysRevB.69.214418} {\bibfield  {journal} {\bibinfo
  {journal} {Phys. Rev. B}\ }\textbf {\bibinfo {volume} {69}}~(\bibinfo
  {number} {21}),\ \bibinfo {eid} {214418}},\ \Eprint
  {https://arxiv.org/abs/cond-mat/0404551} {arXiv:cond-mat/0404551
  [cond-mat.str-el]} \BibitemShut {NoStop}%
\bibitem [{\citenamefont {Tikhanovskaya}\ \emph
  {et~al.}(2021{\natexlab{a}})\citenamefont {Tikhanovskaya}, \citenamefont
  {Guo}, \citenamefont {Sachdev},\ and\ \citenamefont
  {Tarnopolsky}}]{Tikhanovskaya:2020elb}%
  \BibitemOpen
  \bibfield  {author} {\bibinfo {author} {\bibnamefont {Tikhanovskaya},
  \bibfnamefont {Maria}}, \bibinfo {author} {\bibfnamefont {Haoyu}\
  \bibnamefont {Guo}}, \bibinfo {author} {\bibfnamefont {Subir}\ \bibnamefont
  {Sachdev}}, and\ \bibinfo {author} {\bibfnamefont {Grigory}\ \bibnamefont
  {Tarnopolsky}}} (\bibinfo {year} {2021}{\natexlab{a}}),\ \bibfield  {title}
  {\enquote {\bibinfo {title} {{Excitation spectra of quantum matter without
  quasiparticles I: Sachdev-Ye-Kitaev models}},}\ }\href
  {https://doi.org/10.1103/PhysRevB.103.075141} {\bibfield  {journal} {\bibinfo
   {journal} {Phys. Rev. B}\ }\textbf {\bibinfo {volume} {103}}~(\bibinfo
  {number} {7}),\ \bibinfo {pages} {075141}},\ \Eprint
  {https://arxiv.org/abs/2010.09742} {arXiv:2010.09742 [cond-mat.str-el]}
  \BibitemShut {NoStop}%
\bibitem [{\citenamefont {Tikhanovskaya}\ \emph
  {et~al.}(2021{\natexlab{b}})\citenamefont {Tikhanovskaya}, \citenamefont
  {Guo}, \citenamefont {Sachdev},\ and\ \citenamefont
  {Tarnopolsky}}]{Tikhanovskaya:2020zcw}%
  \BibitemOpen
  \bibfield  {author} {\bibinfo {author} {\bibnamefont {Tikhanovskaya},
  \bibfnamefont {Maria}}, \bibinfo {author} {\bibfnamefont {Haoyu}\
  \bibnamefont {Guo}}, \bibinfo {author} {\bibfnamefont {Subir}\ \bibnamefont
  {Sachdev}}, and\ \bibinfo {author} {\bibfnamefont {Grigory}\ \bibnamefont
  {Tarnopolsky}}} (\bibinfo {year} {2021}{\natexlab{b}}),\ \bibfield  {title}
  {\enquote {\bibinfo {title} {{Excitation spectra of quantum matter without
  quasiparticles II: random $t$-$J$ models}},}\ }\href
  {https://doi.org/10.1103/PhysRevB.103.075142} {\bibfield  {journal} {\bibinfo
   {journal} {Phys. Rev. B}\ }\textbf {\bibinfo {volume} {103}}~(\bibinfo
  {number} {7}),\ \bibinfo {pages} {075142}},\ \Eprint
  {https://arxiv.org/abs/2012.14449} {arXiv:2012.14449 [cond-mat.str-el]}
  \BibitemShut {NoStop}%
\bibitem [{\citenamefont {Tikhanovskaya}\ \emph {et~al.}(2022)\citenamefont
  {Tikhanovskaya}, \citenamefont {Sachdev},\ and\ \citenamefont
  {Patel}}]{Tikhanovskaya:2022zqq}%
  \BibitemOpen
  \bibfield  {author} {\bibinfo {author} {\bibnamefont {Tikhanovskaya},
  \bibfnamefont {Maria}}, \bibinfo {author} {\bibfnamefont {Subir}\
  \bibnamefont {Sachdev}}, and\ \bibinfo {author} {\bibfnamefont
  {Aavishkar~A.}\ \bibnamefont {Patel}}} (\bibinfo {year} {2022}),\ \bibfield
  {title} {\enquote {\bibinfo {title} {{Maximal quantum chaos of the critical
  Fermi surface}},}\ }\href@noop {} {\ }\Eprint
  {https://arxiv.org/abs/2202.01845} {arXiv:2202.01845 [cond-mat.str-el]}
  \BibitemShut {NoStop}%
\bibitem [{\citenamefont {Tsuji}\ and\ \citenamefont
  {Werner}(2019)}]{Tsuji_2019}%
  \BibitemOpen
  \bibfield  {author} {\bibinfo {author} {\bibnamefont {Tsuji}, \bibfnamefont
  {Naoto}}, and\ \bibinfo {author} {\bibfnamefont {Philipp}\ \bibnamefont
  {Werner}}} (\bibinfo {year} {2019}),\ \bibfield  {title} {\enquote {\bibinfo
  {title} {{Out-of-time-ordered correlators of the Hubbard model:
  Sachdev-Ye-Kitaev strange metal in the spin-freezing crossover region}},}\
  }\href {https://doi.org/10.1103/PhysRevB.99.115132} {\bibfield  {journal}
  {\bibinfo  {journal} {Phys. Rev. B}\ }\textbf {\bibinfo {volume} {99}},\
  \bibinfo {pages} {115132}}\BibitemShut {NoStop}%
\bibitem [{\citenamefont {Tulipman}\ and\ \citenamefont {Berg}(2021)}]{EB21}%
  \BibitemOpen
  \bibfield  {author} {\bibinfo {author} {\bibnamefont {Tulipman},
  \bibfnamefont {Evyatar}}, and\ \bibinfo {author} {\bibfnamefont {Erez}\
  \bibnamefont {Berg}}} (\bibinfo {year} {2021}),\ \bibfield  {title} {\enquote
  {\bibinfo {title} {{Strongly coupled phonon fluid and Goldstone modes in an
  anharmonic quantum solid: Transport and chaos}},}\ }\href
  {https://doi.org/10.1103/PhysRevB.104.195113} {\bibfield  {journal} {\bibinfo
   {journal} {Phys. Rev. B}\ }\textbf {\bibinfo {volume} {104}}~(\bibinfo
  {number} {19}),\ \bibinfo {pages} {195113}},\ \Eprint
  {https://arxiv.org/abs/2108.01107} {arXiv:2108.01107 [cond-mat.str-el]}
  \BibitemShut {NoStop}%
\bibitem [{\citenamefont {Tyler}\ \emph {et~al.}(1998)\citenamefont {Tyler},
  \citenamefont {Mackenzie}, \citenamefont {NishiZaki},\ and\ \citenamefont
  {Maeno}}]{Tyler1998}%
  \BibitemOpen
  \bibfield  {author} {\bibinfo {author} {\bibnamefont {Tyler}, \bibfnamefont
  {A~W}}, \bibinfo {author} {\bibfnamefont {A.~P.}\ \bibnamefont {Mackenzie}},
  \bibinfo {author} {\bibfnamefont {S.}~\bibnamefont {NishiZaki}}, and\
  \bibinfo {author} {\bibfnamefont {Y.}~\bibnamefont {Maeno}}} (\bibinfo {year}
  {1998}),\ \bibfield  {title} {\enquote {\bibinfo {title} {{High-temperature
  resistivity of ${\mathrm{Sr}}_{2}{\mathrm{RuO}}_{4}:$ Bad metallic transport
  in a good metal}},}\ }\href {https://doi.org/10.1103/PhysRevB.58.R10107}
  {\bibfield  {journal} {\bibinfo  {journal} {Phys. Rev. B}\ }\textbf {\bibinfo
  {volume} {58}},\ \bibinfo {pages} {R10107--R10110}}\BibitemShut {NoStop}%
\bibitem [{\citenamefont {Valla}\ \emph {et~al.}(1999)\citenamefont {Valla},
  \citenamefont {Fedorov}, \citenamefont {Johnson}, \citenamefont {Wells},
  \citenamefont {Hulbert}, \citenamefont {Li}, \citenamefont {Gu},\ and\
  \citenamefont {Koshizuka}}]{Valla99}%
  \BibitemOpen
  \bibfield  {author} {\bibinfo {author} {\bibnamefont {Valla}, \bibfnamefont
  {T}}, \bibinfo {author} {\bibfnamefont {A.~V.}\ \bibnamefont {Fedorov}},
  \bibinfo {author} {\bibfnamefont {P.~D.}\ \bibnamefont {Johnson}}, \bibinfo
  {author} {\bibfnamefont {B.~O.}\ \bibnamefont {Wells}}, \bibinfo {author}
  {\bibfnamefont {S.~L.}\ \bibnamefont {Hulbert}}, \bibinfo {author}
  {\bibfnamefont {Q.}~\bibnamefont {Li}}, \bibinfo {author} {\bibfnamefont
  {G.~D.}\ \bibnamefont {Gu}}, and\ \bibinfo {author} {\bibfnamefont
  {N.}~\bibnamefont {Koshizuka}}} (\bibinfo {year} {1999}),\ \bibfield  {title}
  {\enquote {\bibinfo {title} {{Evidence for Quantum Critical Behavior in the
  Optimally Doped Cuprate Bi$_2$Sr$_2$CaCu$_2$O$_{8+\delta}$}},}\ }\href
  {https://doi.org/10.1126/science.285.5436.2110} {\bibfield  {journal}
  {\bibinfo  {journal} {Science}\ }\textbf {\bibinfo {volume} {285}}~(\bibinfo
  {number} {5436}),\ \bibinfo {pages} {2110--2113}}\BibitemShut {NoStop}%
\bibitem [{\citenamefont {{van der Marel}}\ \emph {et~al.}(2006)\citenamefont
  {{van der Marel}}, \citenamefont {{Carbone}}, \citenamefont {{Kuzmenko}},\
  and\ \citenamefont {{Giannini}}}]{vdM_2006}%
  \BibitemOpen
  \bibfield  {author} {\bibinfo {author} {\bibnamefont {{van der Marel}},
  \bibfnamefont {D}}, \bibinfo {author} {\bibfnamefont {F.}~\bibnamefont
  {{Carbone}}}, \bibinfo {author} {\bibfnamefont {A.~B.}\ \bibnamefont
  {{Kuzmenko}}}, and\ \bibinfo {author} {\bibfnamefont {E.}~\bibnamefont
  {{Giannini}}}} (\bibinfo {year} {2006}),\ \bibfield  {title} {\enquote
  {\bibinfo {title} {{Scaling properties of the optical conductivity of
  Bi-based cuprates}},}\ }\href {https://doi.org/10.1016/j.aop.2006.04.012}
  {\bibfield  {journal} {\bibinfo  {journal} {Annals of Physics}\ }\textbf
  {\bibinfo {volume} {321}}~(\bibinfo {number} {7}),\ \bibinfo {pages}
  {1716--1729}},\ \Eprint {https://arxiv.org/abs/cond-mat/0604037}
  {arXiv:cond-mat/0604037 [cond-mat.str-el]} \BibitemShut {NoStop}%
\bibitem [{\citenamefont {{van Heumen}}\ \emph {et~al.}(2022)\citenamefont
  {{van Heumen}}, \citenamefont {{Feng}}, \citenamefont {{Cassanelli}},
  \citenamefont {{Neubrand}}, \citenamefont {{de Jager}}, \citenamefont
  {{Berben}}, \citenamefont {{Huang}}, \citenamefont {{Kondo}}, \citenamefont
  {{Takeuchi}},\ and\ \citenamefont {{Zaanen}}}]{Heumen22}%
  \BibitemOpen
  \bibfield  {author} {\bibinfo {author} {\bibnamefont {{van Heumen}},
  \bibfnamefont {Erik}}, \bibinfo {author} {\bibfnamefont {Xuanbo}\
  \bibnamefont {{Feng}}}, \bibinfo {author} {\bibfnamefont {Silvia}\
  \bibnamefont {{Cassanelli}}}, \bibinfo {author} {\bibfnamefont {Linda}\
  \bibnamefont {{Neubrand}}}, \bibinfo {author} {\bibfnamefont {Lennart}\
  \bibnamefont {{de Jager}}}, \bibinfo {author} {\bibfnamefont {Maarten}\
  \bibnamefont {{Berben}}}, \bibinfo {author} {\bibfnamefont {Yingkai}\
  \bibnamefont {{Huang}}}, \bibinfo {author} {\bibfnamefont {Takeshi}\
  \bibnamefont {{Kondo}}}, \bibinfo {author} {\bibfnamefont {Tsunehiro}\
  \bibnamefont {{Takeuchi}}}, and\ \bibinfo {author} {\bibfnamefont {Jan}\
  \bibnamefont {{Zaanen}}}} (\bibinfo {year} {2022}),\ \bibfield  {title}
  {\enquote {\bibinfo {title} {{Strange metal dynamics across the phase diagram
  of Bi$_{2}$Sr$_{2}$CuO$_{6+\delta}$ cuprates}},}\ }\href@noop {} {\ }\Eprint
  {https://arxiv.org/abs/2205.00899} {arXiv:2205.00899 [cond-mat.str-el]}
  \BibitemShut {NoStop}%
\bibitem [{\citenamefont {Varma}\ \emph {et~al.}(1989)\citenamefont {Varma},
  \citenamefont {Littlewood}, \citenamefont {Schmitt-Rink}, \citenamefont
  {Abrahams},\ and\ \citenamefont {Ruckenstein}}]{Varma89}%
  \BibitemOpen
  \bibfield  {author} {\bibinfo {author} {\bibnamefont {Varma}, \bibfnamefont
  {C~M}}, \bibinfo {author} {\bibfnamefont {P.~B.}\ \bibnamefont {Littlewood}},
  \bibinfo {author} {\bibfnamefont {S.}~\bibnamefont {Schmitt-Rink}}, \bibinfo
  {author} {\bibfnamefont {E.}~\bibnamefont {Abrahams}}, and\ \bibinfo {author}
  {\bibfnamefont {A.~E.}\ \bibnamefont {Ruckenstein}}} (\bibinfo {year}
  {1989}),\ \bibfield  {title} {\enquote {\bibinfo {title} {{Phenomenology of
  the normal state of Cu-O high-temperature superconductors}},}\ }\href
  {https://doi.org/10.1103/PhysRevLett.63.1996} {\bibfield  {journal} {\bibinfo
   {journal} {Phys. Rev. Lett.}\ }\textbf {\bibinfo {volume} {63}},\ \bibinfo
  {pages} {1996--1999}}\BibitemShut {NoStop}%
\bibitem [{\citenamefont {Varma}(2016)}]{Varma2016}%
  \BibitemOpen
  \bibfield  {author} {\bibinfo {author} {\bibnamefont {Varma}, \bibfnamefont
  {Chandra~M}}} (\bibinfo {year} {2016}),\ \bibfield  {title} {\enquote
  {\bibinfo {title} {Quantum-critical fluctuations in 2d metals: strange metals
  and superconductivity in antiferromagnets and in cuprates},}\ }\href
  {https://doi.org/10.1088/0034-4885/79/8/082501} {\bibfield  {journal}
  {\bibinfo  {journal} {Reports on Progress in Physics}\ }\textbf {\bibinfo
  {volume} {79}}~(\bibinfo {number} {8}),\ \bibinfo {pages}
  {082501}}\BibitemShut {NoStop}%
\bibitem [{\citenamefont {Varma}(2020)}]{Varma2020}%
  \BibitemOpen
  \bibfield  {author} {\bibinfo {author} {\bibnamefont {Varma}, \bibfnamefont
  {Chandra~M}}} (\bibinfo {year} {2020}),\ \bibfield  {title} {\enquote
  {\bibinfo {title} {{Colloquium: Linear in temperature resistivity and
  associated mysteries including high temperature superconductivity}},}\ }\href
  {https://doi.org/10.1103/RevModPhys.92.031001} {\bibfield  {journal}
  {\bibinfo  {journal} {Rev. Mod. Phys.}\ }\textbf {\bibinfo {volume} {92}},\
  \bibinfo {pages} {031001}}\BibitemShut {NoStop}%
\bibitem [{\citenamefont {Ma\ifmmode~\check{c}\else \v{c}\fi{}ek}\ \emph
  {et~al.}(2020)\citenamefont {Ma\ifmmode~\check{c}\else \v{c}\fi{}ek},
  \citenamefont {Dumitrescu}, \citenamefont {Bertrand}, \citenamefont {Triggs},
  \citenamefont {Parcollet},\ and\ \citenamefont {Waintal}}]{QQMCPRL}%
  \BibitemOpen
  \bibfield  {author} {\bibinfo {author} {\bibnamefont
  {Ma\ifmmode~\check{c}\else \v{c}\fi{}ek}, \bibfnamefont {Marjan}}, \bibinfo
  {author} {\bibfnamefont {Philipp~T.}\ \bibnamefont {Dumitrescu}}, \bibinfo
  {author} {\bibfnamefont {Corentin}\ \bibnamefont {Bertrand}}, \bibinfo
  {author} {\bibfnamefont {Bill}\ \bibnamefont {Triggs}}, \bibinfo {author}
  {\bibfnamefont {Olivier}\ \bibnamefont {Parcollet}}, and\ \bibinfo {author}
  {\bibfnamefont {Xavier}\ \bibnamefont {Waintal}}} (\bibinfo {year} {2020}),\
  \bibfield  {title} {\enquote {\bibinfo {title} {{Quantum Quasi-Monte Carlo
  Technique for Many-Body Perturbative Expansions}},}\ }\href
  {https://doi.org/10.1103/PhysRevLett.125.047702} {\bibfield  {journal}
  {\bibinfo  {journal} {Phys. Rev. Lett.}\ }\textbf {\bibinfo {volume} {125}},\
  \bibinfo {pages} {047702}}\BibitemShut {NoStop}%
\bibitem [{\citenamefont {Vishveshwara}(1970)}]{Vishveshwara}%
  \BibitemOpen
  \bibfield  {author} {\bibinfo {author} {\bibnamefont {Vishveshwara},
  \bibfnamefont {C~V}}} (\bibinfo {year} {1970}),\ \bibfield  {title} {\enquote
  {\bibinfo {title} {{Scattering of Gravitational Radiation by a Schwarzschild
  Black-hole}},}\ }\href {https://doi.org/10.1038/227936a0} {\bibfield
  {journal} {\bibinfo  {journal} {Nature}\ }\textbf {\bibinfo {volume}
  {227}}~(\bibinfo {number} {5261}),\ \bibinfo {pages} {936}}\BibitemShut
  {NoStop}%
\bibitem [{\citenamefont {Vojta}\ \emph {et~al.}(2000)\citenamefont {Vojta},
  \citenamefont {Buragohain},\ and\ \citenamefont {Sachdev}}]{Vojta2000}%
  \BibitemOpen
  \bibfield  {author} {\bibinfo {author} {\bibnamefont {Vojta}, \bibfnamefont
  {Matthias}}, \bibinfo {author} {\bibfnamefont {Chiranjeeb}\ \bibnamefont
  {Buragohain}}, and\ \bibinfo {author} {\bibfnamefont {Subir}\ \bibnamefont
  {Sachdev}}} (\bibinfo {year} {2000}),\ \bibfield  {title} {\enquote {\bibinfo
  {title} {Quantum impurity dynamics in two-dimensional antiferromagnets and
  superconductors},}\ }\href {https://doi.org/10.1103/PhysRevB.61.15152}
  {\bibfield  {journal} {\bibinfo  {journal} {Phys. Rev. B}\ }\textbf {\bibinfo
  {volume} {61}},\ \bibinfo {pages} {15152--15184}}\BibitemShut {NoStop}%
\bibitem [{\citenamefont {{Vojta}}\ and\ \citenamefont
  {{Fritz}}(2004)}]{VojtaFritz2004}%
  \BibitemOpen
  \bibfield  {author} {\bibinfo {author} {\bibnamefont {{Vojta}}, \bibfnamefont
  {Matthias}}, and\ \bibinfo {author} {\bibfnamefont {Lars}\ \bibnamefont
  {{Fritz}}}} (\bibinfo {year} {2004}),\ \bibfield  {title} {\enquote {\bibinfo
  {title} {{Upper critical dimension in a quantum impurity model: Critical
  theory of the asymmetric pseudogap Kondo problem}},}\ }\href
  {https://doi.org/10.1103/PhysRevB.70.094502} {\bibfield  {journal} {\bibinfo
  {journal} {Phys. Rev. B}\ }\textbf {\bibinfo {volume} {70}}~(\bibinfo
  {number} {9}),\ \bibinfo {eid} {094502}},\ \Eprint
  {https://arxiv.org/abs/cond-mat/0309262} {arXiv:cond-mat/0309262
  [cond-mat.str-el]} \BibitemShut {NoStop}%
\bibitem [{\citenamefont {Vollmer}\ \emph {et~al.}(2000)\citenamefont
  {Vollmer}, \citenamefont {Pietrus}, \citenamefont {L\"ohneysen},
  \citenamefont {Chau},\ and\ \citenamefont {Maple}}]{Maple00}%
  \BibitemOpen
  \bibfield  {author} {\bibinfo {author} {\bibnamefont {Vollmer}, \bibfnamefont
  {R}}, \bibinfo {author} {\bibfnamefont {T.}~\bibnamefont {Pietrus}}, \bibinfo
  {author} {\bibfnamefont {H.~v.}\ \bibnamefont {L\"ohneysen}}, \bibinfo
  {author} {\bibfnamefont {R.}~\bibnamefont {Chau}}, and\ \bibinfo {author}
  {\bibfnamefont {M.~B.}\ \bibnamefont {Maple}}} (\bibinfo {year} {2000}),\
  \bibfield  {title} {\enquote {\bibinfo {title} {{Phase transitions and
  non-Fermi-liquid behavior in
  ${\mathrm{UCu}}_{5\ensuremath{-}x}{\mathrm{Pd}}_{x}$ at low temperatures}},}\
  }\href {https://doi.org/10.1103/PhysRevB.61.1218} {\bibfield  {journal}
  {\bibinfo  {journal} {Phys. Rev. B}\ }\textbf {\bibinfo {volume} {61}},\
  \bibinfo {pages} {1218--1222}}\BibitemShut {NoStop}%
\bibitem [{\citenamefont {Wang}\ \emph {et~al.}(2020)\citenamefont {Wang},
  \citenamefont {Chudnovskiy}, \citenamefont {Gorsky},\ and\ \citenamefont
  {Kamenev}}]{kamenevSC}%
  \BibitemOpen
  \bibfield  {author} {\bibinfo {author} {\bibnamefont {Wang}, \bibfnamefont
  {Hanteng}}, \bibinfo {author} {\bibfnamefont {A.~L.}\ \bibnamefont
  {Chudnovskiy}}, \bibinfo {author} {\bibfnamefont {Alexander}\ \bibnamefont
  {Gorsky}}, and\ \bibinfo {author} {\bibfnamefont {Alex}\ \bibnamefont
  {Kamenev}}} (\bibinfo {year} {2020}),\ \bibfield  {title} {\enquote {\bibinfo
  {title} {{Sachdev-Ye-Kitaev superconductivity: Quantum Kuramoto and
  generalized Richardson models}},}\ }\href
  {https://doi.org/10.1103/PhysRevResearch.2.033025} {\bibfield  {journal}
  {\bibinfo  {journal} {Phys. Rev. Research}\ }\textbf {\bibinfo {volume}
  {2}},\ \bibinfo {pages} {033025}}\BibitemShut {NoStop}%
\bibitem [{\citenamefont {Wang}\ \emph {et~al.}(2004)\citenamefont {Wang},
  \citenamefont {Yang}, \citenamefont {Sekharan}, \citenamefont {Ding},
  \citenamefont {Engelbrecht}, \citenamefont {Dai}, \citenamefont {Wang},
  \citenamefont {Kaminski}, \citenamefont {Valla}, \citenamefont {Kidd},
  \citenamefont {Fedorov},\ and\ \citenamefont {Johnson}}]{johnson}%
  \BibitemOpen
  \bibfield  {author} {\bibinfo {author} {\bibnamefont {Wang}, \bibfnamefont
  {S-C}}, \bibinfo {author} {\bibfnamefont {H.-B.}\ \bibnamefont {Yang}},
  \bibinfo {author} {\bibfnamefont {A.~K.~P.}\ \bibnamefont {Sekharan}},
  \bibinfo {author} {\bibfnamefont {H.}~\bibnamefont {Ding}}, \bibinfo {author}
  {\bibfnamefont {J.~R.}\ \bibnamefont {Engelbrecht}}, \bibinfo {author}
  {\bibfnamefont {X.}~\bibnamefont {Dai}}, \bibinfo {author} {\bibfnamefont
  {Z.}~\bibnamefont {Wang}}, \bibinfo {author} {\bibfnamefont {A.}~\bibnamefont
  {Kaminski}}, \bibinfo {author} {\bibfnamefont {T.}~\bibnamefont {Valla}},
  \bibinfo {author} {\bibfnamefont {T.}~\bibnamefont {Kidd}}, \bibinfo {author}
  {\bibfnamefont {A.~V.}\ \bibnamefont {Fedorov}}, and\ \bibinfo {author}
  {\bibfnamefont {P.~D.}\ \bibnamefont {Johnson}}} (\bibinfo {year} {2004}),\
  \bibfield  {title} {\enquote {\bibinfo {title} {Quasiparticle line shape of
  ${\mathrm{sr}}_{2}{\mathrm{ruo}}_{4}$ and its relation to anisotropic
  transport},}\ }\href {https://doi.org/10.1103/PhysRevLett.92.137002}
  {\bibfield  {journal} {\bibinfo  {journal} {Phys. Rev. Lett.}\ }\textbf
  {\bibinfo {volume} {92}},\ \bibinfo {pages} {137002}}\BibitemShut {NoStop}%
\bibitem [{\citenamefont {{Wang}}\ \emph {et~al.}(2021)\citenamefont {{Wang}},
  \citenamefont {{Davis}}, \citenamefont {{Pan}}, \citenamefont {{Wang}},\ and\
  \citenamefont {{Meng}}}]{WangMeng21}%
  \BibitemOpen
  \bibfield  {author} {\bibinfo {author} {\bibnamefont {{Wang}}, \bibfnamefont
  {Wei}}, \bibinfo {author} {\bibfnamefont {Andrew}\ \bibnamefont {{Davis}}},
  \bibinfo {author} {\bibfnamefont {Gaopei}\ \bibnamefont {{Pan}}}, \bibinfo
  {author} {\bibfnamefont {Yuxuan}\ \bibnamefont {{Wang}}}, and\ \bibinfo
  {author} {\bibfnamefont {Zi~Yang}\ \bibnamefont {{Meng}}}} (\bibinfo {year}
  {2021}),\ \bibfield  {title} {\enquote {\bibinfo {title} {{Phase diagram of
  the spin-1/2 Yukawa-Sachdev-Ye-Kitaev model: Non-Fermi liquid, insulator, and
  superconductor}},}\ }\href {https://doi.org/10.1103/PhysRevB.103.195108}
  {\bibfield  {journal} {\bibinfo  {journal} {Phys. Rev. B}\ }\textbf {\bibinfo
  {volume} {103}}~(\bibinfo {number} {19}),\ \bibinfo {eid} {195108}},\ \Eprint
  {https://arxiv.org/abs/2102.10755} {arXiv:2102.10755 [cond-mat.str-el]}
  \BibitemShut {NoStop}%
\bibitem [{\citenamefont {{Wang}}\ and\ \citenamefont {{Berg}}(2019)}]{Berg19}%
  \BibitemOpen
  \bibfield  {author} {\bibinfo {author} {\bibnamefont {{Wang}}, \bibfnamefont
  {Xiaoyu}}, and\ \bibinfo {author} {\bibfnamefont {Erez}\ \bibnamefont
  {{Berg}}}} (\bibinfo {year} {2019}),\ \bibfield  {title} {\enquote {\bibinfo
  {title} {{Scattering mechanisms and electrical transport near an Ising
  nematic quantum critical point}},}\ }\href
  {https://doi.org/10.1103/PhysRevB.99.235136} {\bibfield  {journal} {\bibinfo
  {journal} {Phys. Rev. B}\ }\textbf {\bibinfo {volume} {99}}~(\bibinfo
  {number} {23}),\ \bibinfo {eid} {235136}},\ \Eprint
  {https://arxiv.org/abs/1902.04590} {arXiv:1902.04590 [cond-mat.str-el]}
  \BibitemShut {NoStop}%
\bibitem [{\citenamefont {Wang}(2020{\natexlab{a}})}]{YW}%
  \BibitemOpen
  \bibfield  {author} {\bibinfo {author} {\bibnamefont {Wang}, \bibfnamefont
  {Yuxuan}}} (\bibinfo {year} {2020}{\natexlab{a}}),\ \bibfield  {title}
  {\enquote {\bibinfo {title} {{Solvable Strong-Coupling Quantum-Dot Model with
  a Non-Fermi-Liquid Pairing Transition}},}\ }\href
  {https://doi.org/10.1103/PhysRevLett.124.017002} {\bibfield  {journal}
  {\bibinfo  {journal} {Phys. Rev. Lett.}\ }\textbf {\bibinfo {volume} {124}},\
  \bibinfo {pages} {017002}}\BibitemShut {NoStop}%
\bibitem [{\citenamefont {Wang}(2020{\natexlab{b}})}]{Wang:2019bpd}%
  \BibitemOpen
  \bibfield  {author} {\bibinfo {author} {\bibnamefont {Wang}, \bibfnamefont
  {Yuxuan}}} (\bibinfo {year} {2020}{\natexlab{b}}),\ \bibfield  {title}
  {\enquote {\bibinfo {title} {{Solvable Strong-coupling Quantum Dot Model with
  a Non-Fermi-liquid Pairing Transition}},}\ }\href
  {https://doi.org/10.1103/PhysRevLett.124.017002} {\bibfield  {journal}
  {\bibinfo  {journal} {Phys. Rev. Lett.}\ }\textbf {\bibinfo {volume}
  {124}}~(\bibinfo {number} {1}),\ \bibinfo {pages} {017002}},\ \Eprint
  {https://arxiv.org/abs/1904.07240} {arXiv:1904.07240 [cond-mat.str-el]}
  \BibitemShut {NoStop}%
\bibitem [{\citenamefont {Wang}\ \emph {et~al.}(2016)\citenamefont {Wang},
  \citenamefont {Abanov}, \citenamefont {Altshuler}, \citenamefont
  {Yuzbashyan},\ and\ \citenamefont {Chubukov}}]{Chubukov16}%
  \BibitemOpen
  \bibfield  {author} {\bibinfo {author} {\bibnamefont {Wang}, \bibfnamefont
  {Yuxuan}}, \bibinfo {author} {\bibfnamefont {Artem}\ \bibnamefont {Abanov}},
  \bibinfo {author} {\bibfnamefont {Boris~L.}\ \bibnamefont {Altshuler}},
  \bibinfo {author} {\bibfnamefont {Emil~A.}\ \bibnamefont {Yuzbashyan}}, and\
  \bibinfo {author} {\bibfnamefont {Andrey~V.}\ \bibnamefont {Chubukov}}}
  (\bibinfo {year} {2016}),\ \bibfield  {title} {\enquote {\bibinfo {title}
  {Superconductivity near a quantum-critical point: The special role of the
  first matsubara frequency},}\ }\href
  {https://doi.org/10.1103/PhysRevLett.117.157001} {\bibfield  {journal}
  {\bibinfo  {journal} {Phys. Rev. Lett.}\ }\textbf {\bibinfo {volume} {117}},\
  \bibinfo {pages} {157001}}\BibitemShut {NoStop}%
\bibitem [{\citenamefont {Wang}\ and\ \citenamefont
  {Chubukov}(2020)}]{Wang:2020dtj}%
  \BibitemOpen
  \bibfield  {author} {\bibinfo {author} {\bibnamefont {Wang}, \bibfnamefont
  {Yuxuan}}, and\ \bibinfo {author} {\bibfnamefont {Andrey~V.}\ \bibnamefont
  {Chubukov}}} (\bibinfo {year} {2020}),\ \bibfield  {title} {\enquote
  {\bibinfo {title} {{Quantum Phase Transition in the Yukawa-SYK Model}},}\
  }\href {https://doi.org/10.1103/PhysRevResearch.2.033084} {\bibfield
  {journal} {\bibinfo  {journal} {Phys. Rev. Res.}\ }\textbf {\bibinfo {volume}
  {2}}~(\bibinfo {number} {3}),\ \bibinfo {pages} {033084}},\ \Eprint
  {https://arxiv.org/abs/2005.07205} {arXiv:2005.07205 [cond-mat.str-el]}
  \BibitemShut {NoStop}%
\bibitem [{\citenamefont {Weber}\ and\ \citenamefont
  {Vojta}(2022)}]{Weber:2022ada}%
  \BibitemOpen
  \bibfield  {author} {\bibinfo {author} {\bibnamefont {Weber}, \bibfnamefont
  {Manuel}}, and\ \bibinfo {author} {\bibfnamefont {Matthias}\ \bibnamefont
  {Vojta}}} (\bibinfo {year} {2022}),\ \bibfield  {title} {\enquote {\bibinfo
  {title} {{SU(2)-symmetric spin-boson model: Quantum criticality, fixed-point
  annihilation, and duality}},}\ }\href@noop {} {\ }\Eprint
  {https://arxiv.org/abs/2203.02518} {arXiv:2203.02518 [cond-mat.str-el]}
  \BibitemShut {NoStop}%
\bibitem [{\citenamefont {Wei}\ and\ \citenamefont
  {Sedrakyan}(2021)}]{wei_2021}%
  \BibitemOpen
  \bibfield  {author} {\bibinfo {author} {\bibnamefont {Wei}, \bibfnamefont
  {Chenan}}, and\ \bibinfo {author} {\bibfnamefont {Tigran~A.}\ \bibnamefont
  {Sedrakyan}}} (\bibinfo {year} {2021}),\ \bibfield  {title} {\enquote
  {\bibinfo {title} {Optical lattice platform for the sachdev-ye-kitaev
  model},}\ }\href {https://doi.org/10.1103/PhysRevA.103.013323} {\bibfield
  {journal} {\bibinfo  {journal} {Phys. Rev. A}\ }\textbf {\bibinfo {volume}
  {103}},\ \bibinfo {pages} {013323}}\BibitemShut {NoStop}%
\bibitem [{\citenamefont {Wen}(2017)}]{XGW17}%
  \BibitemOpen
  \bibfield  {author} {\bibinfo {author} {\bibnamefont {Wen}, \bibfnamefont
  {Xiao-Gang}}} (\bibinfo {year} {2017}),\ \bibfield  {title} {\enquote
  {\bibinfo {title} {Colloquium: Zoo of quantum-topological phases of
  matter},}\ }\href {https://doi.org/10.1103/RevModPhys.89.041004} {\bibfield
  {journal} {\bibinfo  {journal} {Rev. Mod. Phys.}\ }\textbf {\bibinfo {volume}
  {89}},\ \bibinfo {pages} {041004}}\BibitemShut {NoStop}%
\bibitem [{\citenamefont {Werman}\ and\ \citenamefont {Berg}(2016)}]{Werman}%
  \BibitemOpen
  \bibfield  {author} {\bibinfo {author} {\bibnamefont {Werman}, \bibfnamefont
  {Yochai}}, and\ \bibinfo {author} {\bibfnamefont {Erez}\ \bibnamefont
  {Berg}}} (\bibinfo {year} {2016}),\ \bibfield  {title} {\enquote {\bibinfo
  {title} {{Mott-Ioffe-Regel limit and resistivity crossover in a tractable
  electron-phonon model}},}\ }\href
  {https://doi.org/10.1103/PhysRevB.93.075109} {\bibfield  {journal} {\bibinfo
  {journal} {Phys. Rev. B}\ }\textbf {\bibinfo {volume} {93}},\ \bibinfo
  {pages} {075109}}\BibitemShut {NoStop}%
\bibitem [{\citenamefont {Werman}\ \emph {et~al.}(2017)\citenamefont {Werman},
  \citenamefont {Kivelson},\ and\ \citenamefont {Berg}}]{Werman2}%
  \BibitemOpen
  \bibfield  {author} {\bibinfo {author} {\bibnamefont {Werman}, \bibfnamefont
  {Yochai}}, \bibinfo {author} {\bibfnamefont {Steven~A.}\ \bibnamefont
  {Kivelson}}, and\ \bibinfo {author} {\bibfnamefont {Erez}\ \bibnamefont
  {Berg}}} (\bibinfo {year} {2017}),\ \bibfield  {title} {\enquote {\bibinfo
  {title} {{Non-quasiparticle transport and resistivity saturation: a view from
  the large-$N$ limit}},}\ }\href {https://doi.org/10.1038/s41535-017-0009-8}
  {\bibfield  {journal} {\bibinfo  {journal} {npj Quantum Materials}\ }\textbf
  {\bibinfo {volume} {2}}~(\bibinfo {number} {1}),\ \bibinfo {pages}
  {7}}\BibitemShut {NoStop}%
\bibitem [{\citenamefont {Werner}\ \emph {et~al.}(2006)\citenamefont {Werner},
  \citenamefont {Comanac}, \citenamefont {de' Medici}, \citenamefont {Troyer},\
  and\ \citenamefont {Millis}}]{cthyb2006}%
  \BibitemOpen
  \bibfield  {author} {\bibinfo {author} {\bibnamefont {Werner}, \bibfnamefont
  {Philipp}}, \bibinfo {author} {\bibfnamefont {Armin}\ \bibnamefont
  {Comanac}}, \bibinfo {author} {\bibfnamefont {Luca}\ \bibnamefont {de'
  Medici}}, \bibinfo {author} {\bibfnamefont {Matthias}\ \bibnamefont
  {Troyer}}, and\ \bibinfo {author} {\bibfnamefont {Andrew~J.}\ \bibnamefont
  {Millis}}} (\bibinfo {year} {2006}),\ \bibfield  {title} {\enquote {\bibinfo
  {title} {{Continuous-Time Solver for Quantum Impurity Models}},}\ }\href
  {https://doi.org/10.1103/PhysRevLett.97.076405} {\bibfield  {journal}
  {\bibinfo  {journal} {Phys. Rev. Lett.}\ }\textbf {\bibinfo {volume} {97}},\
  \bibinfo {pages} {076405}}\BibitemShut {NoStop}%
\bibitem [{\citenamefont {Werner}\ \emph {et~al.}(2018)\citenamefont {Werner},
  \citenamefont {Kim},\ and\ \citenamefont {Hoshino}}]{Werner_2018}%
  \BibitemOpen
  \bibfield  {author} {\bibinfo {author} {\bibnamefont {Werner}, \bibfnamefont
  {Philipp}}, \bibinfo {author} {\bibfnamefont {Aaram~J.}\ \bibnamefont {Kim}},
  and\ \bibinfo {author} {\bibfnamefont {Shintaro}\ \bibnamefont {Hoshino}}}
  (\bibinfo {year} {2018}),\ \bibfield  {title} {\enquote {\bibinfo {title}
  {{Spin-freezing and the Sachdev-Ye model}},}\ }\href
  {https://doi.org/10.1209/0295-5075/124/57002} {\bibfield  {journal} {\bibinfo
   {journal} {Europhysics Letters}\ }\textbf {\bibinfo {volume}
  {124}}~(\bibinfo {number} {5}),\ \bibinfo {pages} {57002}}\BibitemShut
  {NoStop}%
\bibitem [{\citenamefont {Winer}\ \emph {et~al.}(2020)\citenamefont {Winer},
  \citenamefont {Jian},\ and\ \citenamefont {Swingle}}]{Winer20}%
  \BibitemOpen
  \bibfield  {author} {\bibinfo {author} {\bibnamefont {Winer}, \bibfnamefont
  {Michael}}, \bibinfo {author} {\bibfnamefont {Shao-Kai}\ \bibnamefont
  {Jian}}, and\ \bibinfo {author} {\bibfnamefont {Brian}\ \bibnamefont
  {Swingle}}} (\bibinfo {year} {2020}),\ \bibfield  {title} {\enquote {\bibinfo
  {title} {{Exponential Ramp in the Quadratic Sachdev-Ye-Kitaev Model}},}\
  }\href {https://doi.org/10.1103/PhysRevLett.125.250602} {\bibfield  {journal}
  {\bibinfo  {journal} {Phys. Rev. Lett.}\ }\textbf {\bibinfo {volume} {125}},\
  \bibinfo {pages} {250602}}\BibitemShut {NoStop}%
\bibitem [{\citenamefont {Witten}(1998)}]{Witten:1998qj}%
  \BibitemOpen
  \bibfield  {author} {\bibinfo {author} {\bibnamefont {Witten}, \bibfnamefont
  {Edward}}} (\bibinfo {year} {1998}),\ \bibfield  {title} {\enquote {\bibinfo
  {title} {{Anti-de Sitter space and holography}},}\ }\href
  {https://doi.org/10.4310/ATMP.1998.v2.n2.a2} {\bibfield  {journal} {\bibinfo
  {journal} {Adv. Theor. Math. Phys.}\ }\textbf {\bibinfo {volume} {2}},\
  \bibinfo {pages} {253--291}},\ \Eprint {https://arxiv.org/abs/hep-th/9802150}
  {arXiv:hep-th/9802150} \BibitemShut {NoStop}%
\bibitem [{\citenamefont {{Wu}}\ \emph {et~al.}(2021)\citenamefont {{Wu}},
  \citenamefont {{Wang}},\ and\ \citenamefont {{Tremblay}}}]{Tremblay21}%
  \BibitemOpen
  \bibfield  {author} {\bibinfo {author} {\bibnamefont {{Wu}}, \bibfnamefont
  {Wei}}, \bibinfo {author} {\bibfnamefont {Xiang}\ \bibnamefont {{Wang}}},
  and\ \bibinfo {author} {\bibfnamefont {A.~M.~S.}\ \bibnamefont {{Tremblay}}}}
  (\bibinfo {year} {2021}),\ \bibfield  {title} {\enquote {\bibinfo {title}
  {{Non-Fermi liquid phase and linear-in-temperature scattering rate in
  overdoped two dimensional Hubbard model}},}\ }\href@noop {} {\ }\Eprint
  {https://arxiv.org/abs/2109.02635} {arXiv:2109.02635 [cond-mat.str-el]}
  \BibitemShut {NoStop}%
\bibitem [{\citenamefont {Wu}\ \emph {et~al.}(2018)\citenamefont {Wu},
  \citenamefont {Chen}, \citenamefont {Jian}, \citenamefont {You},\ and\
  \citenamefont {Xu}}]{CX18}%
  \BibitemOpen
  \bibfield  {author} {\bibinfo {author} {\bibnamefont {Wu}, \bibfnamefont
  {Xiaochuan}}, \bibinfo {author} {\bibfnamefont {Xiao}\ \bibnamefont {Chen}},
  \bibinfo {author} {\bibfnamefont {Chao-Ming}\ \bibnamefont {Jian}}, \bibinfo
  {author} {\bibfnamefont {Yi-Zhuang}\ \bibnamefont {You}}, and\ \bibinfo
  {author} {\bibfnamefont {Cenke}\ \bibnamefont {Xu}}} (\bibinfo {year}
  {2018}),\ \bibfield  {title} {\enquote {\bibinfo {title} {Candidate theory
  for the strange metal phase at a finite-energy window},}\ }\href
  {https://doi.org/10.1103/PhysRevB.98.165117} {\bibfield  {journal} {\bibinfo
  {journal} {Phys. Rev. B}\ }\textbf {\bibinfo {volume} {98}},\ \bibinfo
  {pages} {165117}}\BibitemShut {NoStop}%
\bibitem [{\citenamefont {{Wu}}\ \emph {et~al.}(2020)\citenamefont {{Wu}},
  \citenamefont {{Abanov}}, \citenamefont {{Wang}},\ and\ \citenamefont
  {{Chubukov}}}]{Chubukov20b}%
  \BibitemOpen
  \bibfield  {author} {\bibinfo {author} {\bibnamefont {{Wu}}, \bibfnamefont
  {Yi-Ming}}, \bibinfo {author} {\bibfnamefont {Artem}\ \bibnamefont
  {{Abanov}}}, \bibinfo {author} {\bibfnamefont {Yuxuan}\ \bibnamefont
  {{Wang}}}, and\ \bibinfo {author} {\bibfnamefont {Andrey~V.}\ \bibnamefont
  {{Chubukov}}}} (\bibinfo {year} {2020}),\ \bibfield  {title} {\enquote
  {\bibinfo {title} {{Interplay between superconductivity and non-Fermi liquid
  at a quantum critical point in a metal. II. The $\gamma$ model at a finite
  $T$ for $0 <\gamma <1$}},}\ }\href
  {https://doi.org/10.1103/PhysRevB.102.024525} {\bibfield  {journal} {\bibinfo
   {journal} {Phys. Rev. B}\ }\textbf {\bibinfo {volume} {102}}~(\bibinfo
  {number} {2}),\ \bibinfo {eid} {024525}},\ \Eprint
  {https://arxiv.org/abs/2006.02968} {arXiv:2006.02968 [cond-mat.supr-con]}
  \BibitemShut {NoStop}%
\bibitem [{\citenamefont {Xu}\ and\ \citenamefont
  {Swingle}(2019)}]{Xu:2018dfp}%
  \BibitemOpen
  \bibfield  {author} {\bibinfo {author} {\bibnamefont {Xu}, \bibfnamefont
  {Shenglong}}, and\ \bibinfo {author} {\bibfnamefont {Brian}\ \bibnamefont
  {Swingle}}} (\bibinfo {year} {2019}),\ \bibfield  {title} {\enquote {\bibinfo
  {title} {{Locality, Quantum Fluctuations, and Scrambling}},}\ }\href
  {https://doi.org/10.1103/PhysRevX.9.031048} {\bibfield  {journal} {\bibinfo
  {journal} {Phys. Rev. X}\ }\textbf {\bibinfo {volume} {9}}~(\bibinfo {number}
  {3}),\ \bibinfo {pages} {031048}},\ \Eprint
  {https://arxiv.org/abs/1805.05376} {arXiv:1805.05376 [cond-mat.str-el]}
  \BibitemShut {NoStop}%
\bibitem [{\citenamefont {Xu}\ and\ \citenamefont {Swingle}(2020)}]{BS20}%
  \BibitemOpen
  \bibfield  {author} {\bibinfo {author} {\bibnamefont {Xu}, \bibfnamefont
  {Shenglong}}, and\ \bibinfo {author} {\bibfnamefont {Brian}\ \bibnamefont
  {Swingle}}} (\bibinfo {year} {2020}),\ \bibfield  {title} {\enquote {\bibinfo
  {title} {Accessing scrambling using matrix product operators},}\ }\href
  {https://doi.org/10.1038/s41567-019-0712-4} {\bibfield  {journal} {\bibinfo
  {journal} {Nature Physics}\ }\textbf {\bibinfo {volume} {16}}~(\bibinfo
  {number} {2}),\ \bibinfo {pages} {199--204}}\BibitemShut {NoStop}%
\bibitem [{\citenamefont {Xu}\ \emph {et~al.}(2019)\citenamefont {Xu},
  \citenamefont {McGehee}, \citenamefont {Morong},\ and\ \citenamefont
  {DeMarco}}]{Marco}%
  \BibitemOpen
  \bibfield  {author} {\bibinfo {author} {\bibnamefont {Xu}, \bibfnamefont
  {W}}, \bibinfo {author} {\bibfnamefont {W.~R.}\ \bibnamefont {McGehee}},
  \bibinfo {author} {\bibfnamefont {W.~N.}\ \bibnamefont {Morong}}, and\
  \bibinfo {author} {\bibfnamefont {B.}~\bibnamefont {DeMarco}}} (\bibinfo
  {year} {2019}),\ \bibfield  {title} {\enquote {\bibinfo {title} {Bad-metal
  relaxation dynamics in a fermi lattice gas},}\ }\href
  {https://doi.org/10.1038/s41467-019-09526-x} {\bibfield  {journal} {\bibinfo
  {journal} {Nature Communications}\ }\textbf {\bibinfo {volume}
  {10}}~(\bibinfo {number} {1}),\ \bibinfo {pages} {1588}}\BibitemShut
  {NoStop}%
\bibitem [{\citenamefont {Zaanen}(2004)}]{Zaanen04}%
  \BibitemOpen
  \bibfield  {author} {\bibinfo {author} {\bibnamefont {Zaanen}, \bibfnamefont
  {Jan}}} (\bibinfo {year} {2004}),\ \bibfield  {title} {\enquote {\bibinfo
  {title} {Superconductivity: Why the temperature is high},}\ }\href
  {https://doi.org/10.1038/430512a} {\bibfield  {journal} {\bibinfo  {journal}
  {Nature}\ }\textbf {\bibinfo {volume} {430}}~(\bibinfo {number} {6999}),\
  \bibinfo {pages} {512--513}}\BibitemShut {NoStop}%
\bibitem [{\citenamefont {Zapf}\ \emph {et~al.}(2001)\citenamefont {Zapf},
  \citenamefont {Dickey}, \citenamefont {Freeman}, \citenamefont {Sirvent},\
  and\ \citenamefont {Maple}}]{Maple01}%
  \BibitemOpen
  \bibfield  {author} {\bibinfo {author} {\bibnamefont {Zapf}, \bibfnamefont
  {V~S}}, \bibinfo {author} {\bibfnamefont {R.~P.}\ \bibnamefont {Dickey}},
  \bibinfo {author} {\bibfnamefont {E.~J.}\ \bibnamefont {Freeman}}, \bibinfo
  {author} {\bibfnamefont {C.}~\bibnamefont {Sirvent}}, and\ \bibinfo {author}
  {\bibfnamefont {M.~B.}\ \bibnamefont {Maple}}} (\bibinfo {year} {2001}),\
  \bibfield  {title} {\enquote {\bibinfo {title} {{Magnetic and
  non-Fermi-liquid properties of
  ${\mathrm{U}}_{1\ensuremath{-}x}{\mathrm{La}}_{x}{\mathrm{Pd}}_{2}{\mathrm{Al}}_{3}$}},}\
  }\href {https://doi.org/10.1103/PhysRevB.65.024437} {\bibfield  {journal}
  {\bibinfo  {journal} {Phys. Rev. B}\ }\textbf {\bibinfo {volume} {65}},\
  \bibinfo {pages} {024437}}\BibitemShut {NoStop}%
\bibitem [{\citenamefont {Zhang}\ \emph {et~al.}(2019)\citenamefont {Zhang},
  \citenamefont {Kountz}, \citenamefont {Behnia},\ and\ \citenamefont
  {Kapitulnik}}]{AK19}%
  \BibitemOpen
  \bibfield  {author} {\bibinfo {author} {\bibnamefont {Zhang}, \bibfnamefont
  {Jiecheng}}, \bibinfo {author} {\bibfnamefont {Erik~D.}\ \bibnamefont
  {Kountz}}, \bibinfo {author} {\bibfnamefont {Kamran}\ \bibnamefont {Behnia}},
  and\ \bibinfo {author} {\bibfnamefont {Aharon}\ \bibnamefont {Kapitulnik}}}
  (\bibinfo {year} {2019}),\ \bibfield  {title} {\enquote {\bibinfo {title}
  {Thermalization and possible signatures of quantum chaos in complex
  crystalline materials},}\ }\href {https://doi.org/10.1073/pnas.1910131116}
  {\bibfield  {journal} {\bibinfo  {journal} {Proceedings of the National
  Academy of Sciences}\ }\textbf {\bibinfo {volume} {116}}~(\bibinfo {number}
  {40}),\ \bibinfo {pages} {19869--19874}}\BibitemShut {NoStop}%
\bibitem [{\citenamefont {Zhang}(2017)}]{Zhang17}%
  \BibitemOpen
  \bibfield  {author} {\bibinfo {author} {\bibnamefont {Zhang}, \bibfnamefont
  {Pengfei}}} (\bibinfo {year} {2017}),\ \bibfield  {title} {\enquote {\bibinfo
  {title} {{Dispersive Sachdev-Ye-Kitaev model: Band structure and quantum
  chaos}},}\ }\href {https://doi.org/10.1103/PhysRevB.96.205138} {\bibfield
  {journal} {\bibinfo  {journal} {Phys. Rev. B}\ }\textbf {\bibinfo {volume}
  {96}},\ \bibinfo {pages} {205138}}\BibitemShut {NoStop}%
\bibitem [{\citenamefont {Zhang}(2019)}]{Zhang:2019fcy}%
  \BibitemOpen
  \bibfield  {author} {\bibinfo {author} {\bibnamefont {Zhang}, \bibfnamefont
  {Pengfei}}} (\bibinfo {year} {2019}),\ \bibfield  {title} {\enquote {\bibinfo
  {title} {{Evaporation dynamics of the Sachdev-Ye-Kitaev model}},}\ }\href
  {https://doi.org/10.1103/PhysRevB.100.245104} {\bibfield  {journal} {\bibinfo
   {journal} {Phys. Rev. B}\ }\textbf {\bibinfo {volume} {100}}~(\bibinfo
  {number} {24}),\ \bibinfo {pages} {245104}},\ \Eprint
  {https://arxiv.org/abs/1909.10637} {arXiv:1909.10637 [cond-mat.str-el]}
  \BibitemShut {NoStop}%
\bibitem [{\citenamefont {Zhang}(2021)}]{Zhang:2020szi}%
  \BibitemOpen
  \bibfield  {author} {\bibinfo {author} {\bibnamefont {Zhang}, \bibfnamefont
  {Pengfei}}} (\bibinfo {year} {2021}),\ \bibfield  {title} {\enquote {\bibinfo
  {title} {{More on Complex Sachdev-Ye-Kitaev Eternal Wormholes}},}\ }\href
  {https://doi.org/10.1007/JHEP03(2021)087} {\bibfield  {journal} {\bibinfo
  {journal} {JHEP}\ }\textbf {\bibinfo {volume} {03}},\ \bibinfo {pages}
  {087}},\ \Eprint {https://arxiv.org/abs/2011.10360} {arXiv:2011.10360
  [hep-th]} \BibitemShut {NoStop}%
\bibitem [{\citenamefont {Zhang}(2022)}]{Zhang:2022yaw}%
  \BibitemOpen
  \bibfield  {author} {\bibinfo {author} {\bibnamefont {Zhang}, \bibfnamefont
  {Pengfei}}} (\bibinfo {year} {2022}),\ \bibfield  {title} {\enquote {\bibinfo
  {title} {{Quantum Entanglement in the Sachdev-Ye-Kitaev Model and its
  Generalizations}},}\ }\href@noop {} {\ }\Eprint
  {https://arxiv.org/abs/2203.01513} {arXiv:2203.01513 [cond-mat.str-el]}
  \BibitemShut {NoStop}%
\bibitem [{\citenamefont {Zhang}\ and\ \citenamefont {Sachdev}(2020)}]{YZSS20}%
  \BibitemOpen
  \bibfield  {author} {\bibinfo {author} {\bibnamefont {Zhang}, \bibfnamefont
  {Ya-Hui}}, and\ \bibinfo {author} {\bibfnamefont {Subir}\ \bibnamefont
  {Sachdev}}} (\bibinfo {year} {2020}),\ \bibfield  {title} {\enquote {\bibinfo
  {title} {{Deconfined criticality and ghost Fermi surfaces at the onset of
  antiferromagnetism in a metal}},}\ }\href
  {https://doi.org/10.1103/PhysRevB.102.155124} {\bibfield  {journal} {\bibinfo
   {journal} {Phys. Rev. B}\ }\textbf {\bibinfo {volume} {102}},\ \bibinfo
  {pages} {155124}}\BibitemShut {NoStop}%
\bibitem [{\citenamefont {{Zhao}}\ \emph {et~al.}(2019)\citenamefont {{Zhao}},
  \citenamefont {{Zhang}}, \citenamefont {{Lyu}}, \citenamefont {{Bachus}},
  \citenamefont {{Tokiwa}}, \citenamefont {{Gegenwart}}, \citenamefont
  {{Zhang}}, \citenamefont {{Cheng}}, \citenamefont {{Yang}}, \citenamefont
  {{Chen}}, \citenamefont {{Isikawa}}, \citenamefont {{Si}}, \citenamefont
  {{Steglich}},\ and\ \citenamefont {{Sun}}}]{Sun19}%
  \BibitemOpen
  \bibfield  {author} {\bibinfo {author} {\bibnamefont {{Zhao}}, \bibfnamefont
  {Hengcan}}, \bibinfo {author} {\bibfnamefont {Jiahao}\ \bibnamefont
  {{Zhang}}}, \bibinfo {author} {\bibfnamefont {Meng}\ \bibnamefont {{Lyu}}},
  \bibinfo {author} {\bibfnamefont {Sebastian}\ \bibnamefont {{Bachus}}},
  \bibinfo {author} {\bibfnamefont {Yoshifumi}\ \bibnamefont {{Tokiwa}}},
  \bibinfo {author} {\bibfnamefont {Philipp}\ \bibnamefont {{Gegenwart}}},
  \bibinfo {author} {\bibfnamefont {Shuai}\ \bibnamefont {{Zhang}}}, \bibinfo
  {author} {\bibfnamefont {Jinguang}\ \bibnamefont {{Cheng}}}, \bibinfo
  {author} {\bibfnamefont {Yi-feng}\ \bibnamefont {{Yang}}}, \bibinfo {author}
  {\bibfnamefont {Genfu}\ \bibnamefont {{Chen}}}, \bibinfo {author}
  {\bibfnamefont {Yosikazu}\ \bibnamefont {{Isikawa}}}, \bibinfo {author}
  {\bibfnamefont {Qimiao}\ \bibnamefont {{Si}}}, \bibinfo {author}
  {\bibfnamefont {Frank}\ \bibnamefont {{Steglich}}}, and\ \bibinfo {author}
  {\bibfnamefont {Peijie}\ \bibnamefont {{Sun}}}} (\bibinfo {year} {2019}),\
  \bibfield  {title} {\enquote {\bibinfo {title} {{Quantum-critical phase from
  frustrated magnetism in a strongly correlated metal}},}\ }\href
  {https://doi.org/10.1038/s41567-019-0666-6} {\bibfield  {journal} {\bibinfo
  {journal} {Nature Physics}\ }\textbf {\bibinfo {volume} {15}}~(\bibinfo
  {number} {12}),\ \bibinfo {pages} {1261--1266}},\ \Eprint
  {https://arxiv.org/abs/1907.04255} {arXiv:1907.04255 [cond-mat.str-el]}
  \BibitemShut {NoStop}%
\bibitem [{\citenamefont {Zhou}\ \emph {et~al.}(2021)\citenamefont {Zhou},
  \citenamefont {Pan}, \citenamefont {Chen}, \citenamefont {Zhang},\ and\
  \citenamefont {Zhai}}]{Zhou:2020kxb}%
  \BibitemOpen
  \bibfield  {author} {\bibinfo {author} {\bibnamefont {Zhou}, \bibfnamefont
  {Tian-Gang}}, \bibinfo {author} {\bibfnamefont {Lei}\ \bibnamefont {Pan}},
  \bibinfo {author} {\bibfnamefont {Yu}~\bibnamefont {Chen}}, \bibinfo {author}
  {\bibfnamefont {Pengfei}\ \bibnamefont {Zhang}}, and\ \bibinfo {author}
  {\bibfnamefont {Hui}\ \bibnamefont {Zhai}}} (\bibinfo {year} {2021}),\
  \bibfield  {title} {\enquote {\bibinfo {title} {{Disconnecting a Traversable
  Wormhole: Universal Quench Dynamics in Random Spin Models}},}\ }\href
  {https://doi.org/10.1103/PhysRevResearch.3.L022024} {\bibfield  {journal}
  {\bibinfo  {journal} {Phys. Rev. Res.}\ }\textbf {\bibinfo {volume}
  {3}}~(\bibinfo {number} {2}),\ \bibinfo {pages} {022024}},\ \Eprint
  {https://arxiv.org/abs/2009.00277} {arXiv:2009.00277 [cond-mat.quant-gas]}
  \BibitemShut {NoStop}%
\bibitem [{\citenamefont {Zhou}\ and\ \citenamefont
  {Zhang}(2020)}]{Zhou:2020wgh}%
  \BibitemOpen
  \bibfield  {author} {\bibinfo {author} {\bibnamefont {Zhou}, \bibfnamefont
  {Tian-Gang}}, and\ \bibinfo {author} {\bibfnamefont {Pengfei}\ \bibnamefont
  {Zhang}}} (\bibinfo {year} {2020}),\ \bibfield  {title} {\enquote {\bibinfo
  {title} {{Tunneling through an Eternal Traversable Wormhole}},}\ }\href
  {https://doi.org/10.1103/PhysRevB.102.224305} {\bibfield  {journal} {\bibinfo
   {journal} {Phys. Rev. B}\ }\textbf {\bibinfo {volume} {102}},\ \bibinfo
  {pages} {224305}},\ \Eprint {https://arxiv.org/abs/2009.02641}
  {arXiv:2009.02641 [cond-mat.str-el]} \BibitemShut {NoStop}%
\bibitem [{\citenamefont {Zhu}\ \emph {et~al.}(2003)\citenamefont {Zhu},
  \citenamefont {Grempel},\ and\ \citenamefont {Si}}]{SiIsing2}%
  \BibitemOpen
  \bibfield  {author} {\bibinfo {author} {\bibnamefont {Zhu}, \bibfnamefont
  {Jian-Xin}}, \bibinfo {author} {\bibfnamefont {D.~R.}\ \bibnamefont
  {Grempel}}, and\ \bibinfo {author} {\bibfnamefont {Qimiao}\ \bibnamefont
  {Si}}} (\bibinfo {year} {2003}),\ \bibfield  {title} {\enquote {\bibinfo
  {title} {{Continuous Quantum Phase Transition in a Kondo Lattice Model}},}\
  }\href {https://doi.org/10.1103/PhysRevLett.91.156404} {\bibfield  {journal}
  {\bibinfo  {journal} {Phys. Rev. Lett.}\ }\textbf {\bibinfo {volume} {91}},\
  \bibinfo {pages} {156404}}\BibitemShut {NoStop}%
\bibitem [{\citenamefont {Zhu}\ \emph {et~al.}(2007)\citenamefont {Zhu},
  \citenamefont {Kirchner}, \citenamefont {Bulla},\ and\ \citenamefont
  {Si}}]{SiIsing3}%
  \BibitemOpen
  \bibfield  {author} {\bibinfo {author} {\bibnamefont {Zhu}, \bibfnamefont
  {Jian-Xin}}, \bibinfo {author} {\bibfnamefont {Stefan}\ \bibnamefont
  {Kirchner}}, \bibinfo {author} {\bibfnamefont {Ralf}\ \bibnamefont {Bulla}},
  and\ \bibinfo {author} {\bibfnamefont {Qimiao}\ \bibnamefont {Si}}} (\bibinfo
  {year} {2007}),\ \bibfield  {title} {\enquote {\bibinfo {title}
  {{Zero-Temperature Magnetic Transition in an Easy-Axis Kondo Lattice
  Model}},}\ }\href {https://doi.org/10.1103/PhysRevLett.99.227204} {\bibfield
  {journal} {\bibinfo  {journal} {Phys. Rev. Lett.}\ }\textbf {\bibinfo
  {volume} {99}},\ \bibinfo {pages} {227204}}\BibitemShut {NoStop}%
\bibitem [{\citenamefont {{Zhu}}\ and\ \citenamefont {{Si}}(2002)}]{ZhuSi}%
  \BibitemOpen
  \bibfield  {author} {\bibinfo {author} {\bibnamefont {{Zhu}}, \bibfnamefont
  {Lijun}}, and\ \bibinfo {author} {\bibfnamefont {Qimiao}\ \bibnamefont
  {{Si}}}} (\bibinfo {year} {2002}),\ \bibfield  {title} {\enquote {\bibinfo
  {title} {{Critical local-moment fluctuations in the Bose-Fermi Kondo
  model}},}\ }\href {https://doi.org/10.1103/PhysRevB.66.024426} {\bibfield
  {journal} {\bibinfo  {journal} {Phys. Rev. B}\ }\textbf {\bibinfo {volume}
  {66}}~(\bibinfo {number} {2}),\ \bibinfo {eid} {024426}},\ \Eprint
  {https://arxiv.org/abs/cond-mat/0204121} {arXiv:cond-mat/0204121
  [cond-mat.str-el]} \BibitemShut {NoStop}%
\bibitem [{\citenamefont {Ziman}(1960)}]{ziman}%
  \BibitemOpen
  \bibfield  {author} {\bibinfo {author} {\bibnamefont {Ziman}, \bibfnamefont
  {John~M}}} (\bibinfo {year} {1960}),\ \href@noop {} {\emph {\bibinfo {title}
  {Electrons and phonons: the theory of transport phenomena in solids}}}\
  (\bibinfo  {publisher} {Oxford university press})\BibitemShut {NoStop}%
\bibitem [{\citenamefont {{Zondiner}}\ \emph {et~al.}(2020)\citenamefont
  {{Zondiner}}, \citenamefont {{Rozen}}, \citenamefont {{Rodan-Legrain}},
  \citenamefont {{Cao}}, \citenamefont {{Queiroz}}, \citenamefont
  {{Taniguchi}}, \citenamefont {{Watanabe}}, \citenamefont {{Oreg}},
  \citenamefont {{von Oppen}}, \citenamefont {{Stern}}, \citenamefont {{Berg}},
  \citenamefont {{Jarillo-Herrero}},\ and\ \citenamefont
  {{Ilani}}}]{Zondiner2020}%
  \BibitemOpen
  \bibfield  {author} {\bibinfo {author} {\bibnamefont {{Zondiner}},
  \bibfnamefont {U}}, \bibinfo {author} {\bibfnamefont {A.}~\bibnamefont
  {{Rozen}}}, \bibinfo {author} {\bibfnamefont {D.}~\bibnamefont
  {{Rodan-Legrain}}}, \bibinfo {author} {\bibfnamefont {Y.}~\bibnamefont
  {{Cao}}}, \bibinfo {author} {\bibfnamefont {R.}~\bibnamefont {{Queiroz}}},
  \bibinfo {author} {\bibfnamefont {T.}~\bibnamefont {{Taniguchi}}}, \bibinfo
  {author} {\bibfnamefont {K.}~\bibnamefont {{Watanabe}}}, \bibinfo {author}
  {\bibfnamefont {Y.}~\bibnamefont {{Oreg}}}, \bibinfo {author} {\bibfnamefont
  {F.}~\bibnamefont {{von Oppen}}}, \bibinfo {author} {\bibfnamefont {Ady}\
  \bibnamefont {{Stern}}}, \bibinfo {author} {\bibfnamefont {E.}~\bibnamefont
  {{Berg}}}, \bibinfo {author} {\bibfnamefont {P.}~\bibnamefont
  {{Jarillo-Herrero}}}, and\ \bibinfo {author} {\bibfnamefont {S.}~\bibnamefont
  {{Ilani}}}} (\bibinfo {year} {2020}),\ \bibfield  {title} {\enquote {\bibinfo
  {title} {{Cascade of phase transitions and Dirac revivals in magic-angle
  graphene}},}\ }\href {https://doi.org/10.1038/s41586-020-2373-y} {\bibfield
  {journal} {\bibinfo  {journal} {Nature}\ }\textbf {\bibinfo {volume}
  {582}}~(\bibinfo {number} {7811}),\ \bibinfo {pages} {203--208}},\ \Eprint
  {https://arxiv.org/abs/1912.06150} {arXiv:1912.06150 [cond-mat.mes-hall]}
  \BibitemShut {NoStop}%
\bibitem [{\citenamefont {Zou}\ and\ \citenamefont {Chowdhury}(2020)}]{LZDC20}%
  \BibitemOpen
  \bibfield  {author} {\bibinfo {author} {\bibnamefont {Zou}, \bibfnamefont
  {Liujun}}, and\ \bibinfo {author} {\bibfnamefont {Debanjan}\ \bibnamefont
  {Chowdhury}}} (\bibinfo {year} {2020}),\ \bibfield  {title} {\enquote
  {\bibinfo {title} {{Deconfined metallic quantum criticality: A $U(2)$
  gauge-theoretic approach}},}\ }\href
  {https://doi.org/10.1103/PhysRevResearch.2.023344} {\bibfield  {journal}
  {\bibinfo  {journal} {Phys. Rev. Research}\ }\textbf {\bibinfo {volume}
  {2}},\ \bibinfo {pages} {023344}}\BibitemShut {NoStop}%
\end{thebibliography}%
\end{document}